\documentclass[twocolumn]{article}

\usepackage[dvipsnames]{xcolor}
\usepackage{amsmath}
\usepackage{amssymb}
\usepackage{rotating}
\usepackage{array}
\usepackage{mathtools}
\usepackage[ruled]{algorithm2e}
\usepackage{algorithmic}
\usepackage{bm}
\usepackage{breqn}
\usepackage{comment}
\usepackage{enumitem}
\usepackage{graphics}
\usepackage{graphicx}
\usepackage{latexsym}
\usepackage{mathrsfs}
\usepackage{nicefrac}
\usepackage{times}
\usepackage{subcaption}
\usepackage{morefloats}
\usepackage{todonotes}
\usepackage[framemethod=TikZ]{mdframed}
\usepackage{fullpage}

\mdfdefinestyle{MyFrame}{%
	linecolor=black,
	outerlinewidth=0.5pt,
	innertopmargin=\baselineskip,
	innerbottommargin=\baselineskip,
	innerrightmargin=10pt,
	innerleftmargin=10pt,
	backgroundcolor=white!20!white,
    nobreak=true}

\newtheorem{definition}{Definition}

\newtheorem{theorem}{Theorem}

\newcommand{\N}{\mathbb{N}}

\def\R{\mathbb{R}}
\def\FA{\hat{A}}
\def\FR{\hat{R}}
\def\inf{\mathit{inf}}
\def\globalReb{\phi}
\def\S{N^{\mathit{succ}}}
\def\P{N^{\mathit{pred}}}

\newcommand{\alg}{\mathit{alg}}
\newcommand{\CS}{\mathit{CS}}
\newcommand{\CC}{C^{\dagger}}
\newcommand{\mc}{\mu}
\newcommand{\floor}[1]{\left\lfloor #1 \right\rfloor}
\newcommand{\ceil}[1]{\left\lceil #1 \right\rceil}

\DeclareMathOperator*{\argmax}{arg\,max}
\DeclareMathOperator*{\argmin}{arg\,min}

\begin{document}

\title{Hiding Individuals and Communities in a Social Network}

\author{
Marcin Waniek \\ \small{University of Warsaw} \\ \small{\texttt{vua@mimuw.edu.pl}}
\and
Tomasz Michalak \\ \small{University of Oxford} \\ \small{\& University of Warsaw} \\ \small{\texttt{tomasz.michalak@cs.ox.ac.uk}}
\and
Talal Rahwan \\ \small{Masdar Institute} \\ \small{of Science and Technology} \\ \small{\texttt{trahwan@gmail.com}}
\and
Michael Wooldridge \\ \small{University of Oxford} \\ \small{\texttt{michael.wooldridge@cs.ox.ac.uk}}
}

\date{}

\twocolumn[
\begin{@twocolumnfalse}

\maketitle

\begin{abstract}
The Internet and social media have fueled enormous interest in social network analysis. New tools continue to be developed and used to analyse our personal connections, with particular emphasis on detecting communities or identifying key individuals in a social network. This raises privacy concerns that are likely to exacerbate in the future. With this in mind, we ask the question: \emph{Can individuals or groups actively manage their connections to evade social network analysis tools?}
By addressing this question, the general public may better protect their privacy, oppressed activist groups may better conceal their existence, and security agencies may better understand how terrorists escape detection. We first study how an individual can evade ``network centrality'' analysis without compromising his or her influence within the network. We prove that an optimal solution to this problem is hard to compute. Despite this hardness, we demonstrate that even a simple heuristic, whereby attention is restricted to the individual's immediate neighbourhood, can be surprisingly effective in practice. For instance, it could disguise Mohamed Atta's leading position within the WTC terrorist network, and that is by rewiring a strikingly-small number of connections. Next, we study how a community can increase the likelihood of being overlooked by community-detection algorithms. We propose a measure of concealment, expressing how well a community is hidden, and use it to demonstrate the effectiveness of a simple heuristic, whereby members of the community either  ``unfriend'' certain other members, or ``befriend'' some non-members, in a coordinated effort to camouflage their community.
\end{abstract}

\vspace{2cm}

\end{@twocolumnfalse}
]

\section{Introduction}

The on-going process of datafication continues to turn many aspects of our lives into computerised data \cite{Mayer-Schnberger:2013}. This data is being collected and analysed for various diverse applications by public and private institutions alike.
One particular type of data that has received significant attention over the past decade concerns our social connections. To this end, a number of tools have been advocated for social network analysis, with particular emphasis on the detection of communities or the identification of key individuals within a network.
%
%Although the process of datafication has undeniably enhanced our lives in various ways, it has given rise to many legitimate privacy concerns 
For all their benefits, the widespread use of such tools raises legitimate privacy concerns.
%as it contributes towards the ever growing digital surveillance of our lives by both business enterprises and public institutions  \ref{}.
For instance, Mislove et al. \cite{Mislove:2010} demonstrated how, by analysing Facebook's social network structure, as well as the attributes of some users, it is possible to infer otherwise-private information about other Facebook users.
%
%This highlighted the need to hide one's friend list, as it may divulge substantial information about ourselves.
%

To tackle such privacy issues, various countermeasures have been proposed, ranging from strict legal controls \cite{EU:2015}, through algorithmic solutions \cite{Kearns:et:al:2016}, to market-like mechanisms that allow participants to monetize their personal information \cite{Lane:et:al:2014}. However, to date only few such countermeasures have been implemented, leaving the privacy issue largely unresolved, e.g., as is evident from the very recent release of Facebook's ``\emph{Global Government Requests Report}'' \cite{FacebookReport:2016}, which revealed a global increase in government requests to secretly access user data.
%
%Through its social mapping program, the US government's National Security Agency (NSA) reportedly created a sophisticated web of social connections of some US citizens \cite{NSA:socialNetworks:2013}.
%
Furthermore, it is unlikely that effective legal mechanisms will be introduced in countries with authoritarian regimes, where  social networking sites and other internet content is policed, and anti-governmental blogs and activities are censored \cite{king2013censorship,king2014reverse}.

Against this background, we ask the question: can individuals or communities proactively manage their social connections so that their privacy is less exposed to the workings of network analysis tools? To put it differently, can we disguise our standing in the network to escape detection? This matters because, on one hand, it assists the general public in protecting their privacy against intrusion from government and corporate interests; on the other hand, it assists counterterrorism units and law-enforcement agencies in understanding how criminals and terrorists could escape detection, especially given their increasing reliance of social-media survival strategies \cite{Nordrum:2016,Johnson:etal:2016}. To date, however, this fundamental question has received little attention in the literature, as most research efforts have focused on developing ever more sophisticated network analysis tools, rather than considering how to evade them.

To address the above question from an individual's viewpoint, we focus on three main centrality measures, namely \emph{degree}, \emph{closeness}, and \emph{betweenness}, and study how one can avoid being highlighted by those measures without compromising his or her influence. Since, from a graph-theoretic perspective, this is fundamentally an optimization problem, we analyse its computational complexity to illuminate the theoretical limits of such capability as disguising oneself. Although we show that an optimal solution is often hard to compute, we demonstrate the effectiveness of a surprisingly simple heuristic, whereby the rewiring of social connections is restricted to the individual's immediate network  neighbourhood. Specifically, it involves two actions that are already available on popular social-media platforms: (i) ``unfriending'' certain neighbours; (ii) introducing certain neighbours to each other.

From a group's viewpoint, we study how a community can conceal itself to increase the likelihood of being overlooked by community-detection algorithms. To this end, we propose a measure of concealment, designed to quantify the degree to which a group of individuals is hidden. Using this measure, we demonstrate the effectiveness of yet another simple heuristic, whereby members of the community either ``unfriend'' certain other members, or ``befriend'' some non-members to blend into the surrounding web of social connections.

\section{The Model} This section presents the basic concepts and objectives; all formal definitions can be found in the Supporting Information. 
\ \smallskip\\ \noindent\noindent\textbf{Centrality Measures:}
A measure of centrality reflects the importance of any given node in the network. Arguably, the standard centrality measures are: \textit{degree}, \textit{closeness} and \textit{betweenness}~\cite{freeman1979centrality}. In particular, for any given node $v$, the degree centrality focuses on the number of neighbours that $v$ has (the more neighbours the better). In contrast, the closeness centrality quantifies the importance of $v$ based on its average distance to other nodes (the closer the better). Finally, the betweenness centrality focuses on the number of shortest paths on which $v$ lies (the more paths the better).
\ \smallskip\\ \noindent\textbf{Models of Influence:}
The best established mathematical models of influence are the \emph{Independent Cascade} model~\cite{goldenberg2001using} and the \emph{Linear Threshold} model~\cite{kempe2003maximizing}. Basically, both models start with some ``active'' subset of nodes, called the \emph{seed set}.\footnote{\footnotesize An \emph{active} node can be thought of as an \emph{infected} person who influences, but not necessarily infects, his or her neighbours. Analogously, an \emph{inactive} node can be a \emph{healthy} person who is influenced by any infected neighbours he or she may have; stronger influence corresponds to stronger chances of infection.} Then, as time passes (in discrete rounds), new nodes become activated due to the influence from other previously-activated nodes. The two models differ in the way influence propagates through the network. Specifically, in the Independent Cascade model, an active node activates each of its neighbours with some pre-defined probability. In contrast, with the Linear Threshold model, each node has some random predefined threshold, and gets activated when the number of its active neighbors exceeds that threshold. Under either model, the influence of a node, $v$, on another, $w$, is measured as the probability that $w$ gets activated when the seed set is $\{v\}$.
%
%%%%%%%%%%%%%%%%%%%%%%%%%%%%%%%%%%%%%%%%%%%%%%%%%%%%%%%%%%%%%%%%%%%%%%%
\ \smallskip\\ \noindent\textbf{First Objective:}
Given a network and a \textit{source node}, $v^{\dagger}$, our objective is to conceal the importance of $v^{\dagger}$ by decreasing its centrality (according to each of the aforementioned measures of centrality) without compromising its influence over the network (according to the aforementioned models of influence). We do so by rewiring the links of the network, without exceeding a certain \emph{budget}---the maximum number of links allowed to be modified (i.e., added or removed). To simplify our analysis, we divide the process of disguising $v^{\dagger}$ into two consecutive phases. In the first phase, part of the budget is spent on minimizing the three centrality measures, during which the influence of $v^{\dagger}$ is likely to decrease---we call this the \textit{centrality minimization} problem. The second phase involves spending the remaining budget to recover as much as possible of the influence of $v^{\dagger}$ while avoiding the addition of any links that were removed during the centrality minimization phase. Here, we consider two variants of this latter problem: (i) the \textit{individual influence recovery} problem, where the goal is to recover the influence of $v^{\dagger}$ over every single node, and (ii) the \textit{global influence recovery} problem, where the goal is to recover the \emph{sum of influences} of $v^{\dagger}$ over all nodes.
\ \smallskip\\ \noindent\textbf{Second Objective:}
Given a community, i.e., subset of nodes, $\CC$, our goal is the conceal the identity of $\CC$ by hiding its existence within the network. Recall that a community structure is a partition of the set of nodes into disjoint and exhaustive subsets, or ``communities''. As such, $\CC$ is exposed if a community-detection algorithm is able to return a community structure, $\CS$, such that $\CC\in\CS$. We hide $\CC$ by rewiring the links of the network, again according to some budget, i.e., maximum number of permitted modifications.

%%%%%%%%%%%%%%%%%%%%%%%%%%%%%%%%%%%%%%%%%%%%%%%%%%%%%%%%%%%%%%%
\section{Disguising Individuals}
\subsection{Hardness Results}
Our main theoretical results are summarized in  Table~\ref{tab:theoretical} (for more details, see theorems 1 through 4 in the \emph{Supporting Information}). As shown in the table, all the problems under consideration turn out to be NP-complete, with the exception of minimizing degree centrality. To put it differently, finding an optimal way to disguise one's importance in a social network is extremely difficult (from a computational point of view), not to mention the fact that it requires knowing the entire network structure, and may also require adding or removing links that are far from the source node.

%Despite this rather bleak result, the situation may not be entirely hopeless, provided that we are content with a reasonable, albeit not optimal, solution to the problem. With this in mind, we propose in the following subsection a simple heuristic that can be applied in practice, before evaluating its effectiveness in the Experiments section.

%%%%%%%%%%%%%%%%%%%%%%%%%%%%%%%%%%%%%%%%%%%%%%%%%%%%%%%%%%%%%%%%%%%
%\vspace*{-0.6cm}
\begin{table}[ht]
\centering
\begin{tabular}{ l  c }
\hline
\vspace*{-0.22cm}\fontsize{1}{1}\selectfont{\ }&\fontsize{2}{2}\selectfont{\ }\\
\normalsize{Disguising centrality (Degree)} & \normalsize{P} \vspace*{0.05cm}\\
\normalsize{Disguising centrality (Closeness)} & \normalsize{NPC} \vspace*{0.05cm}\\
\normalsize{Disguising centrality (Betweenness)} & \normalsize{NPC} \vspace*{0.05cm}\\
\normalsize{Individual influence recovery (LT)} & \normalsize{NPC} \vspace*{0.05cm}\\
\normalsize{Individual influence recovery (IC)} & \normalsize{NPC} \vspace*{0.08cm}\\
\normalsize{Global influence recovery (LT)}  & \normalsize{NPC} \vspace*{0.05cm}\\
\normalsize{Global influence recovery (IC)} & \normalsize{NPC} \vspace*{0.05cm}\\
\hline
\end{tabular}
\caption{Summary of our computational-hardness results.}
\label{tab:theoretical}
\end{table}
\vspace*{-0.6cm}
%%%%%%%%%%%%%%%%%%%%%%%%%%%%%%%%%%%%%%%%%%%%%%%%%%%%%%%%%%%%%%%%%%%%
%
%
%%%%%%%%%%%%%%%%%%%%%%%%%%%%%%%%%%%%%%%%%%%%%%%%%%%%%%%%%%%%%%%
\begin{figure*}[th]
 \centering
 \includegraphics[width=1.0\linewidth]{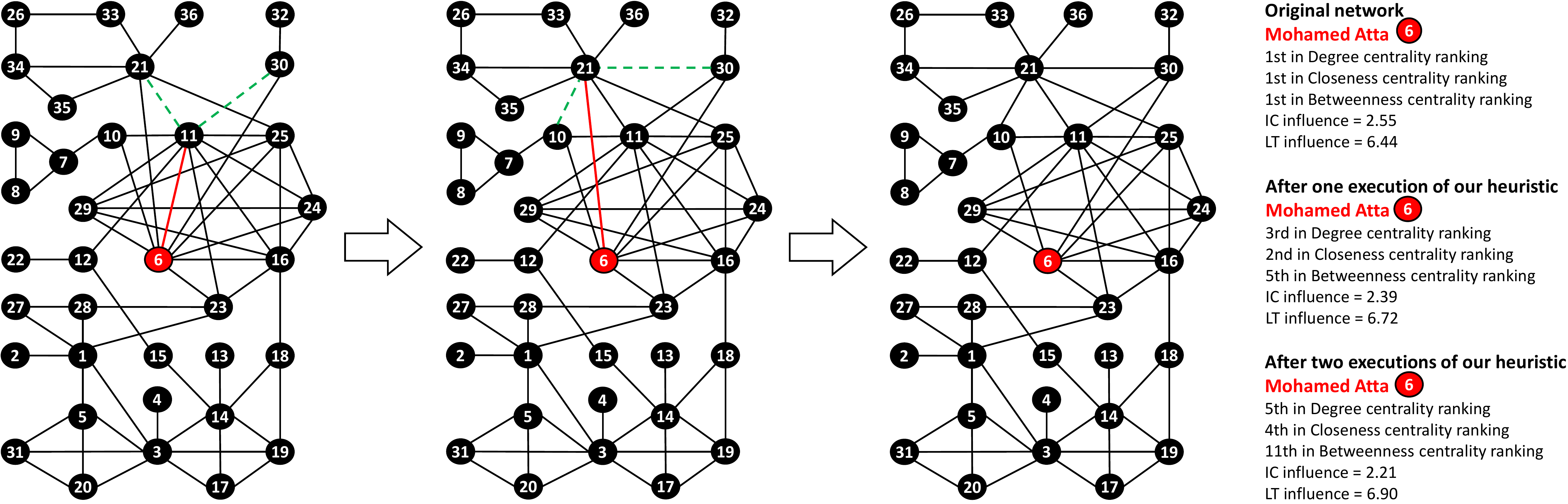}
\caption{Executing the ROAM heuristic twice on the 9/11 terrorist network to hide Mohamed Atta---one of the ringleaders of the attack \cite{Krebs:2002a}. The red link is the one to be to removed by the algorithm, and the dashed links are the ones to be added.}
 \label{fig:wtc}
\end{figure*}
%%%%%%%%%%%%%%%%%%%%%%%%%%%%%%%%%%%%%%%%%%%%%%%%%%%%%%%%%%%%%%%%

\subsection{A Scalable Heuristic} 
Typically, one has very limited knowledge of the social ties beyond his or her immediate friends, or maybe friends of friends. However, even if one was able to somehow acquire information about the entire network structure, our theoretical results from the previous subsection suggest that it is extremely unlikely for such an individual to have the necessary computational power to \emph{optimally} disguise himself or herself. Against this background, we investigate the possibility of disguising one's centrality adequately (albeit not optimally) while restricting one's attention to only his or her immediate neighbourhood, and without requiring massive computational power nor expertise in sophisticated optimization techniques. With this in mind, we propose a heuristic whose instructions are simple enough for an average user of social-networking services to understand and use, regardless of their technical background. Our heuristic, called ROAM---Remove One, Add Many---is detailed in the box below, and an illustration of how it works is presented in Figure~\ref{fig:wtc}.

%%%%%%%%%%%%%%%%%%%%%%%%%%%%%%%%%%%%%%%%%%%%%%%%%%%%%%%%%%%%%%%
\begin{mdframed}[style=MyFrame]
\vspace*{-0.1cm}
\textbf{The ROAM heuristic} given a budget $b$:
\vspace*{-0.18cm}
\begin{itemize}\itemsep0.3em
\item \textbf{Step~1:} Remove the link between the source node, $v^{\dagger}$, and its neighbour of choice, $v_0$;
\item \textbf{Step~2:} Connect $v_0$ to $b-1$ nodes of choice, who are neighbours of $v^{\dagger}$ but not of $v_0$ (if there are fewer than $b-1$ such neighbours, connect $v_0$ to all of them).
\vspace*{-0.18cm}
\end{itemize}
\end{mdframed}
%%%%%%%%%%%%%%%%%%%%%%%%%%%%%%%%%%%%%%%%%%%%%%%%%%%%%%%%%%%%%%%

Let us now comment on this heuristic, starting with \textbf{Step~1}. As far as the  centrality of $v^{\dagger}$ is concerned, this step can only be beneficial. More specifically, cutting off $v^{\dagger}$ from one of its neighbours is the only way to reduce the degree  of $v^{\dagger}$. Likewise, \textbf{Step~1} can only decrease the closeness of $v^{\dagger}$ (this happens when all shortest paths between $v^{\dagger}$ and some other node run through the removed link), and can only decrease the betweenness of $v^{\dagger}$ (this happens when some of the shortest paths going through $v^{\dagger}$ contain the removed link). However, as far as the influence of $v^{\dagger}$ is concerned, \textbf{Step~1} may be detrimental, as it deprives $v^{\dagger}$ from its direct influence over $v_0$.

Moving on to \textbf{Step~2}, this step is primarily designed to compensate for any influence that $v^{\dagger}$ may have lost during the previous step. Specifically, it creates new, indirect connections between $v^{\dagger}$ and $v_0$ to compensate for the direct one that was removed earlier. As far as the centrality of $v^{\dagger}$ is concerned, while \textbf{Step~2} does not affect the degree of $v^{\dagger}$, it increases the degrees of some of its neighbours, which in turn contributes towards concealing the relative importance of $v^{\dagger}$ within the network. Furthermore, the addition of a link, $(v_0,v_i)$---where $v_i$ is some neighbour of $v^{\dagger}$---cannot increase the closeness centrality of $v^{\dagger}$ beyond its original state, i.e., its state before running the ROAM heuristic altogether. This is because any path containing $(v_0,v_i)$ and $(v_i,v^{\dagger})$ is certainly longer than an original path in which $(v_0,v_i)$ and $(v_i,v^{\dagger})$ were replaced with $(v_0,v^{\dagger})$. Likewise, the addition of this link cannot increase the betweenness centrality of $v^{\dagger}$ beyond its original state, because replacing a direct connection between $v^{\dagger}$ and $v_0$ with an indirect one cannot increase the percentage of shortest paths going through $v^{\dagger}$. 

Finally, let us comment on the how to choose $v_0$, and how to choose the neighbours of $v^{\dagger}$ to connect to $v_0$. Based on the simulation study reported in the Supporting Information, we choose $v_0$ to be the neighbour of $v^{\dagger}$ with the most connections, and we connect $v_0$ to the $b-1$ neighbours of $v^{\dagger}$ with the least connections. With such choices, it is relatively straightforward to execute the ROAM heuristic on existing social-networking services. On Facebook, for example, one can typically view the number of friends that each of his friends has (even if some of them make this information private, one can still choose among those that do not). Once the nodes are chosen, \textbf{Step~1} simply requires 	$v^{\dagger}$ to ``unfriend'' $v_0$, whereas \textbf{Step~2} requires $v^{\dagger}$ to ``suggest'' the friendship of $v_0$ to the other chosen nodes. Note that, on Facebook, $v^{\dagger}$ can only introduce two individuals to each other if they were both $v^{\dagger}$'s friends. As such, \textbf{Step~2} must be executed before \textbf{Step~1}, that is, $v^{\dagger}$ must end the friendship with $v_0$ \emph{after} introducing $v_0$ to the other nodes.

%%%%%%%%%%%%%%%%%%%%%%%%%%%%%%%%%%%%%%%%%%%%%%%%%%%%%%%%%%%%%%%%
\section{Disguising Communities}
\subsection{A Measure of Concealment}\label{sec:concealment-measure}

\noindent We propose a measure of how well a community, $\CC$, is hidden in a community structure, $\CS$. Note that $\CC$ is not necessarily a member of $\CS$. To put it differently, when describing $\CC$ as a ``community'', we mean to use this term in its broader sense, where $\CC$ is essentially just a subset of nodes. As such, when measuring how well $\CC$ is hidden in $\CS$, it may well be the case that the members of $\CC$ are spread out across multiple communities in $\CS$.

To this end, we start by proposing two measures, denoted by $\mc'$ and $\mc''$, which capture different aspects of concealment. In particular, $\mc'$ is defined for every community $\CC\subseteq V$ and every community structure $\CS$ as follows:
$$
\mc'(\CC,\CS) = \frac{|\{ C_i \in \CS : C_i \cap \CC \neq \emptyset\}|-1}{\max(|\CS|-1,1) \max_{C_i \in \CS}(|C_i \cap \CC|)}.
$$
Basically, this measure focuses on how well the members of $\CC$ are spread out across the communities in $\CS$. In more detail, we have $\mc'(\CC,\CS) \in [0,1]$, and the greater $\mc'(\CC,\CS)$, the greater the concealment of $\CC$ in $\CS$. Note that the numerator grows linearly with the number of communities that $\CC$ is distributed over. Subtracting $1$ from both the numerator and the $|\CS|$ term of the denominator is meant to handle the worst case, where all members of $\CC$ appear in a single (possibly larger) community in $\CS$; in this case, we have: $\mc'(\CC,\CS)=0$. In contrast, the term $\max_{C \in \CS}(|C \cap \CC|)$ is meant to promote community structures in which the members of $\CC$ are more evenly distributed across the communities in $\CS$. As such, the maximum concealment is achieved when the members of $\CC$ are uniformly distributed, with each member appearing in a separate community; in this case: $\mc'(\CC,\CS)=1$. 

Moving on to the second measure, $\mc''$, it is defined as:
$$
\mc''(\CC,\CS) = \sum_{C_i \in \CS} \frac{|C_i \setminus \CC|}{\max(n-|\CC|,1)}.
$$
Intuitively, $\mc''$ focuses on how well $\CC$ is ``hidden in the crowd''; it grows linearly with the number of non-members of $\CC$ that appear with members of $\CC$ in the same community in $\CS$. Note that $\mc''(\CC,\CS) \in [0,1]$, and the greater the value, the greater the concealment of $\CC$ in $\CS$.

Having defined $\mc'$ and $\mc''$, we now use the two as building blocks to construct a single measure whereby the trade-off between $\mc'$ and $\mc''$ is controlled by a parameter, $\alpha \in [0,1]$. More formally, \emph{our proposed measure of concealment of a community $\CC$ in a community structure $\CS$ is}:
$$
\mc(\CC,\CS) = \alpha \mc'(\CC,\CS) + (1 - \alpha) \mc''(\CC,\CS).
$$
%Naturally, for all $\CC$ and $\CS$ we have $\mc(\CC,\CS) \in [0,1]$, where greater values indicate greater levels of concealment.

\begin{figure}[tb]
\centering
    \includegraphics[width=1.0\columnwidth]{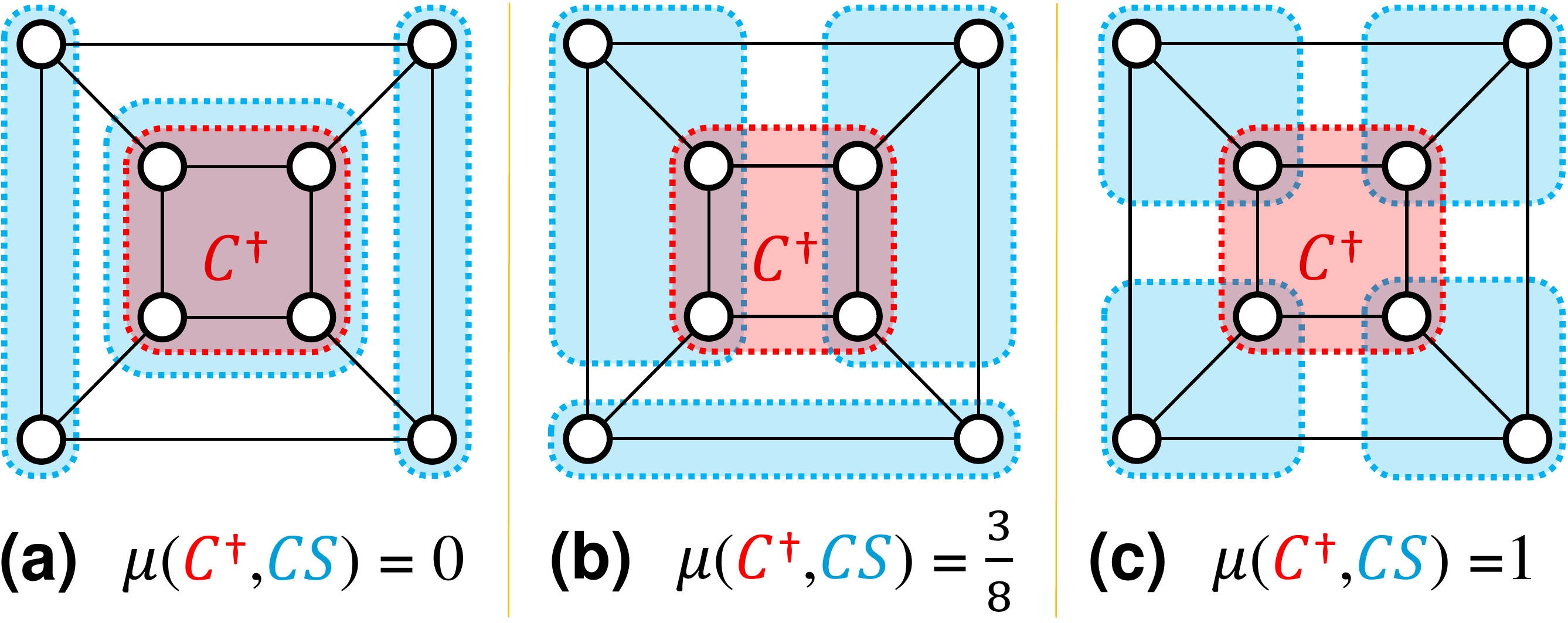}\vspace{-0.25cm}
    \caption{How the concealment of $\CC$ differs from one community structure to another according to $\mu$ where $\alpha=0.5$.}
    \label{fig:measure-example}
\end{figure}

Figure~\ref{fig:measure-example} presents a sample network with three different community structures, and highlights the community that we wish to conceal, namely $\CC$. For every such community structure, we measure the concealment of $\CC$ using our measure $\mc$ with $\alpha = 0.5$. In particular,  Figure~\ref{fig:measure-example}(a) presents one extreme where $\mc(\CC,\CS) = 0$, reflecting the fact that $\CC$ is completely exposed as a community. Figure~\ref{fig:measure-example}(b) presents the other extreme where $\mc(\CC,\CS) = 1$, reflecting the fact that $\CC$ is completely hidden, since every member appears in a separate community along some non-member of $\CC$. Finally, a case between the two extremes is presented in Figure~\ref{fig:measure-example}(c), where $\mc(\CC,\CS) = \frac{3}{8}$.

%%%%%%%%%%%%%%%%%%%%%%%%%%%%%%%%%%%%%%%%%%%%%%%%%%%%%%%%%%%%%%%
%%%%%%%%%%%%%%%%%%%%%%%%%%%%%%%%%%%%%%%%%%%%%%%%%%%%%%%%%%%%%%%

%%%%%%%%%%%%%%%%%%%%%%%%%%%%%%%%%%%%%%%%%%%%%%%%%%%%%%%%%%%%%%%
\subsection{A Scalable Heuristic}
We set to develop a simple heuristic that can be applied by any group of people regardless of their technical background or their knowledge of the network topology. After all, it is of little use to have an exact algorithm that can only be understood or applied by optimization experts armed with enormous processing power. Likewise, exact algorithms that require knowing the entire network topology may prove useless, since such knowledge is rarely available.

Our heuristic, called DICE---Disconnect Internally, Connect Externally---is detailed in the box below.

%%%%%%%%%%%%%%%%%%%%%%%%%%%%%%%%%%%%%%%%%%%%%%%%%%%%%%%%%%%%%%%
\begin{mdframed}[style=MyFrame]
\vspace*{-0.1cm}
\textbf{The DICE heuristic} given a budget $b$:
\vspace*{-0.18cm}
\begin{itemize}\itemsep0.3em
\item \textbf{Step~1:} Disconnect $d\leq b$ links from within $\CC$;
\item \textbf{Step~2:} Connect $b-d$ nodes from within $\CC$ to $b-d$ nodes from outside of $\CC$.
\vspace*{-0.18cm}
\end{itemize}
\end{mdframed}
%%%%%%%%%%%%%%%%%%%%%%%%%%%%%%%%%%%%%%%%%%%%%%%%%%%%%%%%%%%%%%%

This heuristic is inspired by \emph{modularity} \cite{newman2004finding}---a widely used index for measuring the quality of any given community structure. Specifically, it promotes structures that have dense connections \emph{within} communities and sparse connections \emph{between} them. As such, community-detection algorithms are typically designed to search for a structure that maximizes modularity. With this in mind, \textbf{Step~1} of our heuristic decreases the density of the connections within $\CC$, whereas \textbf{Step~2}  increases the connections between $\CC$ and other communities. In doing so, a community-detection algorithm is more likely to overlook $\CC$, i.e., it would fail to recognize $\CC$ as a community, and instead assign its members to multiple communities.

Finally, let us comment on how DICE can be applied in practice. On Facebook, for example, \textbf{Step~1} requires some members to ``unfriend'' other members, which is rather straightforward. As for \textbf{Step~2}, members must send a friendship request to non-members; these could be classmates, coworkers, neighbours living next door, or even random people (it is possible to try multiple random friendship requests, hoping that some of them would be successful).

%%%%%%%%%%%%%%%%%%%%%%%%%%%%%%%%%%%%%%%%%%%%%%%%%%%%%%%%%%%%%%%
\section{Experiments}
%To prove the effectiveness of presented heuristic in disguising centrality problem we perform series of simulations.
%We test the ROAM heuristic on both real-life and randomly-generated networks. We also perform a series of simulations to evaluate the properties of the Lieutenant Network.
%We experiment with randomly-generated networks, as well as real-life networks.

%%%%%%%%%%%%%%%%%%%%%%%%%%%%%%%%%%%%%%%%%%%%%%%%%%%%%%%%%%%%%%%
\begin{figure*}[tbh!]
\centering
\setlength\tabcolsep{1pt}
\renewcommand{\arraystretch}{2.0}
\begin{tabular}{m{.01\textwidth}m{.195\textwidth}m{.195\textwidth}m{.195\textwidth}m{.195\textwidth}m{.195\textwidth}}
& \multicolumn{1}{c}{\small{Degree Ranking}}
& \multicolumn{1}{c}{\small{Closeness Ranking}}
& \multicolumn{1}{c}{\small{Betweenness Ranking}}
& \multicolumn{1}{c}{\small{IC Influence}}
& \multicolumn{1}{c}{\small{LT Influence}} \\
\rotatebox{90}{\small{Madrid bombing}} &
\includegraphics[width=0.95\linewidth]{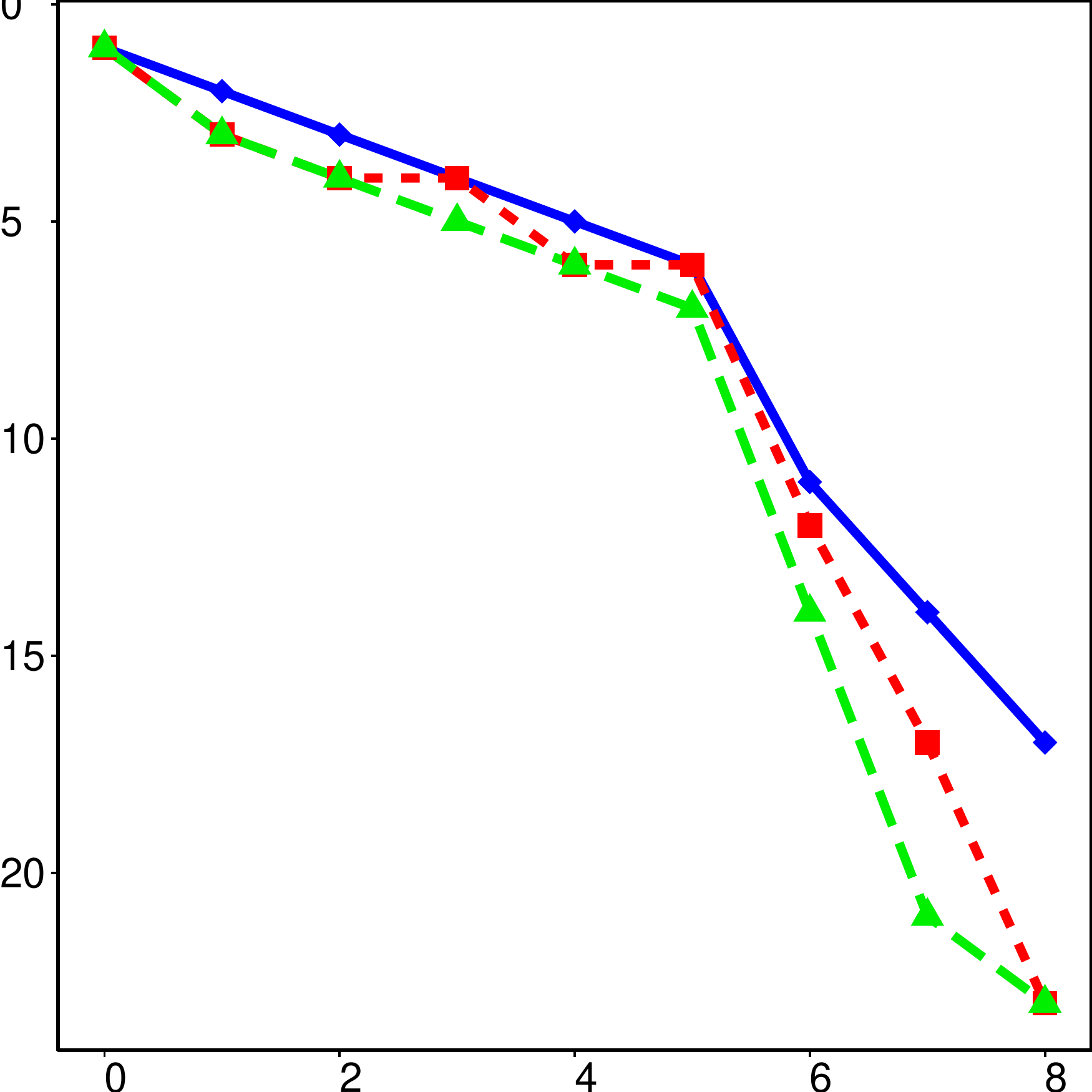} &
\includegraphics[width=0.95\linewidth]{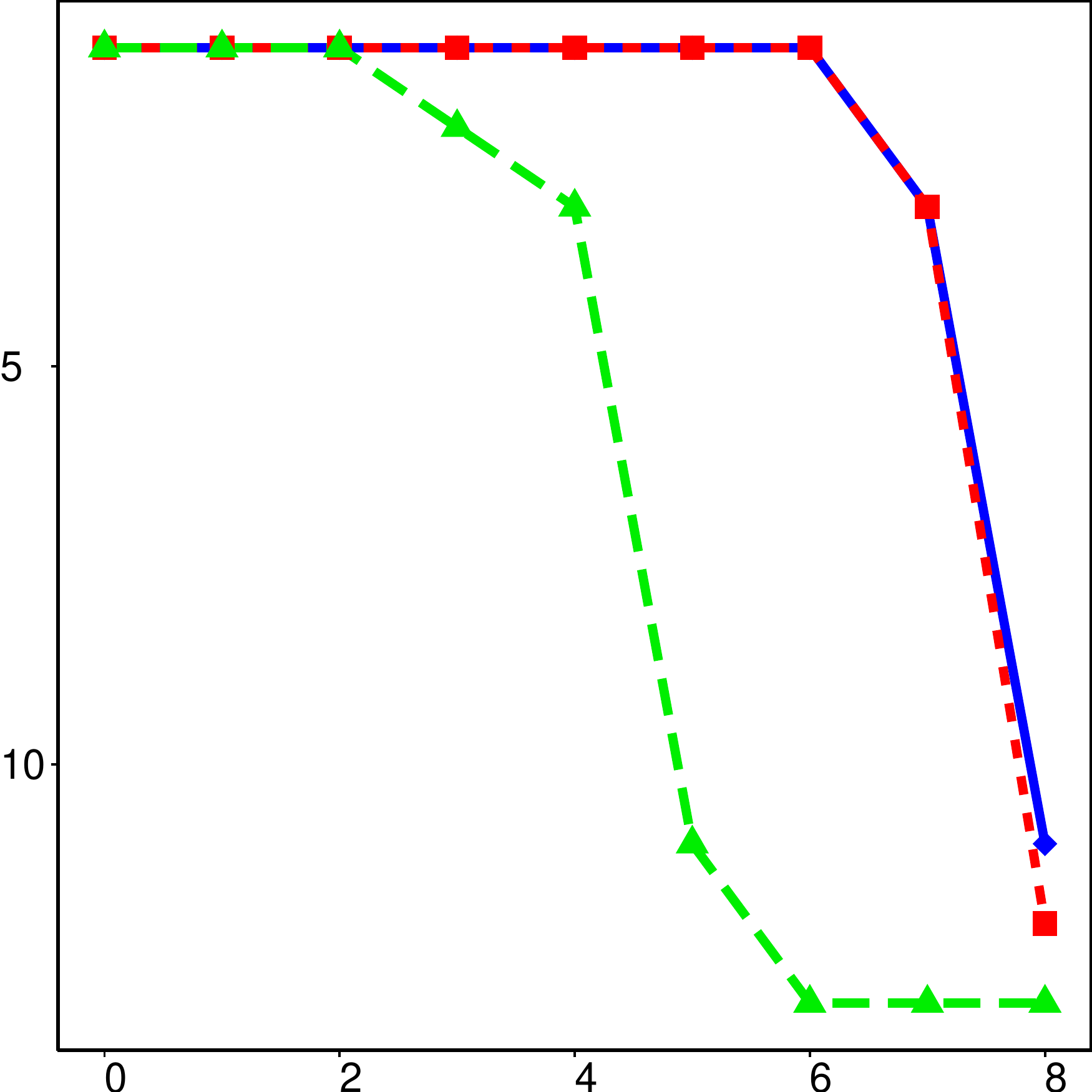} &
\includegraphics[width=0.95\linewidth]{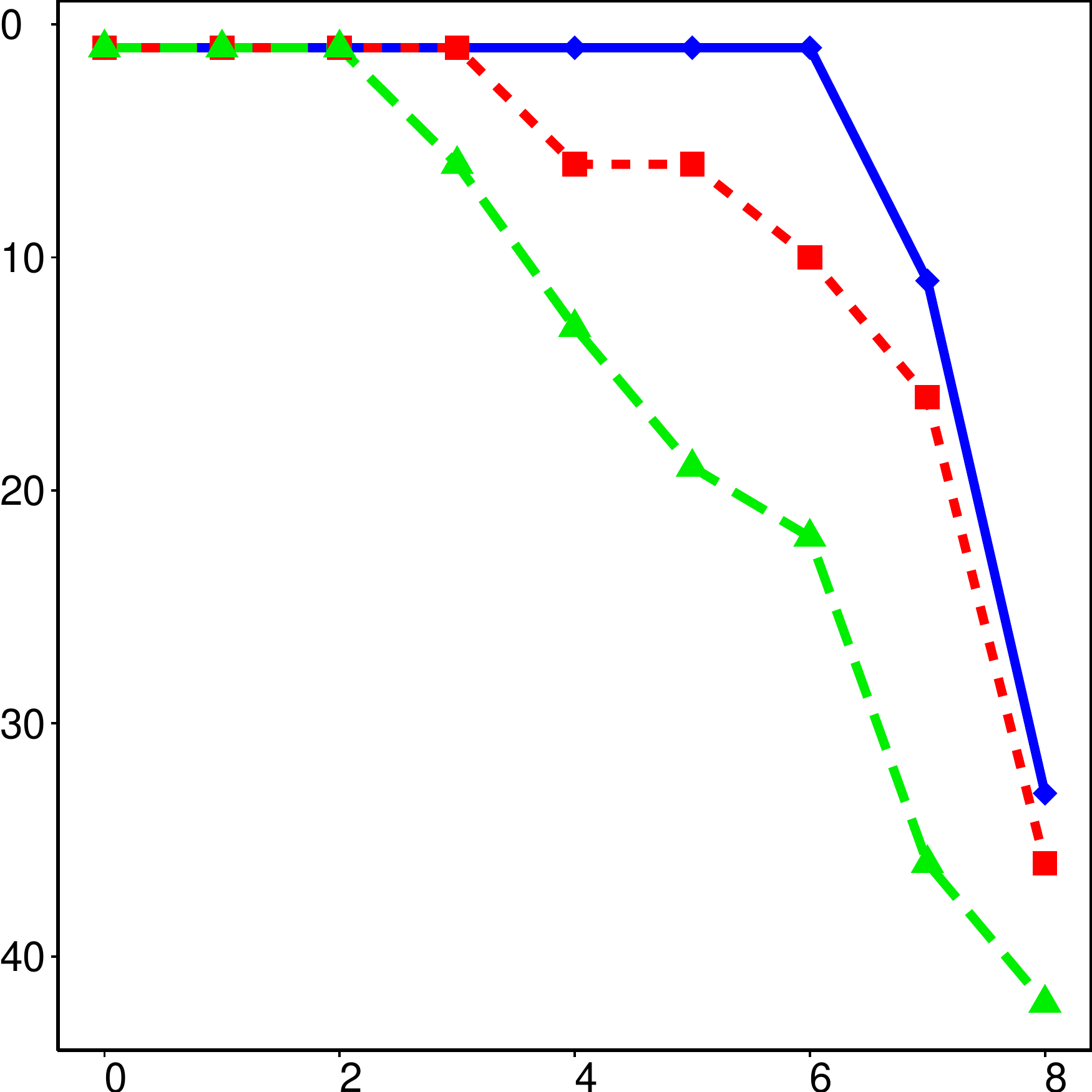} &
\includegraphics[width=0.95\linewidth]{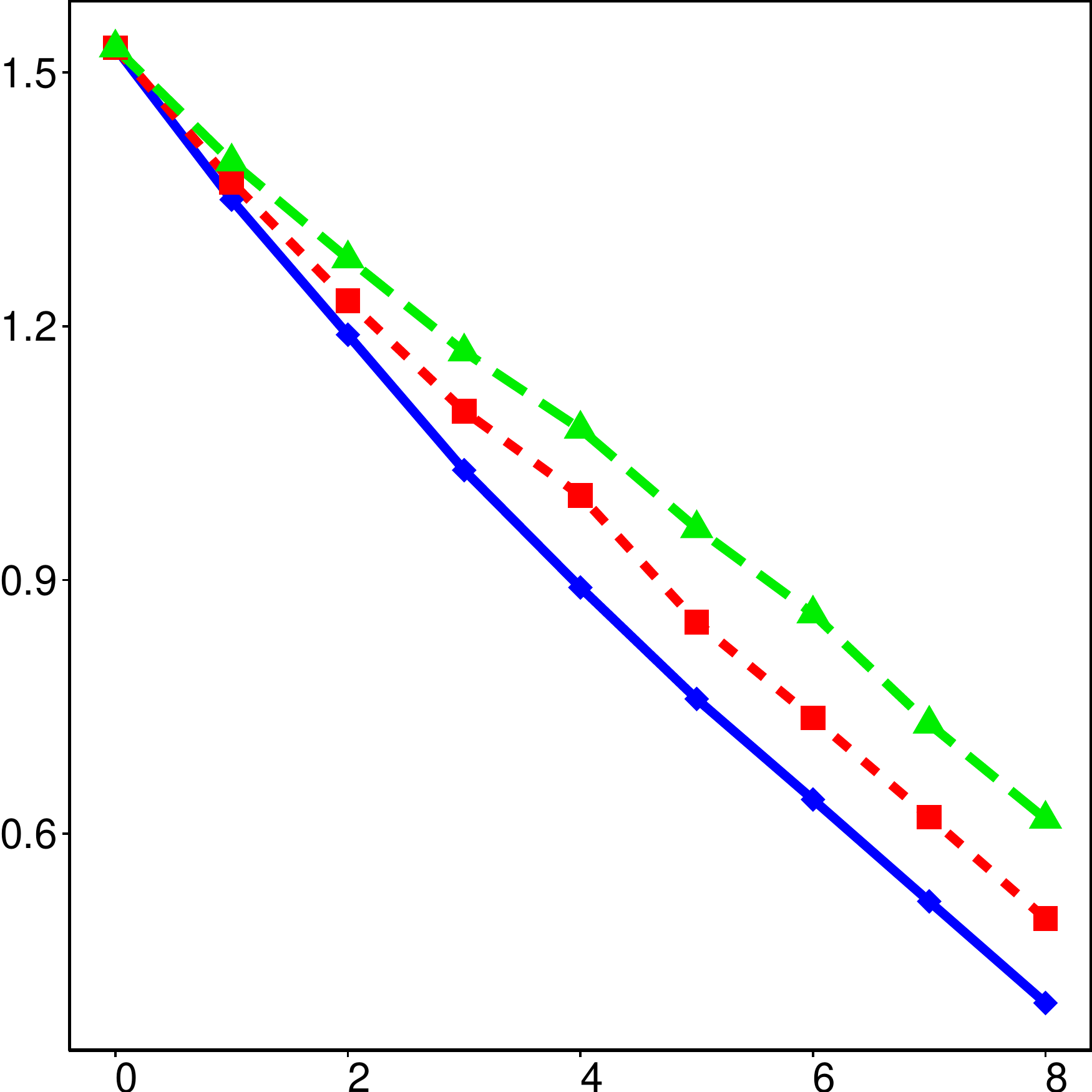} &
\includegraphics[width=0.95\linewidth]{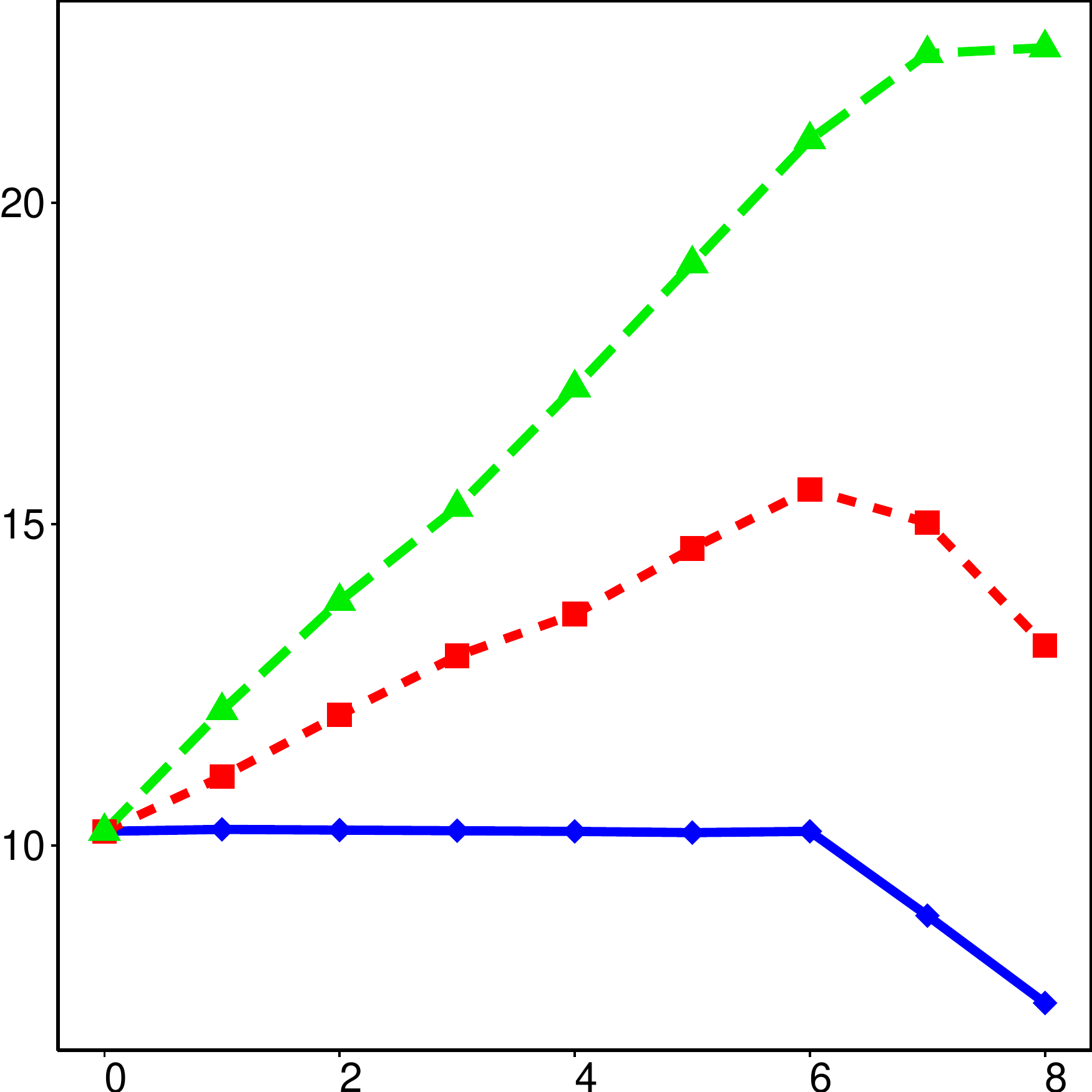} \\
\rotatebox{90}{\small{$\mathit{ScaleFree}(100,3)$}} &
\includegraphics[width=0.95\linewidth]{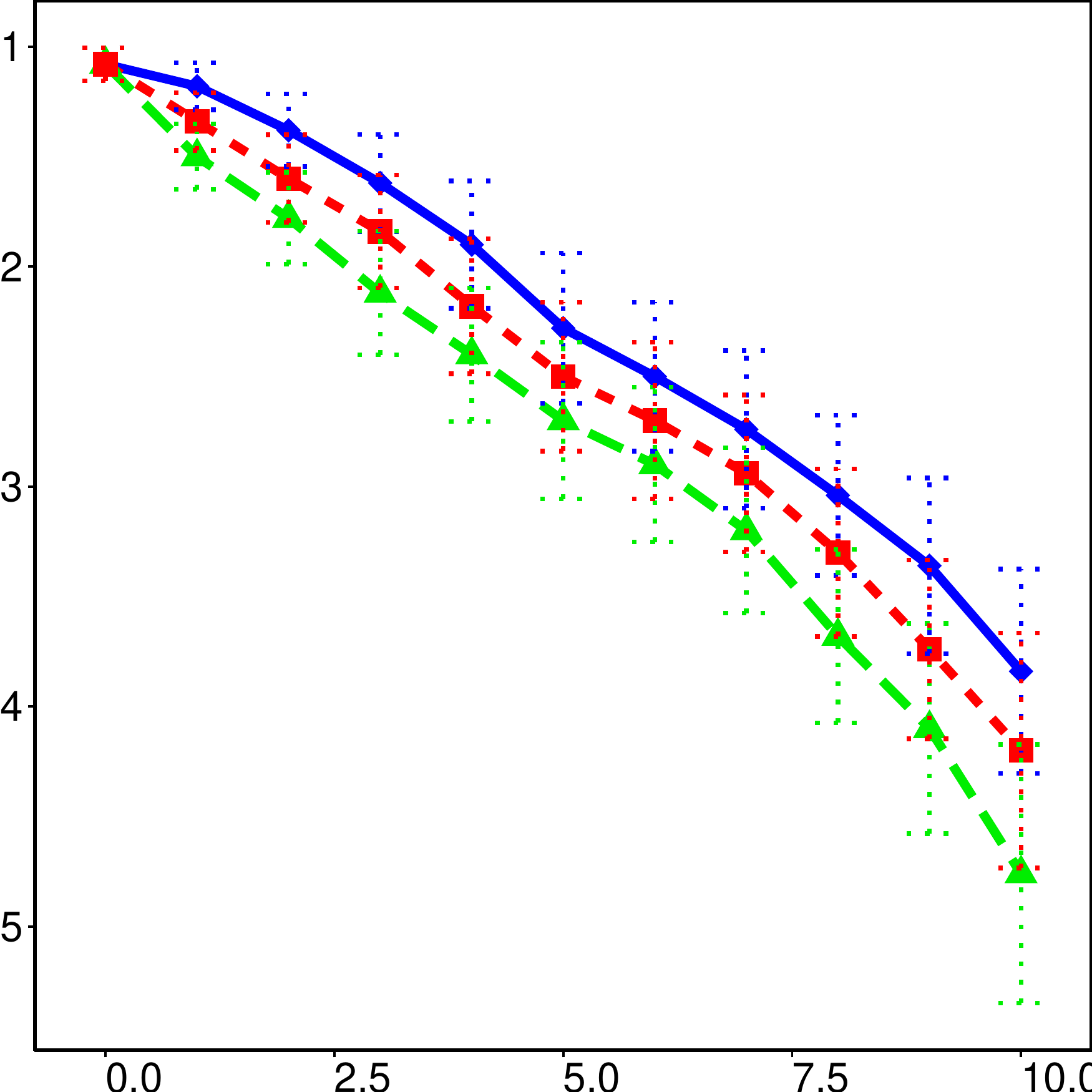} &
\includegraphics[width=0.95\linewidth]{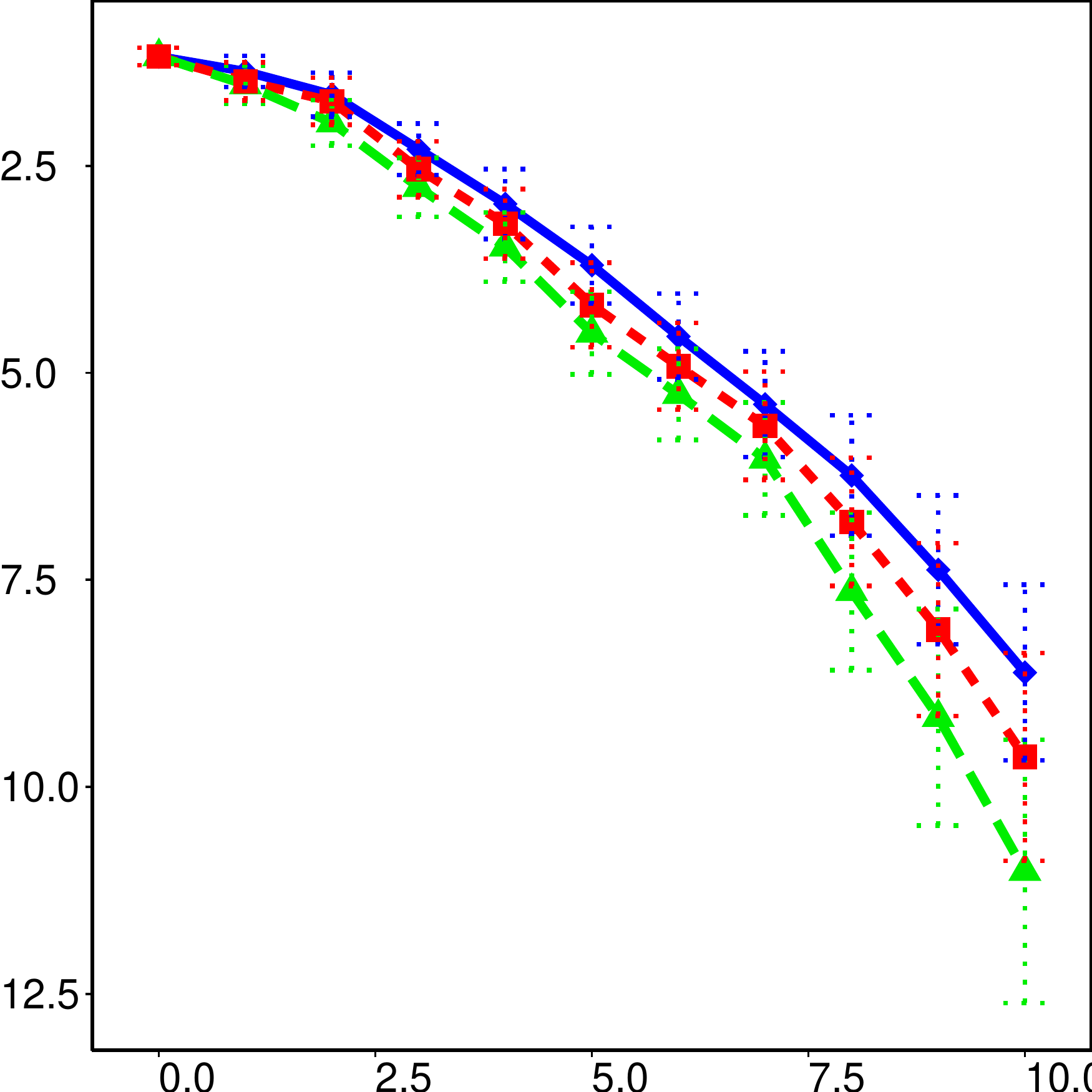} &
\includegraphics[width=0.95\linewidth]{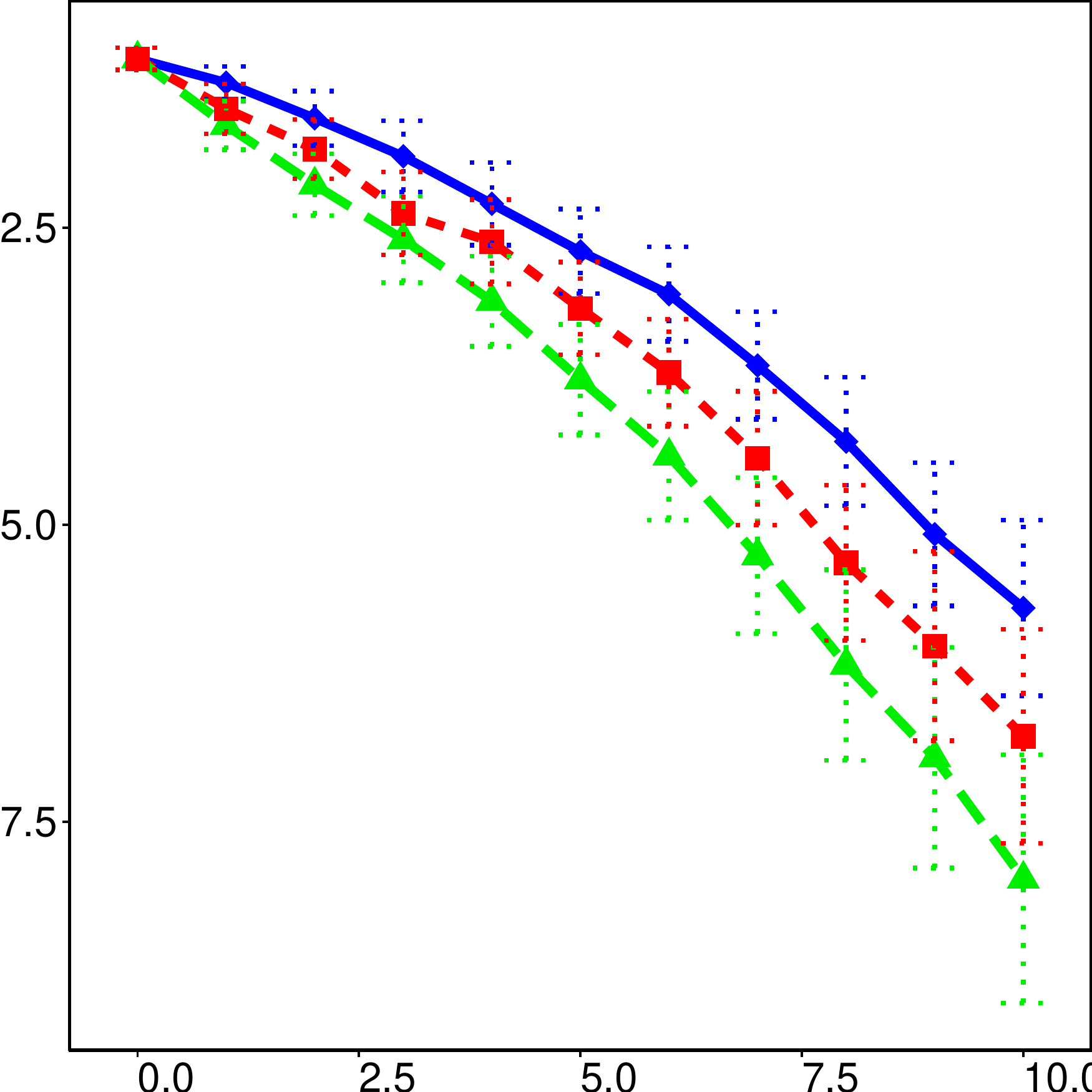} &
\includegraphics[width=0.95\linewidth]{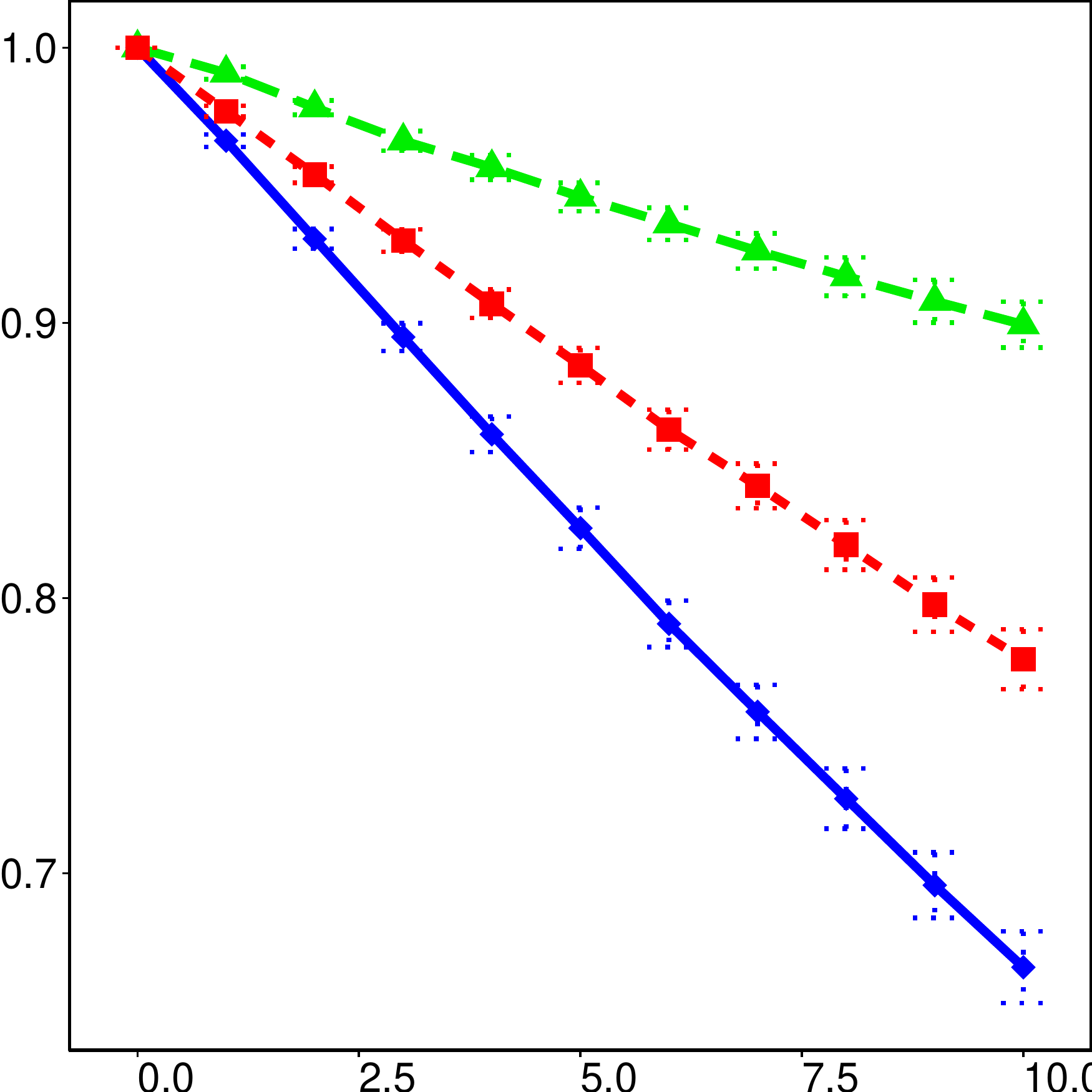} &
\includegraphics[width=0.95\linewidth]{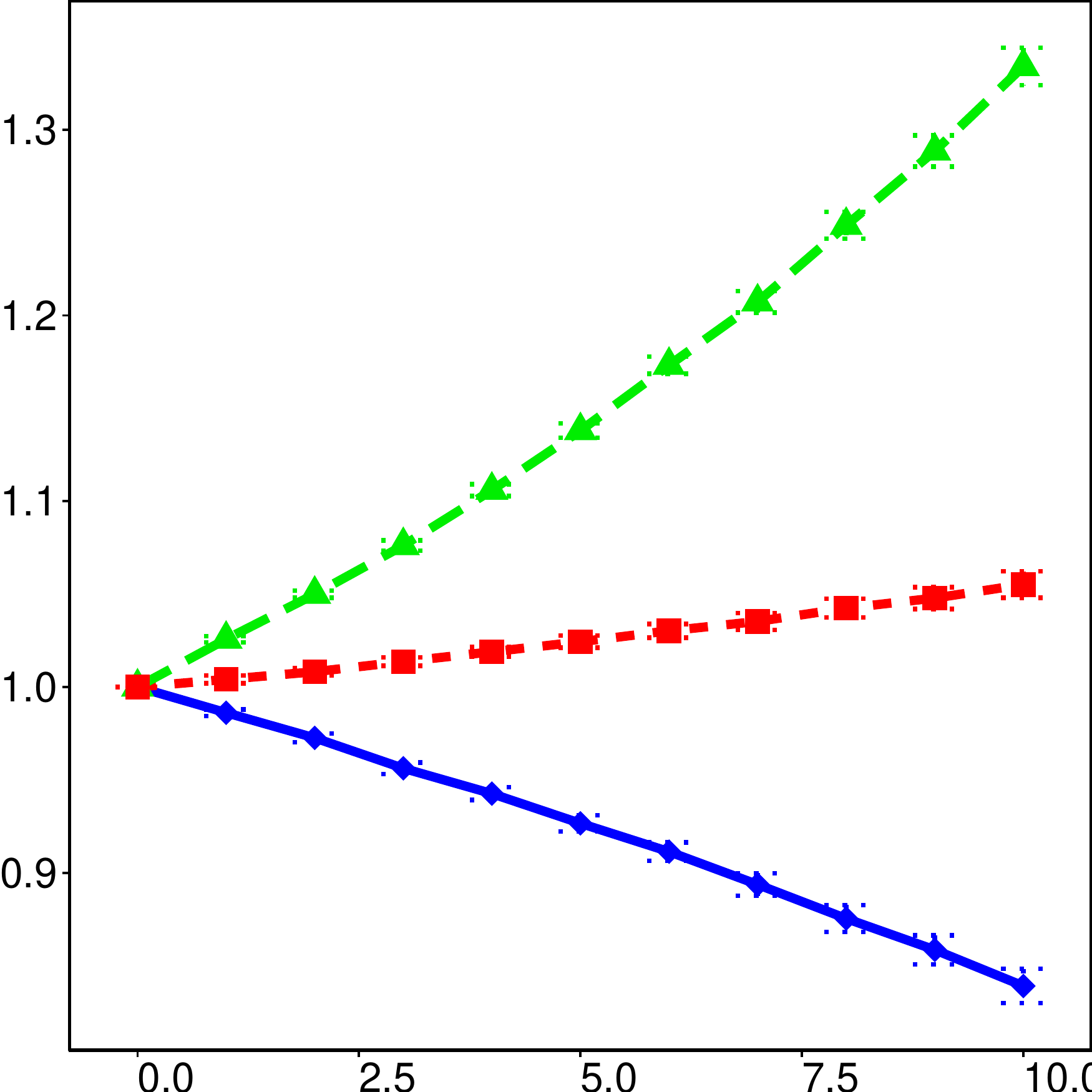} \\
\rotatebox{90}{\small{Facebook (medium)}} &
\includegraphics[width=0.95\linewidth]{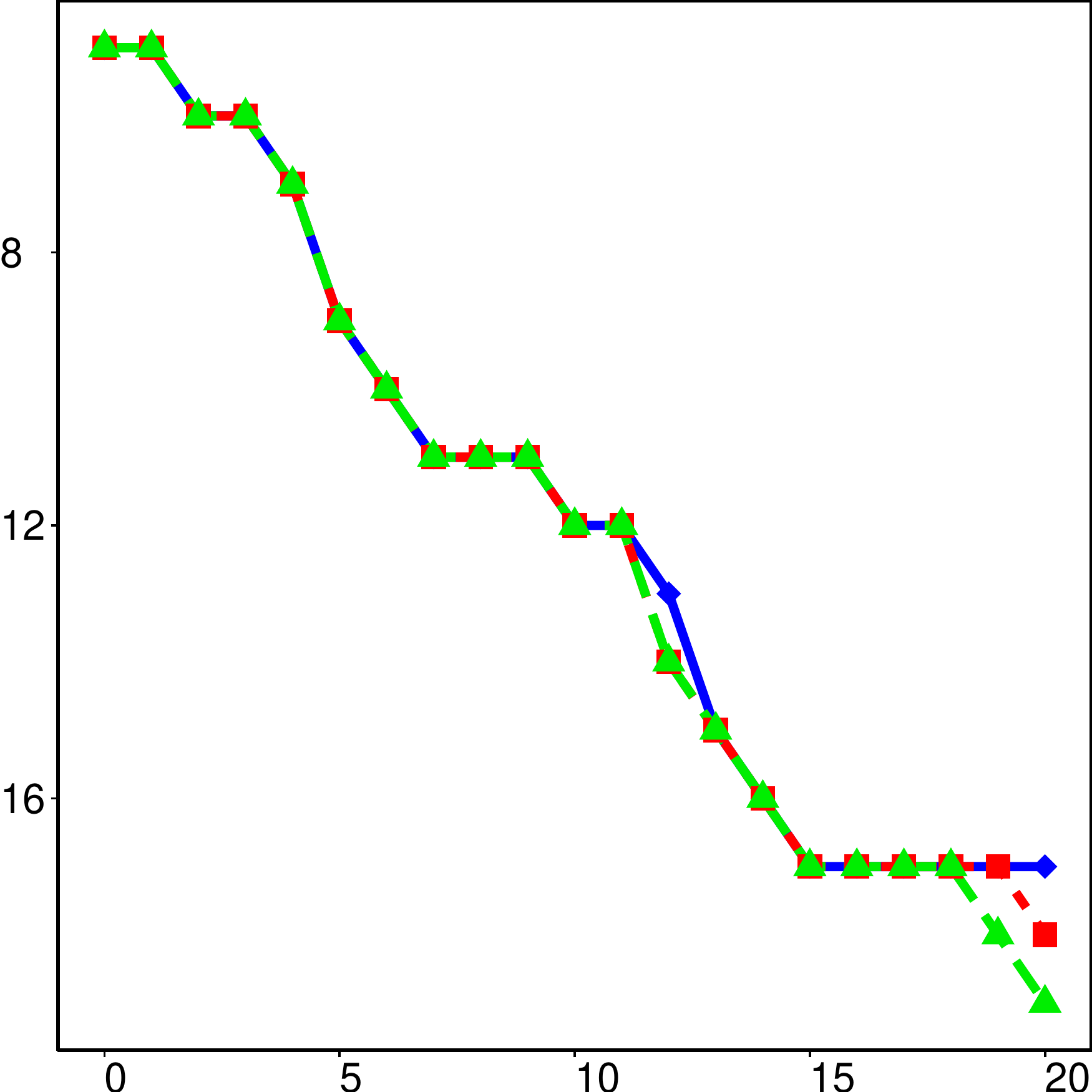} &
\includegraphics[width=0.95\linewidth]{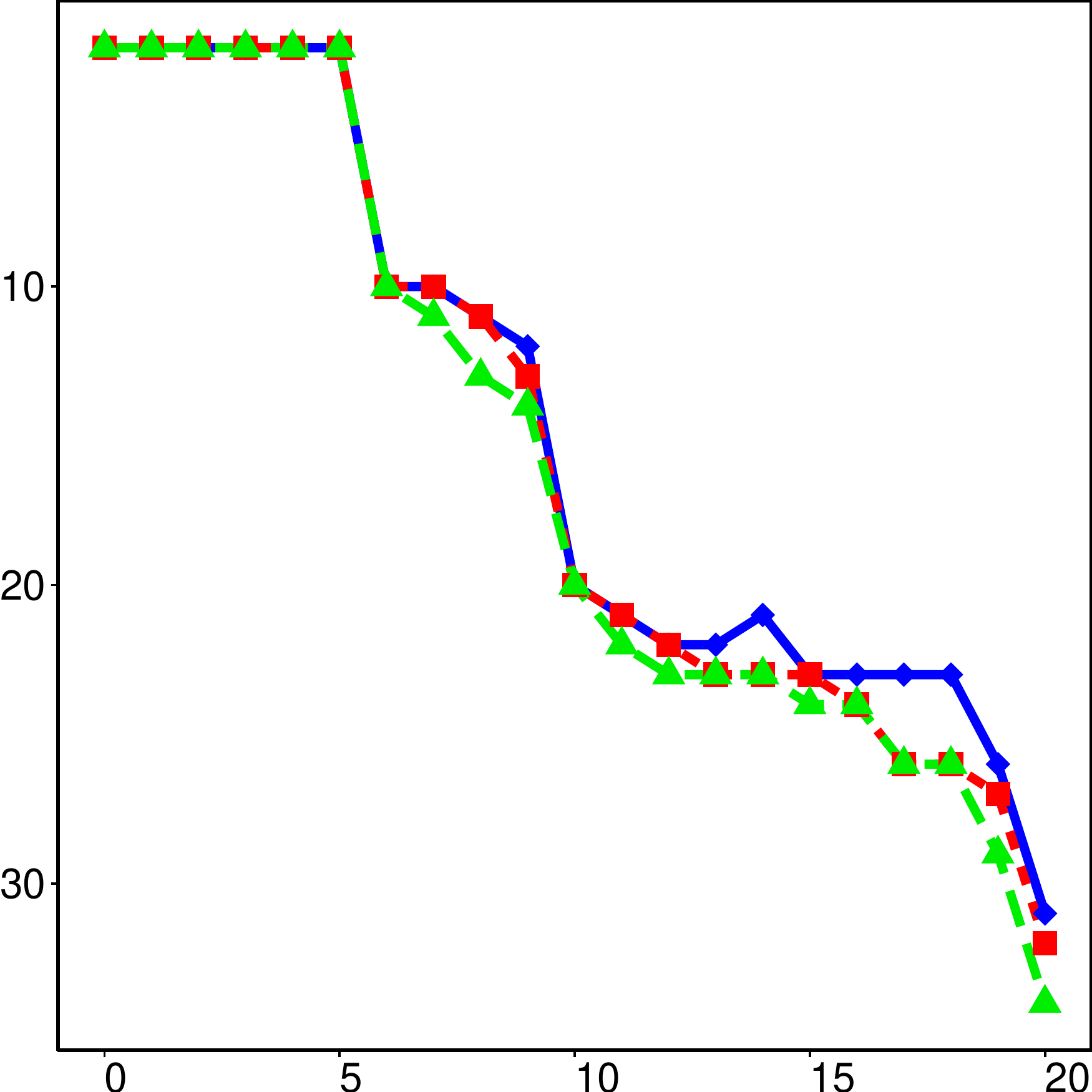} &
\includegraphics[width=0.95\linewidth]{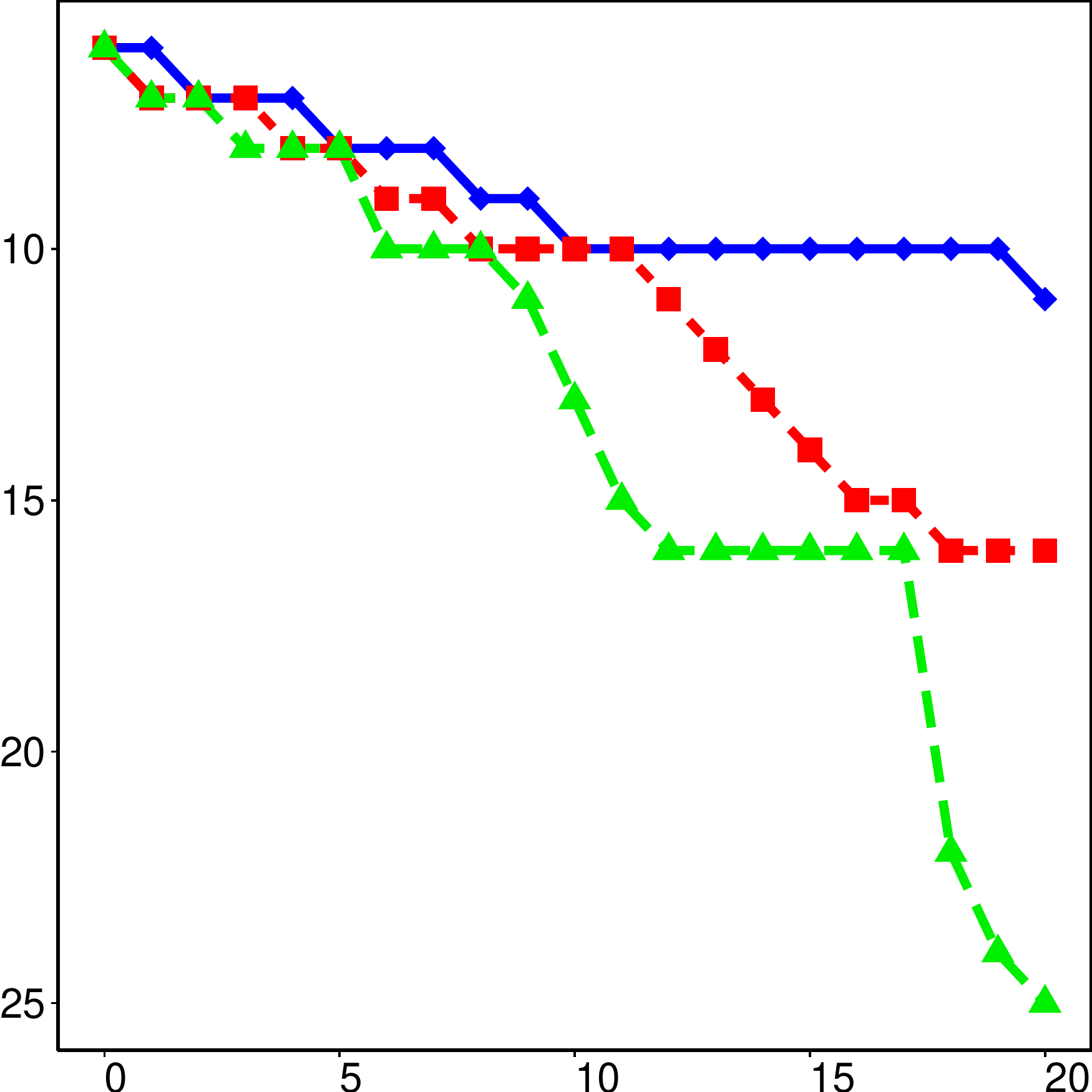} &
\includegraphics[width=0.95\linewidth]{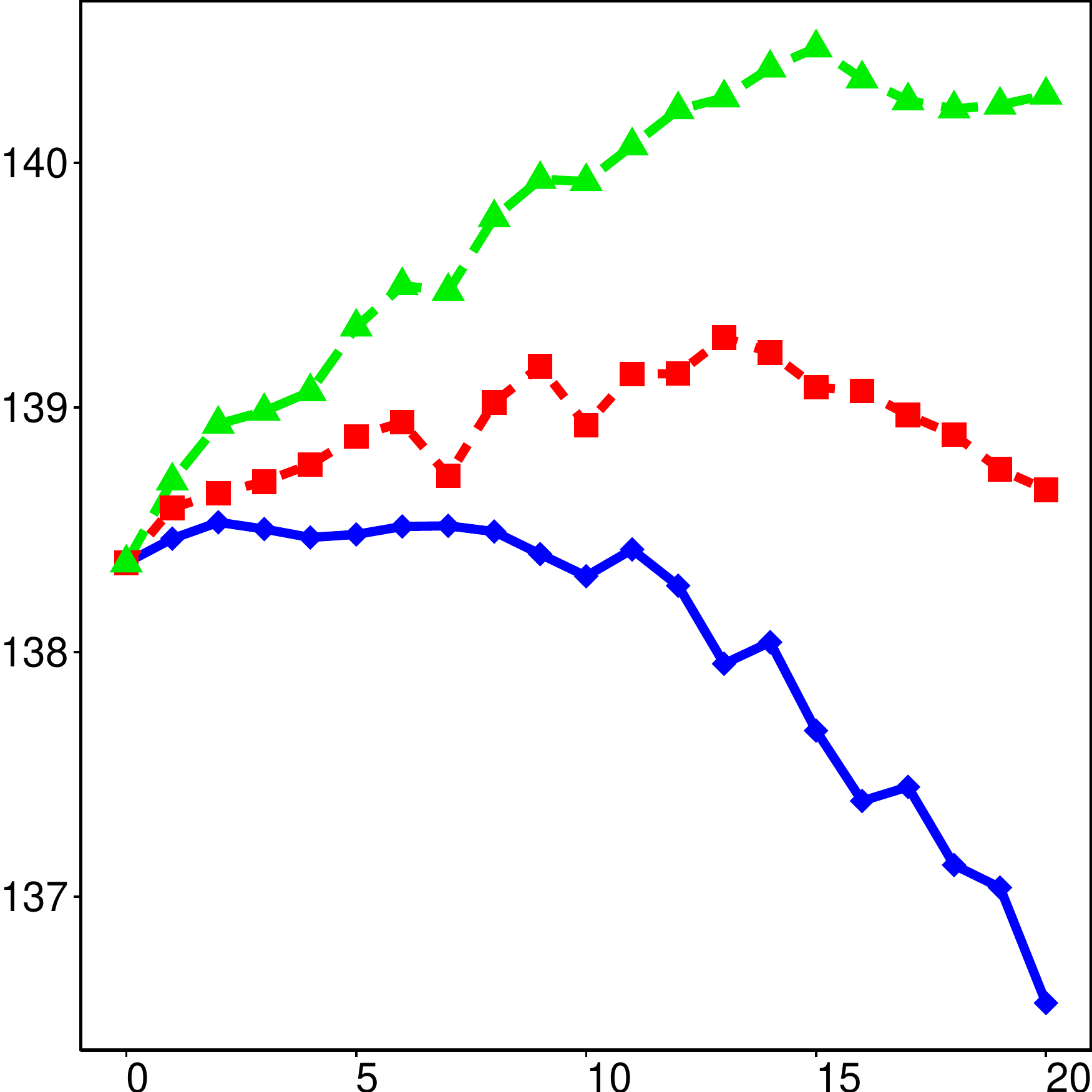} &
\includegraphics[width=0.95\linewidth]{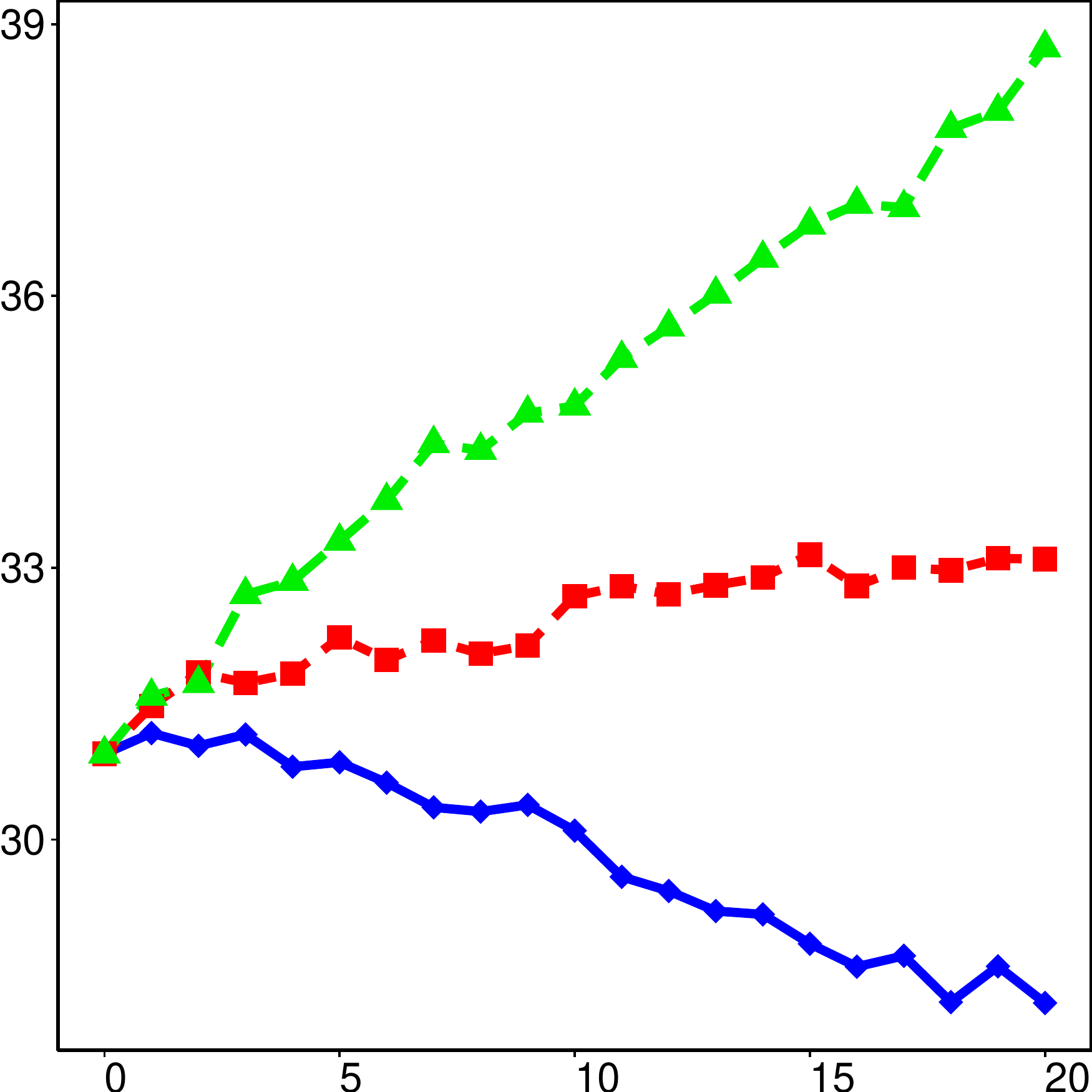} \\
& \multicolumn{5}{c}{\includegraphics[width=0.75\linewidth]{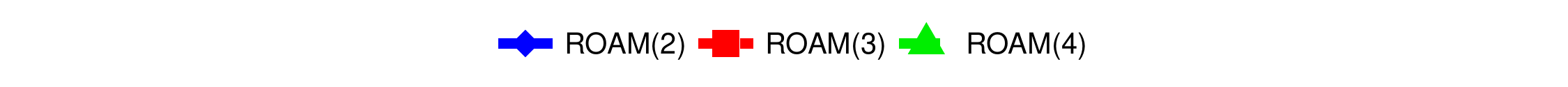}}
\end{tabular}
\vspace*{-0.2cm}
\caption{Executing ROAM multiple, consecutive times (the $x$-axis represents the number of executions). Given the Madrid-attack network, 50 scale-free networks, and a medium-sized fragment of Facebook's network (333 nodes, 5038 edges), the subfigures show the source node's ranking (according to different centrality measures), and the relative change in its influence value (according to different influence models). Results are for $\textnormal{ROAM}(b):b=2, 3, 4$, where $b$ is the budget in each execution.}
\label{fig:centrality-figures}
\end{figure*}
%%%%%%%%%%%%%%%%%%%%%%%%%%%%%%%%%%%%%%%%%%%%%%%%%%%%%%%%%%%%%%%

\subsection{Data sets} 
%We experiment with three types of randomly-generated networks, namely: \emph{scale-free} networks generated using the Barabasi-Albert model \cite{barabasi1999emergence}; \emph{small-world} networks generated using the Watts-Strogatz model \cite{watts1998collective}; and \emph{random graphs} generated using the Erdos-Renyi model \cite{erdds1959random}.
%
We experiment with two types of real-life networks:\vspace*{-0.12cm}
\begin{enumerate}
\item \emph{Covert organizations}: we consider three terrorist network, responsible for the WTC 9/11 attacks \cite{Krebs:2002a}; the 2002 Bali attack \cite{hayes2006connecting}; and the 2004 Madrid train bombings \cite{hayes2006connecting};
\item \emph{Social networks}: we study anonymized fragments of three social networks, namely Facebook, Twitter and Google+. These fragments are taken from SNAP---the Stanford Network Analysis Platform \cite{leskovec2012learning}.
\end{enumerate}
We also study randomly-generated networks, namely:\vspace*{-0.12cm}
\begin{enumerate}
\item \emph{Scale-free} networks using the Barabasi-Albert model \cite{barabasi1999emergence}. We write $\mathit{ScaleFree}(x,y)$ where $x$ is the number of nodes; $y$ is the number of links added with each node;
\item \emph{Small-world} networks using the Watts-Strogatz model \cite{watts1998collective}. We write $\mathit{SmallWorld}(x,y,z)$ where $x$ is the number of nodes; $y$ is the average degree; $z$ is the rewiring probability;
\item \emph{Random graphs} generated using the Erdos-Renyi model \cite{erdds1959random}. We write $\mathit{RandomGraph}(x,y)$ where $x$ is the number of nodes; $y$ is the expected average degree.
\end{enumerate}
For each type of randomly-generated networks, we report the average result taken over 50 such networks, with the error bars representing the 95\% confidence intervals.
%

%%%%%%%%%%%%%%%%%%%%%%%%%%%%%%%%%%%%%%%%%%%%%%%%%%%%%%%%%%%%%%%
\subsection{Experimenting with ROAM} Each of our experiments consists of a network, a budget, a source node, and an influence model. More specifically, we experiment with a budget of 2, 3, and 4. The source node is assumed to be the one with the lowest sum of centrality rankings (ties are broken uniformly at random). Whenever the \emph{Independent Cascade} model is used, an activation probability of $0.15$ is assumed on each link. On the other hand, whenever the \emph{Linear Threshold} model is used, a uniform distribution of thresholds is assumed (see the Supporting Information for more details). For both models, the influence values are approximated using the Monte-Carlo method. In each of these experiment, the ROAM heuristic is executed multiple, consecutive times.

Figure~\ref{fig:centrality-figures} shows the results of some of our experiments (the remaining results are provided in the Supporting Information). The centrality plots depict the \emph{ranking} of the source node, whereas the influence plots depict its \emph{relative} influence value (compared to the \emph{original} influence value before executing the heuristic altogether). As can be seen, the heuristic is effective in decreasing the source node's ranking, and this effectiveness increases with the budget spent on rewiring the network. As for influence, the performance of the heuristic varies depending on the network, the influence model, and the budget. Overall, the greater the budget, the greater the influence, e.g., a budget of 4 manages to maintain (or even increase) the influence in 4 out of 6 cases.

%%%%%%%%%%%%%%%%%%%%%%%%%%%%%%%%%%%%%%%%%%%%%%%%%%%%%%%%%%%%%%%

%%%%%%%%%%%%%%%%%%%%%%%%%%%%%%%%%%%%%%%%%%%%%%%%%%%%%%%%%%%%%%%
\begin{figure}[tb!]
\centering
\setlength\tabcolsep{1pt}
\renewcommand{\arraystretch}{2.0}
\begin{tabular}{m{.05\linewidth}m{.39\linewidth}m{.39\linewidth}}
&
\multicolumn{1}{c}{\small{(($b=4$, $d=0$)}} &
\multicolumn{1}{c}{\small{($b=4$, $d=4$)}} \\
\rotatebox{90}{\small{Madrid bombing}} &
\includegraphics[width=0.95\linewidth]{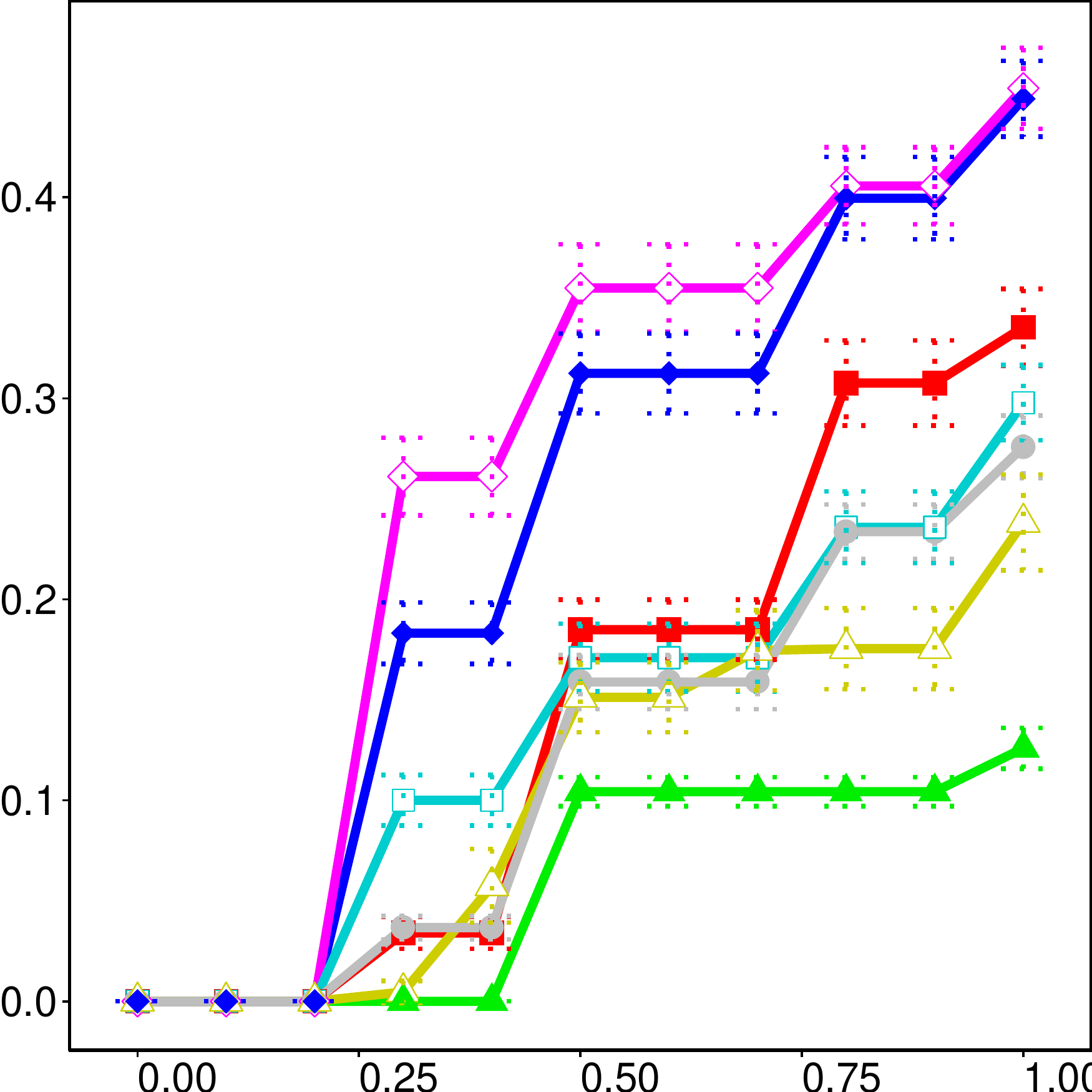} &
\includegraphics[width=0.95\linewidth]{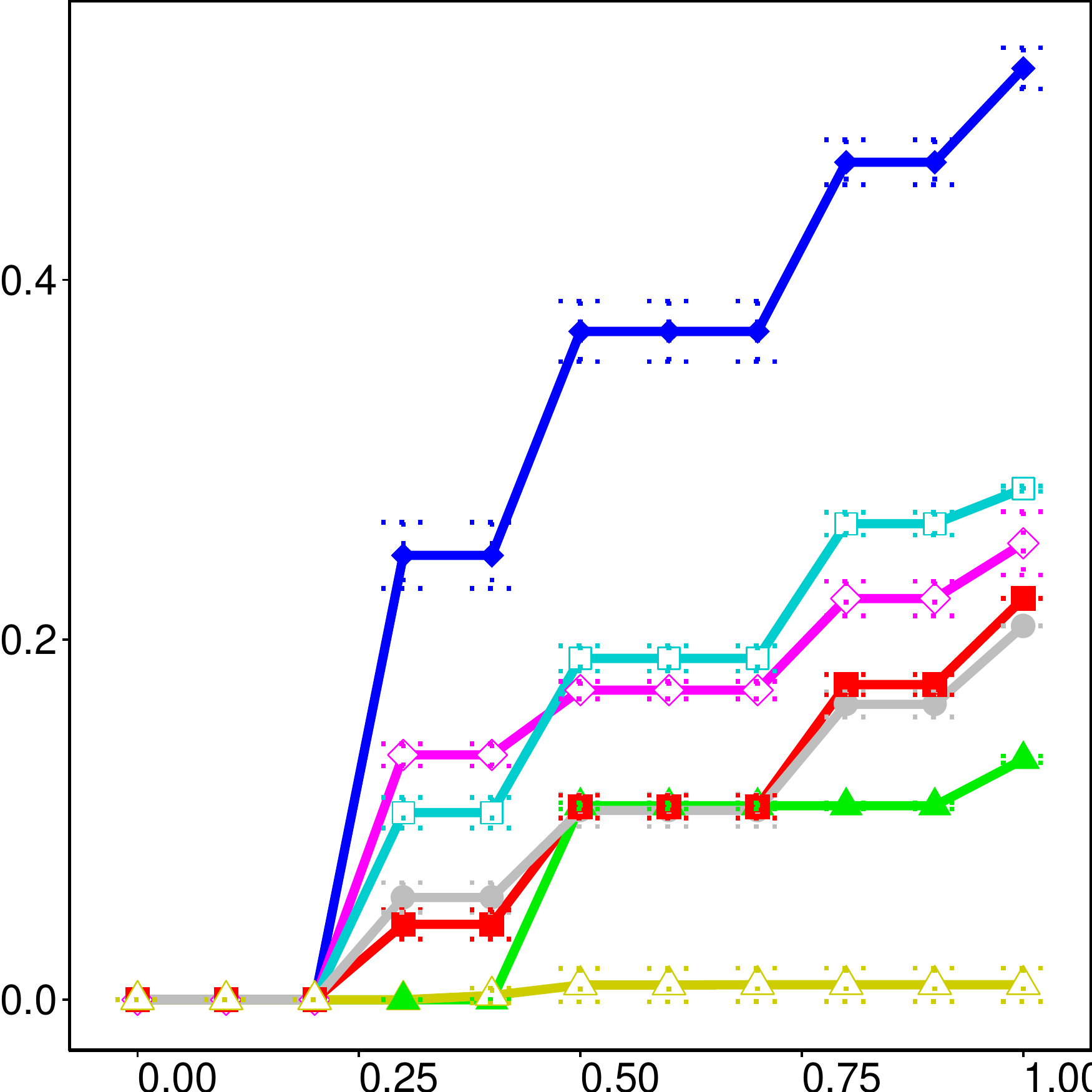} \\
\rotatebox{90}{\small{$\mathit{ScaleFree}(100,3)$}} &
\includegraphics[width=0.95\linewidth]{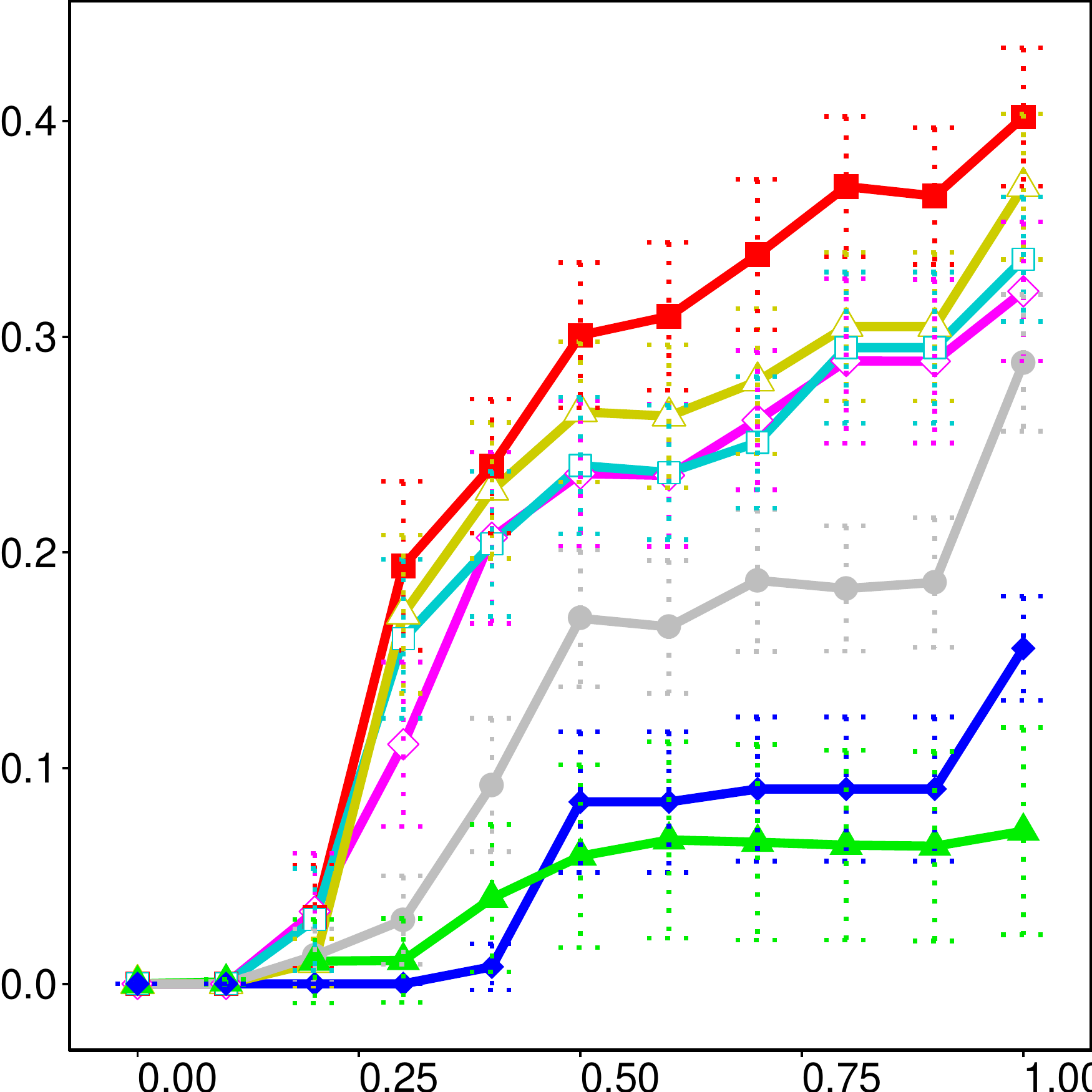} &
\includegraphics[width=0.95\linewidth]{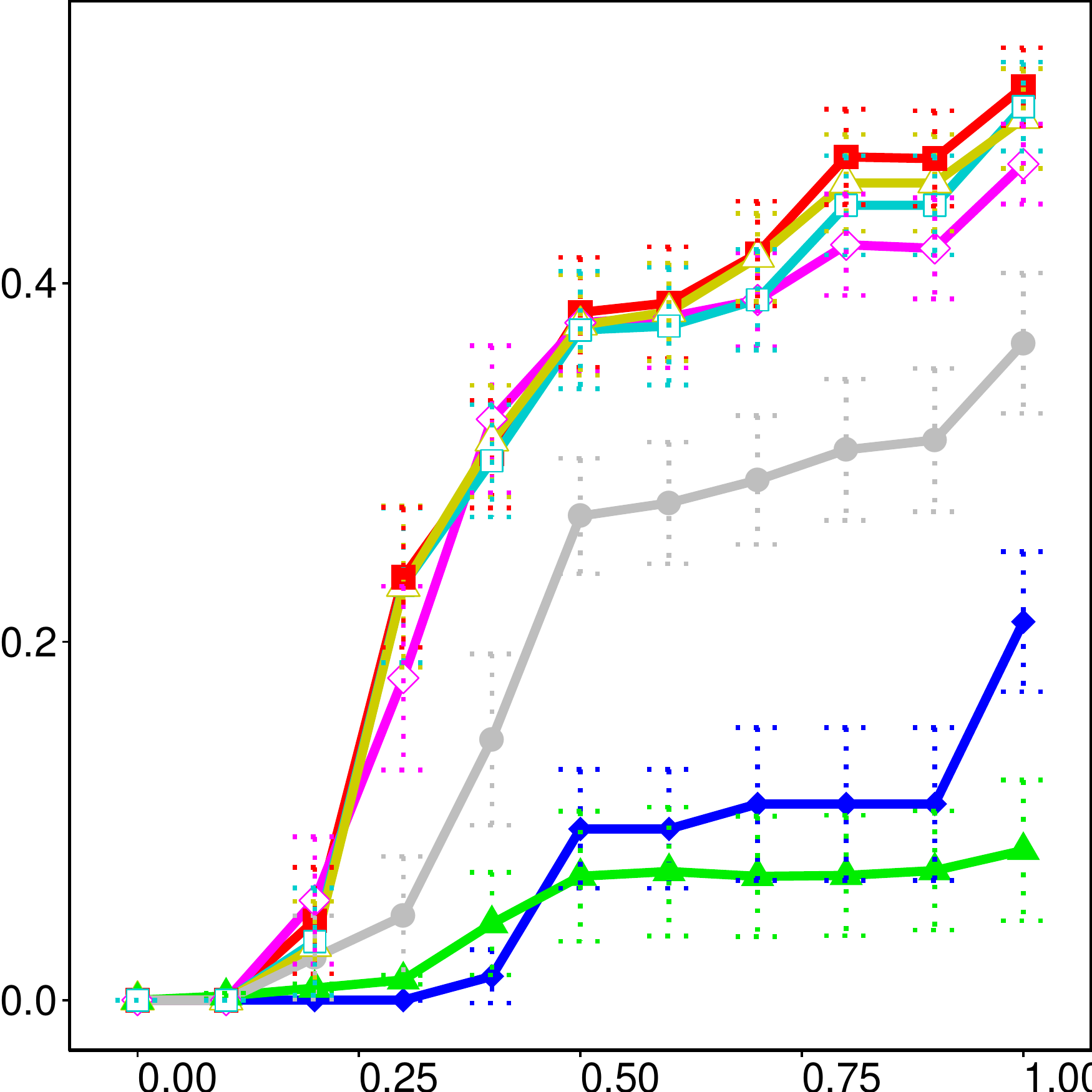} \\
\rotatebox{90}{\small{Facebook (medium)}} &
\includegraphics[width=0.95\linewidth]{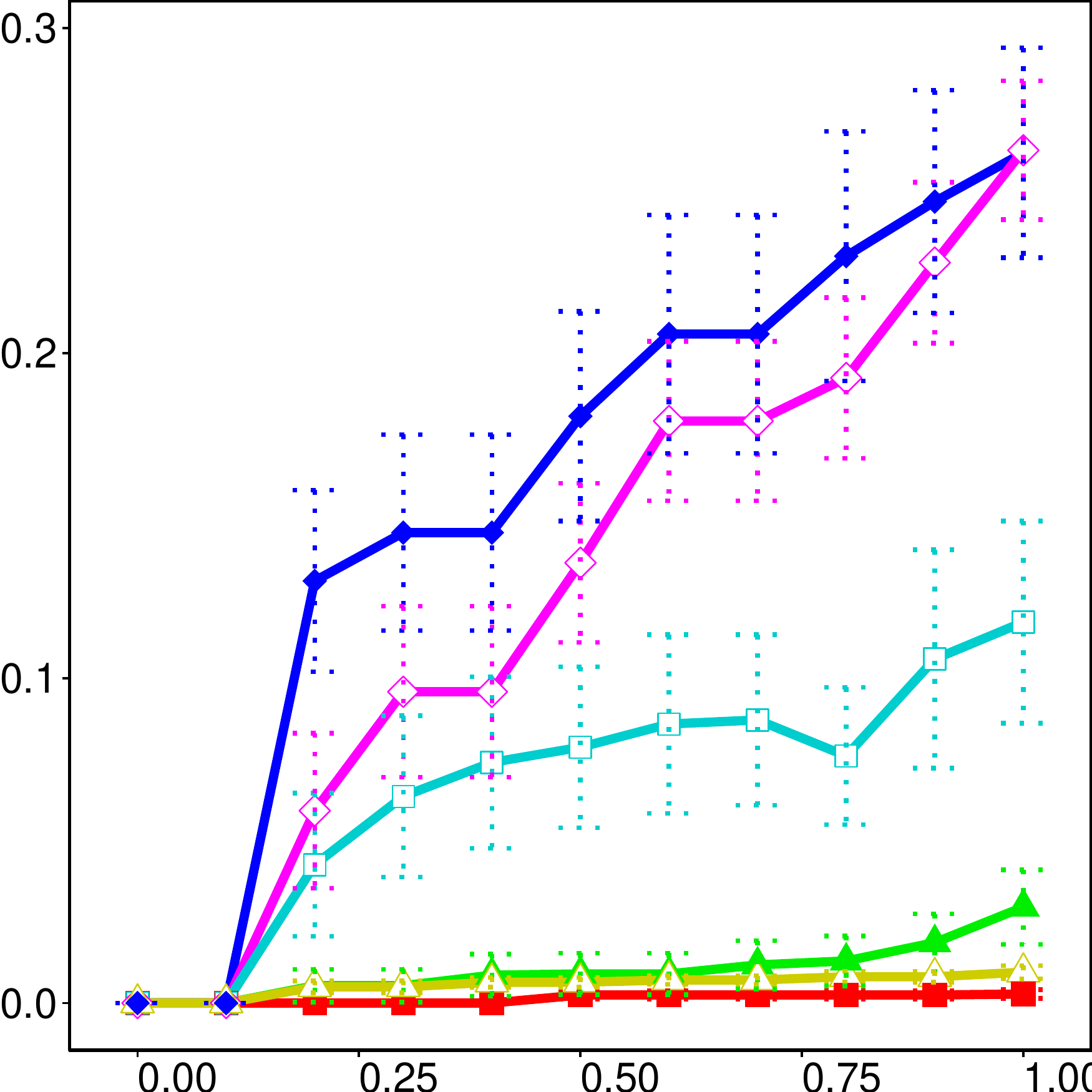} &
\includegraphics[width=0.95\linewidth]{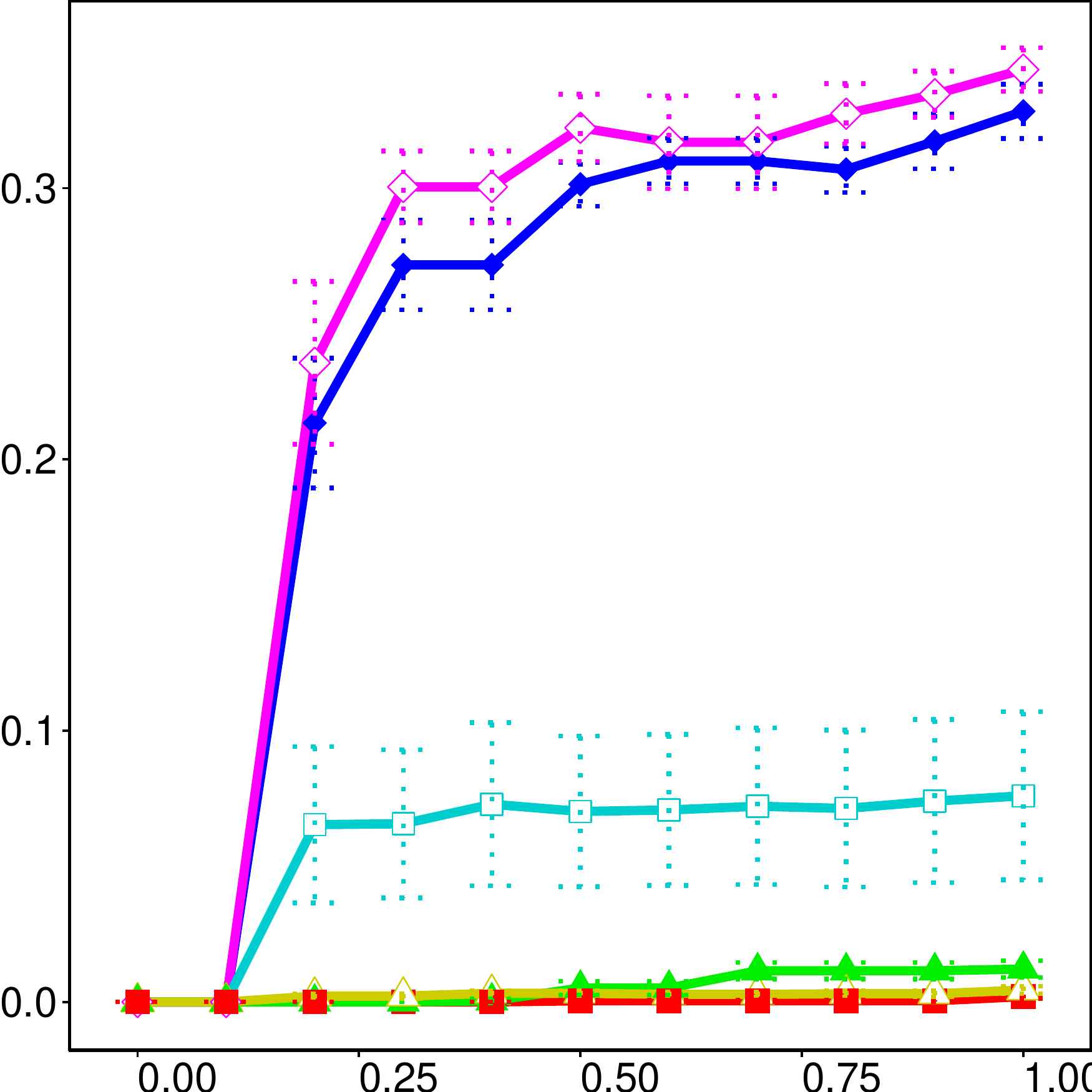} \\
& \multicolumn{2}{c}{\includegraphics[width=0.75\linewidth]{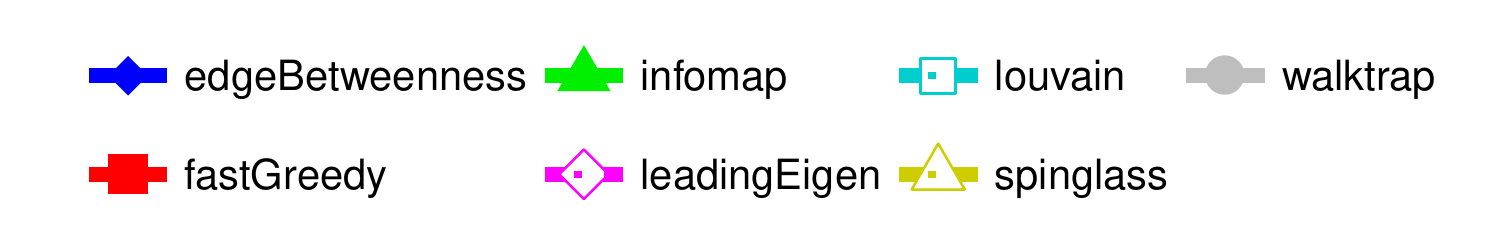}}
\end{tabular}
\vspace*{-0.2cm}
\caption{Executing DICE multiple, consecutive rounds (the $x$-axis represents the percentage of completed rounds), given the Madrid-attack network, scale-free networks, and a fragment of Facebook's network (consisting of 333 nodes, 5038 edges).}
\label{fig:community-figures}
\end{figure}
%%%%%%%%%%%%%%%%%%%%%%%%%%%%%%%%%%%%%%%%%%%%%%%%%%%%%%%%%%%%%%%

%%%%%%%%%%%%%%%%%%%%%%%%%%%%%%%%%%%%%%%%%%%%%%%%%%%%%%%%%%%%%%%%
\begin{figure}[tb]
\centerline{\includegraphics[width=\linewidth]{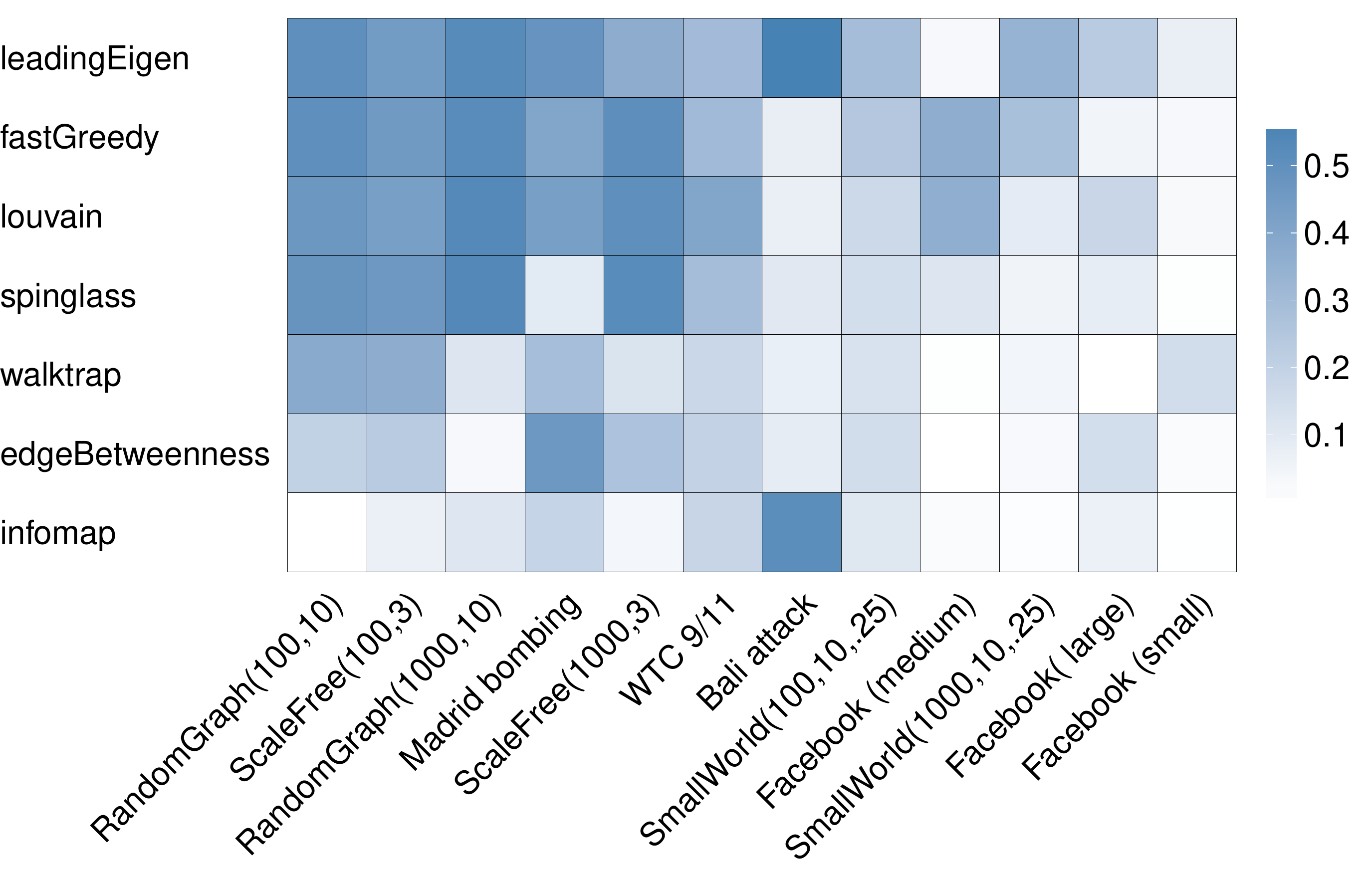}}
\vspace*{-0.06cm}
\caption{Avg. concealment-measure value in each experiment.}
\label{fig:concealment}
\end{figure}
%%%%%%%%%%%%%%%%%%%%%%%%%%%%%%%%%%%%%%%%%%%%%%%%%%%%%%%%%%%%%%%

\subsection{Experimenting with DICE}
For each network, we experiment with seven community-detection algorithms implemented in the \textit{igraph}
%[uncomment if given more space]~\cite{igraph}
package of the R language (version $1.0.1$), namely: Eigenvector \cite{newman2006finding}, Betweenness \cite{newman2004finding}, Walktrap \cite{pons2005computing}, Louvain \cite{blondel2008fast}, Greedy \cite{clauset2004finding}, Infomap \cite{rosvall2010map}  and Spinglass \cite{reichardt2006statistical}. As such, every experiment consists of a community-detection algorithm and a network. The experiment starts by running the algorithm to obtain a community structure, $\CS$. After that, the community to be hidden, i.e., $\CC$, is chosen to be the element in $\CS$ whose size is the median of the sizes of all communities in $\CS$ (ties are broken uniformly at random). Although $\CC$ does not necessary have to be an element of $\CS$, we choose it this way in order to study the worst case scenario in which $\CC$ is initially exposed completely. The experiment then proceeds in rounds, each involving the execution of DICE followed by the execution of the community-detection algorithm, to measure how well $\CC$ is hidden in the new outcome of the algorithm (this measurement is done using $\mc$ with $\alpha=0.5$). We set the number of rounds to be $\ceil{|\CC|/ b}$. In each round, we disconnect $d$ links from within $\CC$ (chosen uniformly at random), and then connect $b-d$ members of $\CC$ to $b-d$ non-members of $\CC$ (again chosen uniformly at random). Due to this randomness in our implementation, DICE may yield different results in different executions. Therefore, we repeat each experiment multiple times, and report the 95\% confidence interval.

Figure~\ref{fig:community-figures} shows the results of some of our experiments (for the remaining results see the Supporting Information). As can be seen, DICE is able to hide the community, $\CC$ with varying levels of success, depending on the community-detection algorithm being used. Importantly, the performance does not appear to be overly-sensitive to the parameter $d$. This is important because it provides the members of $\CC$ with the ability to control this parameter as needed (i.e., control the trade-off between the number of internal edges being removed, and the number of external edges being added). For example, the members of $\CC$ might be interested in hiding their community as much as possible, while removing as few internal links as possible (after all, the added external links are fake, serving no purpose other than disguising the community, whereas the removed internal links are real; they existed in the community for a reason). In such a case, since the addition of an external link is not entirely under the control of $\CC$ (as it requires the consent of a non-member), the number of newly-added external links may be insufficient for providing a satisfactory level of concealment, in which case the members can compensate for this by sacrificing more internal links, i.e., by increasing the parameter $d$.

Figure~\ref{fig:concealment} illustrates the average value of our concealment measure, $\mc$, in each experiment where $b=4$ and $d=2$. In particular, each row represents a community-detection algorithm, each row represents a network, and the intensity of the colour in each cell represents the average value of $\mc$, taken over $50$ simulations, either by generating a new random network in each simulation, or by re-running the simulation over and over on the same real-life network (recall that our implementation of DICE is non-deterministic, and may yield different results on the same network). As can be seen, the Infomap~\cite{rosvall2010map} algorithm seems to be the most difficult to fool.
%[uncomment if given more space] (see how the row representing Infomap has, on average, the lightest colour shades).

%%%%%%%%%%%%%%%%%%%%%%%%%%%%%%%%%%%%%%%%%%%%%%%%%%%%%%%%%%%%%%%%%%%%%%%%
\section{Discussion}
Our goal was to understand the practical limits of disguising individuals and communities, to increase the likelihood of them being overlooked by social network analysis tools. Our main result is that, despite the hardness of finding an optimal solution, disguise is surprisingly easy in practice using simple heuristics that are readily-implementable even by lay people. Viewed from a different perspective, our work can be seen as an extension of the sensitivity analyses of centrality measures \cite{Correa:et:al:2012} and community detection algorithms \cite{Orman:Labatut:2009}; while such analyses typically consider the effects of small network alterations, we consider changes that are much wider in scope, and  strategic in nature.

On one hand, our findings contribute towards charting the limits of protecting privacy in social networks. On the other hand, they expose implications for using generic social network analysis tools in security applications; the fact that such tools can be easily misled underlines the need for developing specialized tools that account for the nature of links and nodes in the network and not just the topology \textit{per se}.

% \cite{Chakraborty:et:al:2013}

%In our simulations, apart from synthetic networks we used real-life  datasets to test our heuristics.

%In this final section, we review our results and their significance relative to previous work in social network theory. 
%We then compare ...., discuss potential criticisms of the model, and finish with some intriguing connections between .....

%Second, our work is limited by our
Despite these findings, our understanding of how to evade social network analysis tools is still limited, with many research questions yet to be answered. For instance, we still do not know
%how to handle setting where communities may overlap \cite{xie2013overlapping}
how a relationship can be hidden from the eyes of link-prediction algorithms \cite{Lu:Zhou:2011}, or how an individual can evade detection by Eigenvector centrality---the backbone of Google's search engine.
%[uncomment if given more space] \cite{ilprints422}. 

{\fontsize{8}{8}\selectfont{
\bibliographystyle{abbrv}
\bibliography{bibliography_final,bibliography_supp}
}}

%%%%%%%%%%%%%%%%%%%%%%%%%%%%%%%%%%%%%%%%%%%%%%%%%%%%%%%%%%%%%%%%%%%%%%%%

\appendix

\onecolumn

\section{Organization of the Appendix}

In this document, we formally define the relevant centrality measures and influence models, before defining our optimization problems (Section~\ref{sec:preliminaries}). After that, we present the proofs of our theoretical results (Section~\ref{sec:proofs}), followed by a discussion of various experimental results (Section~\ref{sec:experiments}). Finally, we study the problem of constructing a network from scratch, designed for the sole purpose of concealing the identity of the leader while ensuring that it is a highly influential node in the network (Section~\ref{sec:lieutenant}).

%%%%%%%%%%%%%%%%%%%%%%%%%%%%%%%%%%%%%%%%%%%%%%%%%%%%%%%%%%%%%%%%%%%%%%%%%%%%%%%

\section{Definitions}\label{sec:preliminaries}
\noindent\textbf{Basic Notation:} Let $G = (V, E) \in \mathbb{G}$ denote a network, where $V=\{v_1,\ldots,v_n\}$ is the set of $n$ nodes and $E \subseteq V \times V$ is the set of edges. A path is a sequence of distinct nodes, $\langle v_l, \ldots, v_k\rangle$, such that every two consecutive nodes are connected by an edge. The length of a path is considered to be the number of edges in that path. For any pair of nodes, $v_i,v_j$ in $G$, the set of all shortest paths between them is denoted by $sp_G(v_i,v_j)$, and the distance between them is denoted by $d_G(v_i,v_j)$, where distance is defined as the length of a shortest path between the two. In case of an \textit{undirected} network $G$ we do not discern between edges $(v_i,v_j)$ and $(v_j,v_i)$; otherwise the network is said to be \textit{directed}. Furthermore, $G$ is said to be \textit{connected} (\textit{strongly connected} for directed networks) if there exists a path between every pair of nodes in $G$.

We denote by $\P_G(v_i)$ the set of \emph{predecessors} of $v_i$ in $G$, that is, $\P_G(v_i) = \{v_j \in V : (v_j,v_i) \in E\}$. On the other hand, we denote by $\S_G(v_i)$ the set of \emph{successors} of $v_i$ in $G$, i.e., $\S_G(v_i) = \{v_j \in V : (v_i,v_j) \in E\}$. Finally, we denote by $N_G(v_i)$ the set of neighbours of $v_i$ in $G$, i.e., $N_G(v_i) = \P_G(v_i) \cup \S_G(v_i)$. For the case of undirected graph, we will assume that $N_G(v_i) = \P_G(v_i) = \S_G(v_i)$.

To make the notation more readable, we will often denote two arbitrary nodes by $v$ and $w$, instead of $v_i$ and $v_j$. Moreover, we will often omit the network itself from the notation whenever it is clear from the context, e.g., by	 writing $d(v,w)$ instead of $d_G(v,w)$; this applies not only to the notation presented thus far, but to all notation.

We consider a \textit{community structure}, $\CS = \{C_1, \ldots, C_k\}$, to be a partition of the set of nodes into disjoint and exhaustive subsets, or communities.\footnote{\footnotesize Some works have considered overlapping community structures~\cite{xie2013overlapping}. However, as common in the literature, we restrict our attention to disjoint communities.} Formally, it satisfies the following three conditions: $\forall_{C_i \in \CS} C_i \subseteq V$, $\bigcup_{C_i \in \CS}C_i = V$, and $\forall_{C_i,C_j \in \CS} C_i \cap C_j = \emptyset$.

\ \\\noindent\textbf{Centrality Measures:} Formally, a centrality measure \cite{freeman1979centrality} is a function $c : \mathbb{G} \times V \rightarrow \R$. The \textit{degree} centrality \cite{shaw1954group} is denoted by $c_{degr}$, the \textit{closeness} centrality \cite{beauchamp1965improved} is denoted by $c_{clos}$, and the \textit{betweeness} centrality \cite{anthonisse1971rush,freeman1977set} is denoted by $c_{betw}$. Specifically, given a node $v_i \in V$ and an undirected network, we have:
$$
	c_{degr}(G,v_i) = \frac{|N_G(v_i)|}{n - 1}
$$

$$
	c_{clos}(G,v_i) = \frac{n - 1}{\sum_{v_j \in V}d_G(v_i,v_j)}
$$

$$
	c_{betw}(G,v_i) = \frac{2}{(n-1)(n-2)}
		\sum_{v_j,v_k \in V \setminus \{ v_i \}}
			\frac
				{|\{ p \in sp_G(v_j,v_k) : v_i \in p \}|}
				{|sp_G(v_j,v_k)|}
$$
On the other hand, given a directed network, we have:
$$
	c_{degr}(G,v_i) = \frac{|N_G(v_i)|}{2(n - 1)}
$$

$$
	c_{clos}(G,v_i) = \frac{1}{n - 1}\sum_{v_j \in V}\frac{1}{d_G(v_i,v_j)}
$$

$$
	c_{betw}(G,v_i) = \frac{1}{(n-1)(n-2)}
		\sum_{v_j,v_k \in V \setminus \{ v_i \}}
			\frac
				{|\{ p \in sp_G(v_j,v_k) : v_i \in p \}|}
				{|sp_G(v_j,v_k)|}
			+ \frac
				{|\{ p \in sp_G(v_k,v_j) : v_i \in p \}|}
				{|sp_G(v_k,v_j)|}
$$

\ \\\noindent\textbf{Models of Influence:} The propagation of influence through the network is often modeled as follows: when a certain node is sufficiently influenced by its neighbour(s), it becomes ``active'', in which case it starts to influence any ``inactive'' neighbour(s) it may have, and so on. Of course, to initiate this propagation process, a set of nodes needs to be activated right from the start; this set is called the \emph{seed set}. Assuming that time moves in discrete rounds, we denote by $I(t)\subseteq V$ the set of nodes that are active at round $t$, implying that $I(1)$ is the seed set. The way influence propagates from the seed set to the remaining nodes depends on the influence model under consideration. Here, the two main models of influence are:
\begin{itemize}
\item\emph{Independent Cascade} \cite{goldenberg2001using}: In this model, every pair of nodes is assigned an activation probability, $p : V \times V \rightarrow [0,1]$. Then, in every round, $t>1$, every node $v\in V$ that became active in round $t-1$ activates every inactive successor, $w\in \S(v)\setminus I(t-1)$, with probability $p(v,w)$. The process ends when there are no new active nodes, i.e., when $I(t)=I(t-1)$. %Under this model, for evey pair of nodes, $v,w\in V$, the influence of $v$ over $w$ will be denoted by $\inf_{IC}(v,w)$.

\item\emph{Linear Threshold} \cite{kempe2003maximizing}: In this model, every node $v \in V$ is assigned a \textit{threshold value} $t_v$ which is sampled (according to some probability distribution) from the set $\{0, \ldots,|\P(v)|\}$. Then, in every round, $t>1$, every inactive node $v$ becomes active, i.e., becomes a member of $I(t)$, if: $|I(t-1) \cap \P(v)| \geq t_{v}$. The process ends when $I(t)=I(t-1)$. %The influence of a node, $v$, over another, $w$, under this model will be denoted by $\inf_{LT}(v,w)$.
\end{itemize}
In either model, the influence of a node, $v$, on another, $w$, is denoted by $\inf_G(v,w)$ and defined as \emph{the probability that $w$ gets activated given the seed set $\{v\}$} (we make the common assumption that $\inf_G(v,v) = 0$ for all $v\in V$). The influence of $v$ over the entire network $G$ is then: $\inf_G(v) = \sum_{w \in V}\inf_G(v,w)$. 
%
%%%%%%%%%%%%%%%%%%%%%%%%%%%%%%%%%%%%%%%%%%%%%%%%%%%%%%%%%%%%%%%%%%%%%%%%%%%%%%%%
\ \\\\
\noindent\textbf{First Objective (Disguising a Node):}
Roughly speaking, given a \emph{source node}, $v^{\dagger}$, and a limited budget, $b$, specifying the maximum number of edges that are allowed to be added or removed, our goal is to first rewire the network so as to minimize the centrality of $v^{\dagger}$, and then to further rewire the network so as to ``recover'' the influence of $v^{\dagger}$ (in an attempt to compensate for any influence that $v^{\dagger}$ might have lost during the centrality-minimization phase). We consider two variants of the influence-recovery problem; the first focuses on the influence of $v^{\dagger}$ over every single node, whereas the second focuses on the influence of $v^{\dagger}$ over the network as a whole. In both cases, only the addition of edges is considered, since the removal of edges can only decrease the influence of $v^{\dagger}$. Next, we formally define the aforementioned problems.

\begin{definition}[Disguising Centrality]
This problem is defined by a tuple, $(G,v^{\dagger},b,c,\FR,\FA)$, where $G=(V,E)\in\mathbb{G}$ is a network, $v^{\dagger}\in V$ is the source node (whose centrality is to be minimized), $b\in\N$ is a budget specifying the maximum number of edges that can be added or removed, $c : \mathbb{G} \times V \rightarrow \R$ is a centrality measure, $\FR\subseteq E$ is a set of edges whose removal is forbidden, $\FA\subseteq (V\times V)\setminus E$ is a set of edges whose addition is forbidden. The goal is then to identify two sets of edges, $R^*\subseteq (E\setminus\FR)$ and $A^*\subseteq (V\times V)\setminus(E\cup \FA)$, such that: $|A^*|+|R^*| \leq b$ and $G^* = \big(V, (E \cup A^*)\setminus R^*\big)$ is connected (strongly connected if $G$ is directed) and $G^*$ is in:
$$
\argmin_{G'\in\left\{\big(V,(E \cup A)\setminus R\big) : R\subseteq (E\setminus\FR), A\subseteq (V\times V)\setminus(E\cup \FA)\right\}} c\big(G',v^{\dagger}\big).
$$
\end{definition}

\begin{definition}[Individual Influence recovery]
This problem is defined by a tuple, $(G,v^{\dagger},\inf,\FA,f)$, where $G=(V,E)\in\mathbb{G}$ is a network, $v^{\dagger}\in V$ is the source node (whose influence is to be recovered), $\inf: V \times V \rightarrow \R$ is an influence measure, $\FA\subseteq (V\times V)\setminus E$ is a set of edges whose addition is forbidden, and $f: V \rightarrow \R$ specifies the influences to be recovered (i.e., for every $v_i\in V$ we want the influence of $v^{\dagger}$ over $v_i$ to be at least $f(v_i)$). The goal is then to identify a set of edges, $A^*$, that is in:
$$
\argmin_{A\subseteq (V\times V)\setminus(E\cup \FA):\forall_{v_i \in V} \inf_{(V, E \cup A)}(v^{\dagger},v_i) \geq f(v_i)} |A|. 
$$
\end{definition}

\begin{definition}[Global Influence recovery]
This problem is defined by a tuple, $(G,v^{\dagger},\inf,\FA,\globalReb)$, where $G=(V,E)\in\mathbb{G}$ is a network, $v^{\dagger}\in V$ is the source node (whose influence is to be recovered), $\inf: V \times V \rightarrow \R$ is an influence measure, $\FA\subseteq (V\times V)\setminus E$ is a set of edges whose addition is forbidden, and $\globalReb\in\R$ is the total influence to be recovered. The goal is then to identify a set of edges, $A^*$, that is in:
$$
\argmin_{A\subseteq (V\times V)\setminus(E\cup \FA):\inf_{(V, E \cup A)}(v^{\dagger}) \geq \globalReb} |A|. 
$$
\end{definition}

%%%%%%%%%%%%%%%%%%%%%%%%%%%%%%%%%%%%%%%%%%%%%%%%%%%%%%%%%%%%%%%%%%%%%%%%%%%%%%%%
\noindent\textbf{Second Objectives (Disguising a Community):}
Roughly speaking, given a community to be hidden, $\CC$, and a limited budget, $b$, specifying the maximum number of edges that are allowed to be added or removed, our goal is to rewire the network so as to hide $\CC$. To this end, we propose a measure of concealment, $\mc$, defined for every community $\CC\subseteq V$ and every community structure $\CS$, as follows:\footnote{\footnotesize Note that $\CC$ is not necessarily a member of $\CS$. To put it differently, when describing $\CC$ as a ``community'' we use this term in its broader sense, where $\CC$ is essentially just a subset of nodes. As such, when measuring how well $\CC$ is hidden in $\CS$, it may well be the case that the members of $\CC$ are spread out across multiple communities in $\CS$.}
$$
\mc(\CC,\CS) = \alpha \mc'(\CC,\CS) + (1 - \alpha) \mc''(\CC,\CS),
$$
\noindent where $\alpha \in [0,1]$ and:
$$
\mc'(\CC,\CS) = \frac{|\{ C_i \in \CS : C_i \cap \CC \neq \emptyset\}|-1}{\max(|\CS|-1,1) \max_{C_i \in \CS}(|C_i \cap \CC|)}
$$
$$
\mc''(\CC,\CS) = \sum_{C_i \in \CS} \frac{|C_i \setminus \CC|}{\max(n-|\CC|,1)}.
$$
Note that $\mc(\CC,\CS) \in [0,1]$ for all $\CC$ and $\CS$, with greater values indicating greater levels of concealment of $\CC$ in $\CS$. Having presented our concealment measure, we are now ready to formally introduce our problem.

\begin{definition}[Disguising a Community]
This problem is defined by a tuple, $(G,\CC,\alg,b)$, where $G=(V,E)$ is a network, $\CC\subseteq V$ is the community to be hidden, $\alg$ is a community-detection algorithm, and $b \in \N$ is a budget specifying the maximum number of edges that can be added or removed. The goal is then to find a set of edges to be added, $A^* \subseteq (V \times V) \setminus E$, and another to be removed, $R^* \subseteq E$,  such that $|A^*| + |R^*| \leq b$ and $G^* = \big(V, (E \cup A^*)\setminus R^*\big)$ is in:
\begin{equation}
\argmax_{G'\in\left\{\substack{\big(V,(E \cup A)\setminus R\big)\ :\ A\subseteq (V \times V)\setminus E,\ R\subseteq E,\ |A| + |R| \leq b}\right\}}\mc(\CC,\alg(G')),\nonumber
\end{equation}
\noindent where $\alg(G)$ is the community structure returned by the algorithm $\alg$ given the network $G$.
\end{definition}

Note that the above optimization problem requires $\CC$ to know the exact community-detection algorithm that the adversary is using. Since such knowledge is hardly available, we avoid this requirement, and instead aim to develop a general-purpose heuristic, designed for no particular community-detection algorithm.

%%%%%%%%%%%%%%%%%%%%%%%%%%%%%%%%%%%%%%%%%%%%%%%%%%%%%%%%%%%%%%%%%%%%%%%%%%%%%%%

\section{Proofs}\label{sec:proofs}
\noindent  From the computational point of view, disguising the degree centrality of $v^{\dagger}$ is easy, since the only way to decrease this centrality is to remove edges connecting $v^{\dagger}$ to its neighbour(s). Next, we study the problems of disguising closeness centrality and betweenness centrality, followed by the problem of influence recovery under the Independent-Cascade model and under the Linear-Threshold model.

%\begin{lemma}
%Let $G=(V,E)$ be a network, and let $G'=(V,E\cup\{(v_i,v_j)\})$ for some $(v_i,v_j)\in (V\times V)\setminus E$. Then, the following holds: $\forall\ v_k \in V, c_{clos}(v_k,G) \leq  c_{clos}(v_k,G')$.
%\end{lemma}

\begin{theorem}\label{theorem:ClosenessHardness}
Disguising closeness centrality is NP-complete.
\end{theorem}

\noindent\textbf{Proof.}
The decision version of the optimization problem is the following: given a network $G=(V,E)$, a source node $v^{\dagger}$, two sets $\FR\subseteq E$, $\FA\subseteq (V\times V)\setminus E$, a budget $b \in \N$ and a value $x \in \R$, does there exist two sets $R^*\subseteq (E\setminus\FR)$ and $A^*\subseteq (V\times V)\setminus(E\cup \FA)$ such that $|A^*|+|R^*| \leq b$, and the network $(V, (E \cup A^*)\setminus R^*)$ is connected (strongly connected if $G$ is directed) and $c_{clos}((V, (E \cup A^*)\setminus R^*), v^{\dagger}) \leq x$?

This problem is in NP, as given a solution, i.e., two sets $A^*$ and $R^*$, we can verify whether $c_{clos}((V, (E \cup A^*)\setminus R^*), v^{\dagger}) \leq x$ in polynomial time; this only requires computing the closeness centrality of node $v^{\dagger}$ in network $(V, (E \cup A^*)\setminus R^*)$.

We will now show that the decision version is NP-hard. To this end, let us denote by $q\in \R$ the smallest possible closeness centrality of $v^{\dagger}$ in any (strongly) connected network whose set of nodes is $V$. One can see that $q=2/n$ in the case of undirected networks, and $q=(\sum_{i=1}^{n-1}\frac{1}{i})/(n - 1)$ in the case of directed network; this happens if and only if:
\begin{itemize}
\item the network is a path of which $v^{\dagger}$ is an end (when dealing with undirected networks); or
\item the network is a directed cycle (when dealing with directed networks).
\end{itemize}
Let us denote such a network by $Q$; the closeness centrality of $v^{\dagger}$ in $Q$ is then $q$. With this in mind, the proof involves a reduction from the \emph{Hamiltonian cycle} problem (i.e., the problem of determining whether there exists a cycle that visits each node exactly once) to the decision problem of determining whether it is possible to reduce the closeness centrality of $v^{\dagger}$ to a value smaller than, or equal to, $q$.

\begin{figure}[tbhp]
\begin{minipage}{\textwidth}
\centering
\includegraphics[width=0.9\linewidth]{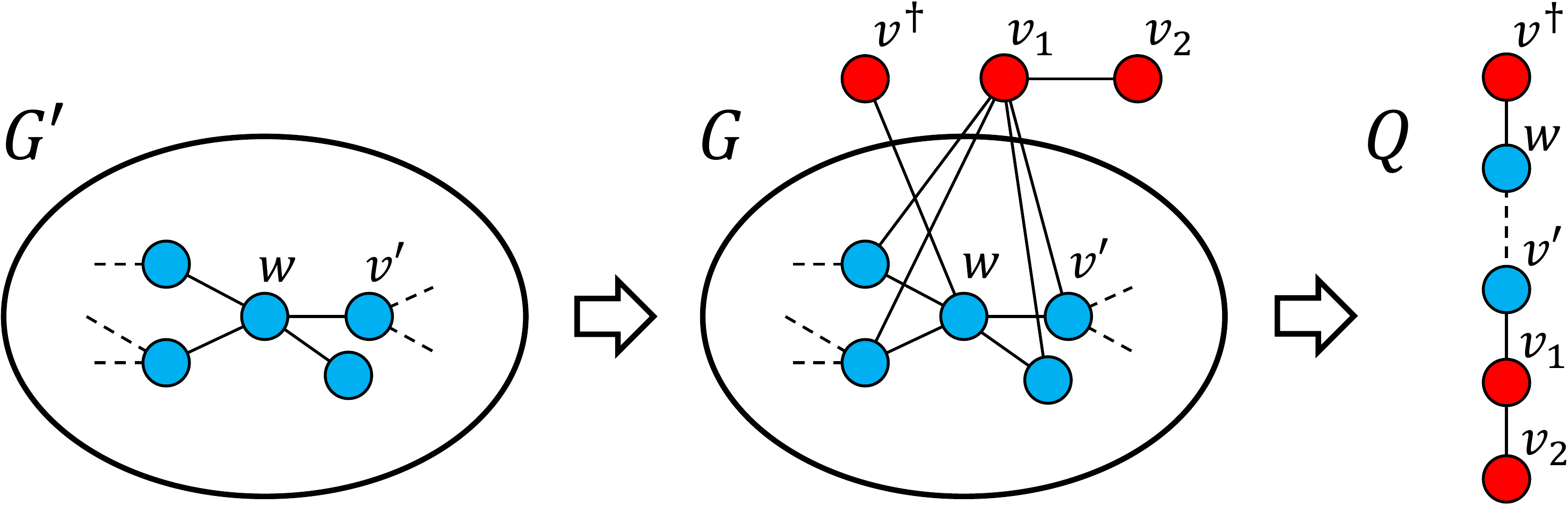}
\caption{\fontsize{10}{10}\selectfont{The main steps of reducing the Hamiltonian cycle problem to the problem of determining whether the closeness centrality of $v^{\dagger}$ can be reduced to a value $\leq q$ in an \textbf{undirected network}.}}
\label{fig:clos_nphard_undir}
\end{minipage}
\hfill\ \\\\
\begin{minipage}{\textwidth}
\centering
\includegraphics[width=0.9\linewidth]{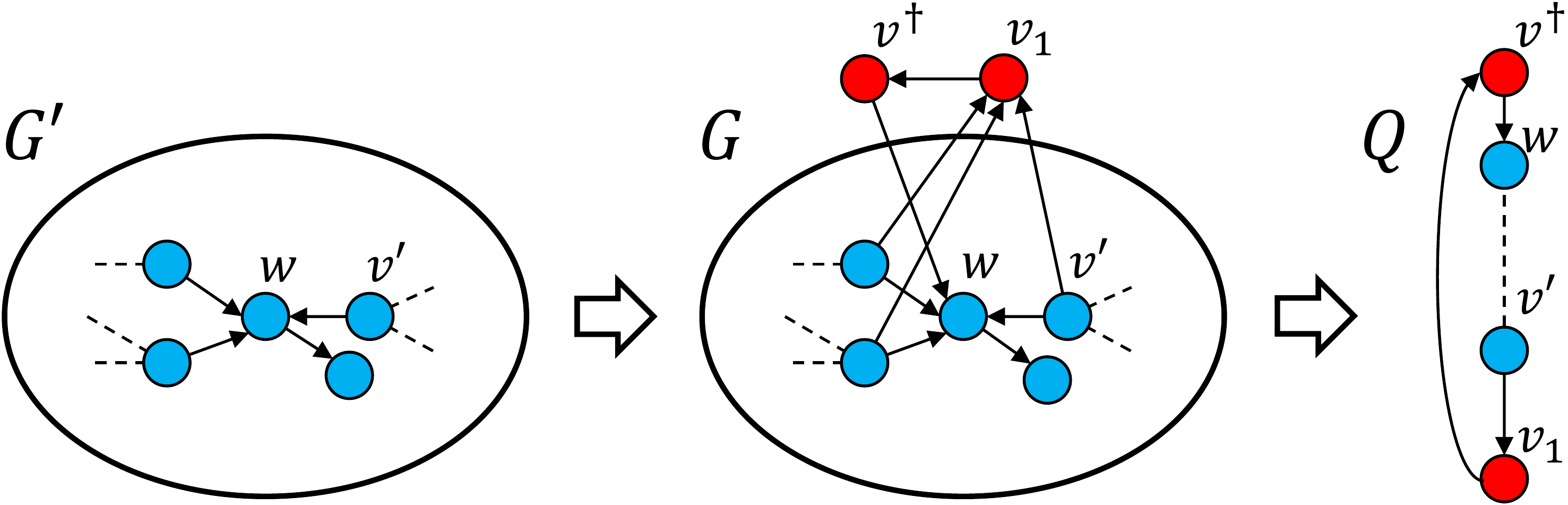}
\caption{\fontsize{10}{10}\selectfont{The main steps of reducing the Hamiltonian cycle problem to the problem of determining whether the closeness centrality of $v^{\dagger}$ can be reduced to a value $\leq q$ in a \textbf{directed network}.}}
\label{fig:clos_nphard_dir}
\end{minipage}
\end{figure}

To this end, given some arbitrary network, $G'=(V',E')$, be it directed or undirected, let us modify $G'$ so as to obtain a new network, $G = (V,E)$, as illustrated in figures \ref{fig:clos_nphard_undir} and \ref{fig:clos_nphard_dir}. Formally, we do so by choosing some arbitrary node, $w \in V'$, and then setting:
\[
V = V' \cup \{v^{\dagger},v_1,v_2\},\ \ \ \ \ 
E = E' \cup \{(v^{\dagger},w),(v_1,v_2)\} \cup \{(v,v_1):v\in V', v \in N_{G'}(w)\},
\]
\noindent in the case of undirected networks, or setting:
\[
V = V' \cup \{v^{\dagger},v_1\},\ \ \ \ \ 
E = E' \cup \{(v^{\dagger},w),(v_1,v^{\dagger})\} \cup \{(v,v_1):v\in V', v \in \P_{G'}(w)\},
\]
\noindent in the case of directed networks.

We will now show that the Hamiltonian cycle problem in $G'$ is equivalent to the following decision problem: Given network $G$ and budget $b = |E'|-|V'|+|\P_{G'}(w)|$, where $\FA = \FR = \emptyset$, determine whether it is possible to reduce the closeness centrality of $v^{\dagger}$ to a value $\leq q$, by removing at most $b$ edges from $G$. Throughout the remainder of the proof, the edges and nodes in $G$ that were in $G'$ will be referred to as ``\textit{original}''. 

%%%%%%%%%%%%%%%%%%%%%%%%%%%%%%%%%%%%%%%%%%%%%%%%%%%%%%%%%%%%%%%%%%%%%%%%%%%%%%%
%%%%%%%%%%%%%%%%%%%%%%%%%%%%%%%%%%%%%%%%%%%%%%%%%%%%%%%%%%%%%%%%%%%%%%%%%%%%%%%

Firstly, we will show that if $G'$ has a Hamiltonian cycle then it is possible to obtain $Q$ by removing $|E'|-|V'|+|\P_{G'}(w)|$ edges from $G$. To this end, fix a Hamiltonian cycle of $G'$, then:
\begin{itemize}
\item remove from $G$ all original edges that are not in the Hamiltonian cycle; there are exactly $|E'| - |V'|$ such edges;
\item in the Hamiltonian cycle, there are exactly two edges of which $w$ is an end; remove any of those edges in the undirected network, or the one pointing to $w$ in the directed network; let us denote the removed edge as $(v',w)$;
\item remove all edges from all predecessors of $w$ to $v_1$, with the exception of $(v', v_1)$; there are exactly $|\P_{G'}(w)| - 1$ such edges.
\end{itemize}
In so doing, we have obtained the network $Q$ by removing a total of $|E'|-|V'|+|\P_{G'}(w)|$ edges from $G$ (see figures \ref{fig:clos_nphard_undir} and \ref{fig:clos_nphard_dir}).

Secondly, we show that if it is possible to obtain $Q$ by removing $|E'|-|V'|+|\P_{G'}(w)|$ edges from $G$, then there exists a Hamiltonian cycle in $G'$. We will first deal with the undirected case, before moving on to the directed case.

In the undirected case, observe that nodes $v^{\dagger}$ and $v_2$ each have a degree of 1 in $G$, since their only neighbours are $w$ and $v_1$, respectively. Now since $Q$ is connected, and since we obtained $Q$ by only removing (rather than adding) edges from $G$, the nodes $v^{\dagger}$ and $v_2$ must each have a degree of 1 in $Q$. Consequently, they must be the two ends of $Q$. This, in turn, implies that $v_1$ must have exactly two neighbours in $Q$; we know that one of them is $v_2$, let us call the other $v'$. This, as well as the fact that $v^{\dagger}$ is only connected to $w$, implies that the segment of $Q$ between $w$ and $v'$ contains all original nodes from $G'$ and only original edges from $G'$ (recall that we did not add any edges between original nodes). Finally, by adding to that segment the original edge between $v'$ and $w$, we obtain a Hamiltonian cycle in $G'$.

As for the directed case, we observe that node $v^{\dagger}$ has only one successor in $Q$, namely $w$, and only one predecessor in $Q$, namely $v_1$. We also know that $v_1$ has only one predecessor in $Q$; let us call that predecessor $v'$. These facts imply that the segment of $Q$ between $w$ and $v'$ contains all original nodes from $G'$ and only original edges from $G'$ (again, recall that we did not add any edges between original nodes). By adding to that segment the original edge between $v'$ and $w$, we obtain a Hamiltonian cycle in $G'$.

We have shown that a Hamiltonian cycle in $G'$ exists if and only if it is possible to reduce the closeness centrality of $v^{\dagger}$ to $q$ by removing exactly $|E'|-|V'|+|\P_{G'}(w)|$ edges from $G$, which concludes the proof. \hspace*{\fill}$\Box$\\

\begin{theorem}\label{theorem:BetweennessHardness}
Disguising betweenness centrality is NP-complete.
\end{theorem}

\noindent\textbf{Proof.}
The decision version of the optimization problem is the following: given a network $G=(V,E)$, a	source node $v^{\dagger}$, two sets $\FR\subseteq E$, $\FA\subseteq (V\times V)\setminus E$, a budget $b \in \N$ and a value $x \in \R$, does there exist two sets $R^*\subseteq (E\setminus\FR)$ and $A^*\subseteq (V\times V)\setminus(E\cup \FA)$ such that $|A^*|+|R^*| \leq b$, and the network $(V, (E \cup A^*)\setminus R^*)$ is connected (strongly connected if $G$ is directed) and $c_{betw}((V, (E \cup A^*)\setminus R^*), v^{\dagger}) \leq x$?

This problem is in NP, as given a solution, i.e., two sets $A^*$ and $R^*$, we can verify whether $c_{betw}((V, (E \cup A^*)\setminus R^*), v^{\dagger}) \leq x$ in polynomial time; this only requires computing the betweenness centrality of node $v^{\dagger}$ in network $(V, (E \cup A^*)\setminus R^*)$.

We will now show that the decision version is NP-hard. To this end, we propose a reduction from the NP-complete \textit{Set cover} problem. The decision version of this problem is defined by a universe $U=\{u_1, \ldots, u_l\}$ and a collection of sets $S = \{S_1, \ldots, S_m\}$ such that $\forall_j{S_j \subset U}$, where the goal is to determine whether there exist $k\leq m$ elements of $S$ the union of which equals $U$.

%%%%%%%%%%%%%%%%%%%%%%%%%%%%%%%%%%%%%%%%%%%%%%%%%%%%%%%%%%%%%%%%%%%%%%%%%%%%%%%
\begin{figure}[tbhp]
\begin{minipage}{0.47\textwidth}
\centering
\includegraphics[width=\linewidth]{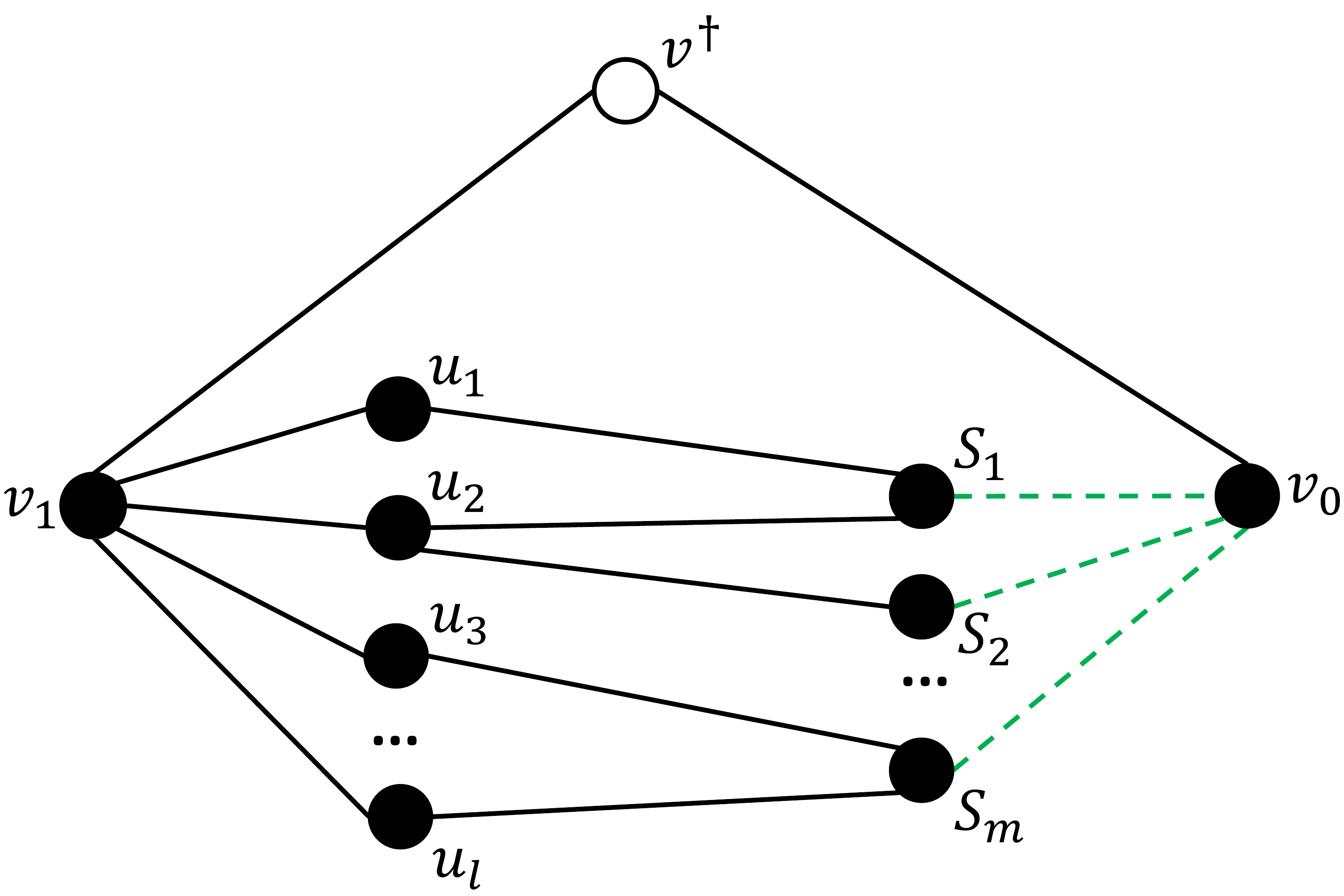}
\caption{\fontsize{10}{10}\selectfont{Undirected network used to reduce the \textit{Set cover} problem to our problem of disguising the betweenness centrality of $v^{\dagger}$. To solve both problems, we consider adding (some of) the dashed edges.}}
\label{fig:betw_nphard_undir}
\end{minipage}
\hfill
\begin{minipage}{0.47\textwidth}
\centering
\includegraphics[width=\linewidth]{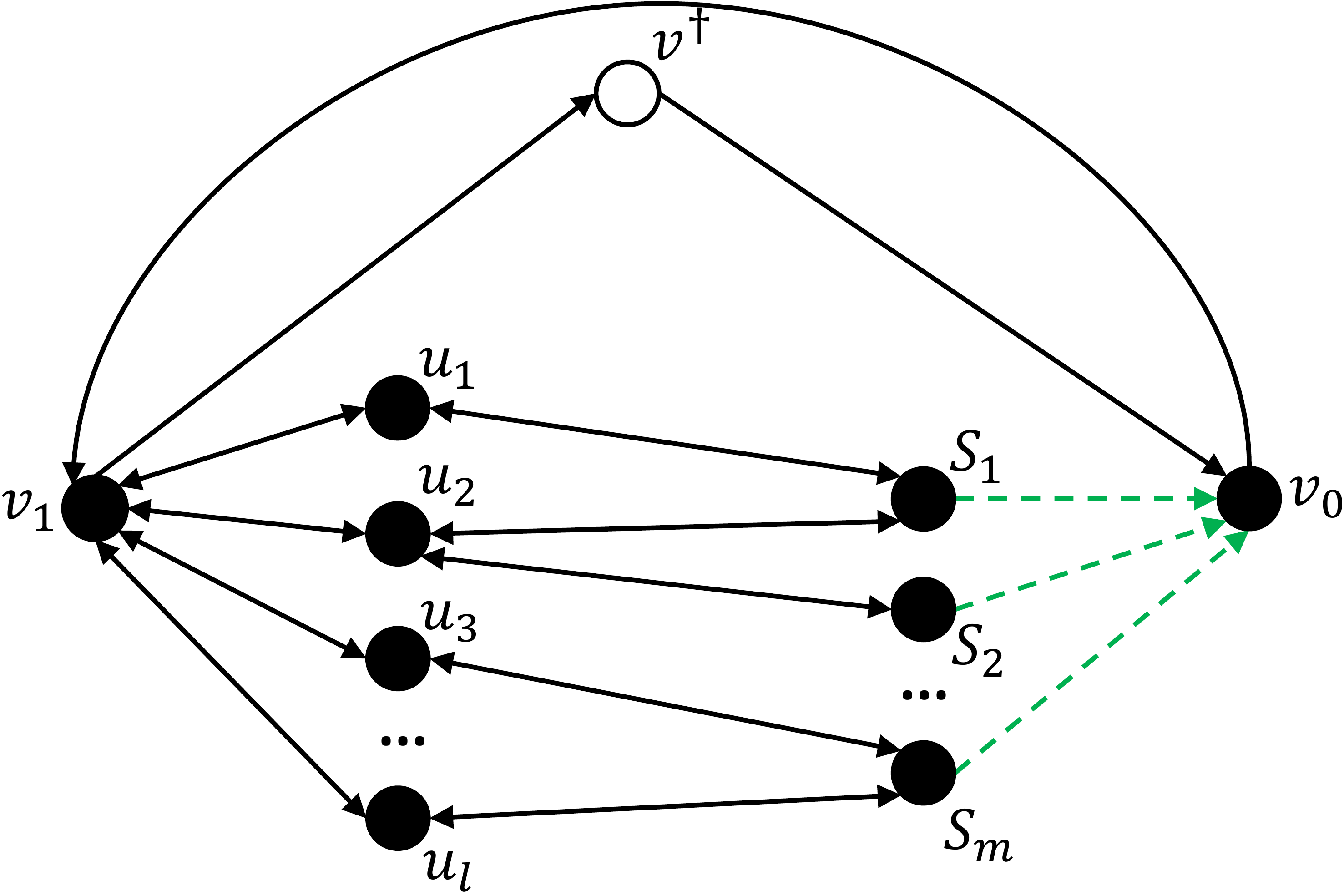}
\caption{\fontsize{10}{10}\selectfont{A directed network used to reduce the \textit{Set cover} problem to our problem of disguising the betweenness centrality of $v^{\dagger}$. To solve both problems, we consider adding (some of) the dashed edges.}}
\label{fig:betw_nphard_dir}
\end{minipage}
\end{figure}
%%%%%%%%%%%%%%%%%%%%%%%%%%%%%%%%%%%%%%%%%%%%%%%%%%%%%%%%%%%%%%%%%%%%%%%%%%%%%%%

First, let us create a network $G$ as shown in figures \ref{fig:betw_nphard_undir} and \ref{fig:betw_nphard_dir}. More specifically, we create one node for every $S_j \in S$, one node for every $u_i \in U$, and three additional nodes, $v^{\dagger}$, $v_0$ and $v_1$. 
Next, we add (either undirected or directed) edges as follows. We add the edges $(v^{\dagger},v_0)$ and $(v_1,v^{\dagger})$, and for every node $u_i \in S_j$ we add the edges $(S_j,u_i)$ and $(u_i,v_1)$. In case of a directed network, we also add the edges $(u_i,S_j)$ and $(v_1,u_i)$ for every $u_i \in S_j$, as well as the edge $(v_0,v_1)$.

Now, consider the problem of disguising the betweenness centrality of $v^{\dagger}$ in $G$ given $\FR = E$ and $\FA = (V \times V) \setminus \{(S_1,v_0), \ldots, (S_m,v_0)\}$. Note that $v^{\dagger}$ ``controls'' (i.e., lies on) every shortest path to $v_0$, and does not control any shortest path between any other pair of nodes. As such, to minimize the betweenness centrality of $v^{\dagger}$, we need to create alternative shortest paths to $v_0$; this should be done by adding (some of) the edges in $\{(S_1,v_0), \ldots, (S_m,v_0)\}$, since no other edge can be added, and no edge can be removed (following the definitions of $\FR$ and $\FA$). To be more precise, we can add at most $b$ edges $\{(S_1,v_0), \ldots, (S_m,v_0)\}$, since we cannot exceed the budget. After this process, the betweenness centrality of $v^{\dagger}$ may drop to as little as $q=\frac{2}{(n-1)(n-2)}$ in the undirected case, or as little as $q=\frac{1}{(n-1)(n-2)}$ in the directed case; this happens when $v^{\dagger}$ no longer controls any of the shortest paths to $v_0$ except for the one from $v_1$ to $v_0$. Note that adding an edge $(S_j,v_0)$ creates a new shortest path from every nodes $u_i \in S_j$ to $v_0$. This implies that the betweenness centrality of $v^{\dagger}$ can be reduced to $q$ if and only if there exists at most $b$ elements of $S$ the union of which equals $U$.

We have just reduced the decision version of the Set Cover problem given $k$ to the following decision problem: Given network $G$ and budget $b = k$, where $\FR = E$ and $\FA = (V \times V) \setminus \{(S_1,v_0), \ldots, (S_m,v_0)\}$, determine whether it is possible to reduce the closeness centrality of $v^{\dagger}$ to some value $\leq q$, by removing at most $b$ edges from $G$. \hspace*{\fill}$\Box$\\
%
%With the forbidden sets as described above, it is only possible to add $k$ edges between nodes $S_1, \ldots, S_m$ and $v_0$. Adding each such edge $(S_j,v_0)$ creates a new shortest path from all nodes $u_i \in S_j$ to $v_0$. Now if there exists a set cover $S': |S'|=x$, then adding the edges $\{(S_j, v_0) : S_j \in S'\}$ creates new shortest paths to $v_0$ from all nodes $u_i$ (and between all nodes $S_j$ and $v_0$ in an undirected case); adding any other set of $x$ edges will not minimize the betweenness of $v^{\dagger}$. Therefore, an optimal solution to the minimizing betweenness centrality problem directly corresponds to the solution of \textit{Set cover} %problem. \hspace*{\fill}$\Box$\\

\begin{theorem}\label{thrm:ic_nphard}
Both the global and the individual influence recovery problems are NP-hard under the Independent Cascade model.
\end{theorem}

\noindent\textbf{Proof.}
We show a reduction from the NP-complete \textit{Set cover} problem, defined by a universe $U=\{u_1, \dots, u_l\}$ and a collection of sets $S = \{S_1, \ldots, S_m\}$ such that $S_1\cup\dots\cup S_m=U$ and $\forall_j{S_j \subseteq U}$, and the goal is to determine whether there exist $k\leq m$ elements of $S$ the union of which equals $U$.

\begin{figure}[tbhp]
\begin{minipage}{0.47\textwidth}
\centering
\includegraphics[width=\linewidth]{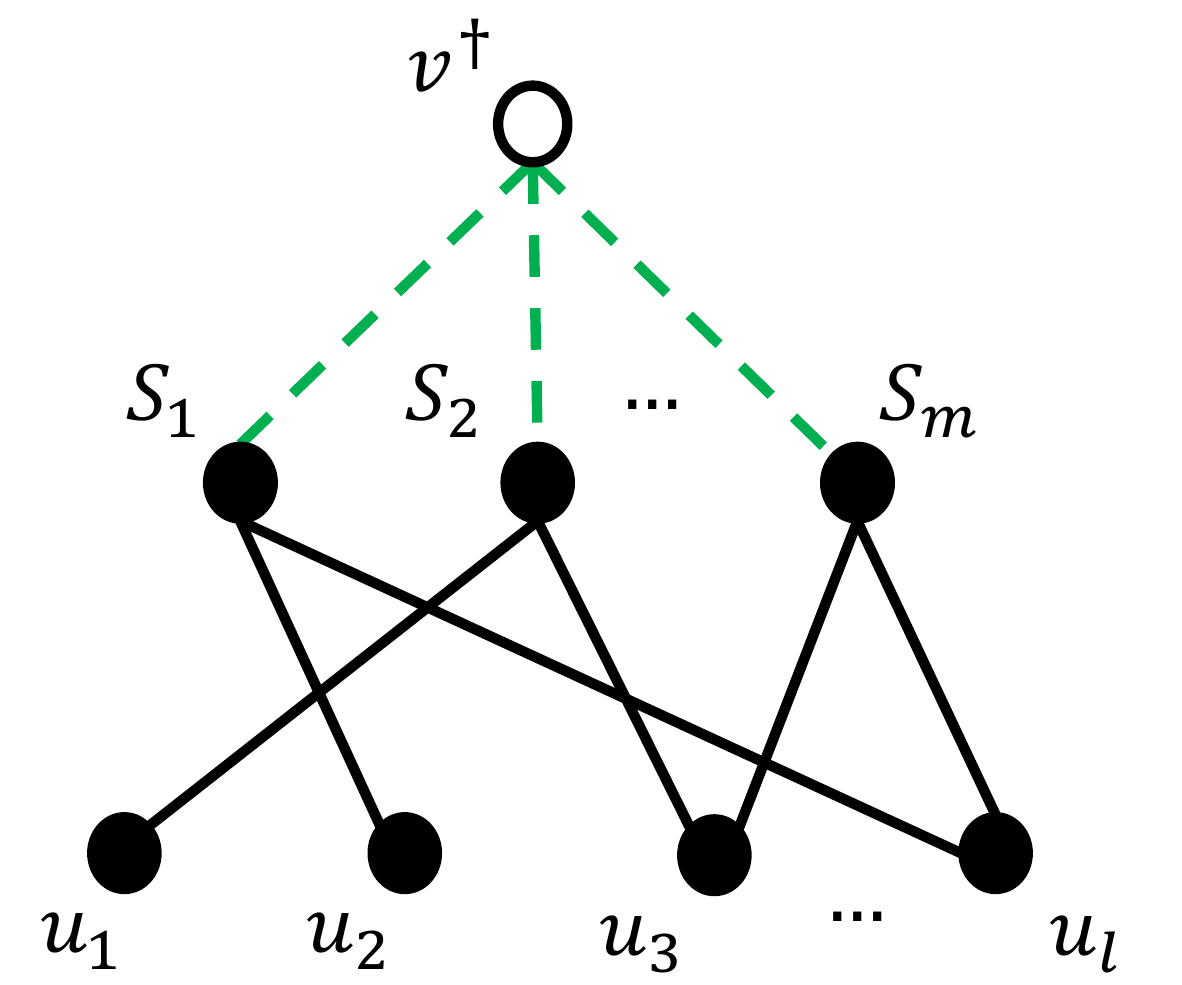}
\caption{\fontsize{10}{10}\selectfont{Undirected network used to reduce the \textit{Set cover} problem to our influence recovery problem. To solve both problems, we consider adding (some of) the dashed edges.}}
\label{fig:ic_nphard_undir}
\end{minipage}
\hfill
\begin{minipage}{0.47\textwidth}
\centering
\includegraphics[width=\linewidth]{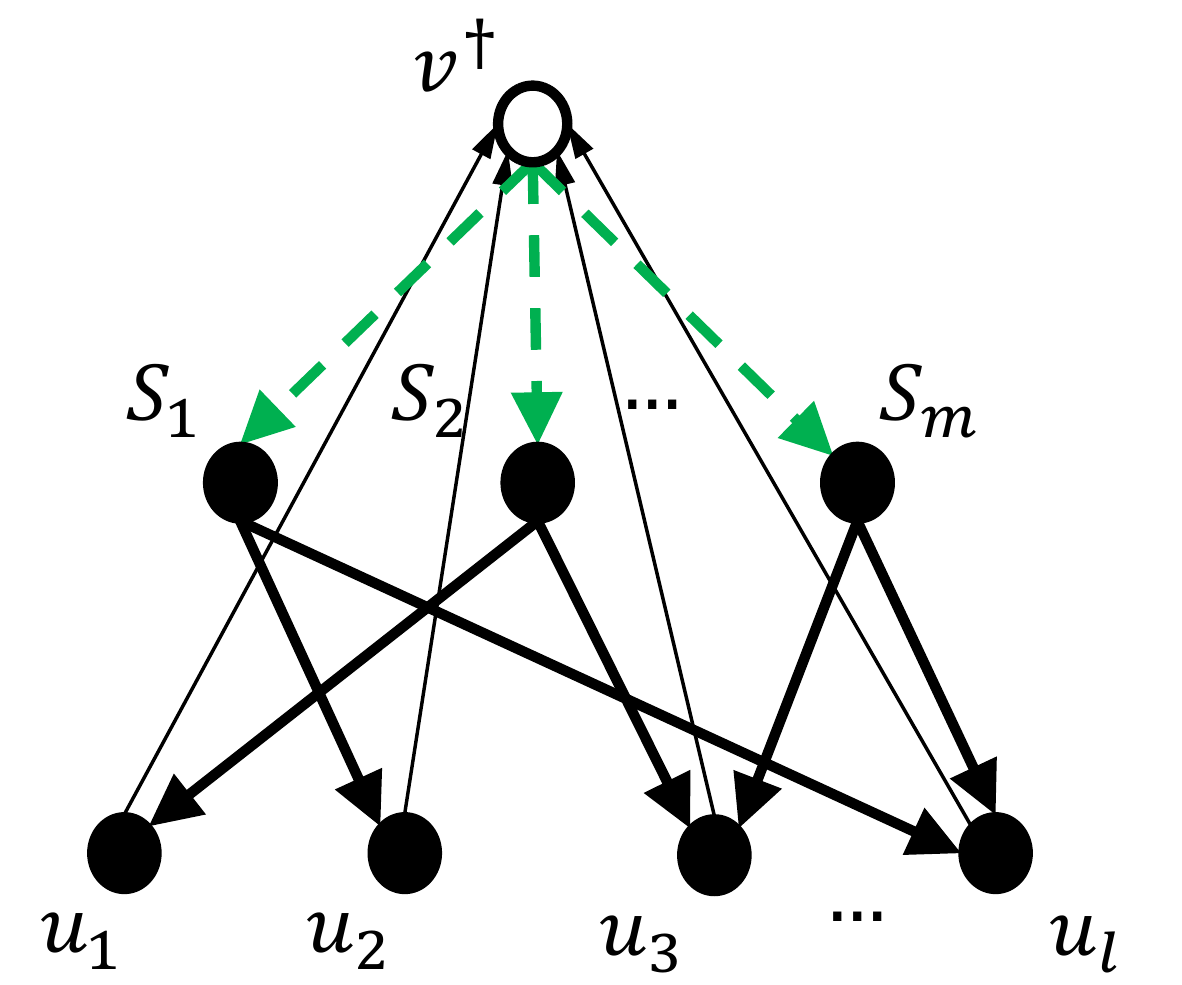}
\caption{\fontsize{10}{10}\selectfont{A directed network used to reduce the \textit{Set cover} problem to our influence recovery problem. To solve both problems, we consider adding (some of) the dashed edges.}}
\label{fig:ic_nphard_dir}
\end{minipage}
\end{figure}

To this end, let us create a network $G$ as illustrated in figures \ref{fig:ic_nphard_undir} and \ref{fig:ic_nphard_dir}. In more detail, we start by creating one node for every $S_j \in S$, one node for every $u_i \in U$, and one additional node $v^{\dagger}$. After that, for every $S_j \in S$ and every $u_i \in S_j$, we add the edge $(S_j, u_i)$ (either directed or undirected). In the directed case we additionally add an edge $(u_i,v^{\dagger})$ for every $u_i \in U$.% to ensure strong connectivity.

%%%%%%%%%%%%%%%%%%%%%%%%%%%%%%%%%%%%%%%%%%%%%%%%%%%%%%%%%%%%%%%%%%%%%%%%%%%%%%%
%%%%%%%%%%%%%%%%%%%%%%%%%%%%%%%%%%%%%%%%%%%%%%%%%%%%%%%%%%%%%%%%%%%%%%%%%%%%%%%

Consider the influence recovery problem in $G$ under the Independent Cascade model, where:
\begin{itemize}
\item $\FA = (V \times V) \setminus \{(v^{\dagger}, S_1), \ldots, (v^{\dagger}, S_m)\}$;
\item $p : V \times V \rightarrow [0,1]$ such that $\forall_{S_j \in S}\ p(v^{\dagger}, S_j) = 1$ and $\forall_{S_j\in S}\forall_{u_i \in S_j}\ p(S_j, u_i) = 1$, and $p(v, w)=0$ for every other pair of nodes;
\item for \emph{individual} influence recovery, $\forall_{u_i \in U} f(u_i) = 1$ and $f(v)=0$ for every other node;
\item for \emph{global} influence recovery, $\globalReb = k + l$.
\end{itemize}

The goal is then to identify the smallest subset of edges to be added to the network, $A\subseteq\{(v^{\dagger}, S_1), \ldots, (v^{\dagger}, S_m) \}$, such that either $\inf_{(V, E \cup A)}(v^{\dagger}) \geq \globalReb$ in the \emph{global} variant of the problem, or $\forall_{v_i \in V} \inf_{(V, E \cup A)}(v^{\dagger},v_i) \geq f(v_i)$ in the \emph{individual} variant of the problem.

Recall that the influence of $v^{\dagger}$ is measured by setting the seed set as $\{v^{\dagger}\}$ and calculating the probability that other nodes get activated. Also recall that under the Independent Cascade model an active node, $v$, activates any of its predecessors, $w$, with probability $p(v,w)$. Importantly, with the $p$ function defined as above, adding an edge $(v^{\dagger},S_j)$ for some $S_i\in S$ makes the influence of $v^{\dagger}$ on every $u_i\in S_j$ equal to 1. Furthermore, the above definitions of $\globalReb$ and $f$ imply that our goal (in both the individual and the global variants of the problem) is achieved if and only if the influence of $v^{\dagger}$ on \emph{every node $u_i \in U$} equals $1$. Consequently, our goal is achieved if and only if we add to $G$ a set of edges, $A\subseteq\{(v^{\dagger}, S_1), \ldots, (v^{\dagger}, S_m) \}$, such that:
$$
\bigcup_{(v^{\dagger}, S_j)\in A} S_j = U.
$$
Since we are interested in finding the smallest such subset, a solution to the above instance of the influence recovery problem gives us a solution to the Set Cover problem. \hspace*{\fill}$\Box$\\
%
%First, let us notice that in both problems the only way to recover the influence is to get it back to $1$ for every node $u_i$. Since we can add a set of $k$ edges $A$ only between $v^{\dagger}$ and $S$, every such choice corresponds to the choice of $k$ sets in the given \textit{Set cover} problem.
%
%If $A$ is a solution to the influence recovery problem, then for every $u_i \in U$ there exist edges $(v^{\dagger}, S_j)$ and $(S_j, u_i)$. However, edge $(S_j, u_i)$ exists only when $u_i \in S_j$ in the given \textit{Set cover} problem. Therefore, a solution to the influence recovery problem constitutes solution to corresponding \textit{Set cover} problem. \hspace*{\fill}$\Box$\\

\begin{theorem}\label{thrm:lt_nphard}
Both the global and the individual influence recovery problems are NP-hard under the Linear Threshold model.
\end{theorem}

\noindent\textbf{Proof.}
We show a reduction from the NP-complete \textit{Set cover} problem, defined by a universe $U=\{u_1, \dots, u_l\}$ and a collection of sets $S = \{S_1, \ldots, S_m\}$ such that $S_1\cup\dots\cup S_m=U$ and $\forall_j{S_j \subseteq U}$, and the goal is to determine whether there exist $k\leq m$ elements of $S$ the union of which equals $U$.

%%%%%%%%%%%%%%%%%%%%%%%%%%%%%%%%%%%%%%%%%%%%%%%%%%%%%%%%%%%%%%%%%%%%%%%%%%%%%%%
\begin{figure}[tbhp]
\centering
\includegraphics[width=0.6\linewidth]{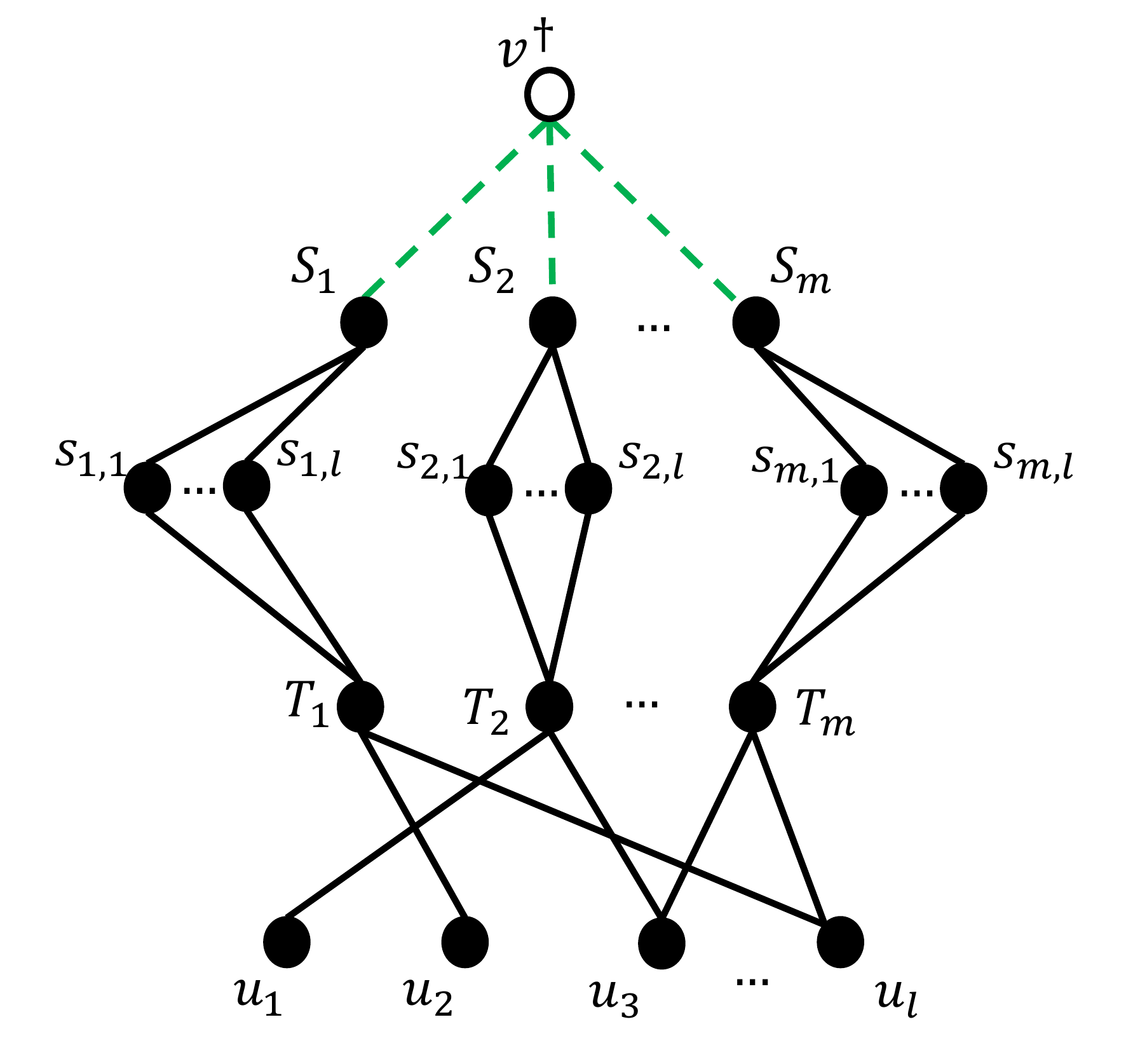}
\vspace*{-0.25cm}
\caption{\fontsize{10}{10}\selectfont{Undirected network used to reduce the \textit{Set cover} problem to our influence recovery problem under the Linear Threshold model. To solve both problems, we consider adding (some of) the dashed edges.}}
\label{fig:lt_nphard_undir}
\end{figure}
%%%%%%%%%%%%%%%%%%%%%%%%%%%%%%%%%%%%%%%%%%%%%%%%%%%%%%%%%%%%%%%%%%%%%%%%%%%%%%%

For the directed case, we create a network $G$ as illustrated earlier in Figure~\ref{fig:ic_nphard_dir}. As for the undirected case, we create $G$ as illustrated in Figure~\ref{fig:lt_nphard_undir}. In more detail, for every $S_j \in S$, we create two nodes, namely $S_j$ and $T_j$, as well as $l$ additional nodes, namely $s_{j,1}, \ldots, s_{j,l}$. We also create one node for every $u_i \in U$, and finally add the source node, $v^{\dagger}$. As for the edges, for every $S_j\in S$ and every $u_i \in S_j$, we add the edge $(T_j, u_i)$. Furthermore, for every node $s_{j,i}$, we add the edges $(S_j,s_{j,i})$ and $(s_{j,i},T_j)$.

Now consider the influence recovery problem in $G$ under the Linear Threshold model, where:
\begin{itemize}
\item $\FA = (V \times V) \setminus \{(v^{\dagger}, S_1), \ldots, (v^{\dagger}, S_m)\}$;
\item $t_v=l$ for every node $v\in \{T_1,\dots,T_m\}$ and $t_v=1$ for every other node;
\item for \emph{individual} influence recovery, $\forall_{u_i \in U} f(u_i) = 1$ and $f(v)=0$ for every other node;
\item for \emph{global} influence recovery, $\globalReb = k + l$ for the directed case, and $\globalReb = k (l + 2) + l$ for the undirected case.
\end{itemize}

The goal is then to identify the smallest subset of edges to be added to the network, $A\subseteq\{(v^{\dagger}, S_1), \ldots, (v^{\dagger}, S_m) \}$, such that either $\inf_{(V, E \cup A)}(v^{\dagger}) \geq \globalReb$ in the \emph{global} variant of the problem, or $\forall_{v_i \in V} \inf_{(V, E \cup A)}(v^{\dagger},v_i) \geq f(v_i)$ in the \emph{individual} variant of the problem.

Recall that the influence of $v^{\dagger}$ is measured by setting the seed set as $\{v^{\dagger}\}$ and calculating the probability that other nodes get activated. Also recall that under the Linear Threshold model a node, $v$, gets activated if the number of its active predecessors exceeds $t_v$. Note that, with $t_v$ defined as above, adding an edge $(v^{\dagger},S_j)$ in the undirected case leads to the activation of nodes $s_{i,j}$ and $T_i$, which in turn leads to the activation of every $u_i \in S_j$ (see Figure~\ref{fig:lt_nphard_undir}). Likewise, in the directed case, adding $(v^{\dagger},S_j)$ leads to the activation of every $u_i \in S_j$ (see Figure~\ref{fig:ic_nphard_dir}). To put it differently, when adding $(v^{\dagger},S_j)$, the influence of $v^{\dagger}$ on every $u_i\in S_j$ equals $1$. Importantly, the above definitions of $\globalReb$ and $f$ imply that our goal (in both the individual and the global variants of the problem) is achieved if and only if the influence of $v^{\dagger}$ on \emph{every node $u_i \in U$} equals 1. Those observations imply that our goal is achieved if and only if we add to $G$ a set of edges, $A\subseteq\{(v^{\dagger}, S_1), \ldots, (v^{\dagger}, S_m) \}$, such that:
$$
\bigcup_{(v^{\dagger}, S_j)\in A} S_j = U.
$$
Since we are interested in finding the smallest such subset, a solution to the above instance of the influence recovery problem gives us a solution to the Set Cover problem. \hspace*{\fill}$\Box$\\

%Firstly, let us notice that with the aforementioned forbidden sets we can choose $k$ edges to be added between $v^{\dagger}$ and $S$. Every such choice corresponds to the choice of $k$ sets in the given \textit{Set cover} problem. This is because adding edge $(v^{\dagger},S_i)$ in an undirected case causes activation of nodes $t_{i,j}$ and $T_i$, and this consecutively causes activation of nodes $u_j$ for $u_j \in S_i$. In a directed case adding edge $(v^{\dagger},S_i)$ causes activation of nodes $u_j$ for $u_j \in S_i$. Both in the global and individual problems the only way to recover the required influence is to get it back to $1$ for every node $u_i$.

%If $A$ is a solution to the influence recovery problem, then for every $u_i \in U$ there exist edge $(v^{\dagger}, S_j)$ such that either $(T_j,u_i) \in E$ in an undirected case or $(S_j,u_i) \in E$ in a directed case. However, such edges exist only when, in the given \textit{Set cover} problem, we have that $u_i \in S_j$. Therefore, solution to the influence recovery problem constitutes solution to the corresponding \textit{Set cover} problem. \hspace*{\fill}$\Box$\\

%%%%%%%%%%%%%%%%%%%%%%%%%%%%%%%%%%%%%%%%%%%%%%%%%%%%%%%%%%%%%%%%%%%%%%%%%%%%%%%

\section{Empirical Evaluation}\label{sec:experiments}

\subsection{Configuring the ROAM Heuristic}
\begin{figure}[tbhp]
	\centering.
	\setlength\tabcolsep{1pt}
	\renewcommand{\arraystretch}{2.0}
	\begin{tabular}{m{.195\textwidth}m{.195\textwidth}m{.195\textwidth}m{.195\textwidth}m{.195\textwidth}}
	\multicolumn{1}{c}{{\fontsize{8.5}{8.5}\selectfont{Degree Ranking}}} &
	\multicolumn{1}{c}{{\fontsize{8.5}{8.5}\selectfont{Closeness Ranking}}} &
	\multicolumn{1}{c}{{\fontsize{8.5}{8.5}\selectfont{Betweenness Ranking}}} &
	\multicolumn{1}{c}{{\fontsize{8.5}{8.5}\selectfont{IC Influence}}} &
	\multicolumn{1}{c}{{\fontsize{8.5}{8.5}\selectfont{LT Influence}}} \\
	\includegraphics[width=0.95\linewidth]{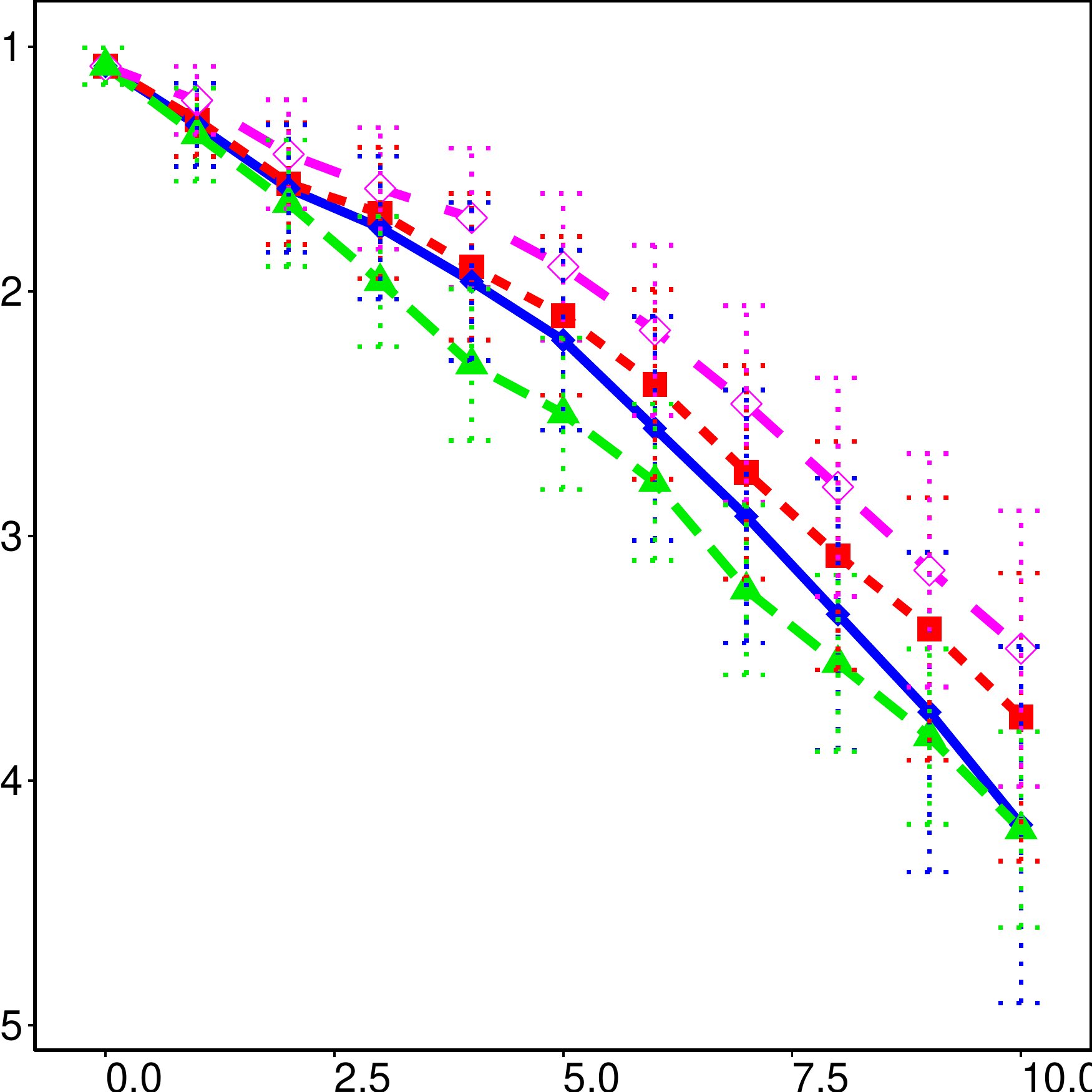} &
	\includegraphics[width=0.95\linewidth]{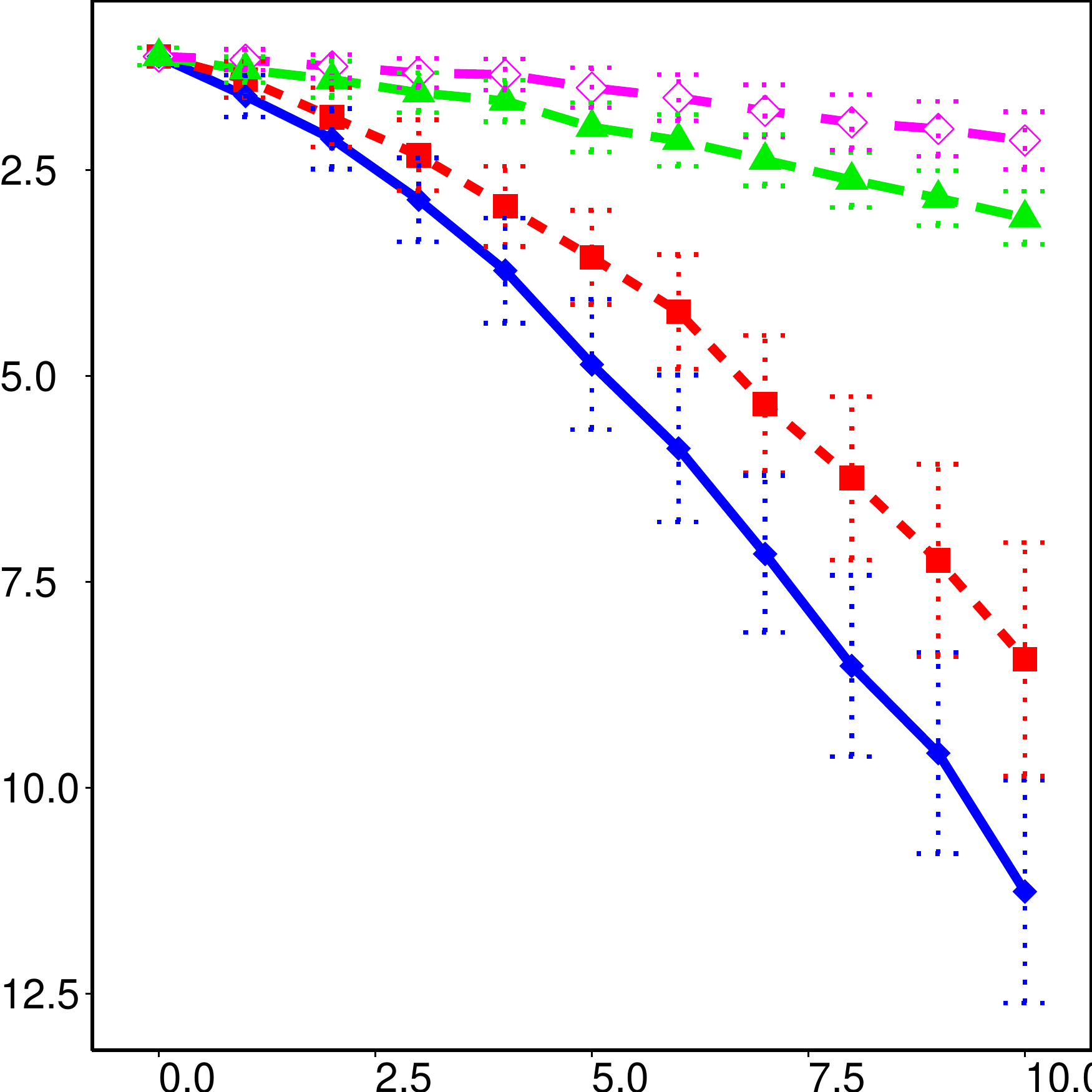} &
	\includegraphics[width=0.95\linewidth]{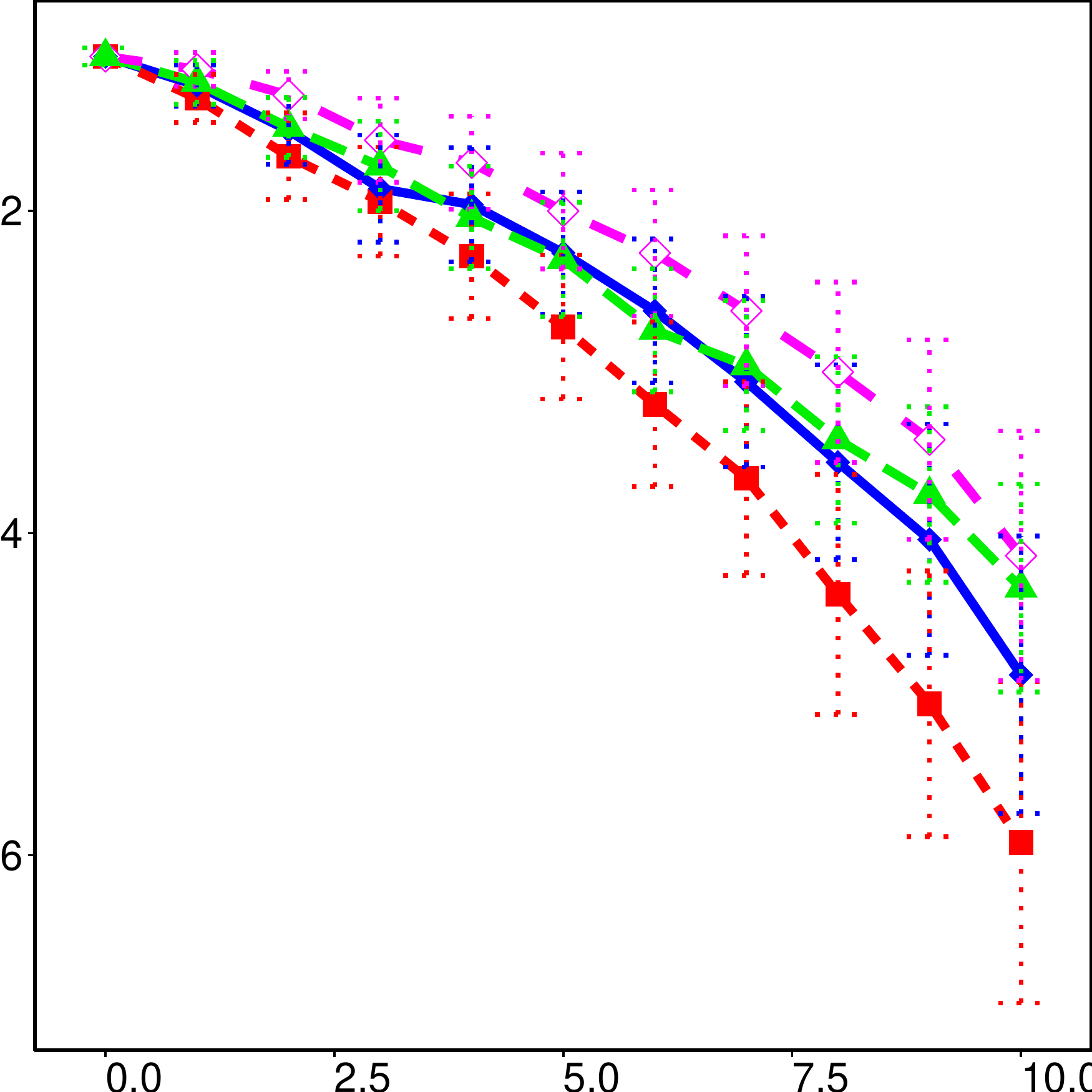} &
	\includegraphics[width=0.95\linewidth]{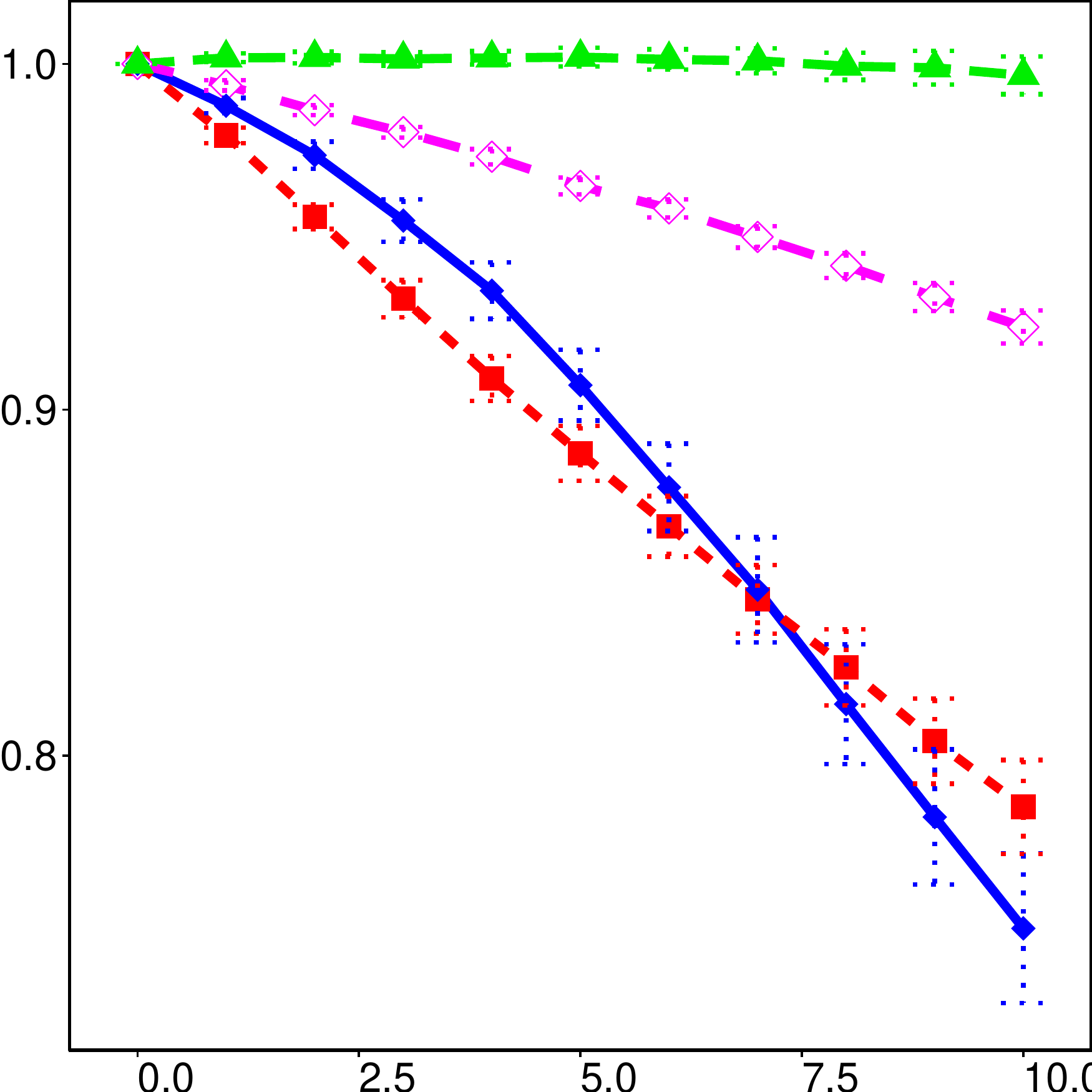} &
	\includegraphics[width=0.95\linewidth]{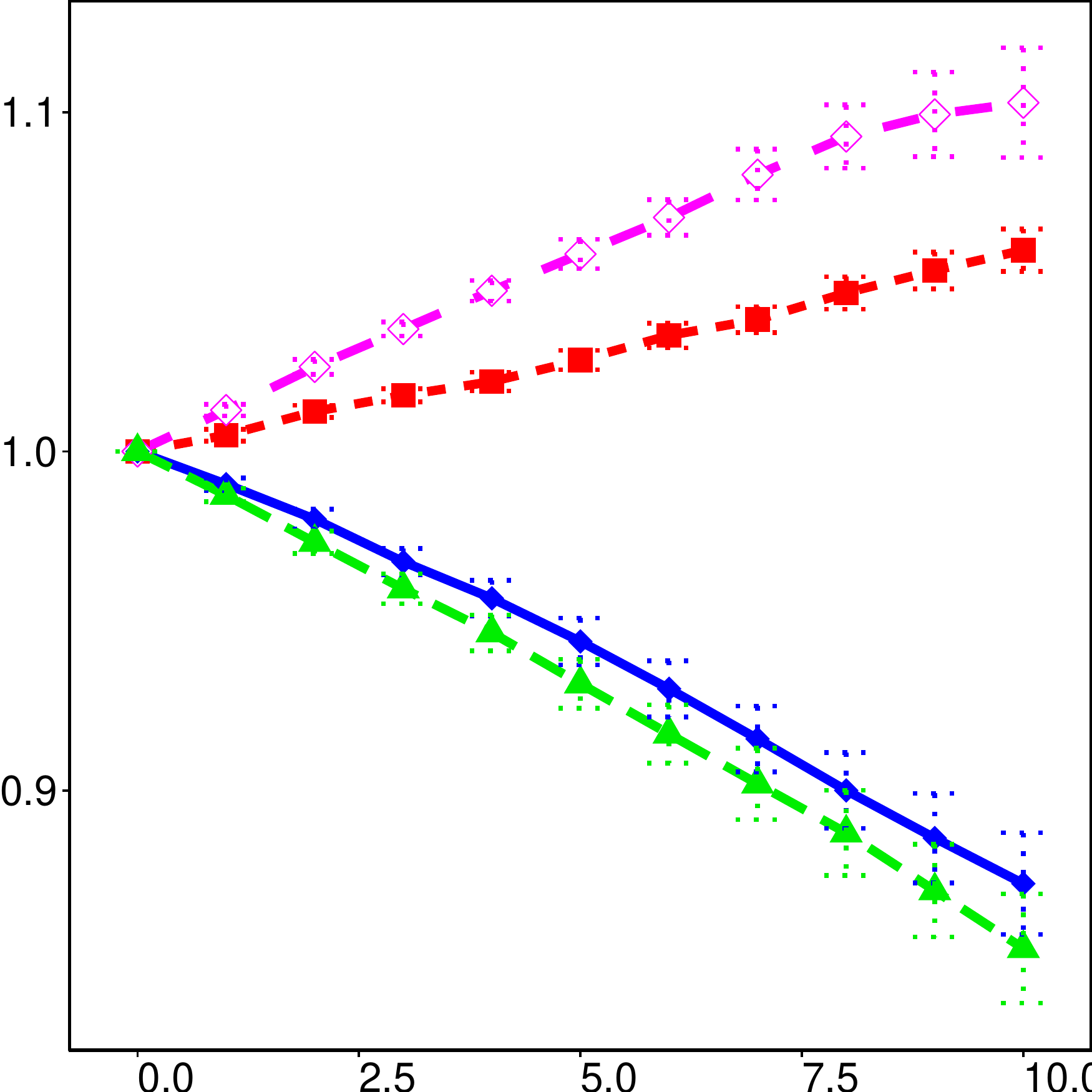} \\
	\multicolumn{5}{c}{\includegraphics[width=0.75\linewidth]{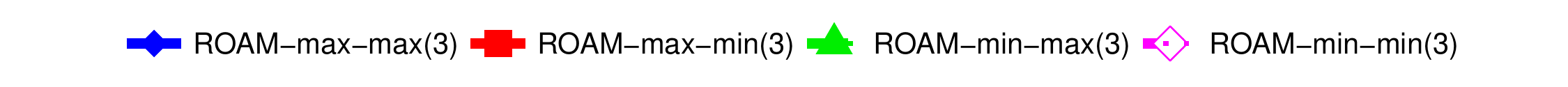}} \\
	\end{tabular}
	\vspace*{-0.25cm}
	\caption{\fontsize{10}{10}\selectfont{Comparing different settings of ROAM on 50 randomly generated scale-free network consisting of 100 nodes, with 3 edges added in each step of the generation process. For each such network, ROAM is executed multiple, consecutive times (the $x$-axis represents the number of executions). The subfigures show the source node's ranking (according to different centrality measures), and the relative change in its influence value (according to different influence models).}}
	\label{fig:minmaxcomp}
\end{figure}

\noindent As mentioned in the main article, the ROAM heuristic involves choosing $v_0$ (the neighbour of $v^{\dagger}$ whom the heuristic will disconnect from $v^{\dagger}$), and choosing the $b-1$ neighbours of $v^{\dagger}$ whom the heuristic will connect to $v_0$. We conducted a number of experiments to determine whether it is more beneficial to choose $v_0$ as the neighbour of $v^{\dagger}$ with the \emph{least} connections or the \emph{most} connections. Likewise, we wanted to determine whether it is more beneficial to choose the $b-1$ neighbours of $v^{\dagger}$ (who will be connected to $v_0$) as the ones with the \emph{least} connections or the \emph{most} connections. In particular, Figure~\ref{fig:minmaxcomp} compares the different settings given $50$ radomly generated scale-free networks consisting of 100 nodes each, where 3 edges are added with each step of the generation process (for more details, see \cite{barabasi1999emergence}); we chose scale-free networks as they resemble real-life networks in many way, e.g., in terms of degree distribution. As for the source node, it is chosen to be the one with the lowest sum of centrality rankings (ties are broken uniformly at random). As for the Independent Cascade model, we set the activation probability to be $p(v,w) = 0.15$ for every pair of nodes, $v,w\in V$. As for the Linear Threshold model, for every node, $v\in V$, the threshold value, $t_v$, is sampled uniformly at random from the set $\{0, \ldots,|\P(v)|\}$. For both models, the influence values are approximated using the Monte-Carlo method. In the figure, we write ROAM-$x$-$y$($b$), where $x$ can either be ``\emph{max}'' or ``\emph{min}'' (indicating that $v_0$ is the neighbour with the \emph{most} connections or the \emph{least} connections, respectively) and $y$ can either be ``\emph{max}'' or ``\emph{min}'' (indicating that the $b-1$ neighbours are chosen to be the ones with the \emph{most} connections or the \emph{least} connections, respectively), whereas $b$ represents the budget (which is set to 3 in this experiment). Since the results are averaged over 50 random networks, the error bars in the figure represent the $95\%$ confidence intervals. For each network, the ROAM heuristic is executed multiple, consecutive times; the $x$-axis in each subfigure represents the number of executions. As can be seen, while there is no setting that dominates the others, the best overall performance seems to be achieved by ROAM-max-min(3). Based on this, in all subsequent experiments on ROAM, we choose $v_0$ as the neighbour of $v^{\dagger}$ with the \emph{most} connections, and we connect $v_0$ to the $b-1$ neighbours of $v^{\dagger}$ with the \emph{least} connections.
%It seems that the best performance is achieved by heuristic, that chooses the high degree neighbour to be cut off (as it has high probability to be on many shortest paths to $v^{\dagger}$, thus being more effective in lowering centrality) and connect her to neighbours with low degree (as the impact of influence increase on them is usually bigger).

%%%%%%%%%%%%%%%%%%%%%%%%%%%%%%%%%%%%%%%%%%%%%%%%%%%%%%%%%%%%%%%%%%%%%%%%%%%%%%%
%%%%%%%%%%%%%%%%%%%%%%%%%%%%%%%%%%%%%%%%%%%%%%%%%%%%%%%%%%%%%%%%%%%%%%%%%%%%%%%

In the main article, due to space constraints, we only specified how the two main steps of ROAM can be applied on \emph{undirected} networks. Next, we specify how these steps are modified to work on \emph{directed} networks. First of all, $v_0$ is not chosen among the \emph{neighbours} of $v^{\dagger}$, but rather among the \emph{successors} of $v^{\dagger}$. This is mainly because removing a successor of $v^{\dagger}$ reduces its closeness centrality, whereas removing a predecessor has no such impact. As for the $b-1$ neighbours of $v^{\dagger}$ to be connected to $v_0$, they are chosen among the \emph{predecessors} of $v^{\dagger}$; for each such predecessor, $v_i$, we add the edge $(v_i,v_0)$. This is mainly because it could potentially rebuild the influence of $v^{\dagger}$ on $v_0$, which was hampered by the removal of the edge $(v^{\dagger},v_0)$. Furthermore, for every shortest path that contains the edge $(v^{\dagger},v_i)$, the addition of $(v_i,v_0)$ could create a new alternative shortest path that does not pass through $v^{\dagger}$, thus further reducing the betweenness centrality of $v^{\dagger}$.

%%%%%%%%%%%%%%%%%%%%%%%%%%%%%%%%%%%%%%%%%%%%%%%%%%%%%%%%%%%%%%%%%%%%%%%%%%%%%%%

\subsection{Experimental Results}
\noindent In the main article, we only presented some of the our experimental results due to space constraints; in this subsection, we present all of our experimental results. Although most of the experimental details can be found in the main article, we add here the only missing detail, which concerns the anonymized fragments of the social networks of Facebook, Twitter and Google+ (note that the fragments of Twitter and Google+ are the only directed networks in our experiments; the remaining networks are all undirected). all anonymized fragments were taken from SNAP---the Stanford Network Analysis Platform \cite{leskovec2012learning}.%\footnote{\footnotesize {\tt http://snap.stanford.edu/data/\#socnets}}. For each of these social networks, we experiment with fragments of three different sizes. In particular:
\begin{itemize}
\item Facebook: the small fragment consists of 61 nodes and 272 edges; the medium one consists of 333 nodes and 2523 edges; the large one consists of 786 nodes and 14027 edges;
\item Twitter: the small fragment consists of 201 nodes and 2503 edges; the medium one consists of 247 nodes and 8041 edges; the large one consists of 235 nodes and 15957 edges;
\item Google+: the small fragment consists of 108 nodes and 2884 edges; the medium one contains 215 nodes and 7132 edges; the large one consists of 338 nodes and 12341 edges.
\end{itemize}

\begin{figure}[p!]
	\centering.
	\setlength\tabcolsep{2pt}
	\renewcommand{\arraystretch}{2.0}
	\begin{tabular}{m{.01\textwidth}m{.195\textwidth}m{.195\textwidth}m{.195\textwidth}m{.195\textwidth}m{.195\textwidth}}
	&
	\multicolumn{1}{c}{\fontsize{8.5}{8.5}\selectfont{Degree Ranking}} &
	\multicolumn{1}{c}{\fontsize{8.5}{8.5}\selectfont{Closeness Ranking}} &
	\multicolumn{1}{c}{\fontsize{8.5}{8.5}\selectfont{Betweenness Ranking}} &
	\multicolumn{1}{c}{\fontsize{8.5}{8.5}\selectfont{IC Influence}} &
	\multicolumn{1}{c}{\fontsize{8.5}{8.5}\selectfont{LT Influence}} \\
	\rotatebox{90}{{\fontsize{8}{8}\selectfont{$\mathit{ScaleFree}(100,3)$}}} &
	\includegraphics[width=0.93\linewidth]{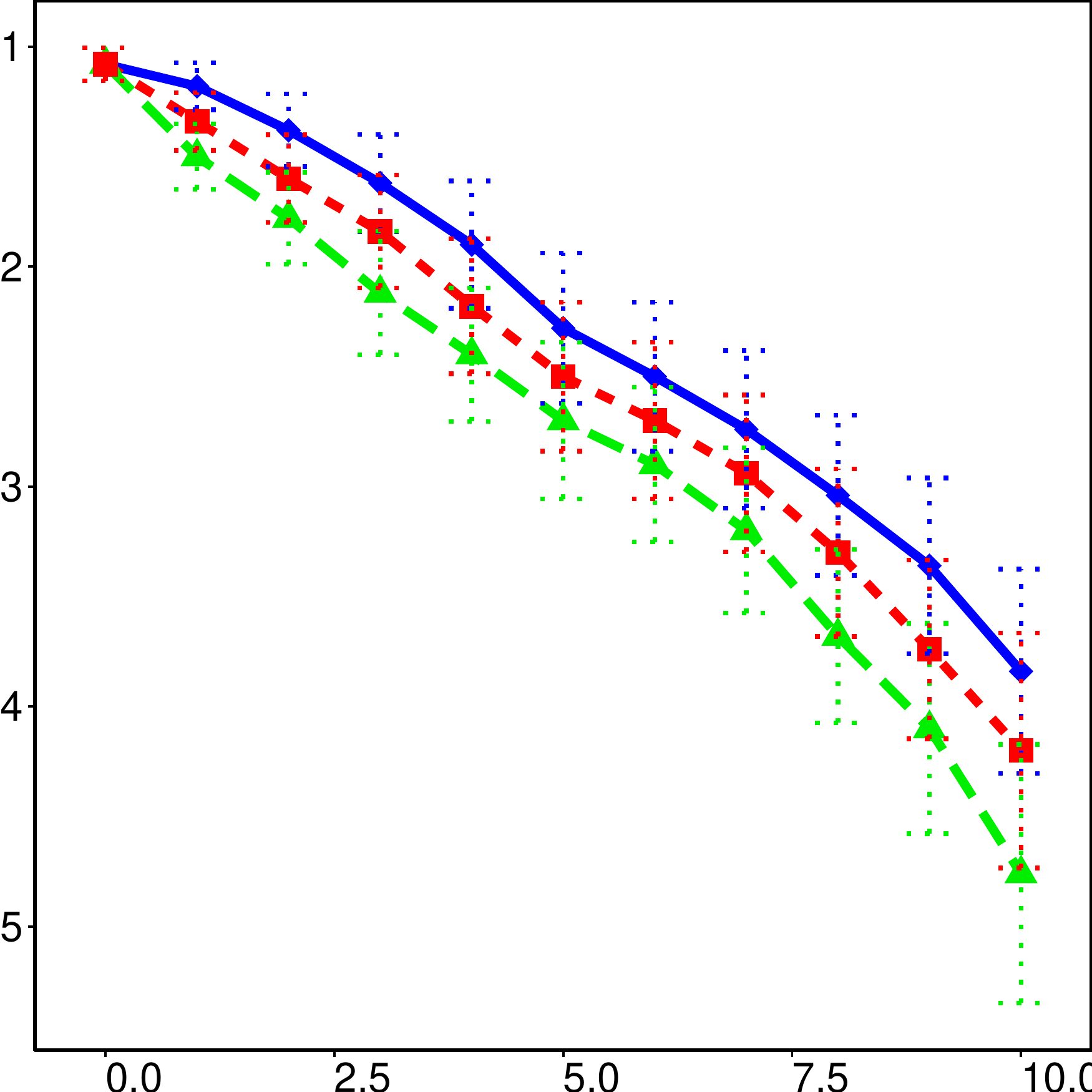} &
	\includegraphics[width=0.93\linewidth]{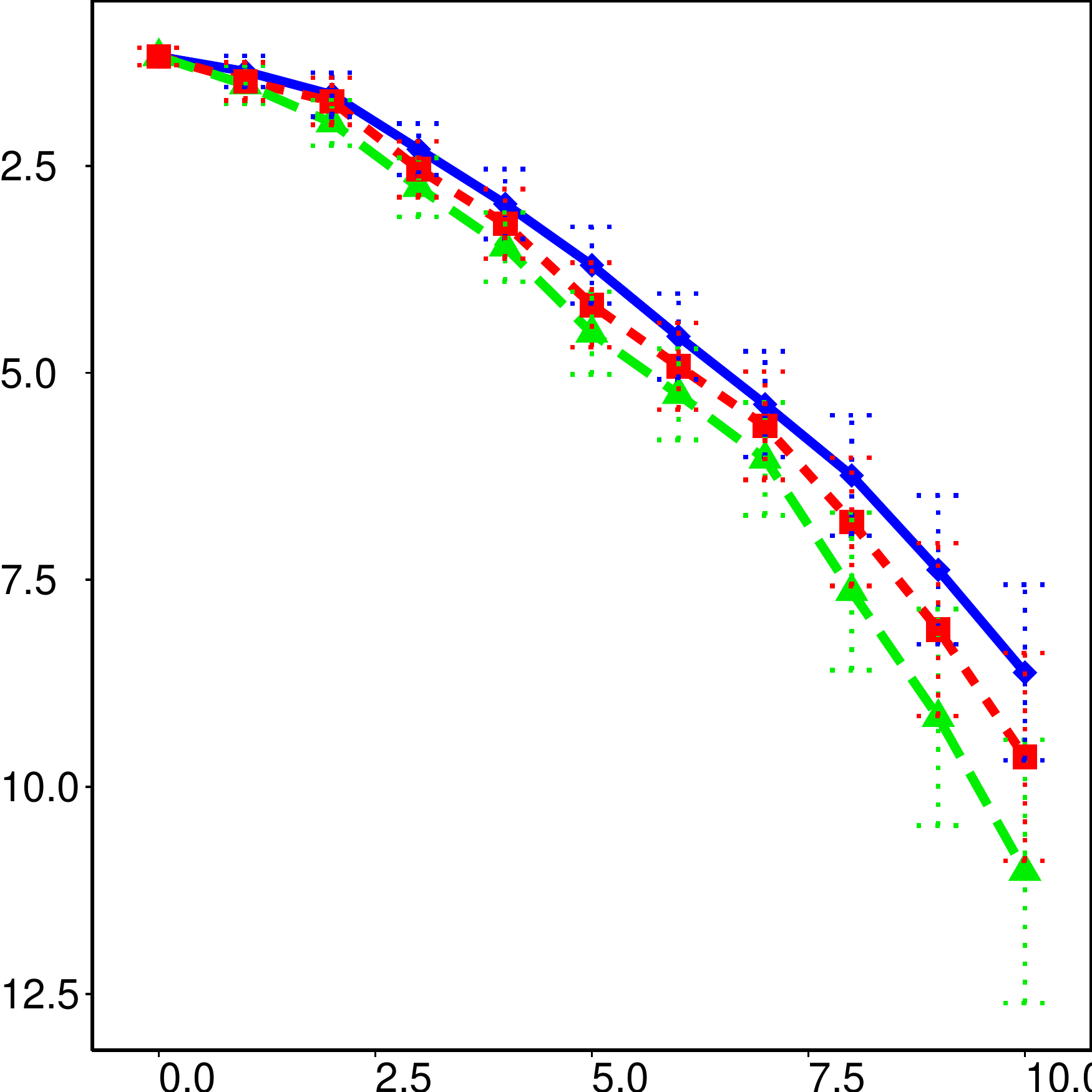} &
	\includegraphics[width=0.93\linewidth]{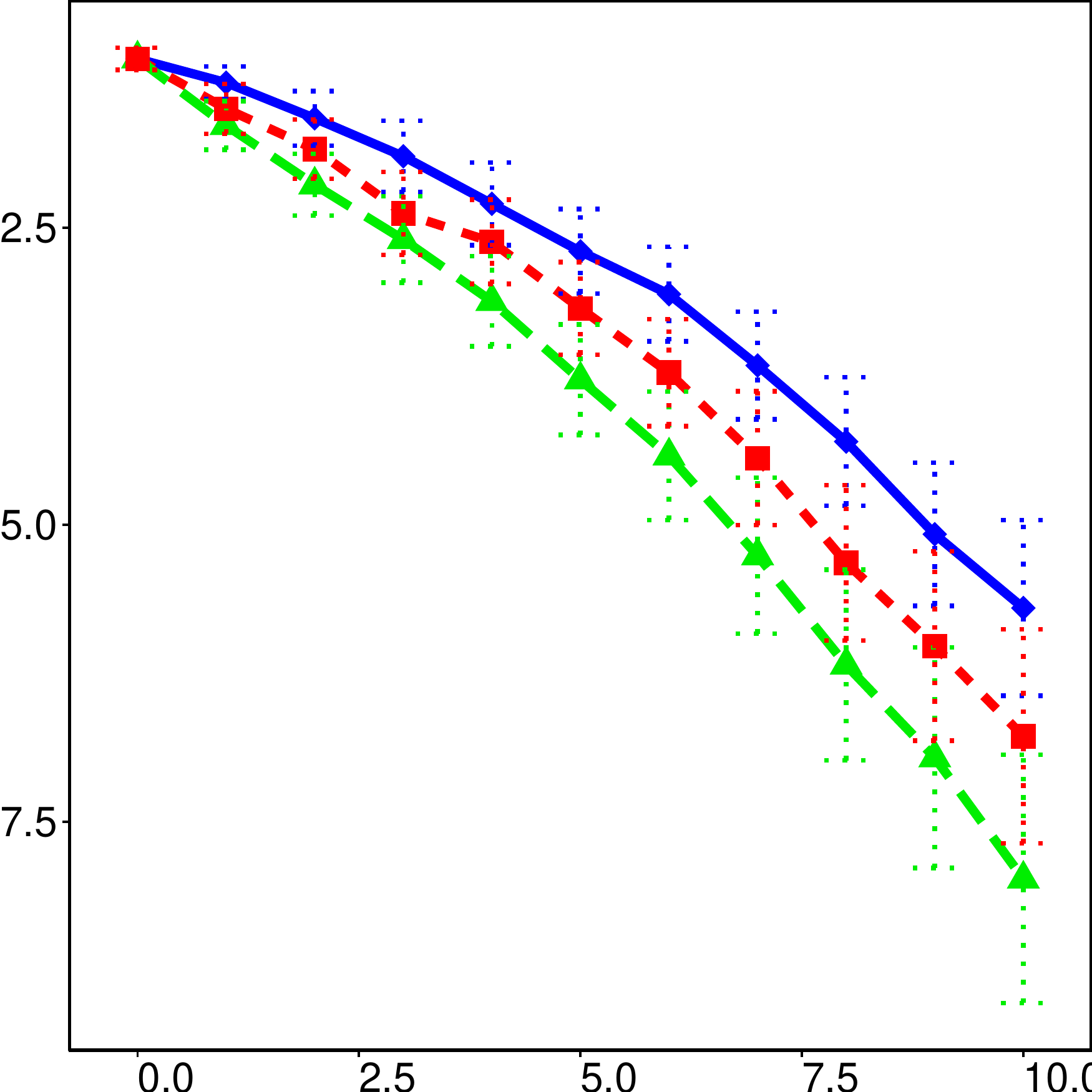} &
	\includegraphics[width=0.93\linewidth]{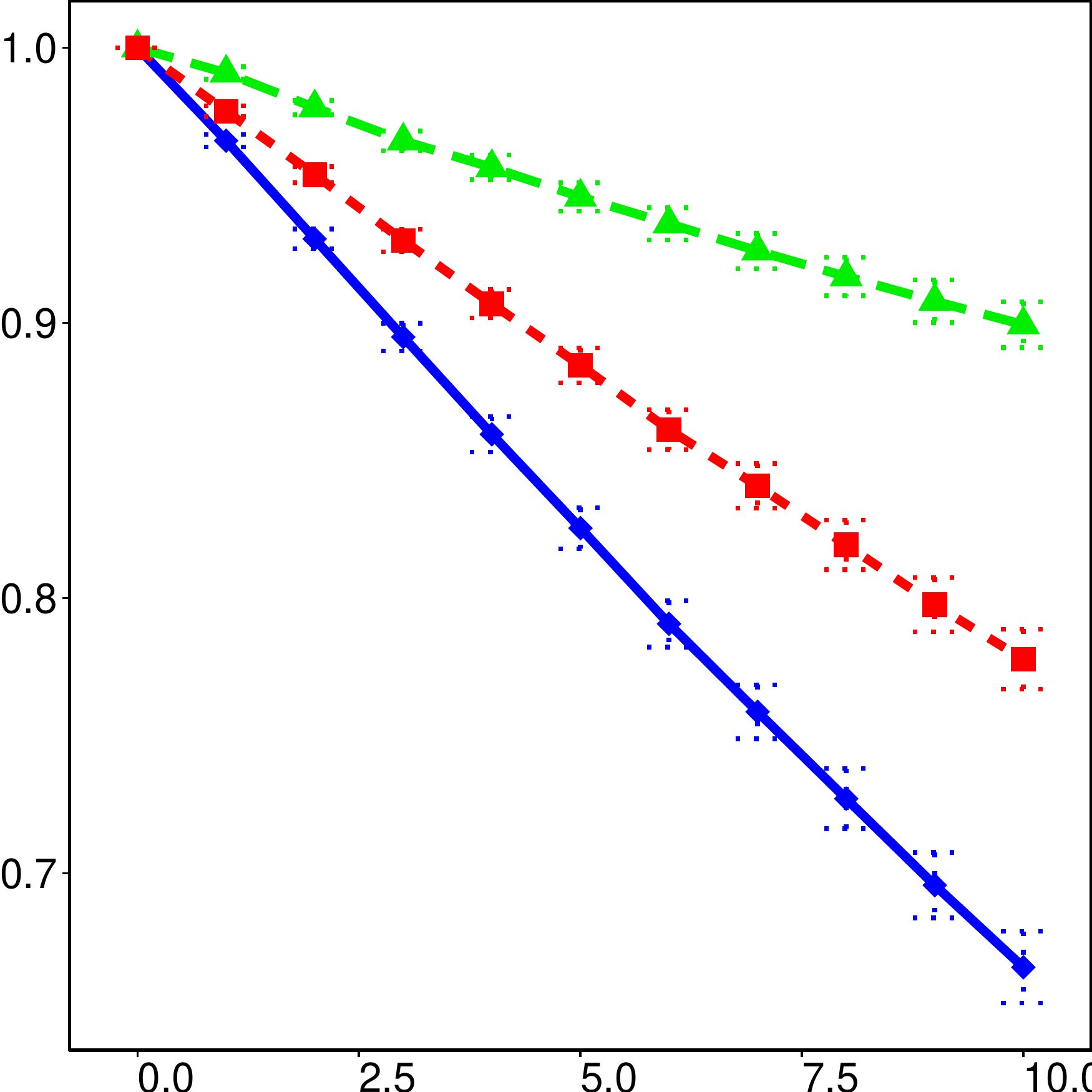} &
	\includegraphics[width=0.93\linewidth]{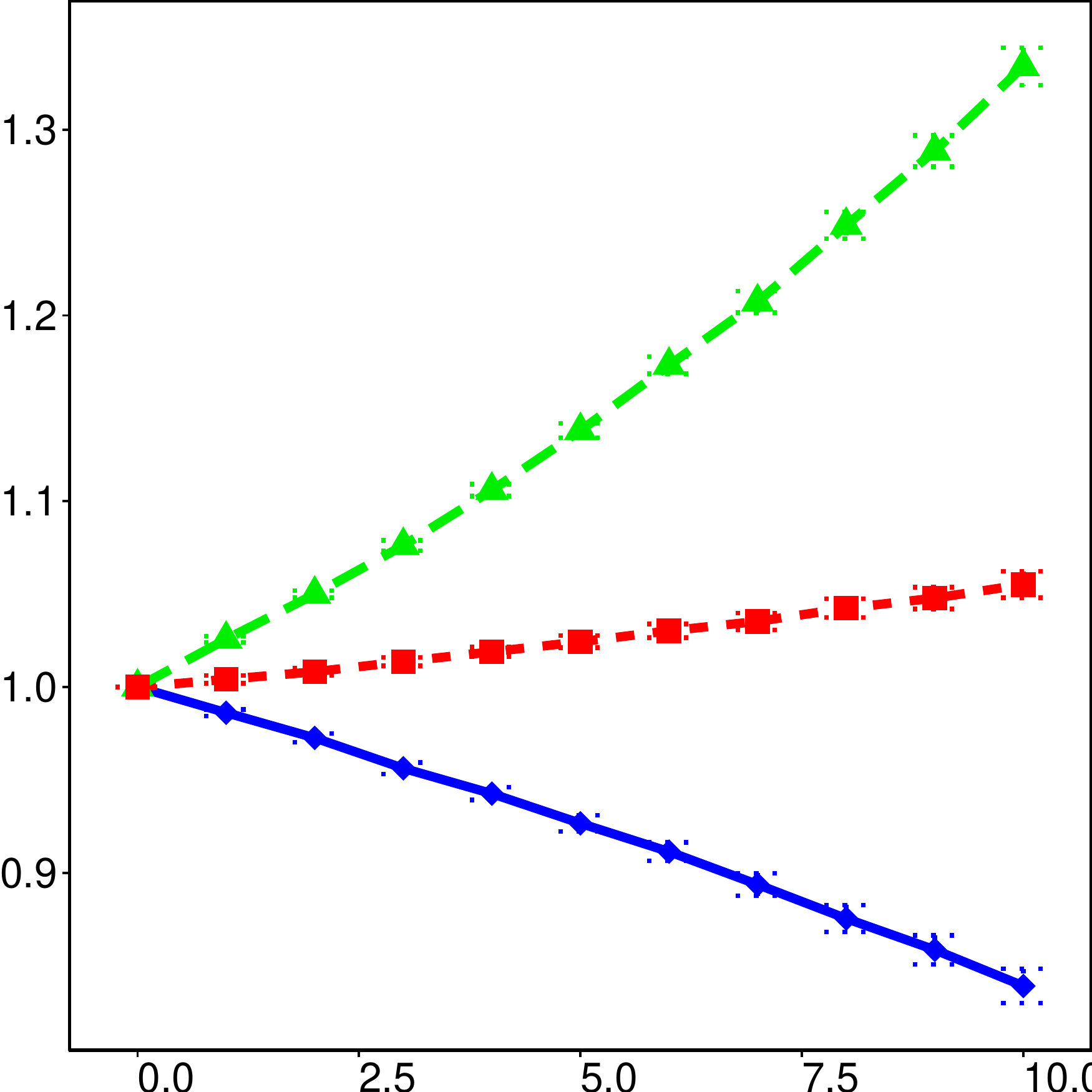} \\
	\rotatebox{90}{{\fontsize{8}{8}\selectfont{$\mathit{ScaleFree}(1000,3)$}}} &
	\includegraphics[width=0.93\linewidth]{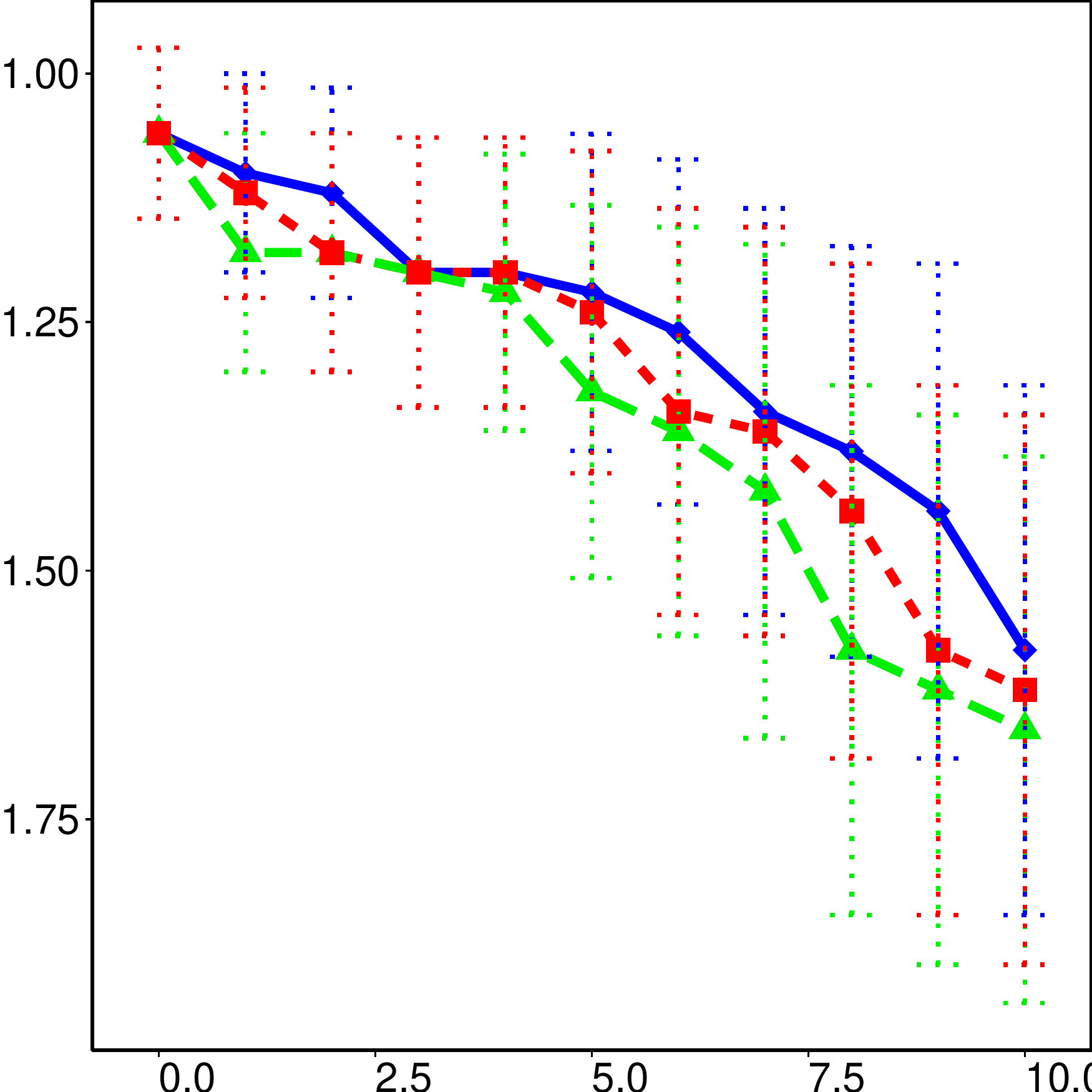} &
	\includegraphics[width=0.93\linewidth]{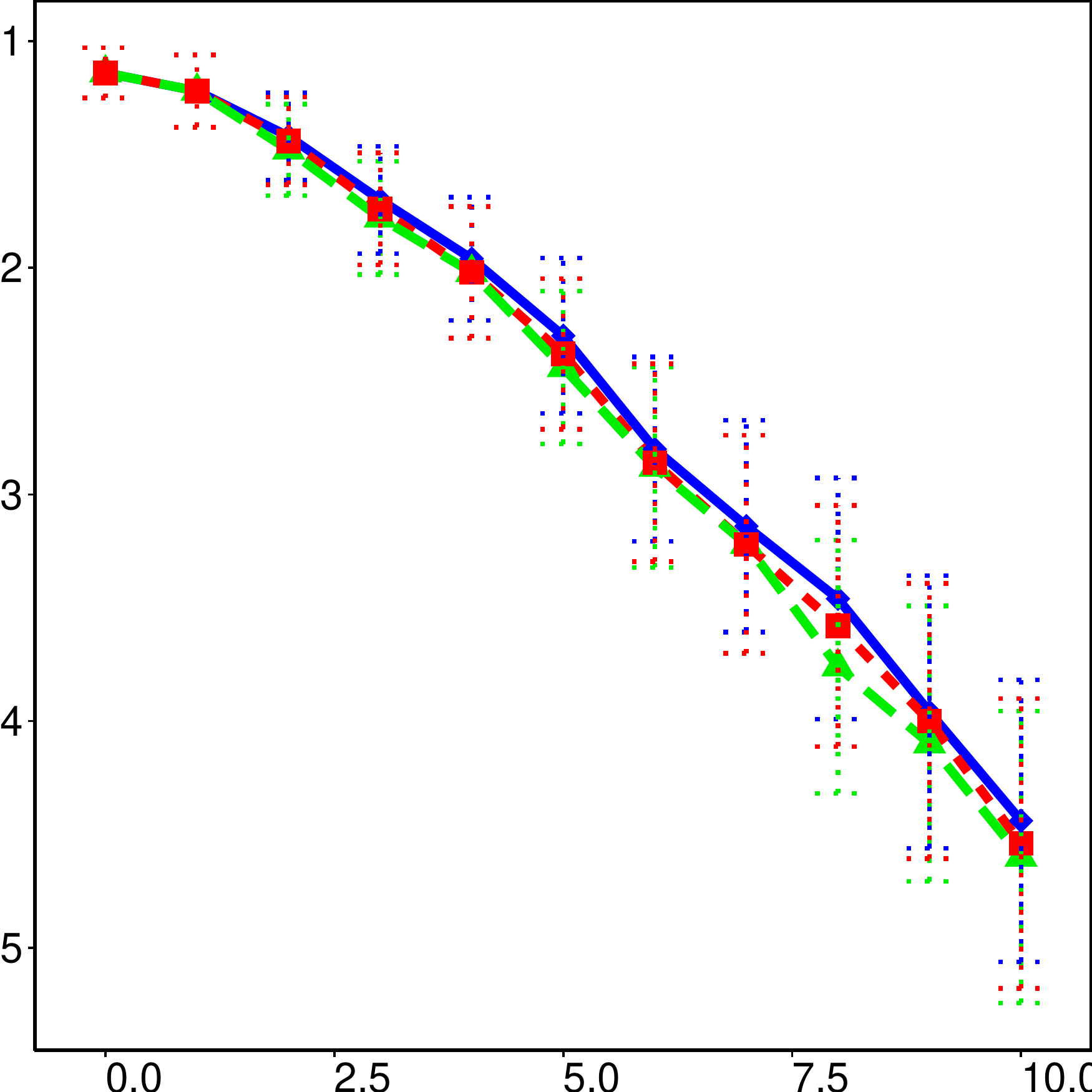} &
	\includegraphics[width=0.93\linewidth]{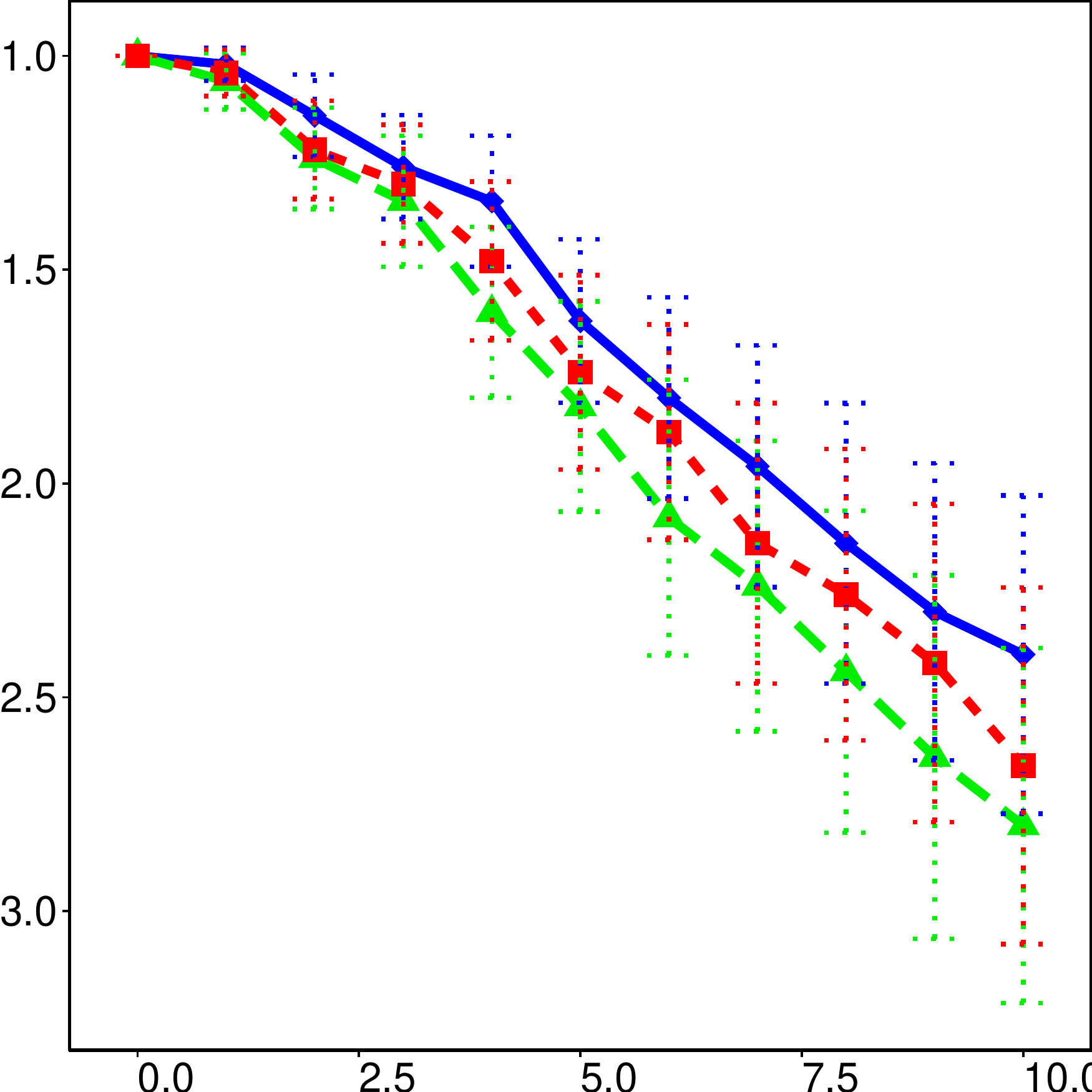} &
	\includegraphics[width=0.93\linewidth]{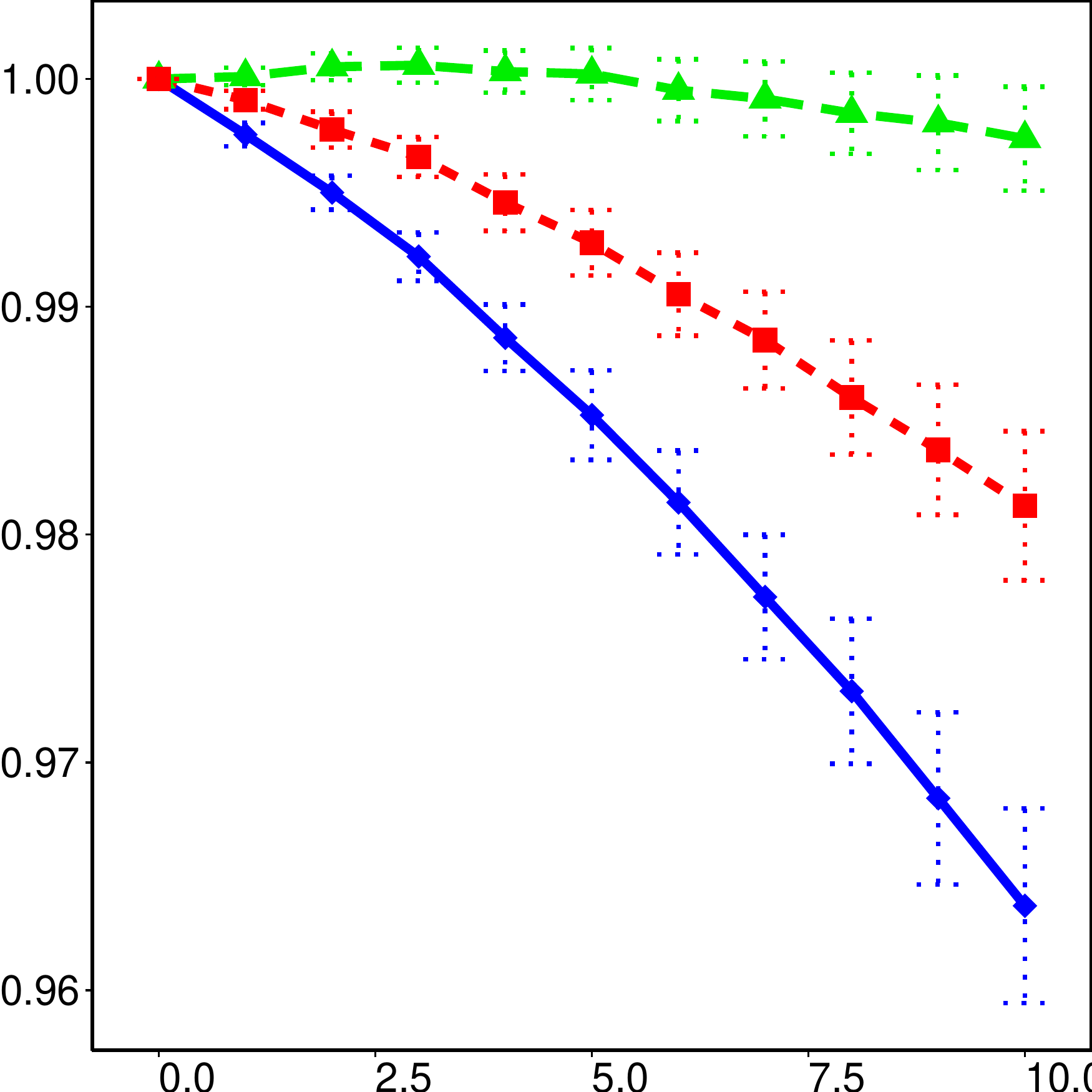} &
	\includegraphics[width=0.93\linewidth]{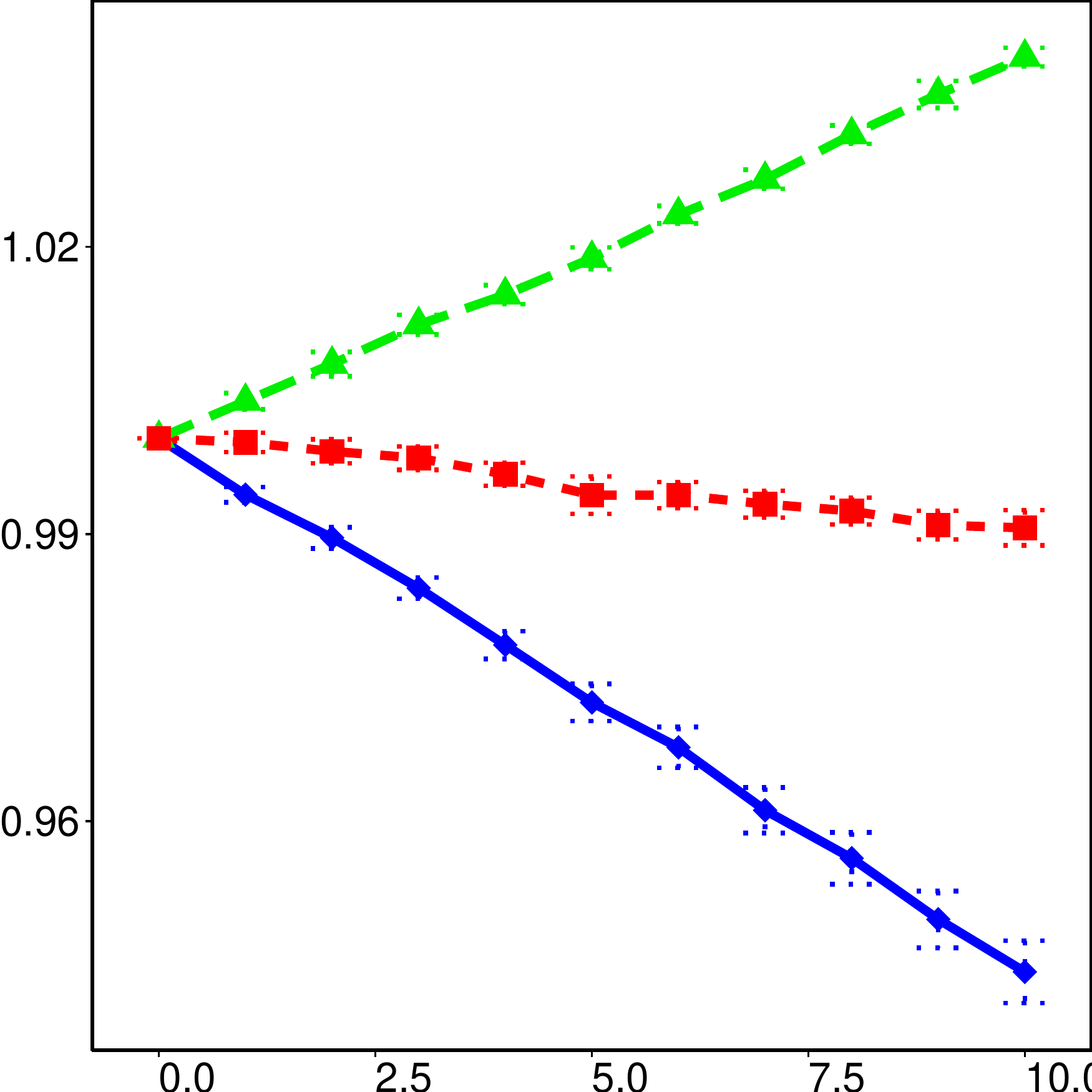} \\
	\rotatebox{90}{{\fontsize{6}{6}\selectfont{$\mathit{RandomGraph}(100,10)$}}} &
	\includegraphics[width=0.93\linewidth]{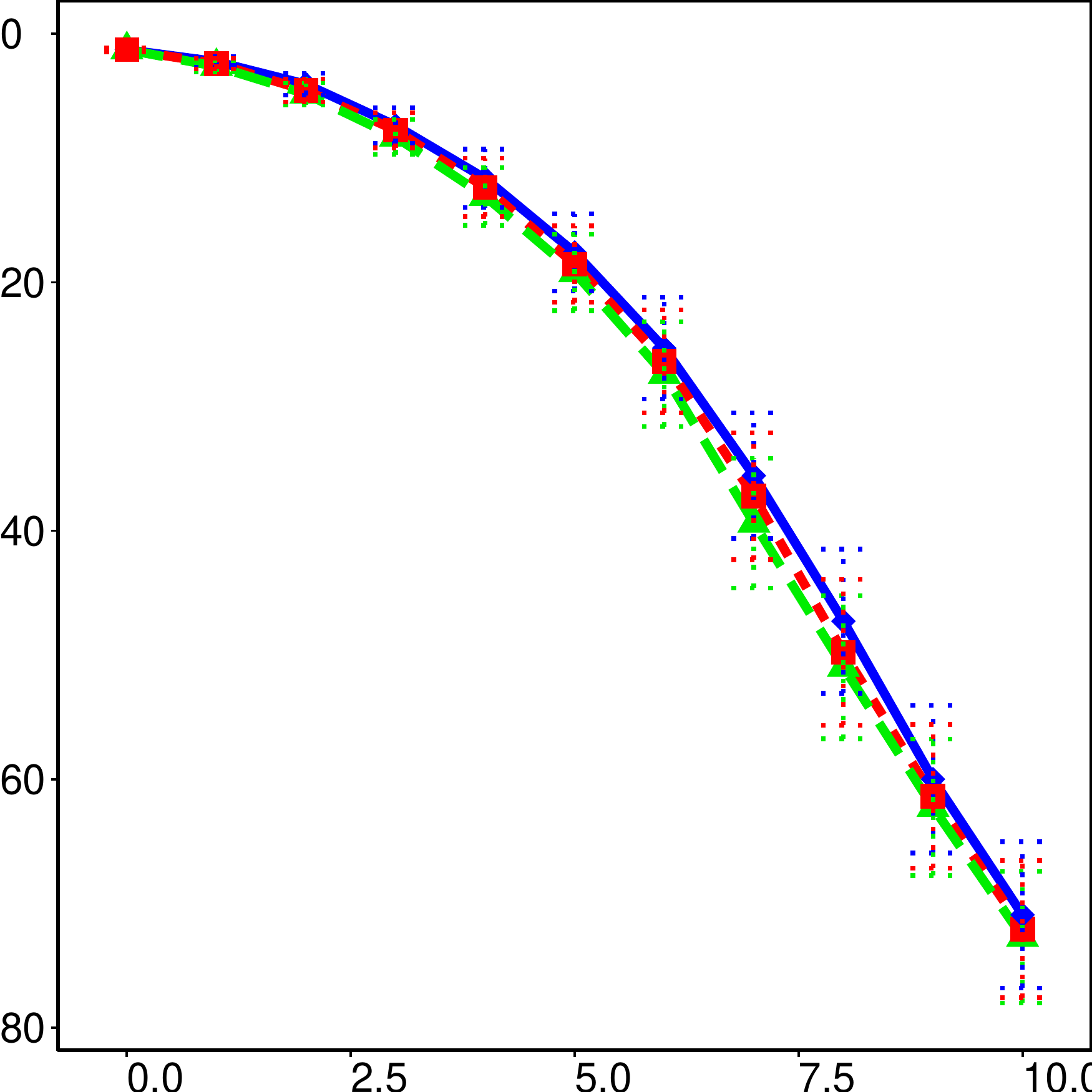} &
	\includegraphics[width=0.93\linewidth]{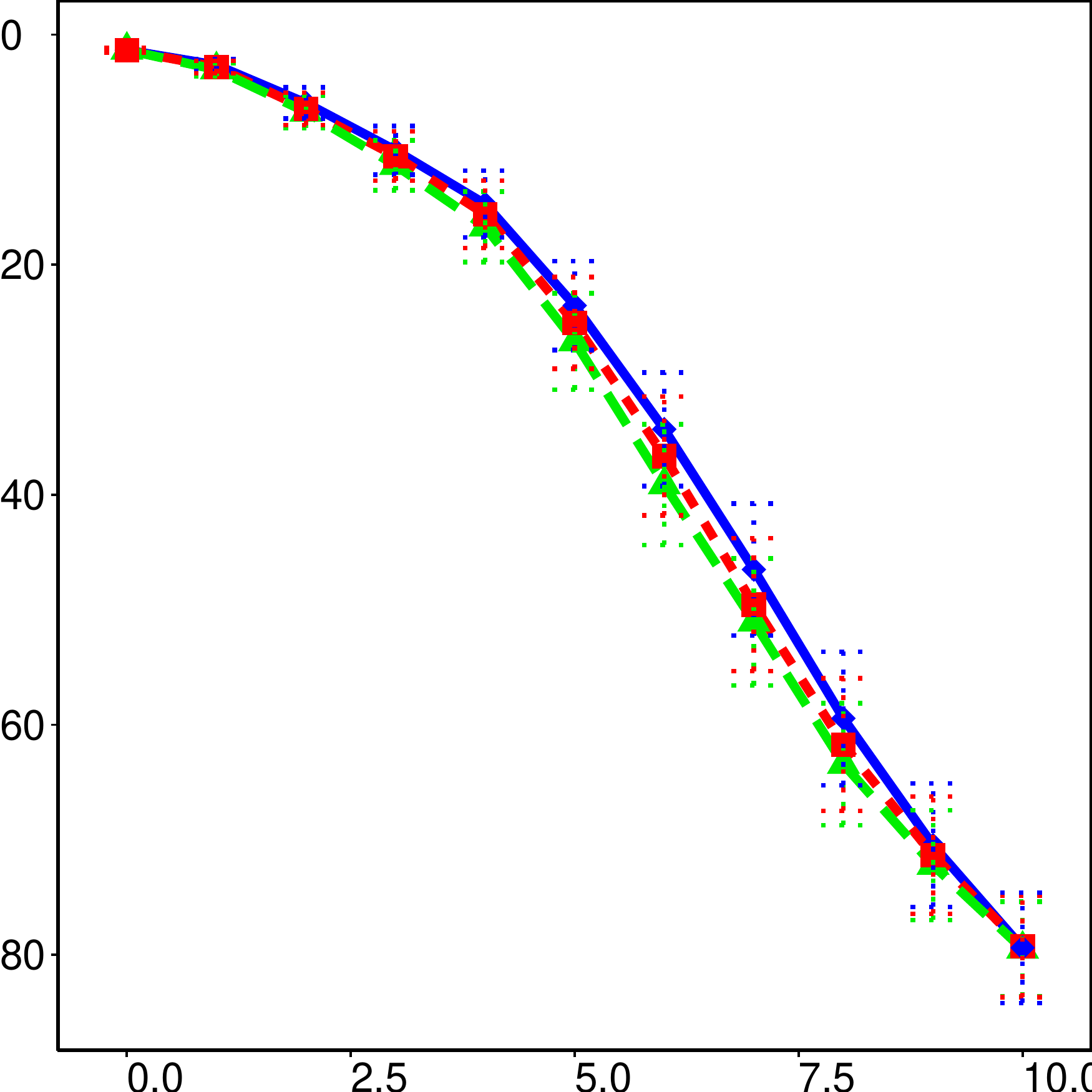} &
	\includegraphics[width=0.93\linewidth]{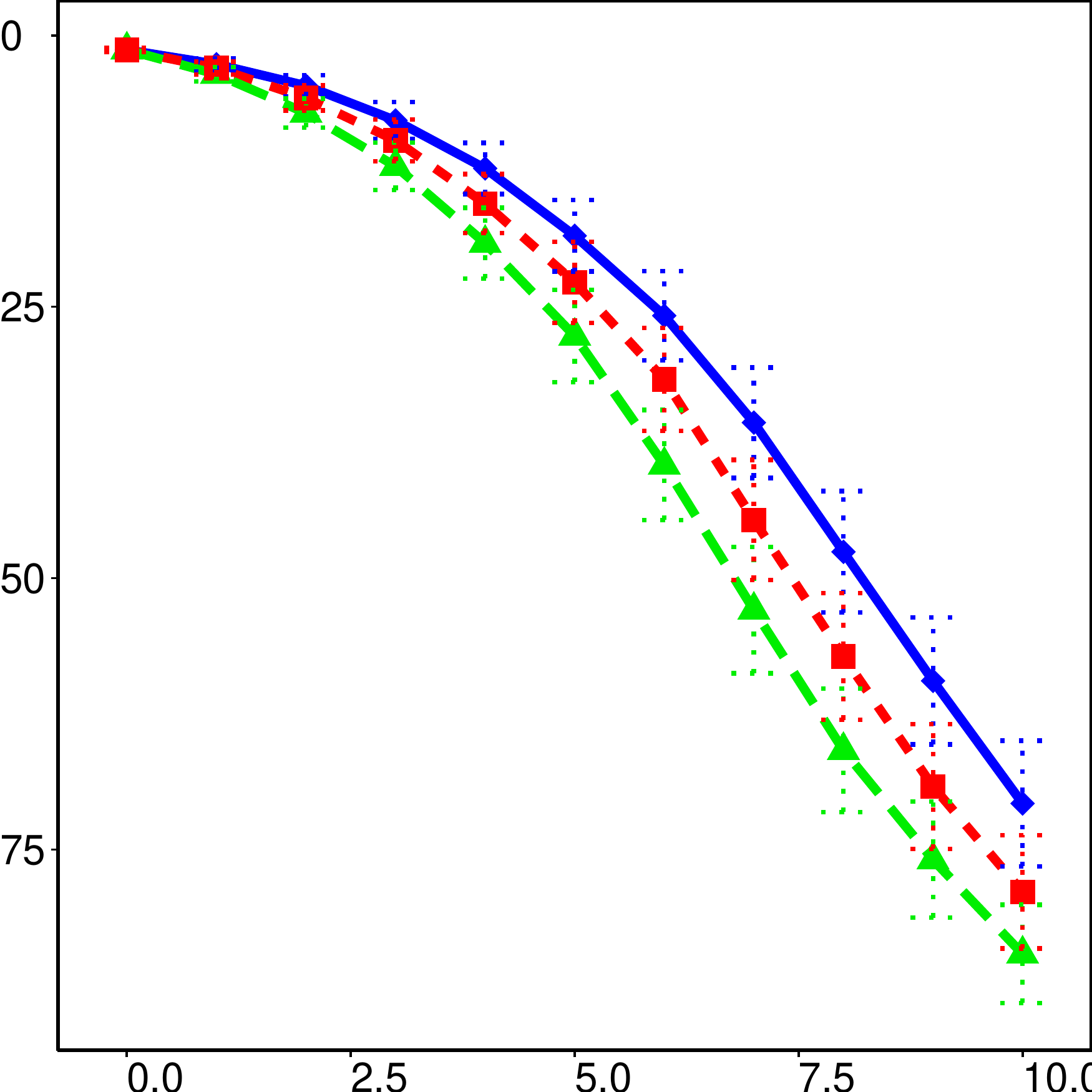} &
	\includegraphics[width=0.93\linewidth]{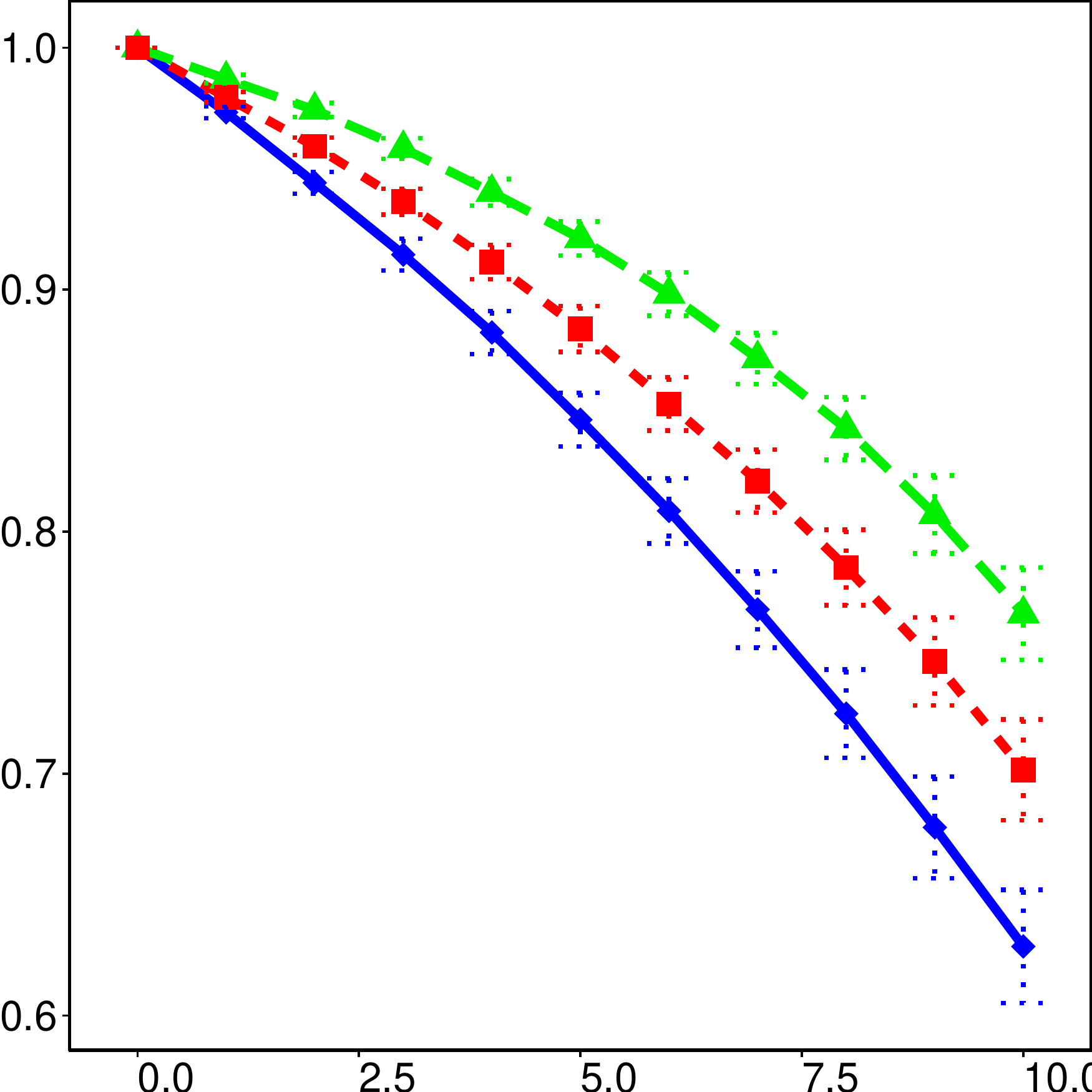} &
	\includegraphics[width=0.93\linewidth]{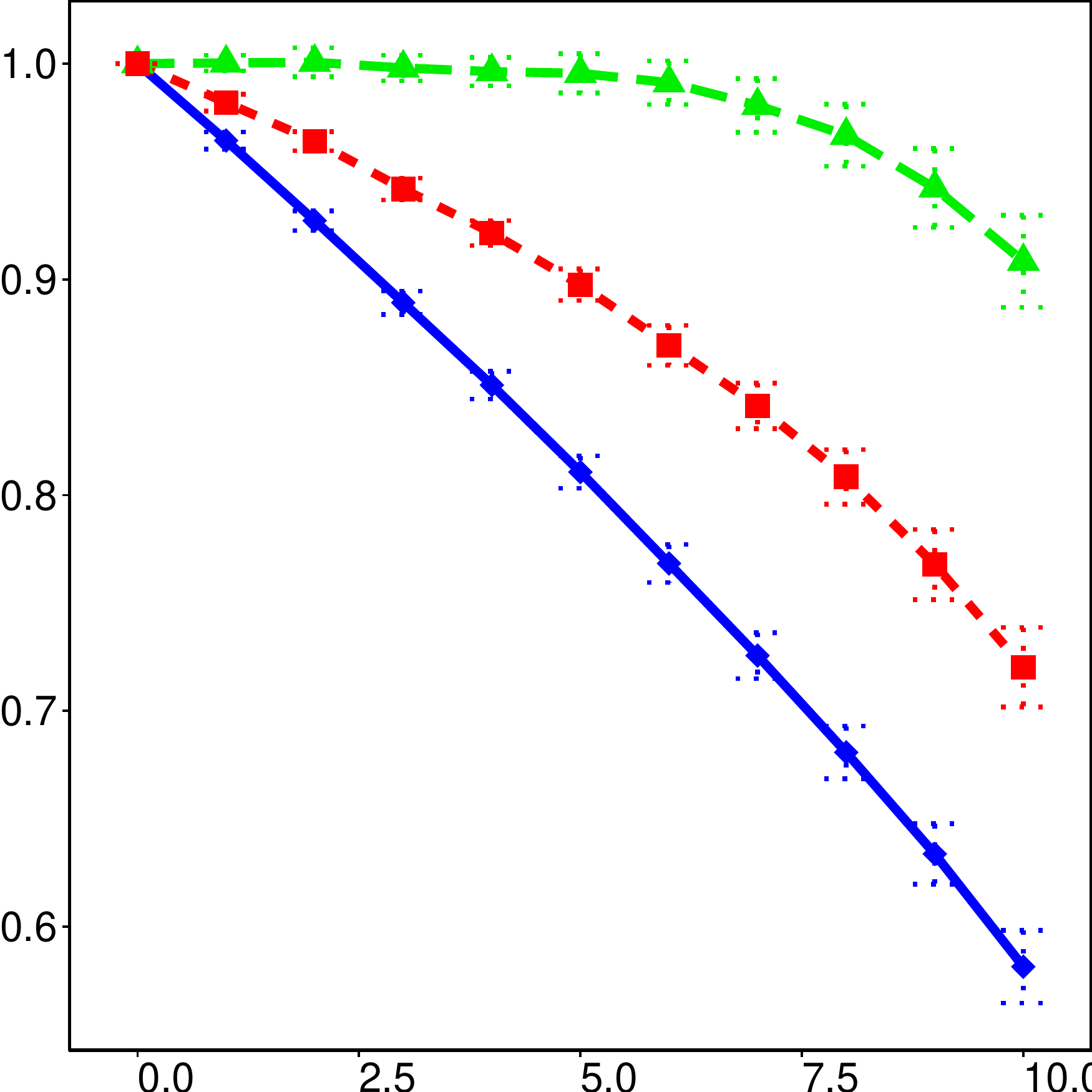} \\
	\rotatebox{90}{{\fontsize{6}{6}\selectfont{$\mathit{RandomGraph}(1000,10)$}}} &
	\includegraphics[width=0.93\linewidth]{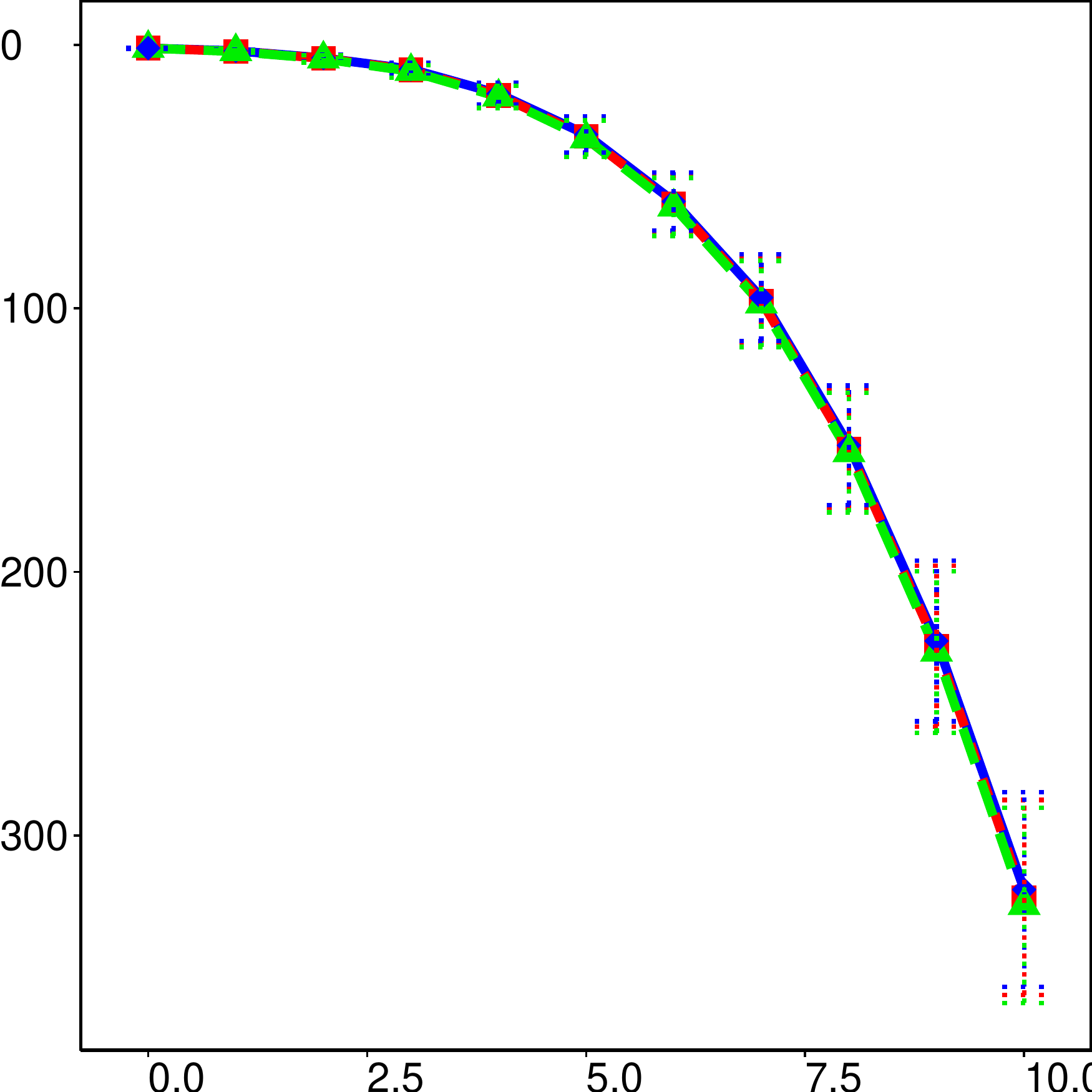} &
	\includegraphics[width=0.93\linewidth]{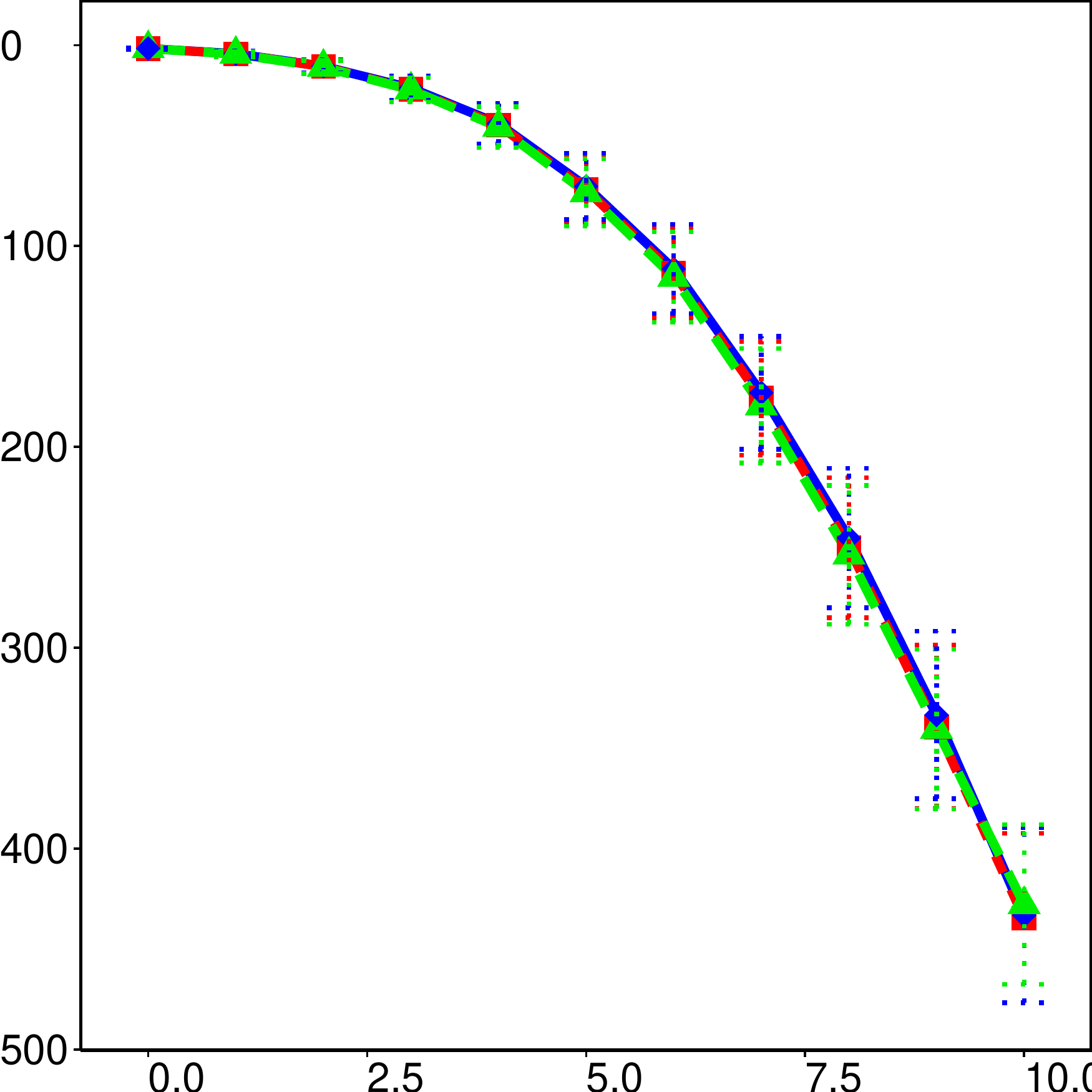} &
	\includegraphics[width=0.93\linewidth]{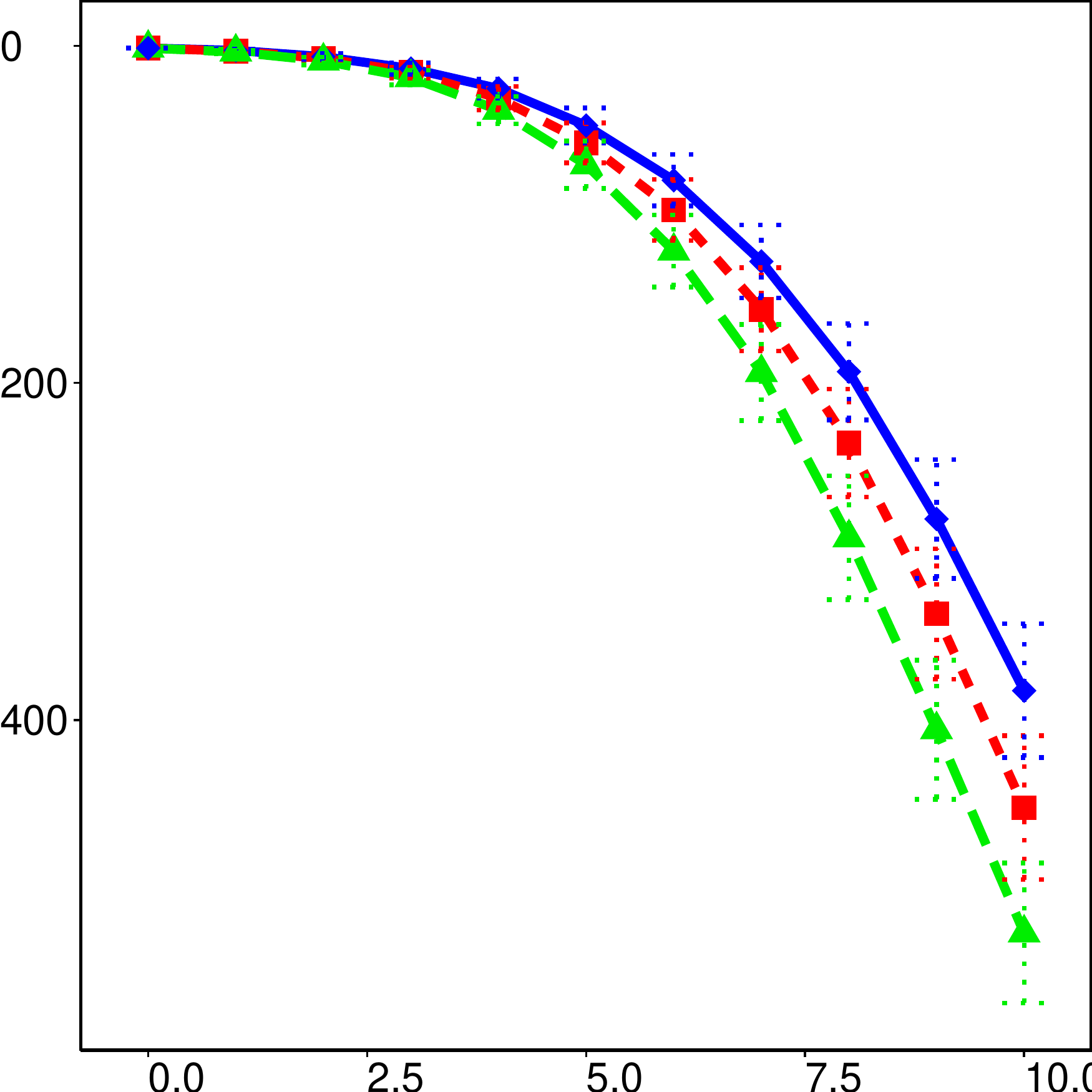} &
	\includegraphics[width=0.93\linewidth]{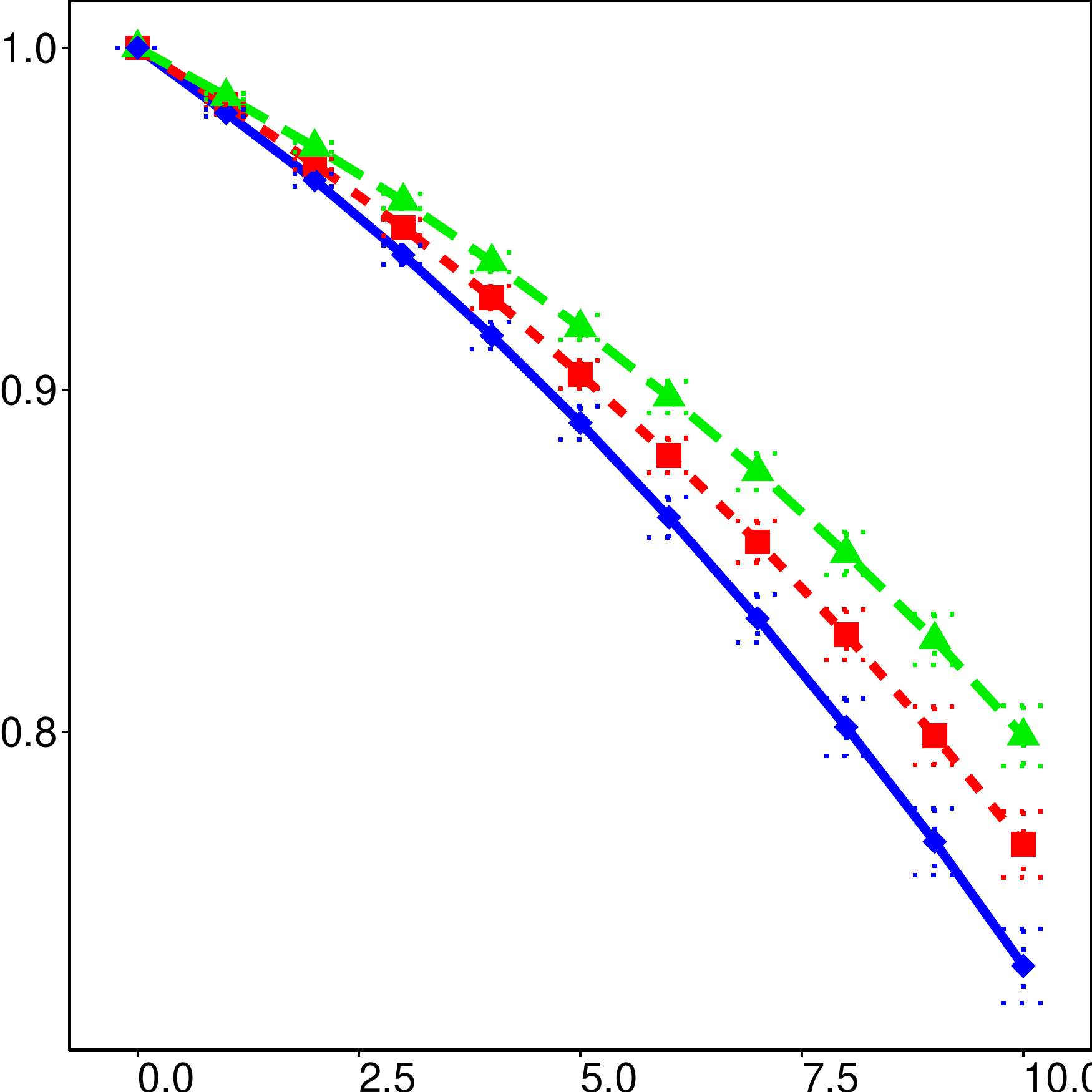} &
	\includegraphics[width=0.93\linewidth]{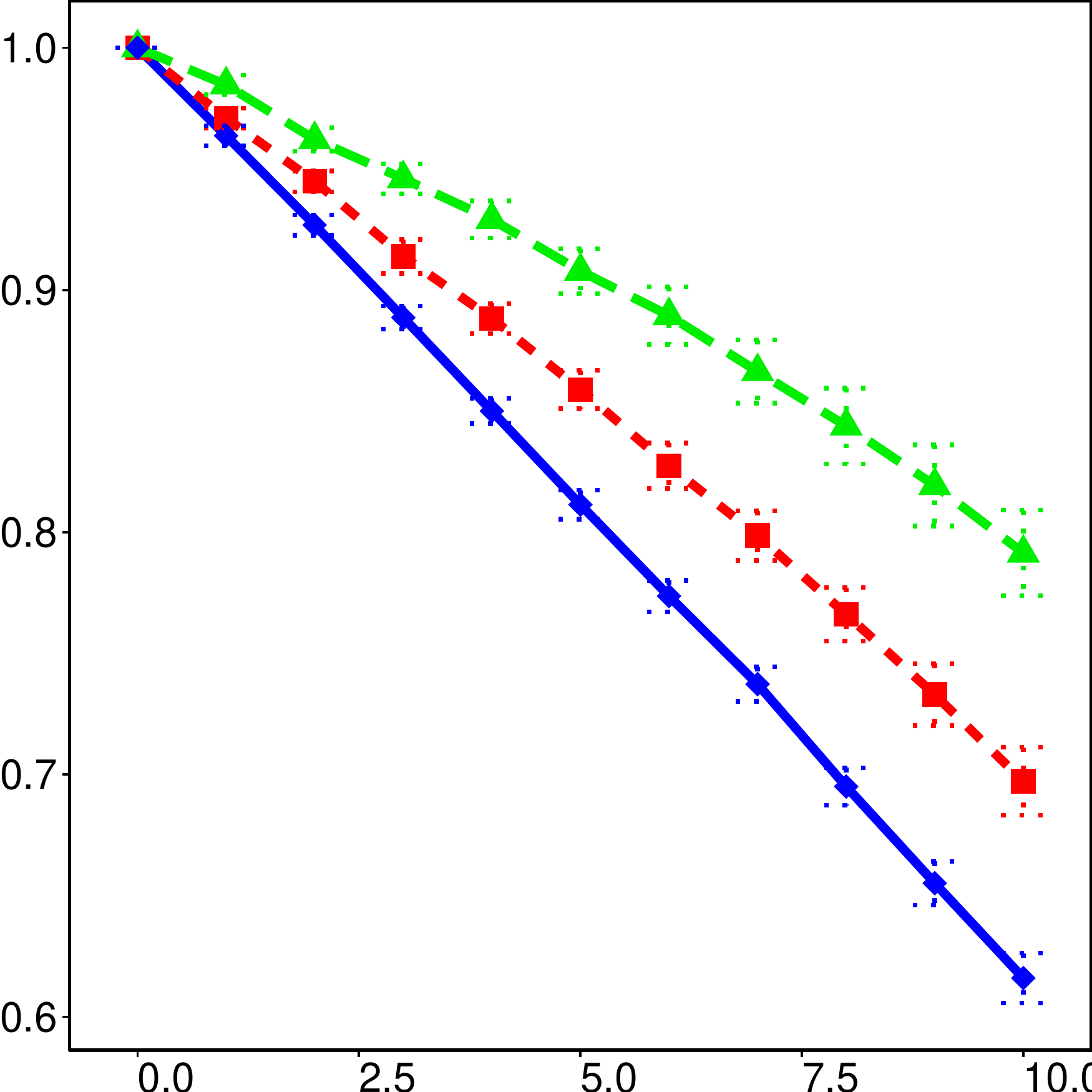} \\
	\rotatebox{90}{{\fontsize{6}{6}\selectfont{$\mathit{SmallWorld}(100,10,.25)$}}} &
	\includegraphics[width=0.93\linewidth]{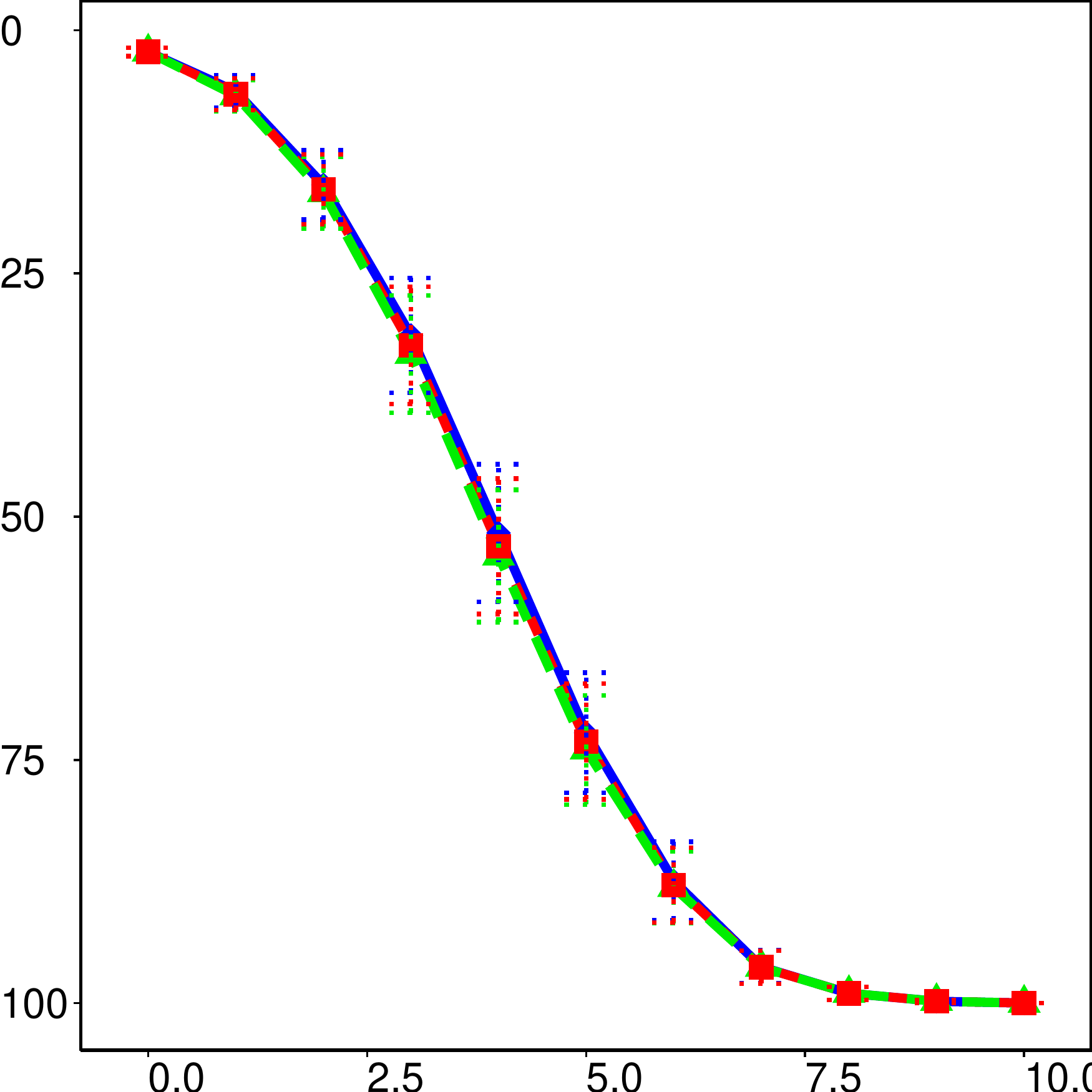} &
	\includegraphics[width=0.93\linewidth]{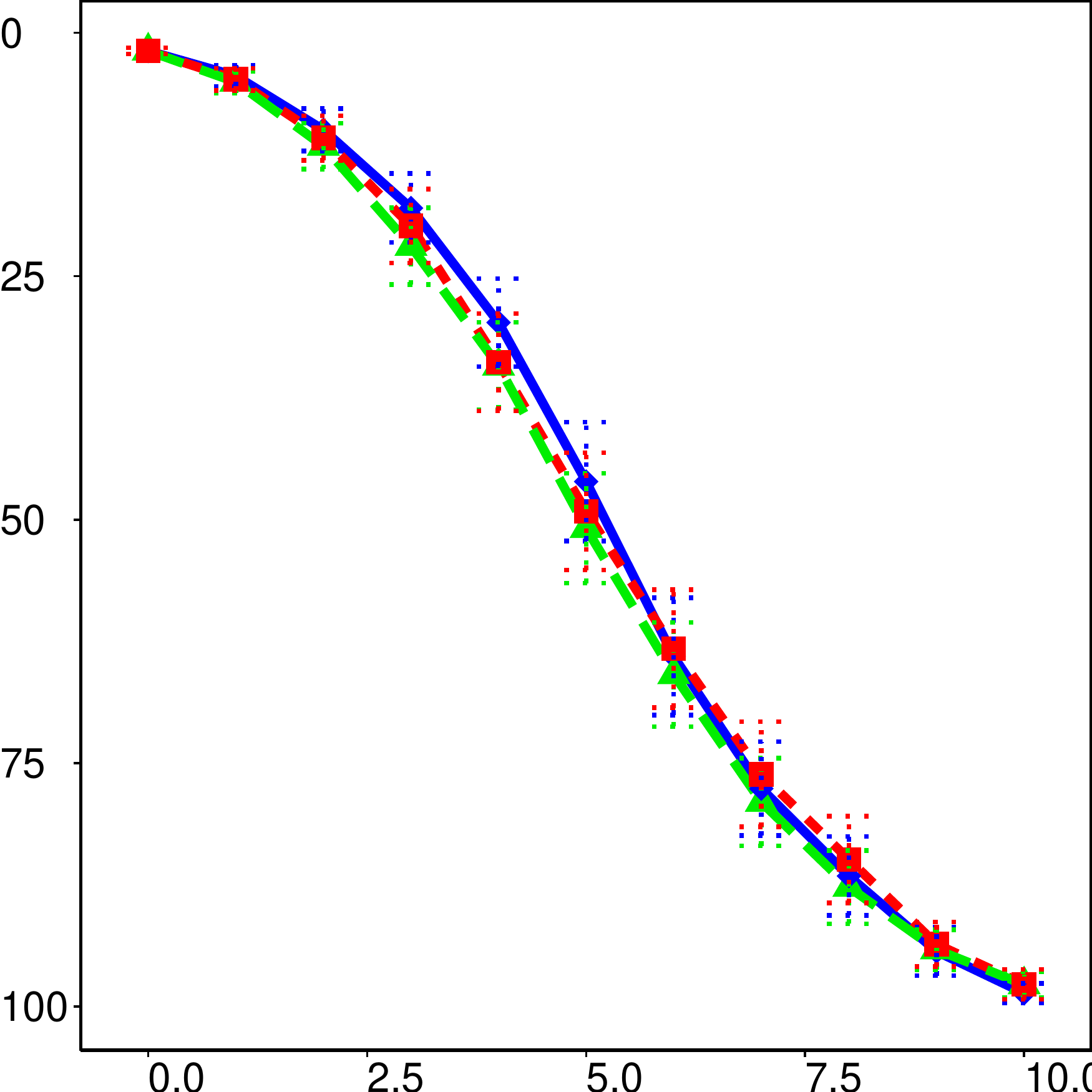} &
	\includegraphics[width=0.93\linewidth]{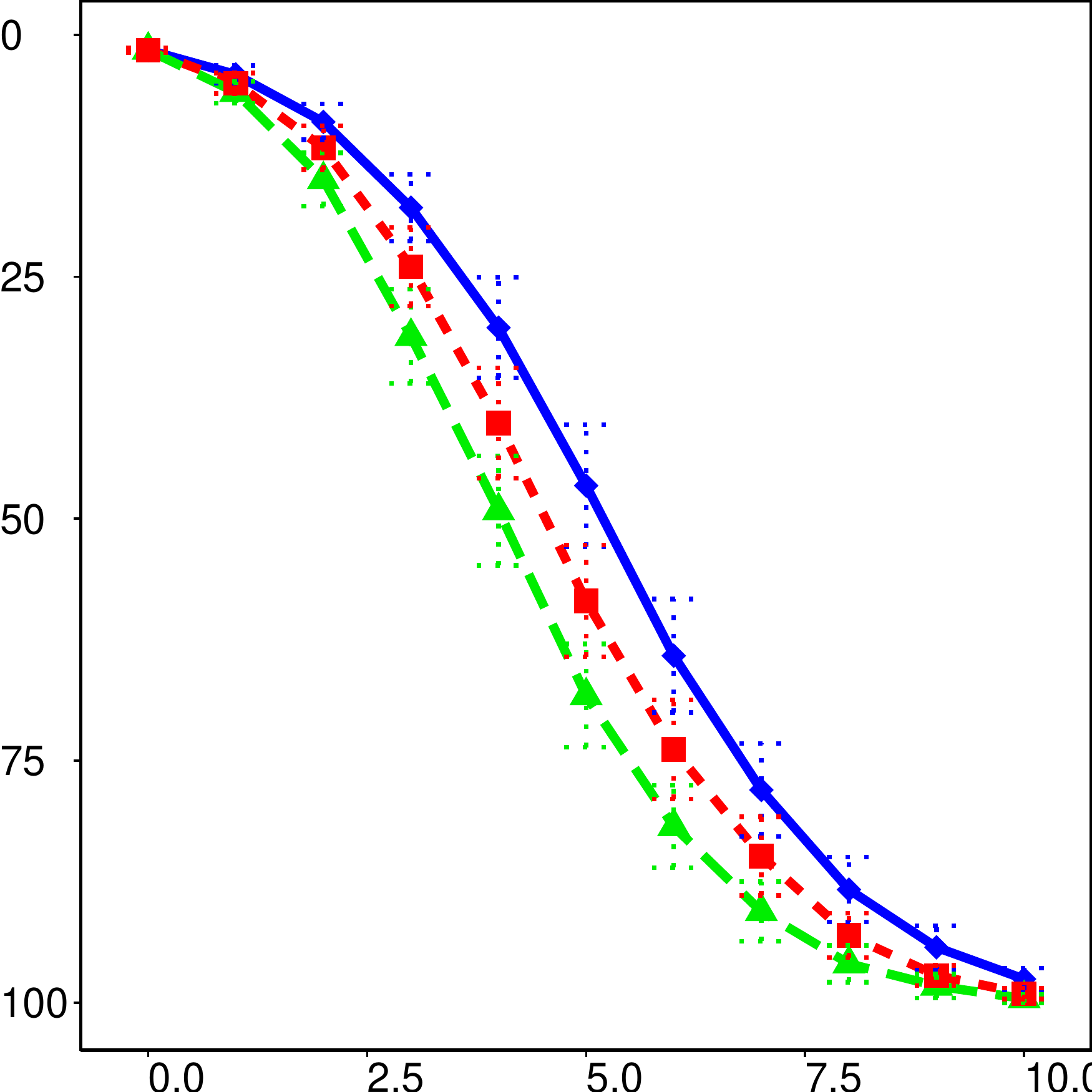} &
	\includegraphics[width=0.93\linewidth]{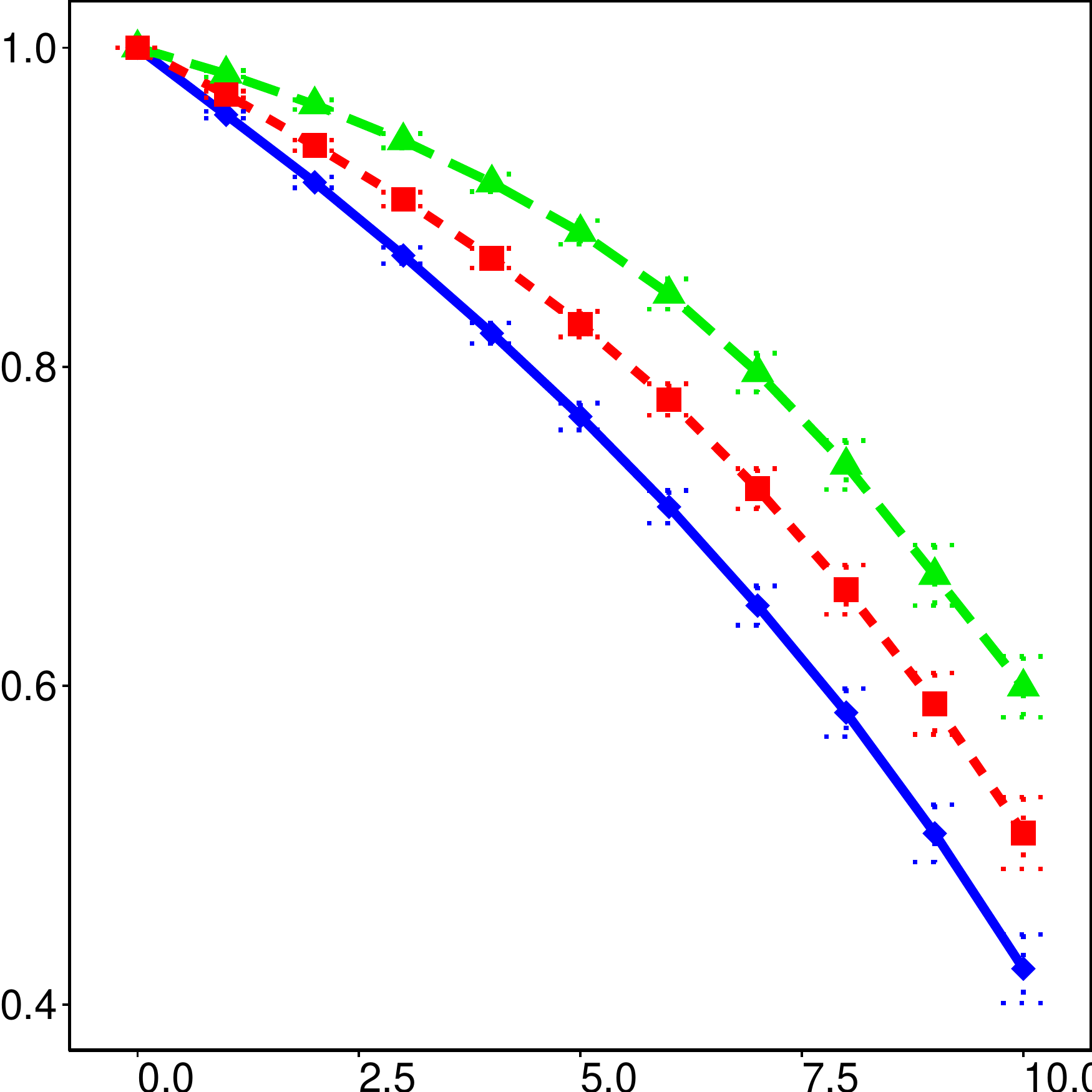} &
	\includegraphics[width=0.93\linewidth]{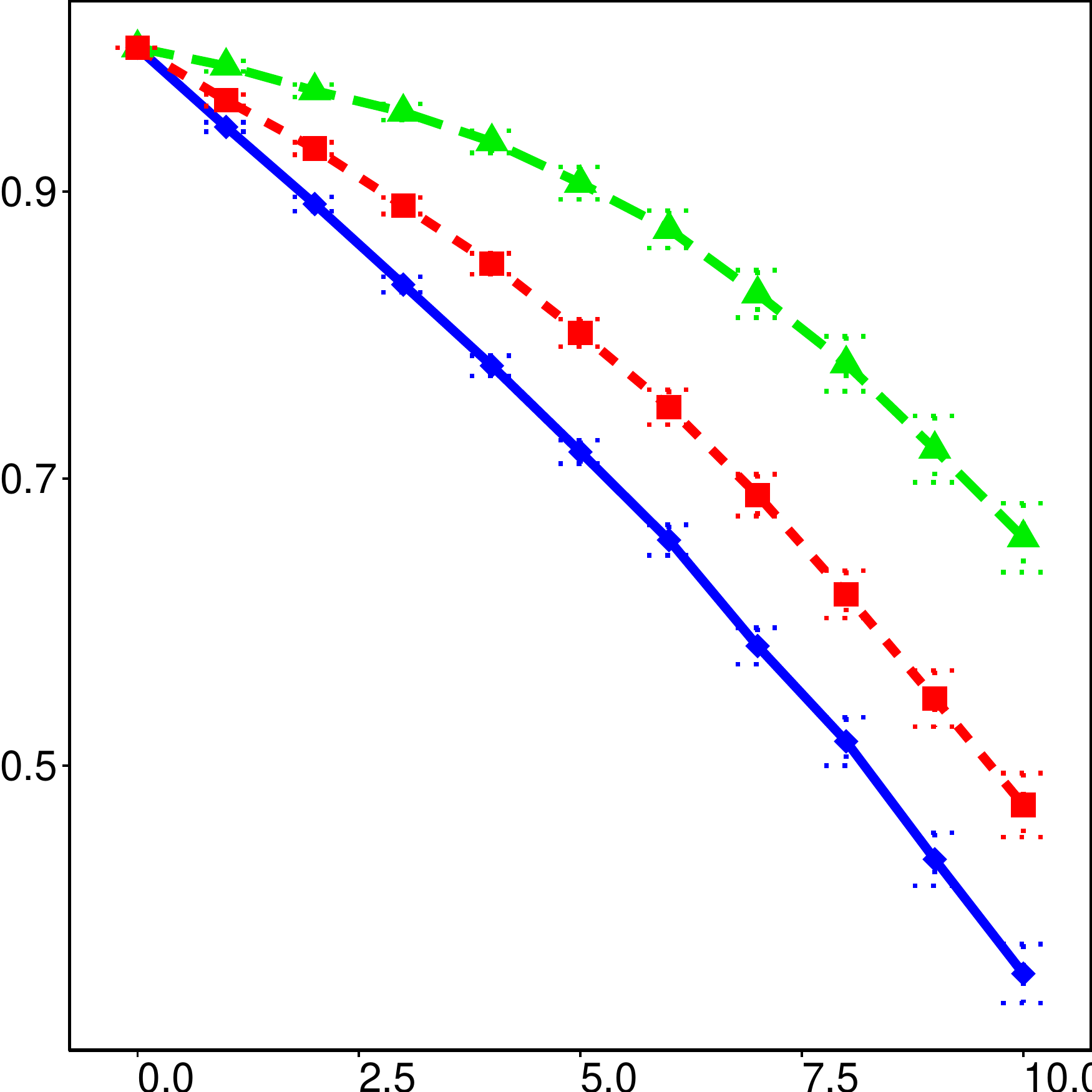} \\
	\rotatebox{90}{{\fontsize{6}{6}\selectfont{$\mathit{SmallWorld}(1000,10,.25)$}}} &
	\includegraphics[width=0.93\linewidth]{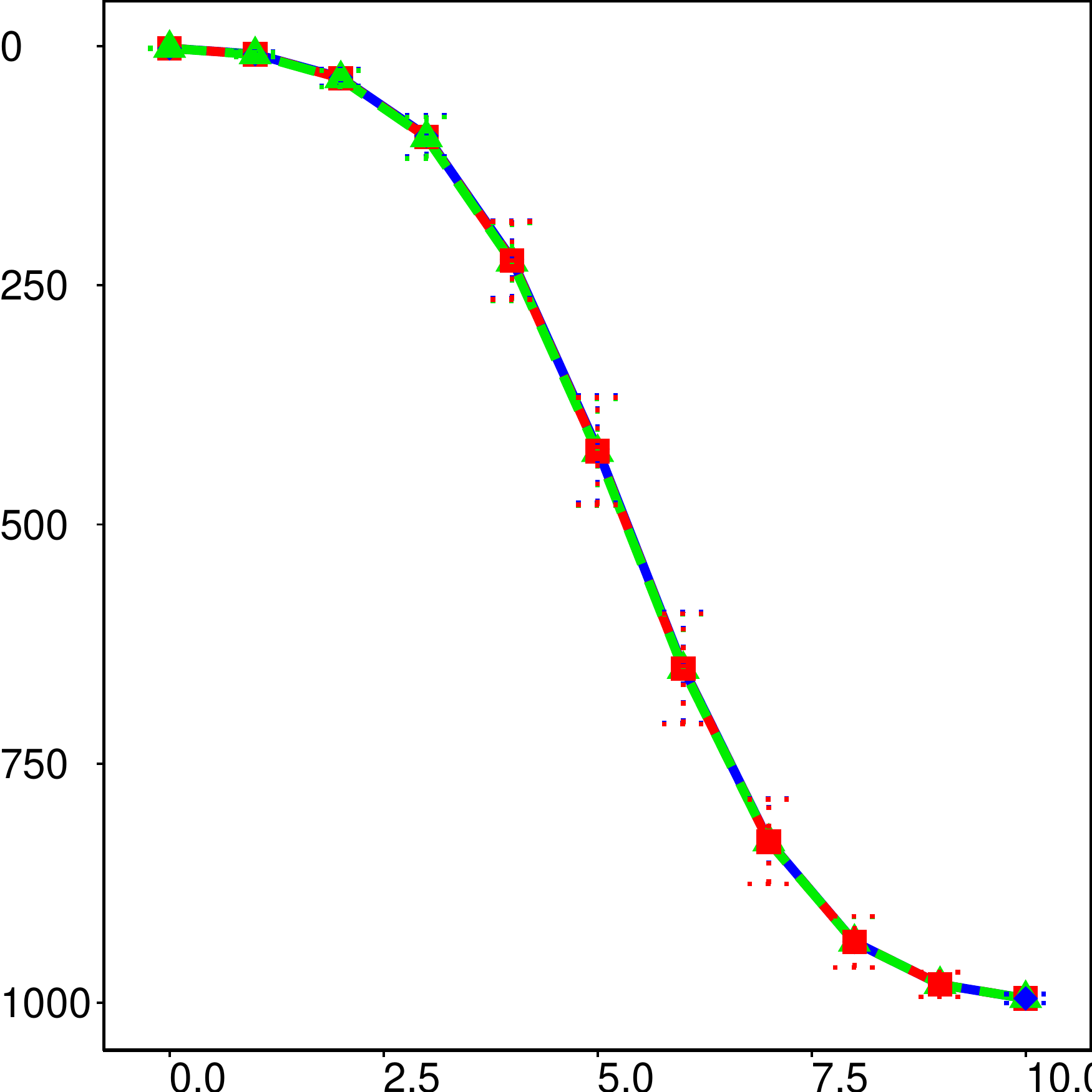} &
	\includegraphics[width=0.93\linewidth]{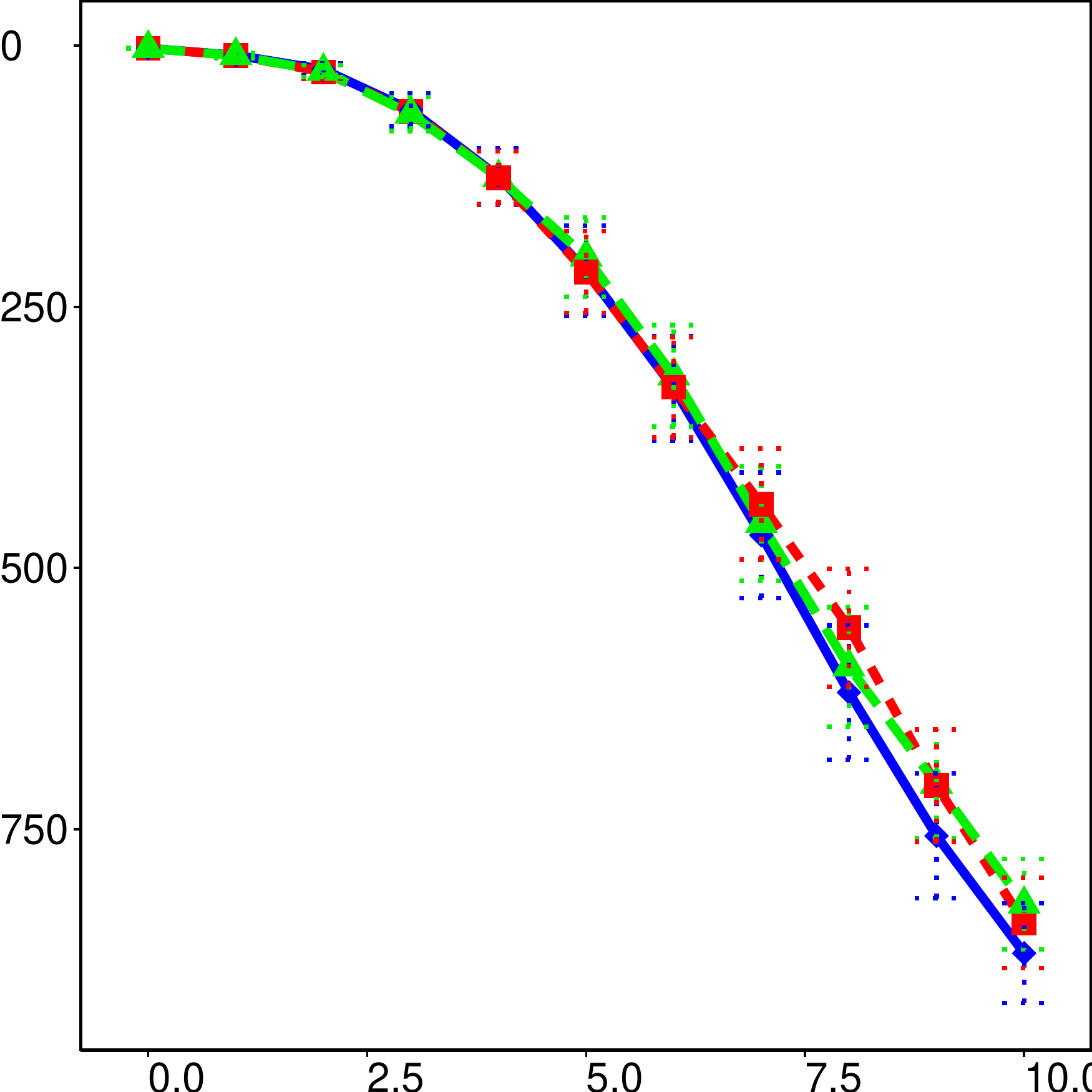} &
	\includegraphics[width=0.93\linewidth]{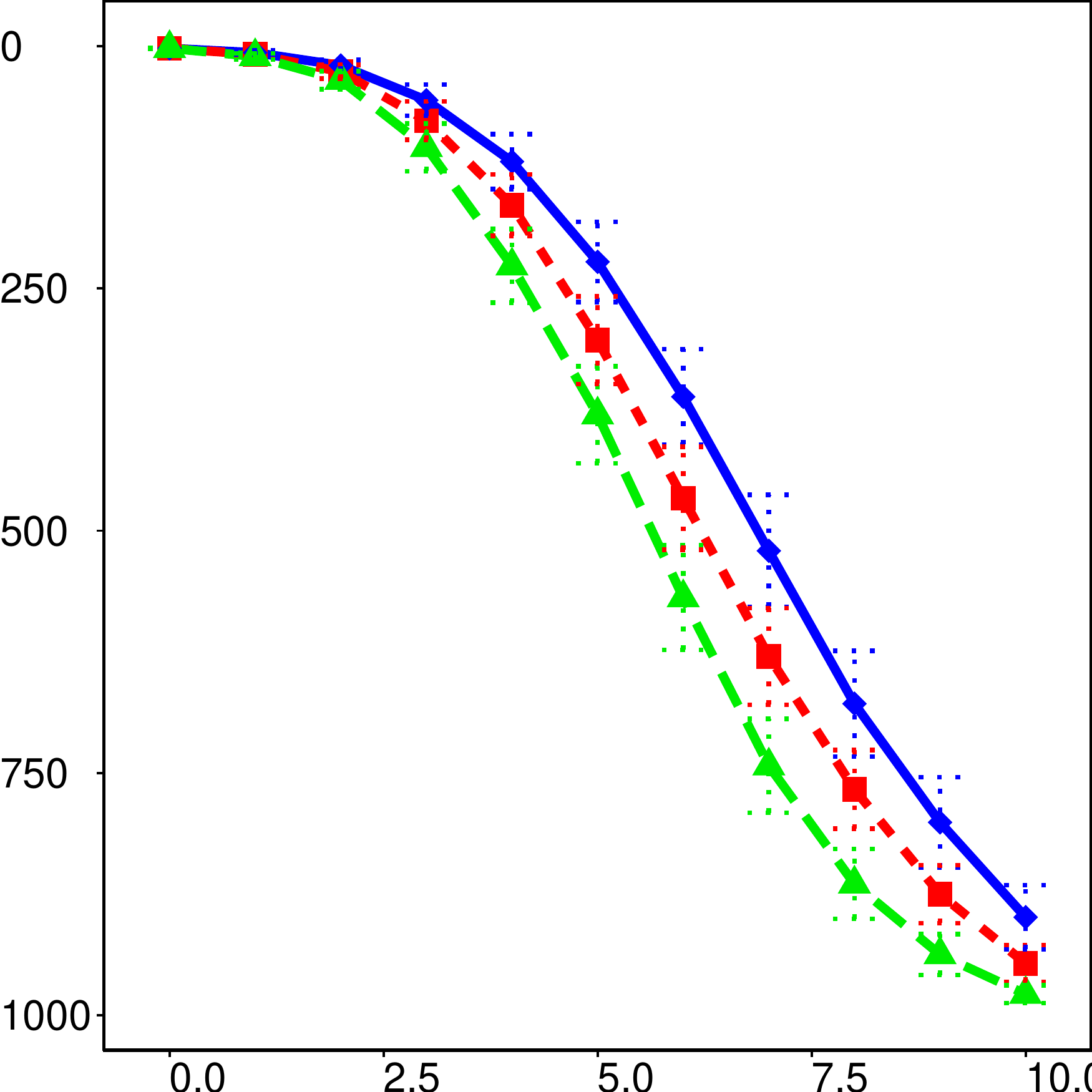} &
	\includegraphics[width=0.93\linewidth]{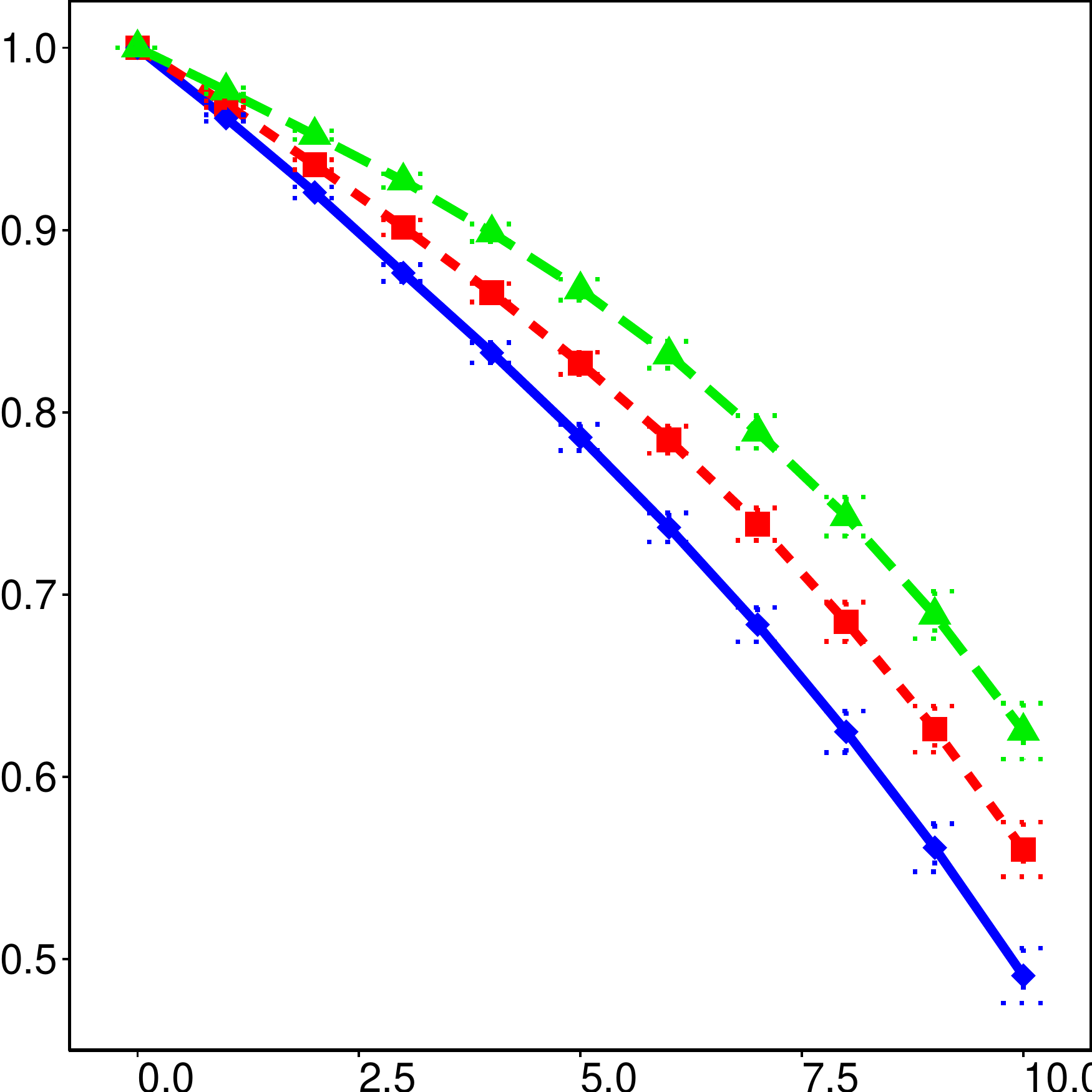} &
	\includegraphics[width=0.93\linewidth]{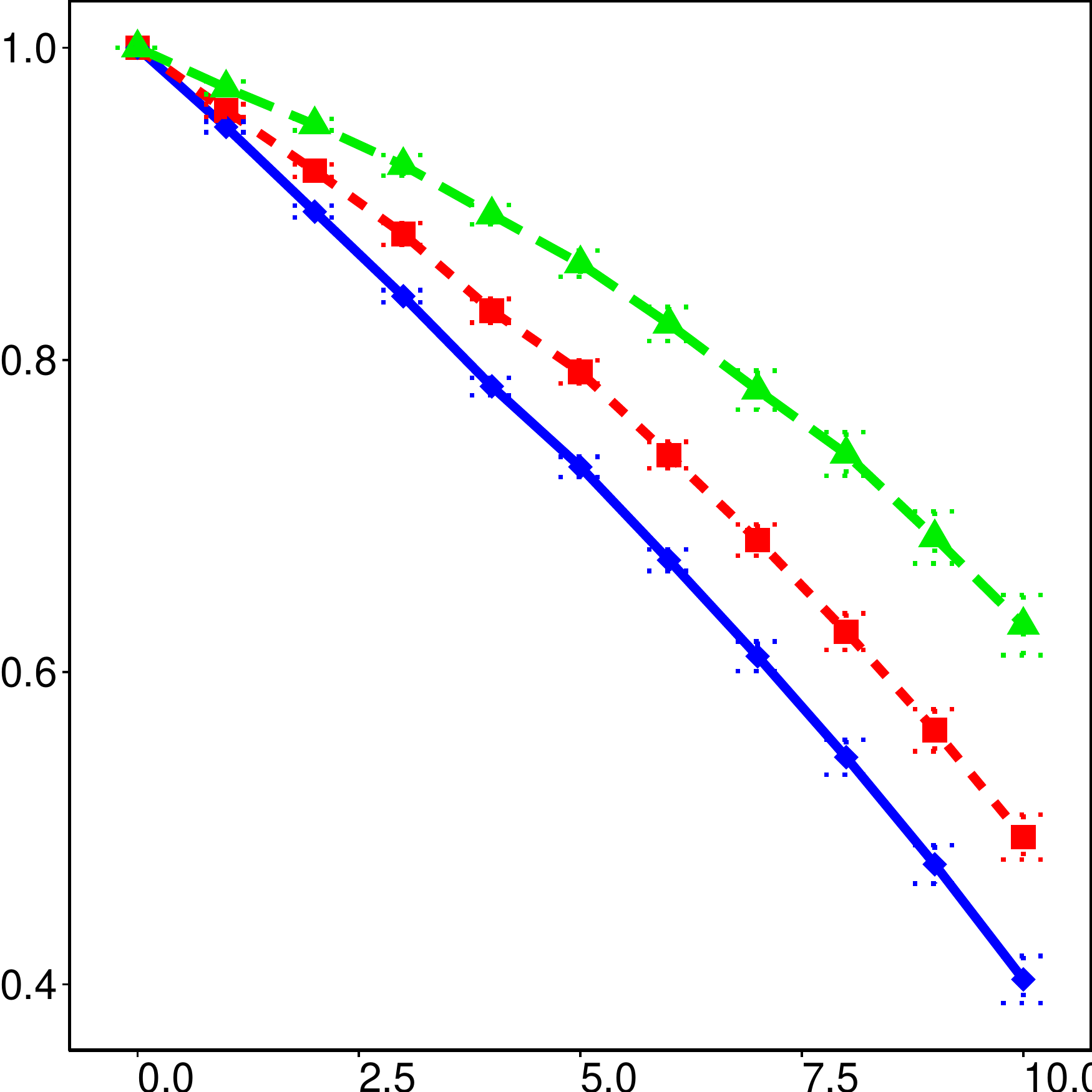} \\
	&
	\multicolumn{5}{c}{\includegraphics[width=0.75\linewidth]{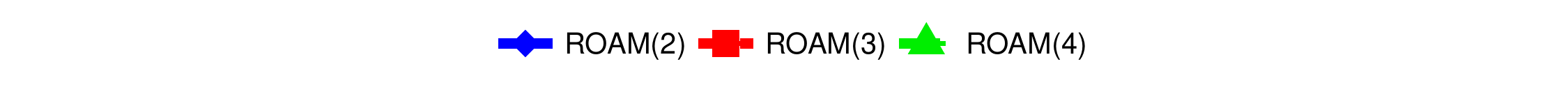}} \\
	\end{tabular}
	\vspace*{-0.5cm}
	\caption{\fontsize{10}{10}\selectfont{Consecutive execution of ROAM (the $x$-axis represents the number of executions). Specifically, given different random networks, the subfigures show the source node's ranking (according to the centrality measures), and the relative change in its influence value (according to the influence models). Results are shown for $\textnormal{ROAM}(b):b=2, 3, 4$, where $b$ is the budget in each execution.}}
	\label{fig:centralityResults:random}
\end{figure}

\begin{figure}[p!]
	\centering.
	\setlength\tabcolsep{2pt}
	\renewcommand{\arraystretch}{2.0}
	\begin{tabular}{m{.01\textwidth}m{.195\textwidth}m{.195\textwidth}m{.195\textwidth}m{.195\textwidth}m{.195\textwidth}}
	&
	\multicolumn{1}{c}{\fontsize{8.5}{8.5}\selectfont{Degree Ranking}} &
	\multicolumn{1}{c}{\fontsize{8.5}{8.5}\selectfont{Closeness Ranking}} &
	\multicolumn{1}{c}{\fontsize{8.5}{8.5}\selectfont{Betweenness Ranking}} &
	\multicolumn{1}{c}{\fontsize{8.5}{8.5}\selectfont{IC Influence}} &
	\multicolumn{1}{c}{\fontsize{8.5}{8.5}\selectfont{LT Influence}} \\
	\rotatebox{90}{\small{Bali attack}} &
	\includegraphics[width=0.95\linewidth]{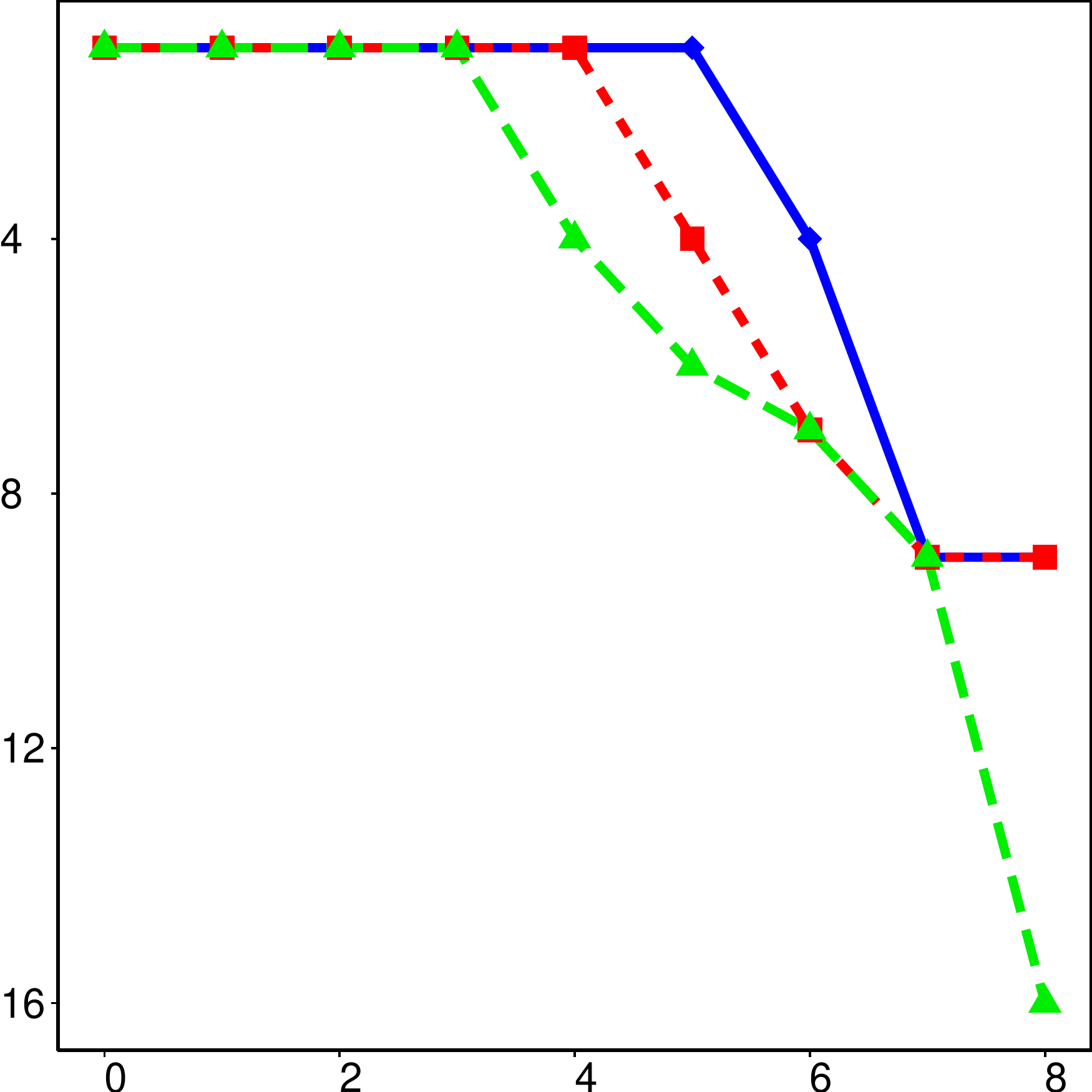} &
	\includegraphics[width=0.93\linewidth]{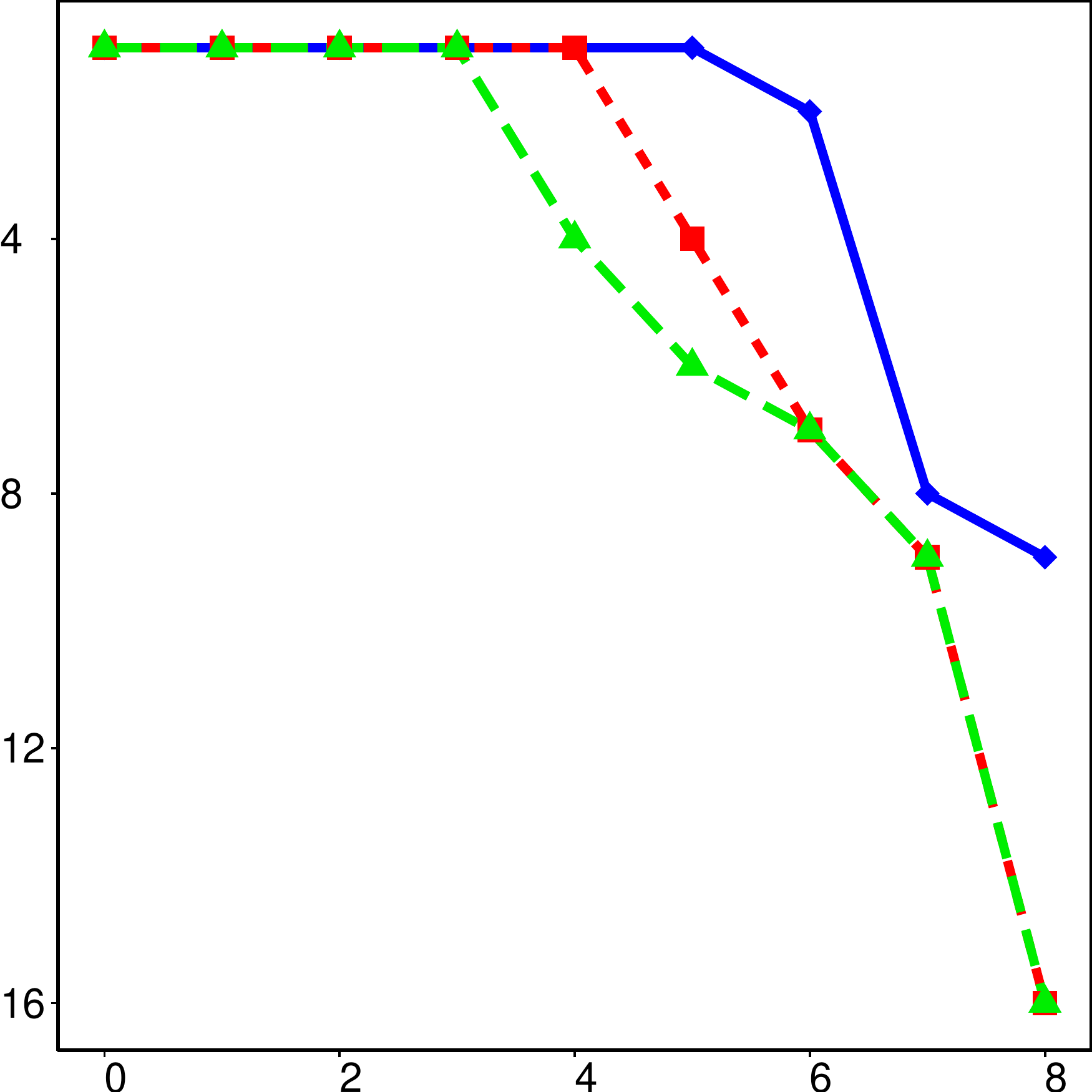} &
	\includegraphics[width=0.93\linewidth]{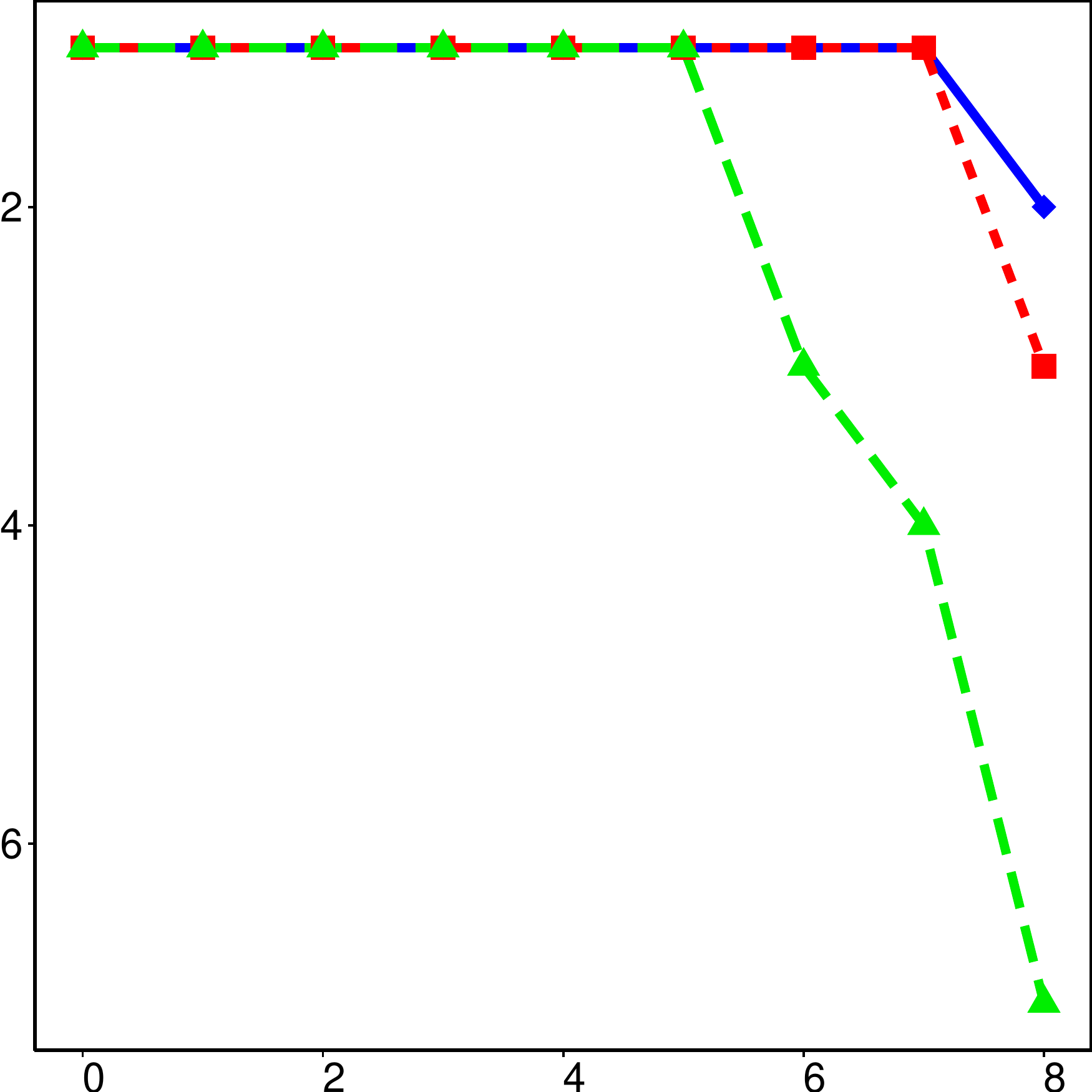} &
	\includegraphics[width=0.93\linewidth]{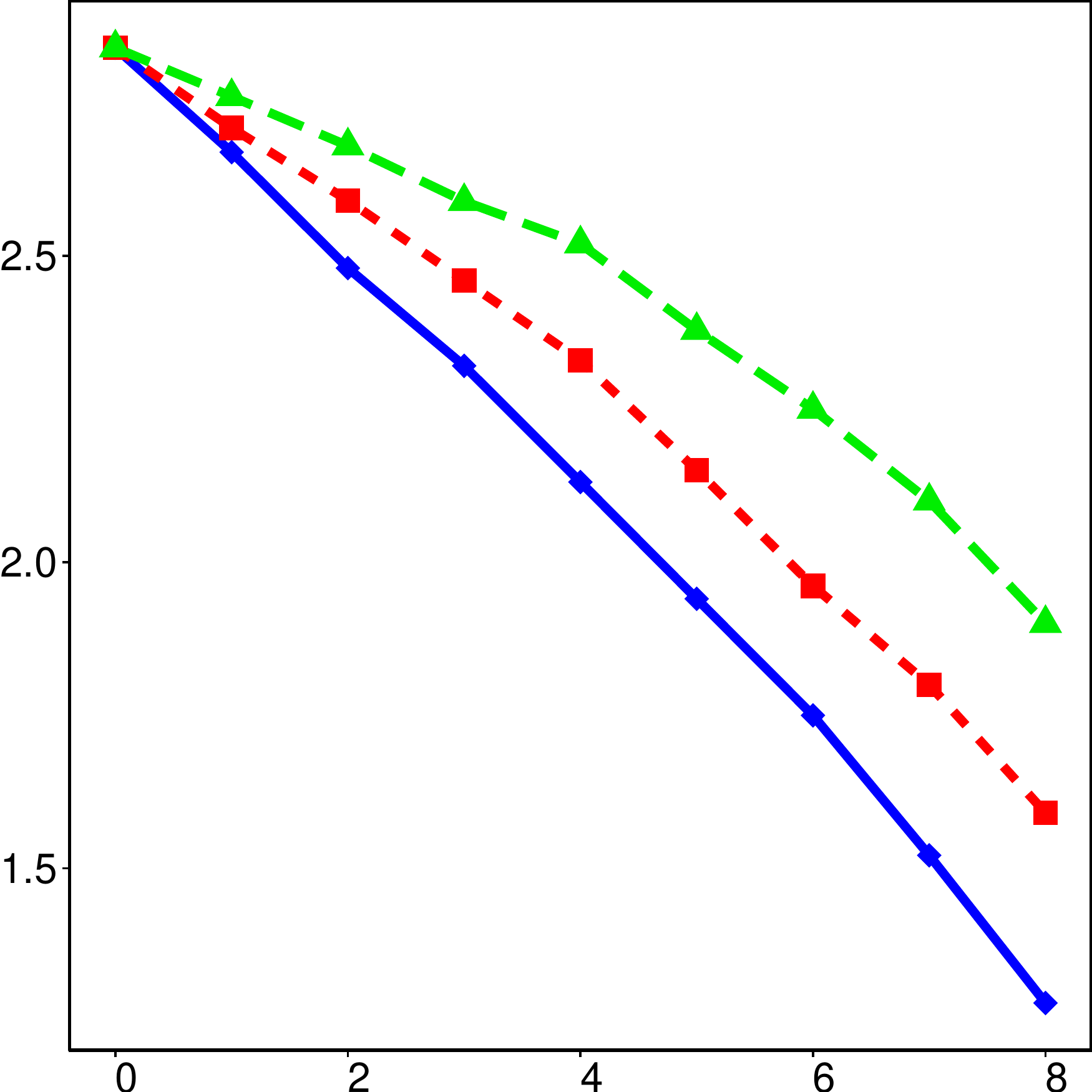} &
	\includegraphics[width=0.93\linewidth]{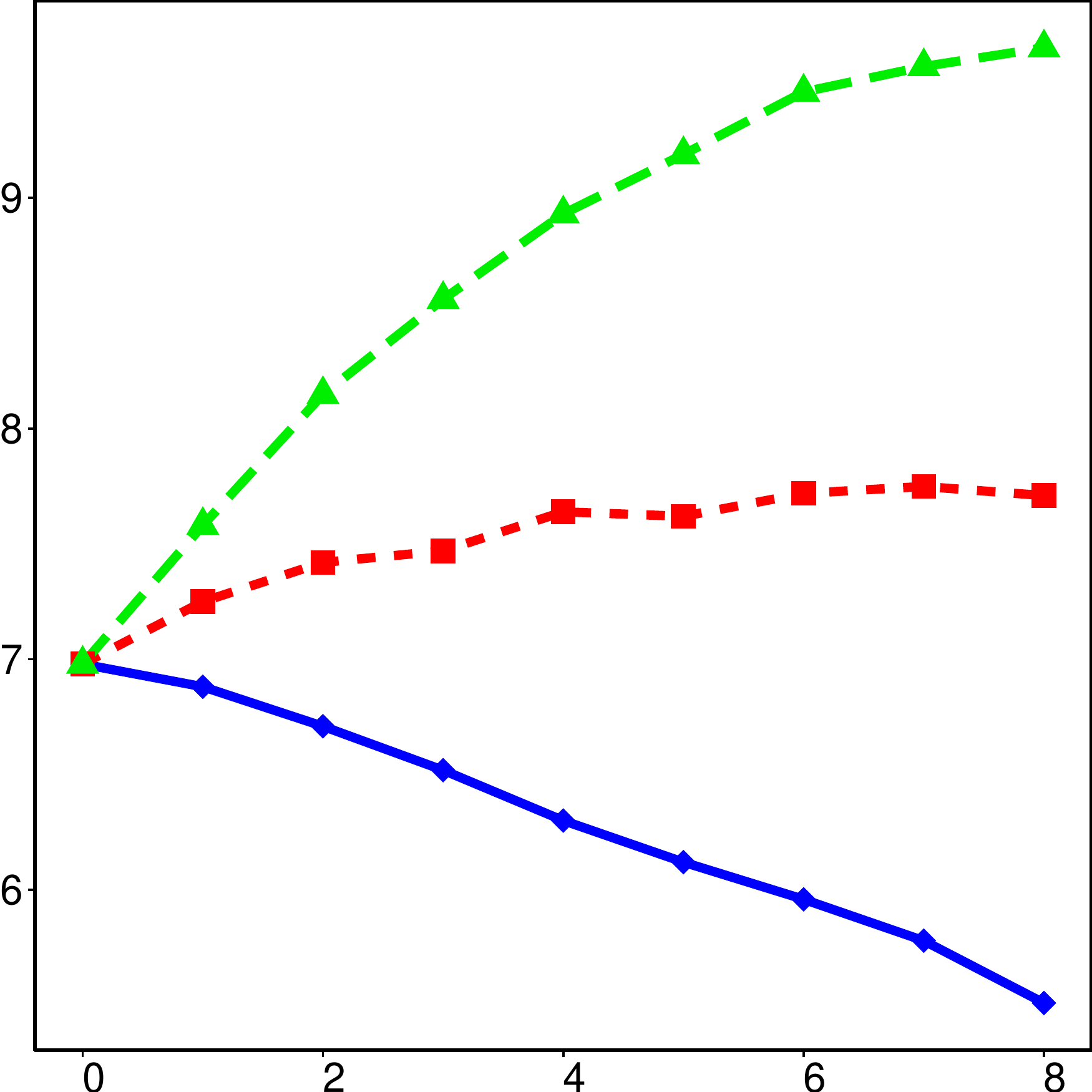} \\
	\rotatebox{90}{\small{Madrid bombing}} &
	\includegraphics[width=0.93\linewidth]{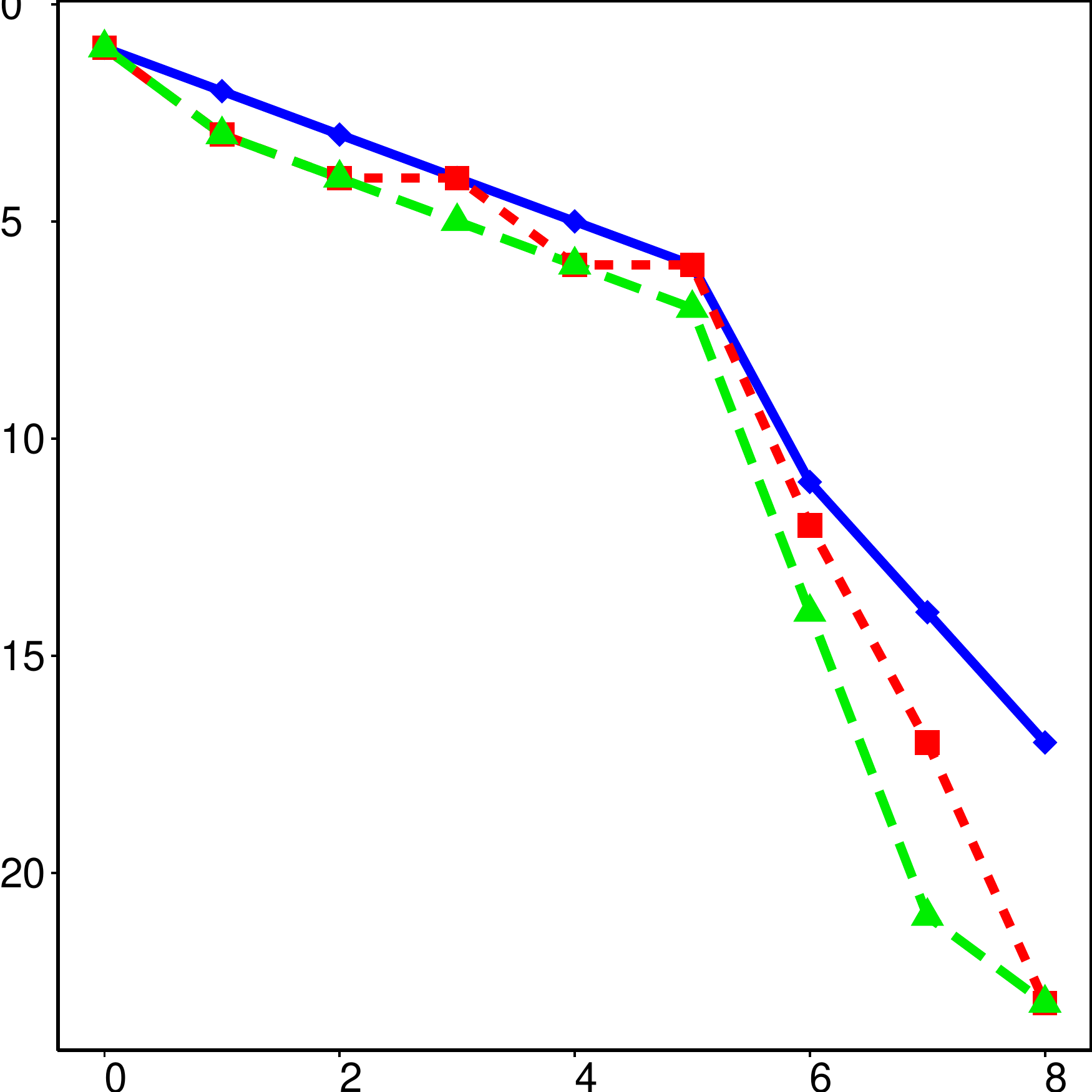} &
	\includegraphics[width=0.93\linewidth]{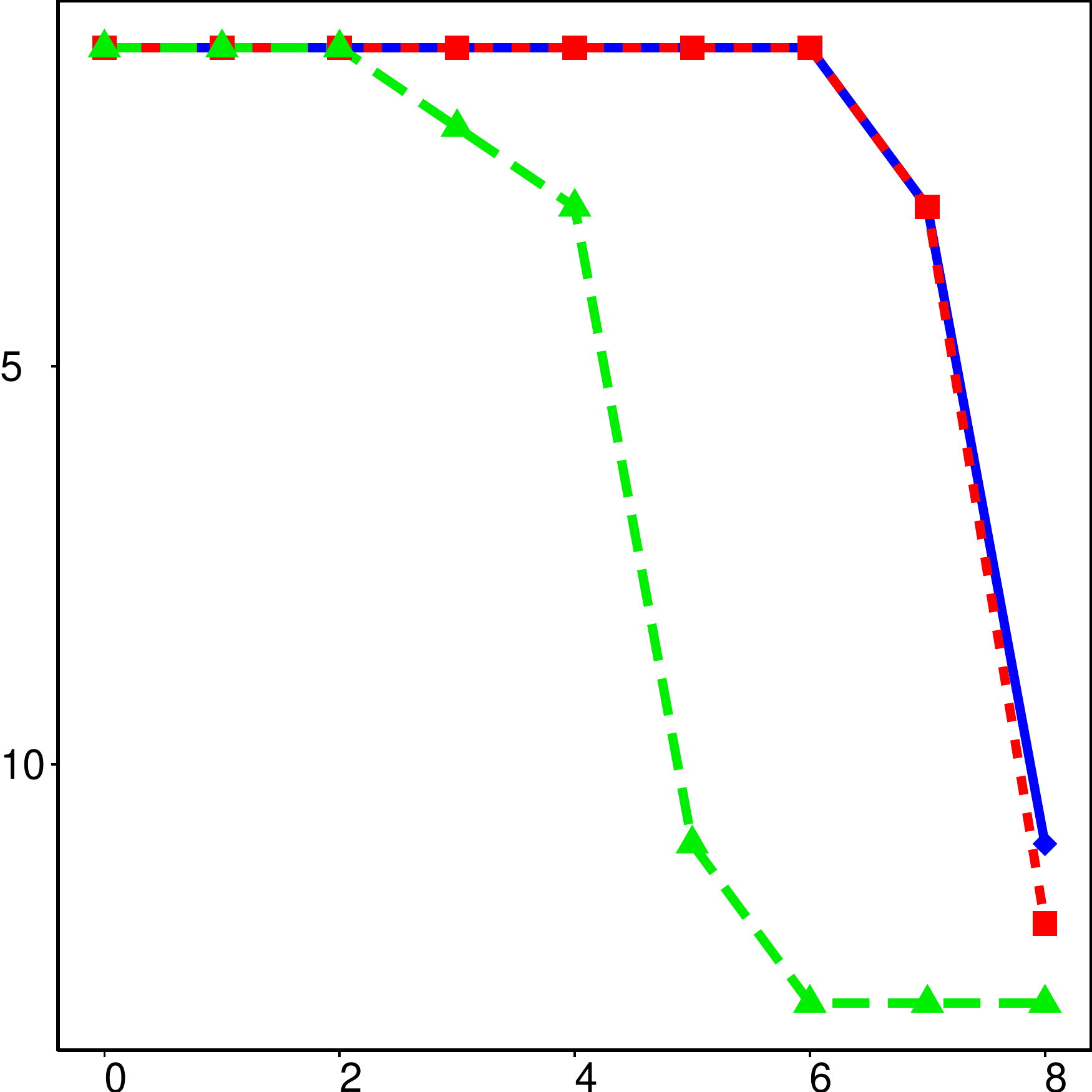} &
	\includegraphics[width=0.93\linewidth]{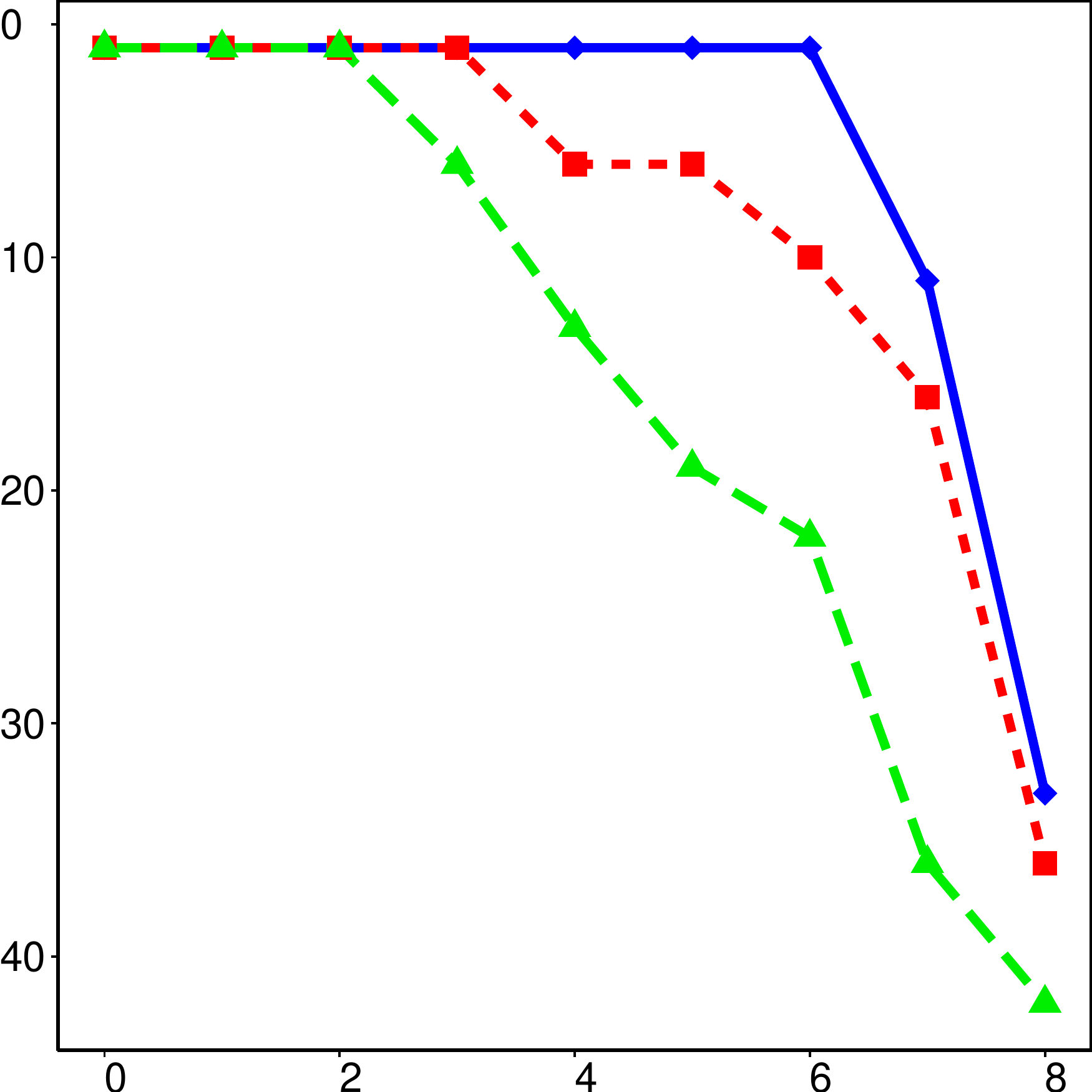} &
	\includegraphics[width=0.93\linewidth]{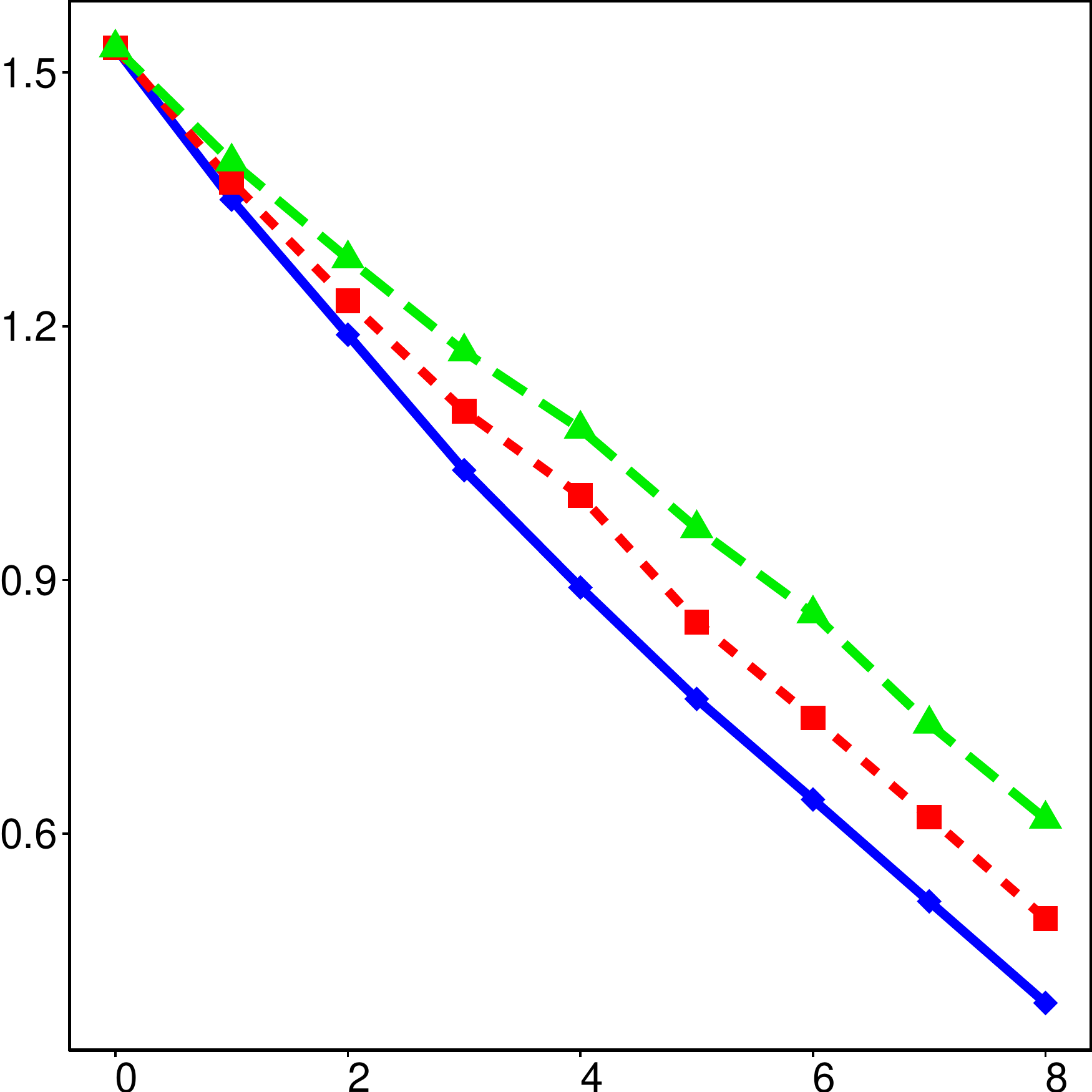} &
	\includegraphics[width=0.93\linewidth]{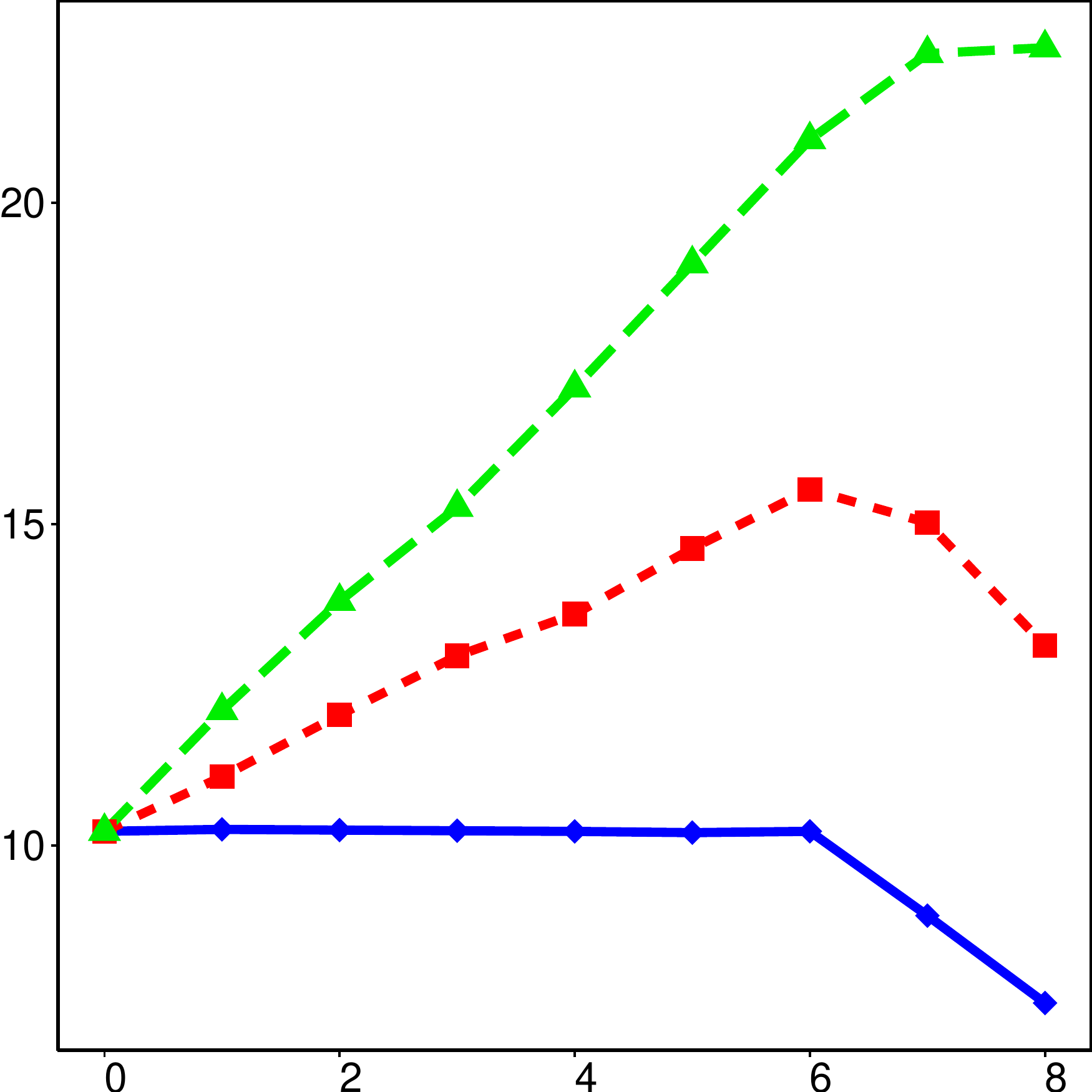} \\
	\rotatebox{90}{\small{WTC 9/11}} &
	\includegraphics[width=0.93\linewidth]{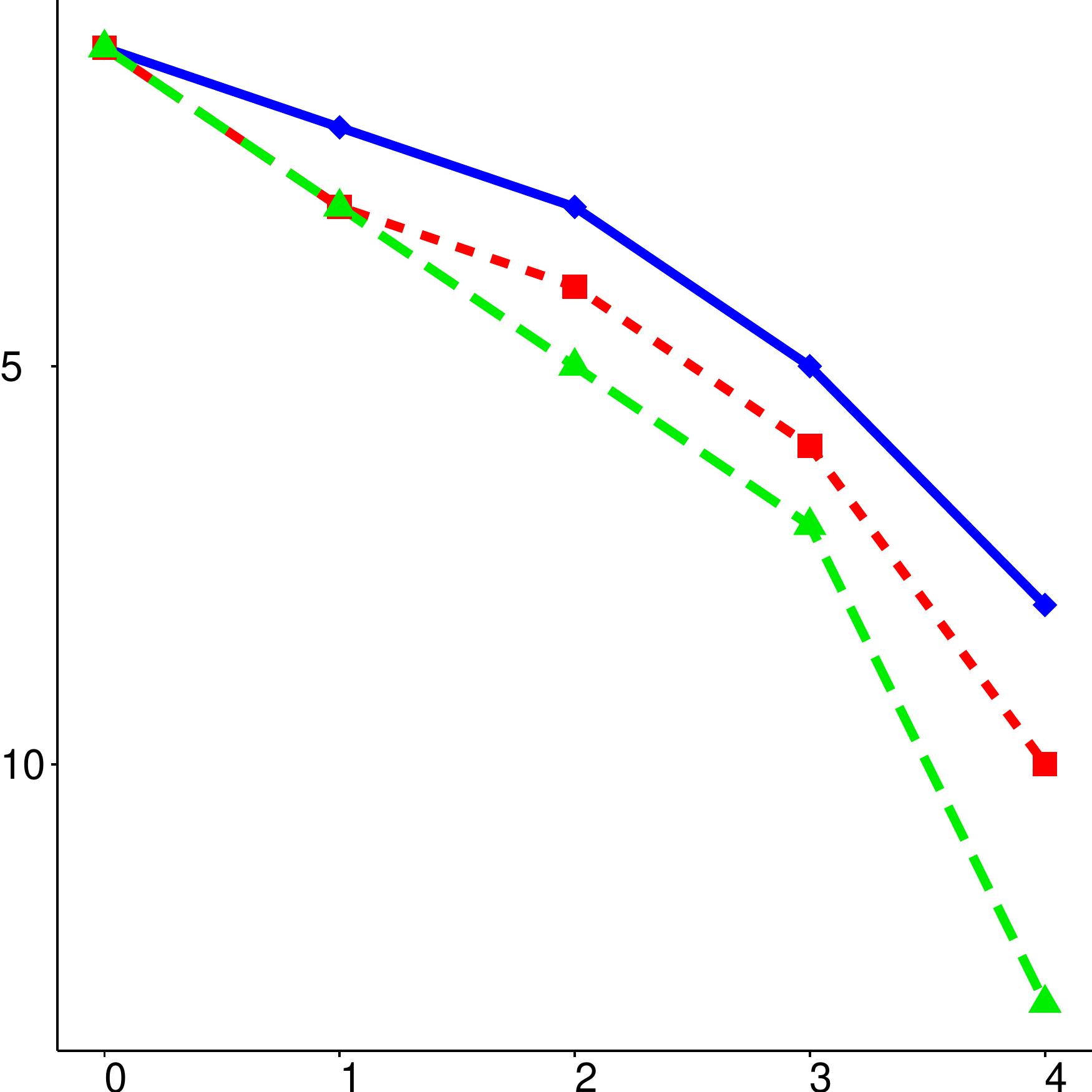} &
	\includegraphics[width=0.93\linewidth]{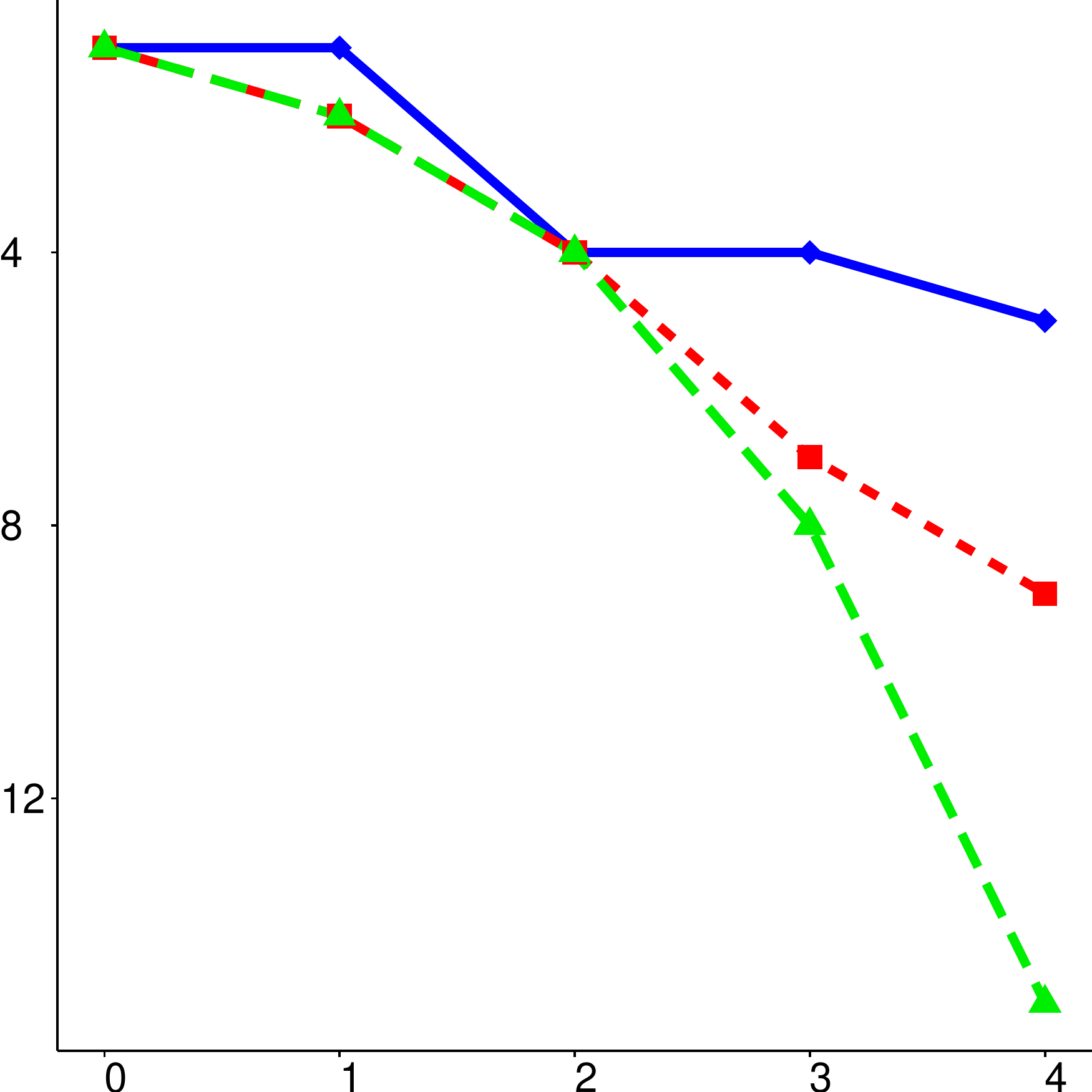} &
	\includegraphics[width=0.93\linewidth]{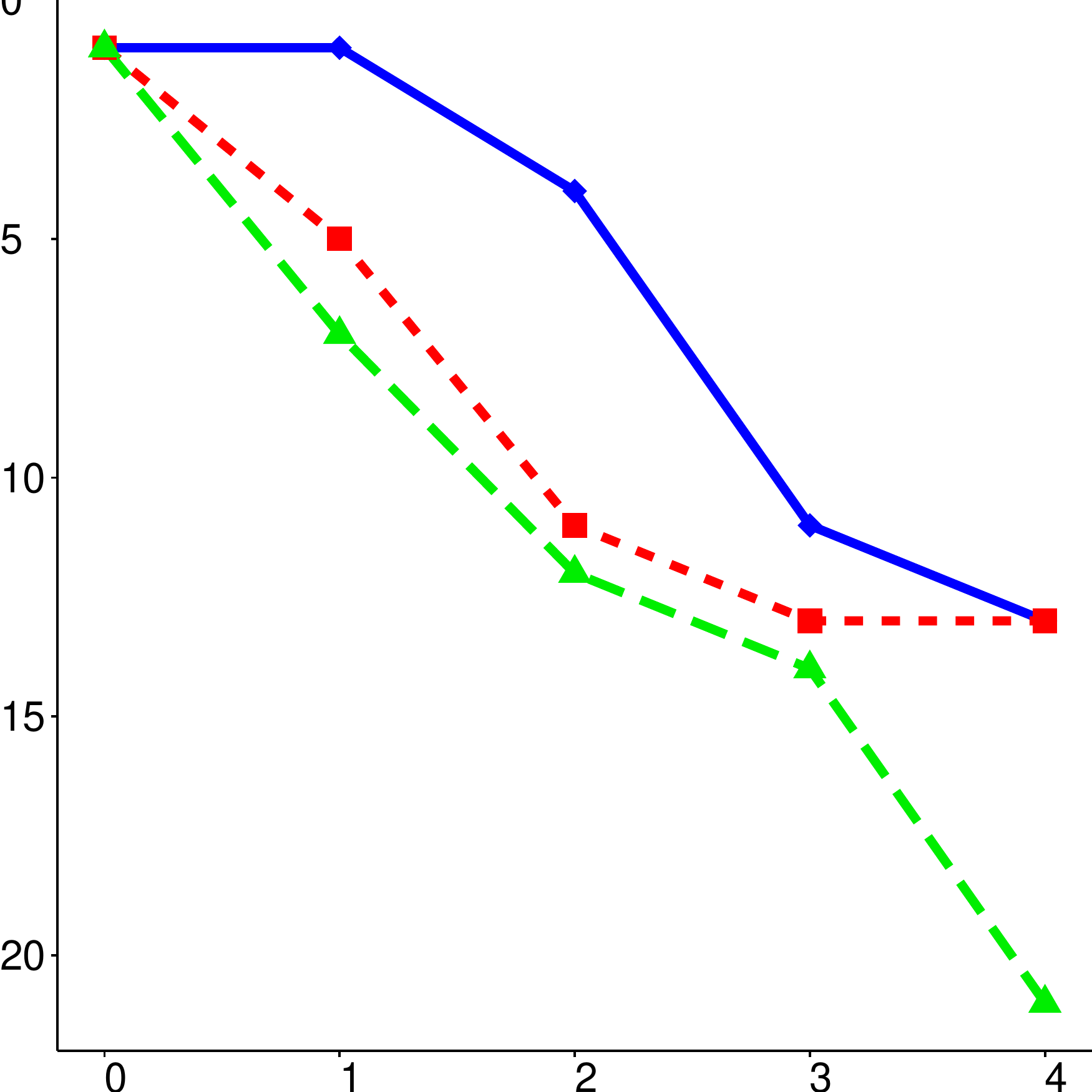} &
	\includegraphics[width=0.93\linewidth]{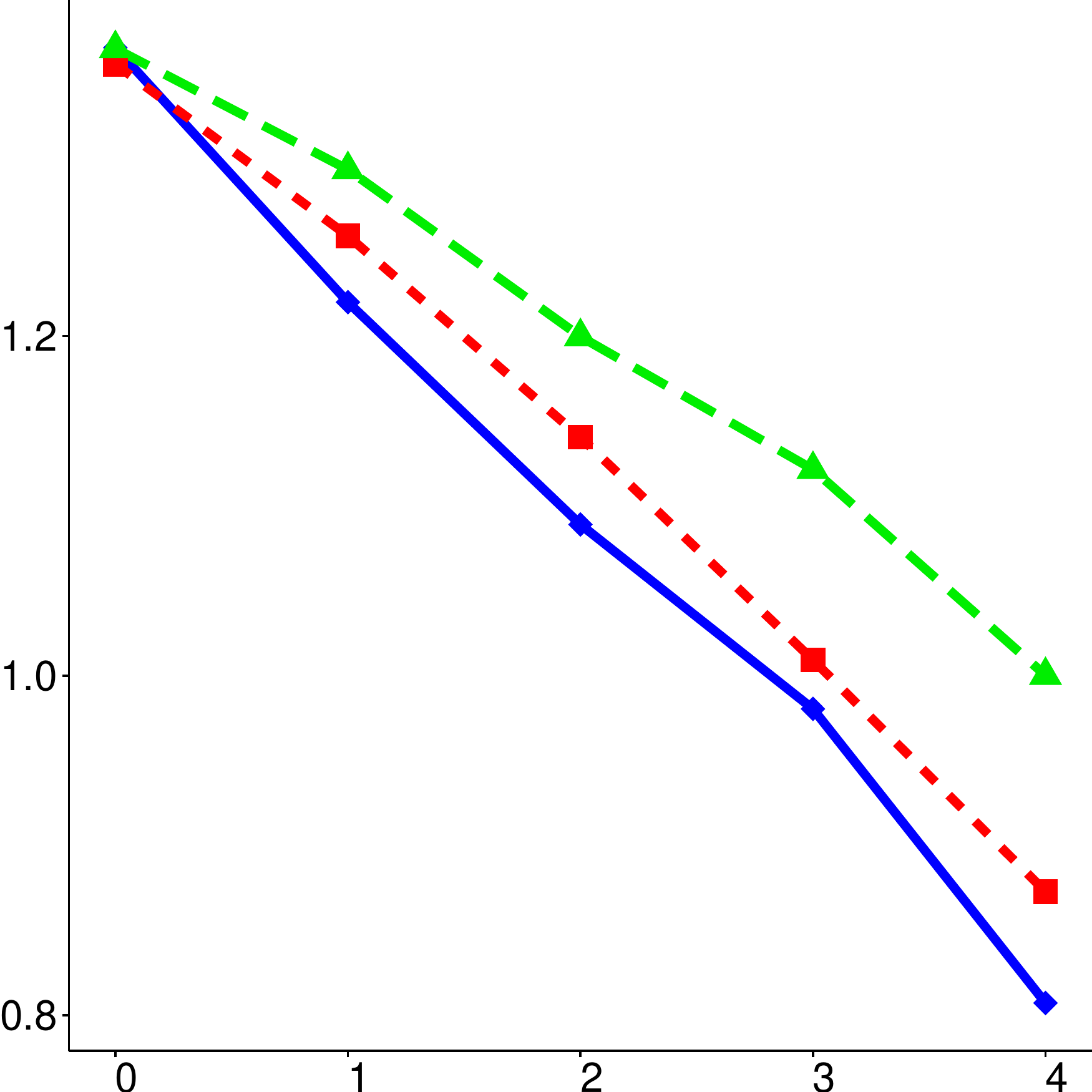} &
	\includegraphics[width=0.93\linewidth]{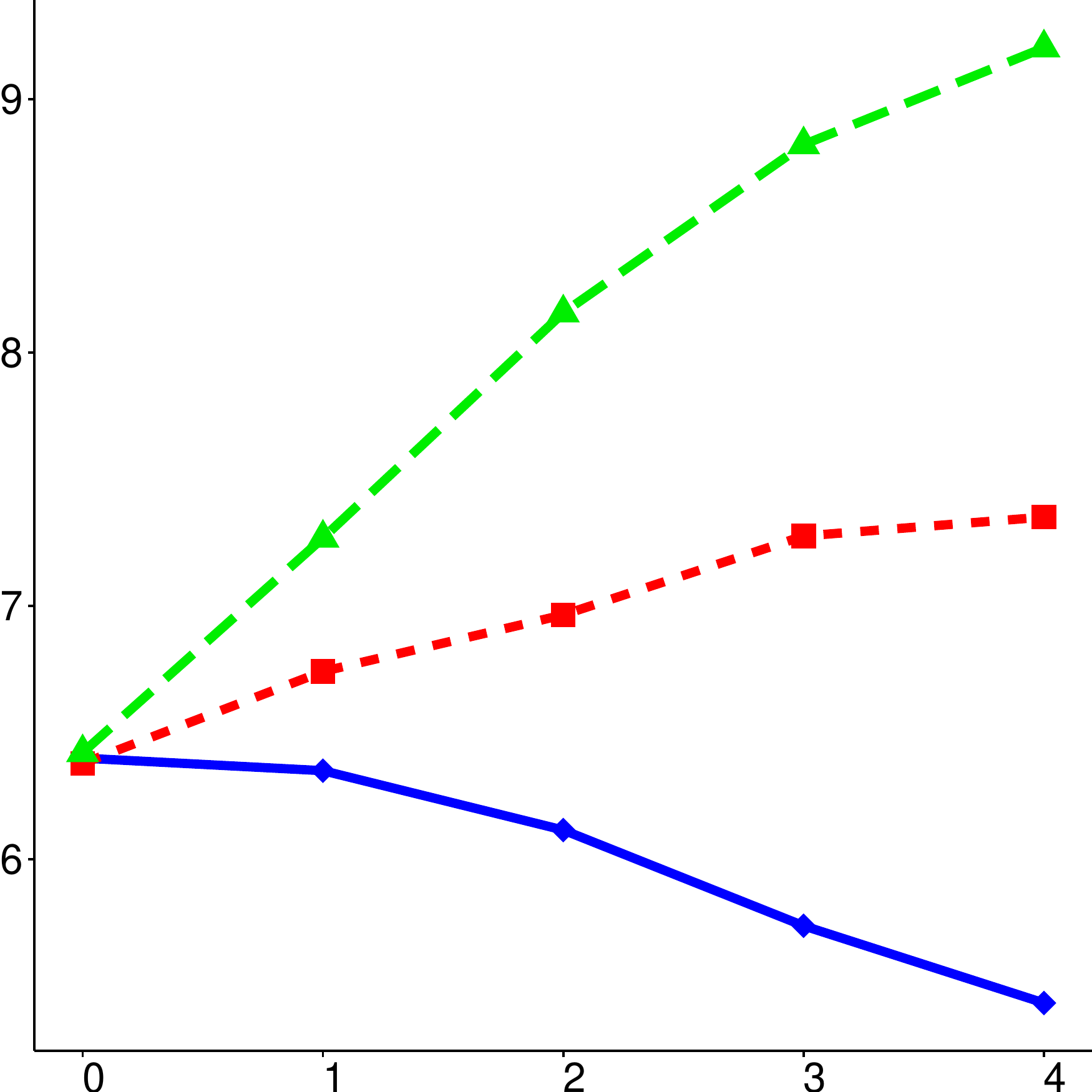} \\
	\rotatebox{90}{\small{Facebook (small)}} &
	\includegraphics[width=0.93\linewidth]{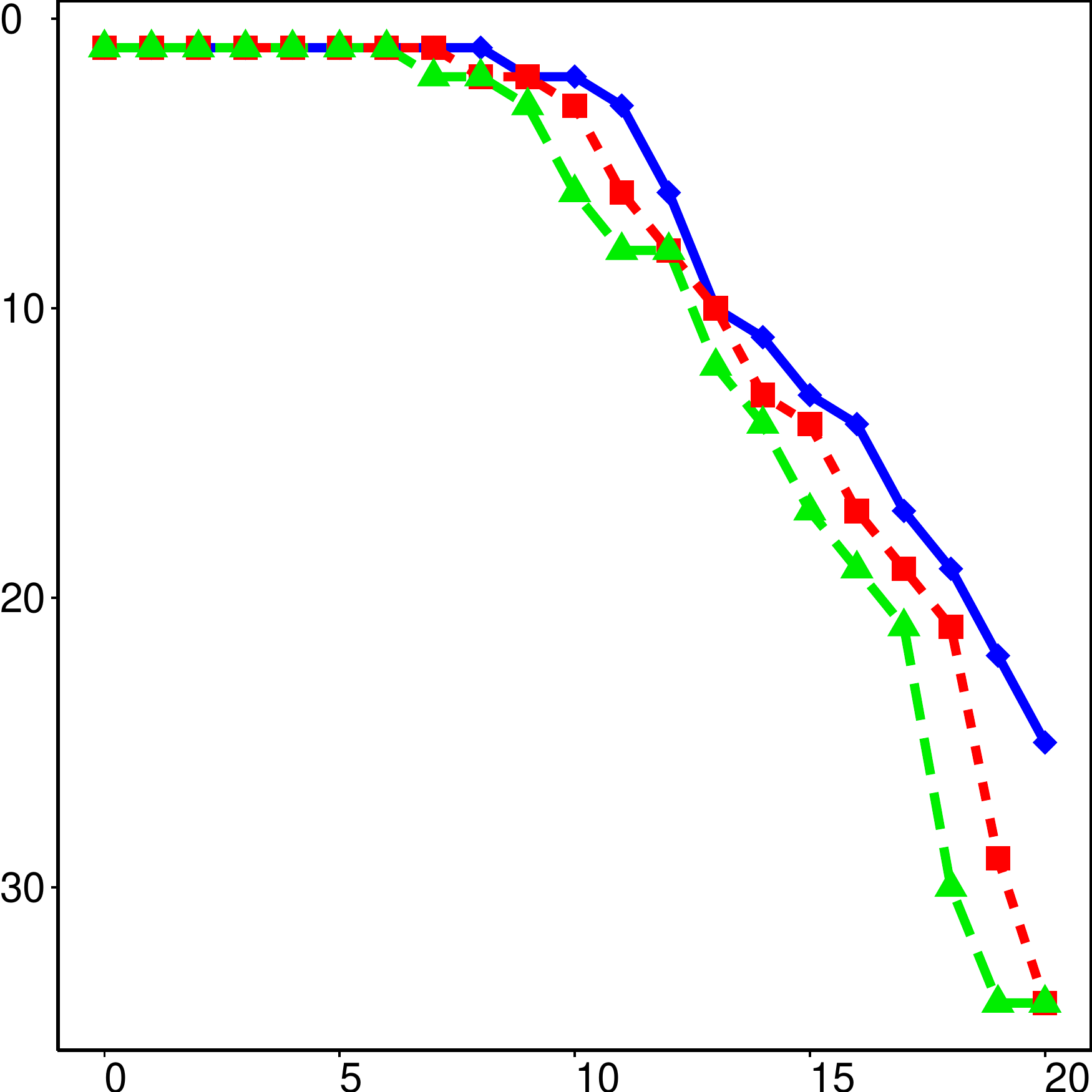} &
	\includegraphics[width=0.93\linewidth]{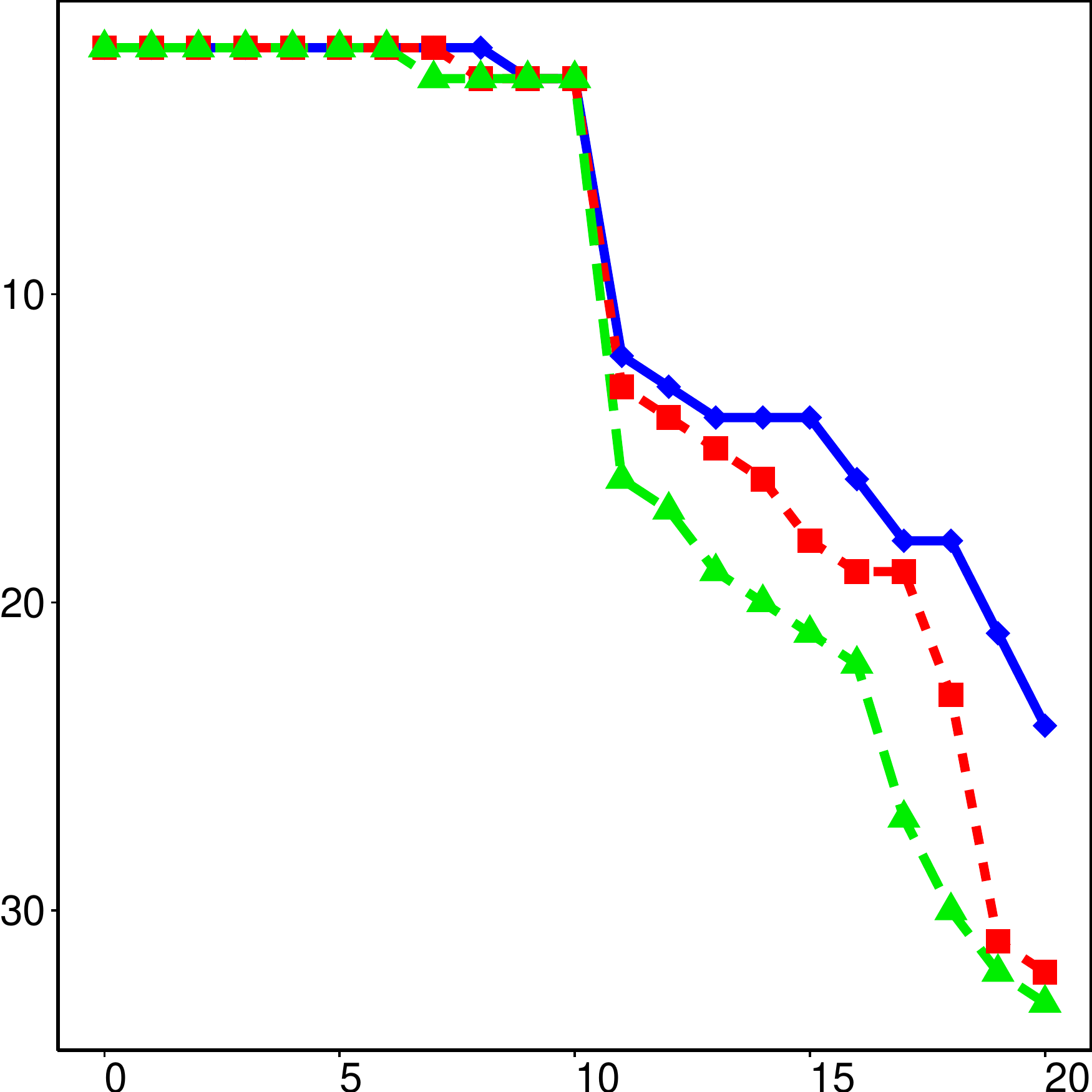} &
	\includegraphics[width=0.93\linewidth]{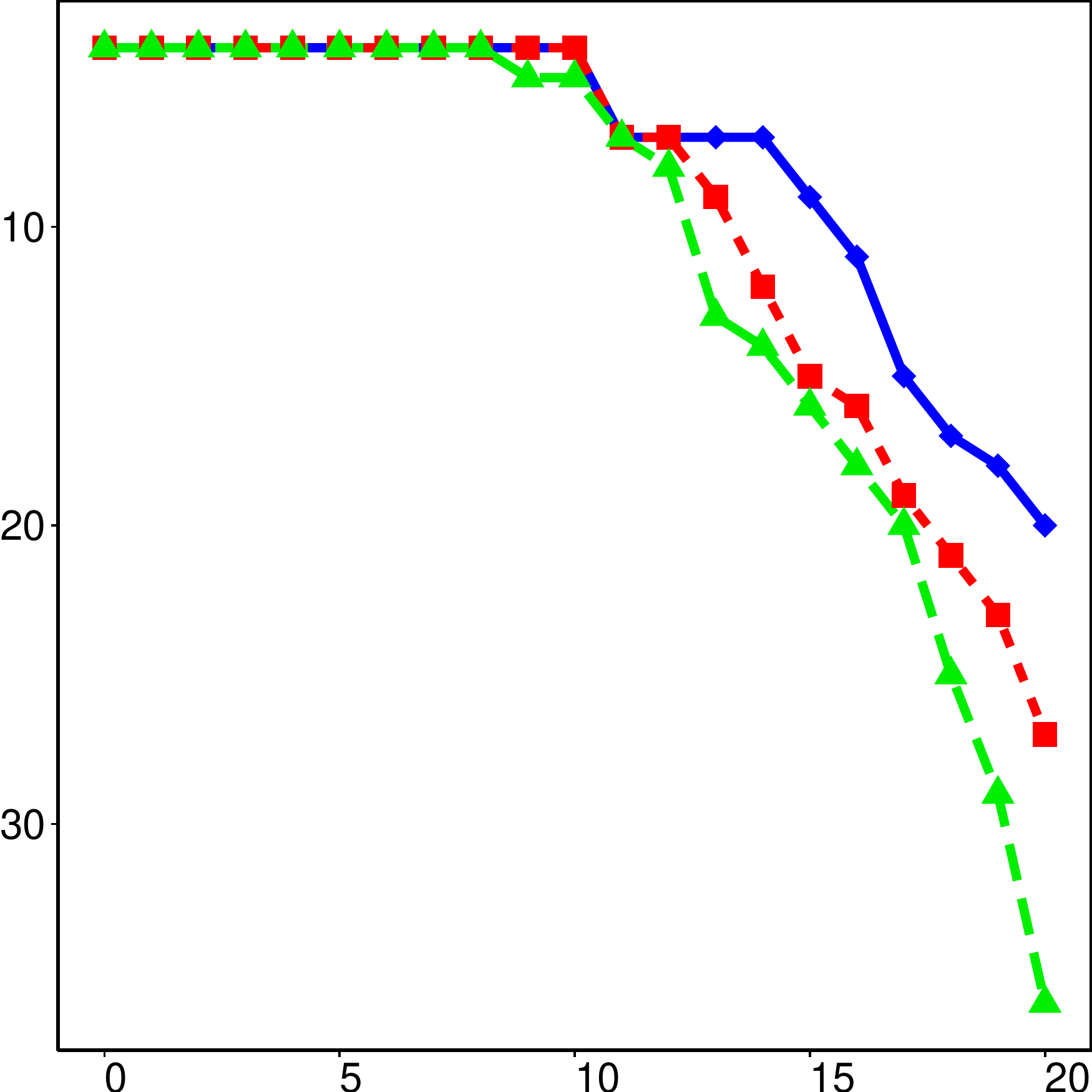} &
	\includegraphics[width=0.93\linewidth]{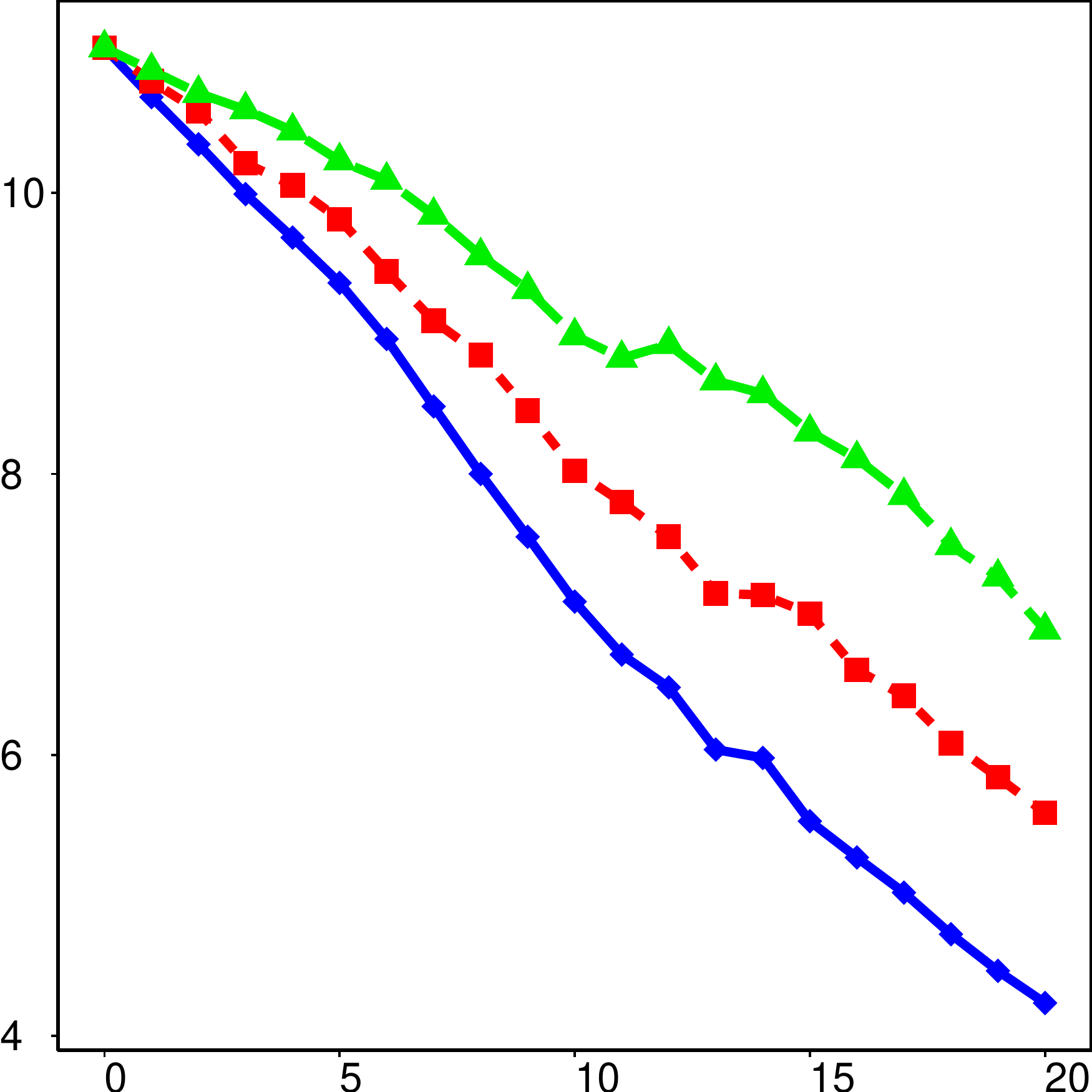} &
	\includegraphics[width=0.93\linewidth]{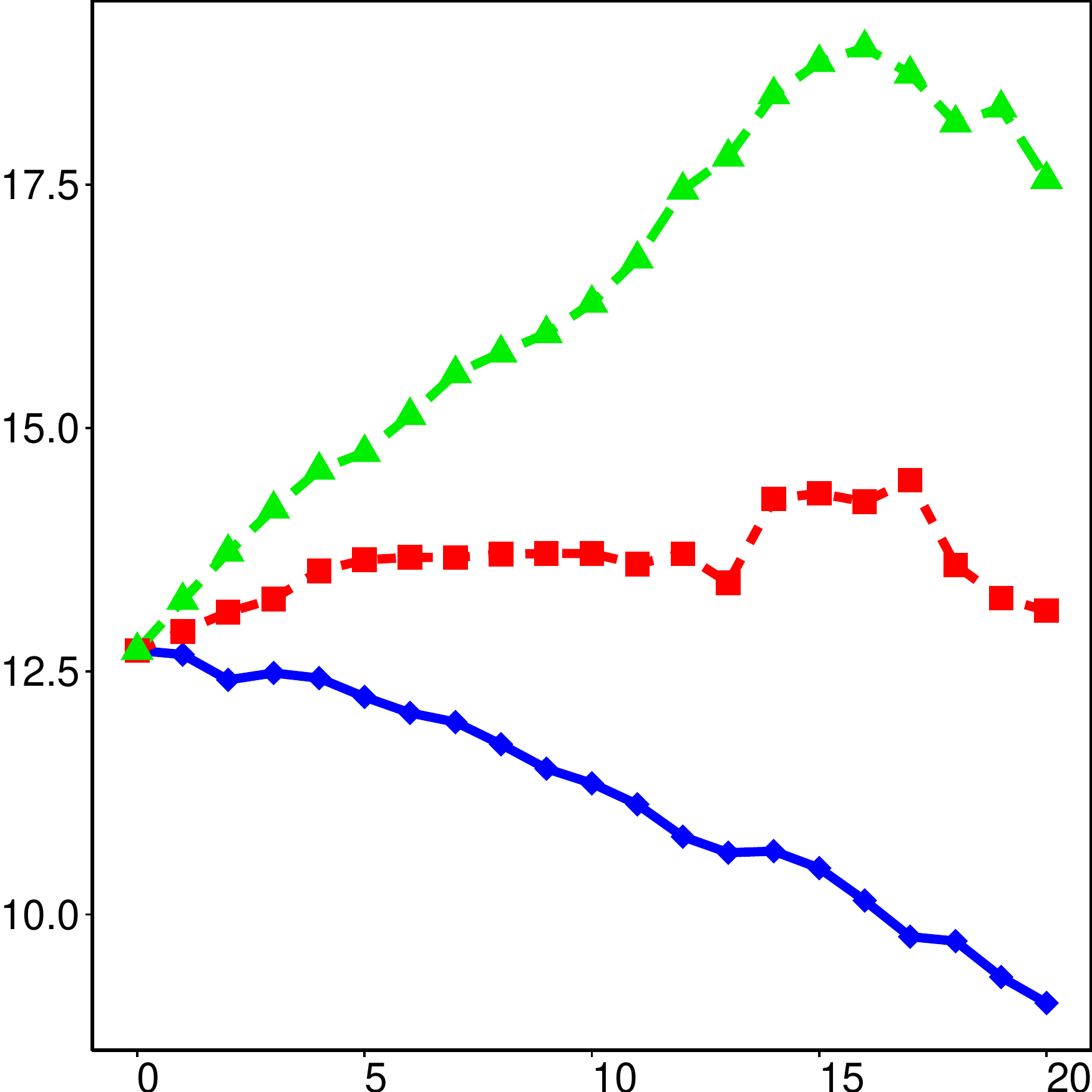} \\
	\rotatebox{90}{\small{Facebook (medium)}} &
	\includegraphics[width=0.93\linewidth]{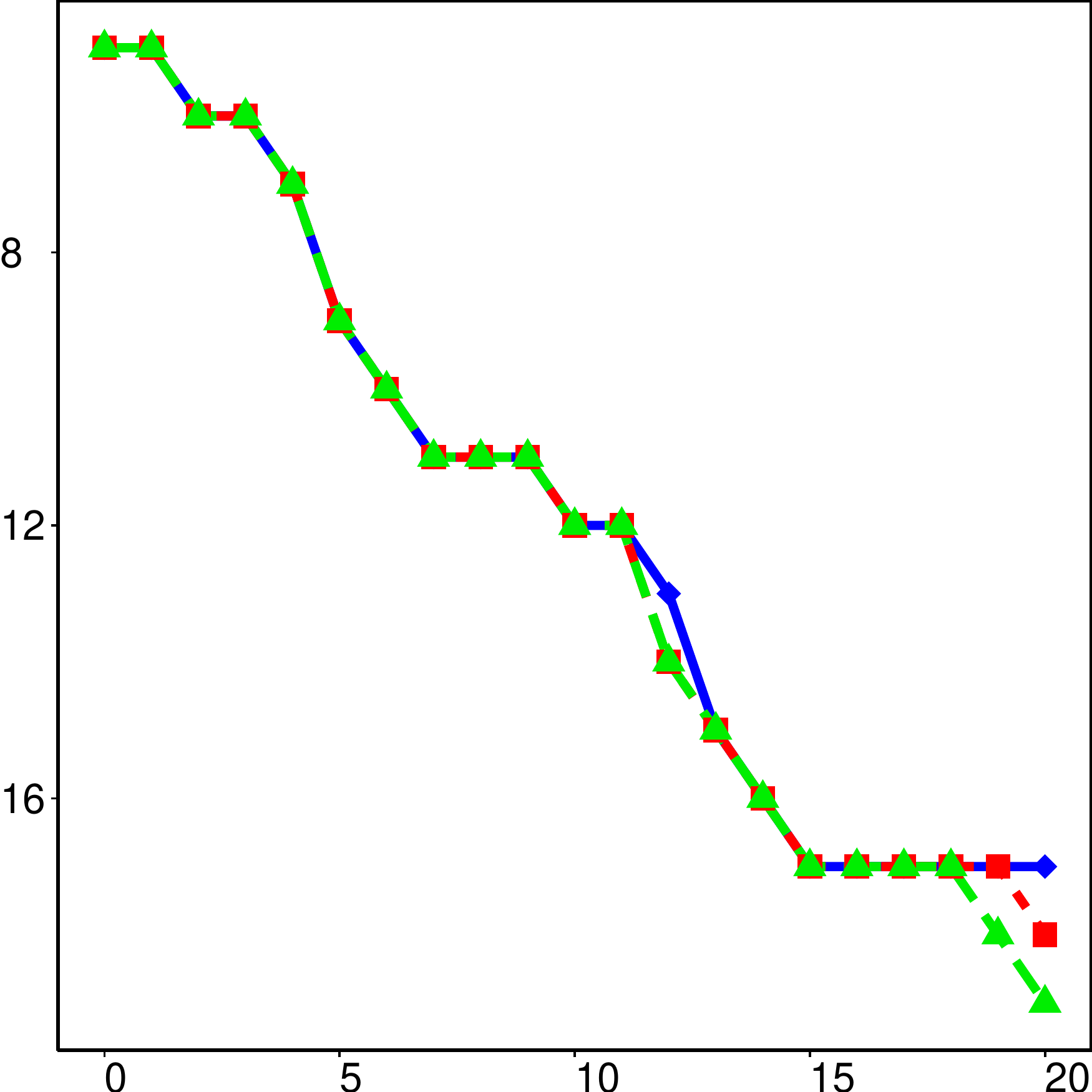} &
	\includegraphics[width=0.93\linewidth]{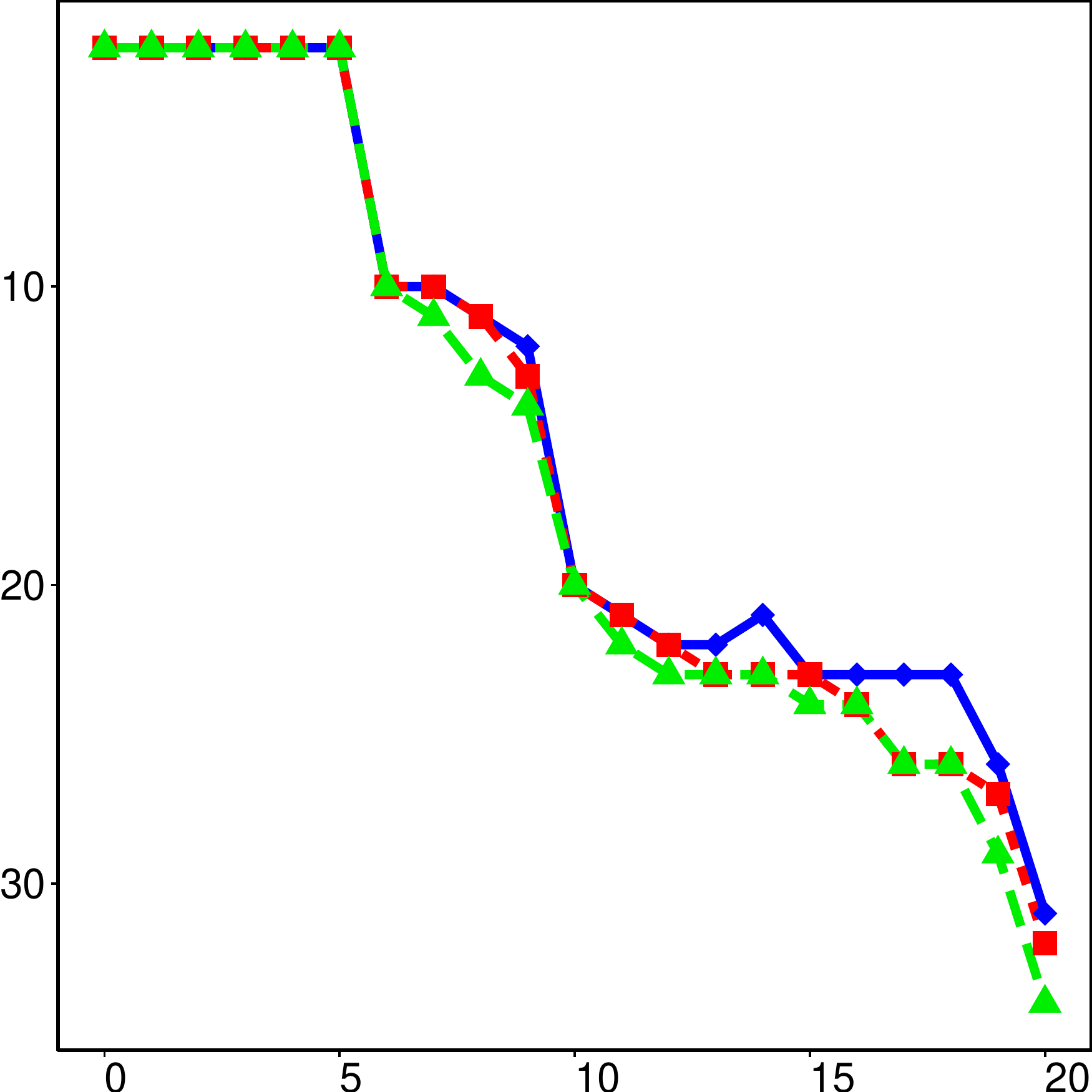} &
	\includegraphics[width=0.93\linewidth]{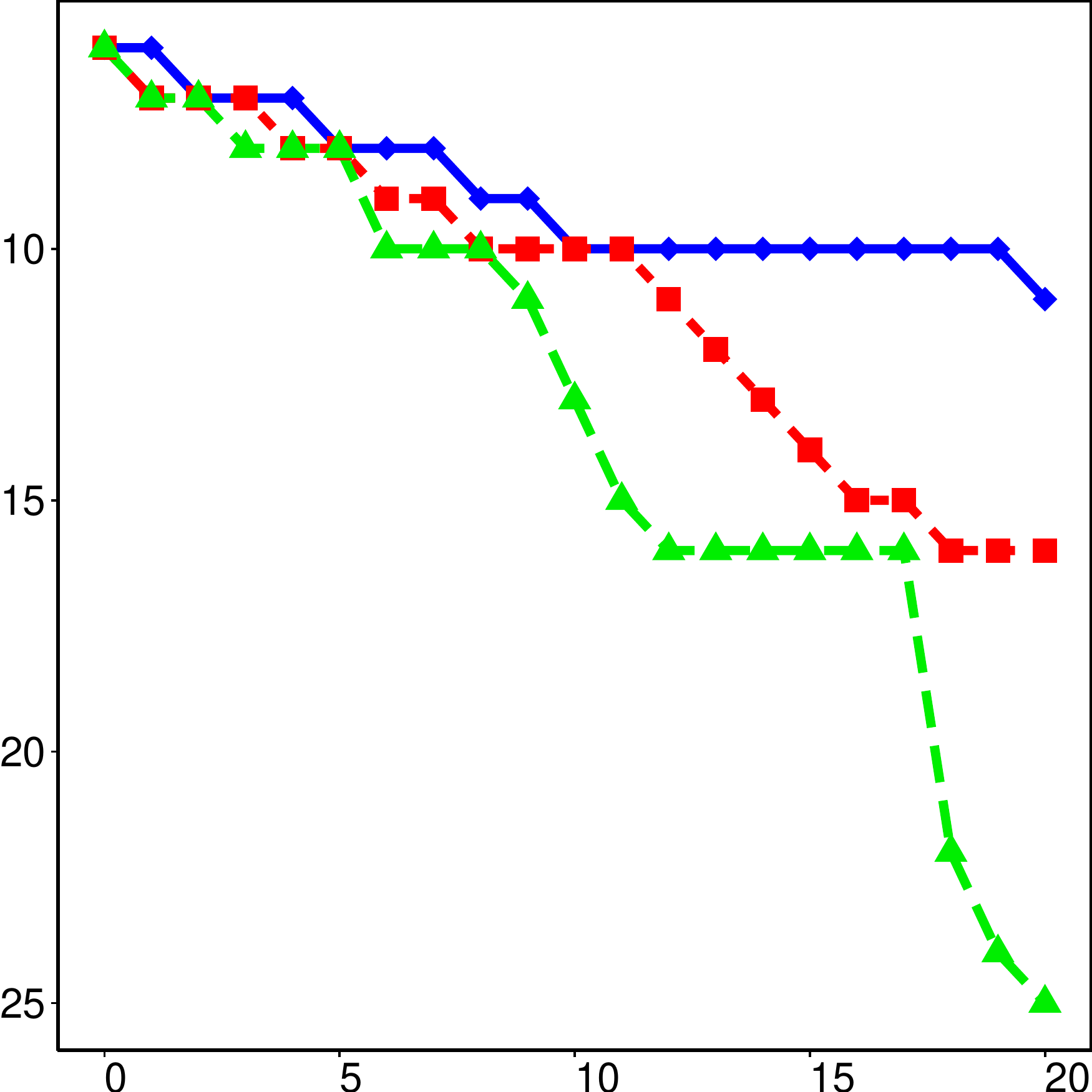} &
	\includegraphics[width=0.93\linewidth]{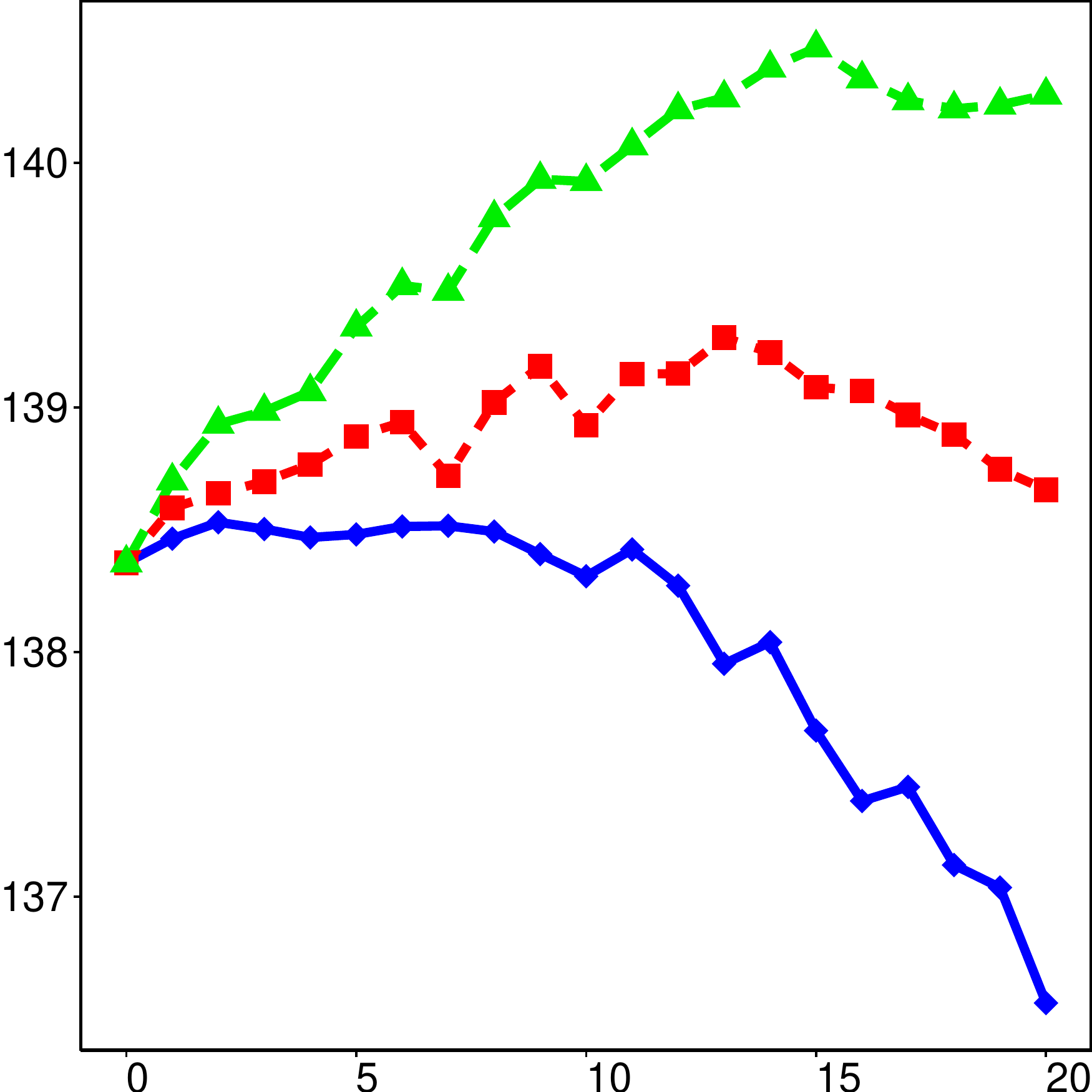} &
	\includegraphics[width=0.93\linewidth]{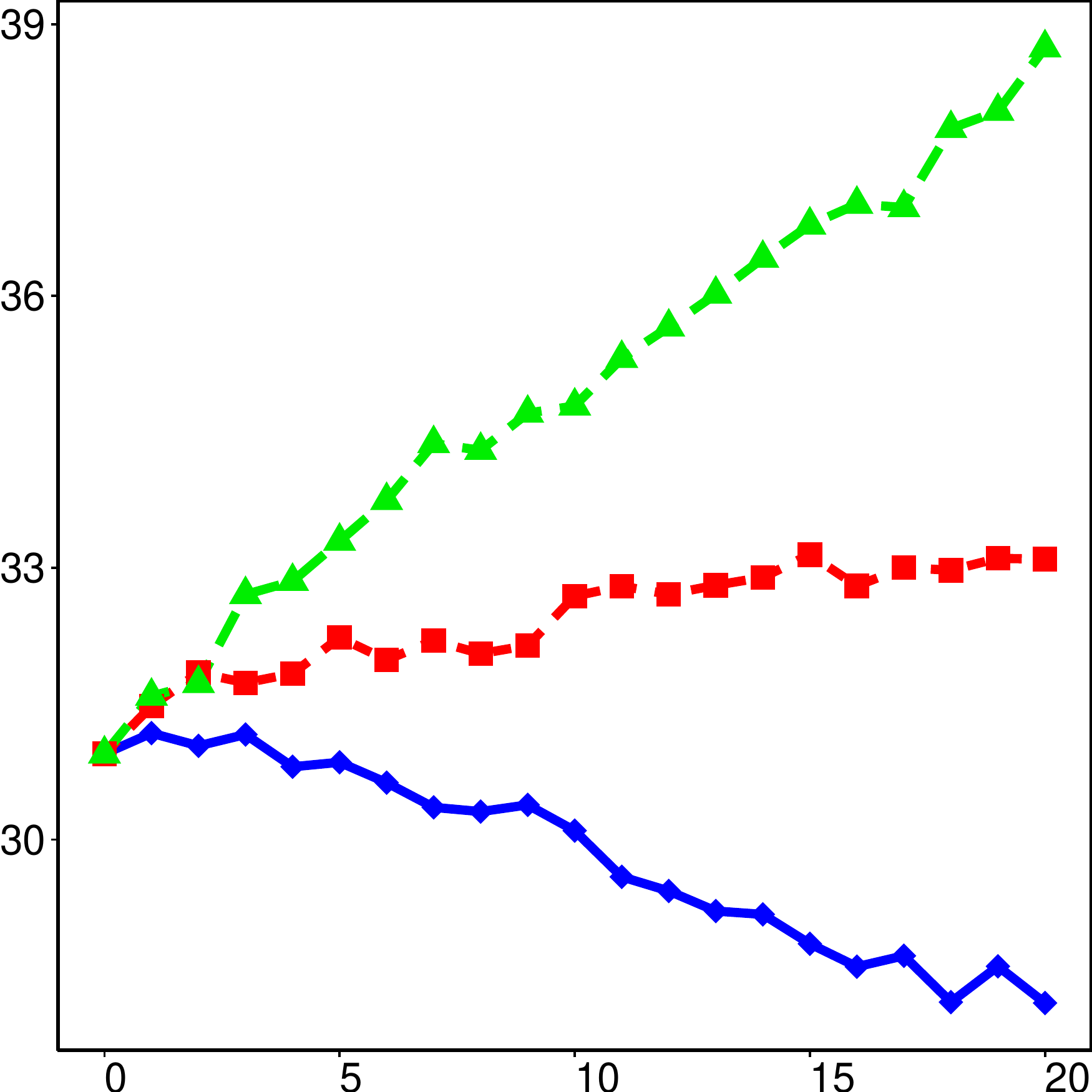} \\
	\rotatebox{90}{\small{Facebook (large)}} &
	\includegraphics[width=0.93\linewidth]{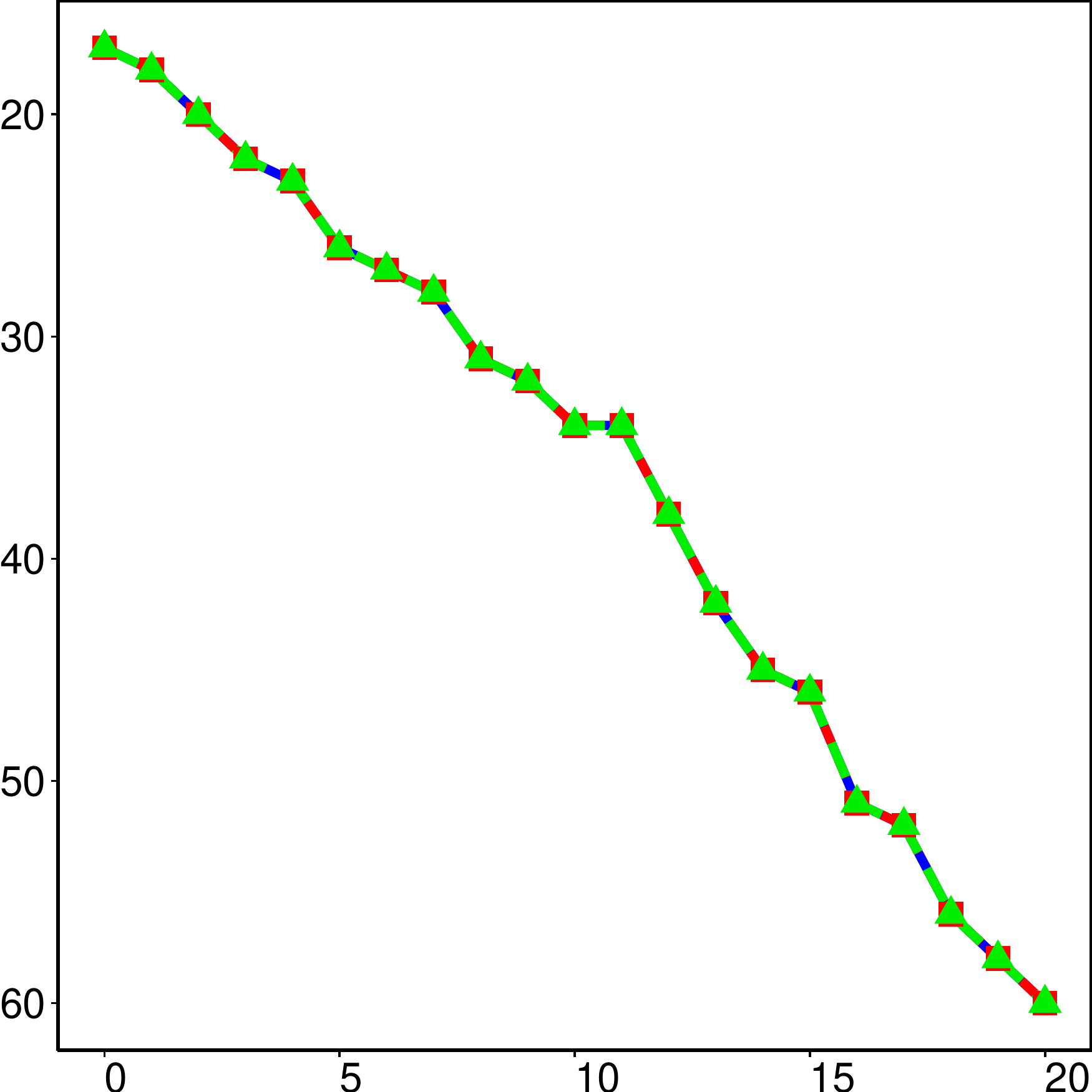} &
	\includegraphics[width=0.93\linewidth]{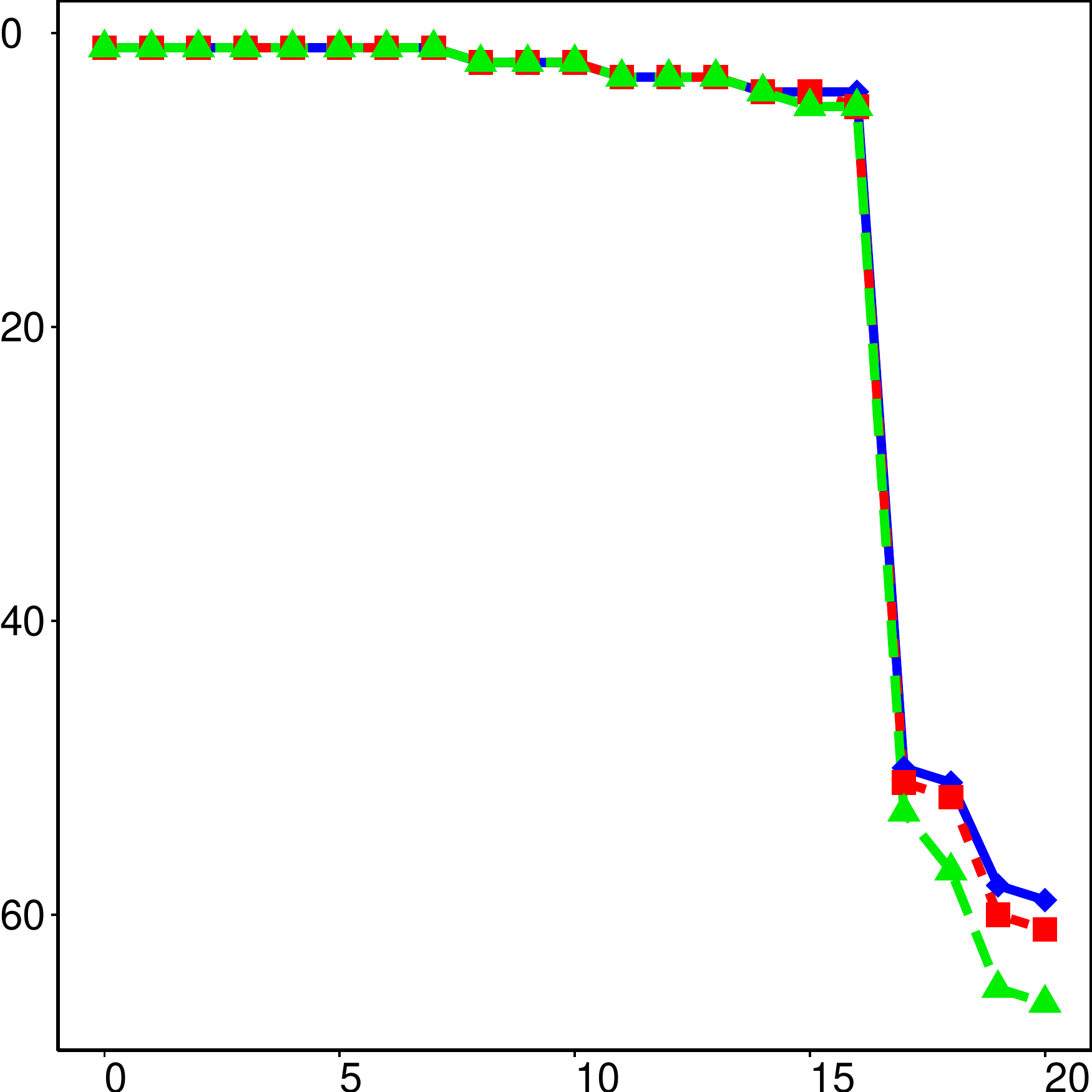} &
	\includegraphics[width=0.93\linewidth]{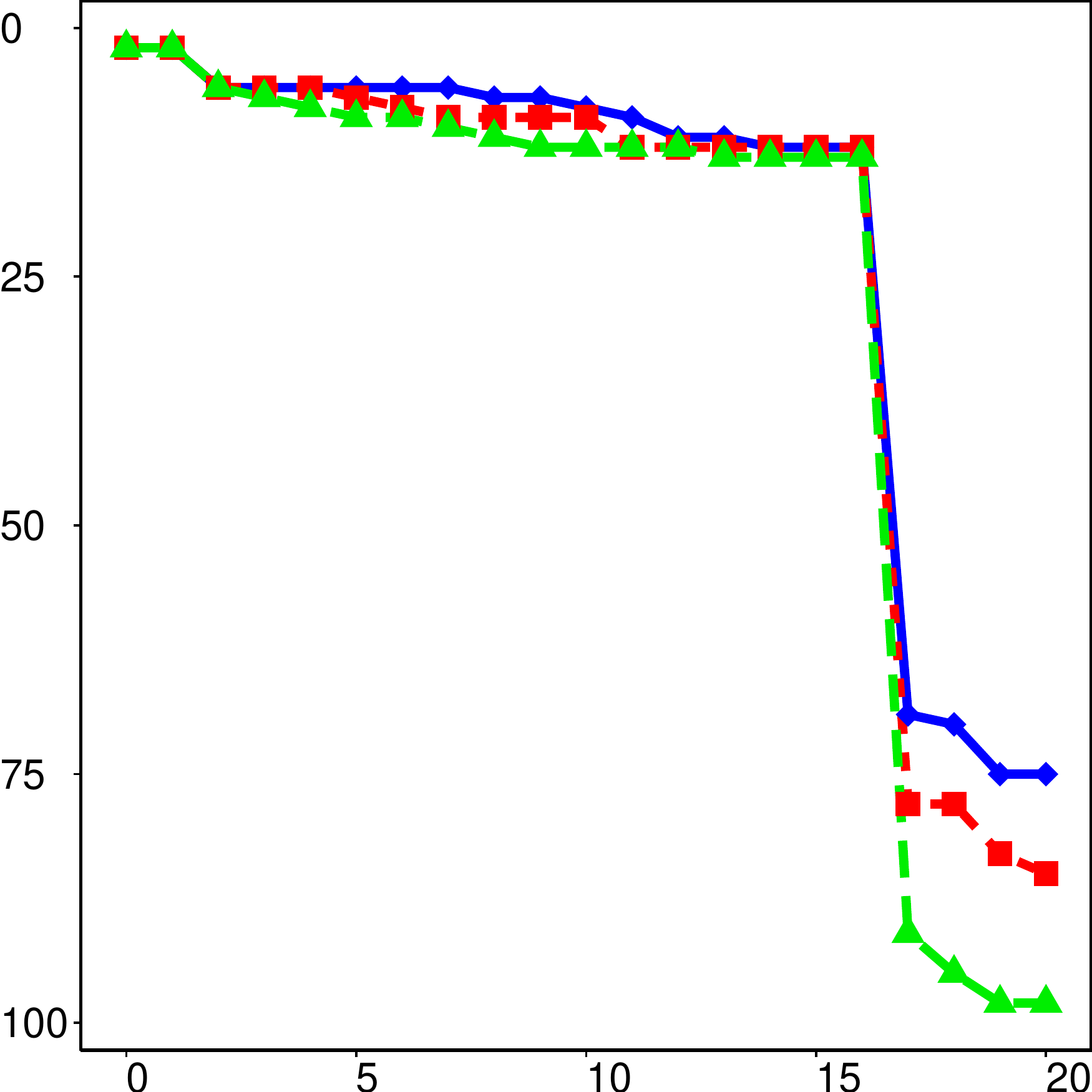} &
	\includegraphics[width=0.93\linewidth]{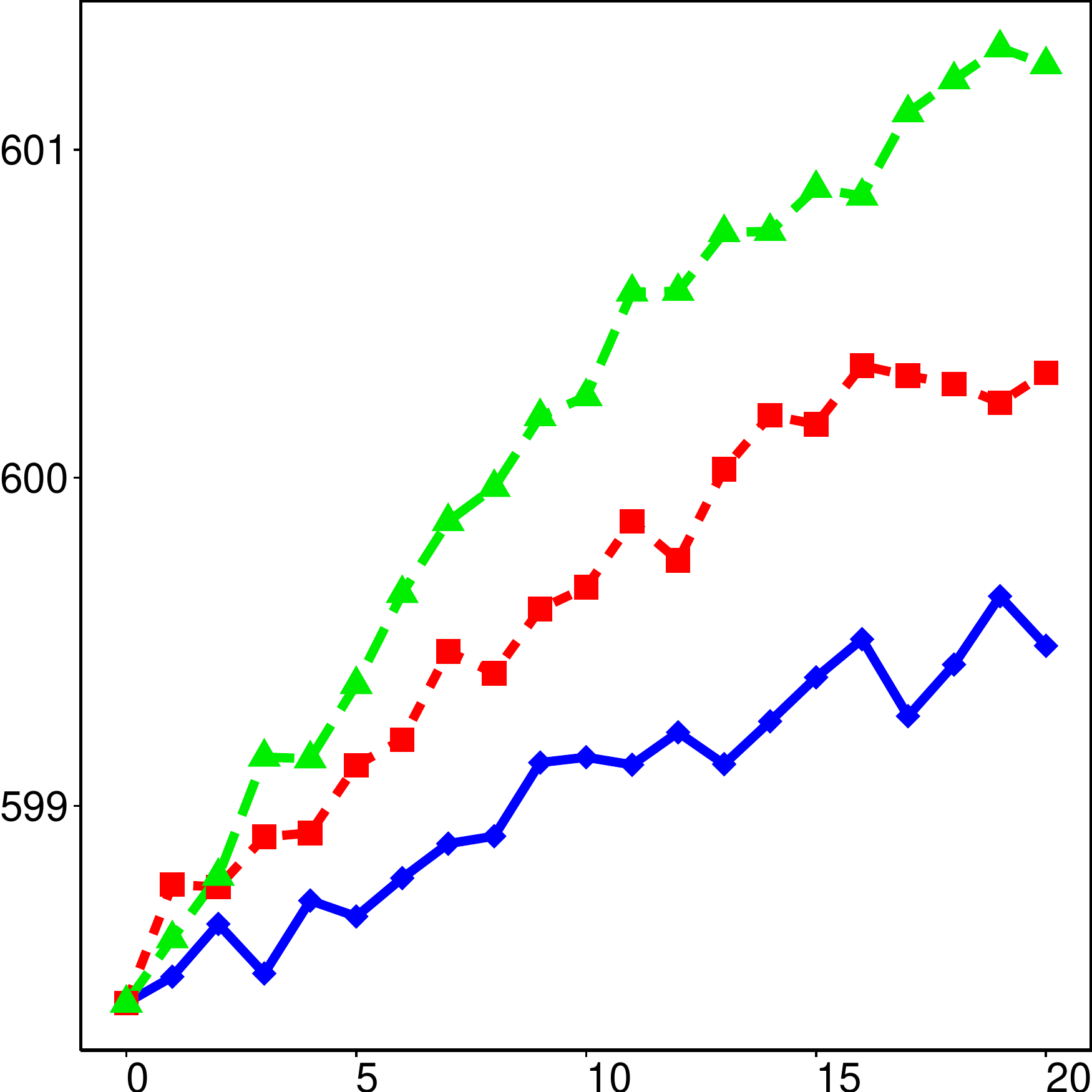} &
	\includegraphics[width=0.93\linewidth]{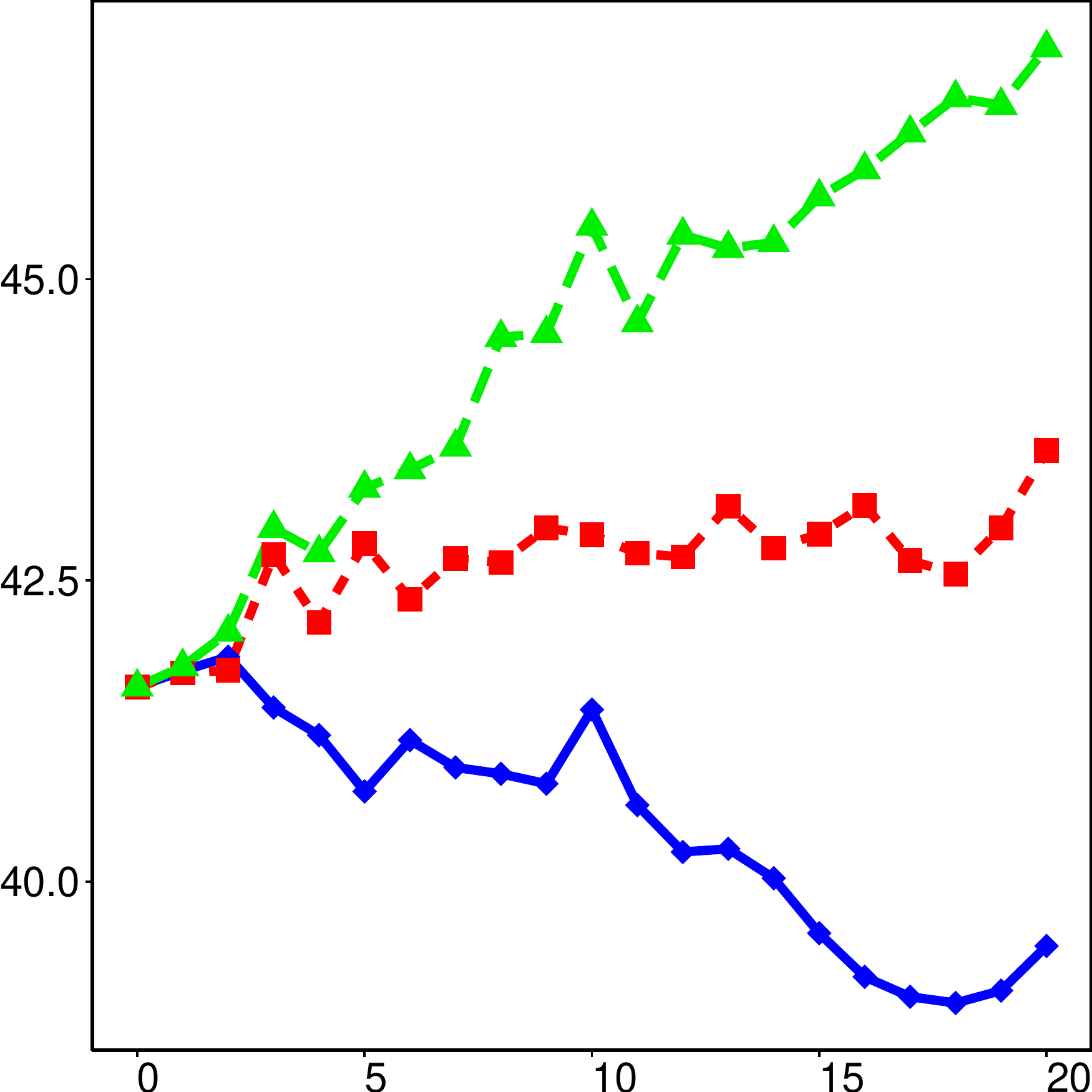} \\
	&
	\multicolumn{5}{c}{\includegraphics[width=0.75\linewidth]{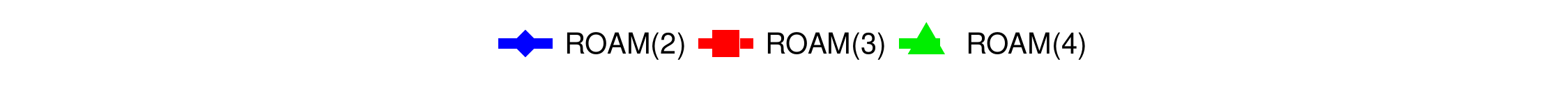}} \\
	\end{tabular}
	\vspace*{-0.5cm}
	\caption{\fontsize{10}{10}\selectfont{Consecutive execution of ROAM (the $x$-axis represents the number of executions). Given three terrorist networks, and different fragments of Facebook's network, the subfigures show the source node's ranking (according to the centrality measures), and the relative change in its influence value (according to the influence models). Results are for $\textnormal{ROAM}(b):b=2, 3, 4$, where $b$ is the budget in each execution.}}
	\label{fig:centralityResults:terroist:Facebook}
\end{figure}

\begin{figure}[p!]
	\centering.
	\setlength\tabcolsep{2pt}
	\renewcommand{\arraystretch}{2.0}
	\begin{tabular}{m{.01\textwidth}m{.195\textwidth}m{.195\textwidth}m{.195\textwidth}m{.195\textwidth}m{.195\textwidth}}
	&
	\multicolumn{1}{c}{\fontsize{8.5}{8.5}\selectfont{Degree Ranking}} &
	\multicolumn{1}{c}{\fontsize{8.5}{8.5}\selectfont{Closeness Ranking}} &
	\multicolumn{1}{c}{\fontsize{8.5}{8.5}\selectfont{Betweenness Ranking}} &
	\multicolumn{1}{c}{\fontsize{8.5}{8.5}\selectfont{IC Influence}} &
	\multicolumn{1}{c}{\fontsize{8.5}{8.5}\selectfont{LT Influence}} \\
	\rotatebox{90}{\small{Twitter (small)}} &
	\includegraphics[width=0.93\linewidth]{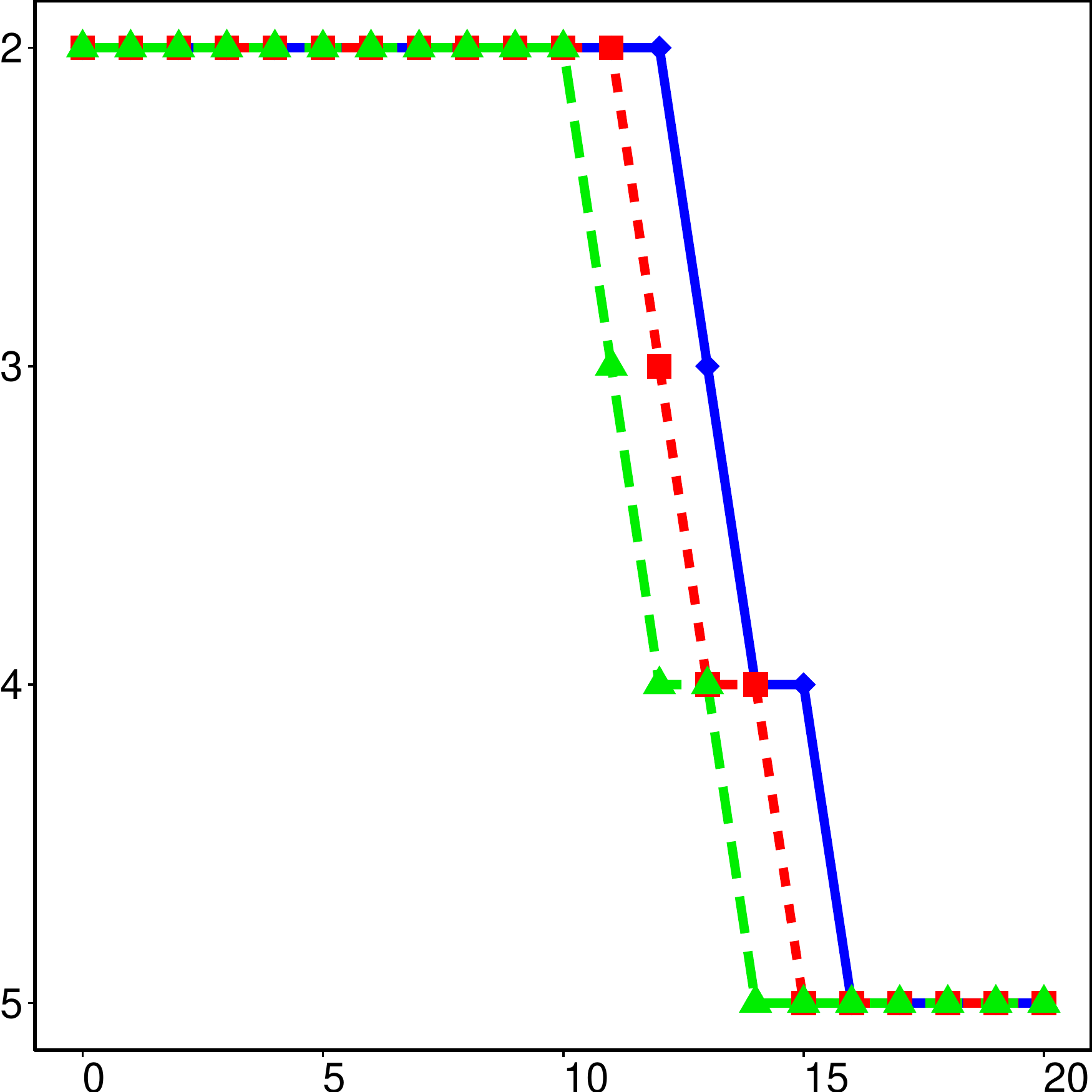} &
	\includegraphics[width=0.93\linewidth]{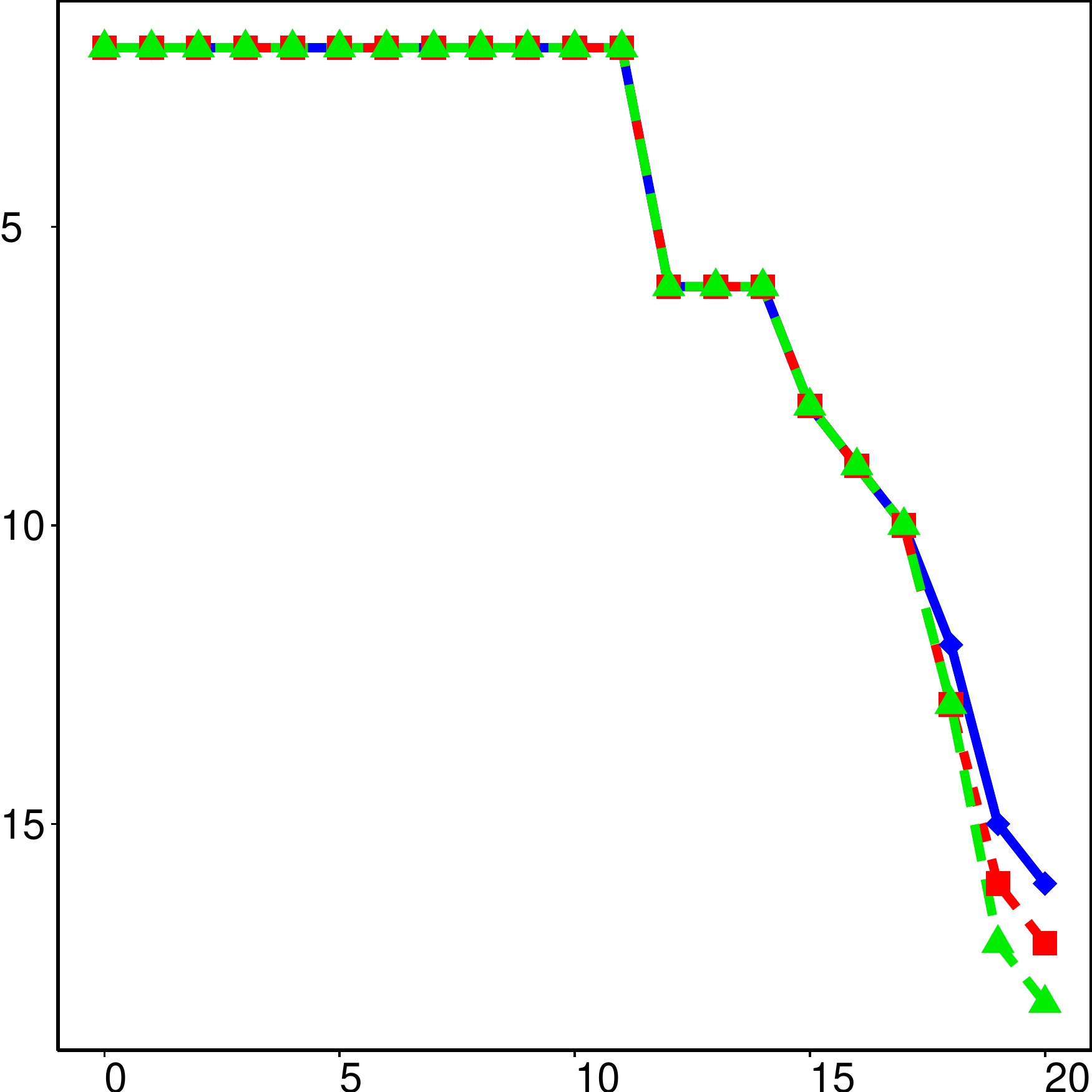} &
	\includegraphics[width=0.93\linewidth]{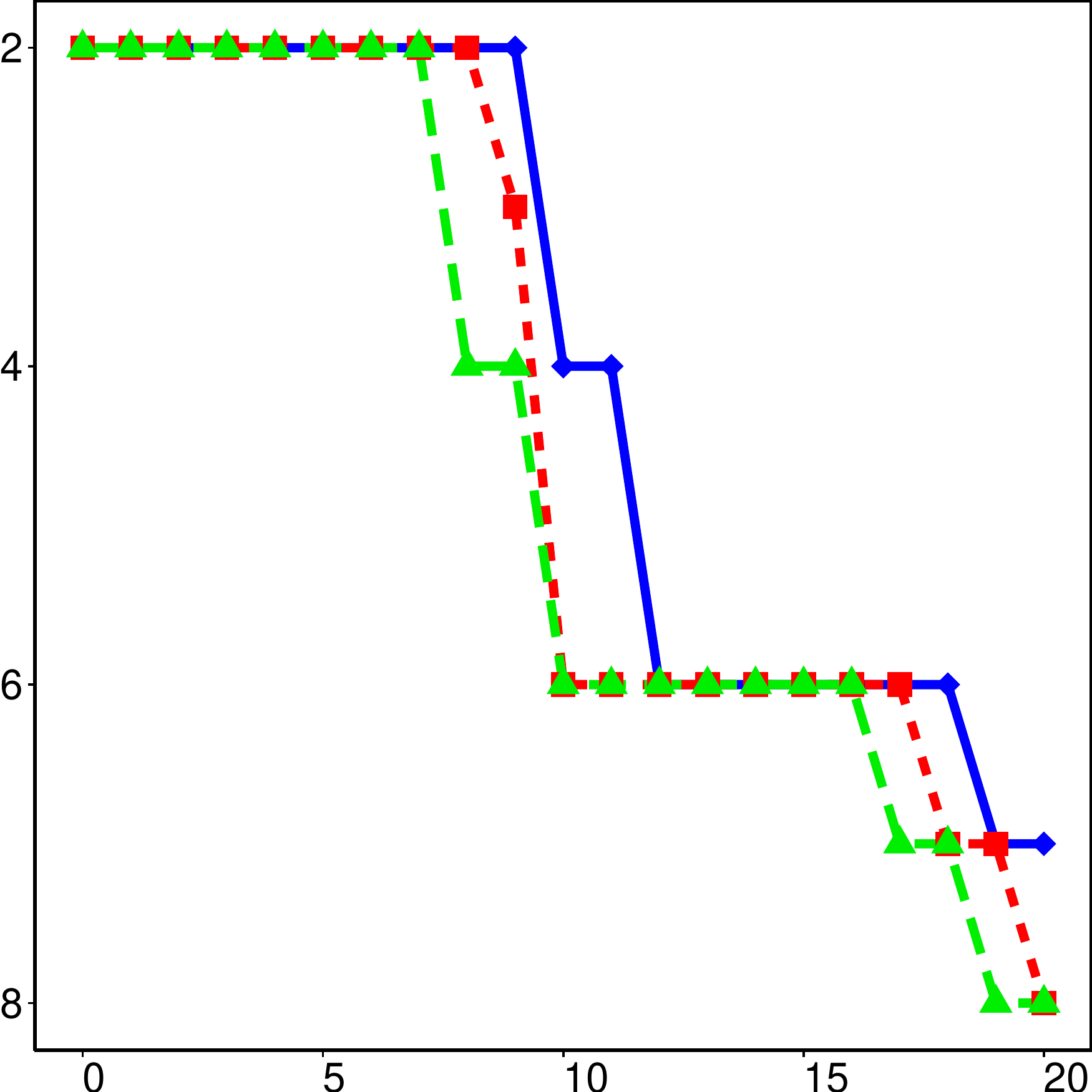} &
	\includegraphics[width=0.93\linewidth]{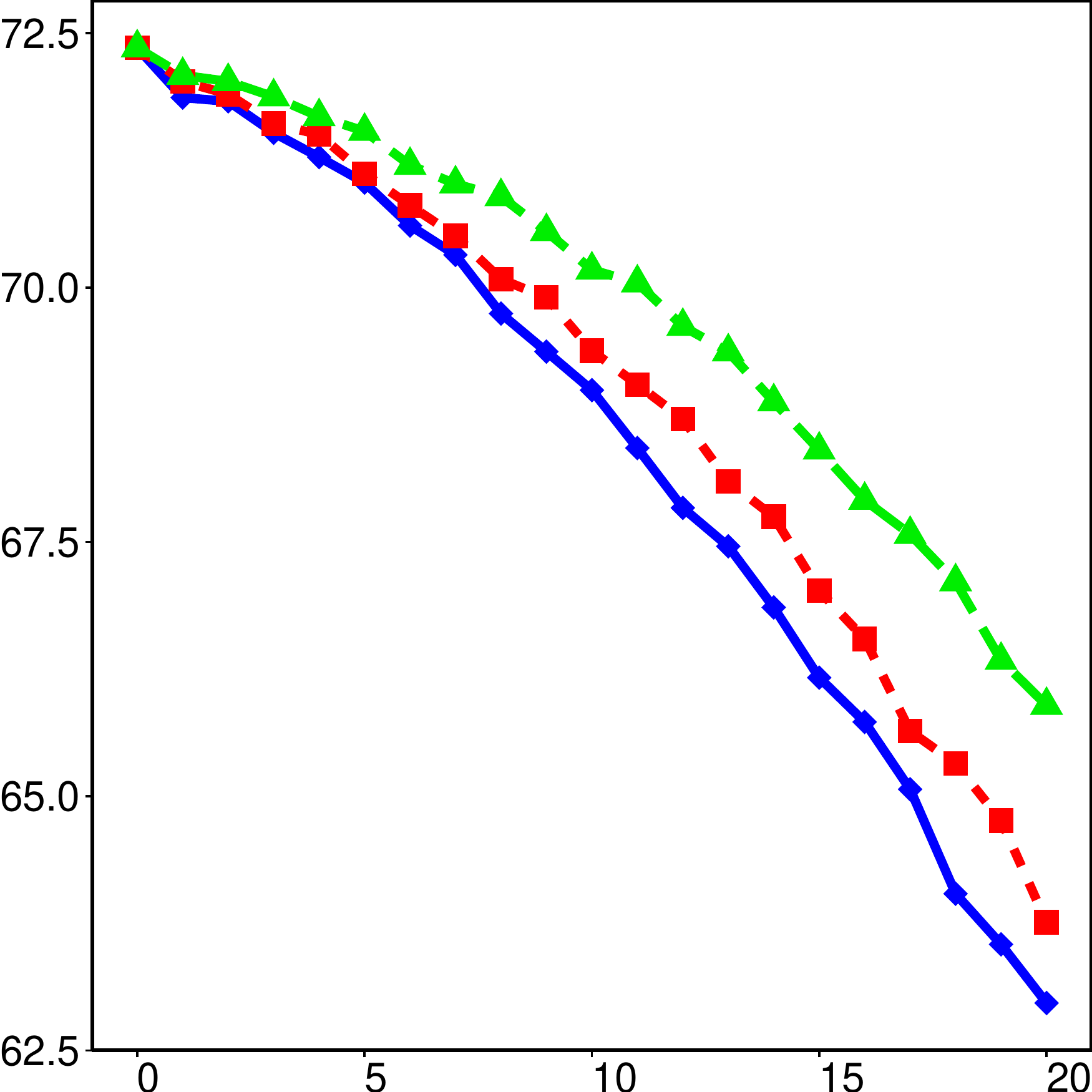} &
	\includegraphics[width=0.93\linewidth]{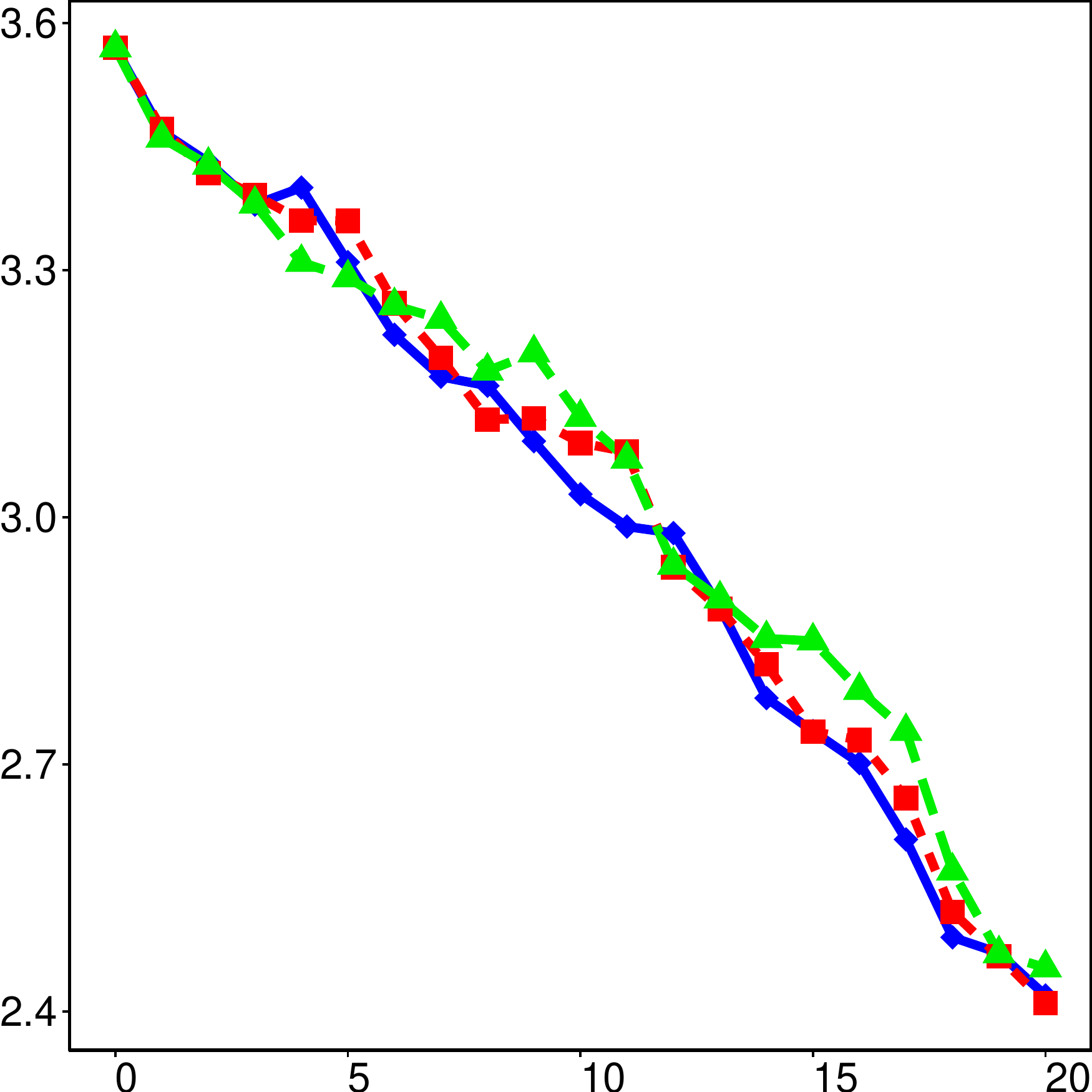} \\
	\rotatebox{90}{\small{Twitter (medium)}} &
	\includegraphics[width=0.93\linewidth]{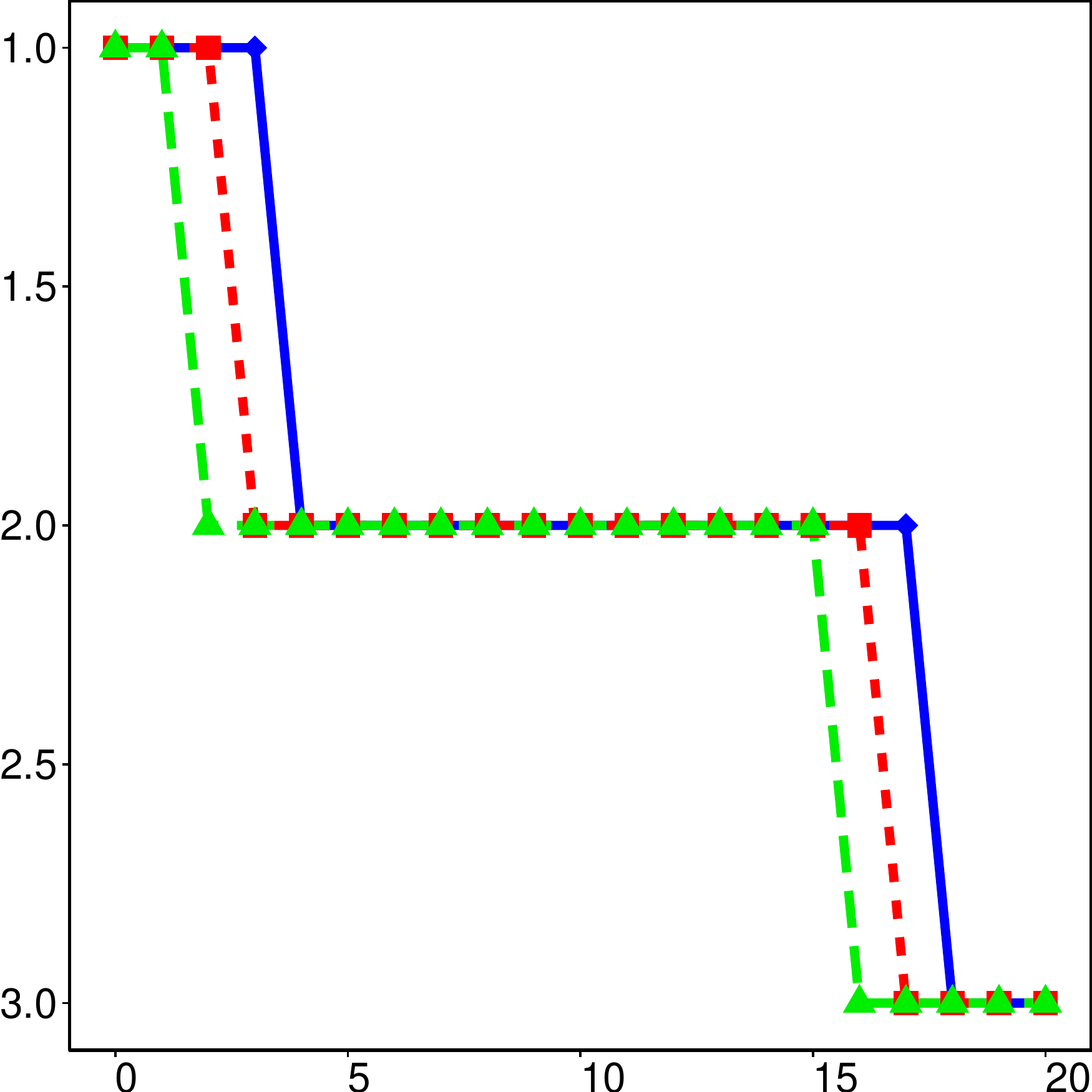} &
	\includegraphics[width=0.93\linewidth]{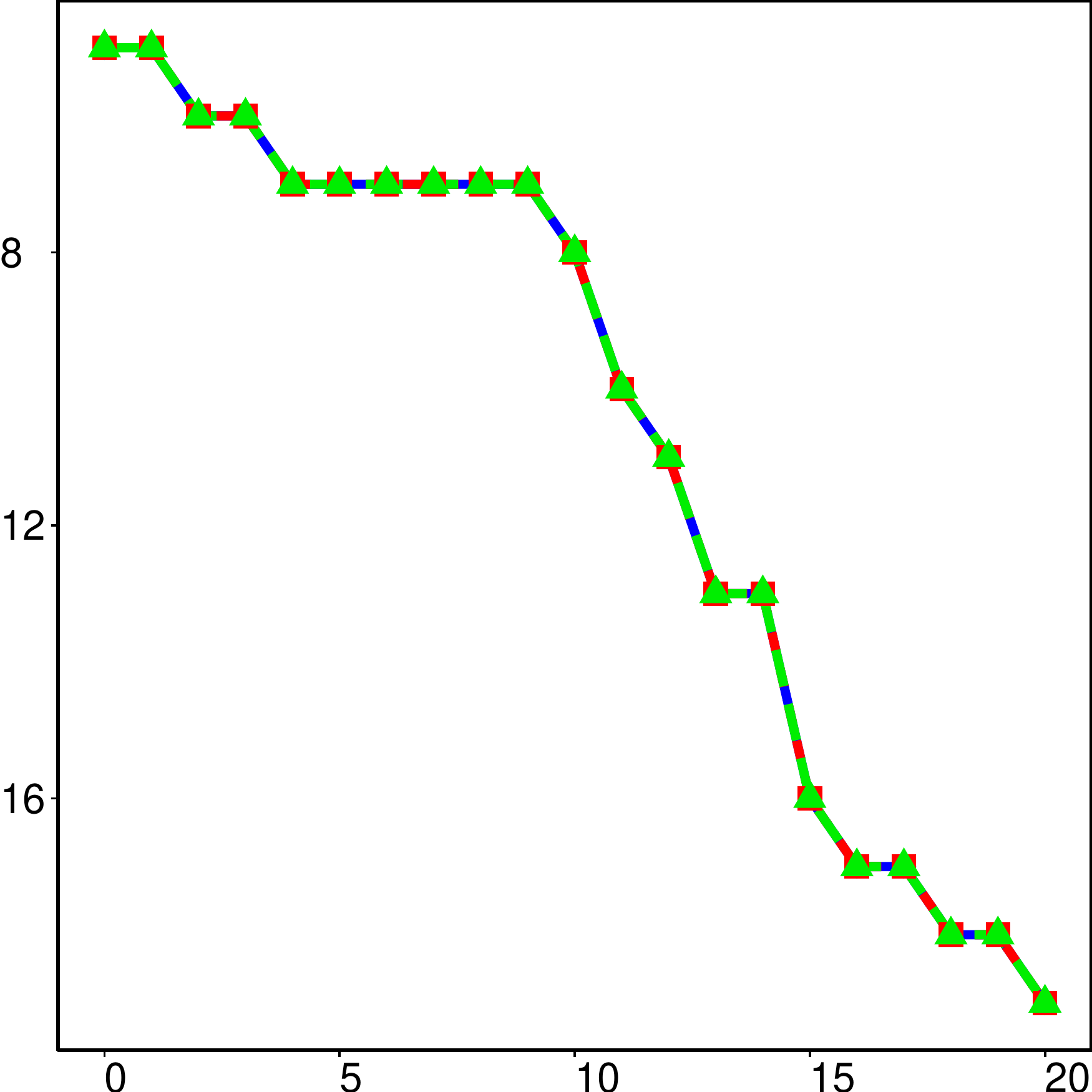} &
	\includegraphics[width=0.93\linewidth]{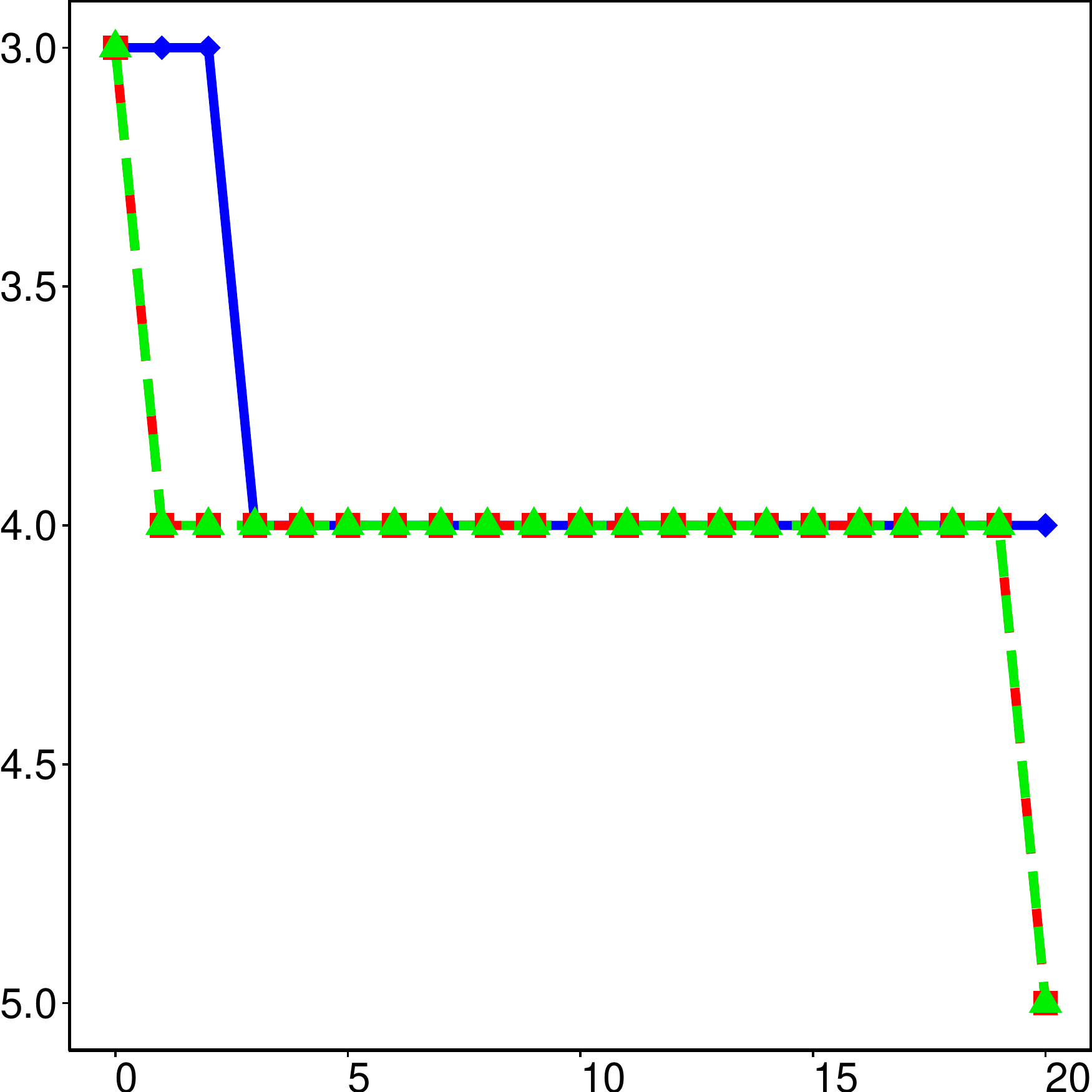} &
	\includegraphics[width=0.93\linewidth]{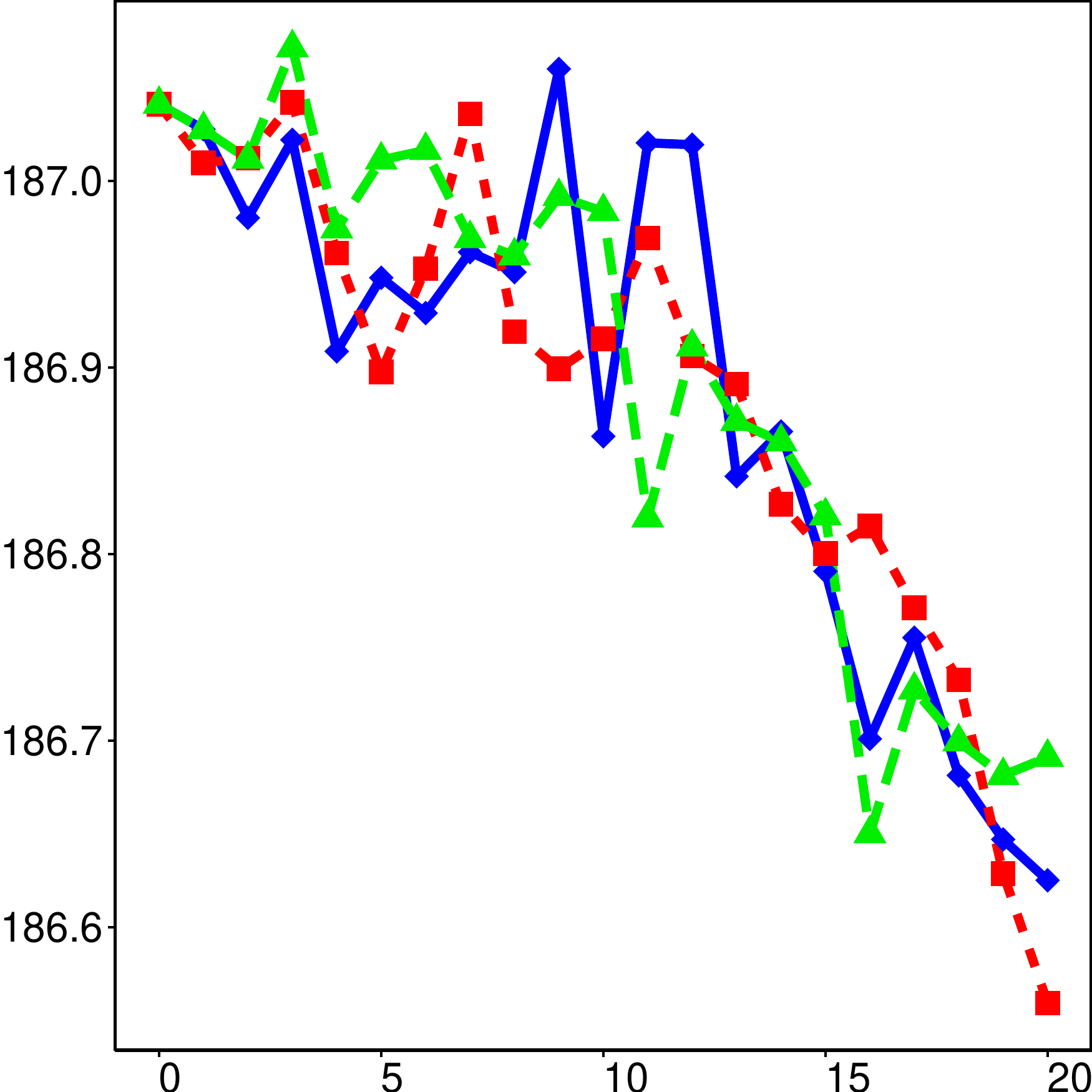} &
	\includegraphics[width=0.93\linewidth]{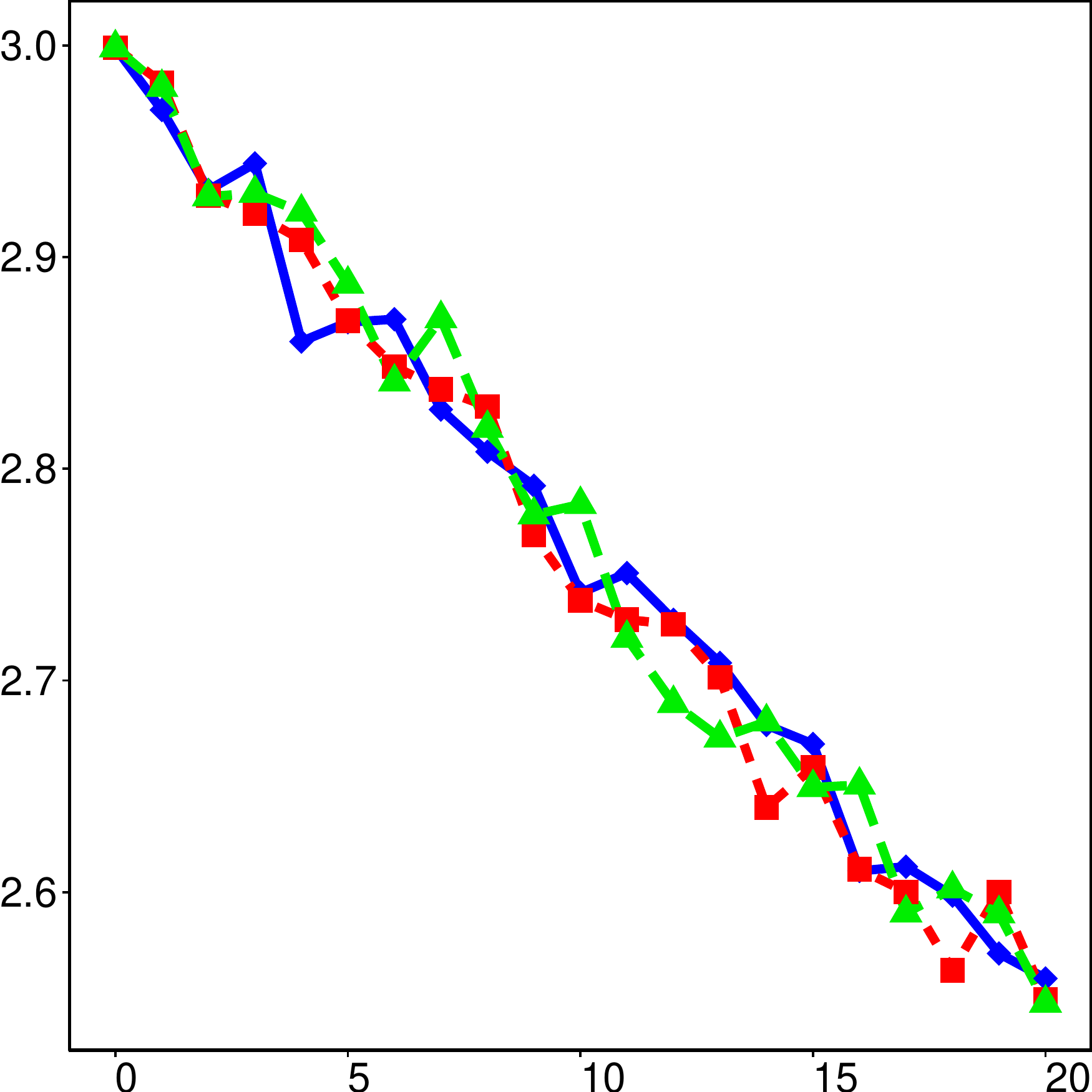} \\
	\rotatebox{90}{\small{Twitter (large)}} &
	\includegraphics[width=0.93\linewidth]{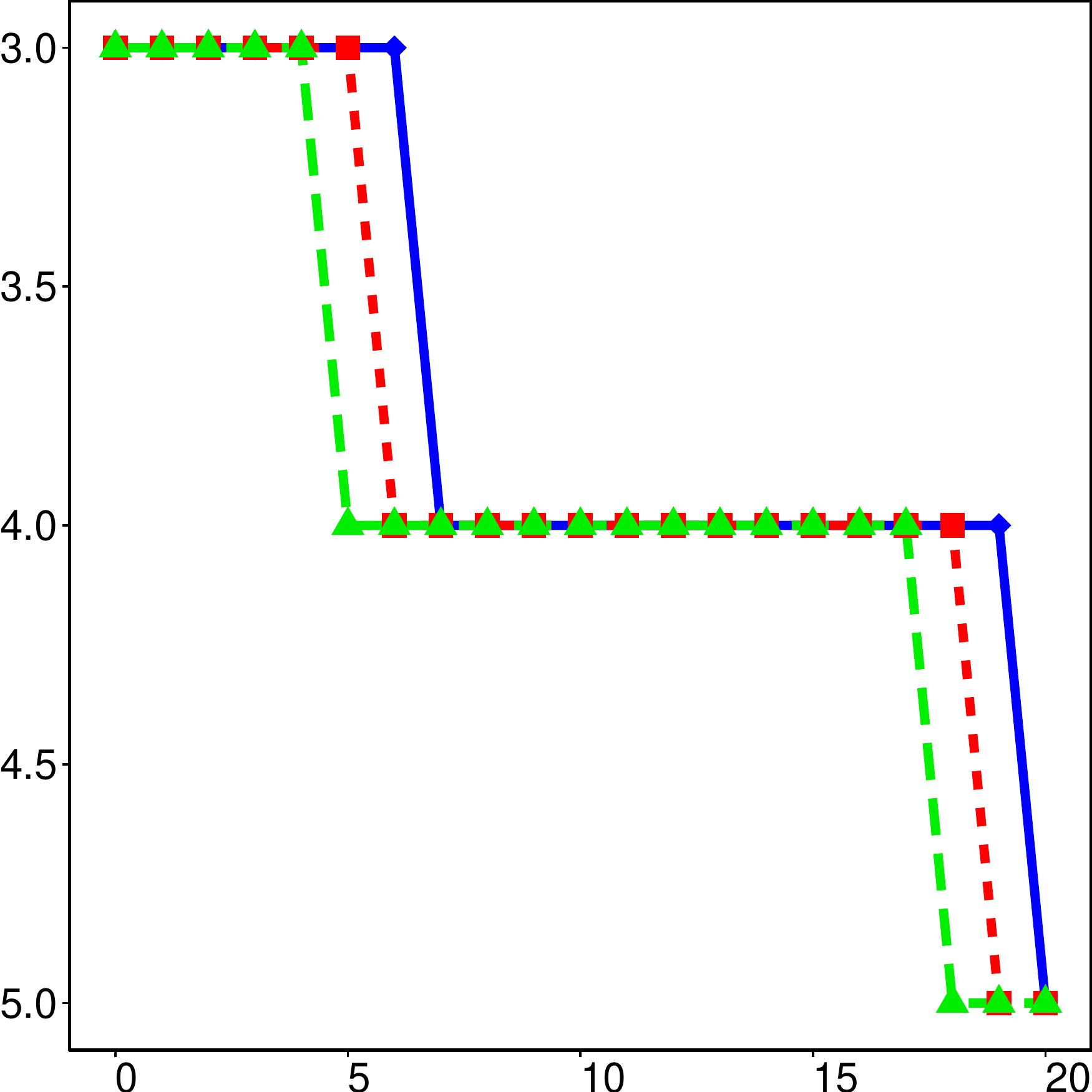} &
	\includegraphics[width=0.93\linewidth]{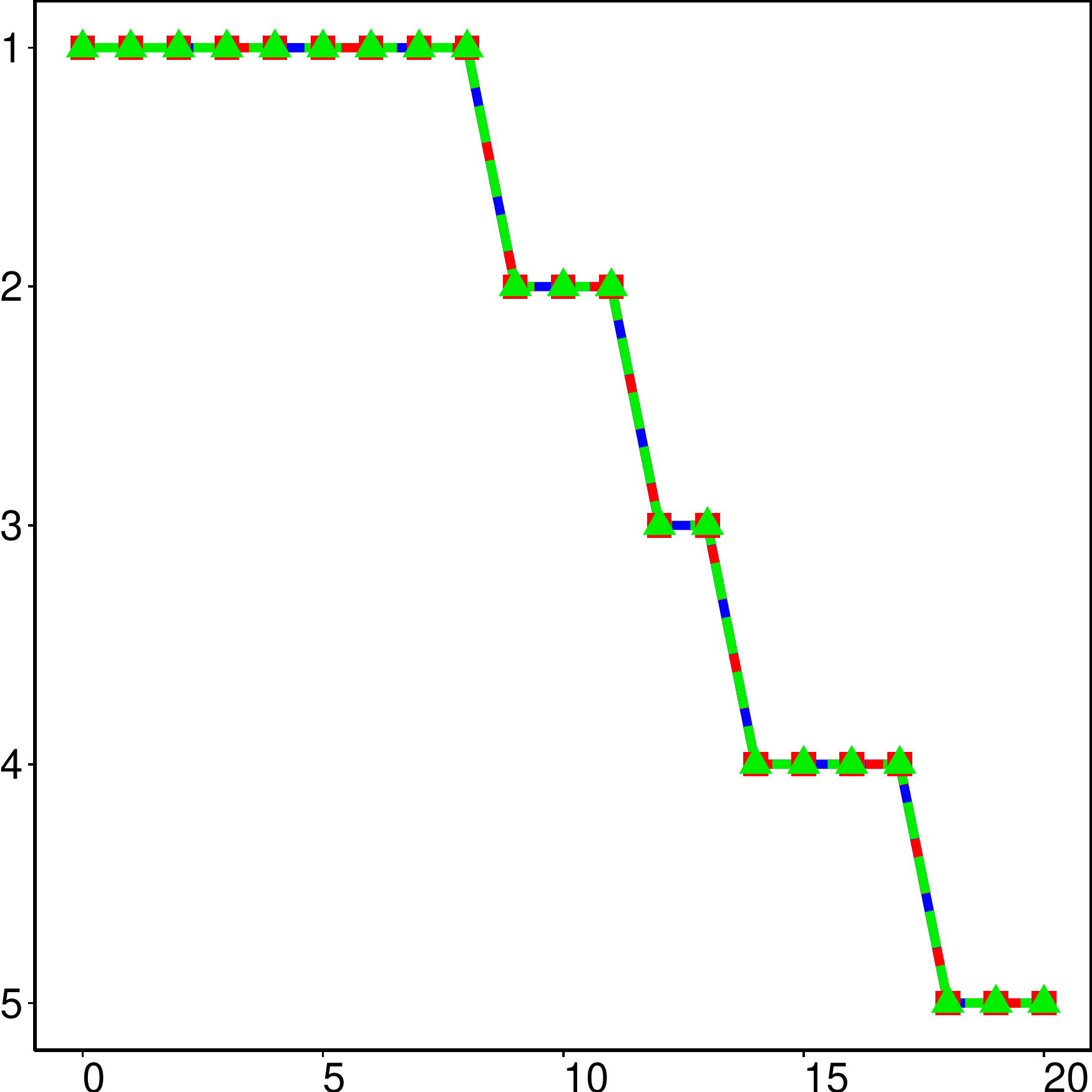} &
	\includegraphics[width=0.93\linewidth]{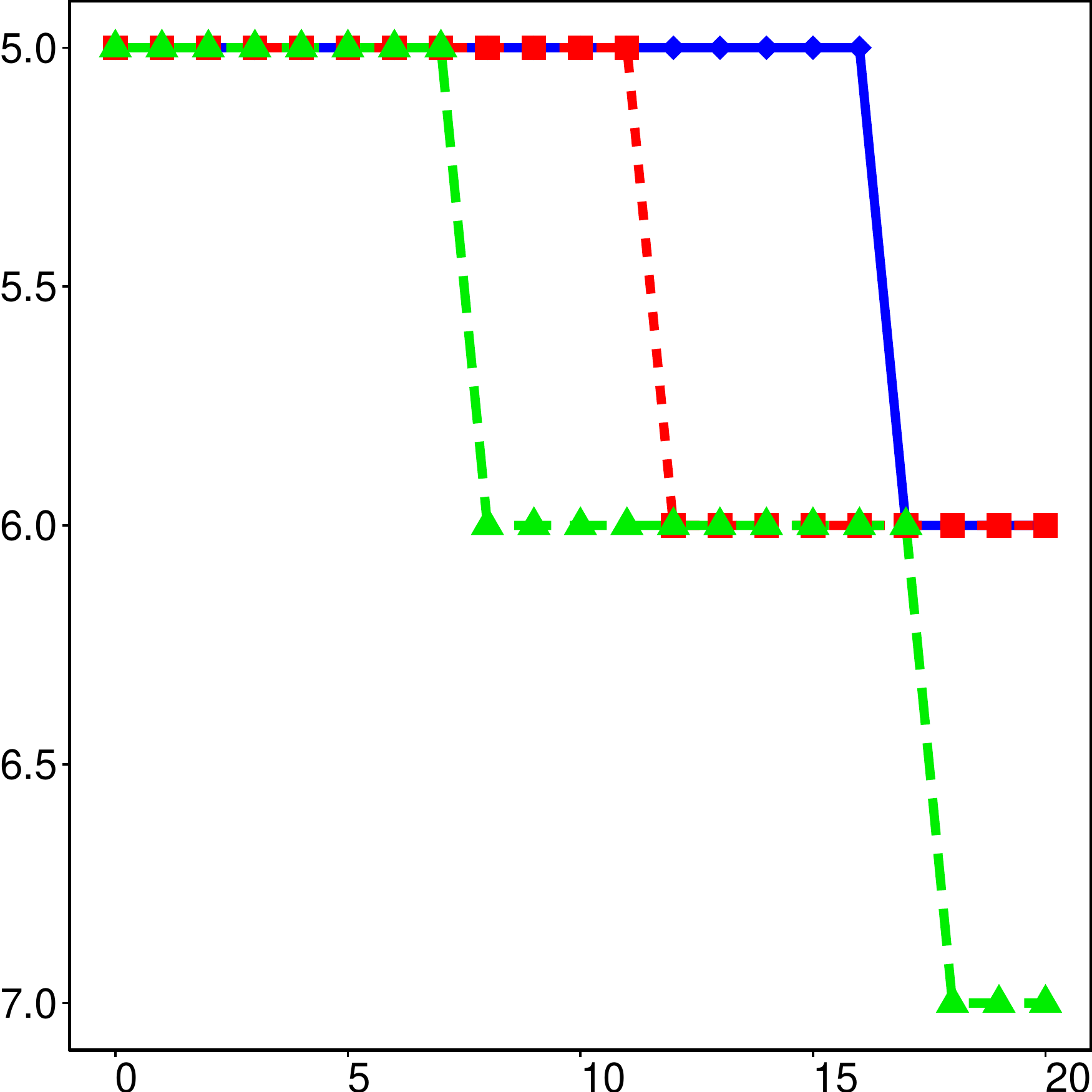} &
	\includegraphics[width=0.93\linewidth]{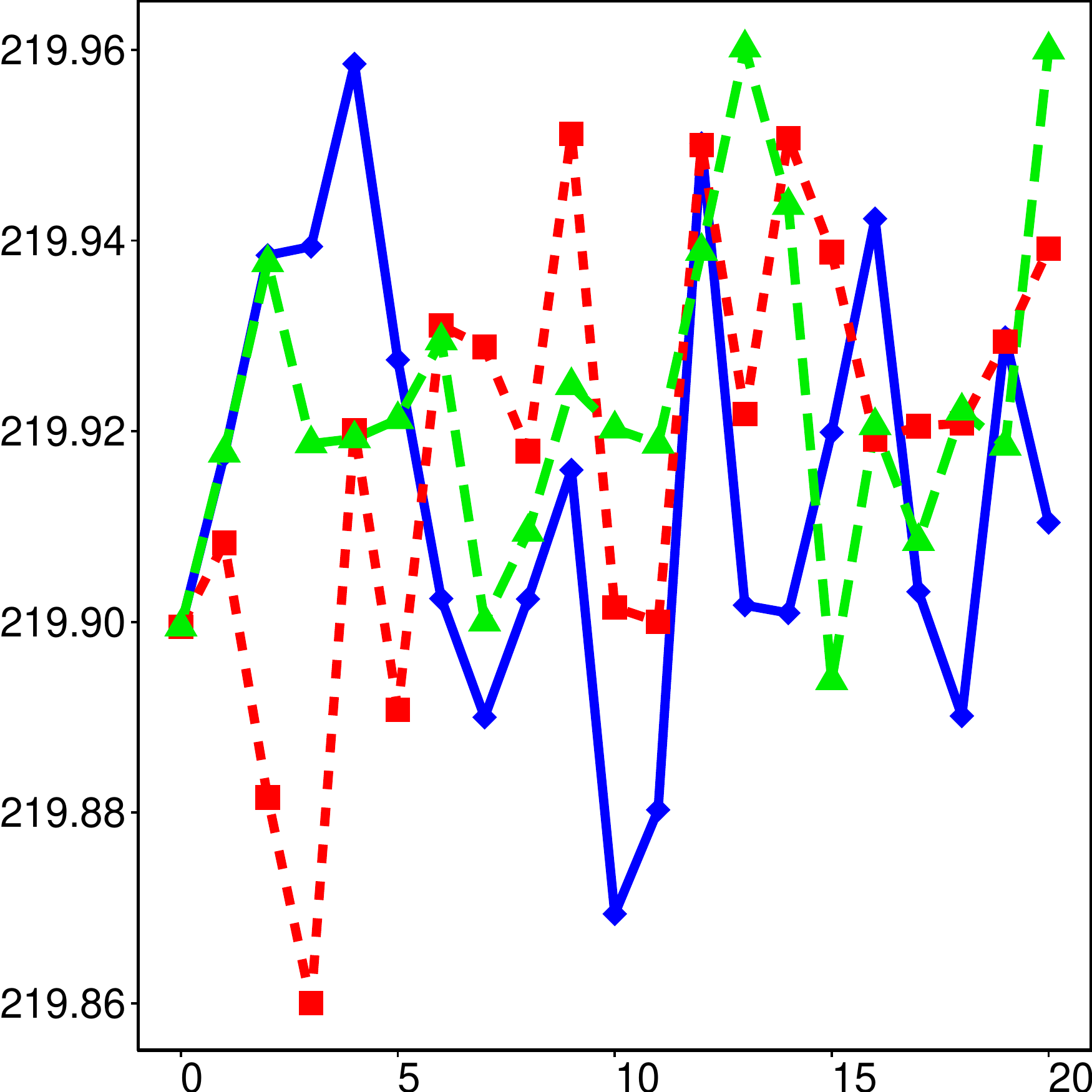} &
	\includegraphics[width=0.93\linewidth]{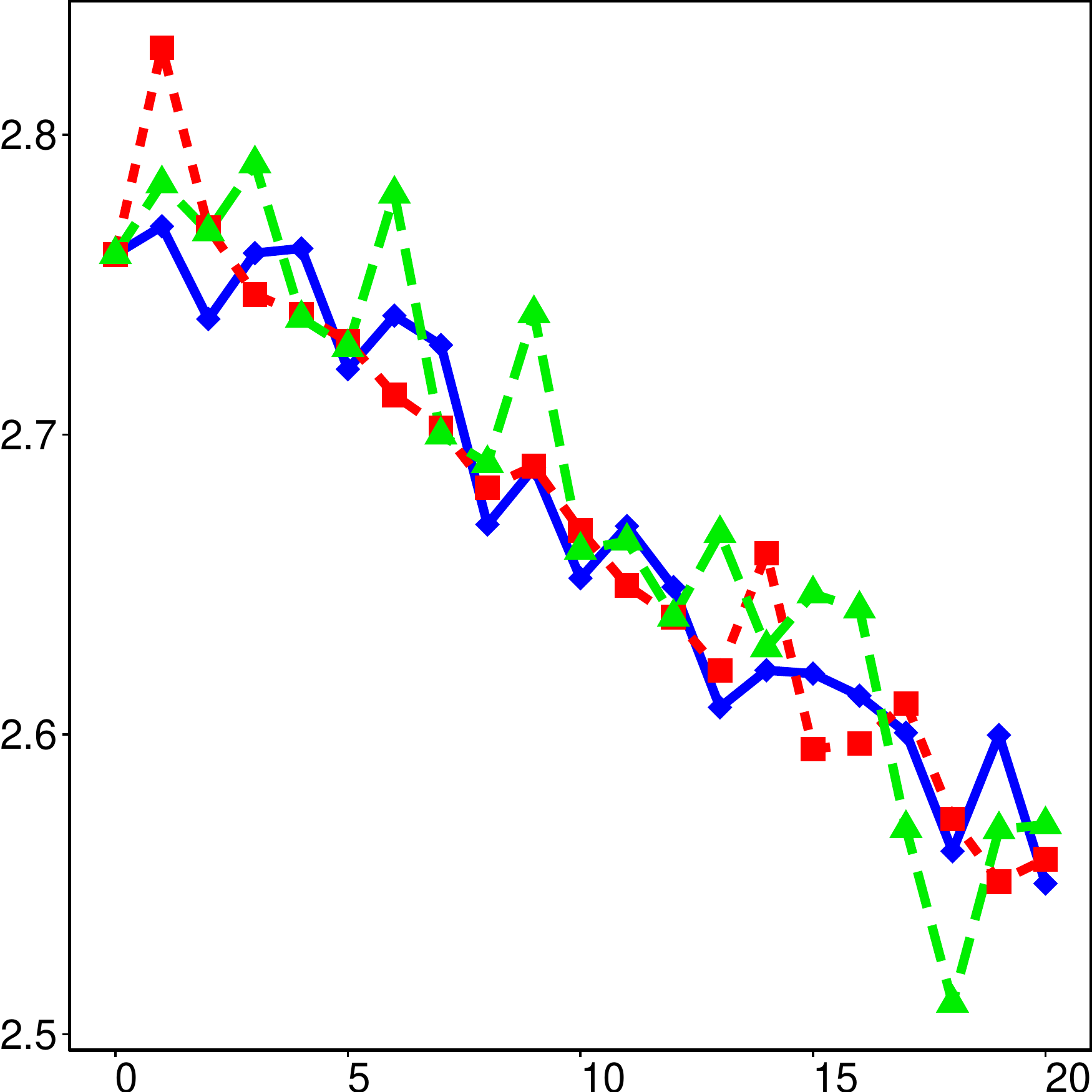} \\
	\rotatebox{90}{\small{Google+ (small)}} &
	\includegraphics[width=0.93\linewidth]{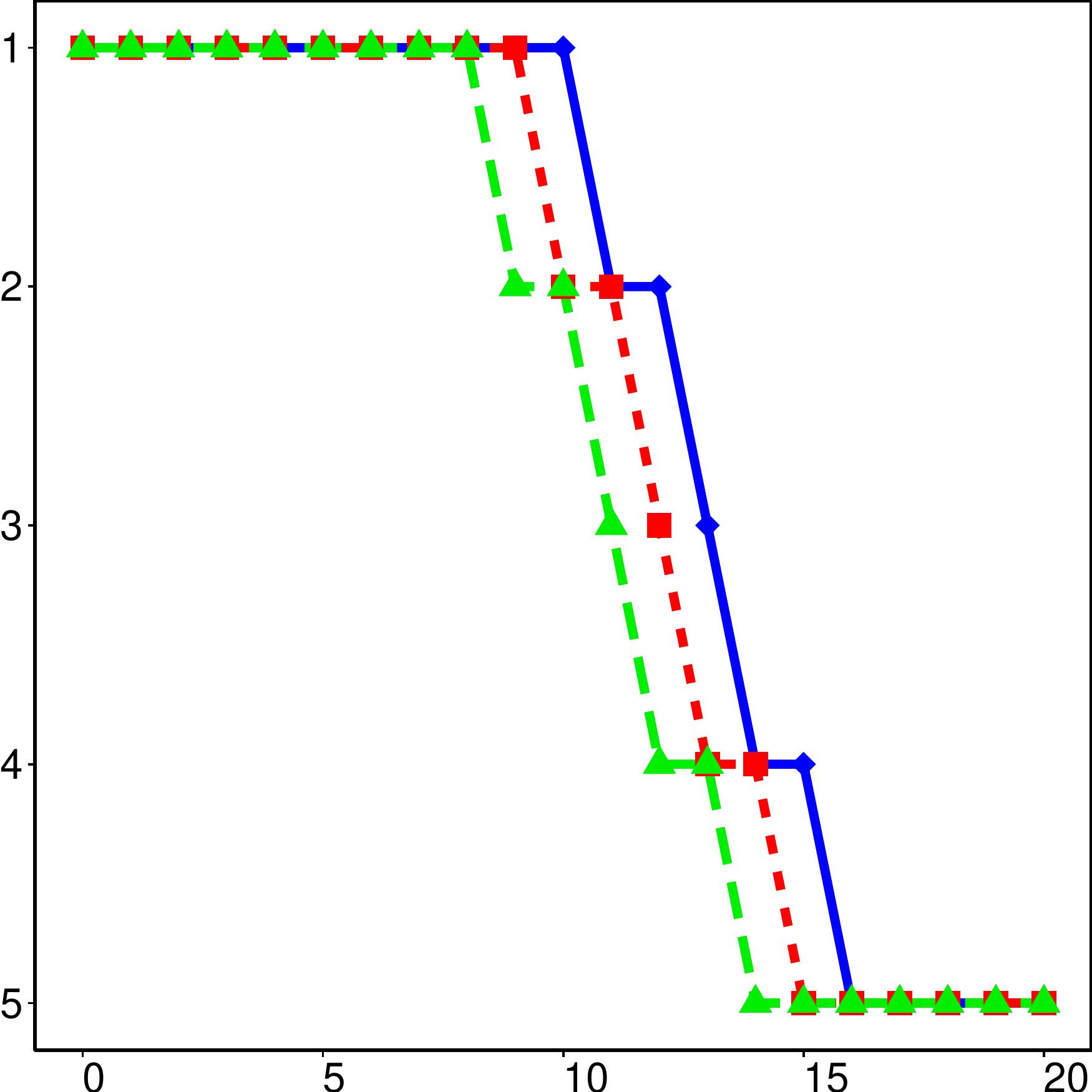} &
	\includegraphics[width=0.93\linewidth]{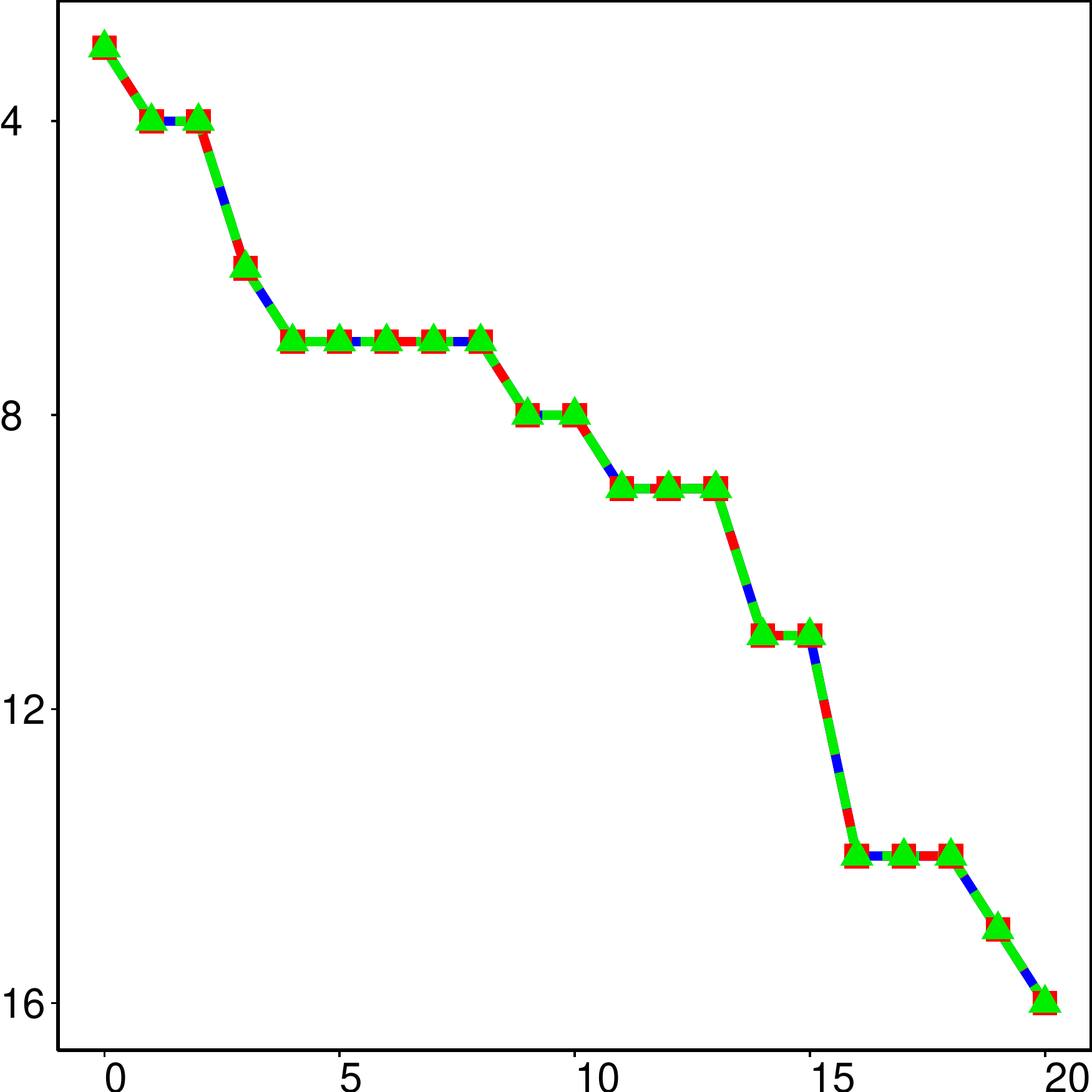} &
	\includegraphics[width=0.93\linewidth]{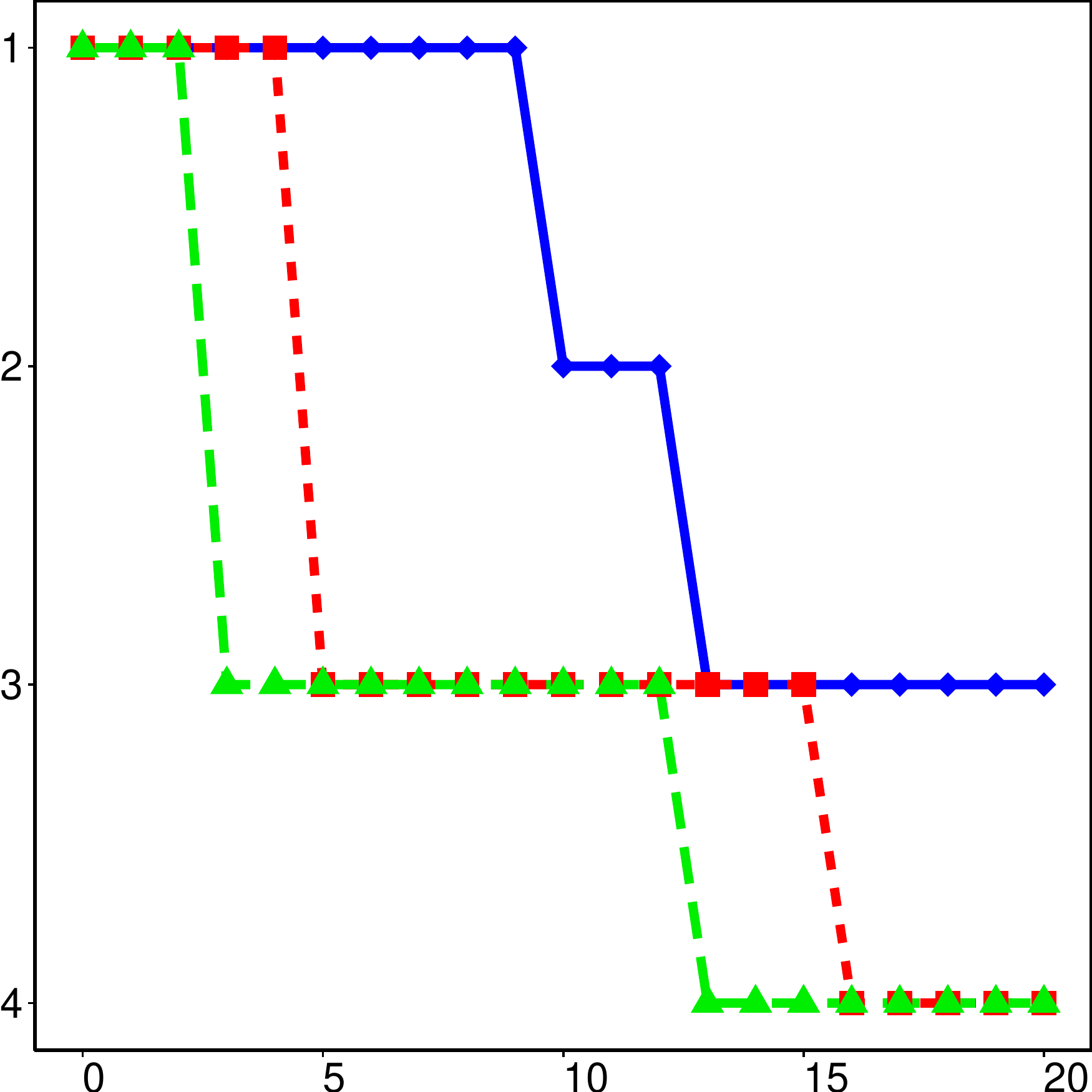} &
	\includegraphics[width=0.93\linewidth]{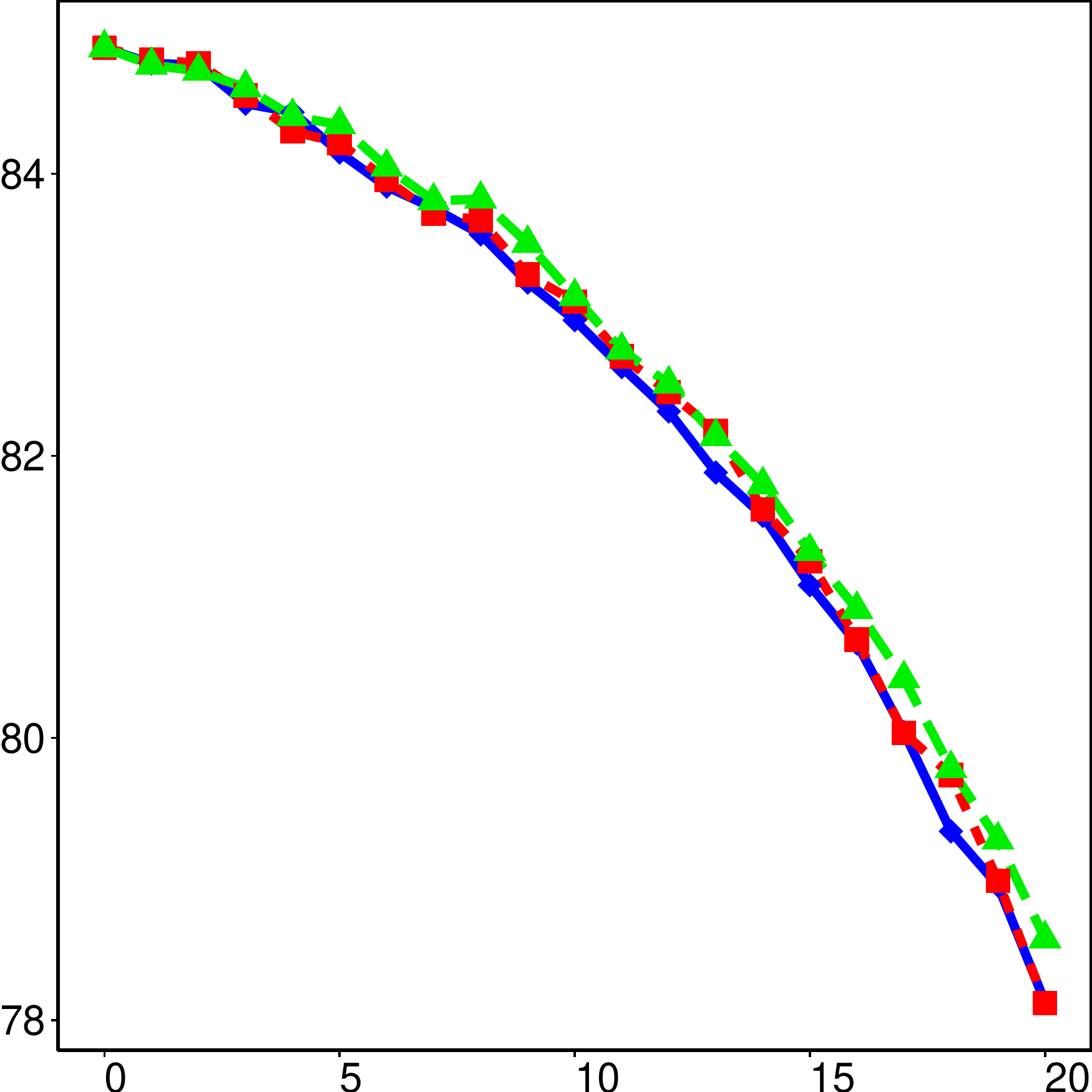} &
	\includegraphics[width=0.93\linewidth]{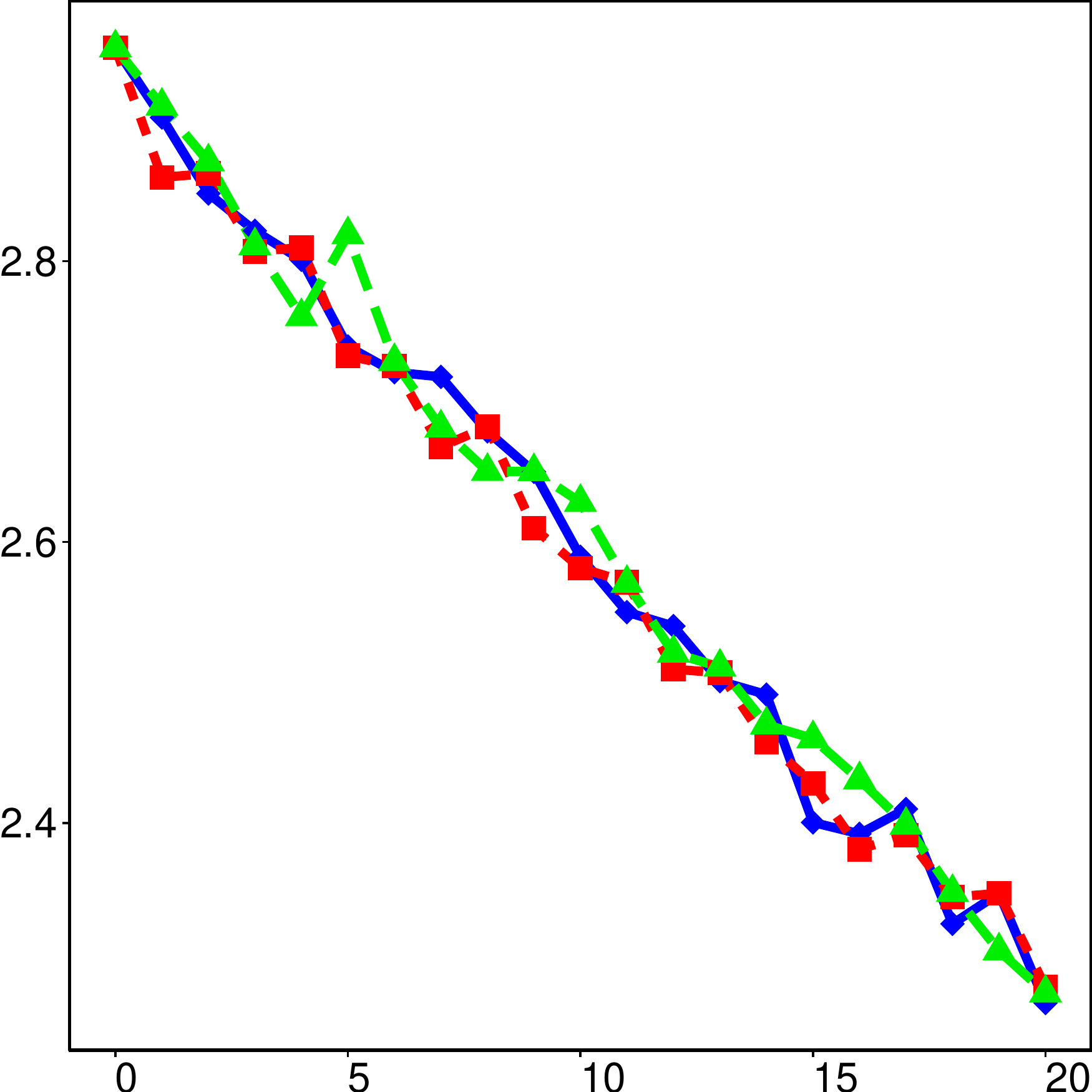} \\
	\rotatebox{90}{\small{Google+ (medium)}} &
	\includegraphics[width=0.93\linewidth]{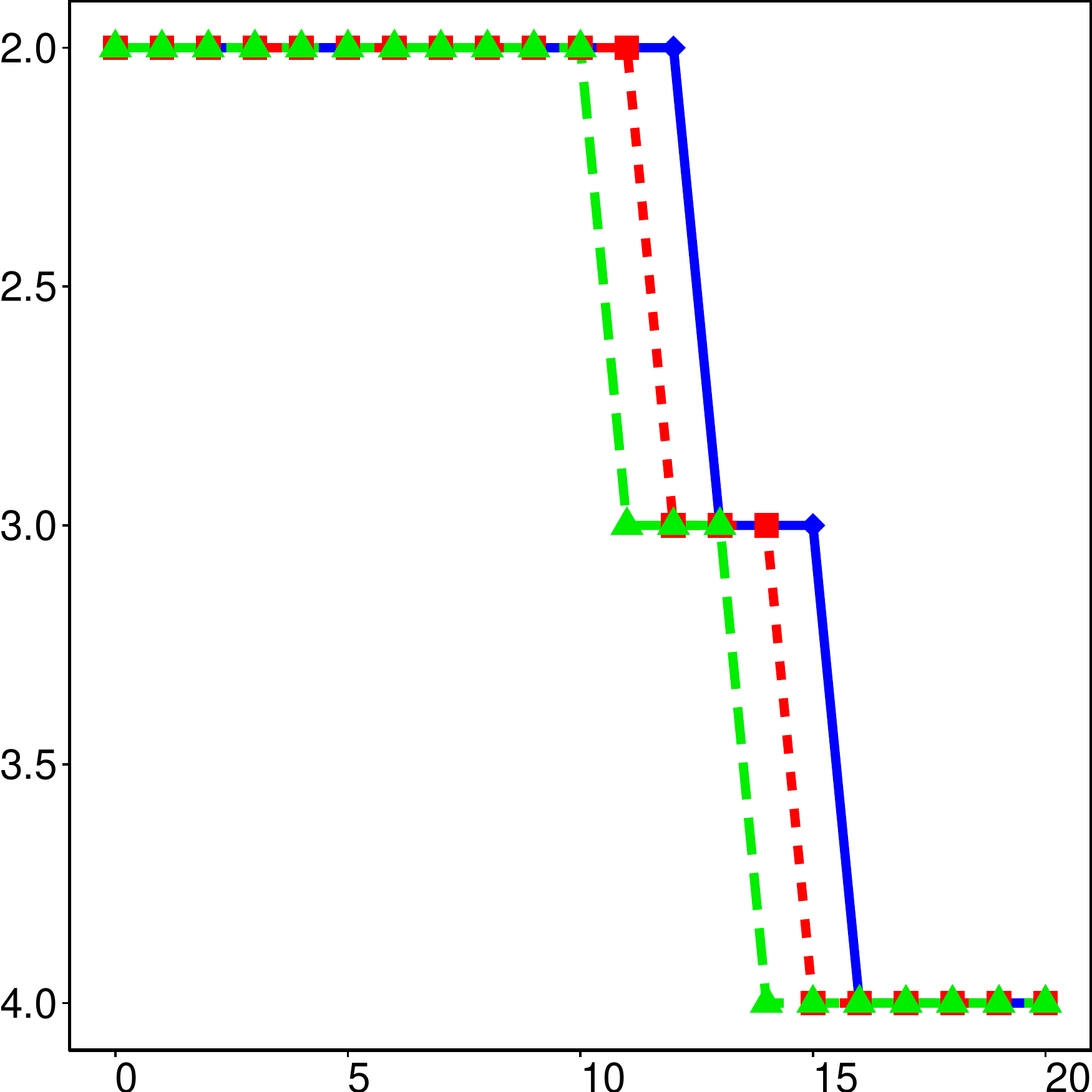} &
	\includegraphics[width=0.93\linewidth]{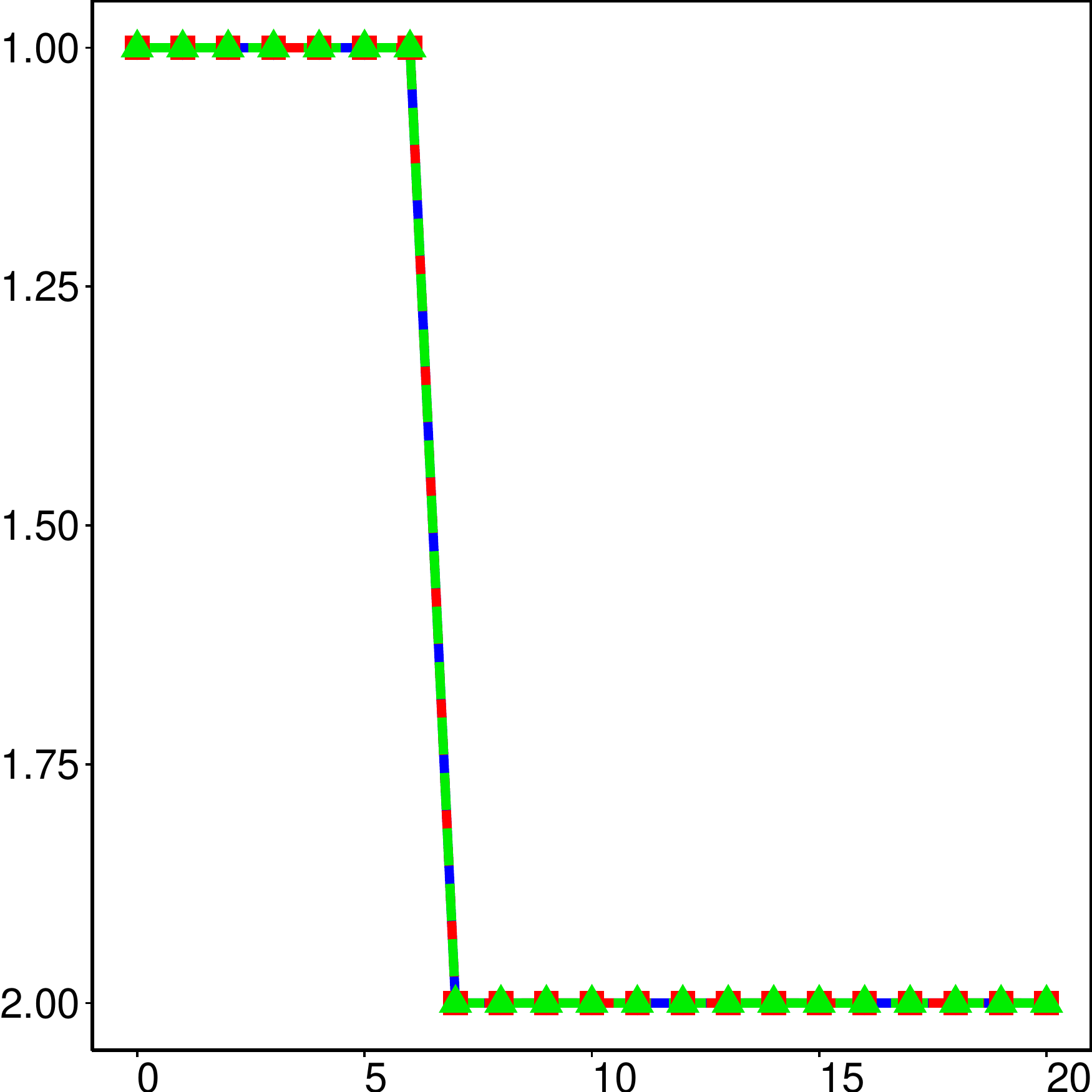} &
	\includegraphics[width=0.93\linewidth]{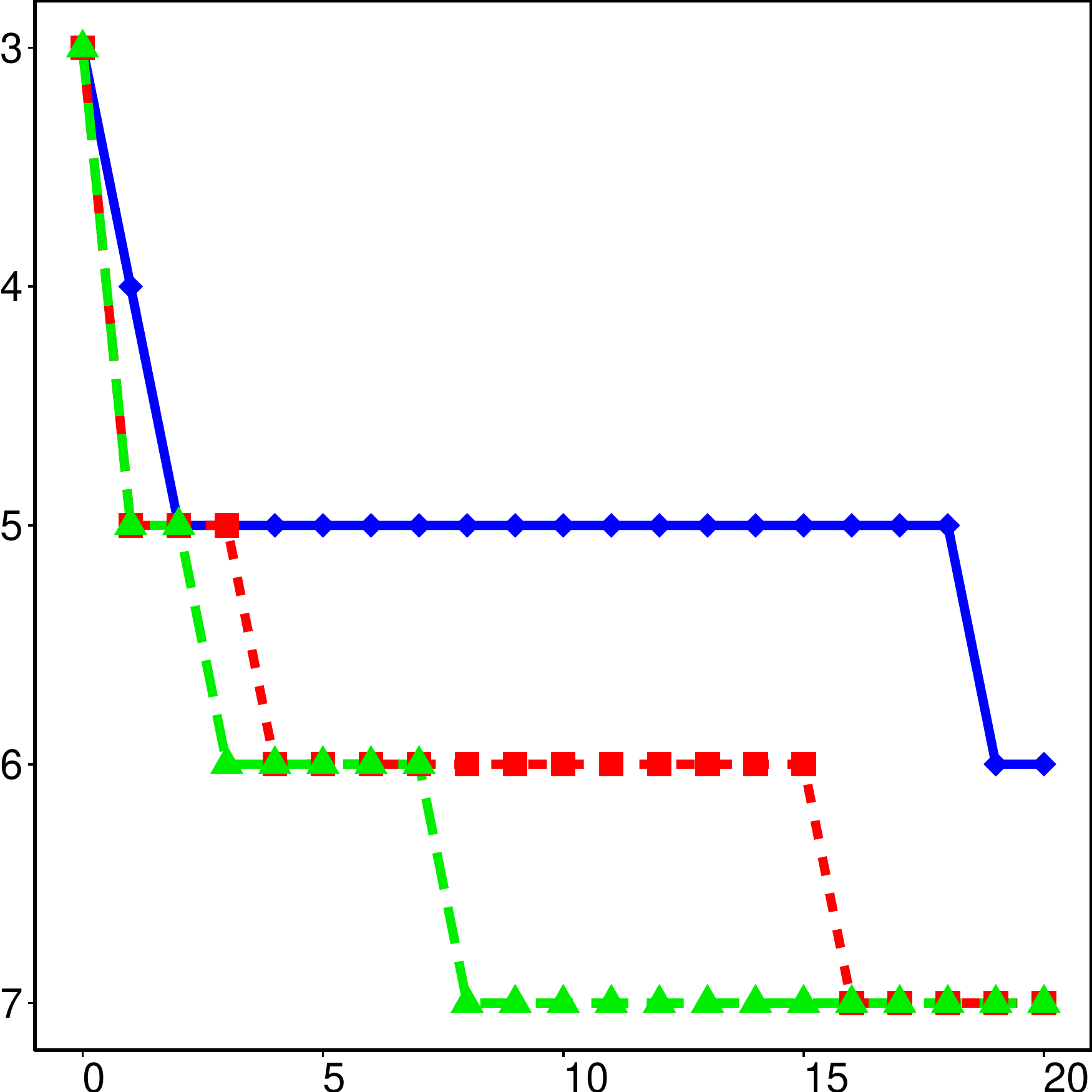} &
	\includegraphics[width=0.93\linewidth]{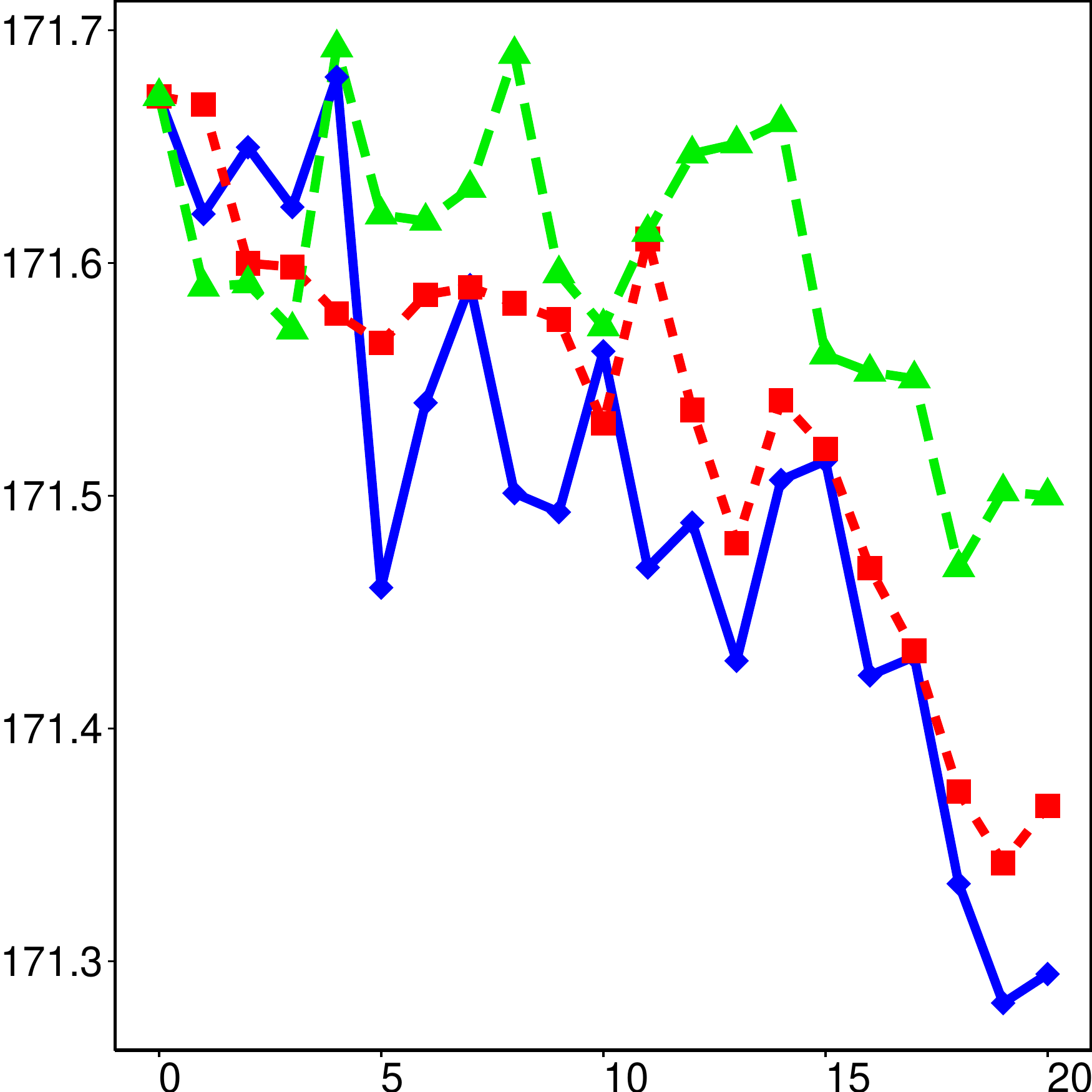} &
	\includegraphics[width=0.93\linewidth]{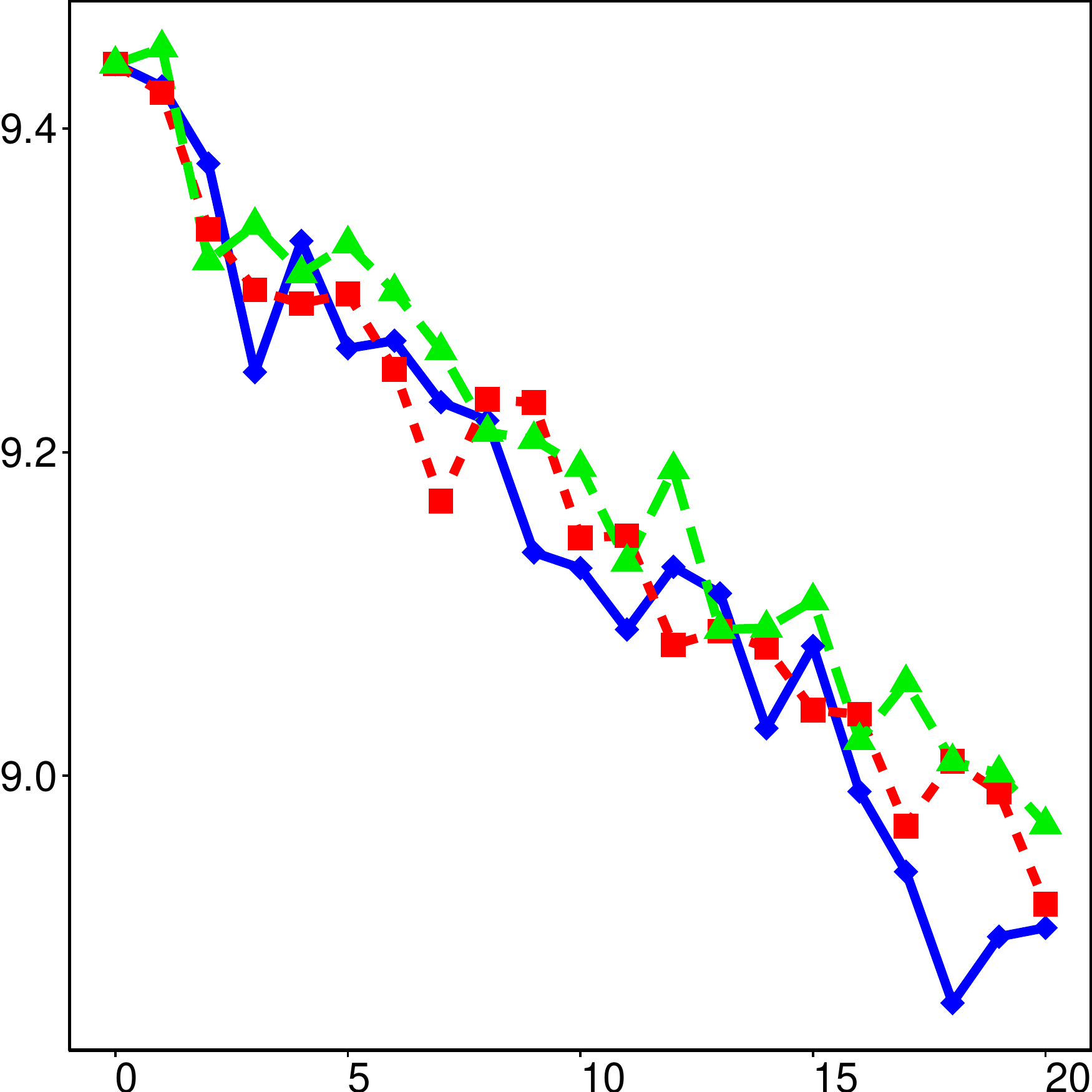} \\
	\rotatebox{90}{\small{Google+ (large)}} &
	\includegraphics[width=0.93\linewidth]{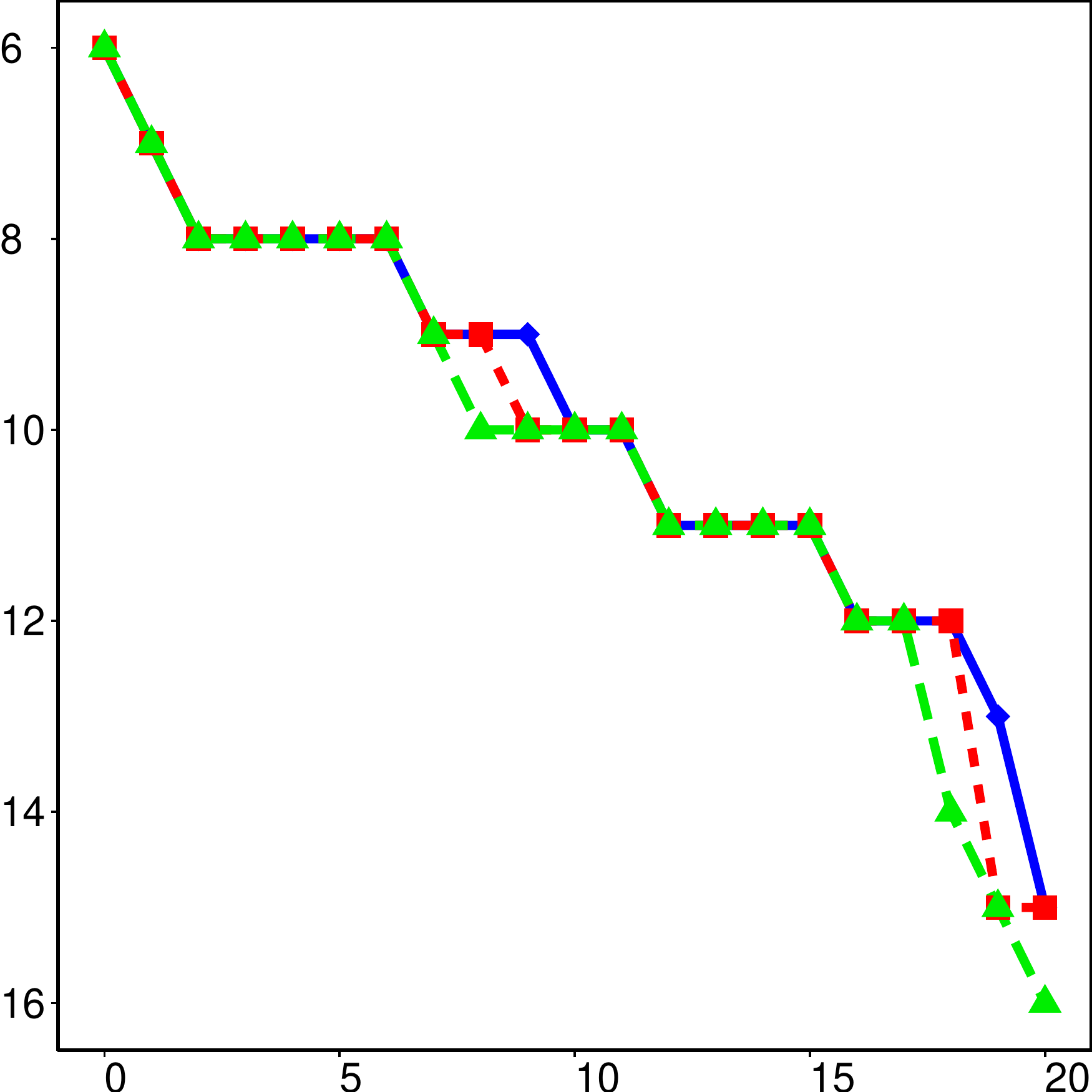} &
	\includegraphics[width=0.93\linewidth]{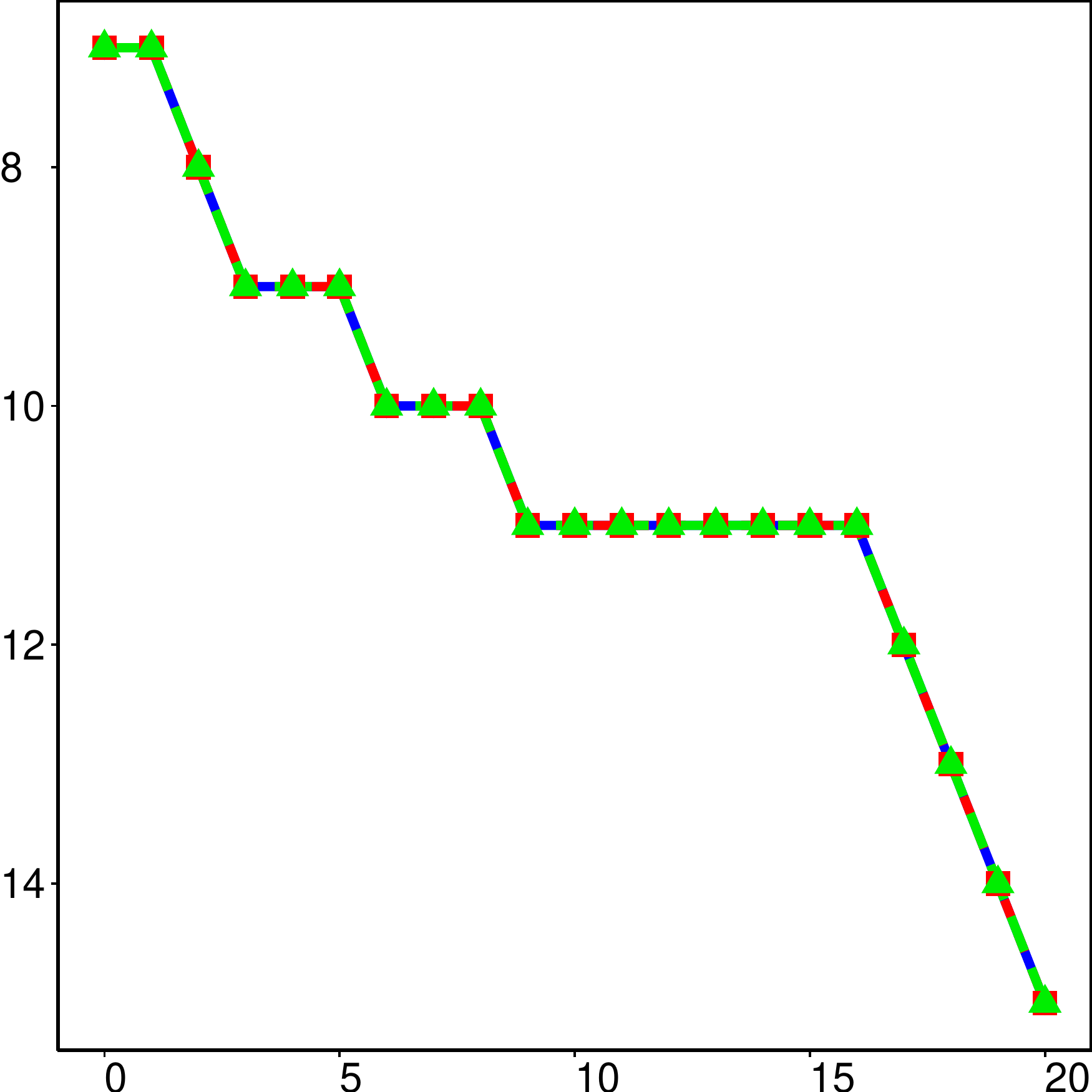} &
	\includegraphics[width=0.93\linewidth]{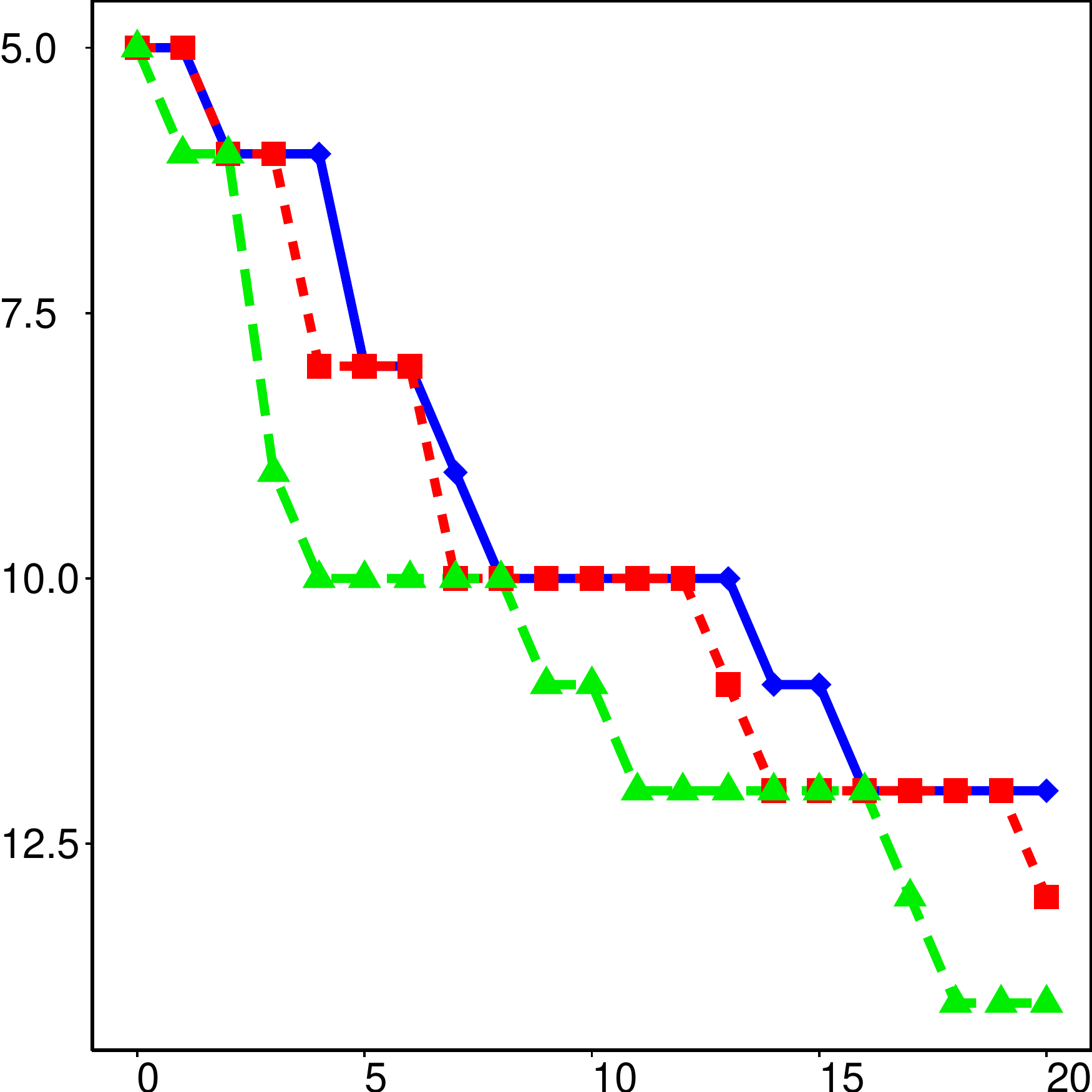} &
	\includegraphics[width=0.93\linewidth]{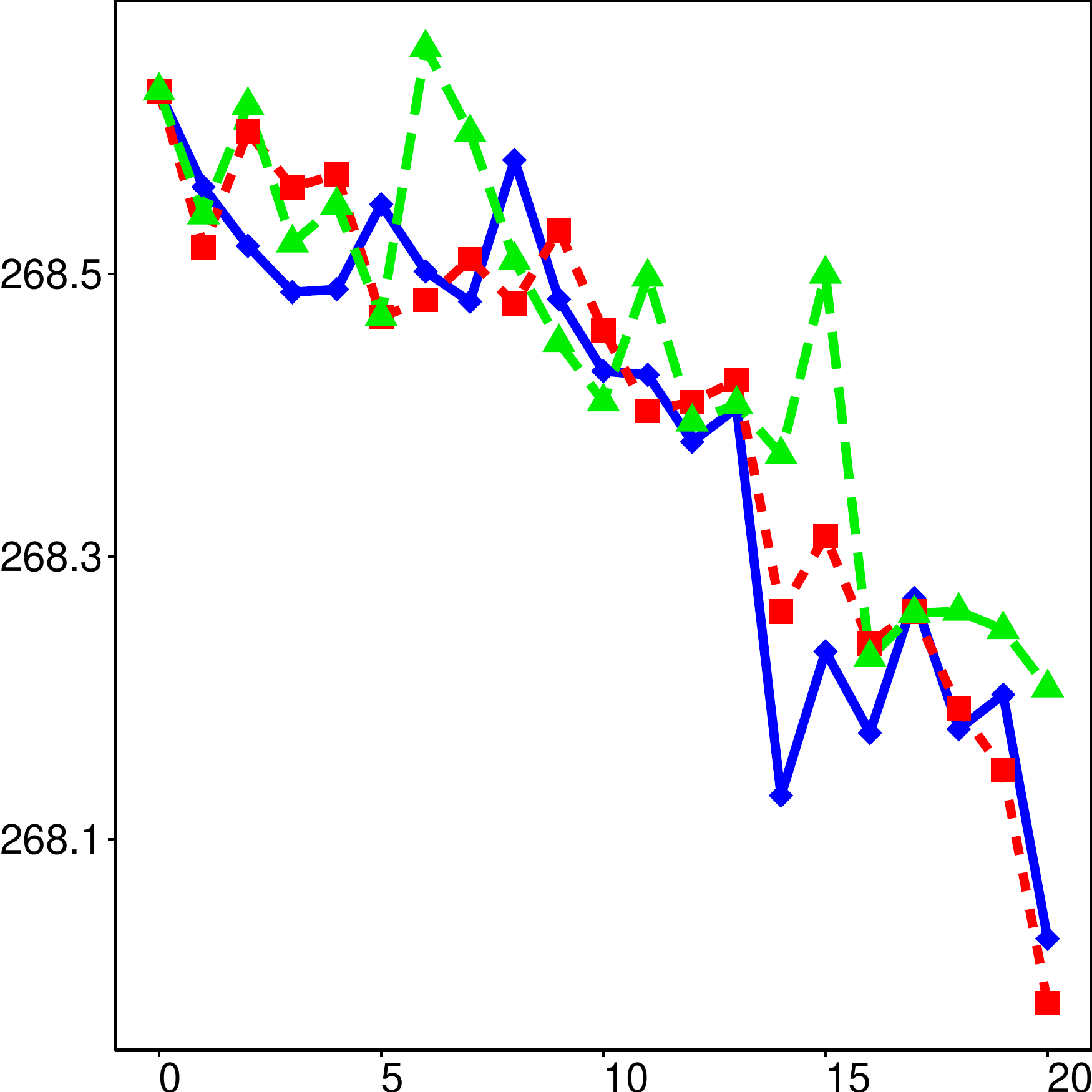} &
	\includegraphics[width=0.93\linewidth]{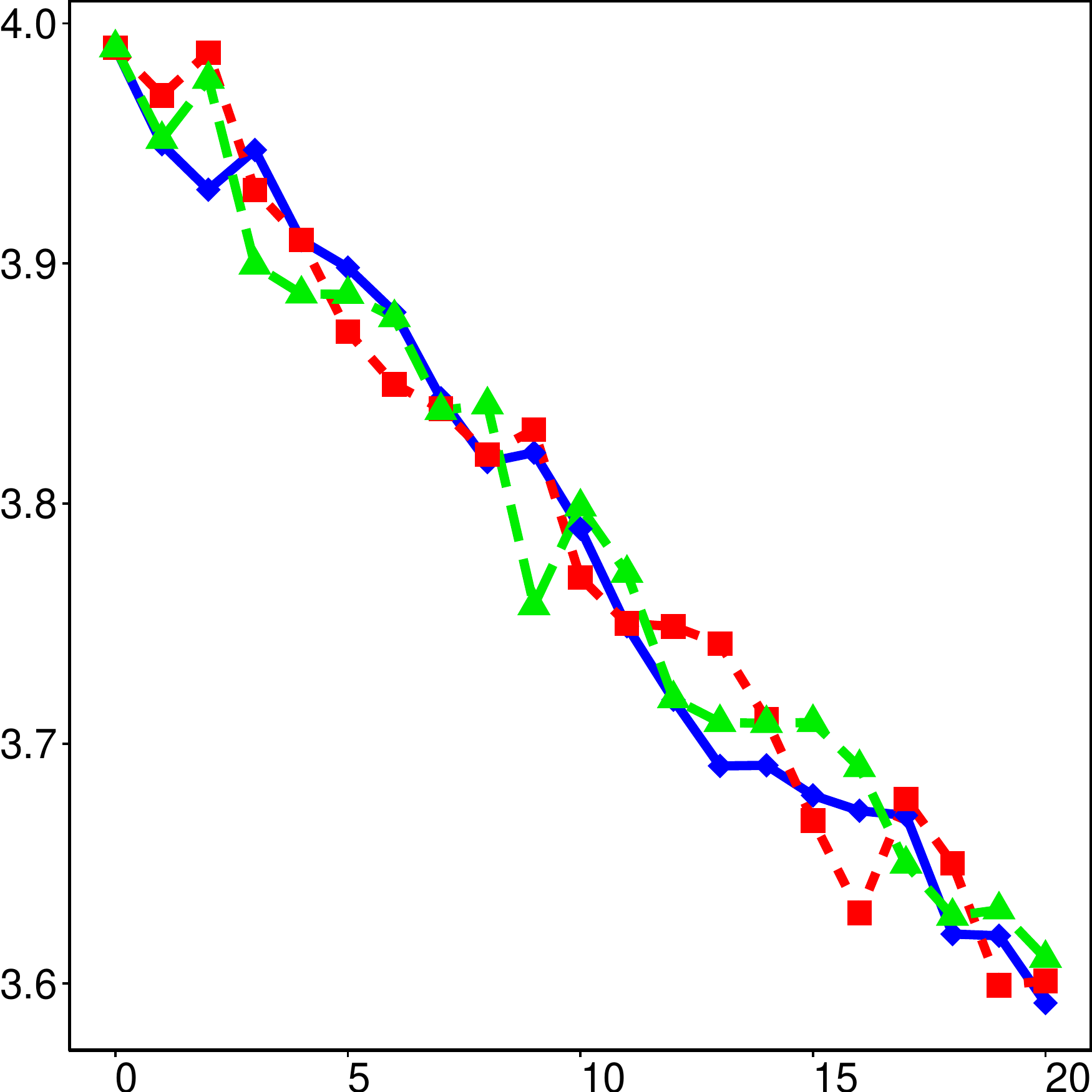} \\
	&
	\multicolumn{5}{c}{\includegraphics[width=0.75\linewidth]{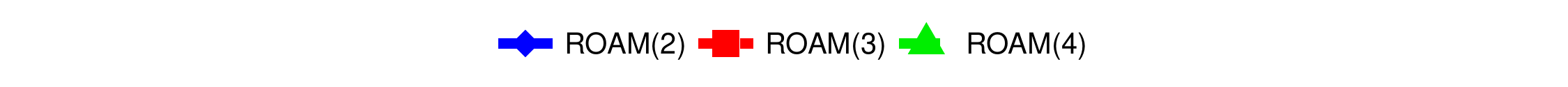}} \\
	\end{tabular}
	\vspace*{-0.5cm}
	\caption{\fontsize{10}{10}\selectfont{Consecutive execution of ROAM (the $x$-axis represents the number of executions). Given different fragments of the social networks of Twitter and Google+, the subfigures show the source node's ranking (according to the centrality measures), and the relative change in its influence value (according to the influence models). Results are for $\textnormal{ROAM}(b):b=2, 3, 4$, where $b$ is the budget in each execution.}}
	\label{fig:centralityResults:Twitter:Google+}
\end{figure}

Our experimental results for the ROAM heuristic are all presented in figures \ref{fig:centralityResults:random}, \ref{fig:centralityResults:terroist:Facebook} and \ref{fig:centralityResults:Twitter:Google+}, which can be found at the end of this document.

%%%%%%%%%%%%%%%%%%%%%%%%%%%%%%%%%%%%%%%%%%%%%%%%%%%%%%%%%%%%%%%%%%%%%%%%%%%%%%%

\section{Constructing a Network from Scratch}\label{sec:lieutenant}

\noindent
%\pagenumbering{gobble} %disable page numbers
Having studied the problem of disguising a node by rewiring an existing network, we now study the same problem but from a different perspective, where the goal is to construct a network from scratch, designed for the sole purpose of concealing the source node, $v^{\dagger}$. In this section, we will restrict our attention to undirected networks. Specifically, given $n$ nodes, our goal is to identify a topology in which $v^{\dagger}$ has a reasonably-high influence, while at the same time ensuring that a certain number of nodes is ranked higher than $v^{\dagger}$ according to each of the three centrality measures. To tackle this problem, we propose what we call a \emph{Lieutenant network}, the structure of which is detailed in the box below.

%%%%%%%%%%%%%%%%%%%%%%%%%%%%%%%%%%%%%%%%%%%%%%%%%%%%%%%%%%%%%%%%%%%%%%%%%%%%%%%
\begin{mdframed}[style=MyFrame]
\textbf{The Lieutenant network} of size $n$: \vspace*{-0.14cm}
\begin{itemize}\itemsep0.3em
\item Label one node as the source node, $v^{\dagger}$;
\item Label two groups of \emph{lieutenants}, containing $k$ nodes each, namely: $L = \{l_1, \ldots, l_k\}$ and $L' = \{l'_1, \ldots, l'_k\}$;
\item Label all remaining nodes as $M = \{m_1, \ldots, m_{\lambda}\}$ where $\lambda = n - 2k - 1$; these are called \emph{members};
	\item Connect the source node to every lieutenant; 
	\item Connect every lieutenant in $L$ to every one in $L'$;
	\item Connect every member to exactly $c$ lieutenants from $L$ and exactly $c$ lieutenants from $L'$ while ensuring that the degrees of lieutenants differ by at most 1.
\end{itemize}
\end{mdframed}
%%%%%%%%%%%%%%%%%%%%%%%%%%%%%%%%%%%%%%%%%%%%%%%%%%%%%%%%%%%%%%%%%%%%%%%%%%%%%%%

\begin{figure}[tbhp]
\centering
\includegraphics[width=0.6\linewidth]{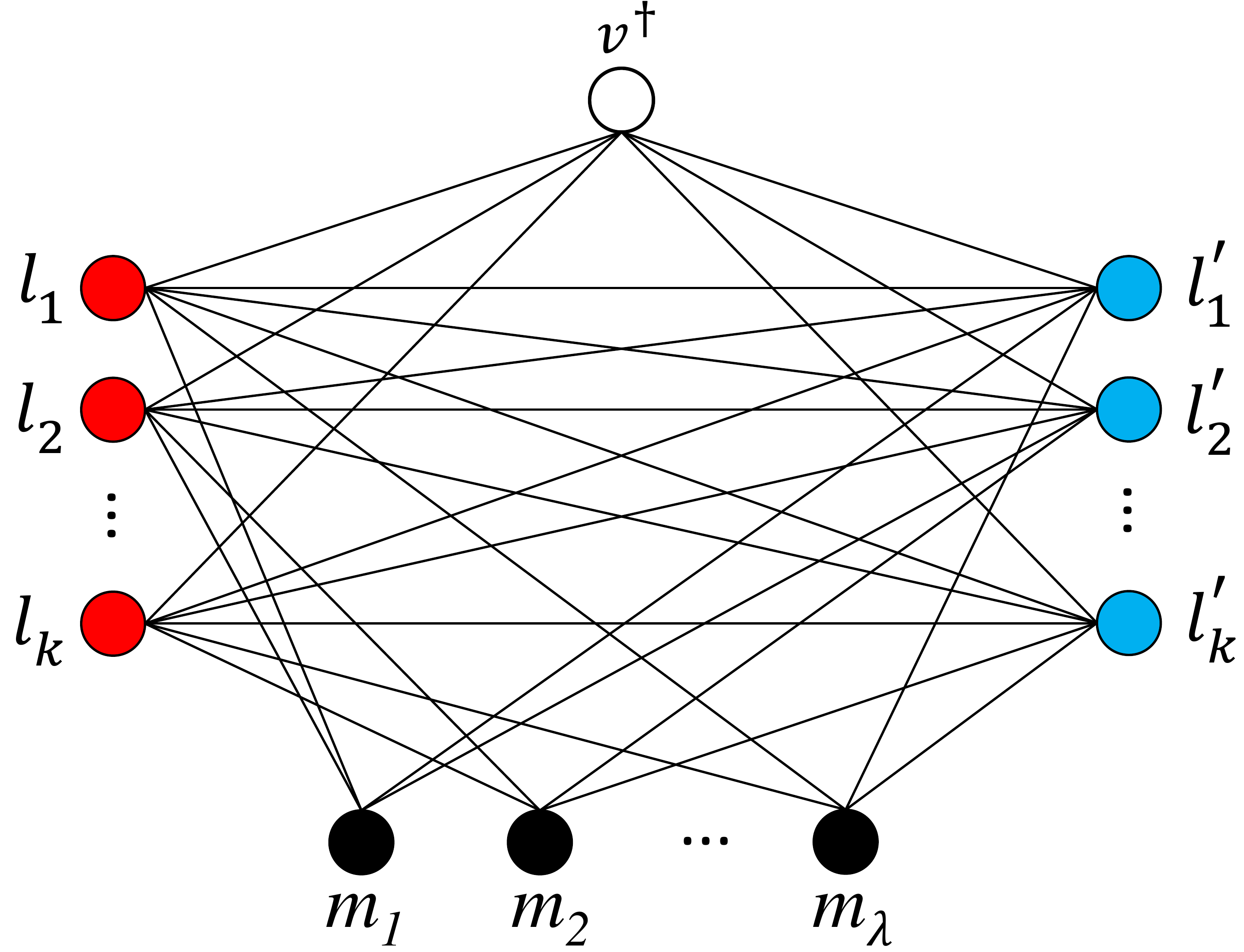}
\caption{\fontsize{10}{10}\selectfont{An illustration of a Lieutenant network with $c = 2$.}}
\label{fig:scratch_lieutenant}
\end{figure}

%\pagenumbering{arabic} %enable page numbers
Here is how the Lieutenant network works. The source node $v^{\dagger}$ only comes into contact with its lieutenants. These are the ones that are supposed to conceal $v^{\dagger}$ by ensuring that they are each ranked higher than $v^{\dagger}$ according to the three standard centrality measures. These are also the nodes that are supposed to pass on the influence of $v^{\dagger}$ to the rest of the network. Figure~\ref{fig:scratch_lieutenant} illustrates a sample Lieutenant network with $c = 2$.

%%%%%%%%%%%%%%%%%%%%%%%%%%%%%%%%%%%%%%%%%%%%%%%%%%%%%%%%%%%%%%%%%%%%%%%%%%%%%%%
%%%%%%%%%%%%%%%%%%%%%%%%%%%%%%%%%%%%%%%%%%%%%%%%%%%%%%%%%%%%%%%%%%%%%%%%%%%%%%%

\begin{theorem}
Let $f$ denote the minimum number of members connected to any single lieutenant, i.e., $f = \floor{\frac{c\lambda}{k}}$. Then, for every Lieutenant network such that $f > k -1$ and $f^2 > 4ck$, all lieutenants have greater degree, closeness and betweenness centrality than the source node $v^{\dagger}$.
\end{theorem}

\noindent\textbf{Proof.}
Starting with degree centrality, the degree of the source node, $v^{\dagger}$, is $c_{degr}(G,v^{\dagger}) = \frac{2 k}{n-1}$, since it is only connected to lieutenants. On the other hand, the degree of a lieutenant, $l_i$, is $c_{degr}(G,l_i) \geq \frac{1 + k + f}{n-1}$, since it is connected to the source node, to all lieutenants from the other group, and to at least $f$ members. As such, we have:
$$
c_{degr}(G,l_i) - c_{degr}(G,v^{\dagger}) \geq \frac{f - k + 1}{n-1}
$$
Therefore, $c_{degr}(G,l_i) > c_{degr}(G,v^{\dagger})$ for all $l_i\in L\cup L'$ when $f > k - 1$.

Moving on to closeness centrality, for any given node, $v$, this centrality depends inversely on the sum of the lengths of shortest paths from $v$ to every other nodes, i.e., $\sum_{u \in V}d_G(v,u)$.
For both the source node and every lieutenant, the distance to every other node is either $1$ or $2$. More precisely, for every $v\in \{v^{\dagger}\}\cup L\cup L'$, we have: $\sum_{u \in V}d_G(v,u) = 1 |N(v)| + 2 (n - |N(v)|) = 2n - |N(v)|$. Consequently, whenever all lieutenants have greater degree centrality than $v^{\dagger}$, they must also have greater closeness centrality than $v^{\dagger}$. This in turn implies that $c_{clos}(G,l_i) > c_{clos}(G,v^{\dagger})$ for all $l_i\in L\cup L'$ when $f > k - 1$.

Finally, regarding betweenness centrality, let $\delta(v)$ denote: $\sum_{u,w \in V \setminus \{ v \}:u\neq w} \frac {|\{ p \in \mathit{sp}_G(u,w) : v \in p \}|} {|\mathit{sp}_G(u,w)|}$. Then the betweenness centrality of a node $v\in V$ can be written as: $c_{betw}(G,v) = \frac{2}{(n-1)(n-2)}\delta(v)$. Furthermore, for any two lieutenants, $u, w\in L\cup L':u\neq w$, let $\gamma_{u,w}$ denote the number of members that are neighbours to both of them, i.e.,  $\gamma_{u,w} = |M \cap N_G(u) \cap N_G(w)|$. Note that, for every pair of lieutenants belonging to the same group, the source node belongs to exactly one of the shortest path between those two lieutenants. Based on this, we have:
$$
\delta(v^{\dagger}) = \sum_{u,w \in L:u\neq w} \frac{1}{k + 1 + \gamma_{u,w}} + \sum_{u,w \in L':u\neq w	} \frac{1}{k + 1 + \gamma_{u,w}}
$$
By observing that for any $a,b > 0$ we have $\frac{1}{a+b} < \frac{1}{a}$, we conclude that:
$$
\delta(v^{\dagger}) < \sum_{u,w \in L:u\neq w} \frac{1}{k + 1} + \sum_{u,w \in L':u\neq w} \frac{1}{k + 1}
$$
Now since the number of pairs of different lieutenants from each group is $\frac{k (k - 1)}{2} < \frac{k (k + 1)}{2}$, then:
$$
\delta(v^{\dagger}) < k
$$
Having analyzed $\delta(v^{\dagger})$, let us now analyze $\delta(l_i)$ for some lieutenant $l_i \in L$ (the same analysis can be done for a lieutenant $l_j \in L'$). In particular, since $l_i$ belongs to shortest paths (i) between every pair of lieutenants from the other group, (ii) between the source node and every member connected to $l_i$, and (iii) between every pair of members connected to $l_i$, we have:
$$
\delta(l_i) = \sum_{u,w \in L':u\neq w} \frac{1}{k + 1 + \gamma_{u,w}} + \sum_{v \in M \cap N(l_i)} \frac{1}{2c} + \sum_{u,w \in M \cap N(l_i):u\neq w} \frac{1}{|\mathit{sp}_G(u,w)|}
$$
By omitting the first term of the right-hand side of the equation and observing that $|M \cap N(l_i)| \geq f$ and $|\{\{u,w\} \subseteq M \cap N(l_i)\}| \geq \frac{f(f-1)}{2}$ and $|\mathit{sp}_G(u,w)| \leq 2c$ for every $u,w \in M \cap N(l_i):u\neq w$, we end up with the following:
$$
\delta(l_i) > \frac{f}{2c} +  \frac{f (f - 1)}{4c} > \frac{f ^ 2}{4c}
$$
Finally, by comparing $\delta(v^{\dagger})$ with $\delta(l_i)$, we find that:
$$
\delta(l_i) - \delta(v^{\dagger}) > \frac{f ^ 2}{4c} - k
$$
Therefore, if $f^2 > 4ck$ every lieutenant has higher betweenness centrality than the source node. \hspace*{\fill}$\Box$\\

As stated in the theorem, a Lieutenant network can indeed conceal its source node as far as centrality is concerned. On the other hand, as far as influence is concerned, we evaluate the network empirically to see how the different parameters affect the influence of the source node. To this end, given a Lieutenant network of 400 nodes, we varied the parameters of the network, namely $k$ (the size of each lieutenant group) and $c$ (the number of lieutenants from each group, connected to any given member). For every pair, $(k,c)$, we measured the difference in centrality between the source node, $v^{\dagger}$, and any given lieutenant (the greater the difference, the more $v^{\dagger}$ is disguised), and measured the influence of $v^{\dagger}$ to see how this influence is affected by the disguising process. 

\begin{figure*}[tbhp]
\centering
\setlength\tabcolsep{1pt}
\begin{tabular}{m{.195\textwidth}m{.195\textwidth}m{.195\textwidth}m{.195\textwidth}m{.195\textwidth}}
  \multicolumn{1}{c}{\footnotesize{Degree}}
& \multicolumn{1}{c}{\footnotesize{Closeness}}
& \multicolumn{1}{c}{\footnotesize{Betweenness}}
& \multicolumn{1}{c}{\footnotesize{IC Influence}}
& \multicolumn{1}{c}{\footnotesize{LT Influence}} \\
\includegraphics[width=\linewidth]{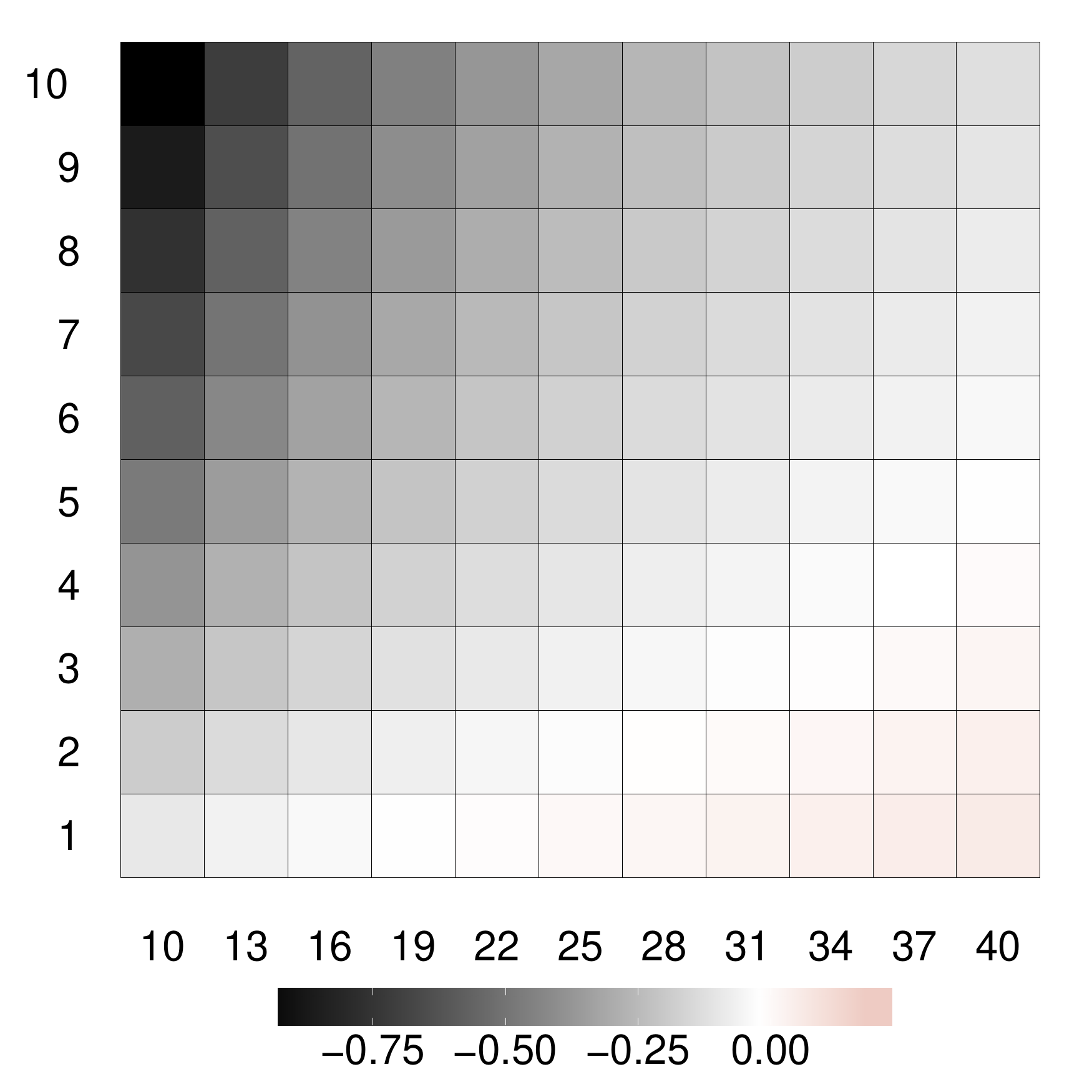} &
\includegraphics[width=\linewidth]{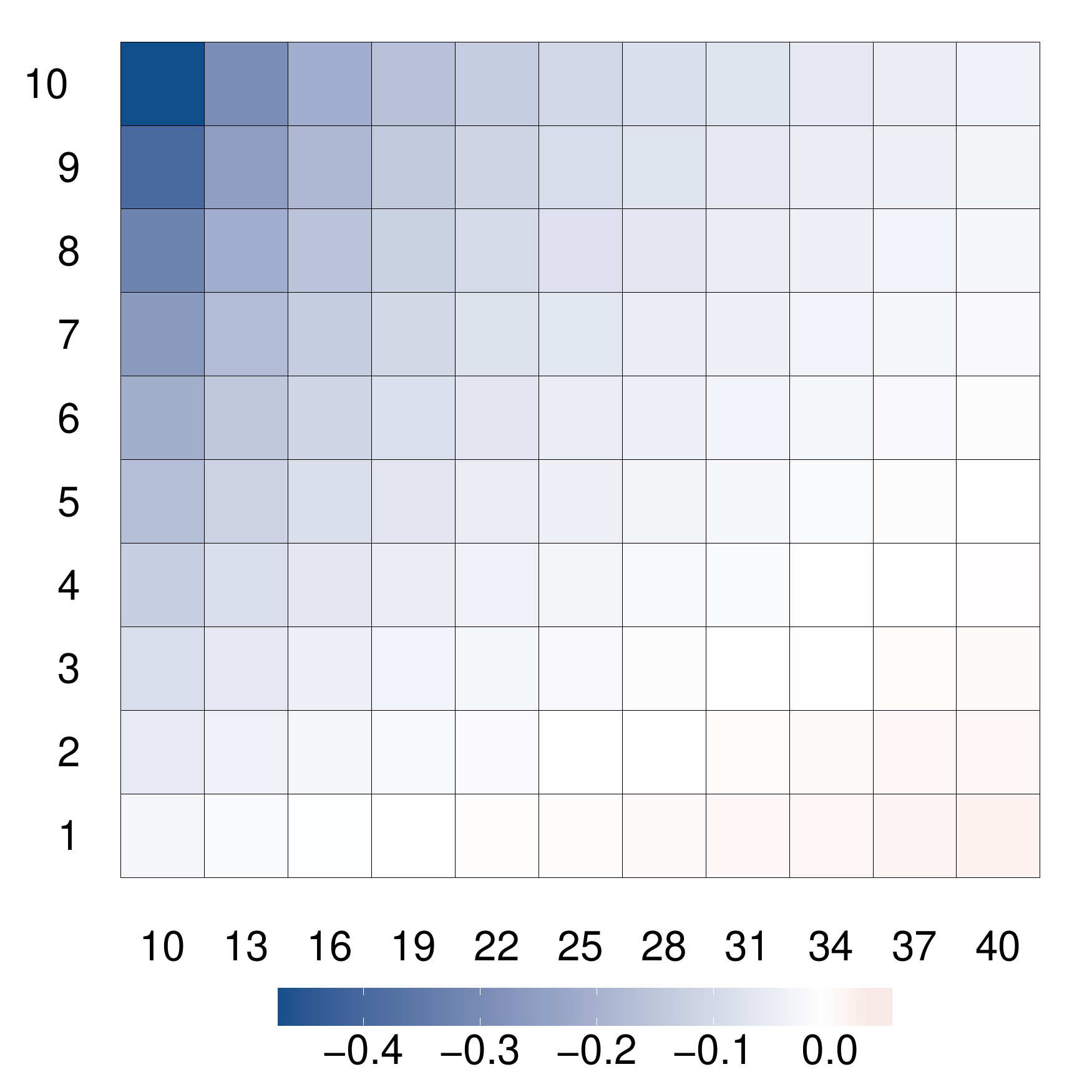} &
\includegraphics[width=\linewidth]{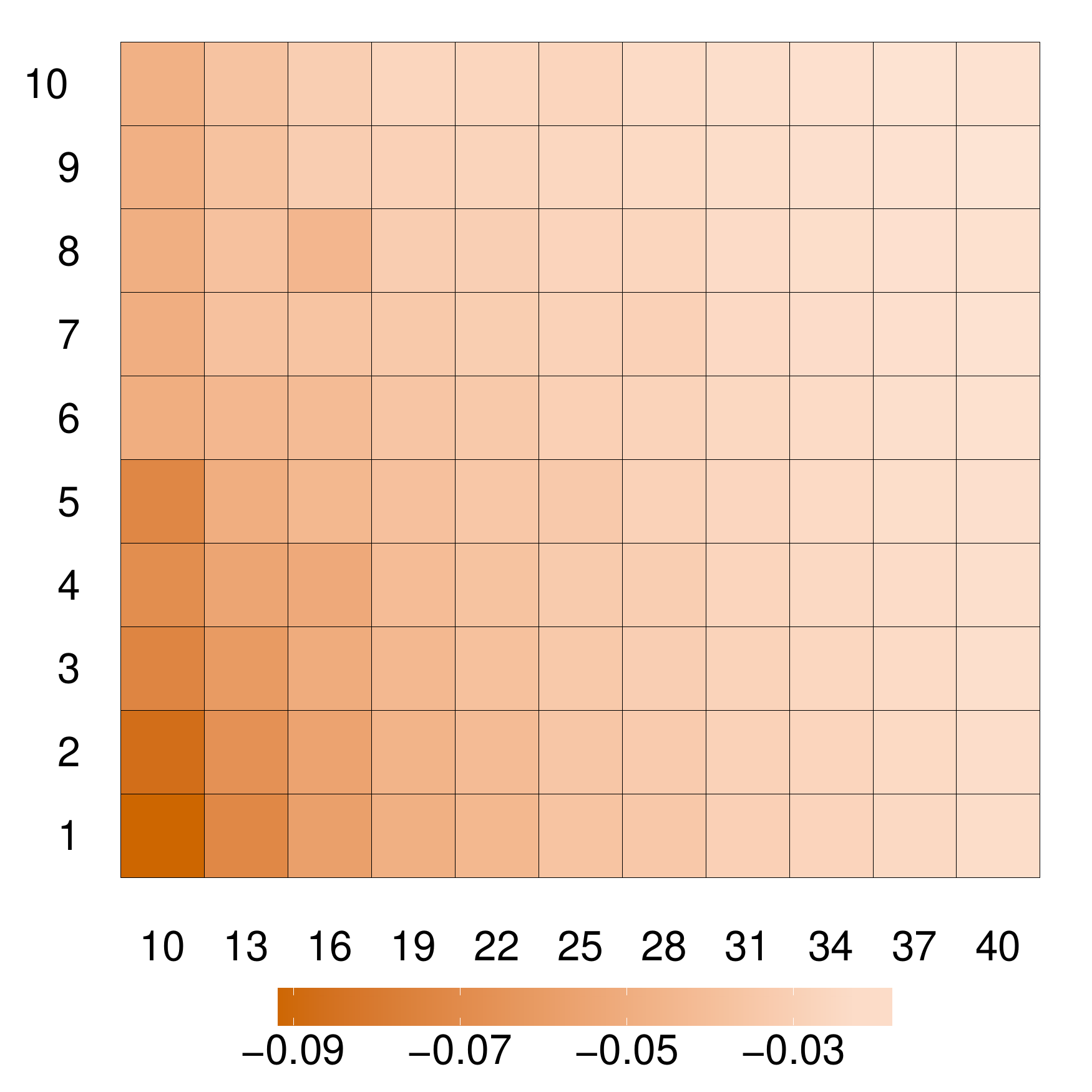} &
\includegraphics[width=\linewidth]{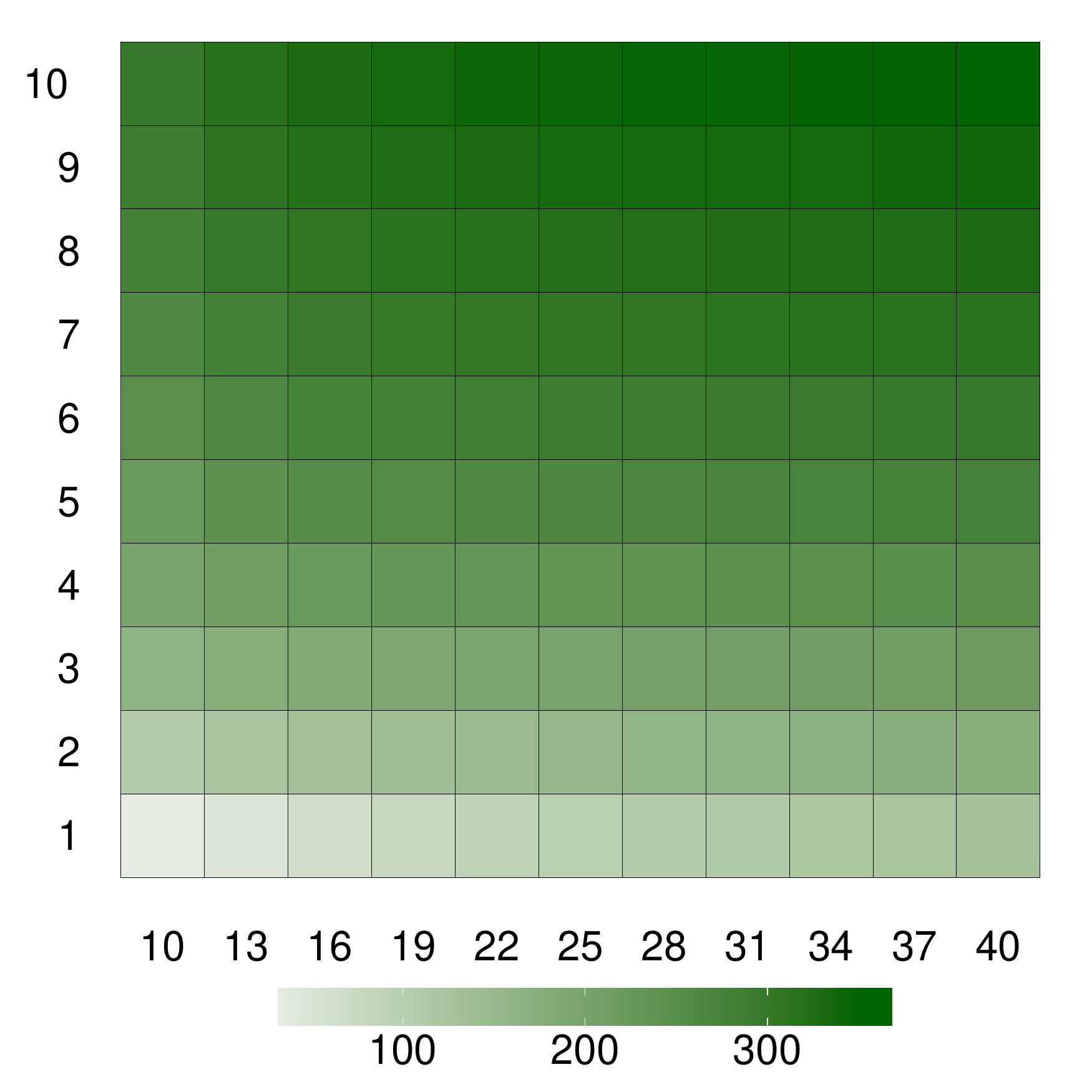} &
\includegraphics[width=\linewidth]{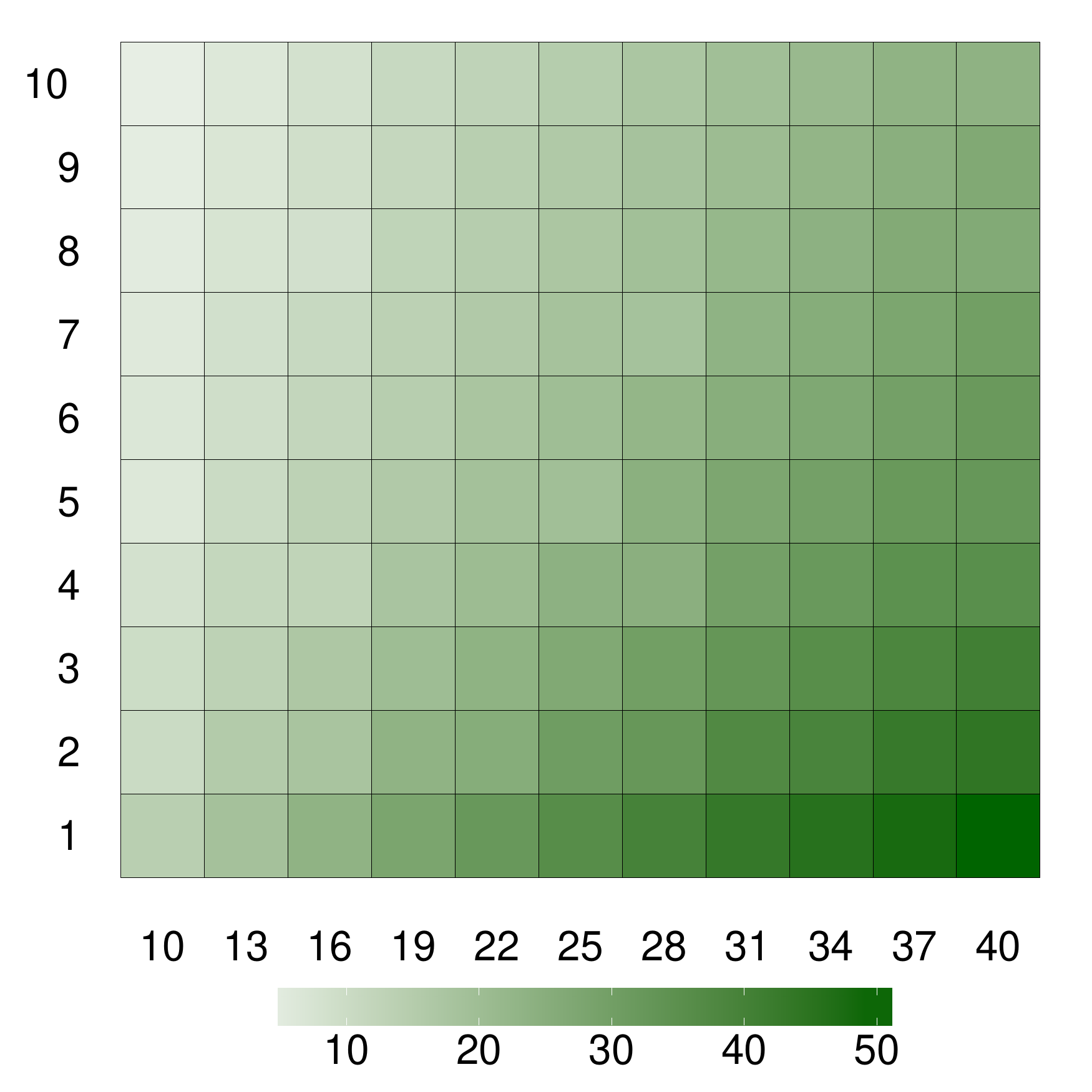} \\
\end{tabular}
\caption{\fontsize{10}{10}\selectfont{Given a Lieutenant netowrk of 400 nodes, with different values of parameter $k$ (the $x$-axis) and parameter $c$ (the $y$-axis), the figure depicts the difference in centrality between the $v^{\dagger}$ and a lieutenant, as well as the influence value of $v^{\dagger}$.}}
\label{fig:lieutenant-heatmaps}
\end{figure*}

The results are depicted in Figure~\ref{fig:lieutenant-heatmaps}, where the $x$-axis represents $k$ and the $y$-axis represents $c$.\footnote{\footnotesize we set $max(c) \leq min(k)$ to ensure that we have a value in every cell of the grid; otherwise some cells would correspond to networks in which the are no sufficient lieutenants to connect to.} Roughly speaking, the results can be categorized into four categories:
\begin{itemize} 
\item \emph{small $k$ and small $c$}: This yields relatively high levels of disguise in terms of betweenness, but not in terms of degree and closeness. On the other hand, it yields rather low levels of Independent-Cascade influence and Linear-Threshold influence;
\item \emph{small $k$ and large $c$}: This yields relatively high levels of disguise in terms of degree and closeness, but not in terms of betweenness. On the other hand, it yields relatively high levels of Independent-Cascade influence, but not Linear-Threshold influence;
\item \emph{large $k$ and small $c$}: This yields relatively low levels of disguise in terms of degree, closeness and betweenness. On the other hand, it yields relatively high levels of Linear-Threshold influence, but not Independent-Cascade influence;
\item \emph{large $k$ and large $c$}: This yields relatively low levels of disguise in terms of degree, closeness and betweenness. On the other hand, it yields relatively high levels of Independent-Cascade influence, but not Linear-Threshold influence.
\end{itemize}
For future work, it would be interesting to identify other network structures that manage to disguise the source node according to all three centrality measures, while at the same time maintaining high levels of influence according to both models of influence.
%
%%%%%%%%%%%%%%%%%%%%%%%%%%%%%%%%%%%%%%%%%%%%%%%%%%%%%%%%%%%%%%%%%%%%%%%%%%%%%%%
%%%%%%%%%%%%%%%%%%%%%%%%%%%%%%%%%%%%%%%%%%%%%%%%%%%%%%%%%%%%%%%%%%%%%%%%%%%%%%%

\newpage
\pagenumbering{gobble}
%%%%%%%%%%%%%%%%%%%%%%%%%%%%%%%%%%%%%%%%%%%%%%%%%%%%%%%%%%%%%%%%%%%%%%%%%%%%%%%
%%%%%%%%%%%%%%%%%%%%%%%%%%%%%%%%%%%%%%%%%%%%%%%%%%%%%%%%%%%%%%%%%%%%%%%%%%%%%%%

%%%%%%%%%%%%%%%%%%%%%%%%%%%%%%%%%%%%%%%%%%%%%%%%%%%%%%%%%%%%%%%%%%%%%%%%%%%%%%%
%%%%%%%%%%%%%%%%%%%%%%%%%%%%%%%%%%%%%%%%%%%%%%%%%%%%%%%%%%%%%%%%%%%%%%%%%%%%%%%

%%%%%%%%%%%%%%%%%%%%%%%%%%%%%%%%%%%%%%%%%%%%%%%%%%%%%%%%%%%%%%%%%%%%%%%%%%%%%%%
%%%%%%%%%%%%%%%%%%%%%%%%%%%%%%%%%%%%%%%%%%%%%%%%%%%%%%%%%%%%%%%%%%%%%%%%%%%%%%%

%%%%%%%%%%%%%%%%%%%%%%%%%%%%%%%%%%%%%%%%%%%%%%%%%%%%%%%%%%%%%%%%%%%%%%%%%%%%%%%
\begin{figure}[p!]
	\centering
	\setlength\tabcolsep{2pt}
	\renewcommand{\arraystretch}{2.0}
	\begin{tabular}{m{.01\textwidth}m{.195\textwidth}m{.195\textwidth}m{.195\textwidth}m{.195\textwidth}m{.195\textwidth}}
	&
	\multicolumn{1}{c}{\small{$b=4$, $d=0$}} &
	\multicolumn{1}{c}{\small{$b=4$, $d=1$}} &
	\multicolumn{1}{c}{\small{$b=4$, $d=2$}} &
	\multicolumn{1}{c}{\small{$b=4$, $d=3$}} &
	\multicolumn{1}{c}{\small{$b=4$, $d=4$}} \\	
	\rotatebox{90}{{\fontsize{8}{8}\selectfont{$\mathit{ScaleFree}(100,3)$}}} &
	\includegraphics[width=0.95\linewidth]{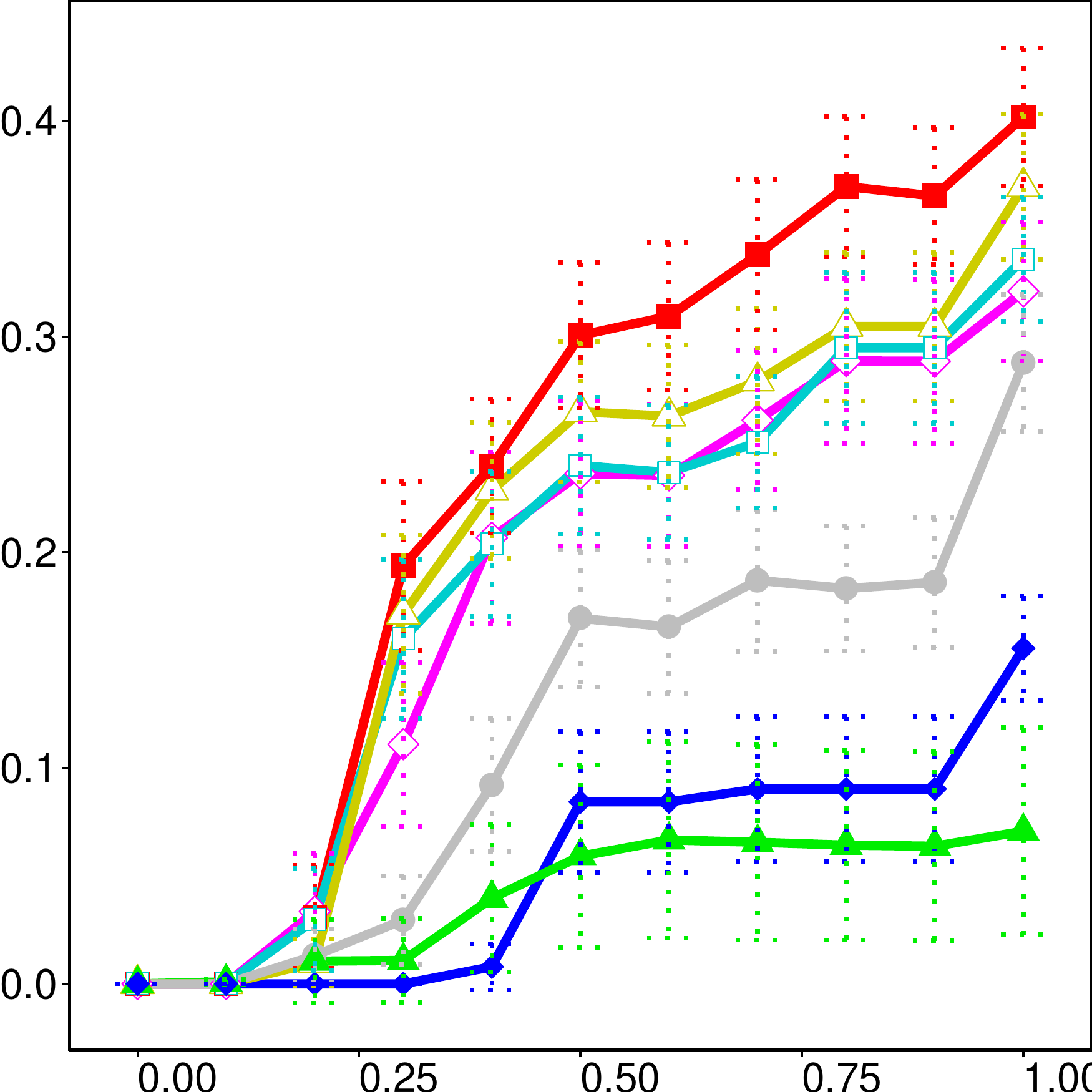} &
	\includegraphics[width=0.95\linewidth]{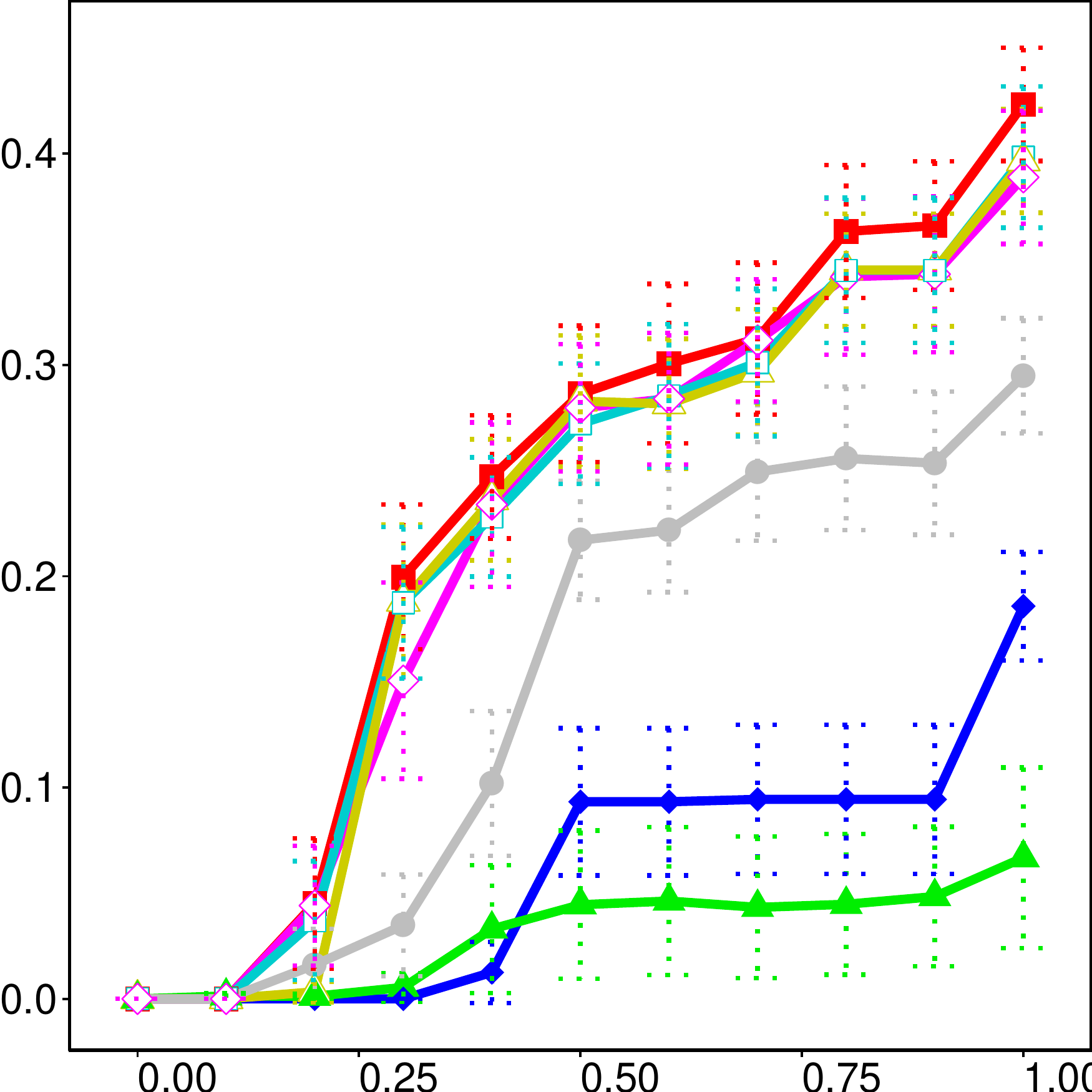} &
	\includegraphics[width=0.95\linewidth]{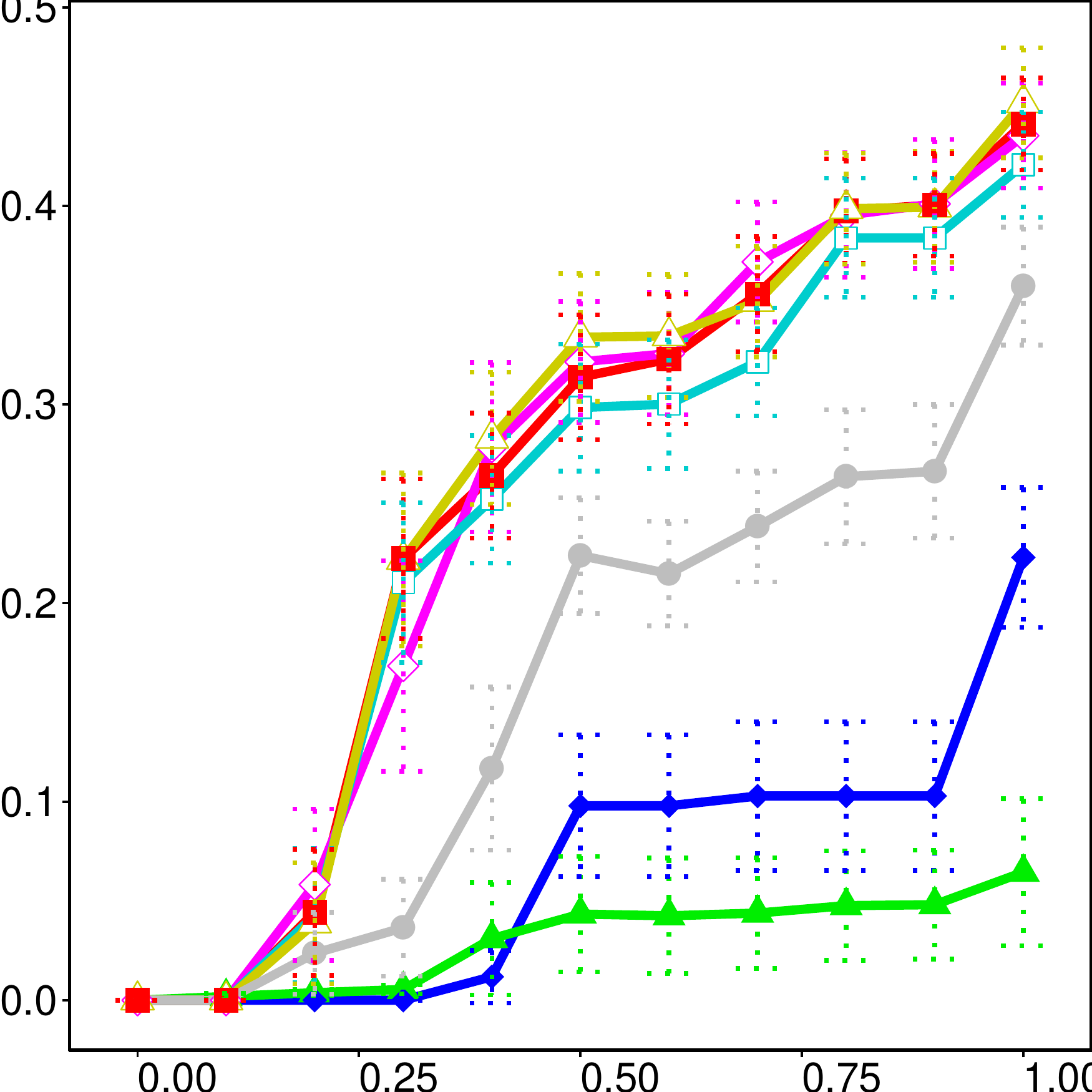} &
	\includegraphics[width=0.95\linewidth]{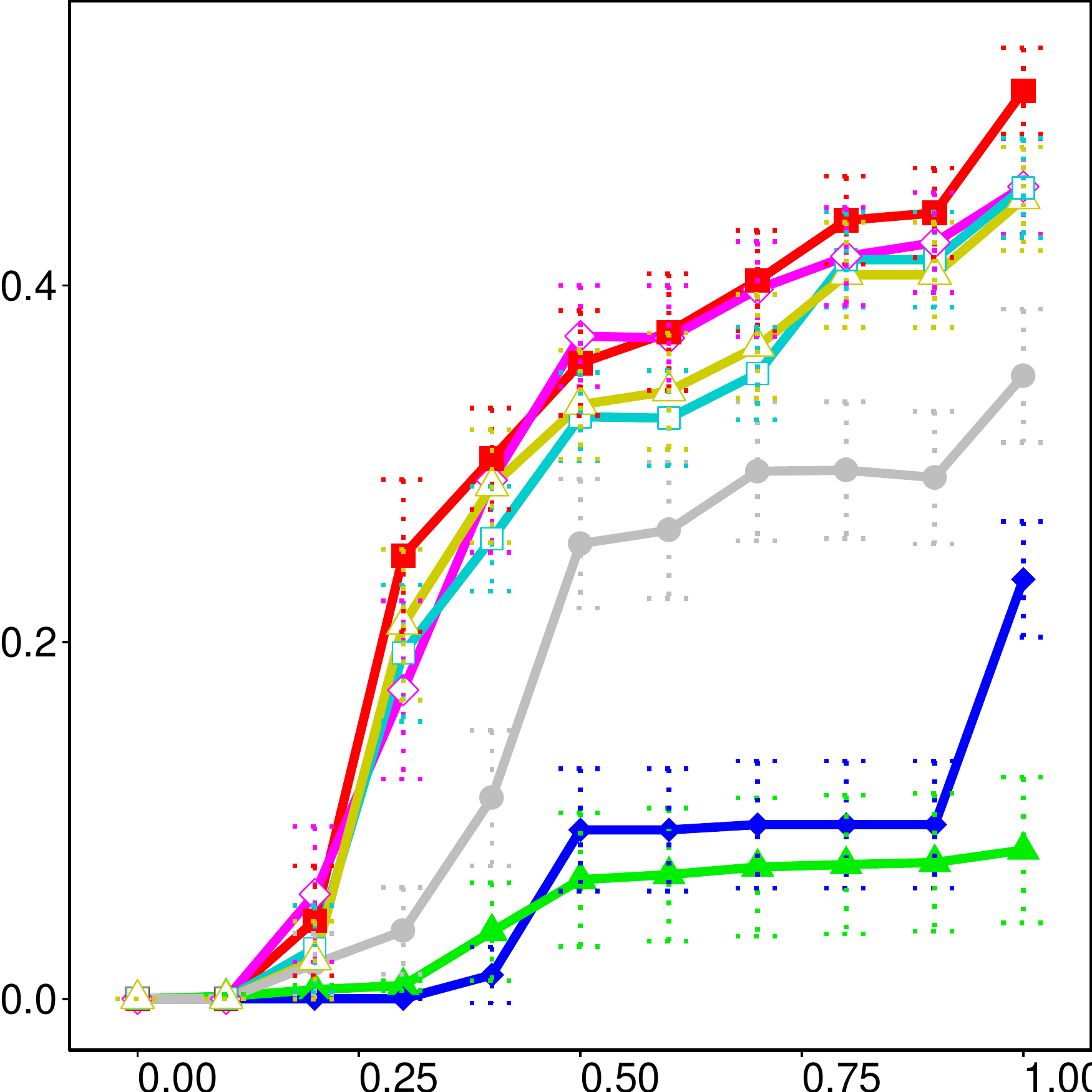} &
	\includegraphics[width=0.95\linewidth]{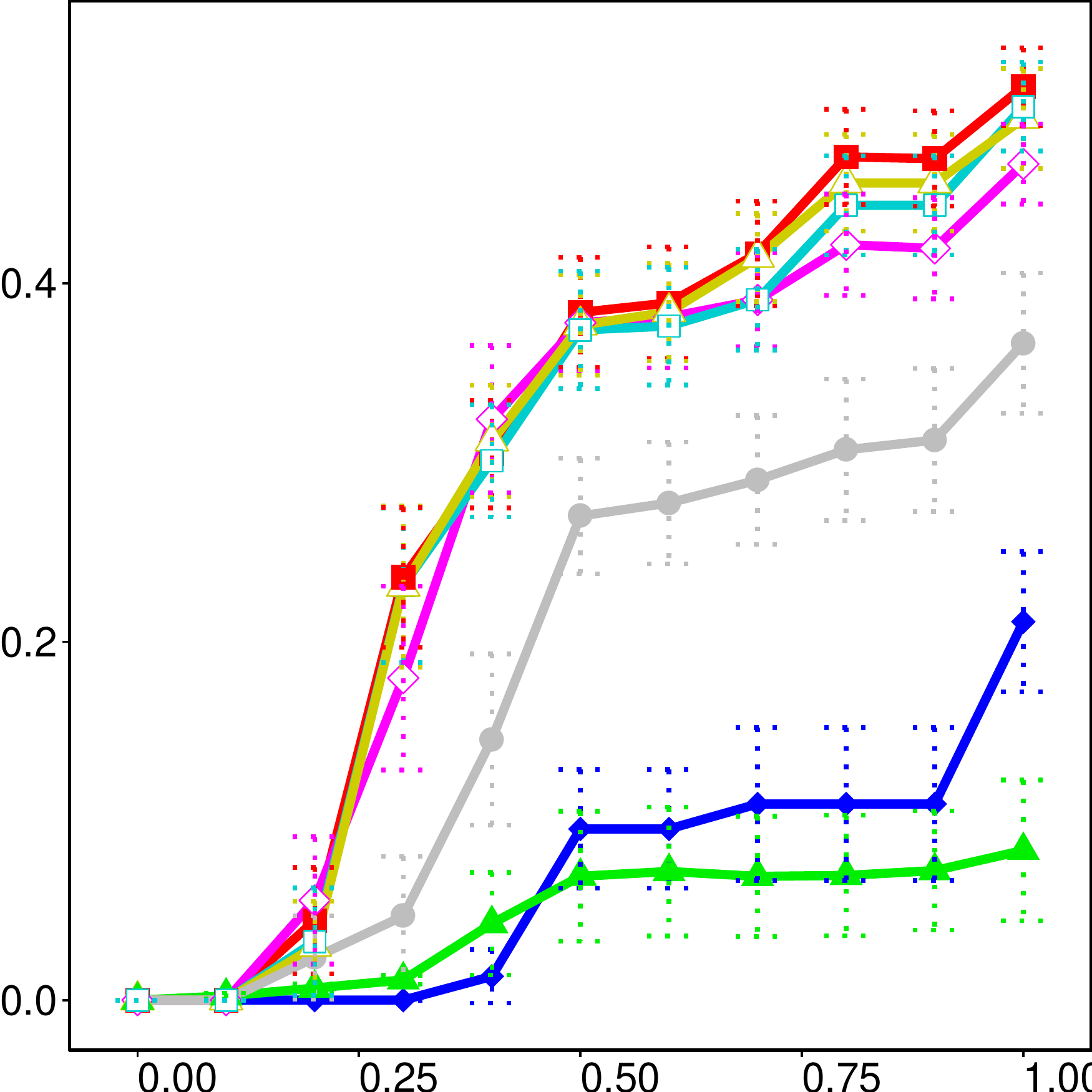} \\
	\rotatebox{90}{{\fontsize{8}{8}\selectfont{$\mathit{ScaleFree}(1000,3)$}}} &
	\includegraphics[width=0.95\linewidth]{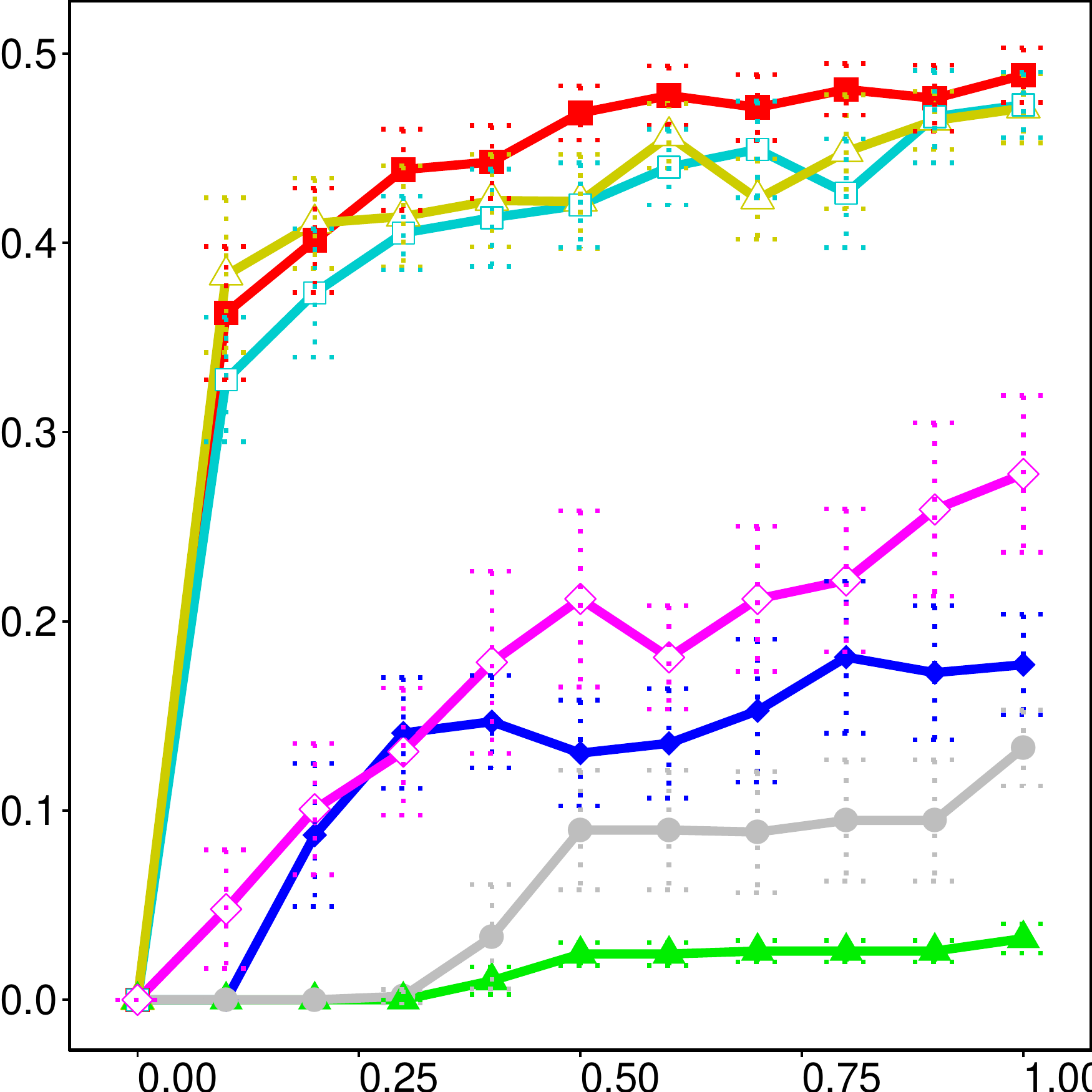} &
	\includegraphics[width=0.95\linewidth]{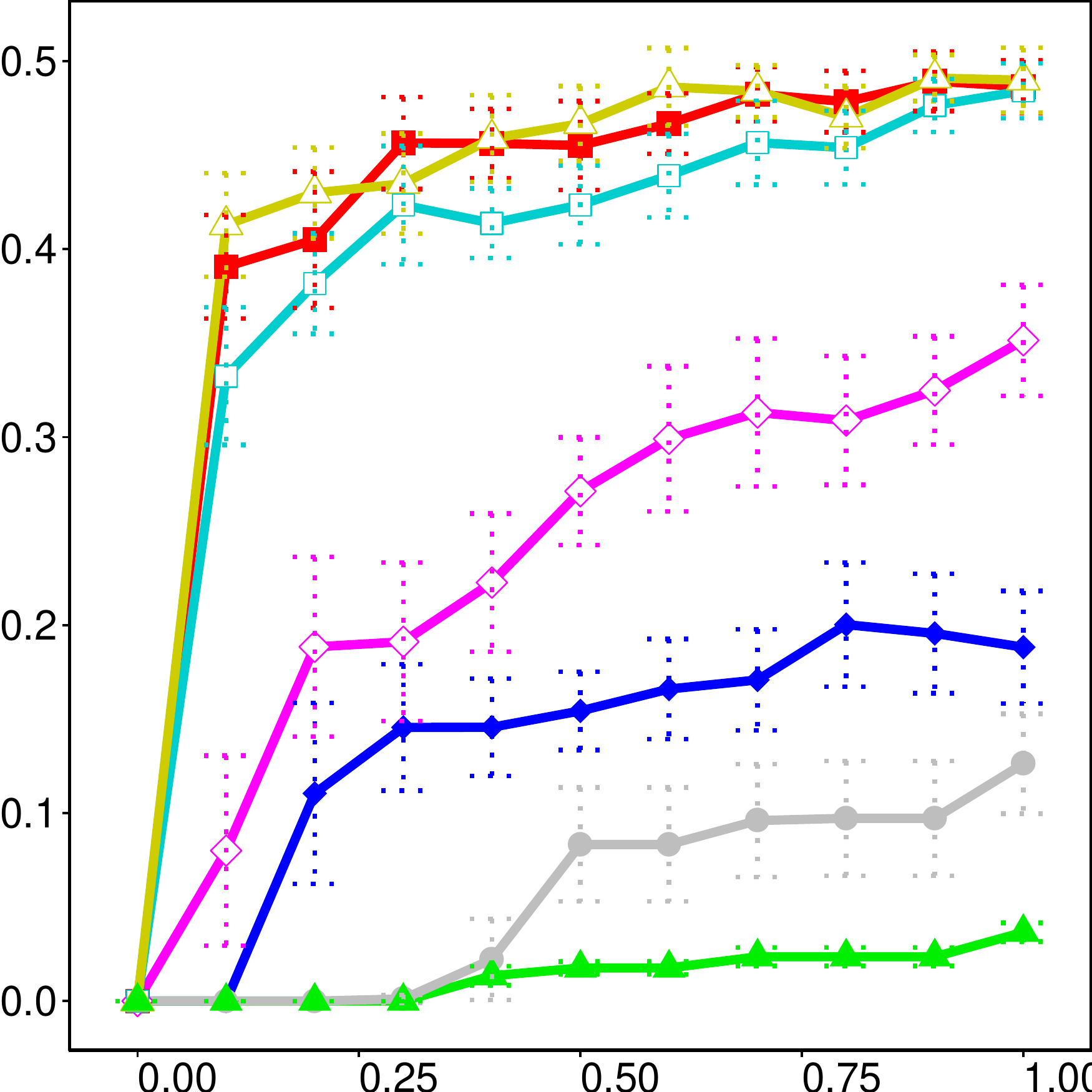} &
	\includegraphics[width=0.95\linewidth]{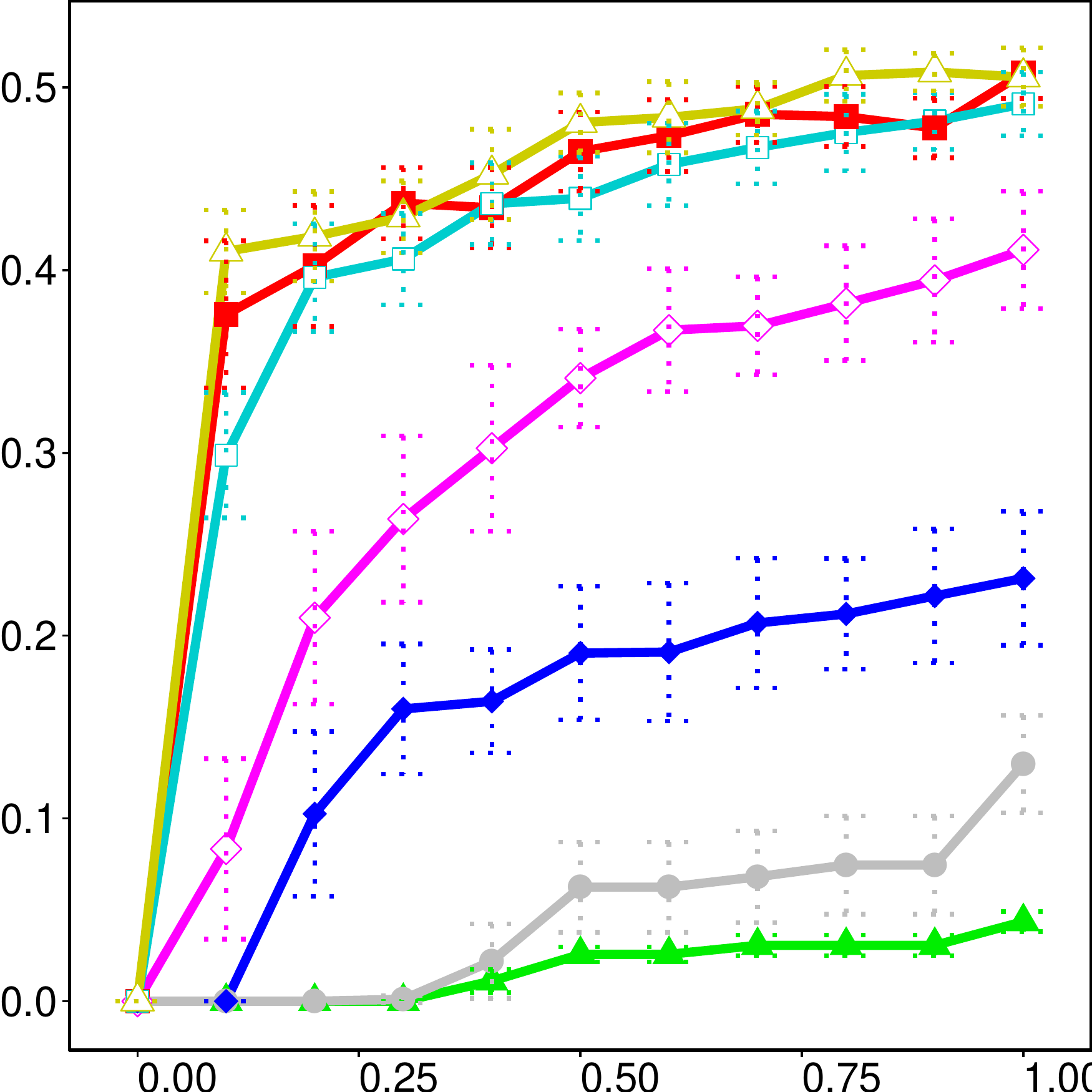} &
	\includegraphics[width=0.95\linewidth]{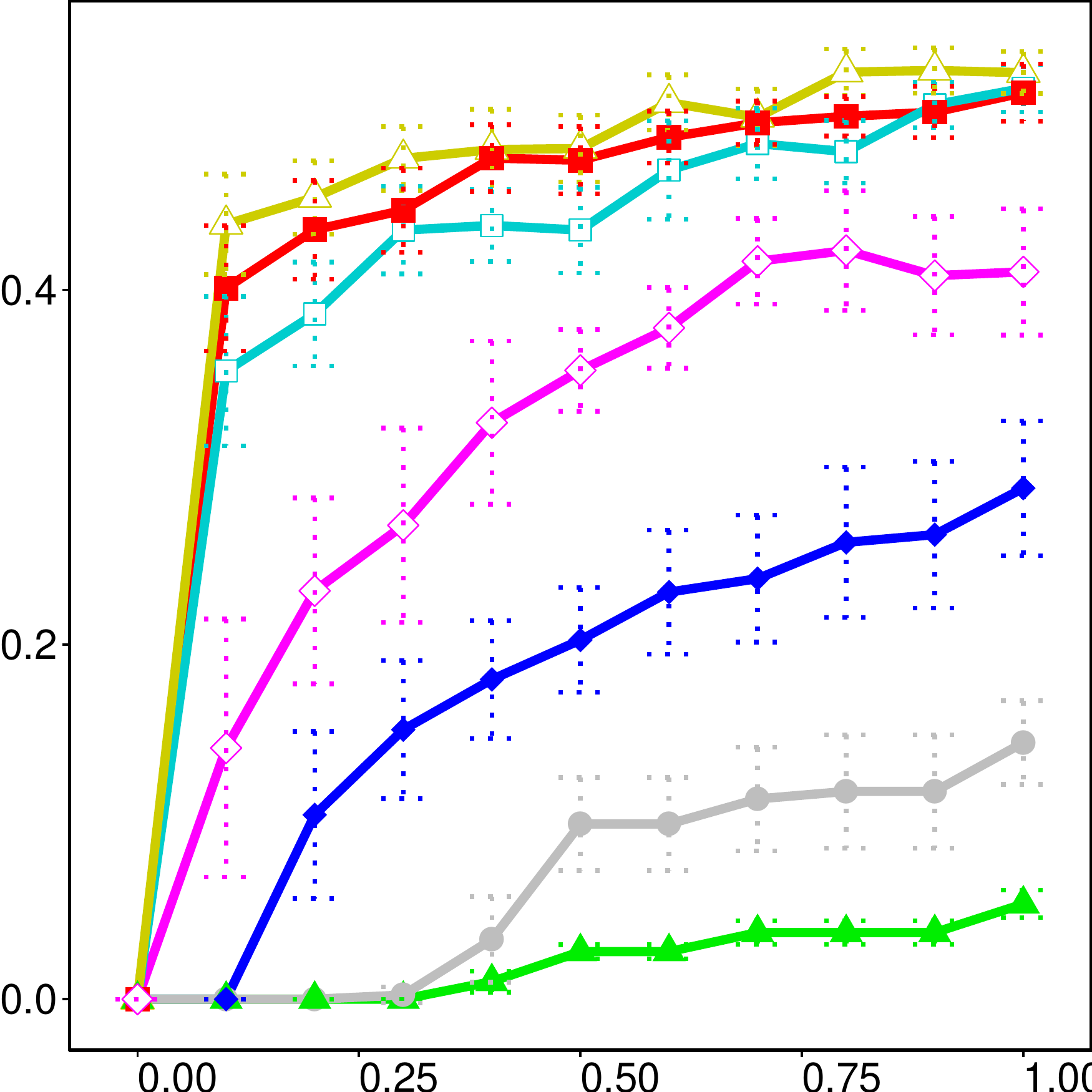} &
	\includegraphics[width=0.95\linewidth]{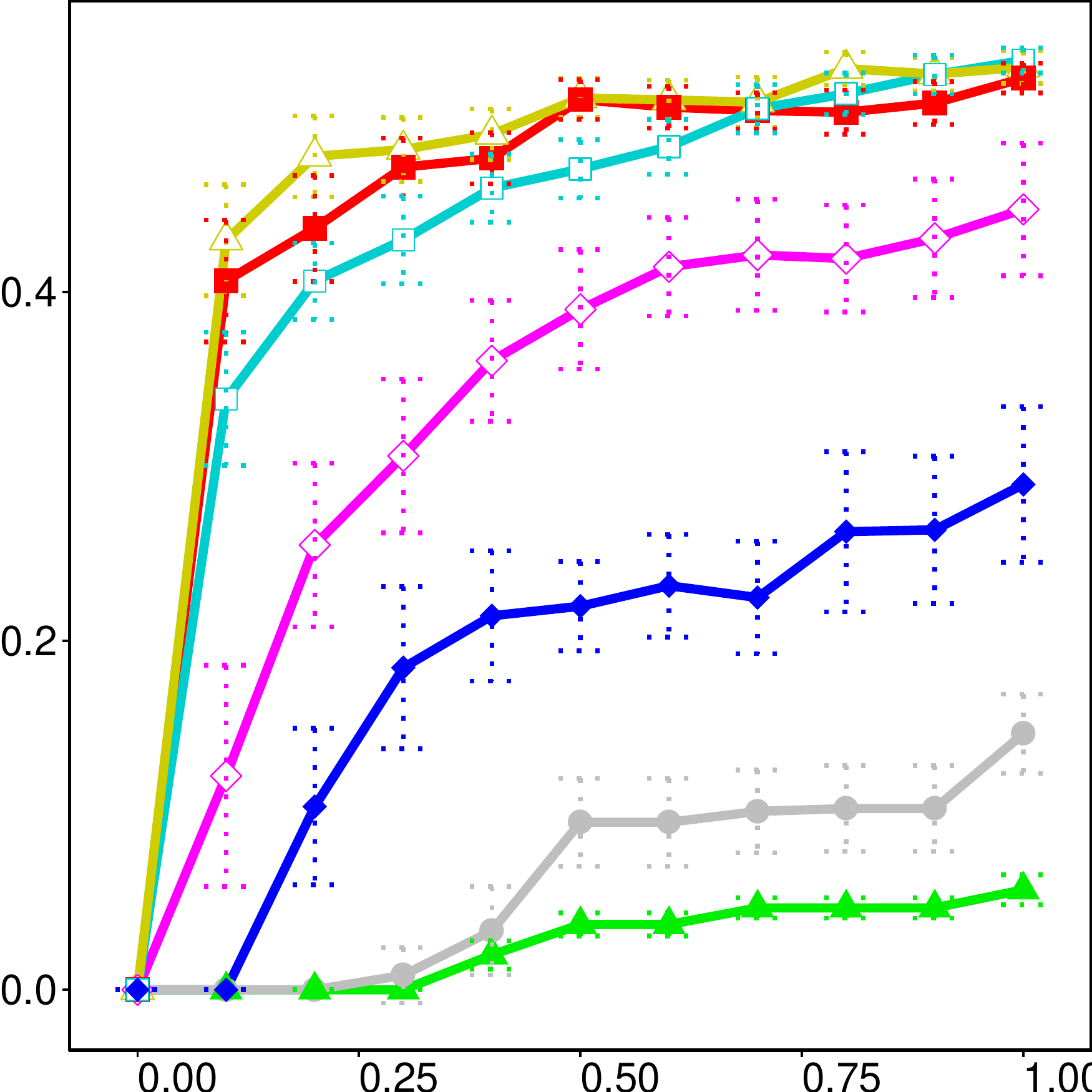} \\
	\rotatebox{90}{{\fontsize{6}{6}\selectfont{$\mathit{RandomGraph}(100,10)$}}} &
	\includegraphics[width=0.95\linewidth]{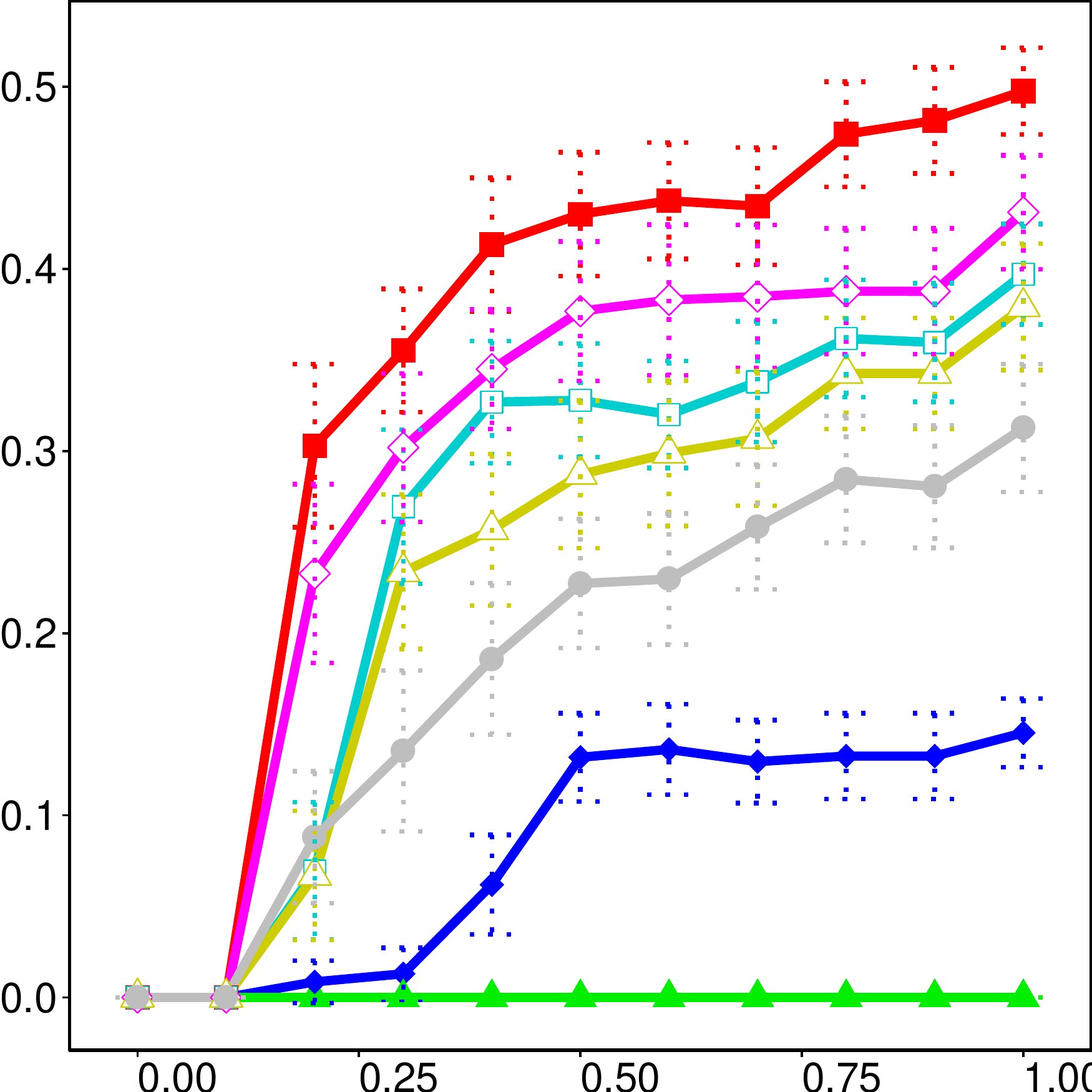} &
	\includegraphics[width=0.95\linewidth]{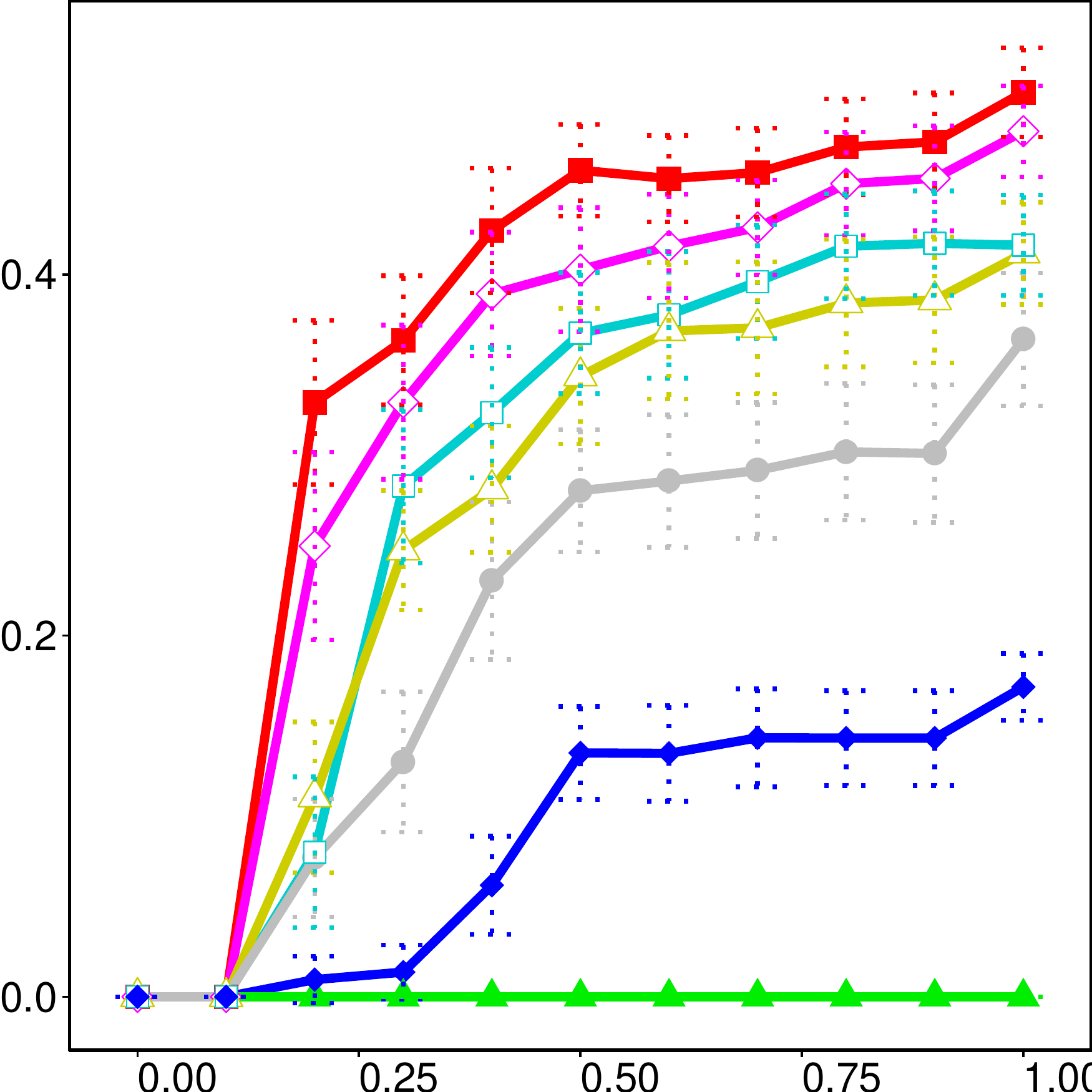} &
	\includegraphics[width=0.95\linewidth]{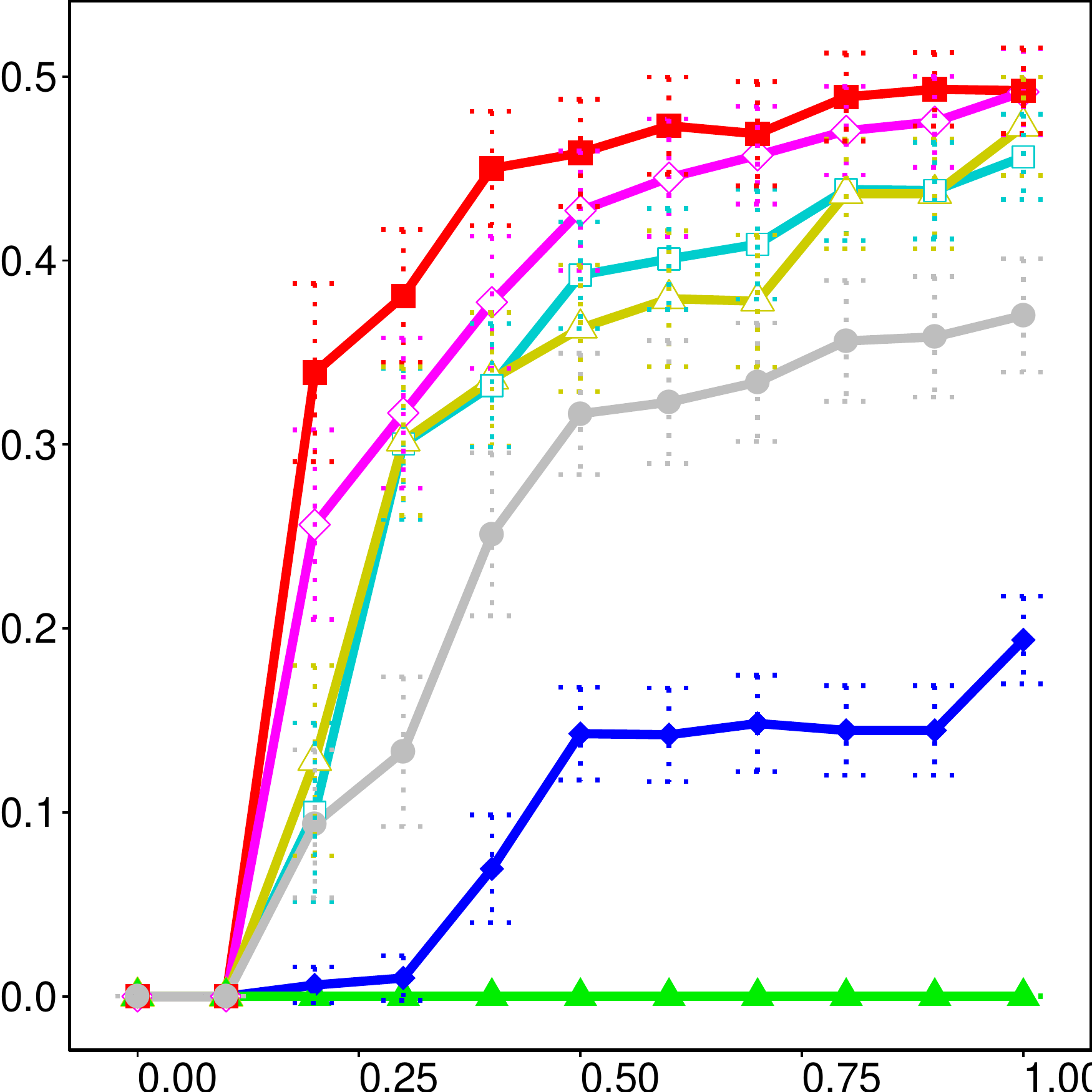} &
	\includegraphics[width=0.95\linewidth]{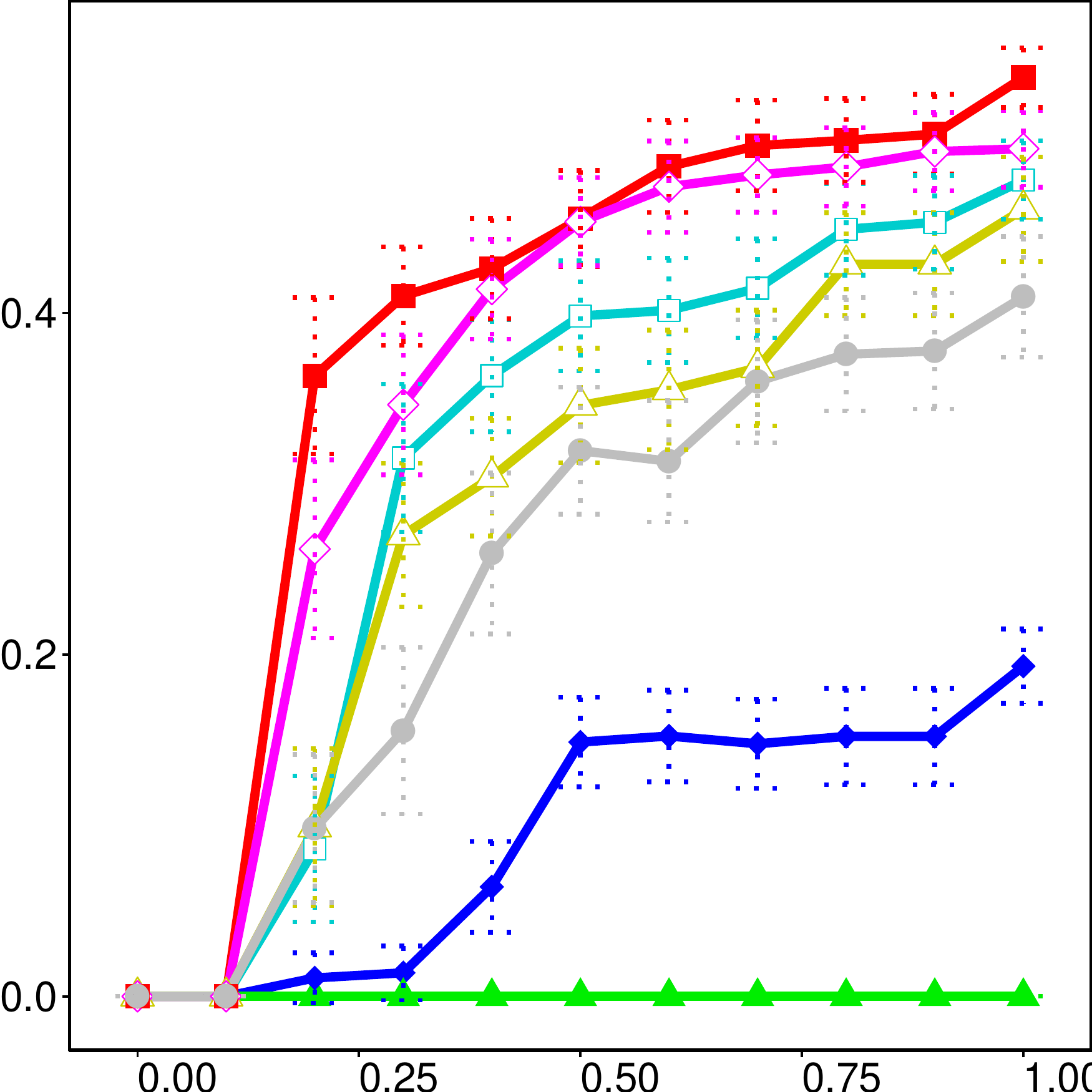} &
	\includegraphics[width=0.95\linewidth]{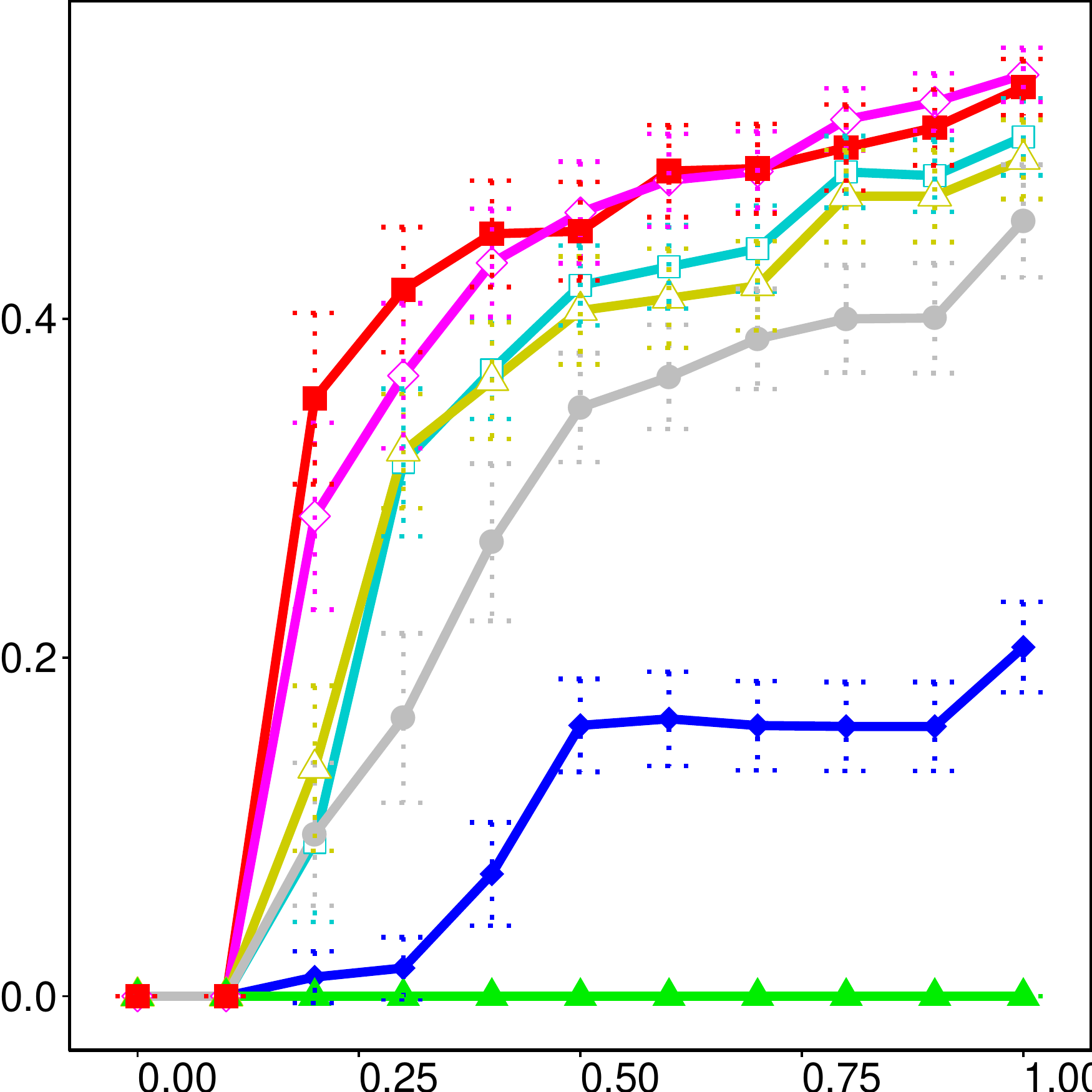} \\
	\rotatebox{90}{{\fontsize{6}{6}\selectfont{$\mathit{RandomGraph}(1000,10)$}}} &
	\includegraphics[width=0.95\linewidth]{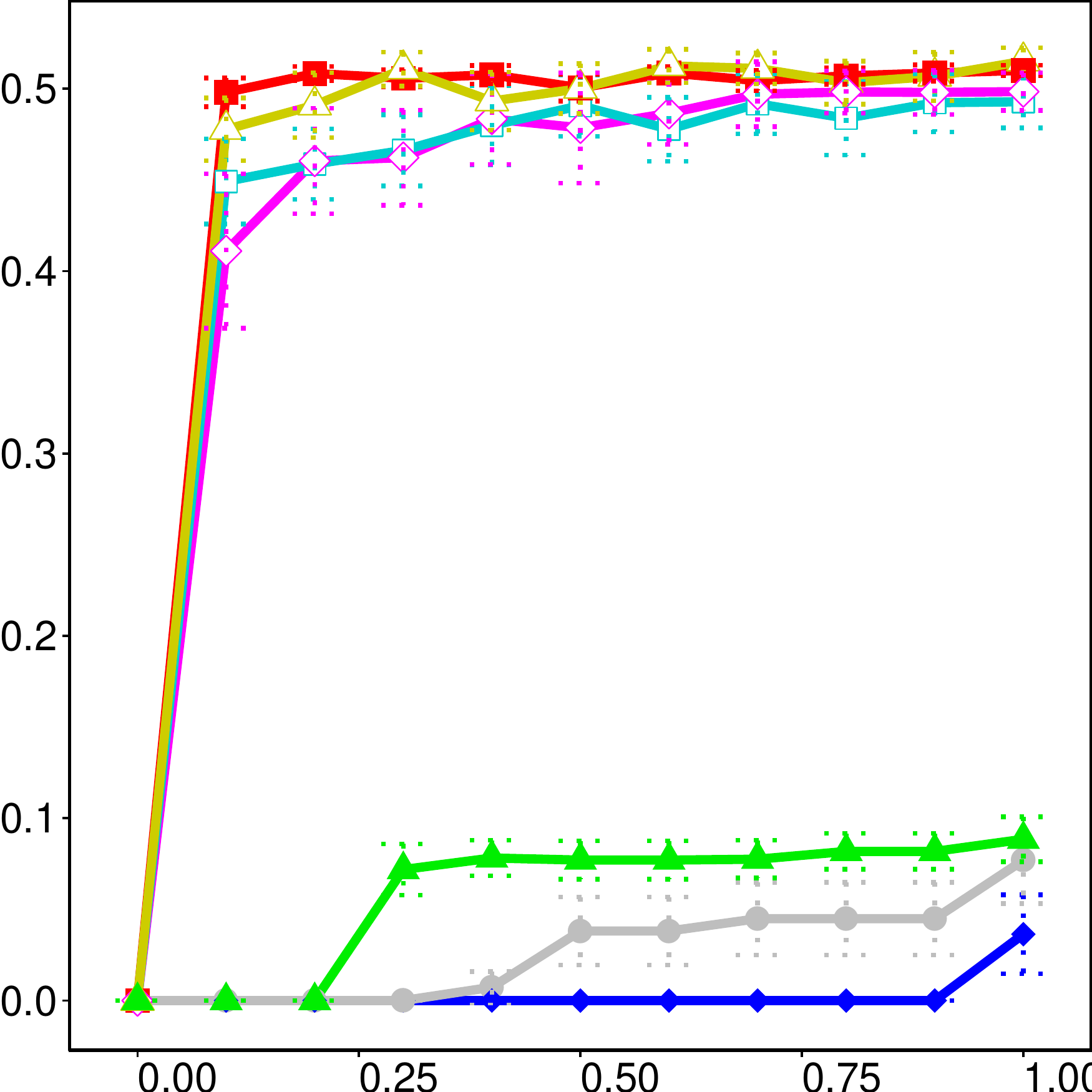} &
	\includegraphics[width=0.95\linewidth]{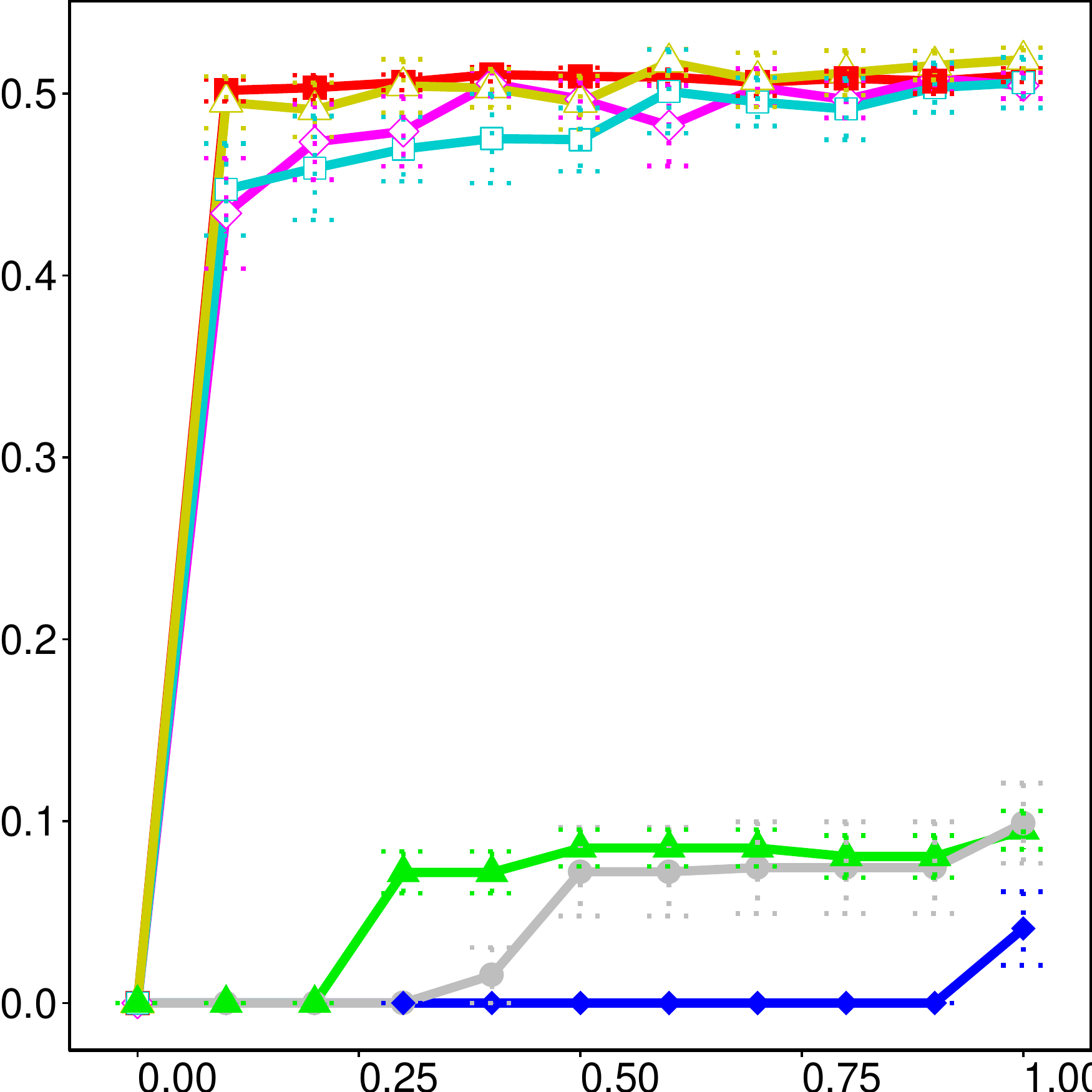} &
	\includegraphics[width=0.95\linewidth]{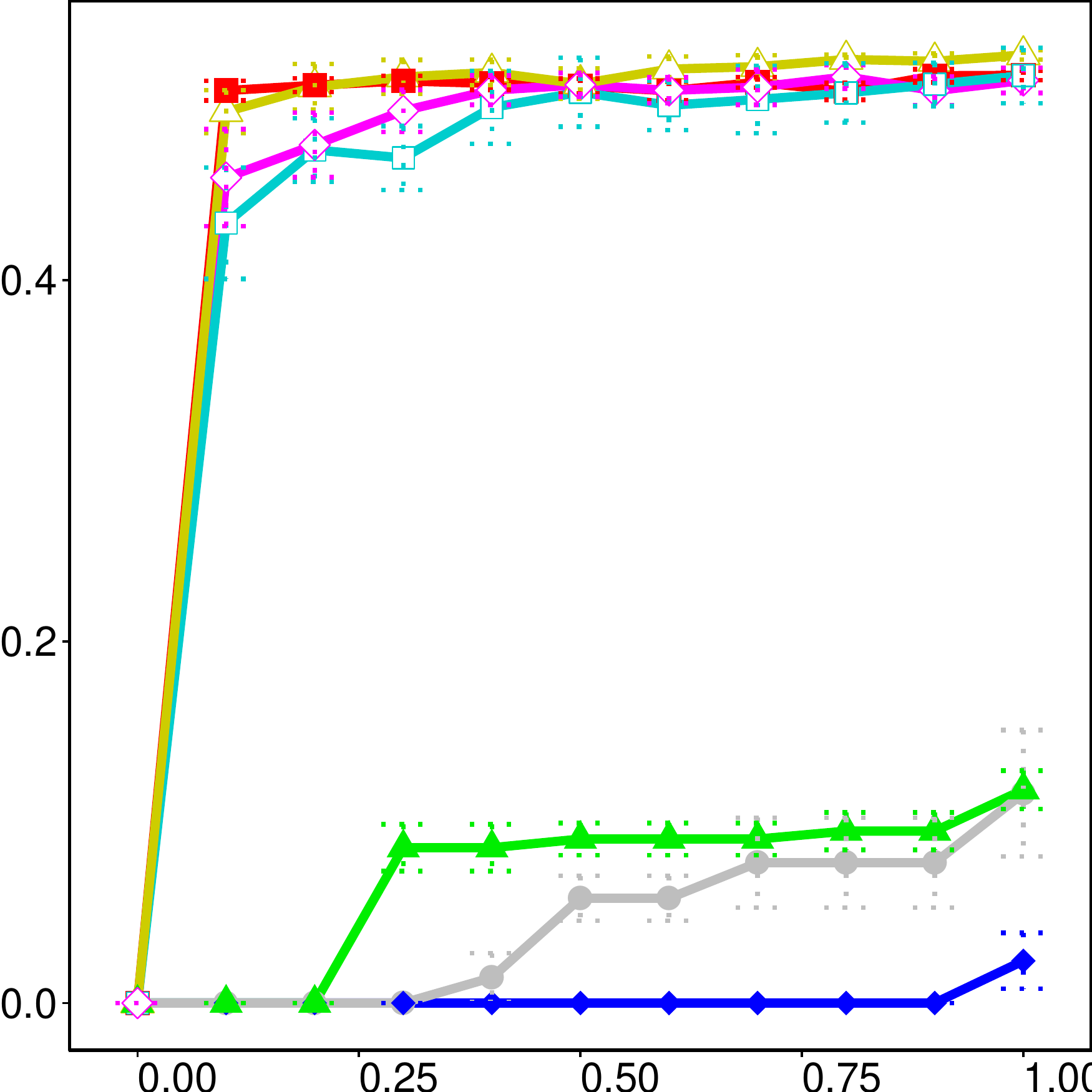} &
	\includegraphics[width=0.95\linewidth]{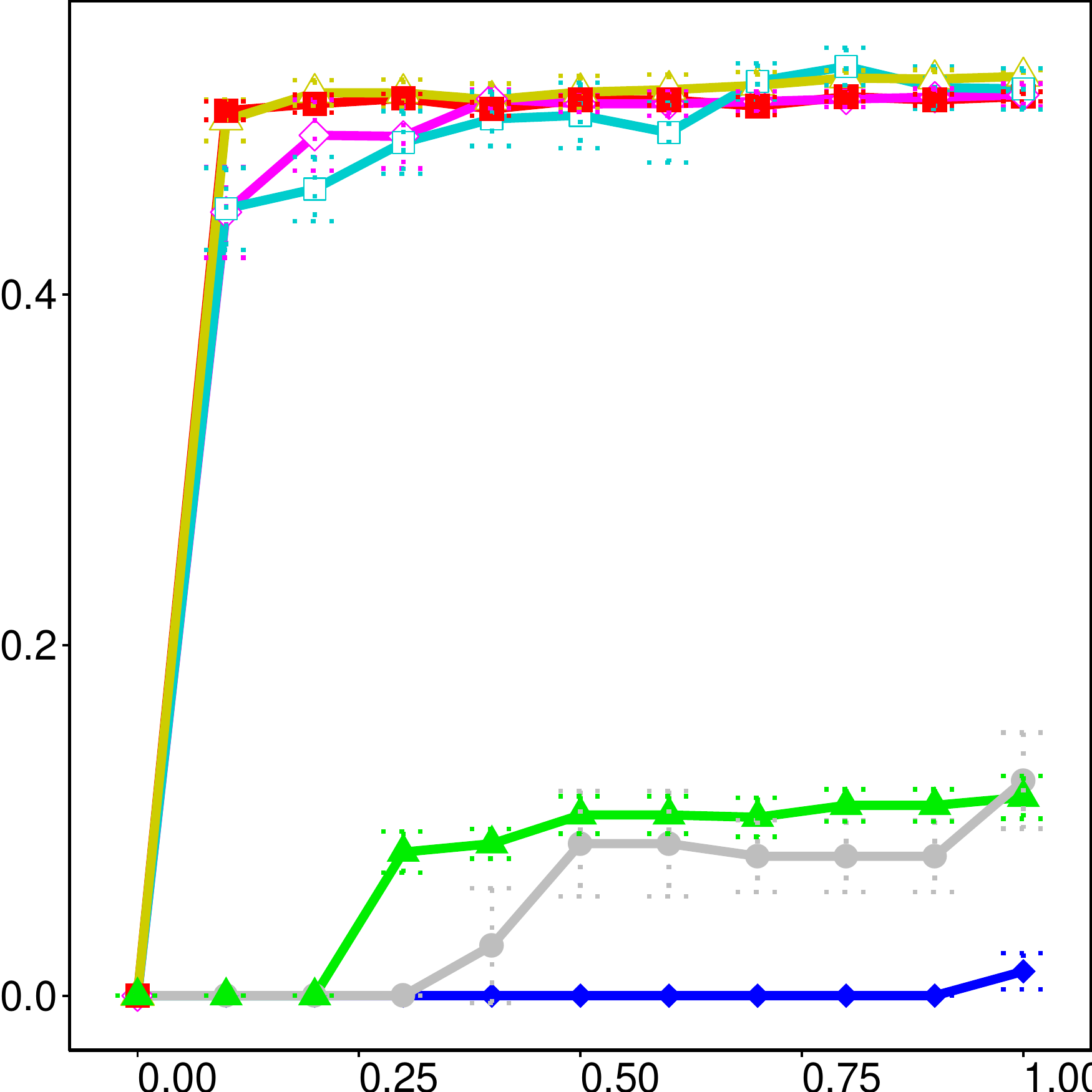} &
	\includegraphics[width=0.95\linewidth]{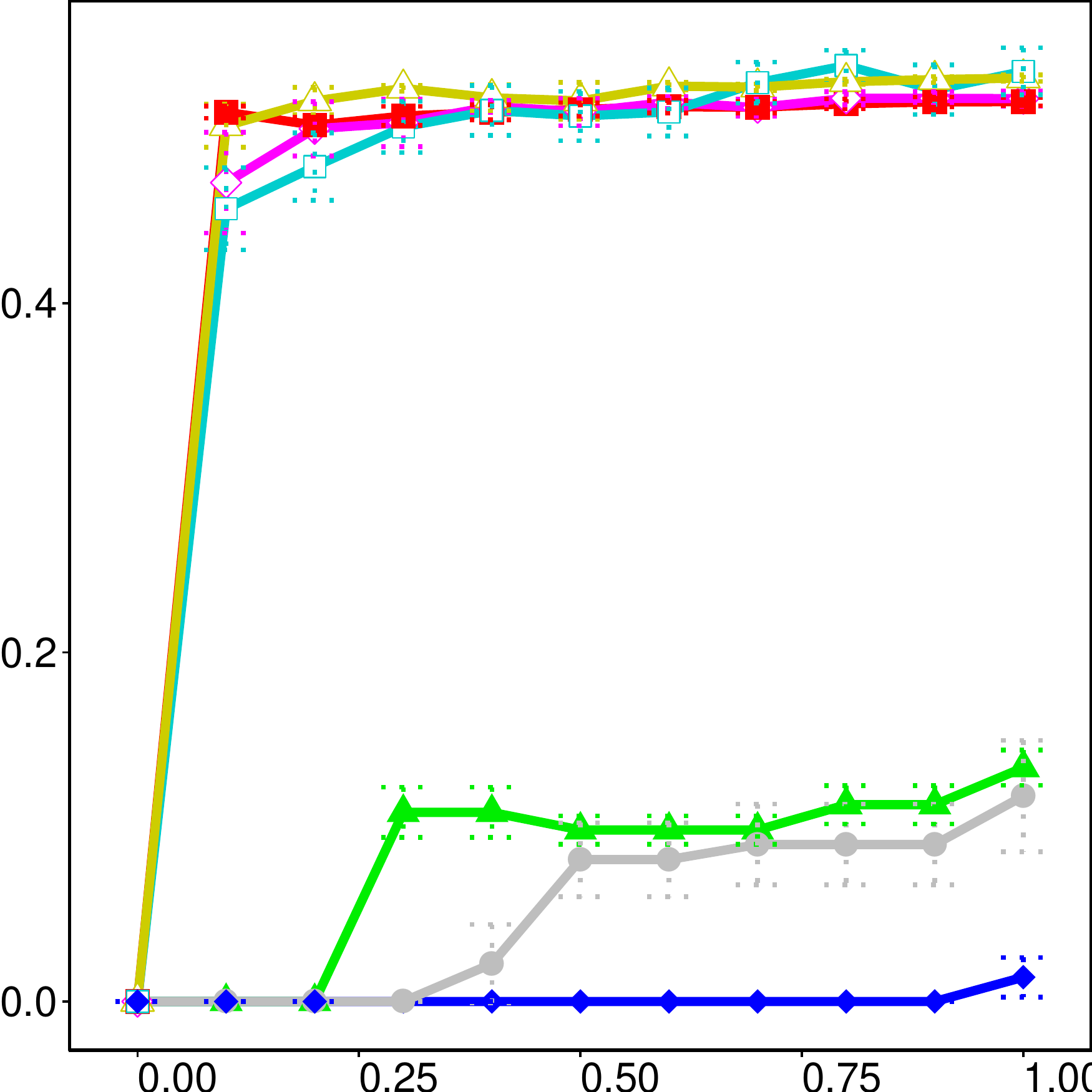} \\
	\rotatebox{90}{{\fontsize{6}{6}\selectfont{$\mathit{SmallWorld}(100,10,.25)$}}} &
	\includegraphics[width=0.95\linewidth]{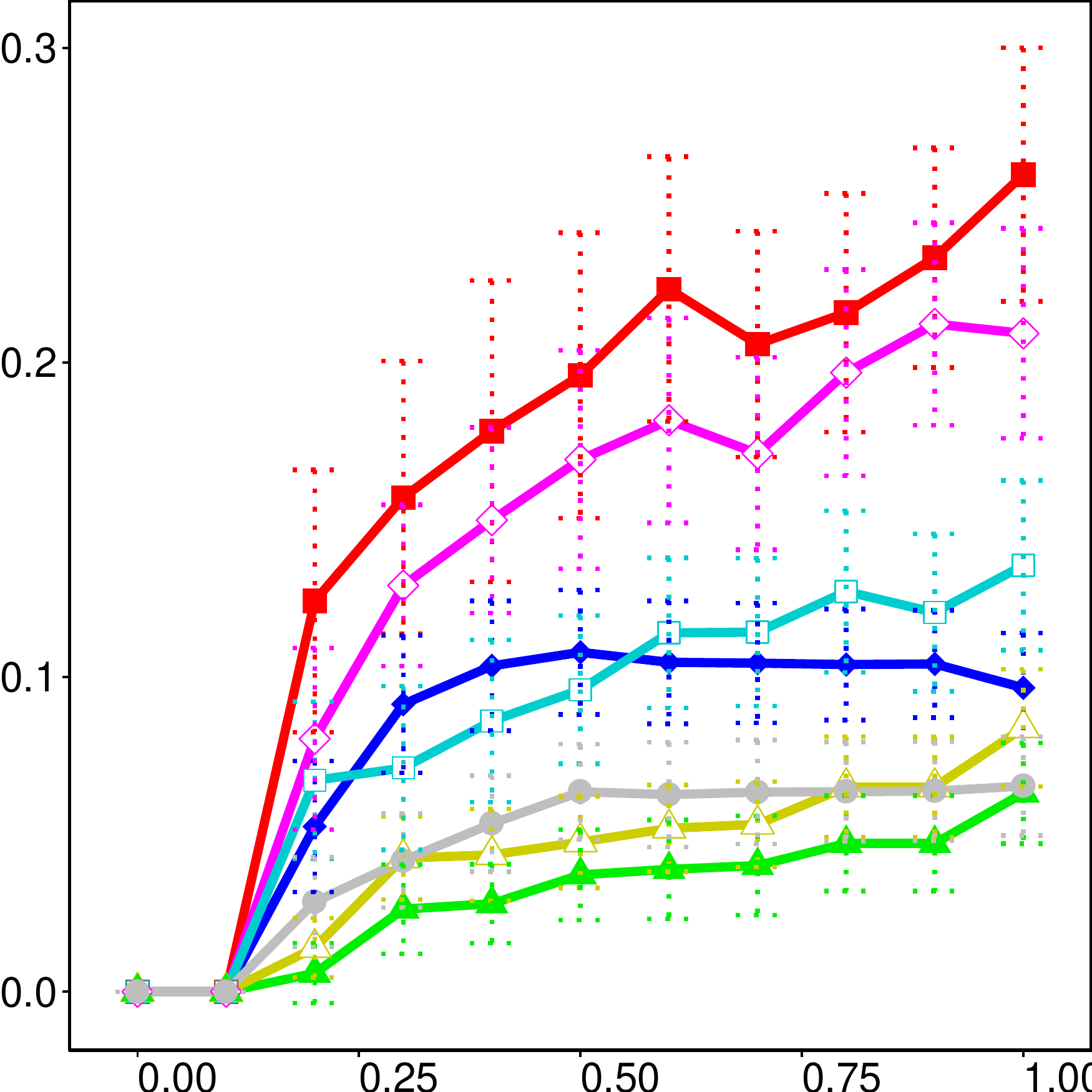} &
	\includegraphics[width=0.95\linewidth]{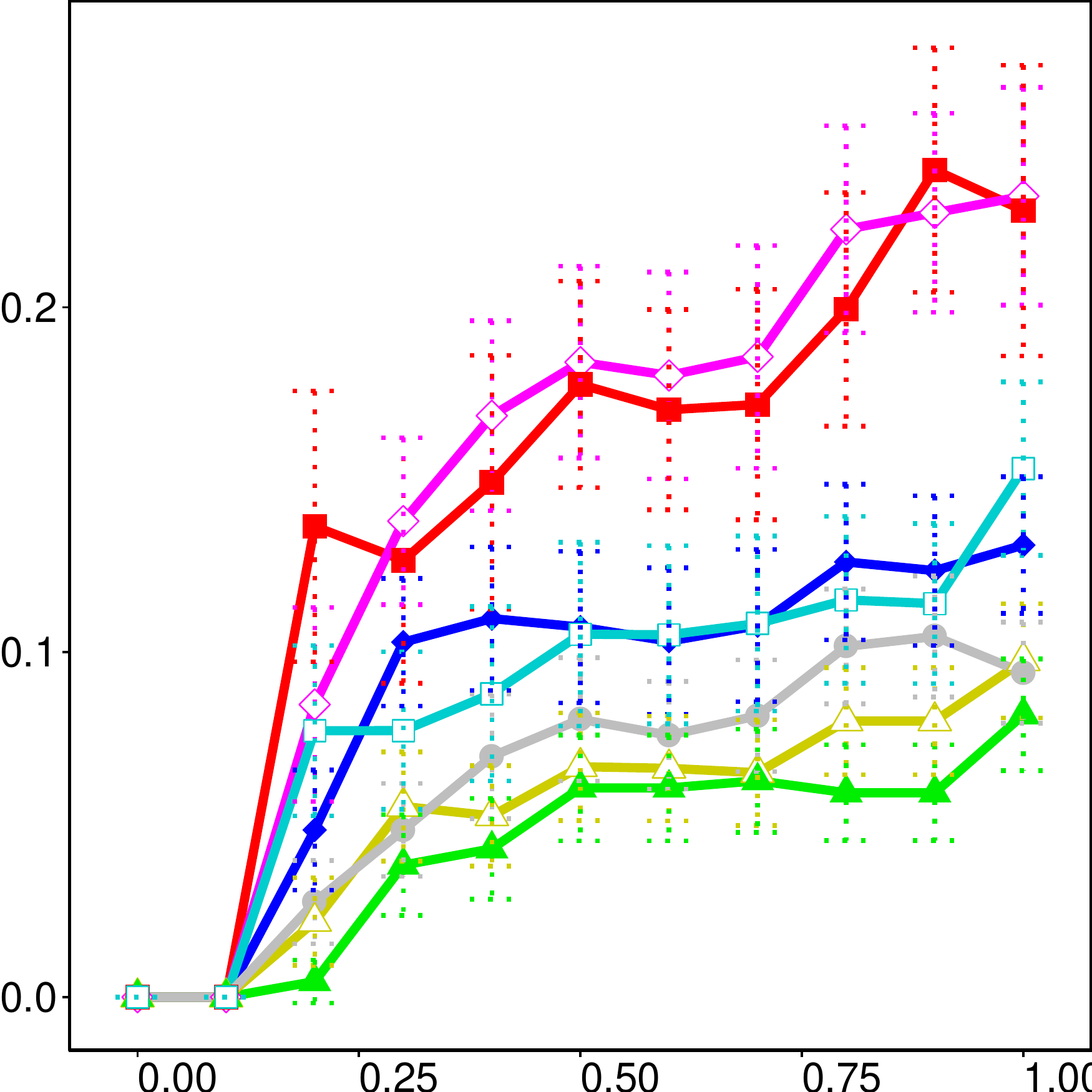} &
	\includegraphics[width=0.95\linewidth]{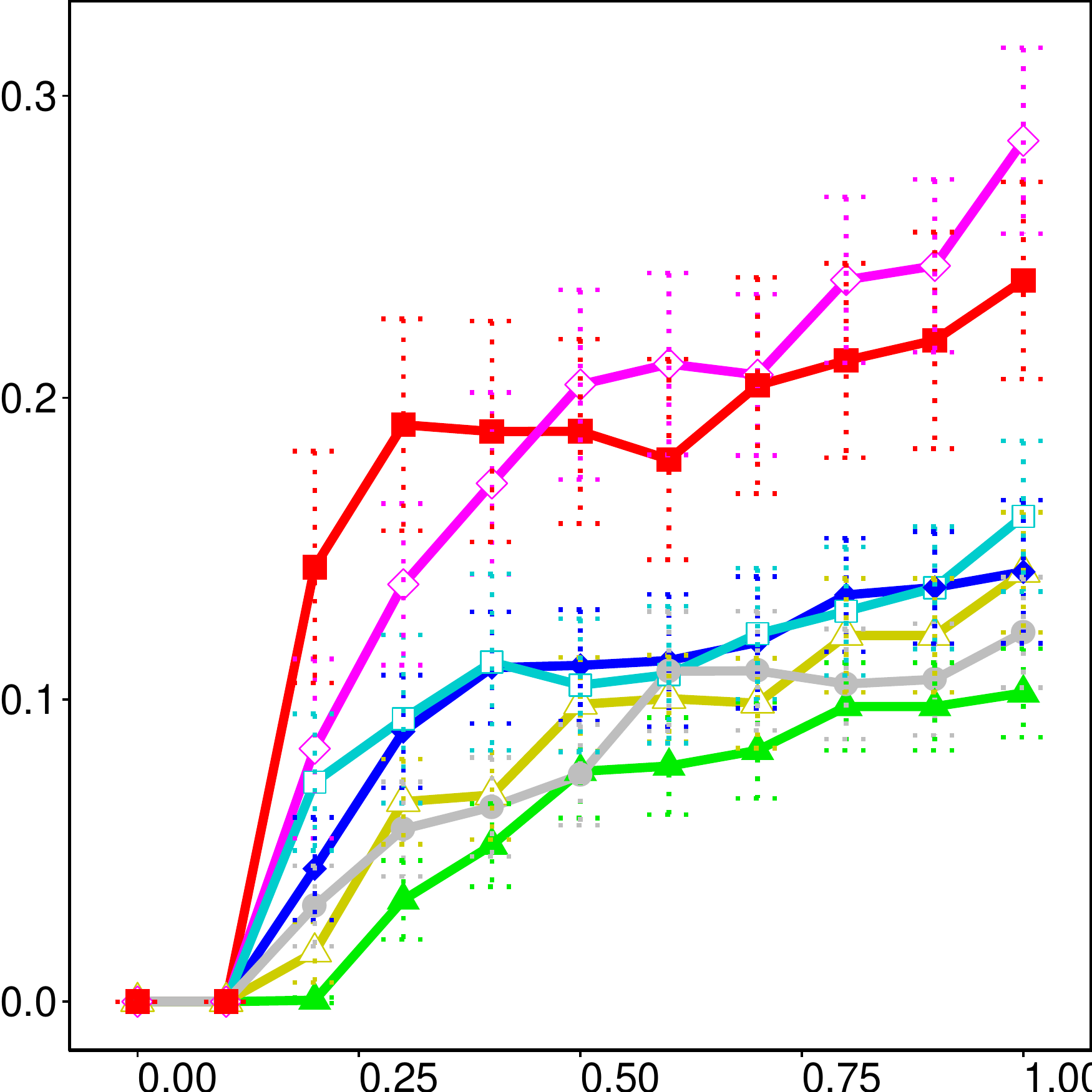} &
	\includegraphics[width=0.95\linewidth]{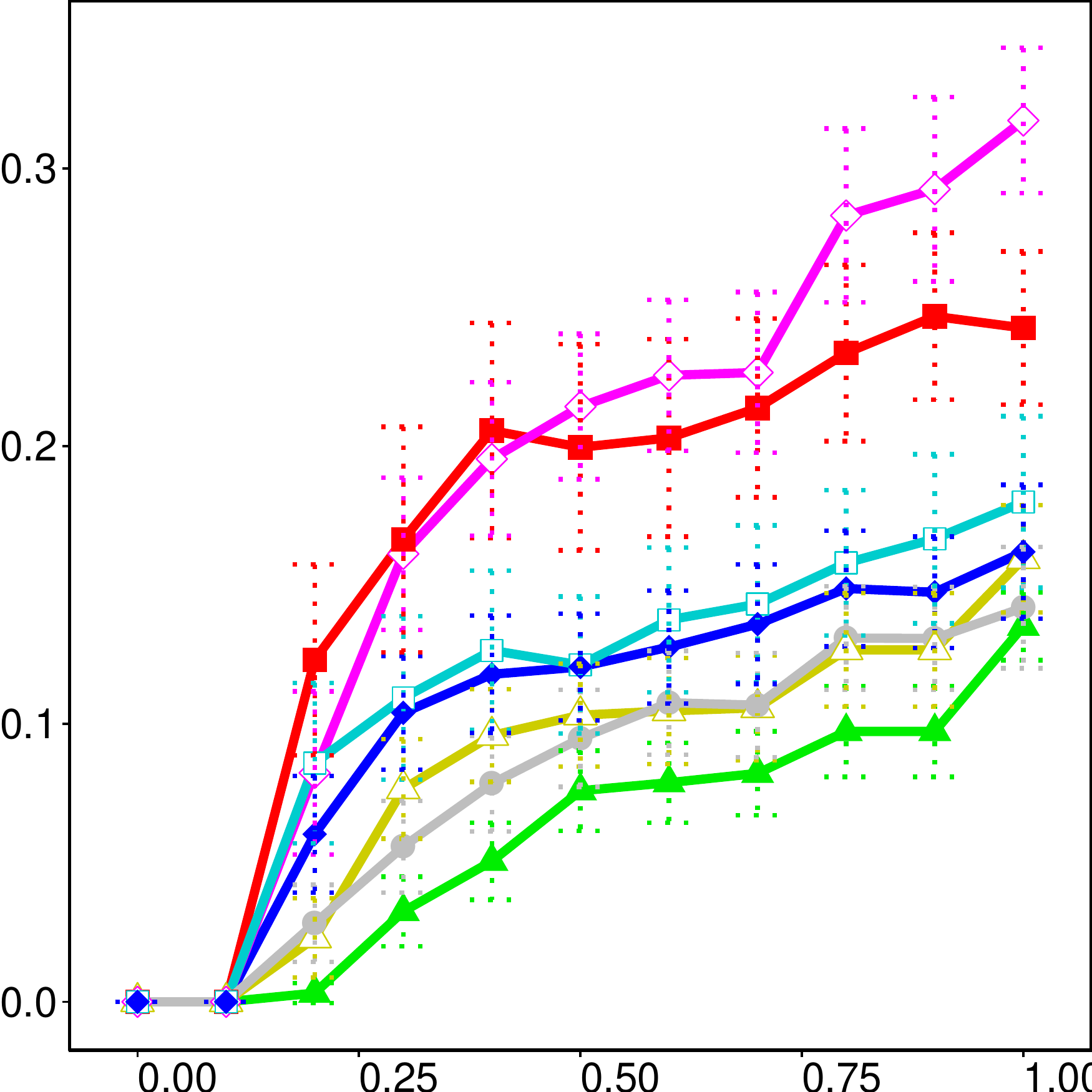} &
	\includegraphics[width=0.95\linewidth]{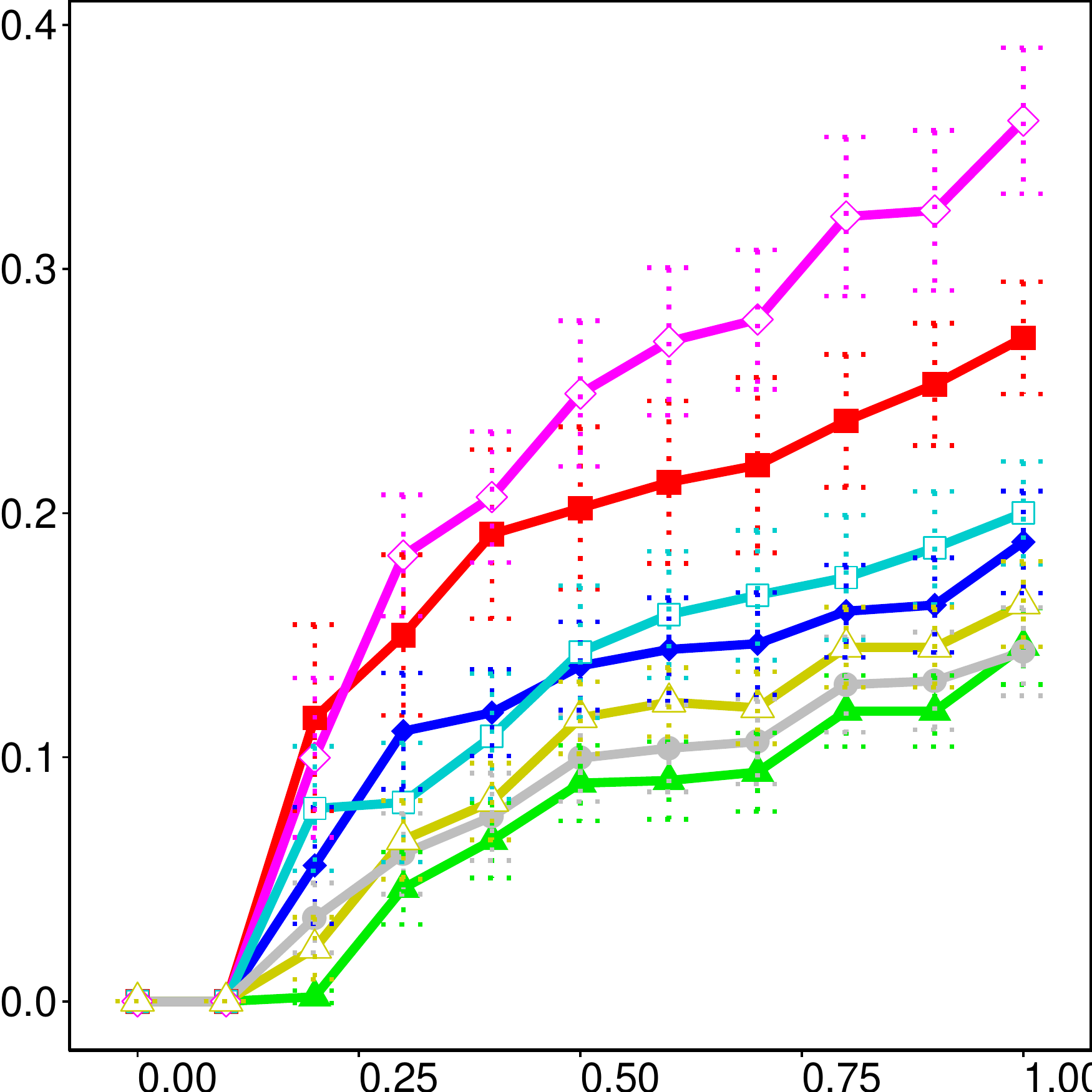} \\
	\rotatebox{90}{{\fontsize{6}{6}\selectfont{$\mathit{SmallWorld}(1000,10,.25)$}}} &
	\includegraphics[width=0.95\linewidth]{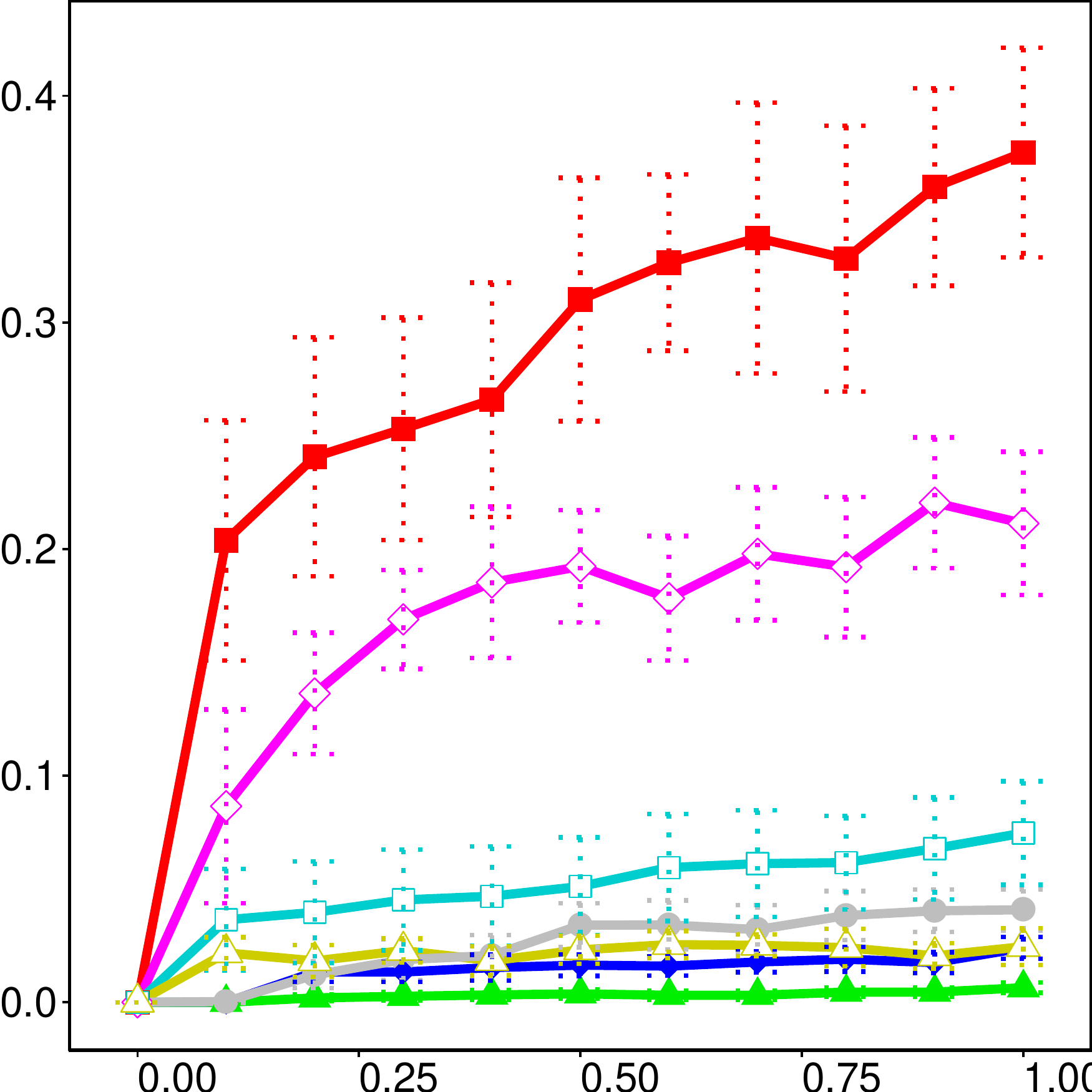} &
	\includegraphics[width=0.95\linewidth]{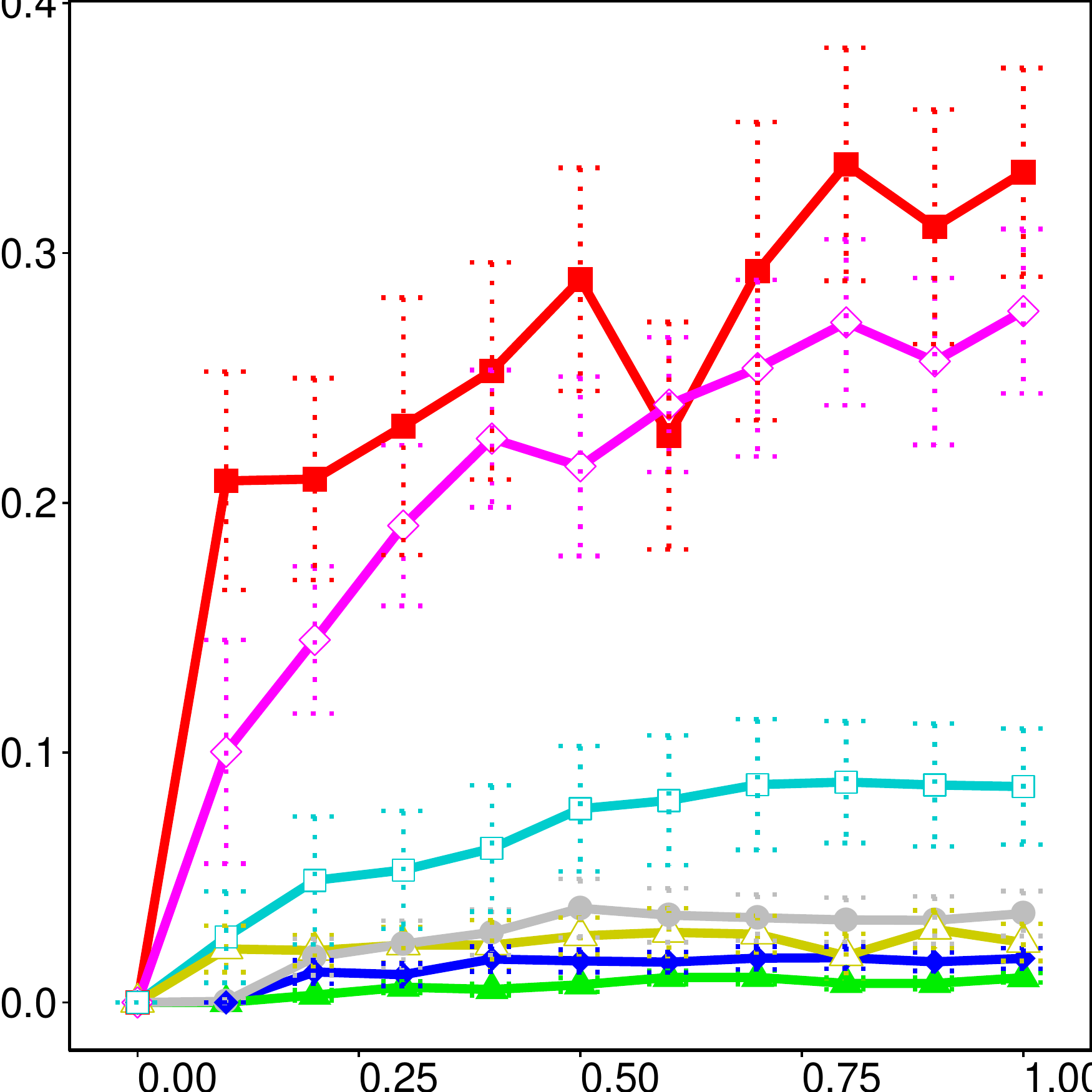} &
	\includegraphics[width=0.95\linewidth]{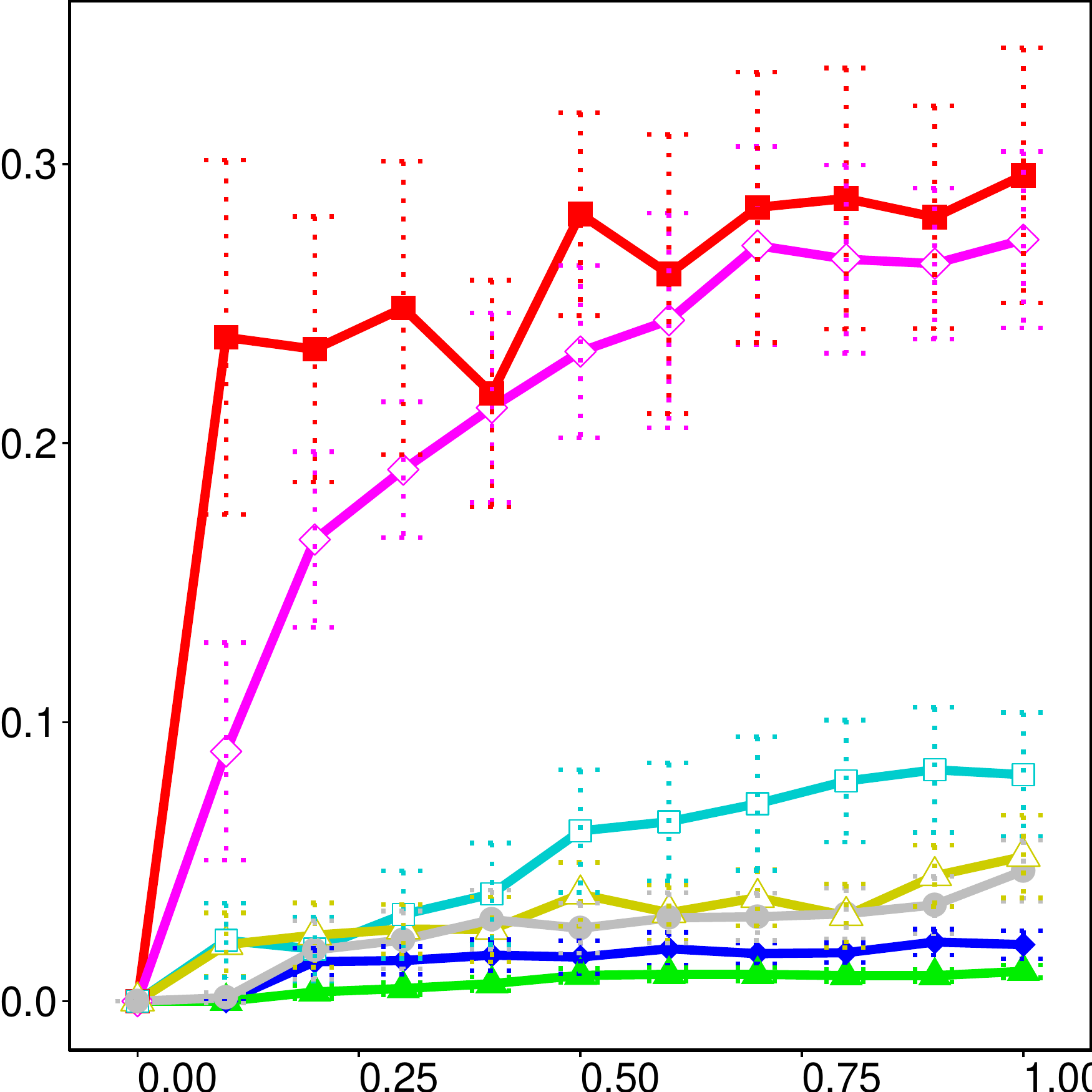} &
	\includegraphics[width=0.95\linewidth]{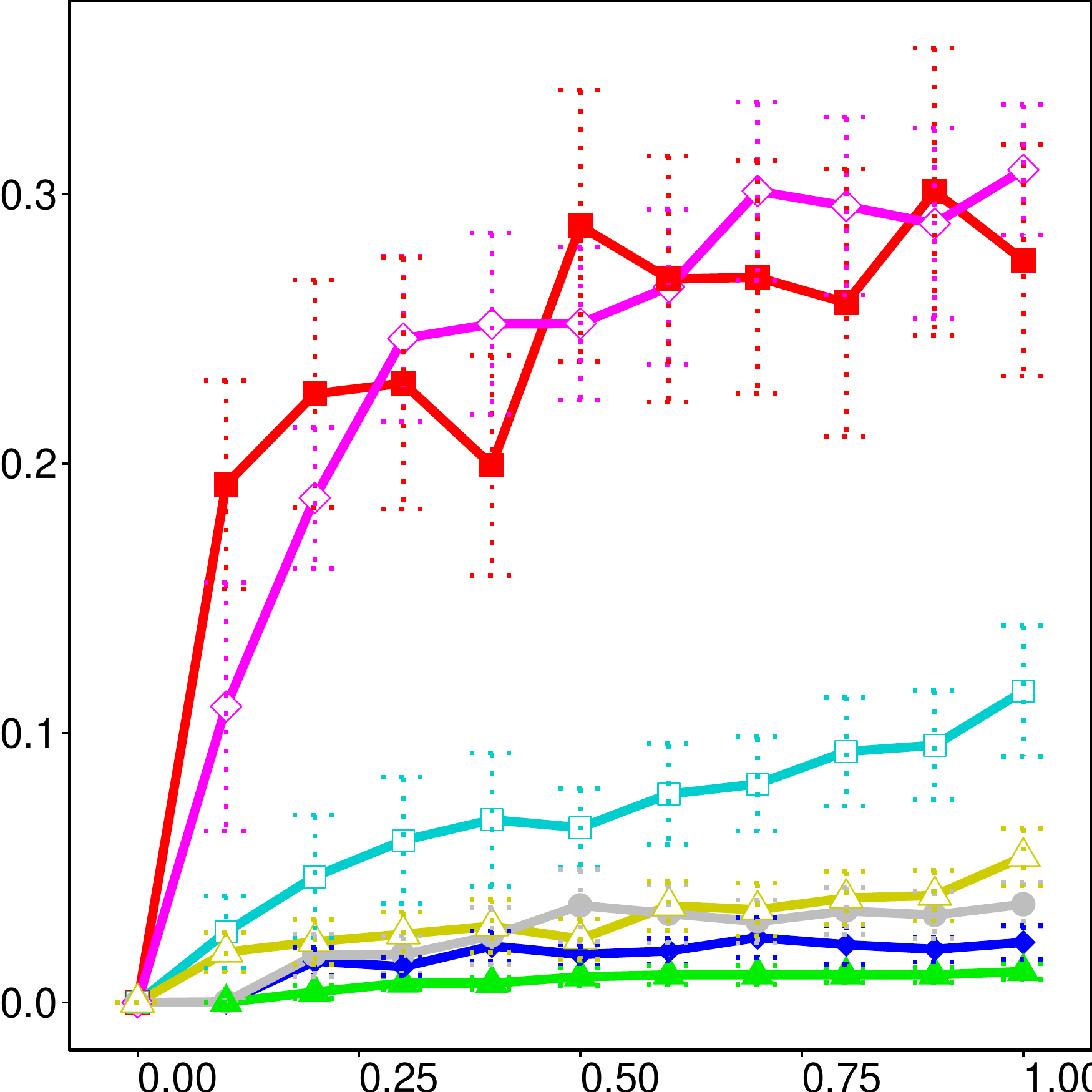} &
	\includegraphics[width=0.95\linewidth]{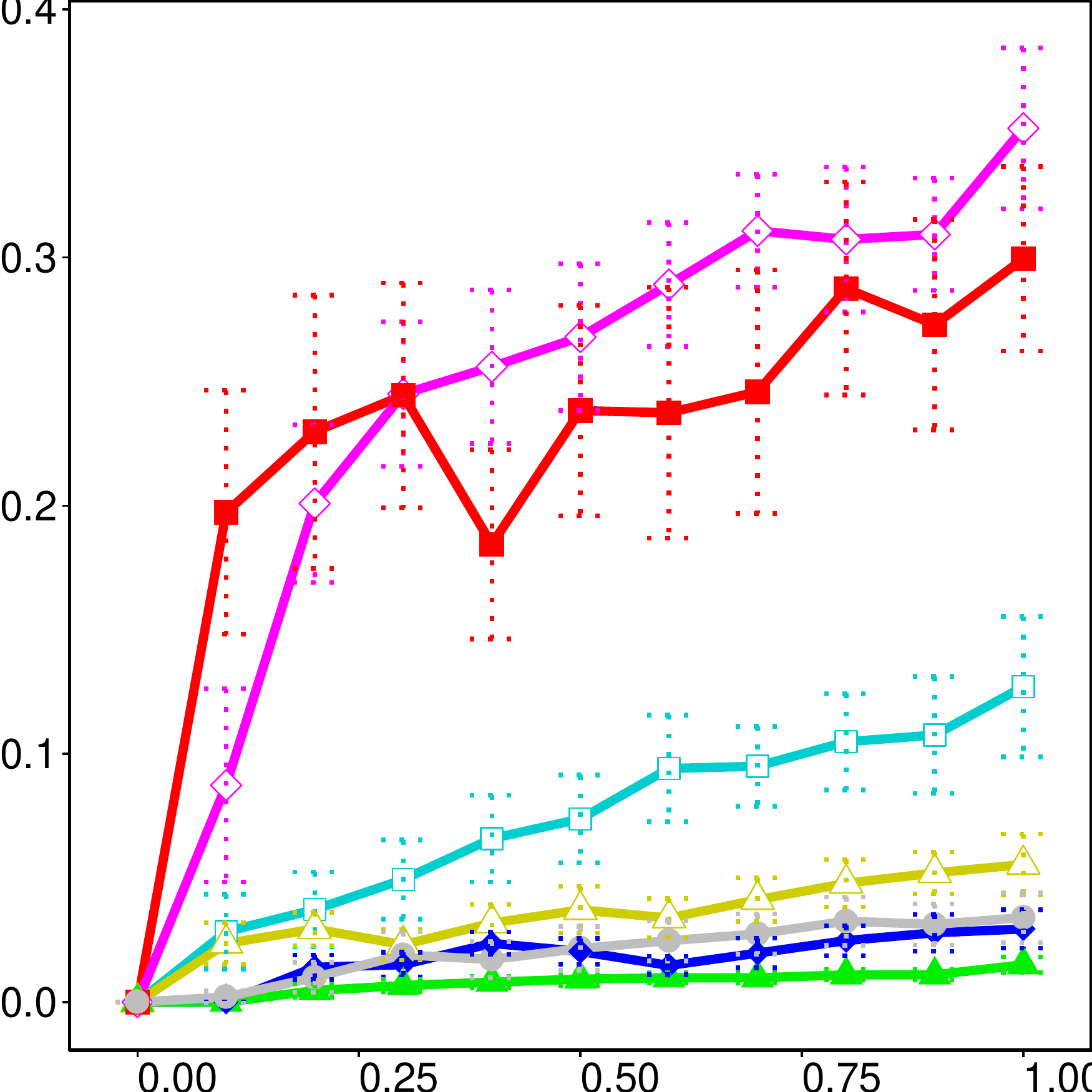} \\
	\end{tabular}
	\caption{Executing DICE multiple, consecutive rounds (the $x$-axis represents the percentage of completed rounds) in undirected random networks.}
	\label{fig:community-plots-undirected-random}
\end{figure}
%%%%%%%%%%%%%%%%%%%%%%%%%%%%%%%%%%%%%%%%%%%%%%%%%%%%%%%%%%%%%%%%%%%%%%%%%%%%%%%

%%%%%%%%%%%%%%%%%%%%%%%%%%%%%%%%%%%%%%%%%%%%%%%%%%%%%%%%%%%%%%%%%%%%%%%%%%%%%%%
\begin{figure}[p!]
	\centering
	\setlength\tabcolsep{2pt}
	\renewcommand{\arraystretch}{2.0}
	\begin{tabular}{m{.01\textwidth}m{.195\textwidth}m{.195\textwidth}m{.195\textwidth}m{.195\textwidth}m{.195\textwidth}}
	&
	\multicolumn{1}{c}{\small{$b=4$, $d=0$}} &
	\multicolumn{1}{c}{\small{$b=4$, $d=1$}} &
	\multicolumn{1}{c}{\small{$b=4$, $d=2$}} &
	\multicolumn{1}{c}{\small{$b=4$, $d=3$}} &
	\multicolumn{1}{c}{\small{$b=4$, $d=4$}} \\
	\rotatebox{90}{\small{Bali attack}} &
	\includegraphics[width=0.95\linewidth]{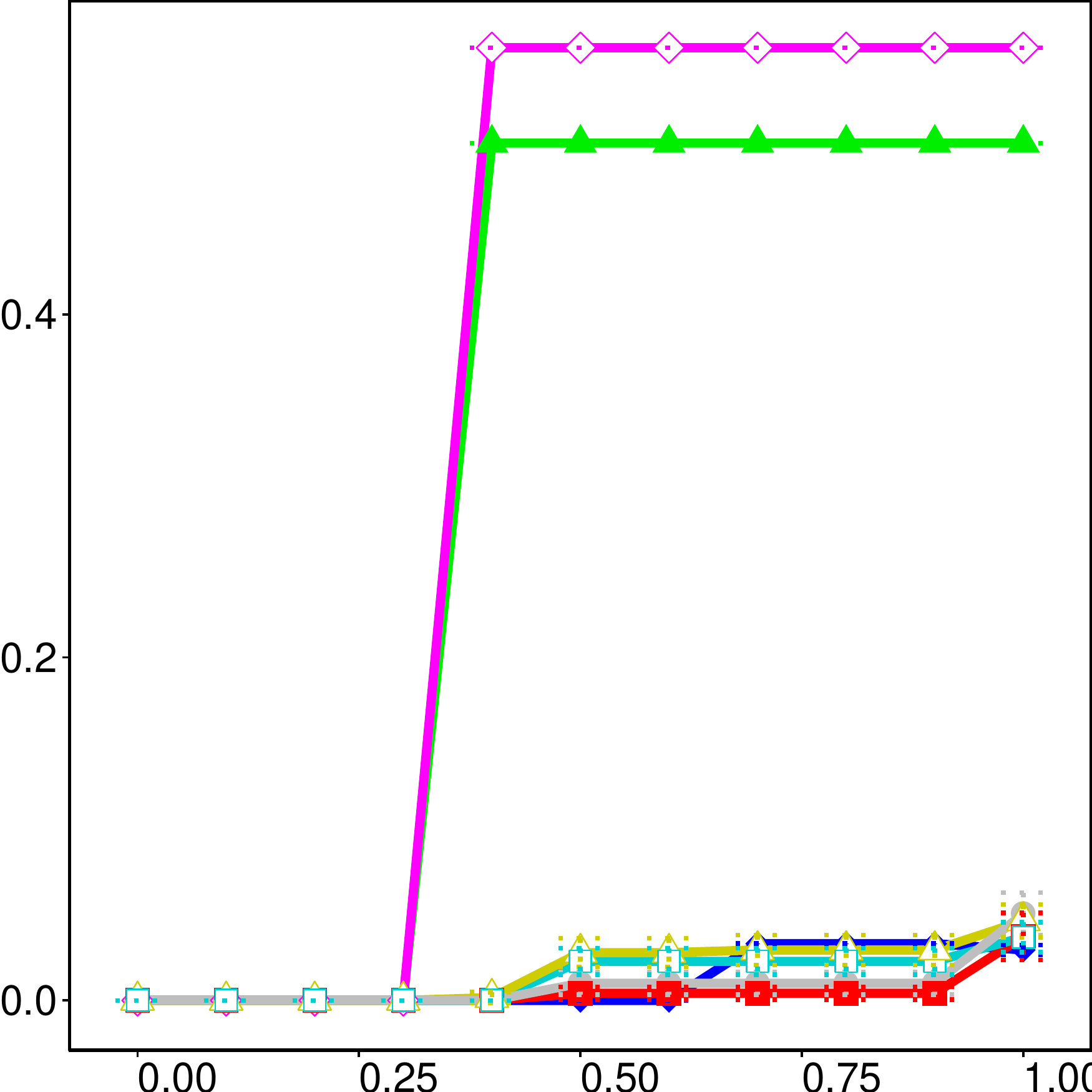} &
	\includegraphics[width=0.95\linewidth]{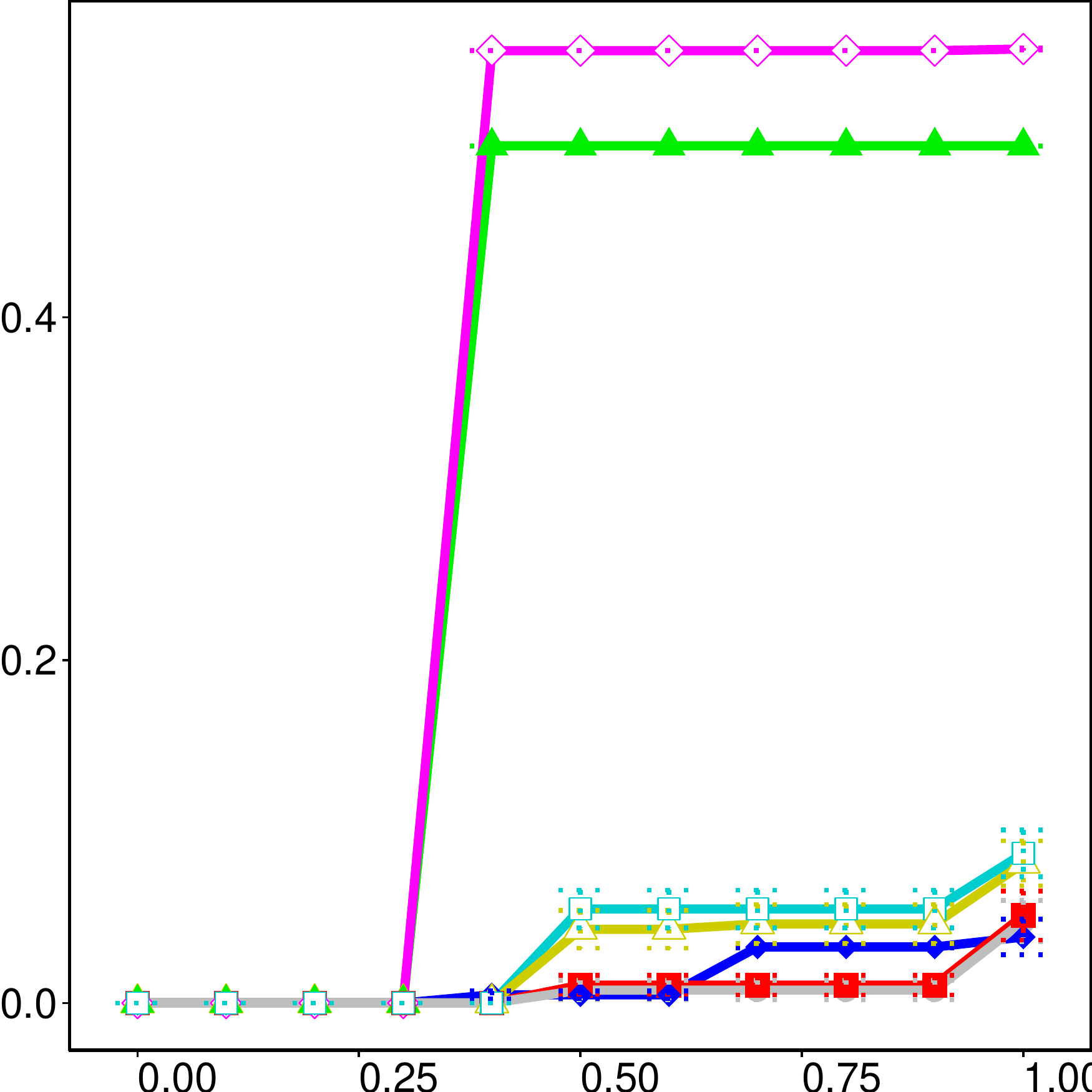} &
	\includegraphics[width=0.95\linewidth]{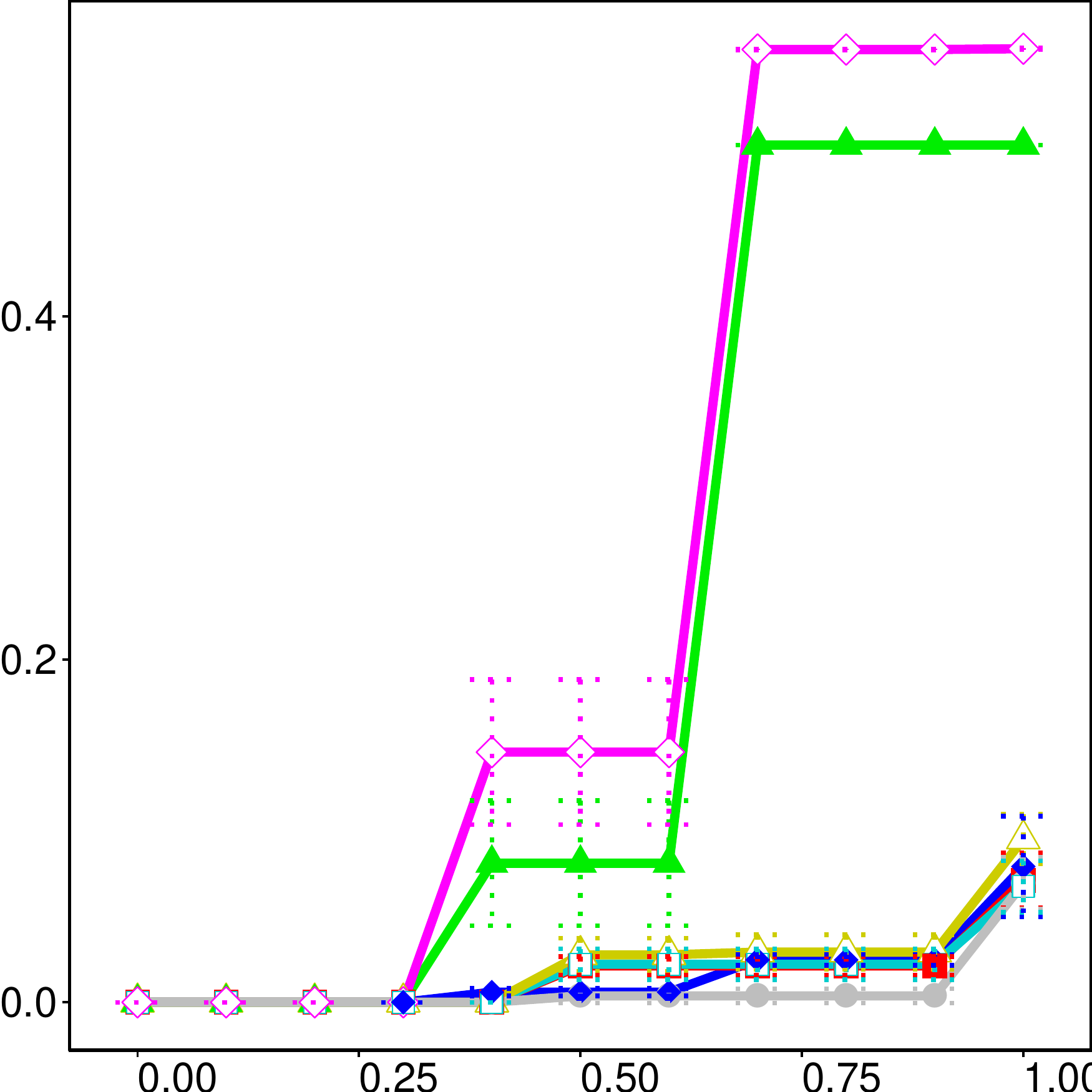} &
	\includegraphics[width=0.95\linewidth]{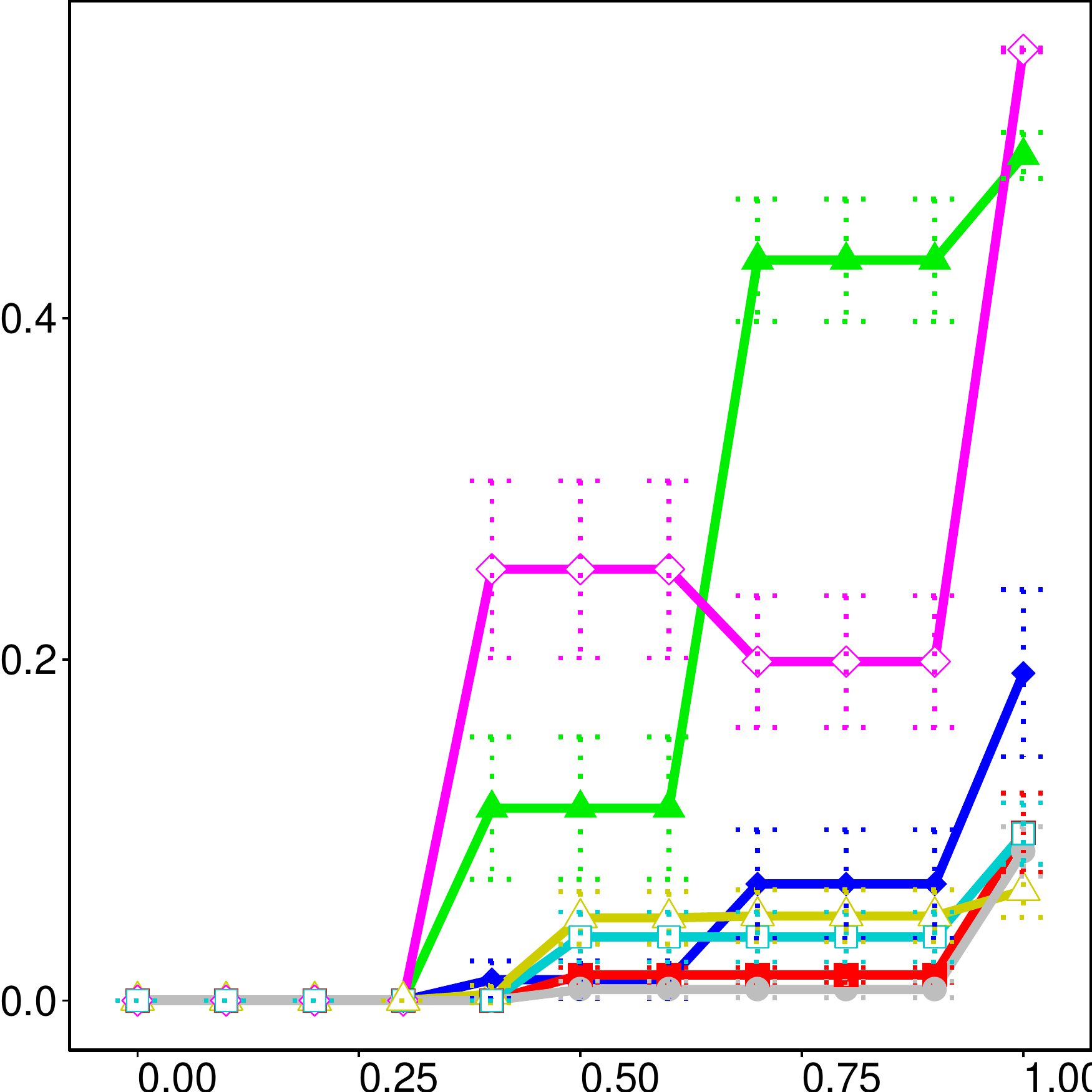} &
	\includegraphics[width=0.95\linewidth]{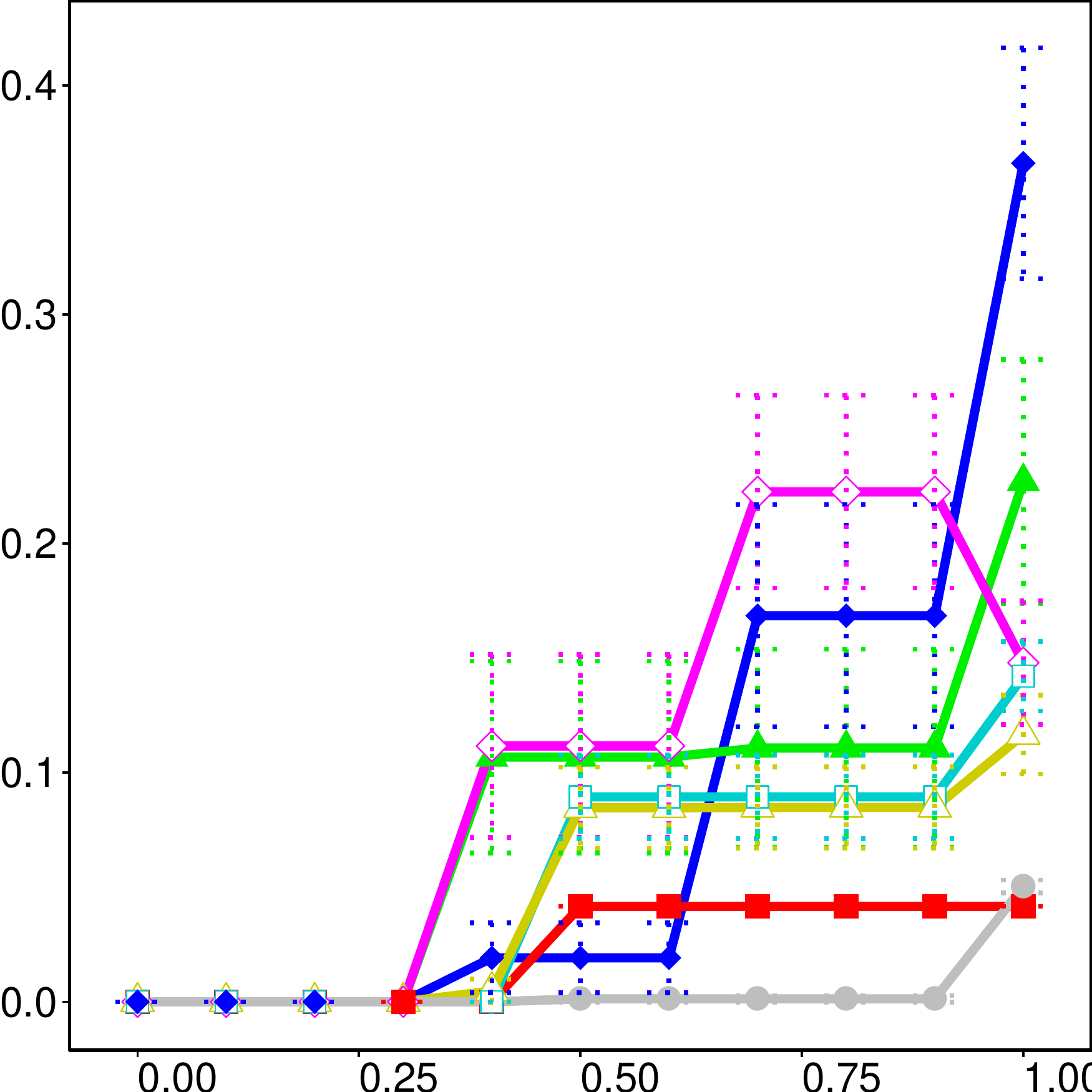} \\
	\rotatebox{90}{\small{Madrid bombing}} &
	\includegraphics[width=0.95\linewidth]{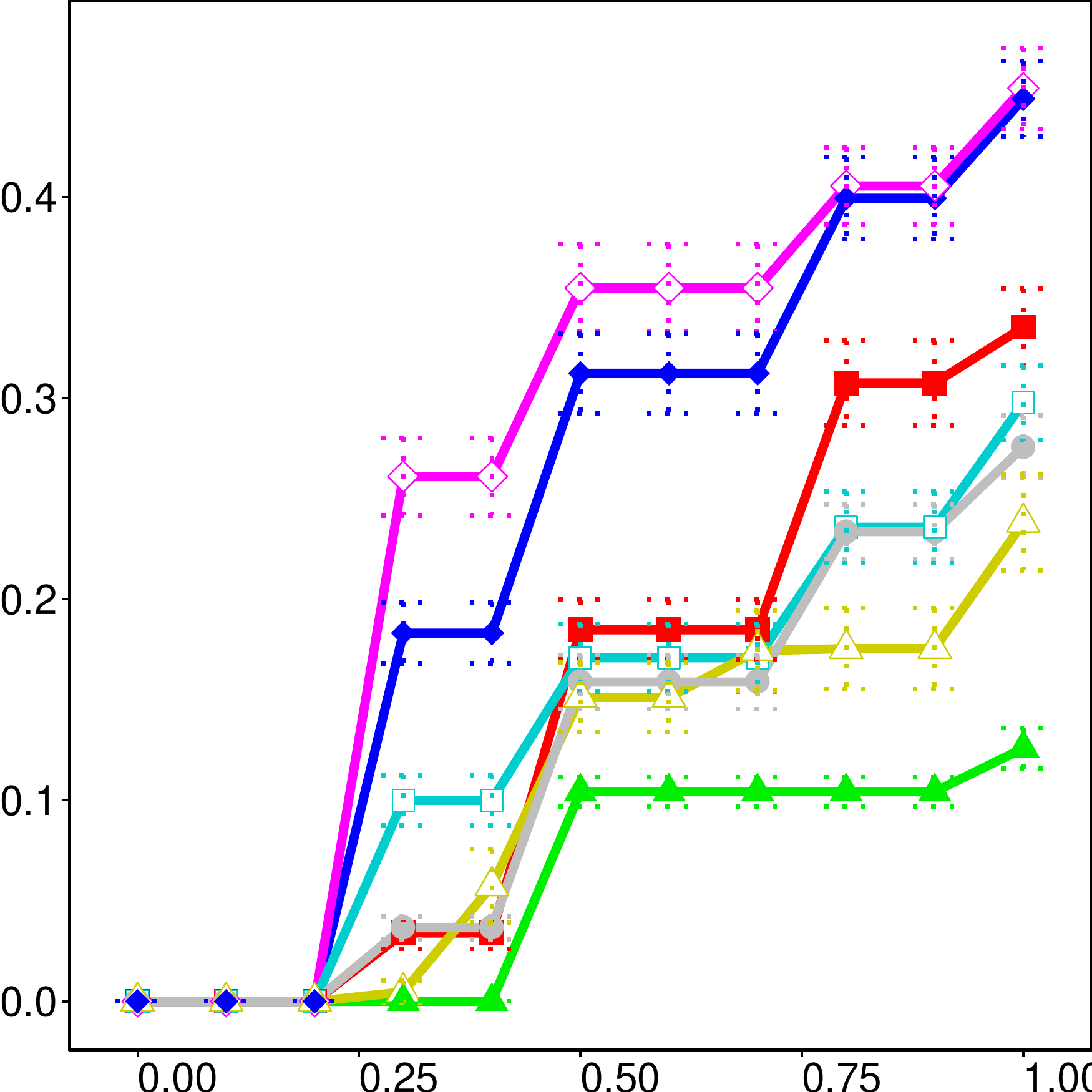} &
	\includegraphics[width=0.95\linewidth]{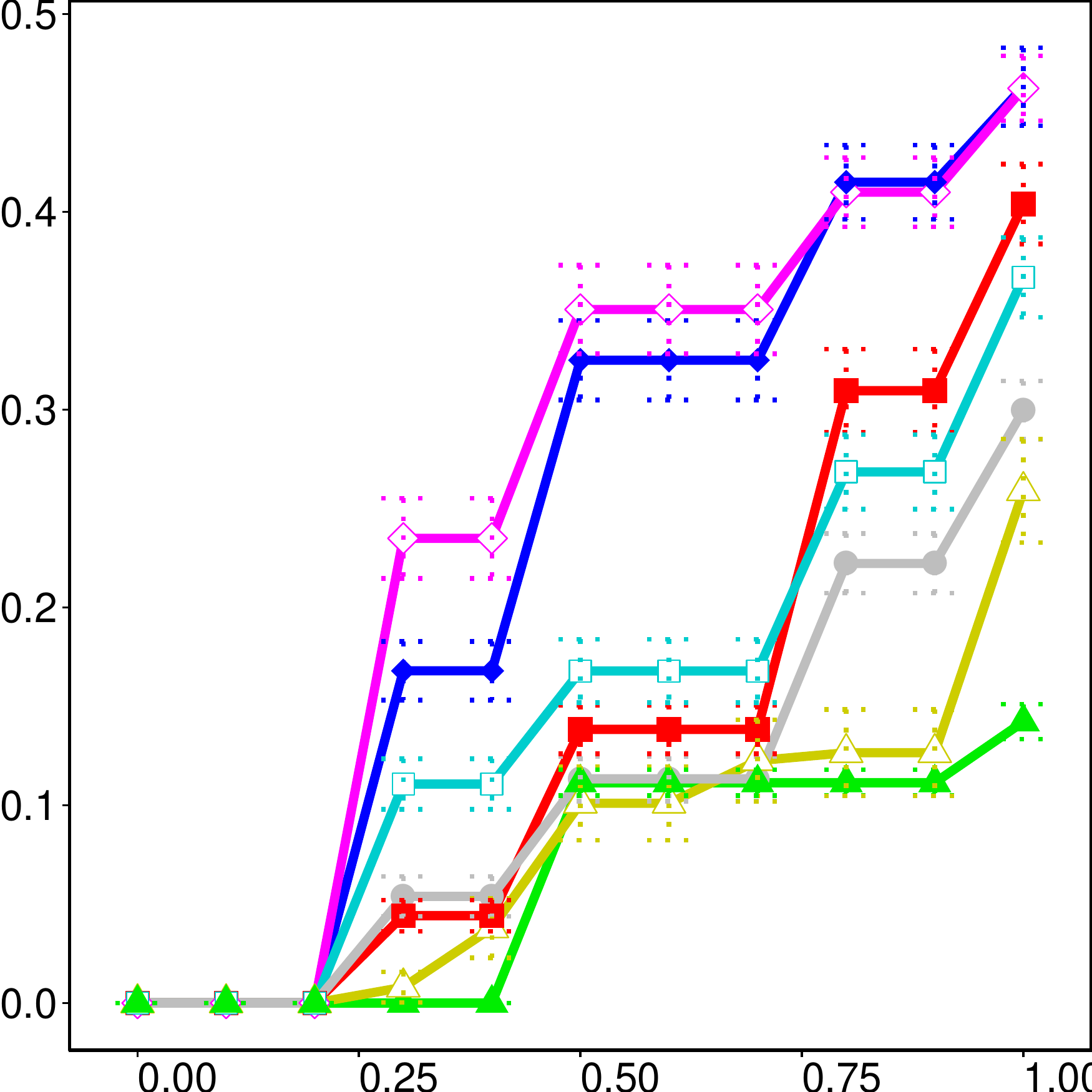} &
	\includegraphics[width=0.95\linewidth]{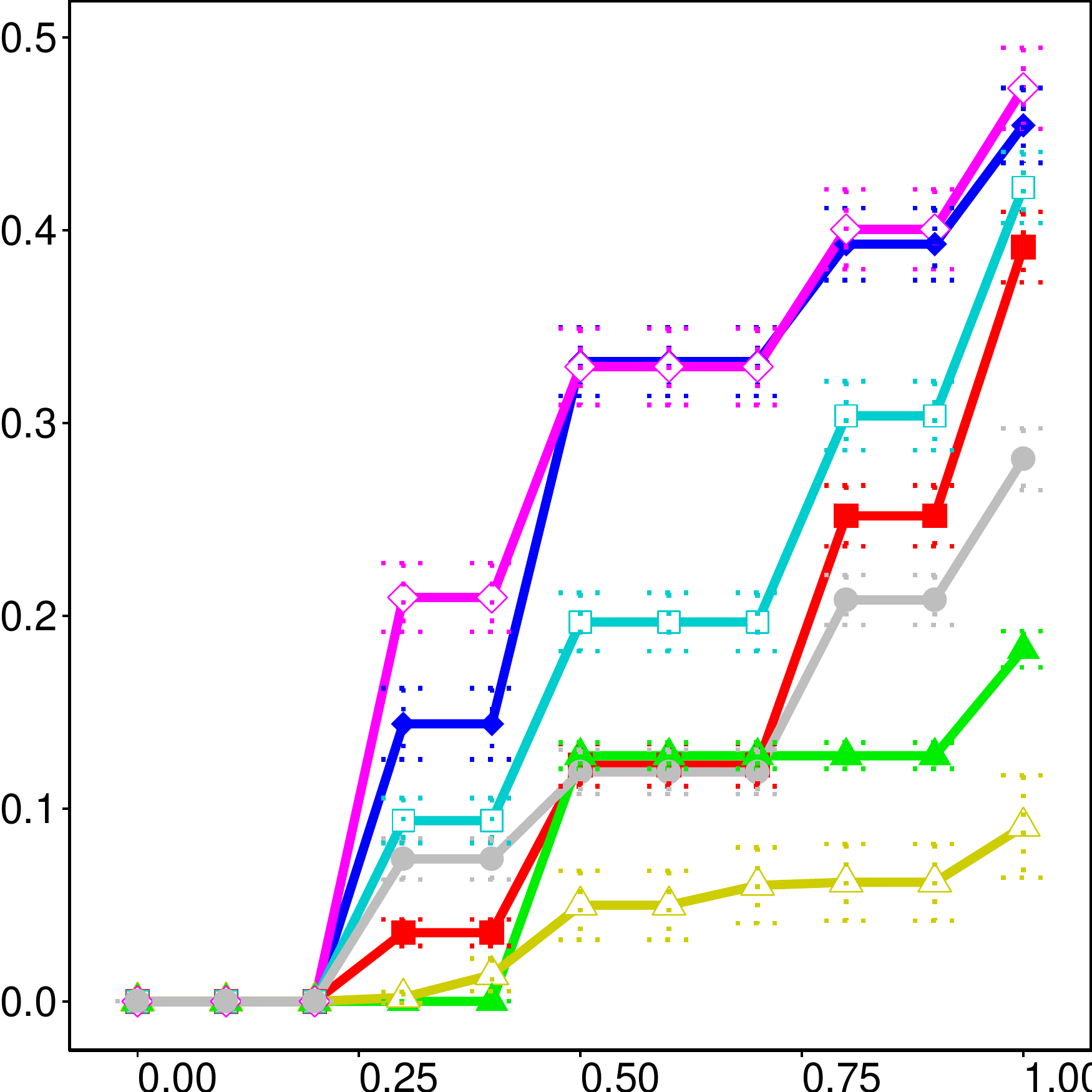} &
	\includegraphics[width=0.95\linewidth]{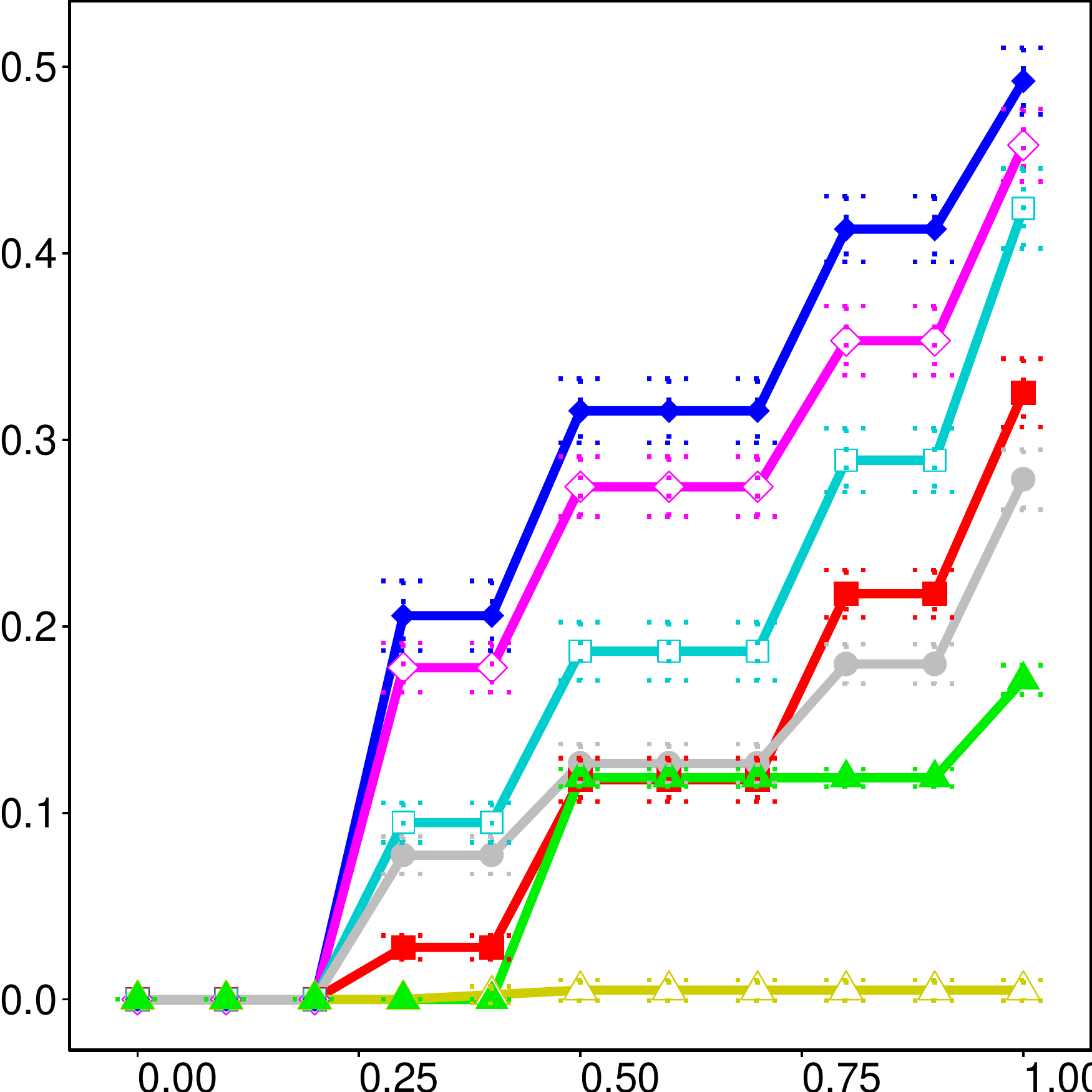} &
	\includegraphics[width=0.95\linewidth]{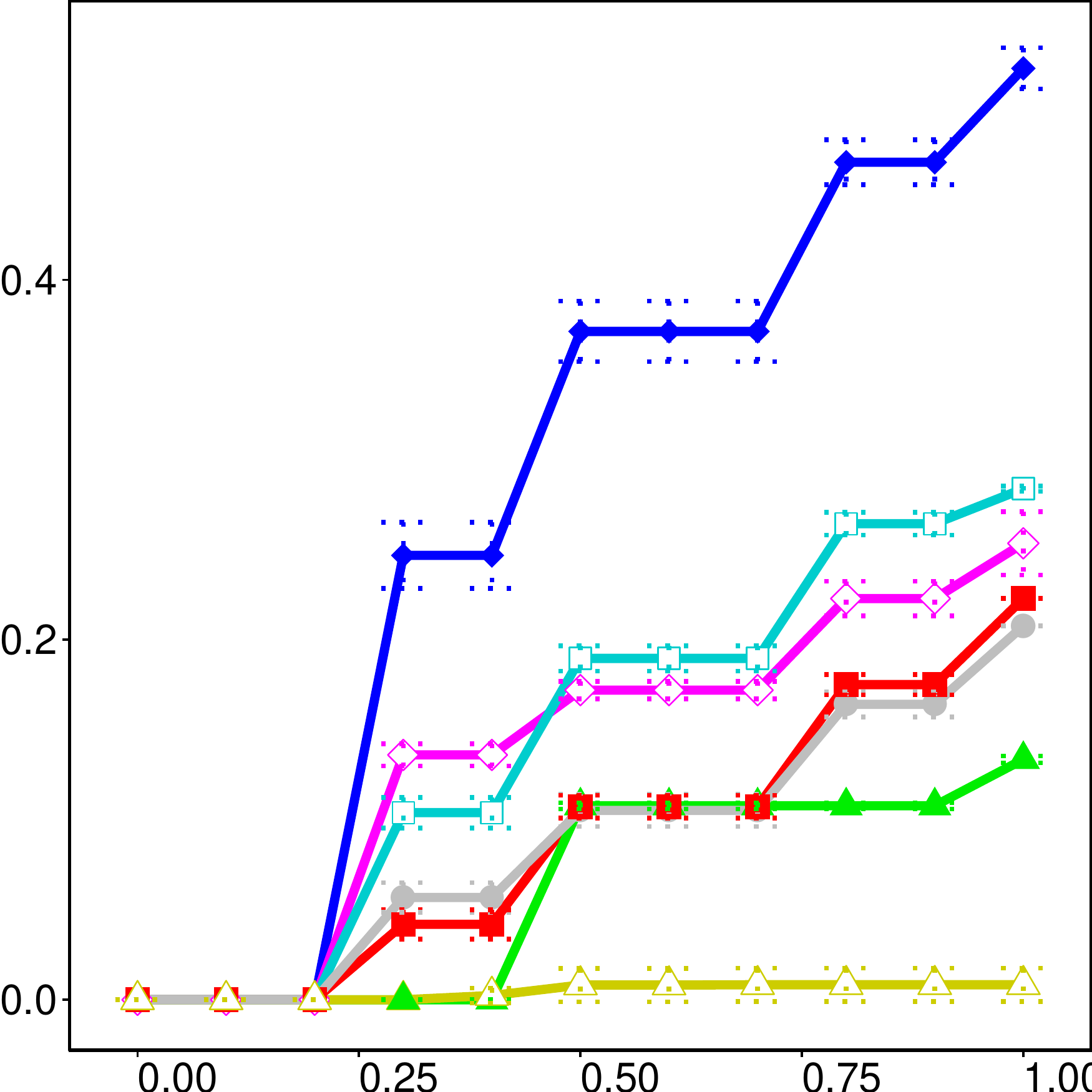} \\
	\rotatebox{90}{\small{WTC 9/11}} &
	\includegraphics[width=0.95\linewidth]{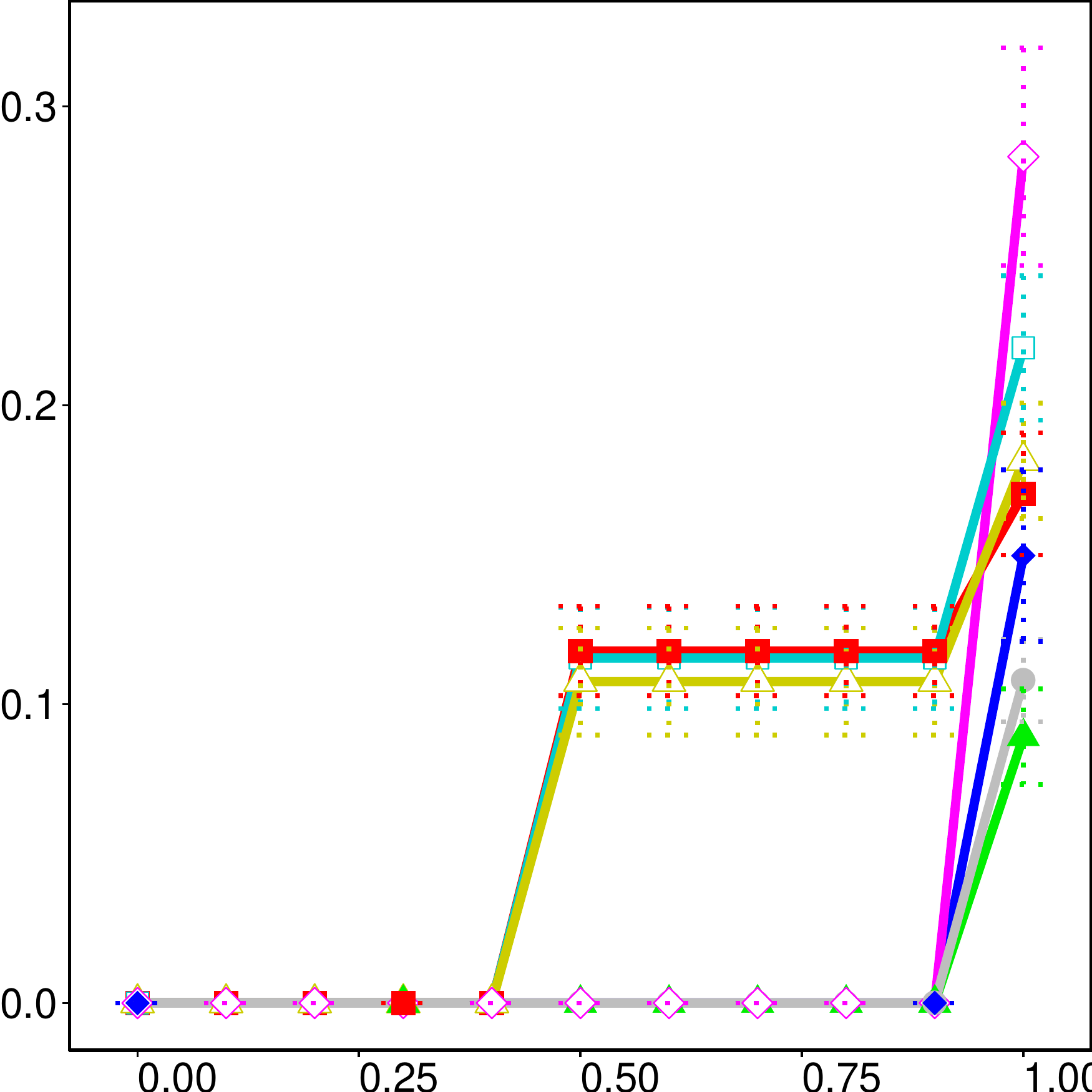} &
	\includegraphics[width=0.95\linewidth]{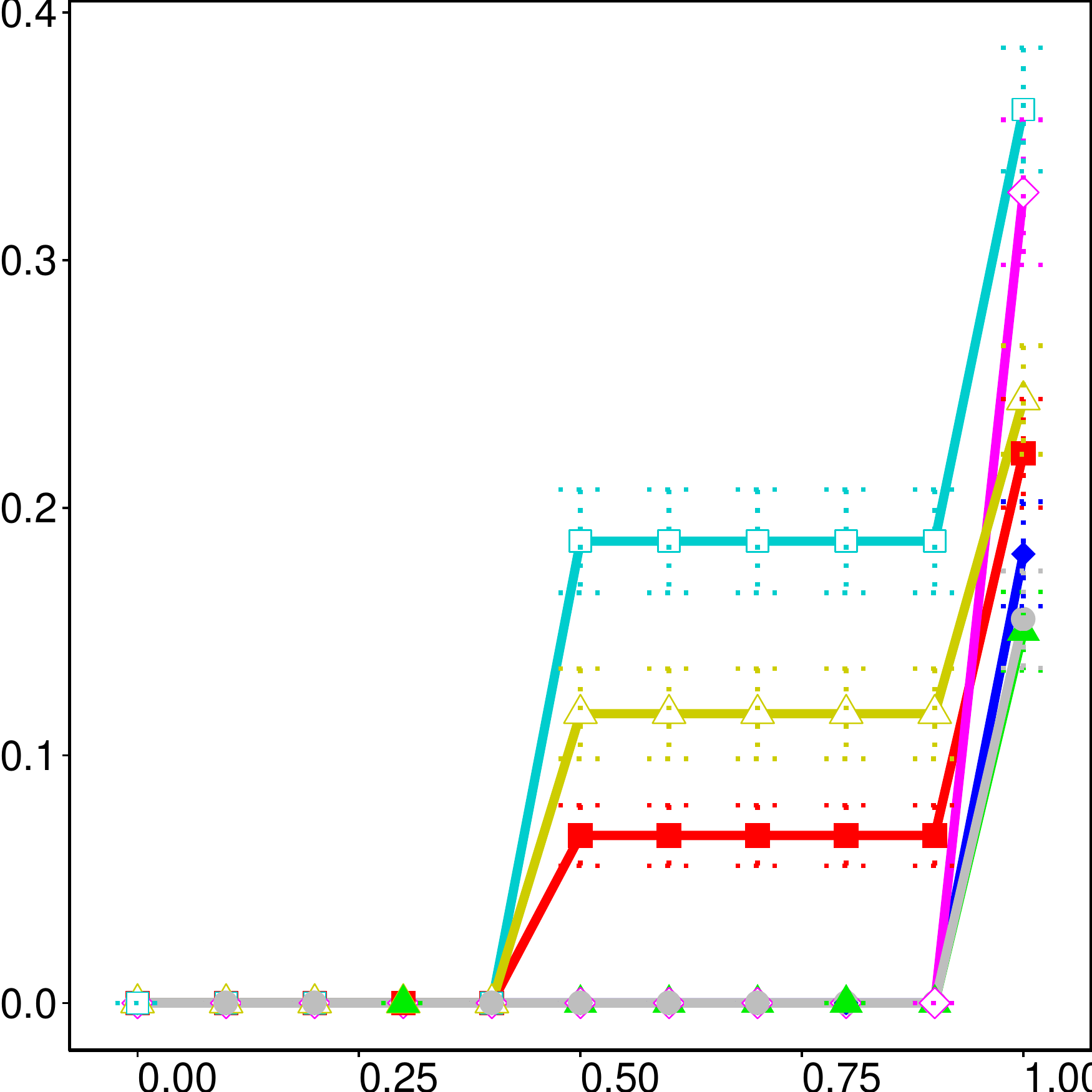} &
	\includegraphics[width=0.95\linewidth]{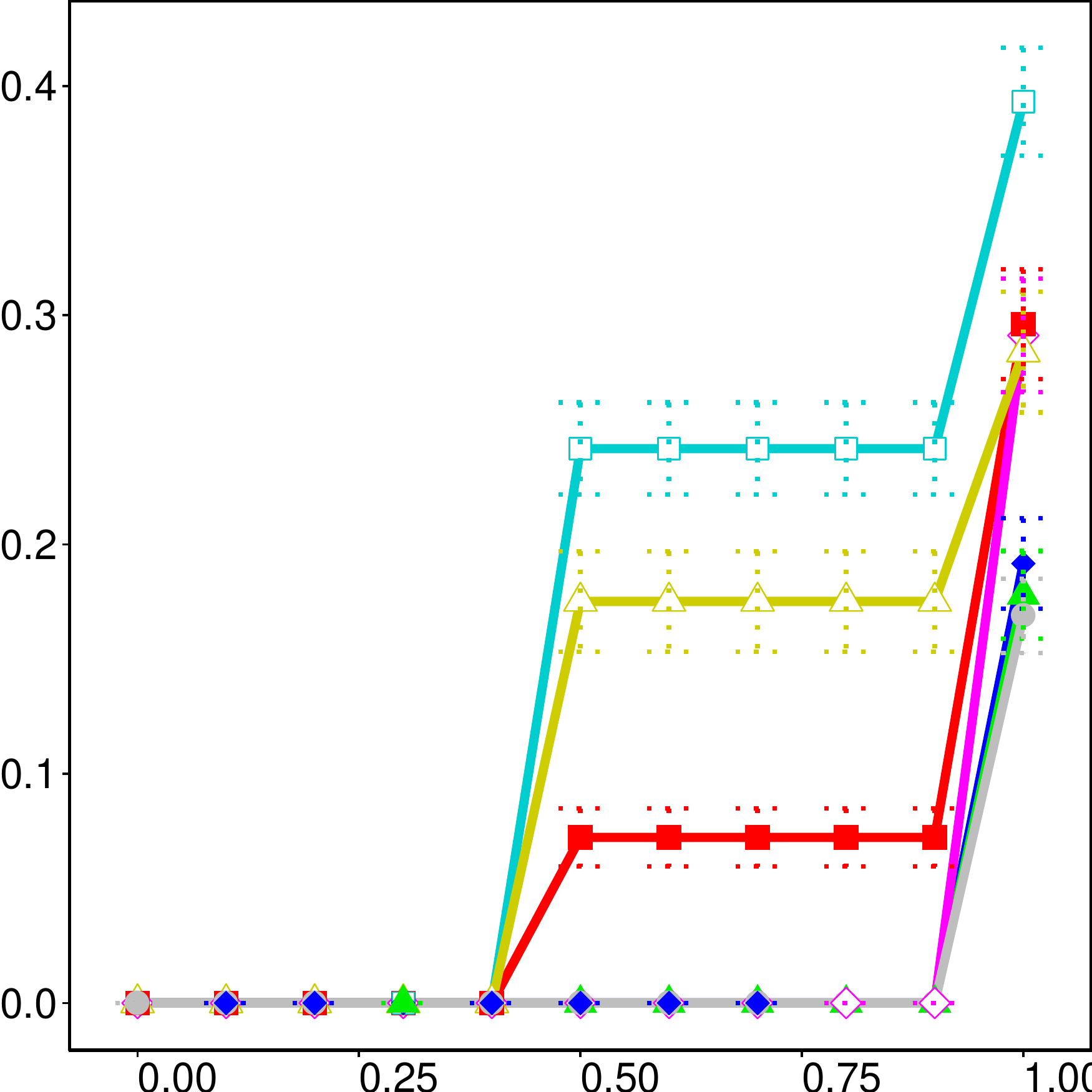} &
	\includegraphics[width=0.95\linewidth]{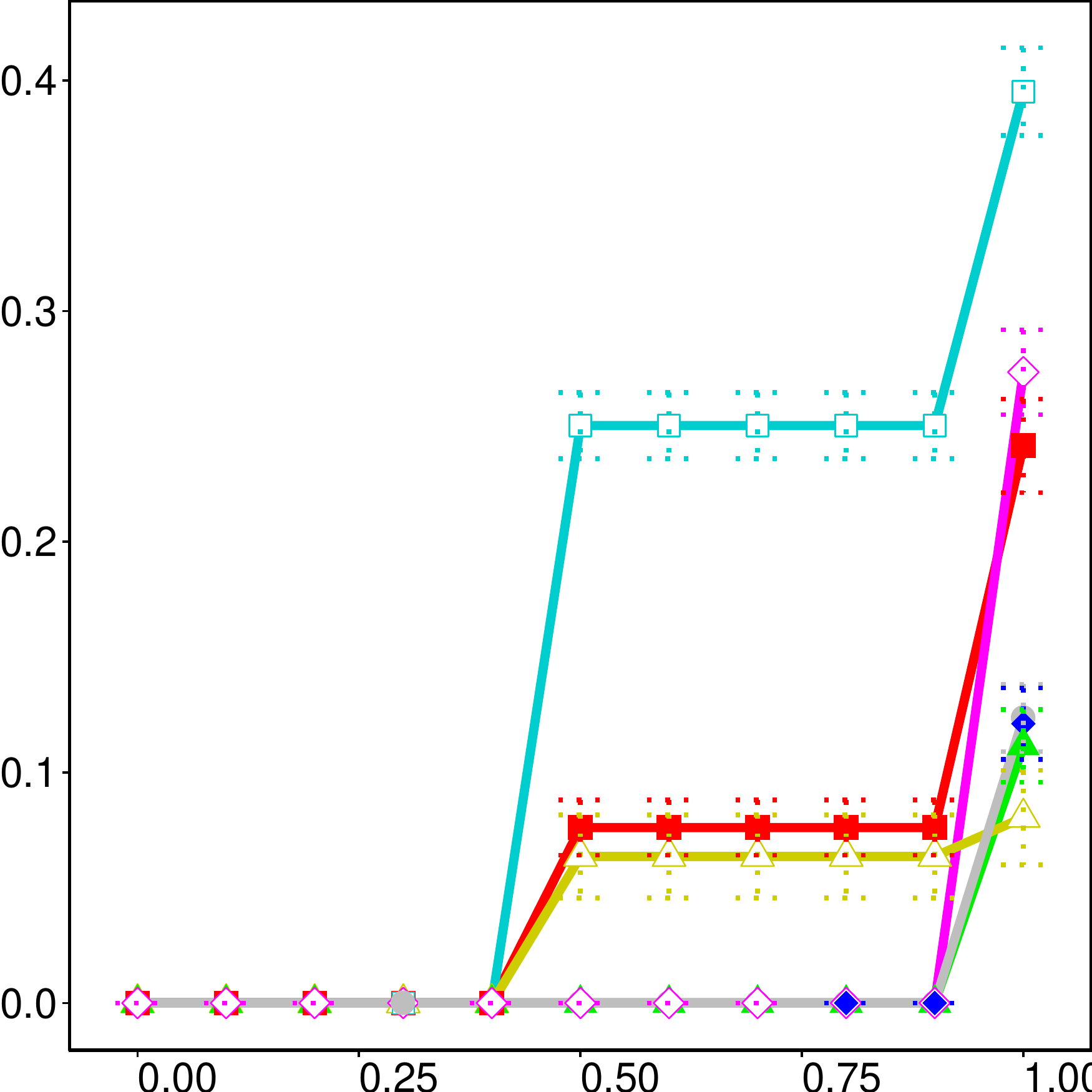} &
	\includegraphics[width=0.95\linewidth]{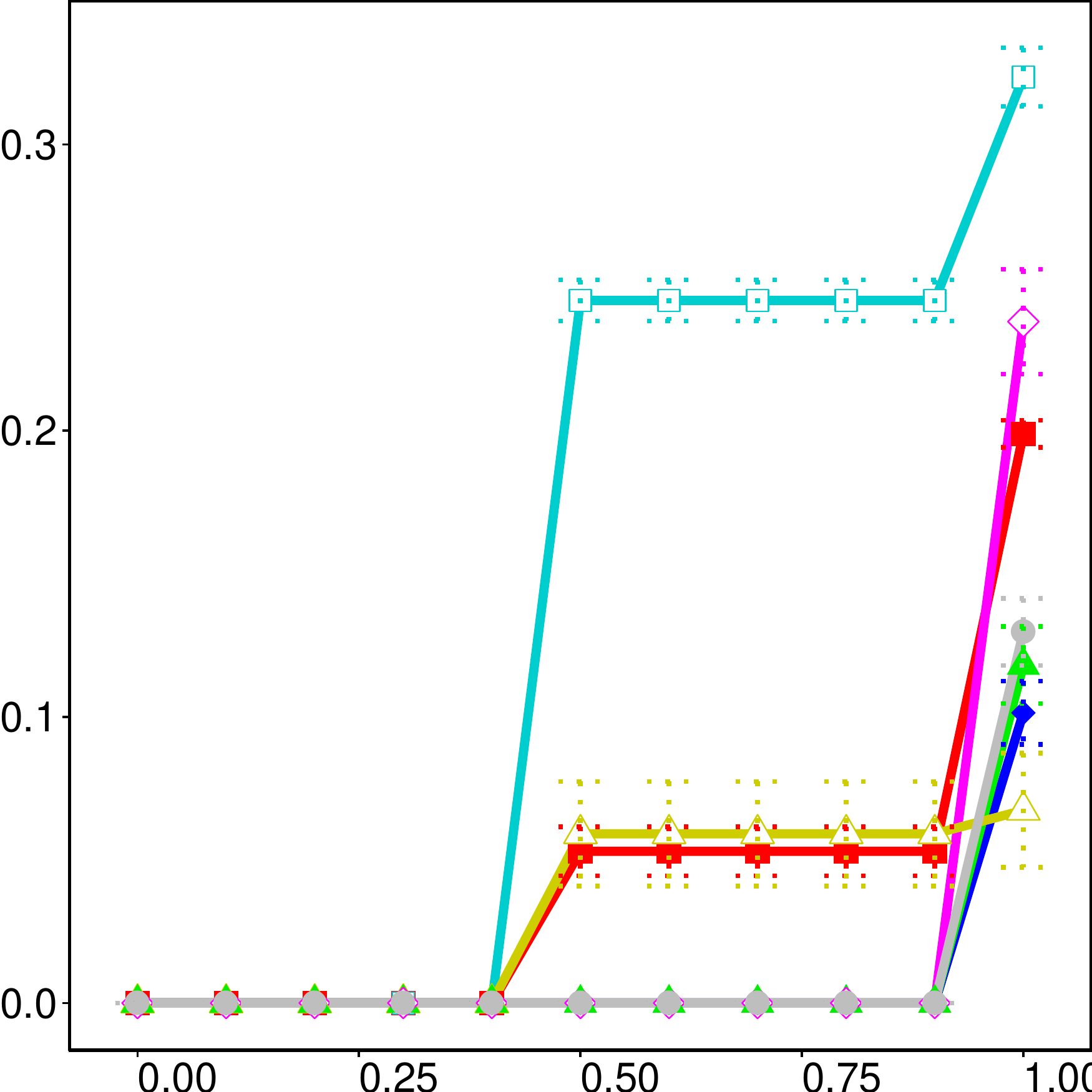} \\
	\rotatebox{90}{\small{Facebook (small)}} &
	\includegraphics[width=0.95\linewidth]{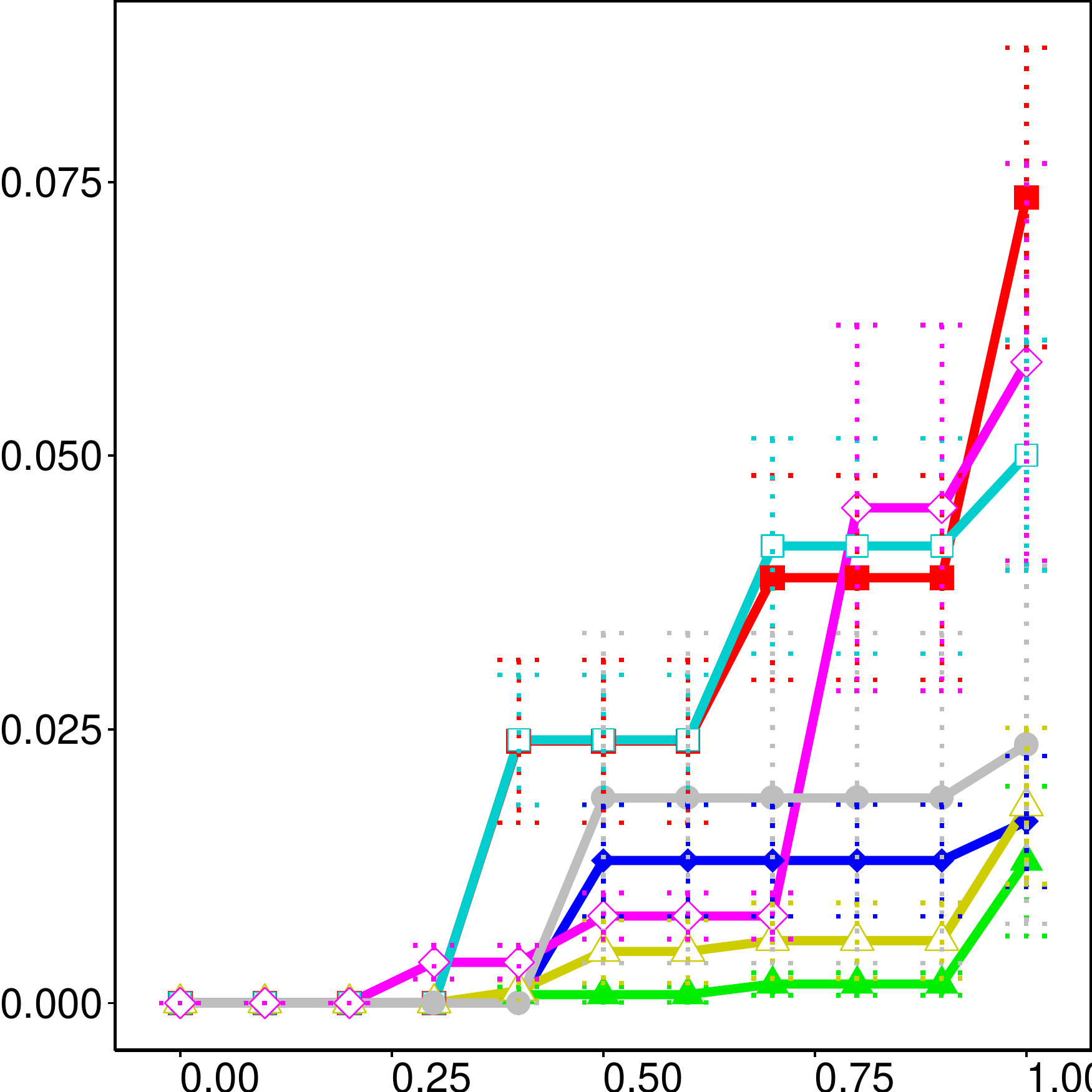} &
	\includegraphics[width=0.95\linewidth]{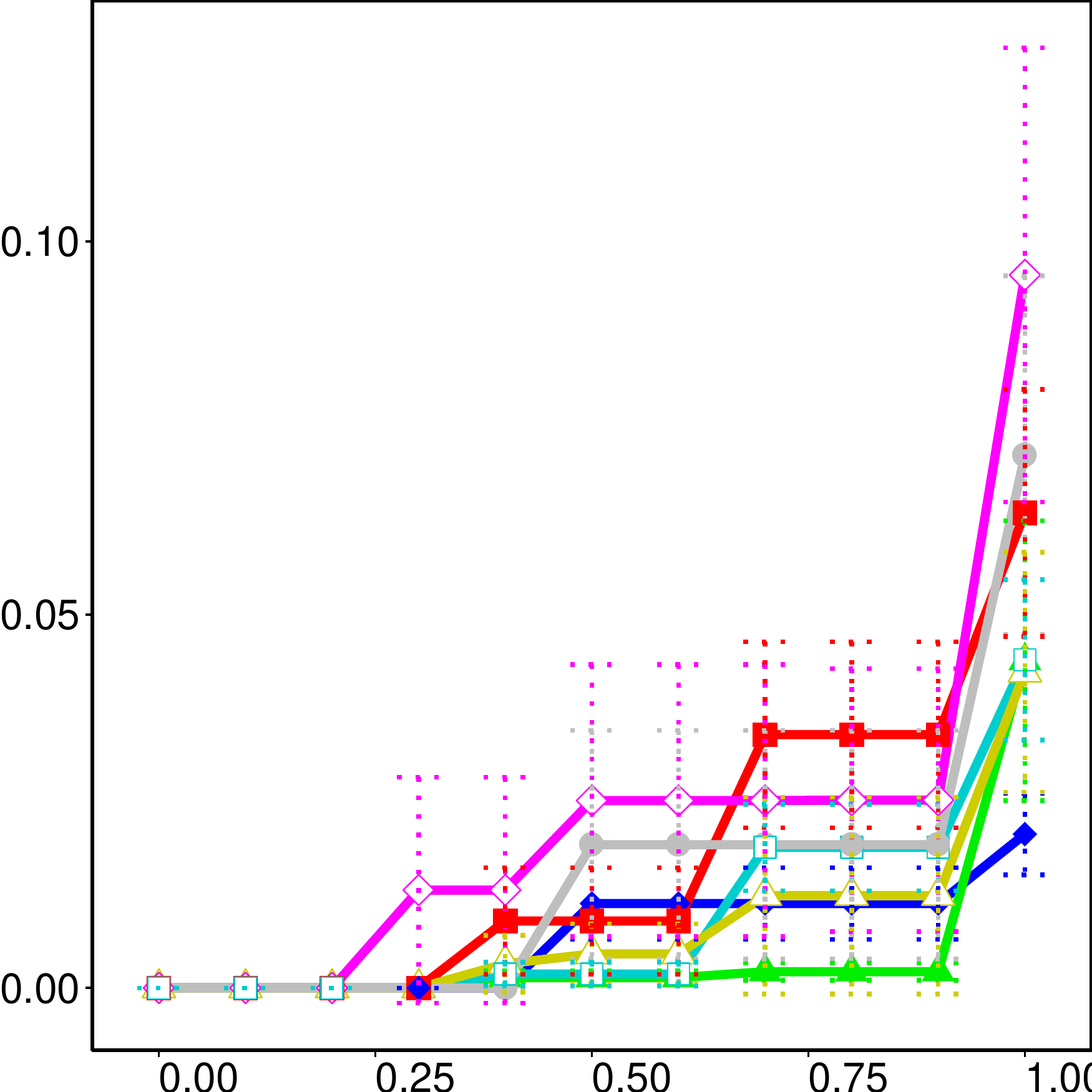} &
	\includegraphics[width=0.95\linewidth]{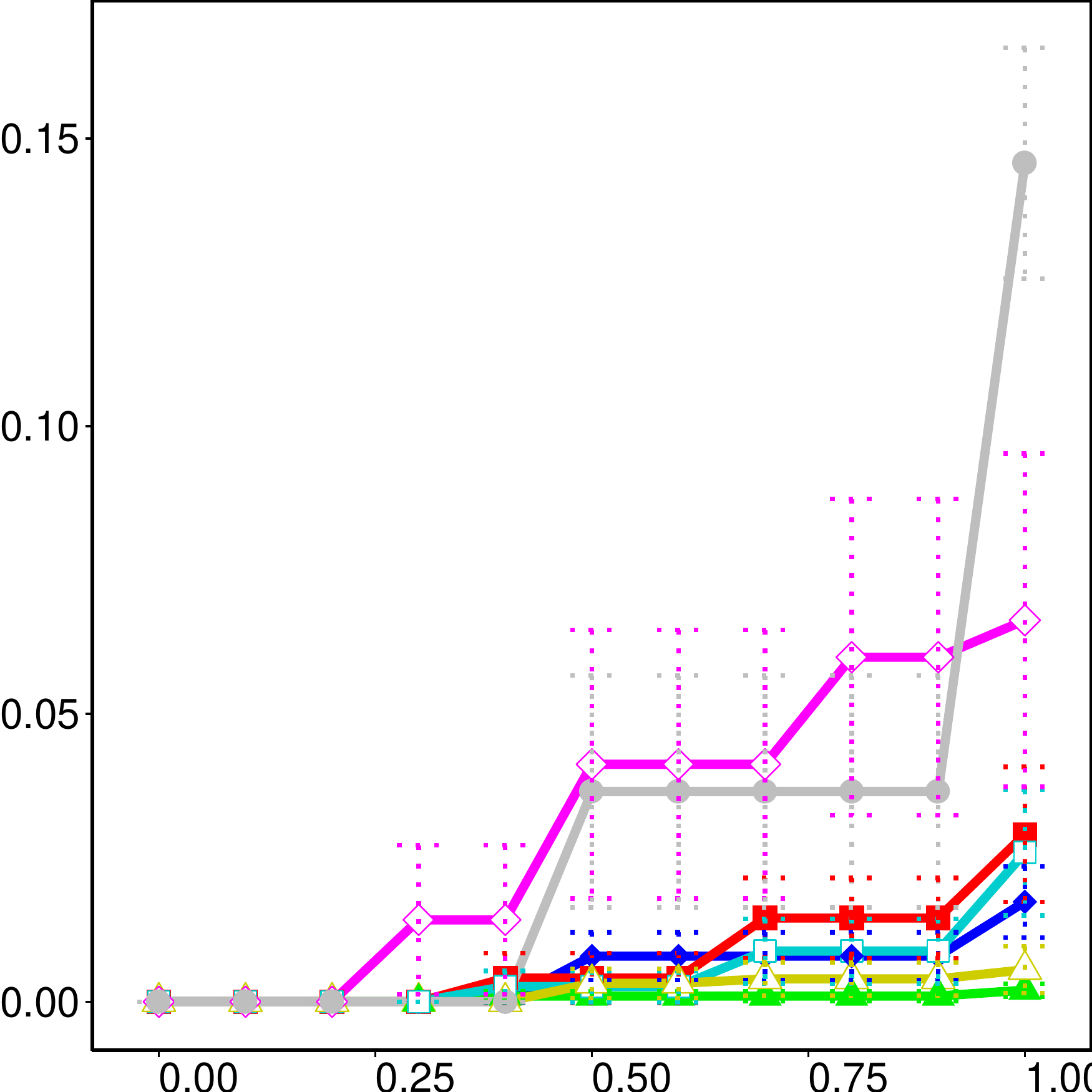} &
	\includegraphics[width=0.95\linewidth]{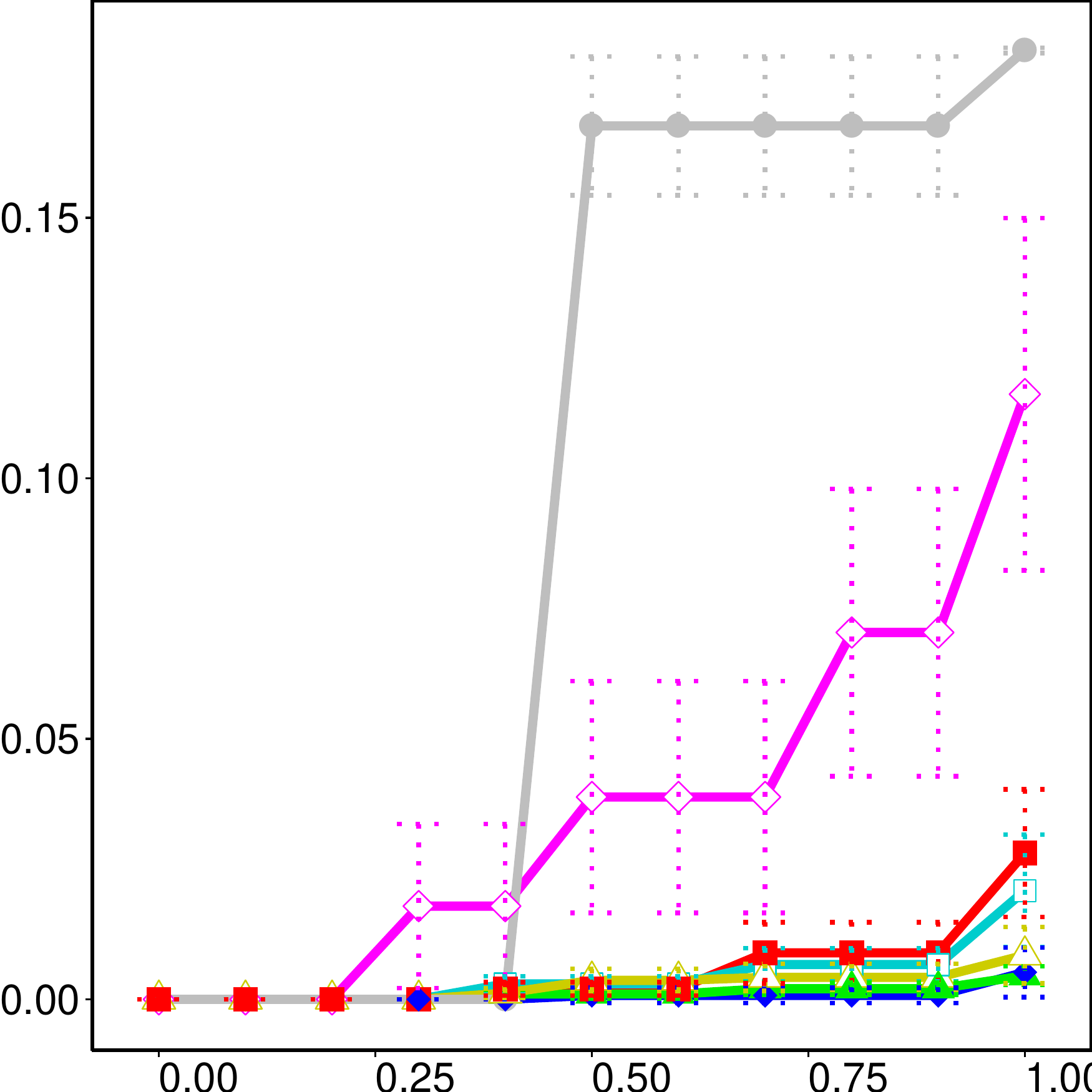} &
	\includegraphics[width=0.95\linewidth]{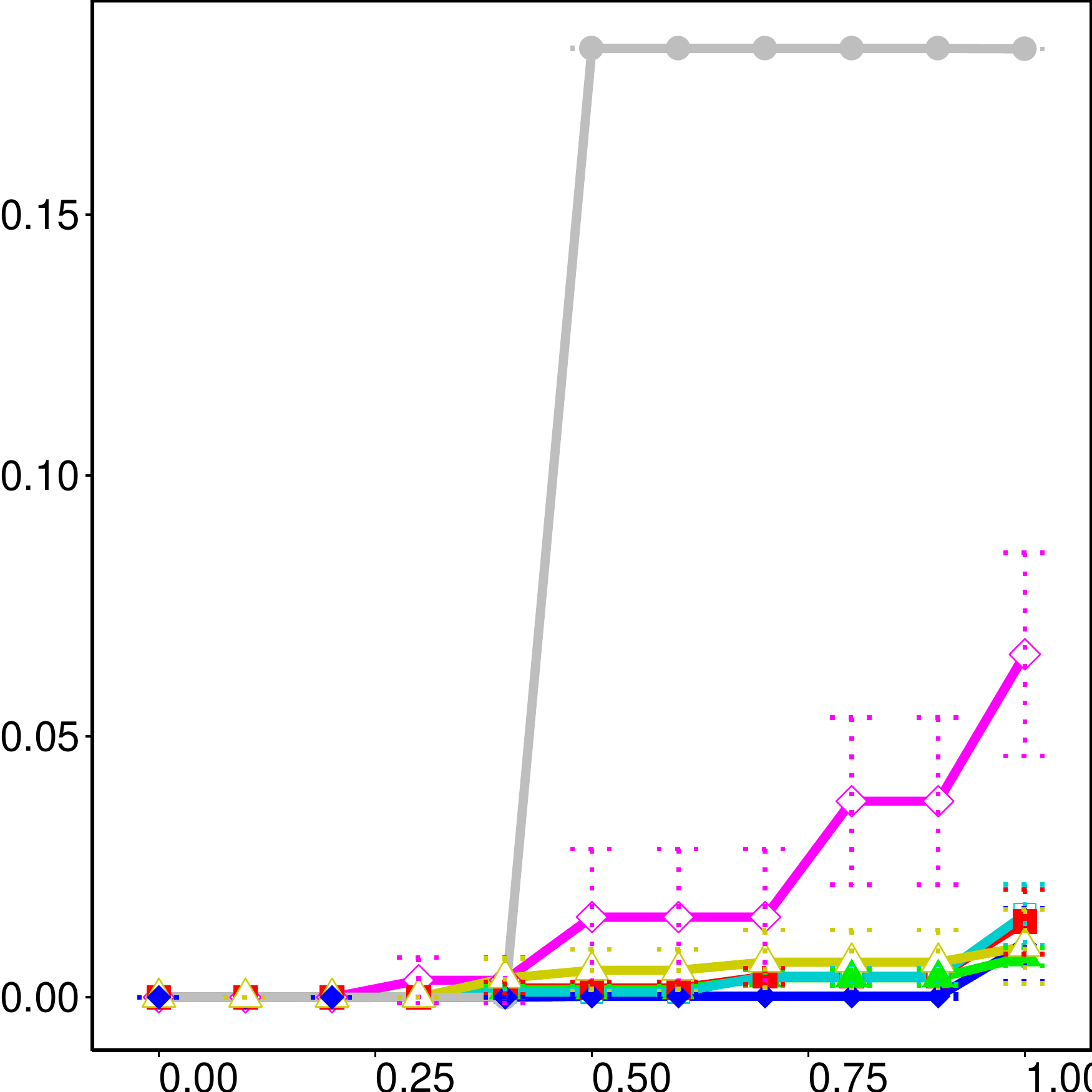} \\
	\rotatebox{90}{\small{Facebook (medium)}} &
	\includegraphics[width=0.95\linewidth]{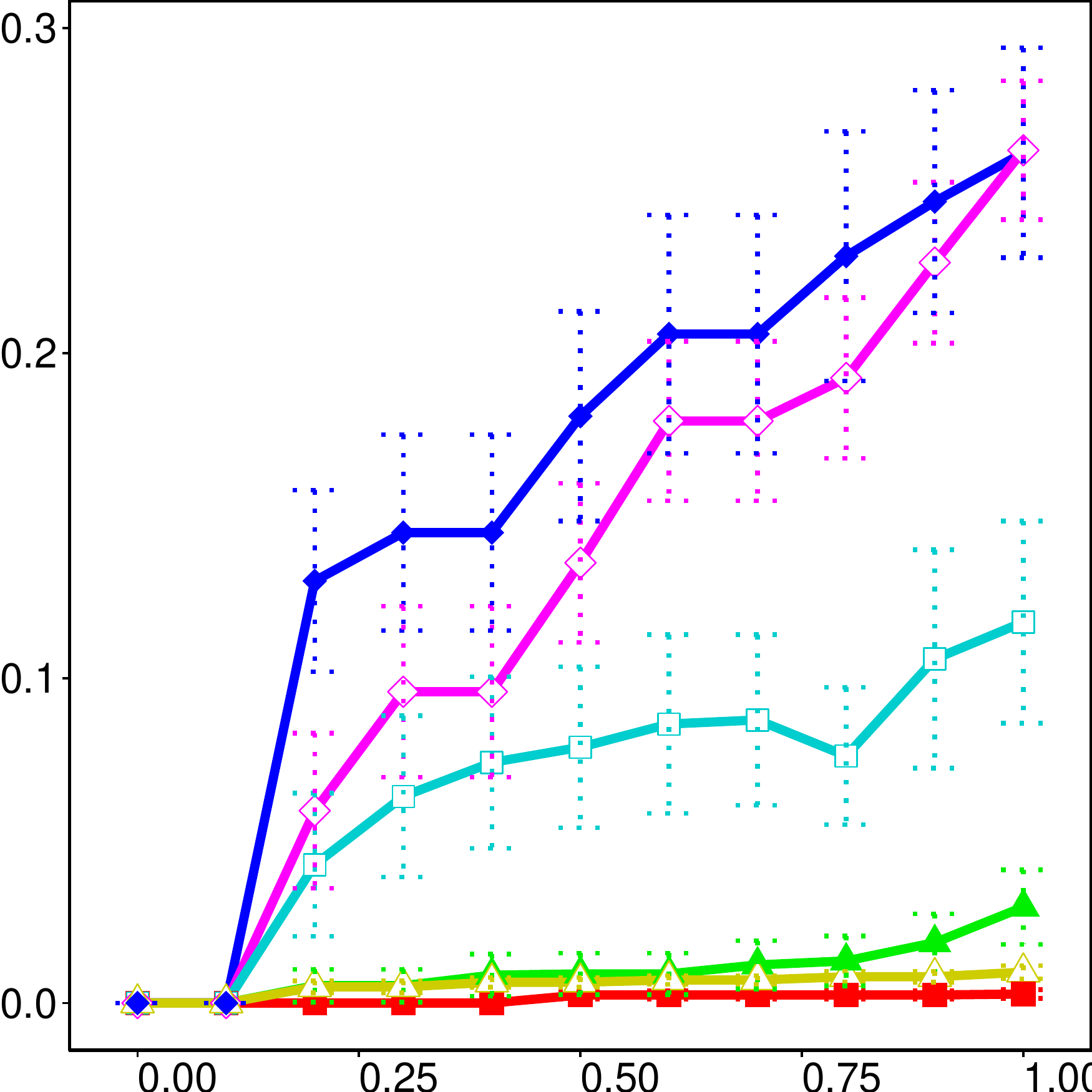} &
	\includegraphics[width=0.95\linewidth]{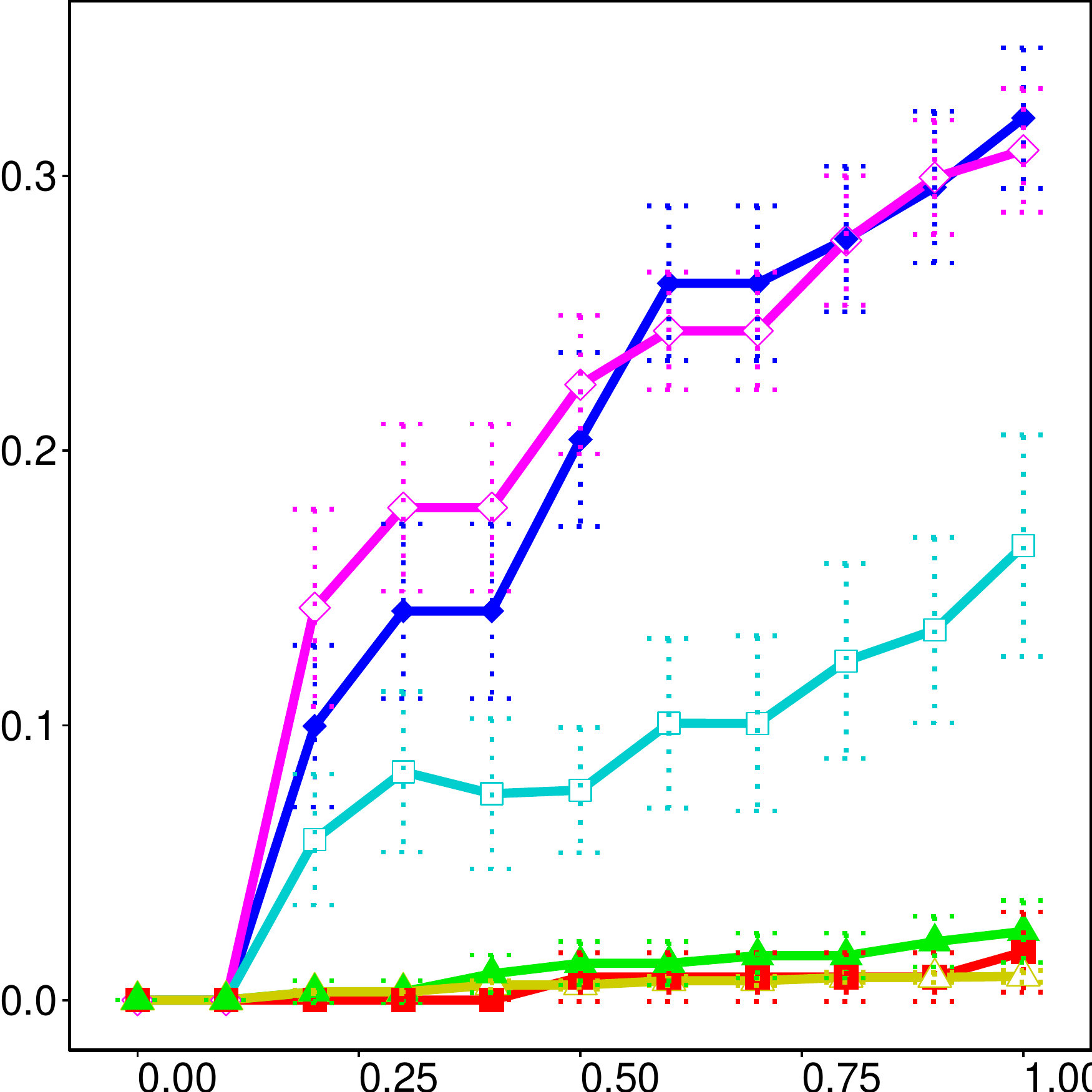} &
	\includegraphics[width=0.95\linewidth]{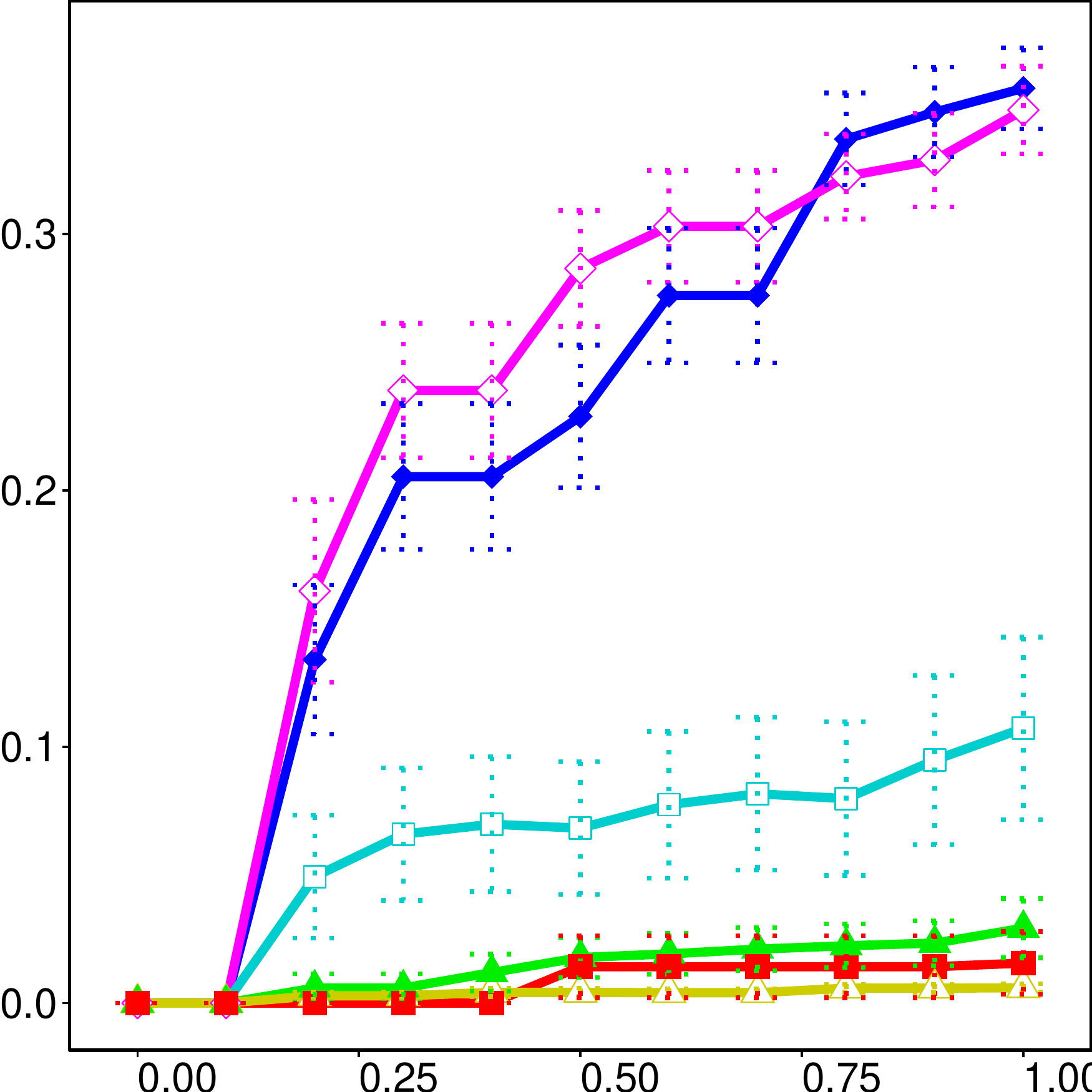} &
	\includegraphics[width=0.95\linewidth]{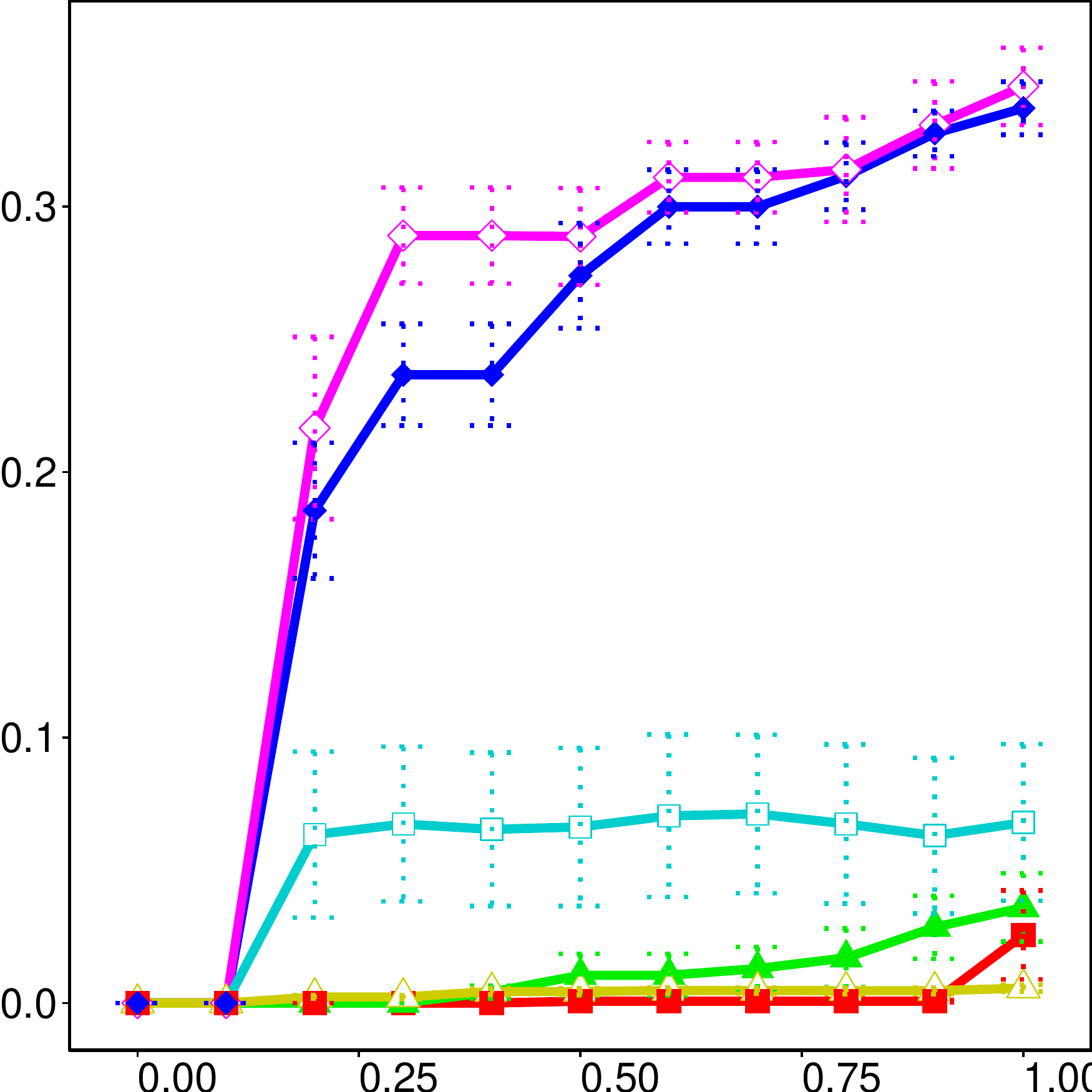} &
	\includegraphics[width=0.95\linewidth]{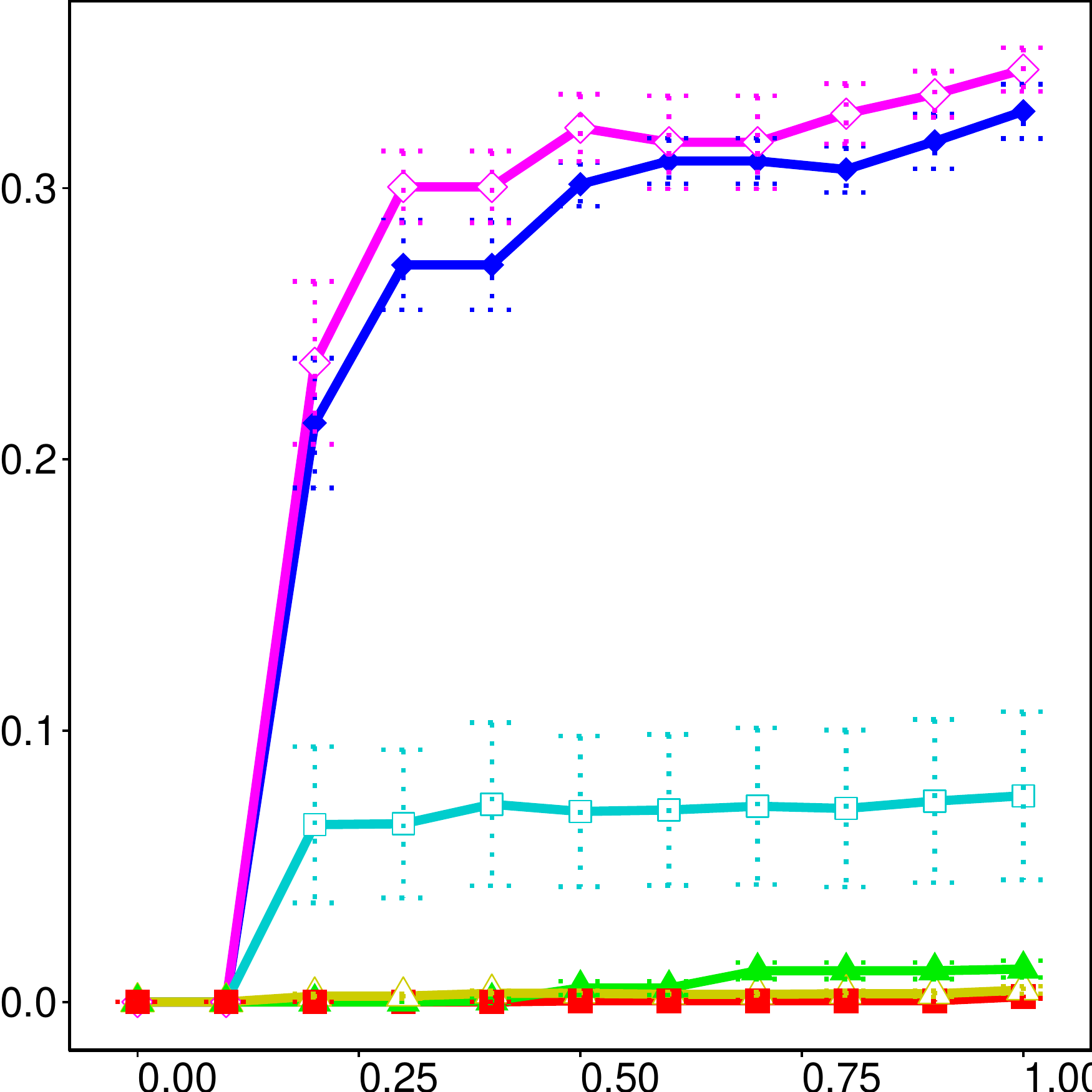} \\
	& \multicolumn{5}{c}{\includegraphics[width=0.75\linewidth]{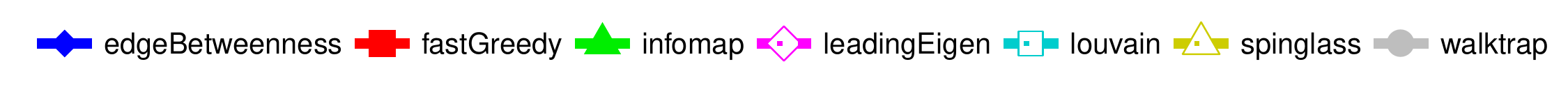}}
	\end{tabular}
	\caption{Executing DICE multiple, consecutive rounds (the $x$-axis represents the percentage of completed rounds) in undirected real-life networks.}
	\label{fig:community-plots-undirected-real}
\end{figure}
%%%%%%%%%%%%%%%%%%%%%%%%%%%%%%%%%%%%%%%%%%%%%%%%%%%%%%%%%%%%%%%%%%%%%%%%%%%%%%%

%%%%%%%%%%%%%%%%%%%%%%%%%%%%%%%%%%%%%%%%%%%%%%%%%%%%%%%%%%%%%%%%%%%%%%%%%%%%%%%
\begin{figure}[p!]
	\centering
	\setlength\tabcolsep{2pt}
	\renewcommand{\arraystretch}{2.0}
	\begin{tabular}{m{.01\textwidth}m{.195\textwidth}m{.195\textwidth}m{.195\textwidth}m{.195\textwidth}m{.195\textwidth}}
	&
	\multicolumn{1}{c}{\small{$b=4$, $d=0$}} &
	\multicolumn{1}{c}{\small{$b=4$, $d=1$}} &
	\multicolumn{1}{c}{\small{$b=4$, $d=2$}} &
	\multicolumn{1}{c}{\small{$b=4$, $d=3$}} &
	\multicolumn{1}{c}{\small{$b=4$, $d=4$}} \\
	\rotatebox{90}{\small{Twitter (small)}} &
	\includegraphics[width=0.95\linewidth]{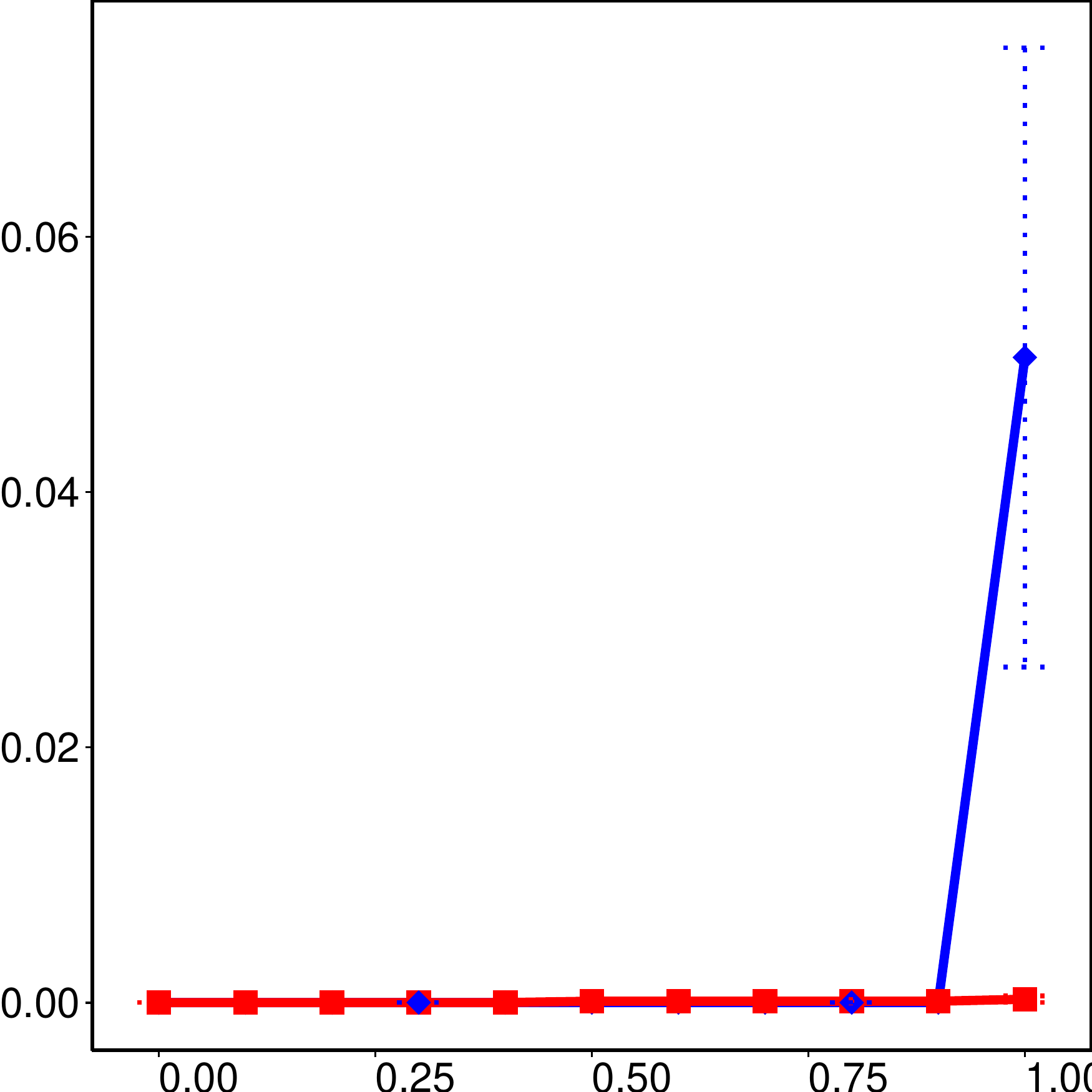} &
	\includegraphics[width=0.95\linewidth]{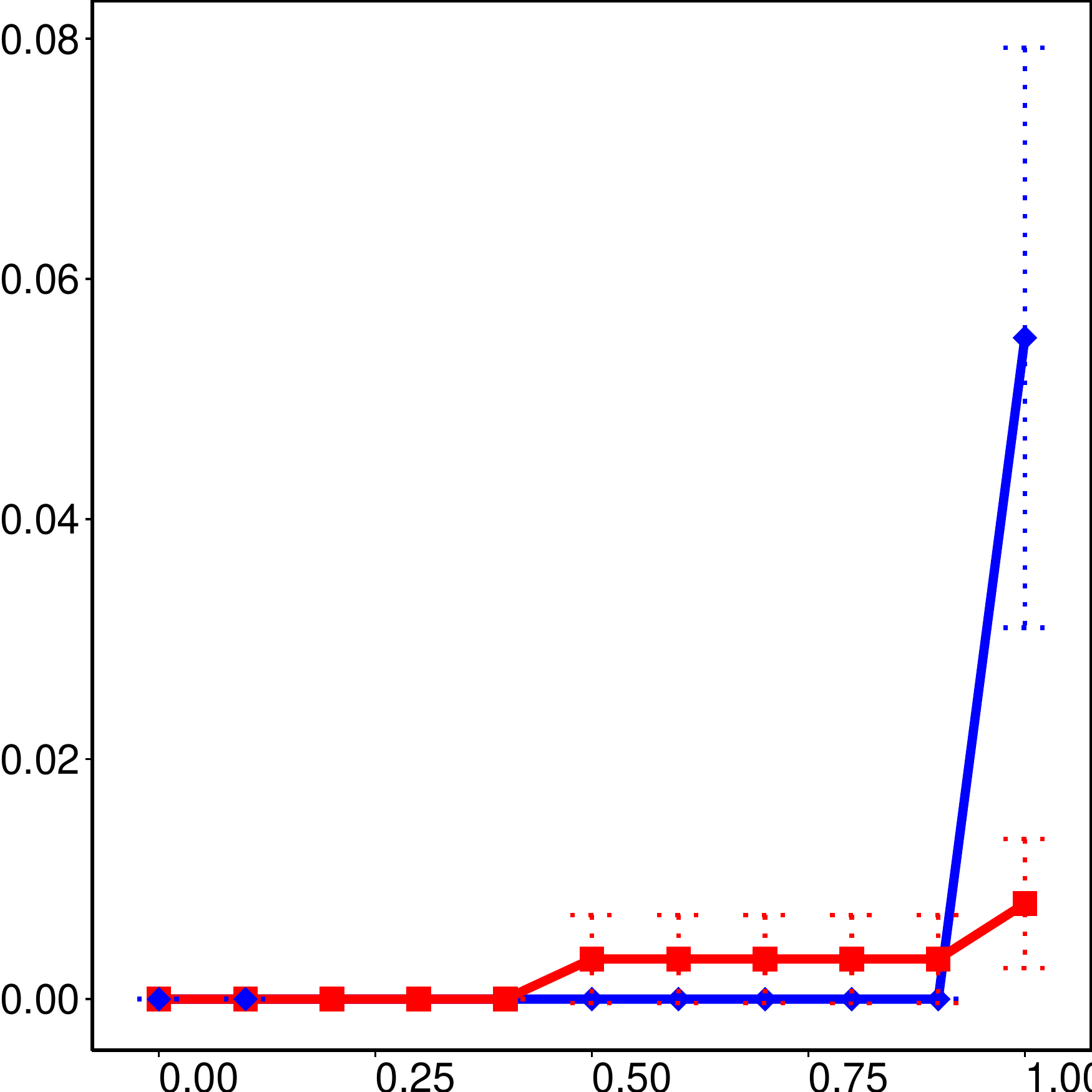} &
	\includegraphics[width=0.95\linewidth]{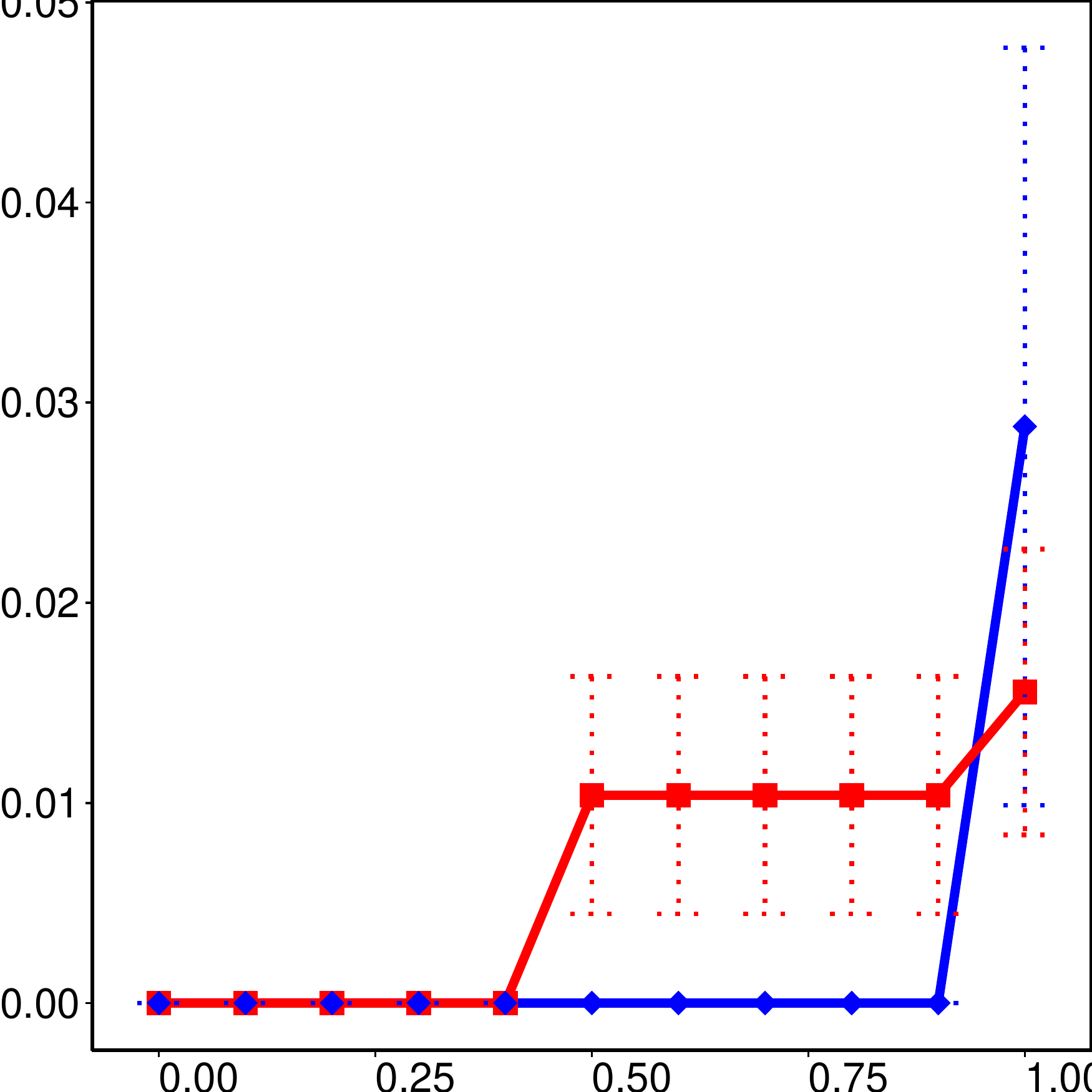} &
	\includegraphics[width=0.95\linewidth]{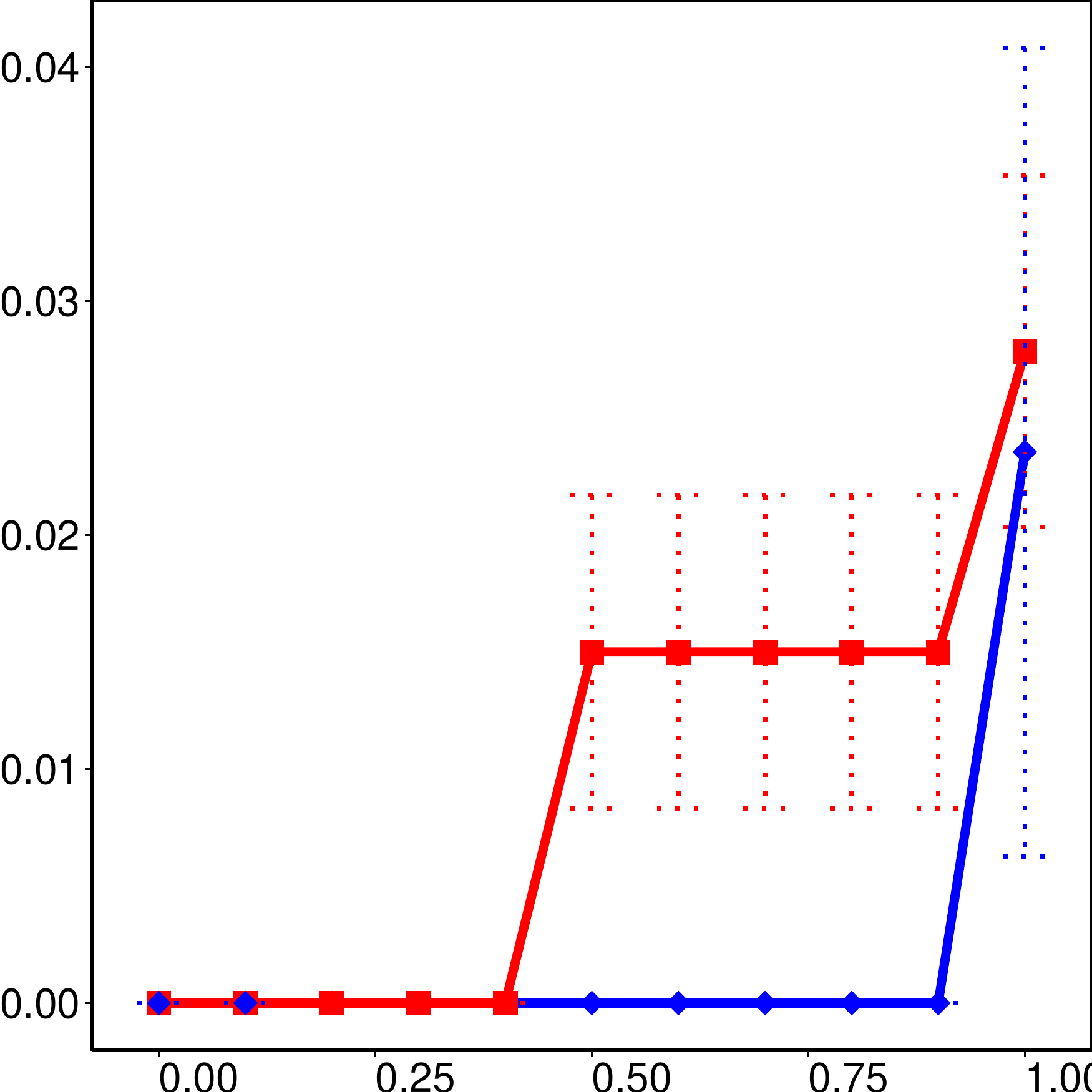} &
	\includegraphics[width=0.95\linewidth]{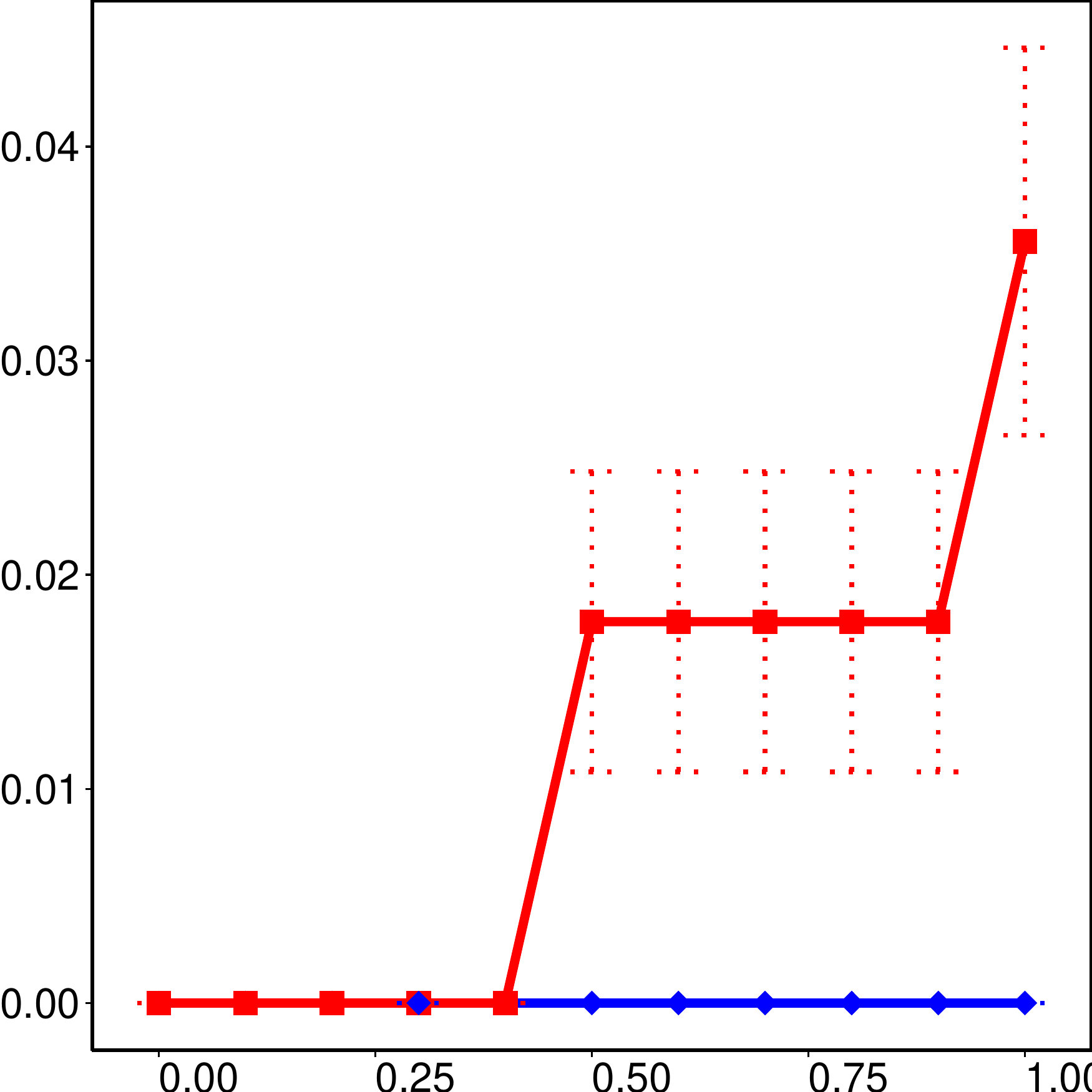} \\
	\rotatebox{90}{\small{Twitter (medium)}} &
	\includegraphics[width=0.95\linewidth]{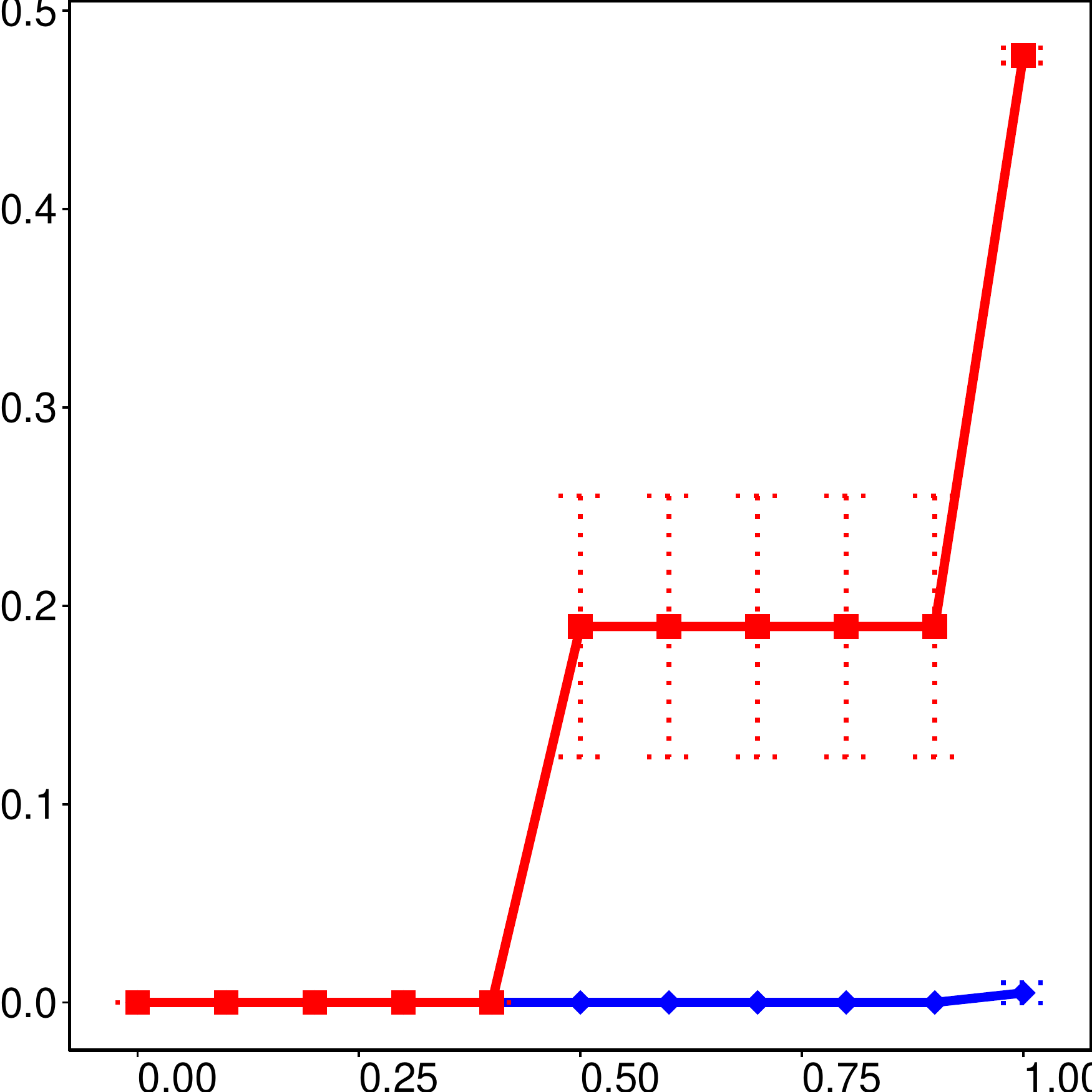} &
	\includegraphics[width=0.95\linewidth]{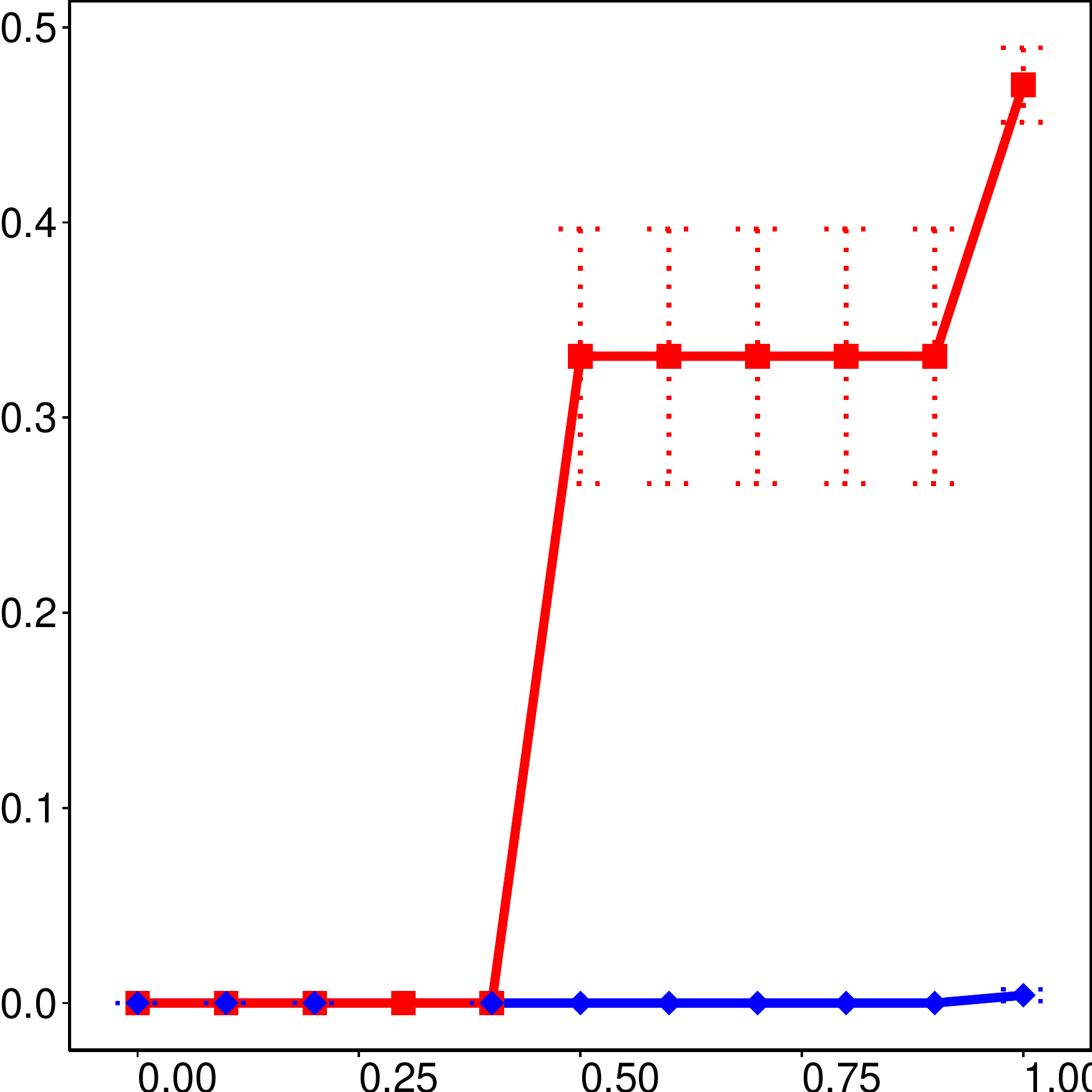} &
	\includegraphics[width=0.95\linewidth]{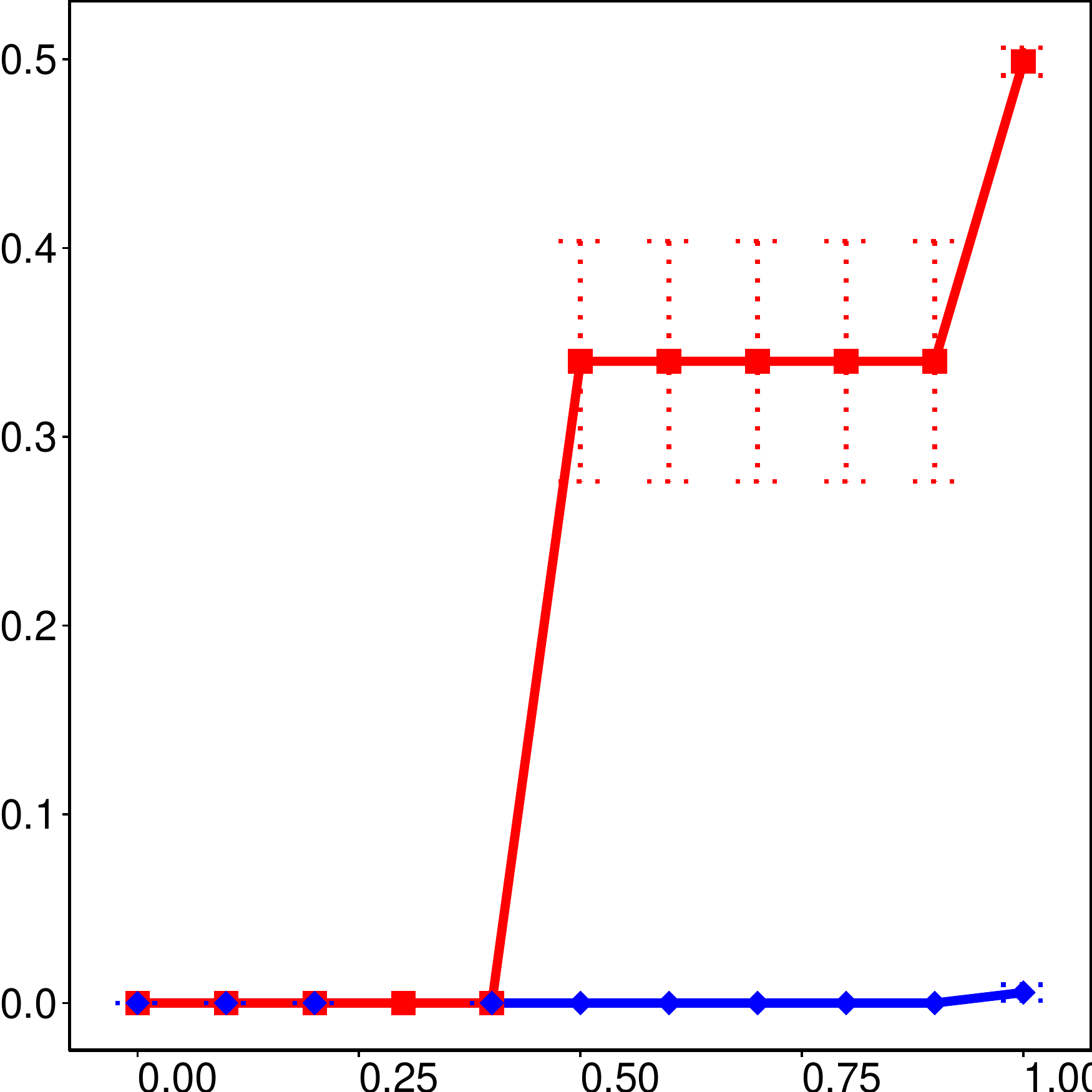} &
	\includegraphics[width=0.95\linewidth]{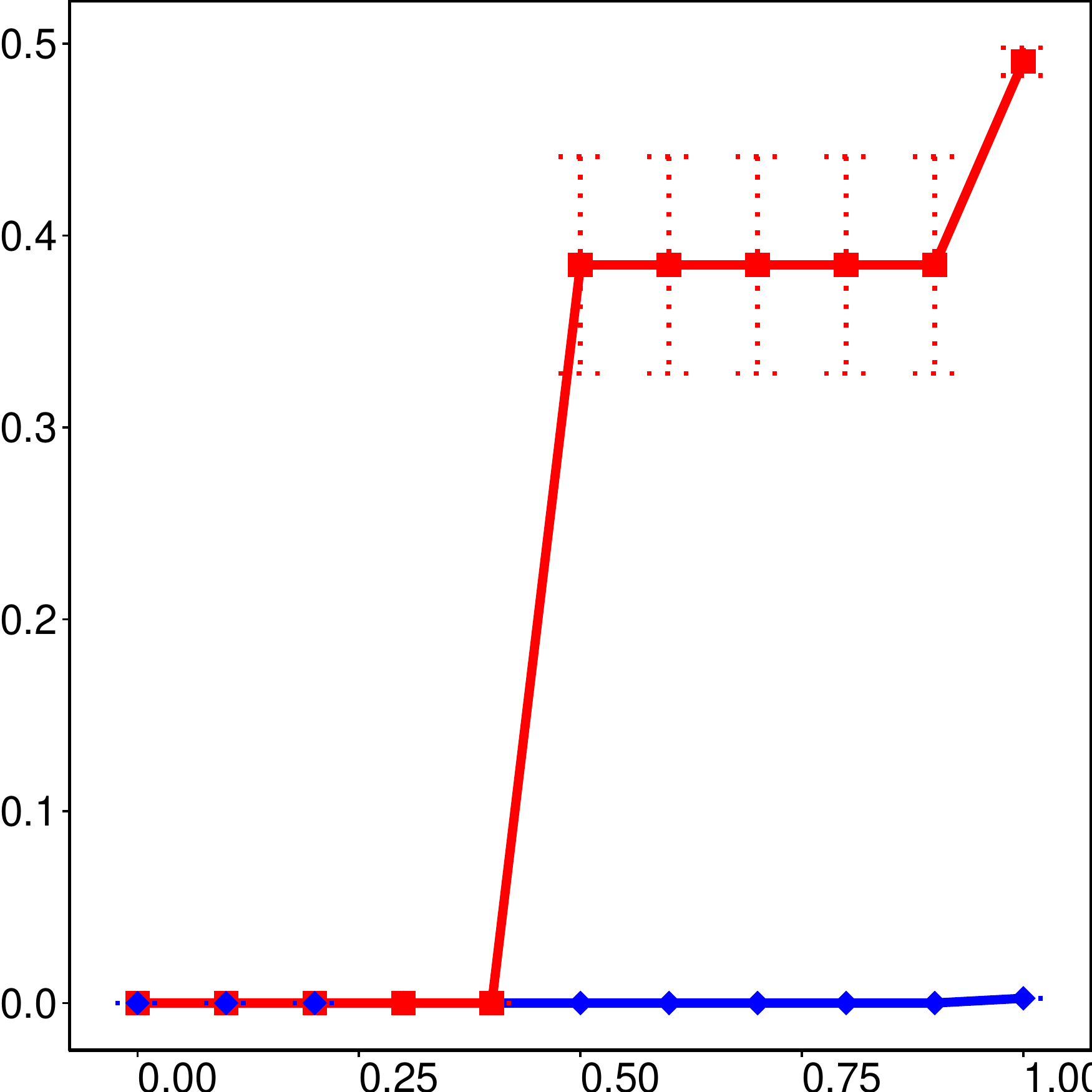} &
	\includegraphics[width=0.95\linewidth]{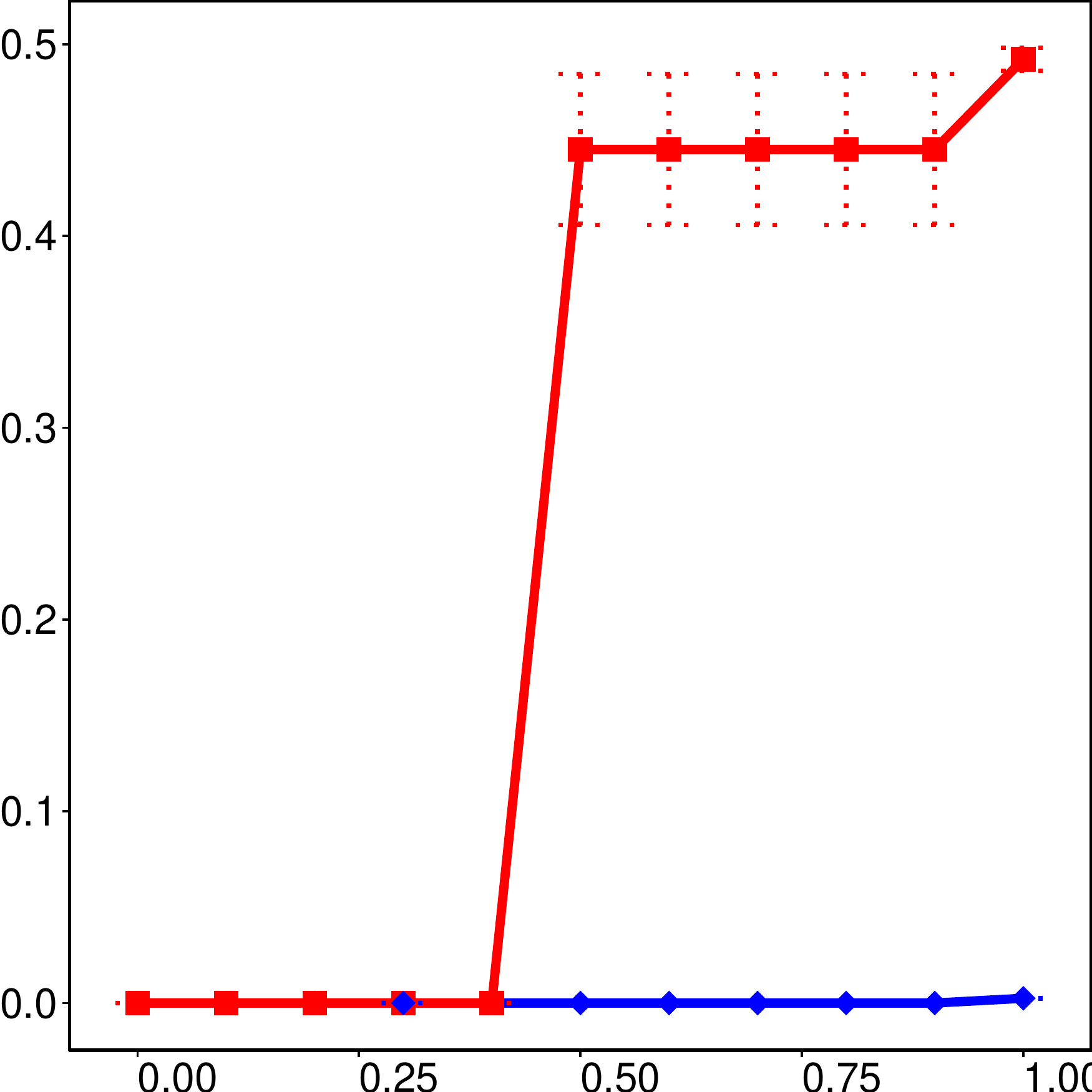} \\
	\rotatebox{90}{\small{Google+ (small)}} &
	\includegraphics[width=0.95\linewidth]{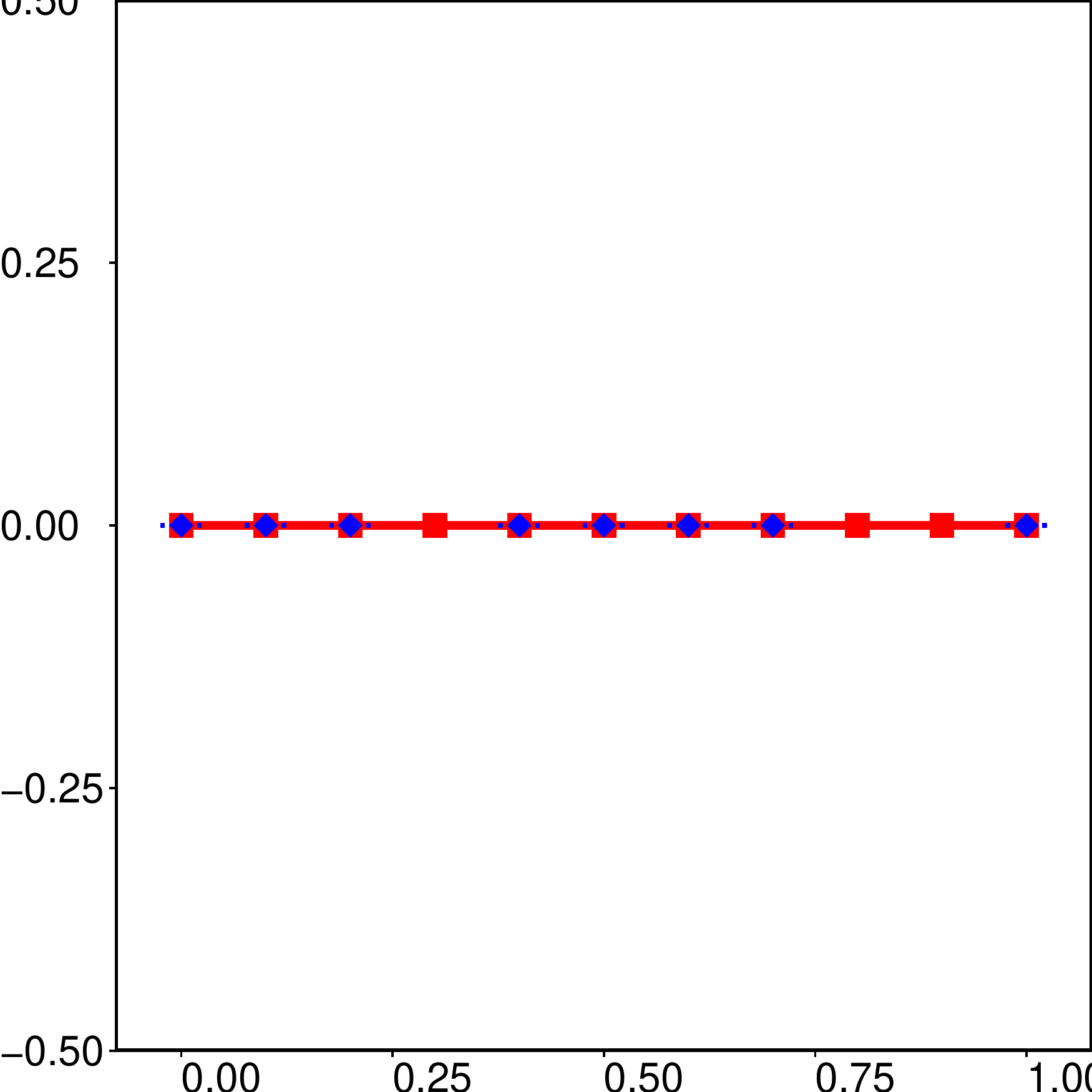} &
	\includegraphics[width=0.95\linewidth]{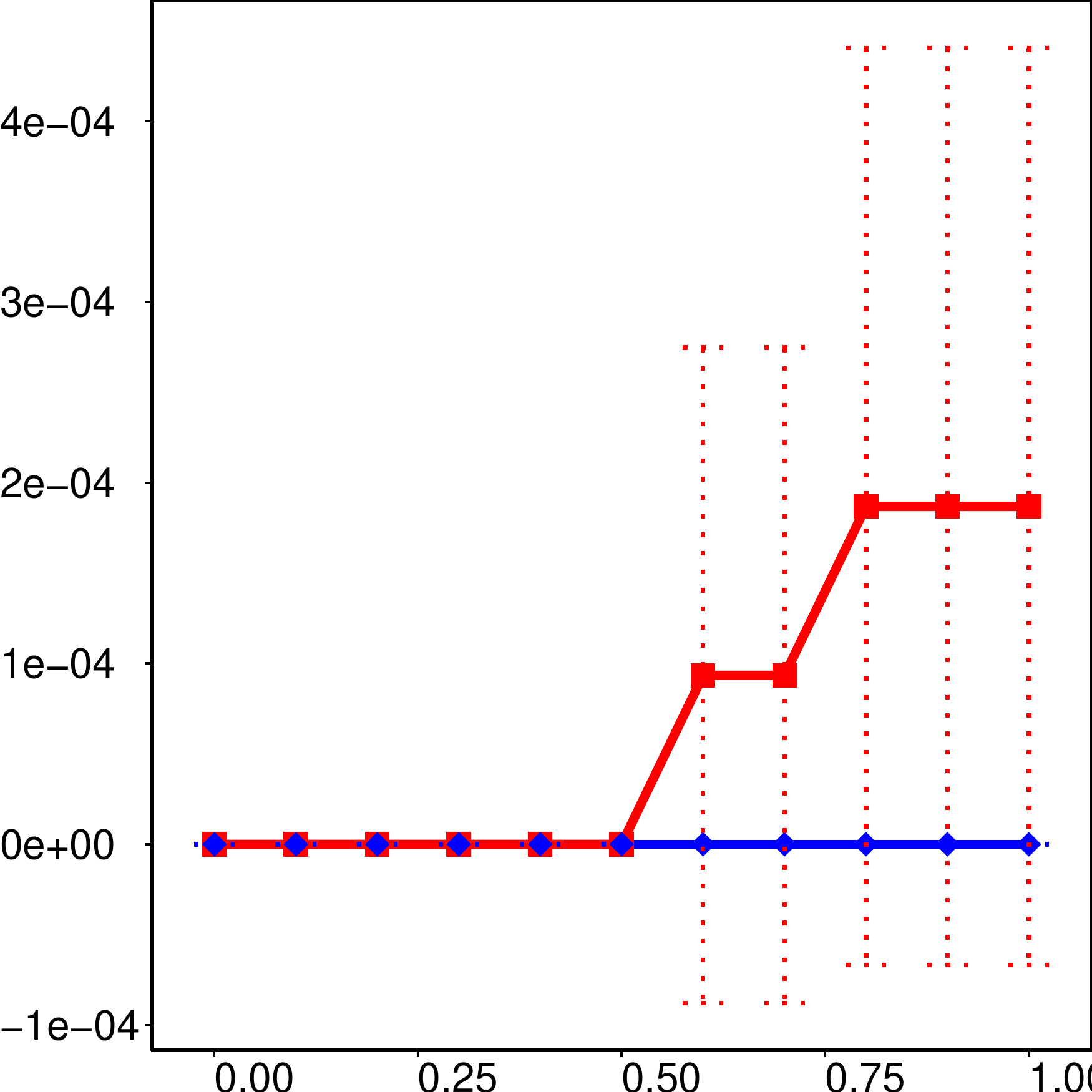} &
	\includegraphics[width=0.95\linewidth]{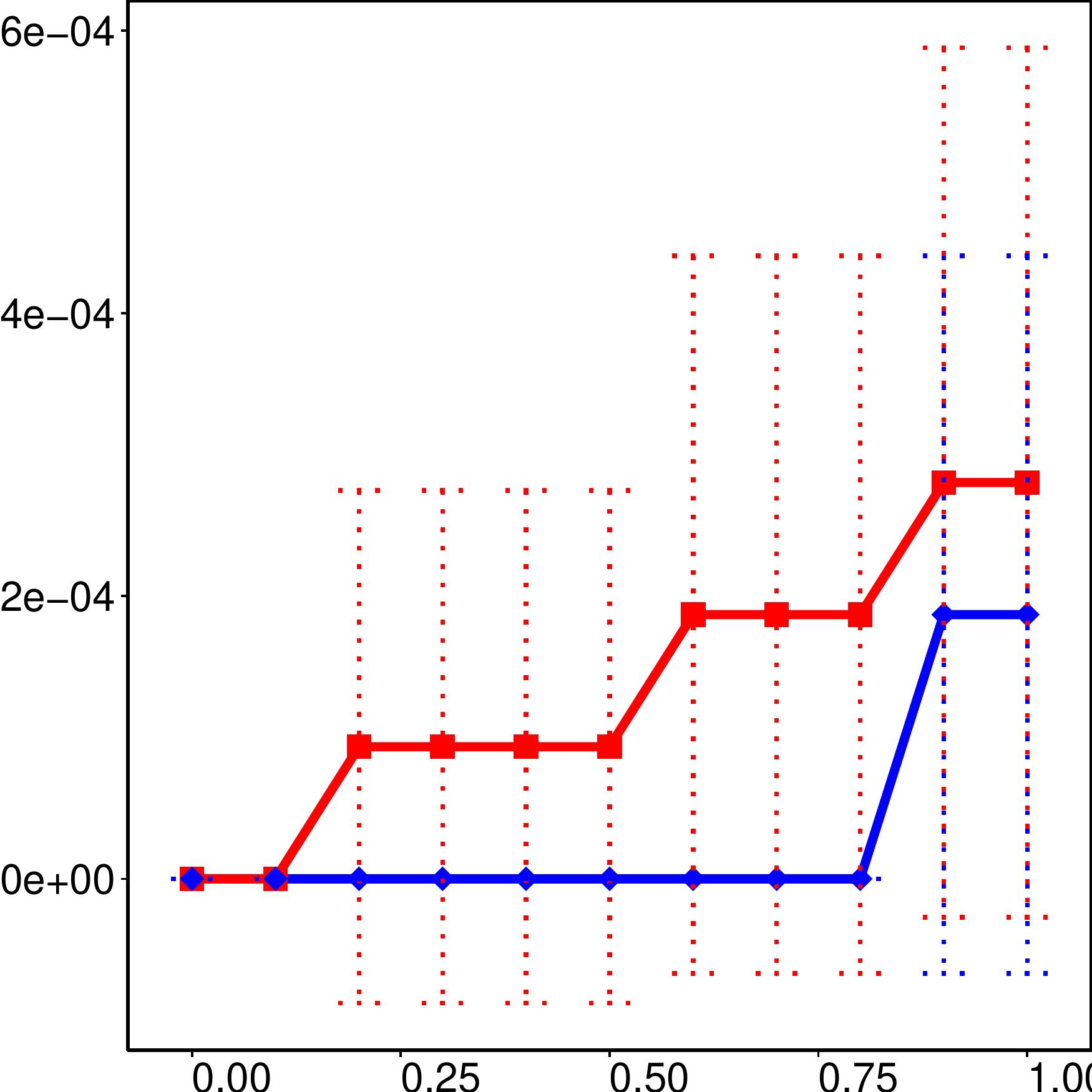} &
	\includegraphics[width=0.95\linewidth]{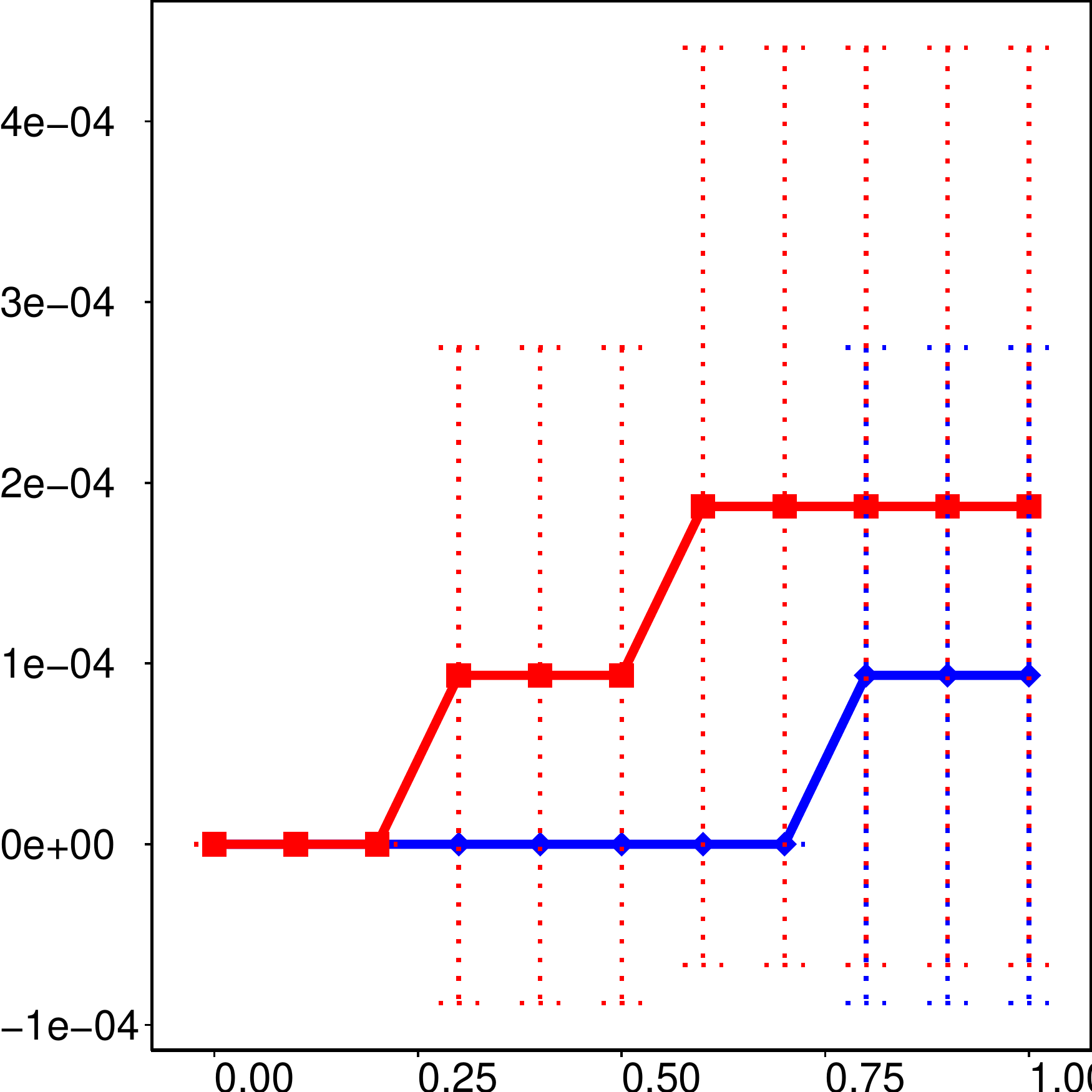} &
	\includegraphics[width=0.95\linewidth]{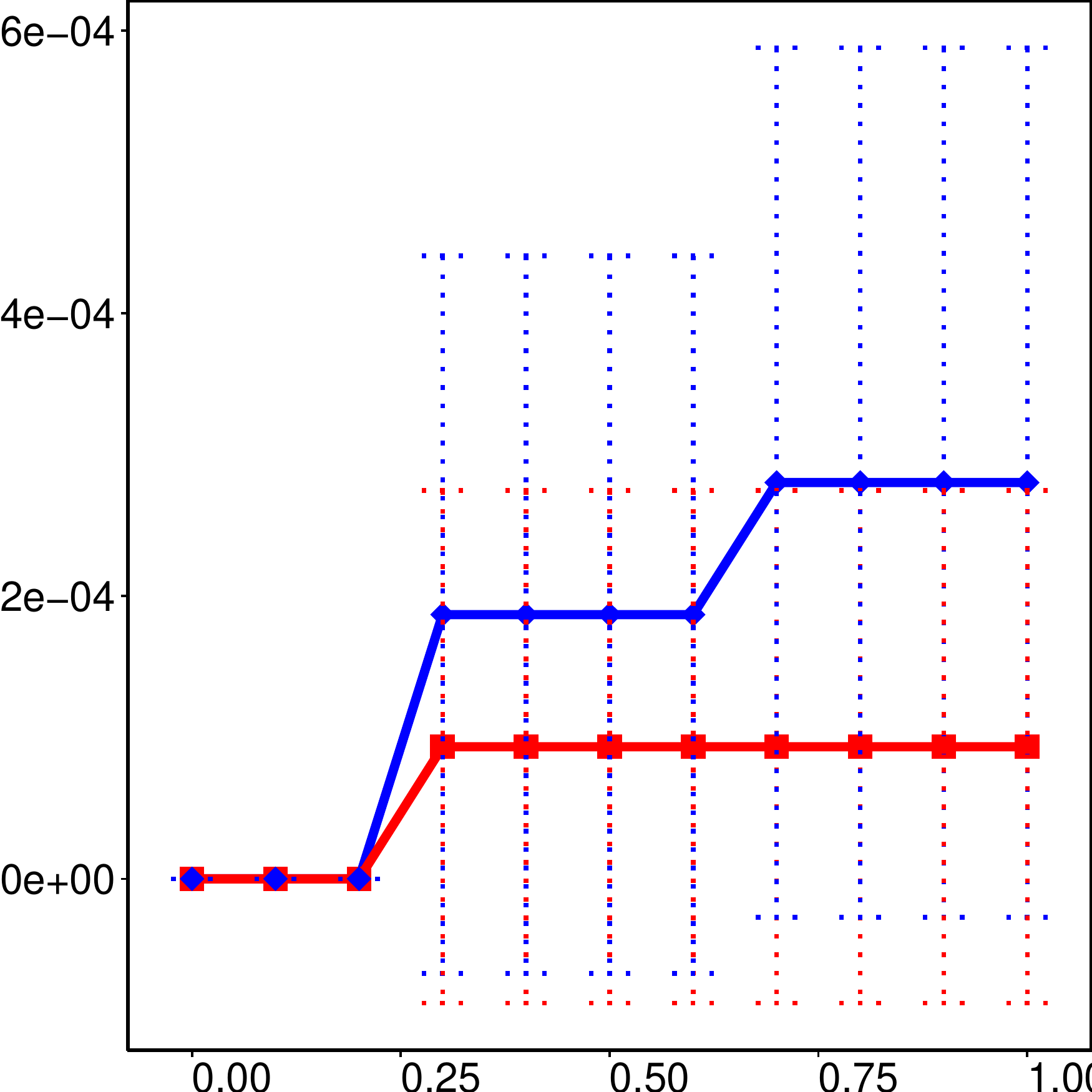} \\
	\rotatebox{90}{\small{Google+ (medium)}} &
	\includegraphics[width=0.95\linewidth]{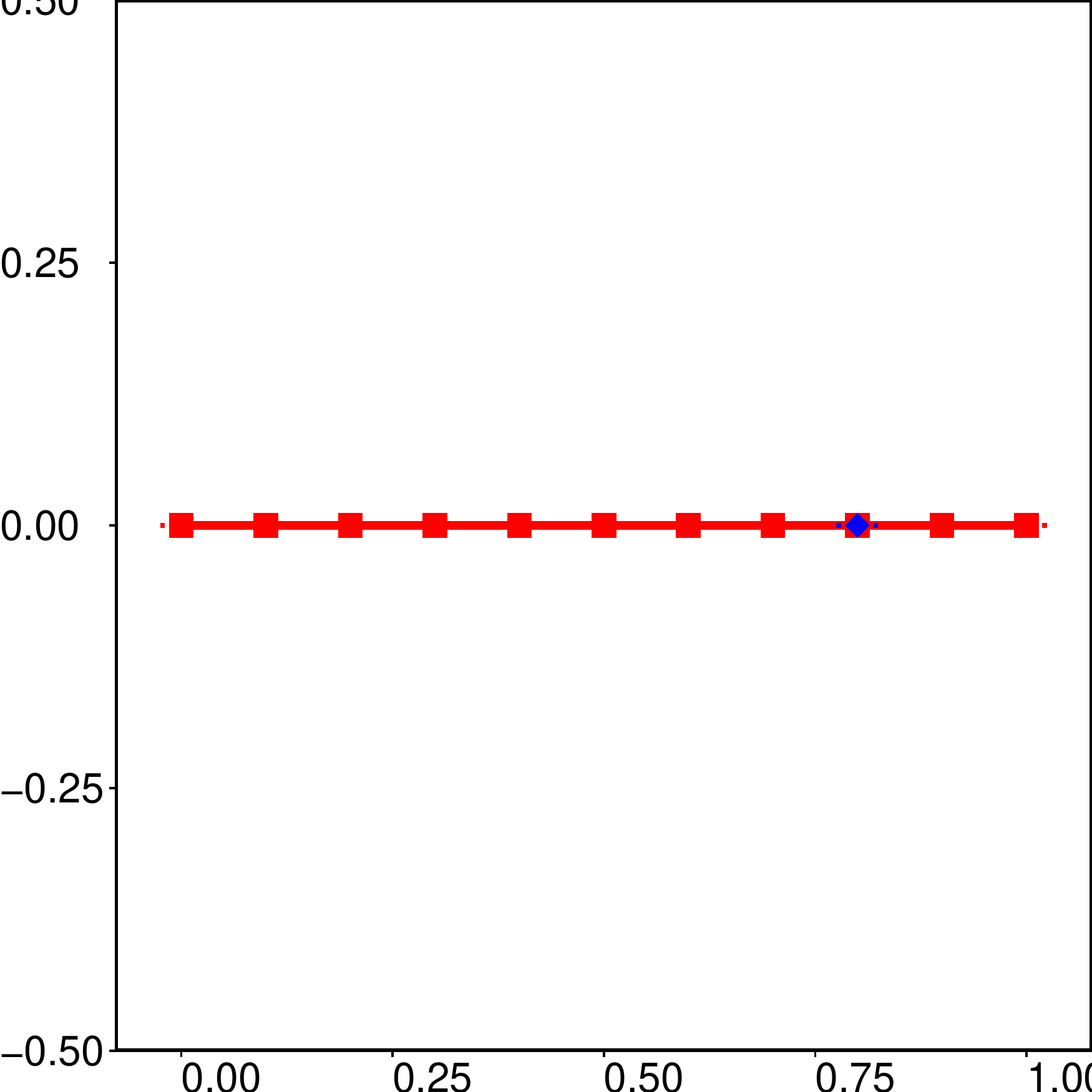} &
	\includegraphics[width=0.95\linewidth]{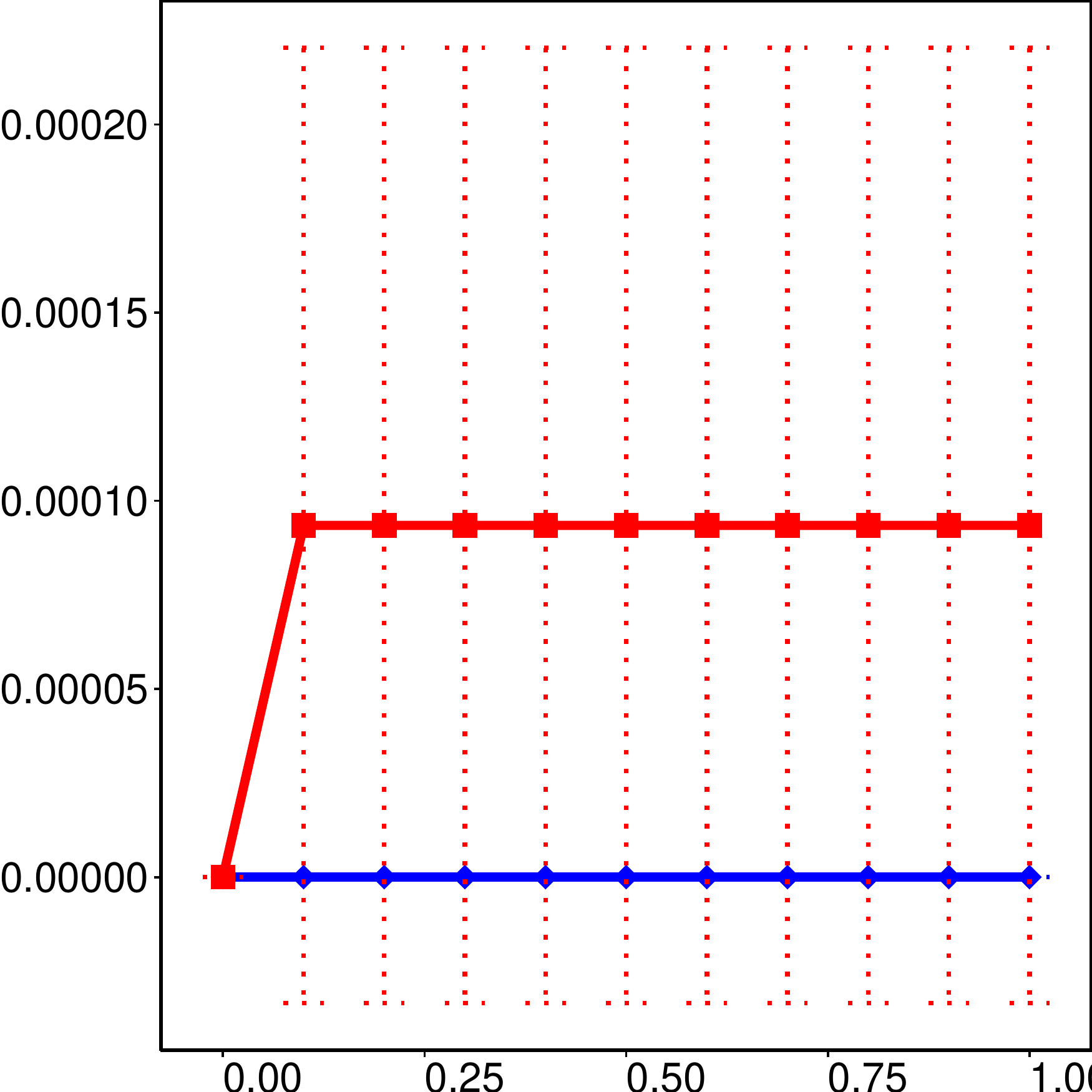} &
	\includegraphics[width=0.95\linewidth]{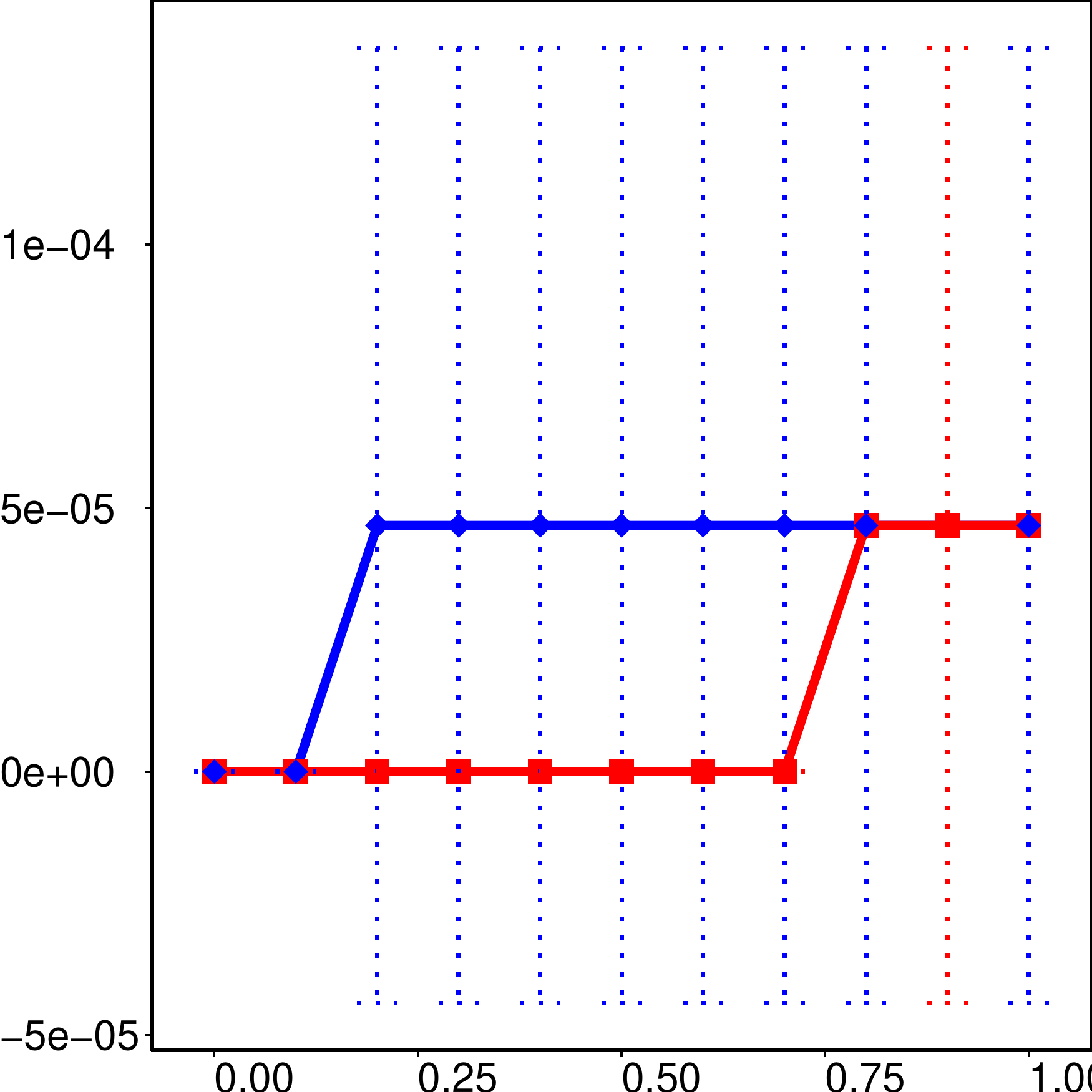} &
	\includegraphics[width=0.95\linewidth]{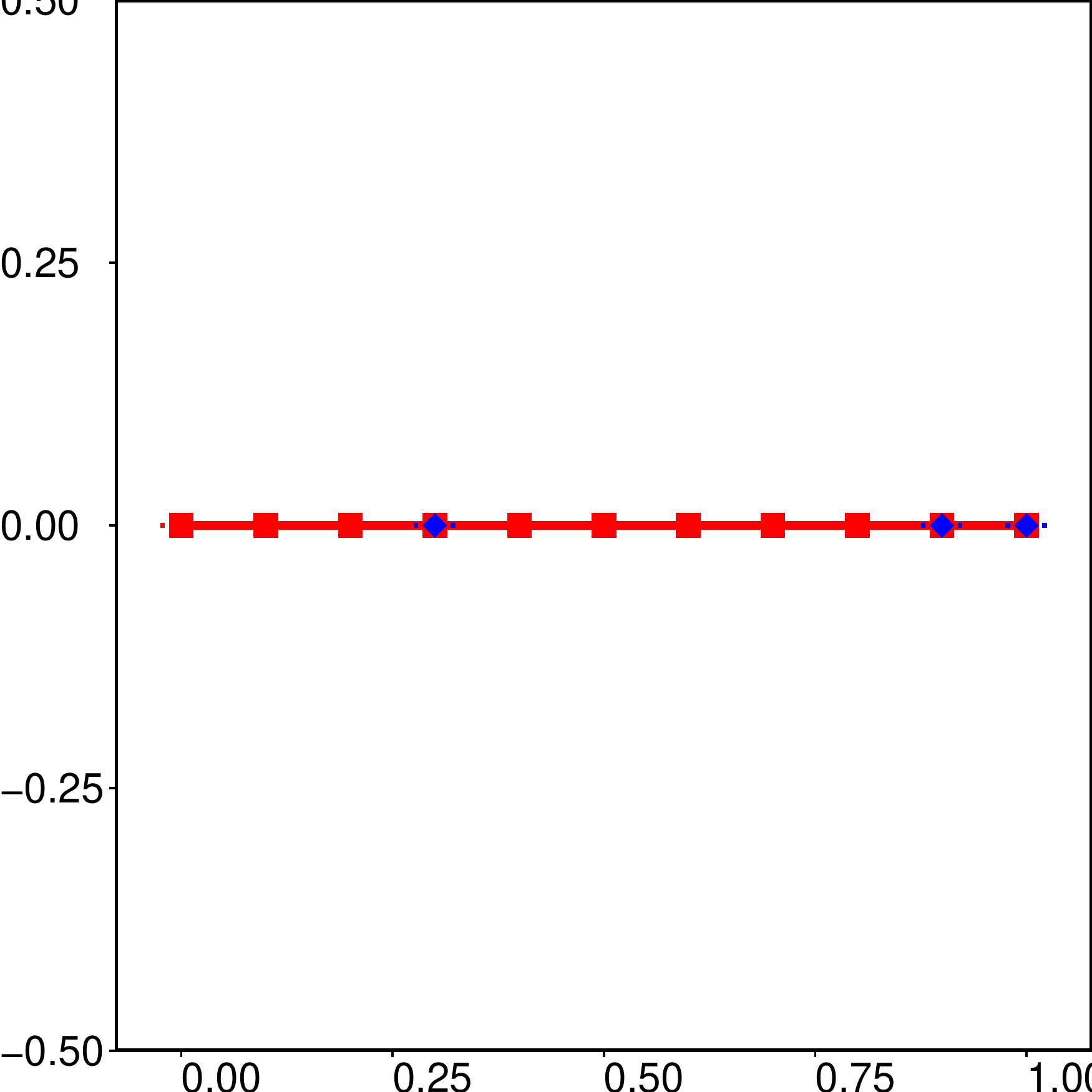} &
	\includegraphics[width=0.95\linewidth]{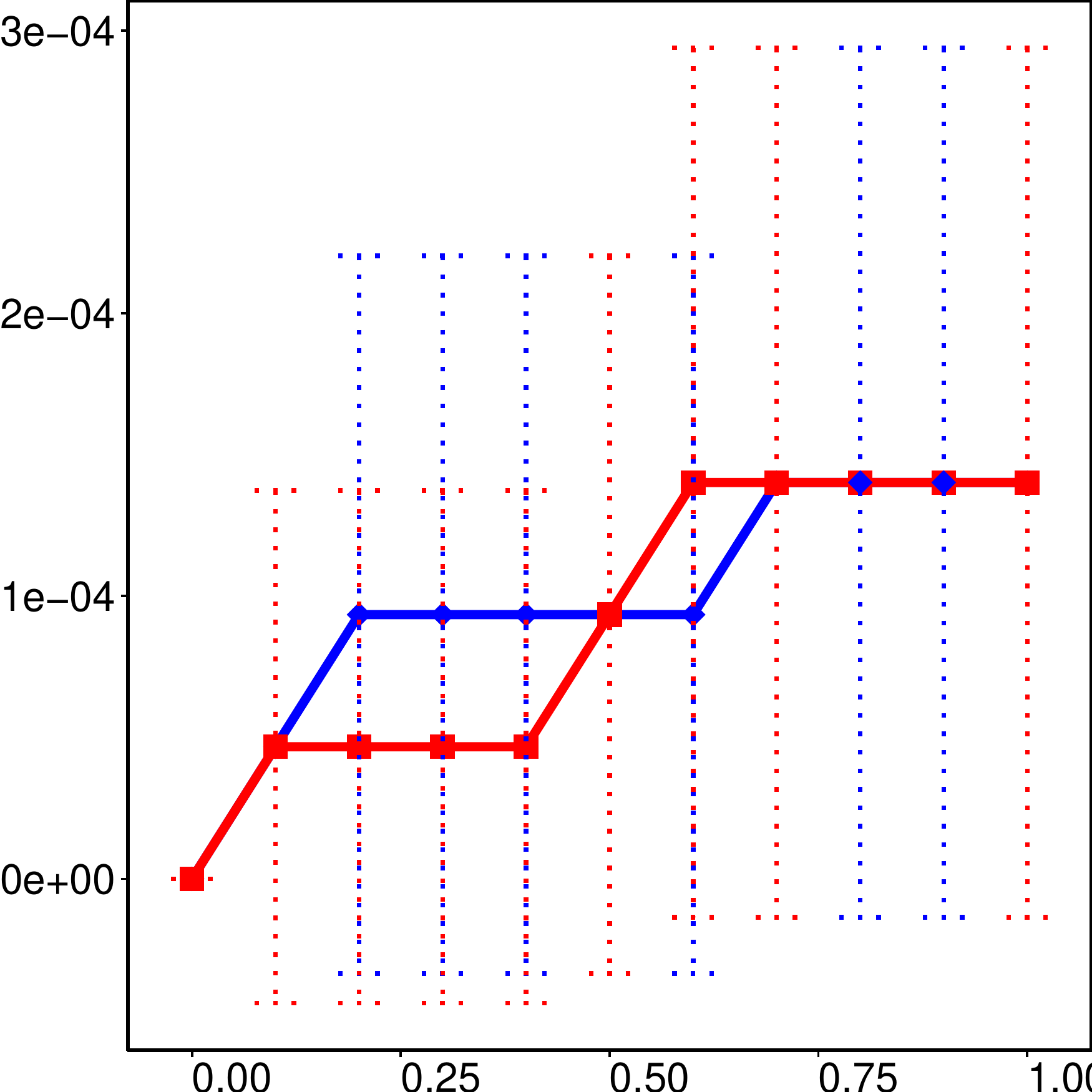} \\
	& \multicolumn{5}{c}{\includegraphics[width=0.75\linewidth]{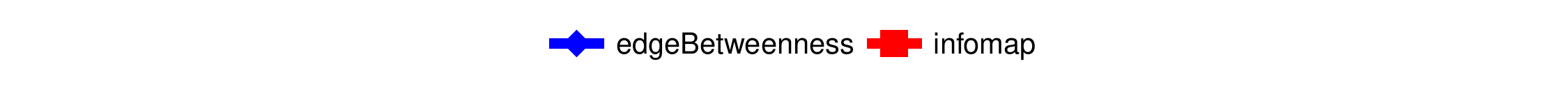}}
	\end{tabular}
	\caption{Executing DICE multiple, consecutive rounds (the $x$-axis represents the percentage of completed rounds) in directed real-life networks.}
	\label{fig:community-plots-directed}
\end{figure}
%%%%%%%%%%%%%%%%%%%%%%%%%%%%%%%%%%%%%%%%%%%%%%%%%%%%%%%%%%%%%%%%%%%%%%%%%%%%%%%

%%%%%%%%%%%%%%%%%%%%%%%%%%%%%%%%%%%%%%%%%%%%%%%%%%%%%%%%%%%%%%%%%%%%%%%%%%%%%%%
\begin{figure}[p!]
	\centering
	\setlength\tabcolsep{2pt}
	\renewcommand{\arraystretch}{2.0}
%	\begin{tabular}{m{.195\textwidth}m{.195\textwidth}m{.195\textwidth}m{.195\textwidth}m{.195\textwidth}}
	\begin{tabular}{m{.33\textwidth}m{.33\textwidth}m{.33\textwidth}}
	\multicolumn{1}{c}{\small{($b=4$, $d=0$)}} &
%	\multicolumn{1}{c}{\small{($b=4$, $d=1$)}} &
	\multicolumn{1}{c}{\small{($b=4$, $d=2$)}} &
%	\multicolumn{1}{c}{\small{($b=4$, $d=3$)}} &
	\multicolumn{1}{c}{\small{($b=4$, $d=4$)}} \\
	\vspace*{-0.25cm}
	\includegraphics[width=0.88\linewidth]{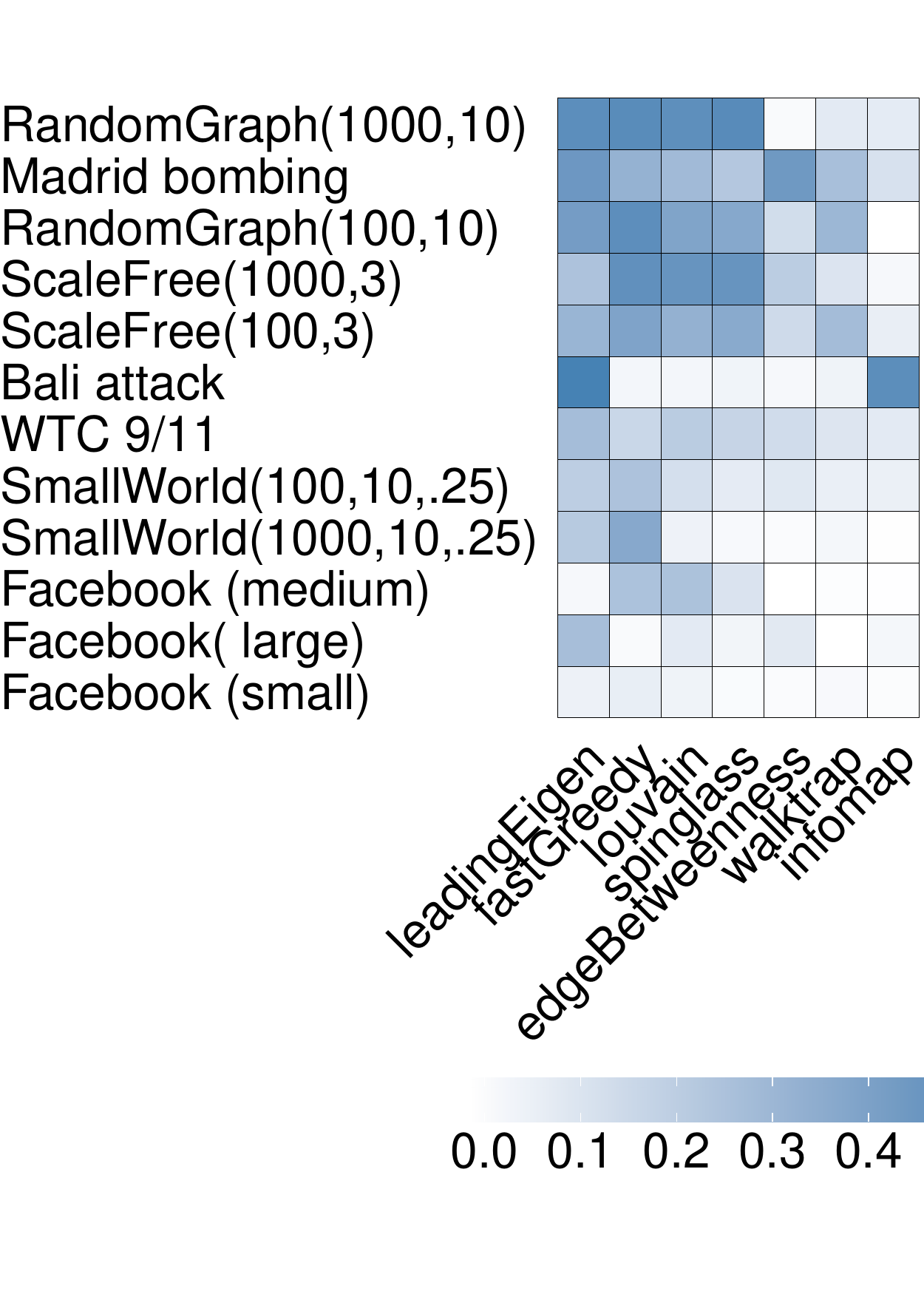} &
%	\vspace*{-0.25cm}
%	\includegraphics[width=0.88\linewidth]{figures/plots/community/undirected/DICE_1,3_-heatmap.pdf} &
	\vspace*{-0.25cm}
	\includegraphics[width=0.88\linewidth]{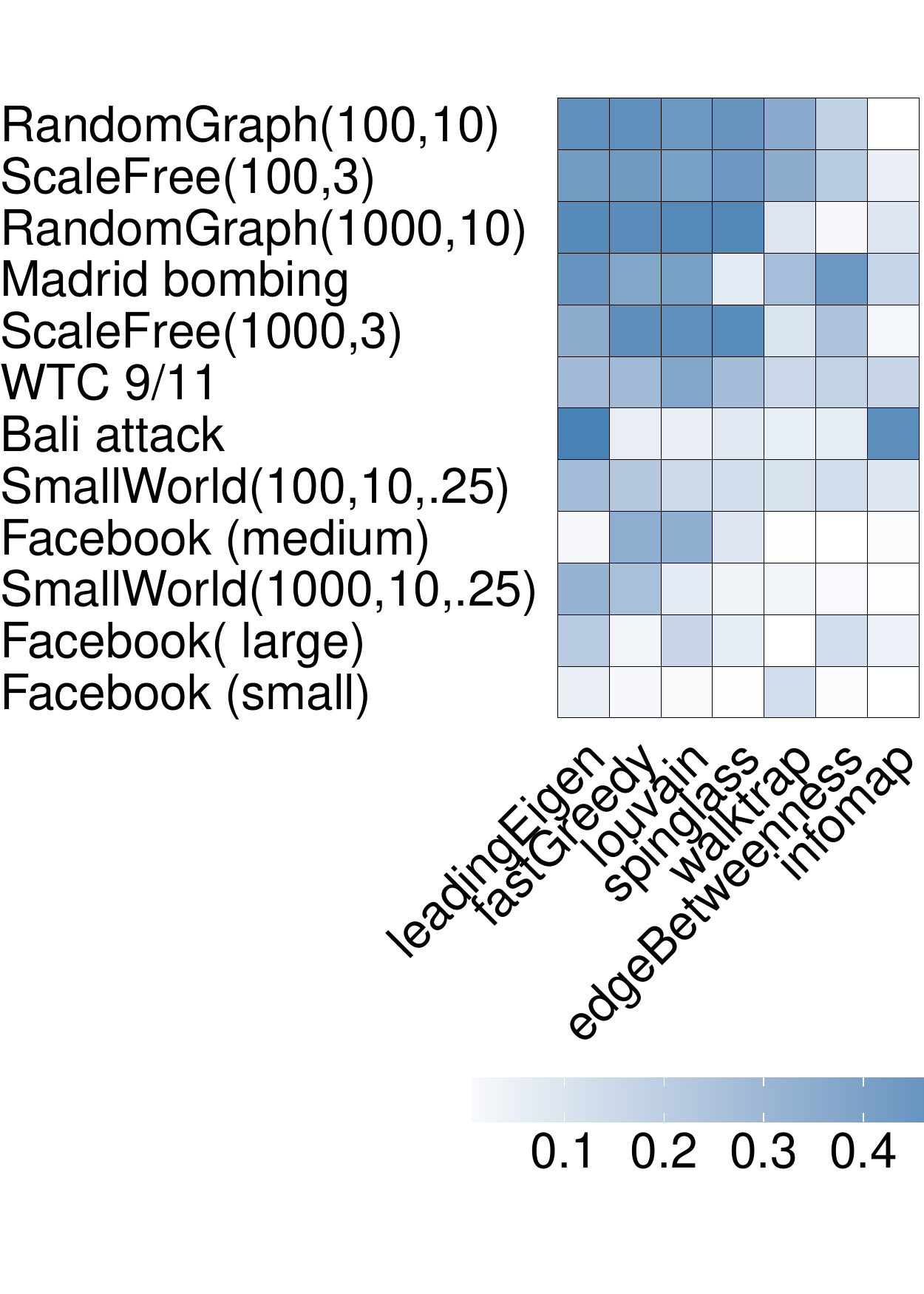} &
%	\vspace*{-0.25cm}
%	\includegraphics[width=0.88\linewidth]{figures/plots/community/undirected/DICE_3,1_-heatmap.pdf} &
	\vspace*{-0.25cm}	
	\includegraphics[width=0.88\linewidth]{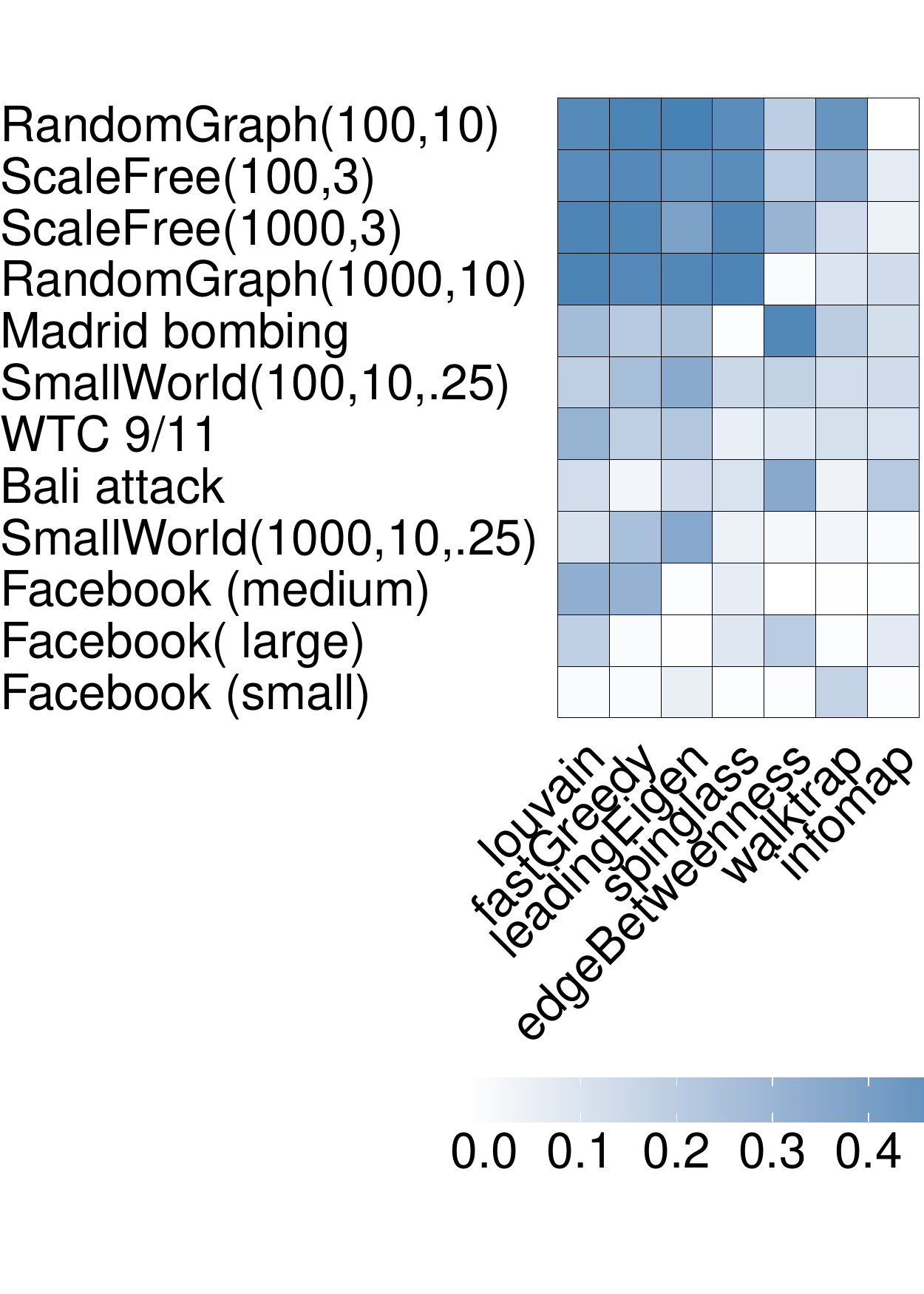} \\
	\end{tabular}
%	\begin{tabular}{m{.195\textwidth}m{.195\textwidth}m{.195\textwidth}m{.195\textwidth}m{.195\textwidth}}
	\begin{tabular}{m{.28\textwidth}m{.28\textwidth}m{.28\textwidth}}
	\multicolumn{1}{c}{\small{($b=4$, $d=0$)}} &
%	\multicolumn{1}{c}{\small{($b=4$, $d=1$)}} &
	\multicolumn{1}{c}{\small{($b=4$, $d=2$)}} &
%	\multicolumn{1}{c}{\small{($b=4$, $d=3$)}} &
	\multicolumn{1}{c}{\small{($b=4$, $d=4$)}} \\
	\includegraphics[width=0.95\linewidth]{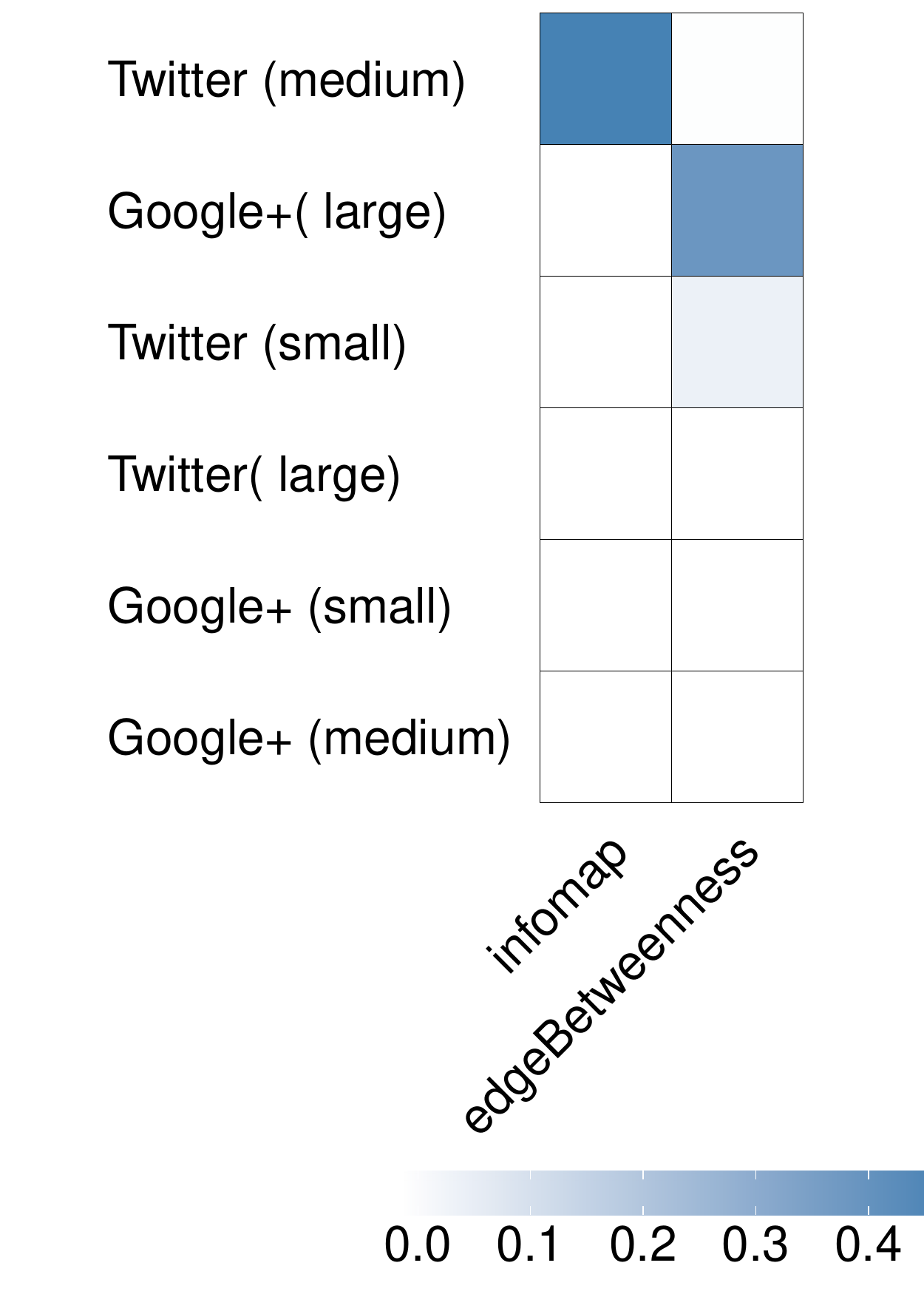} &
	\includegraphics[width=0.95\linewidth]{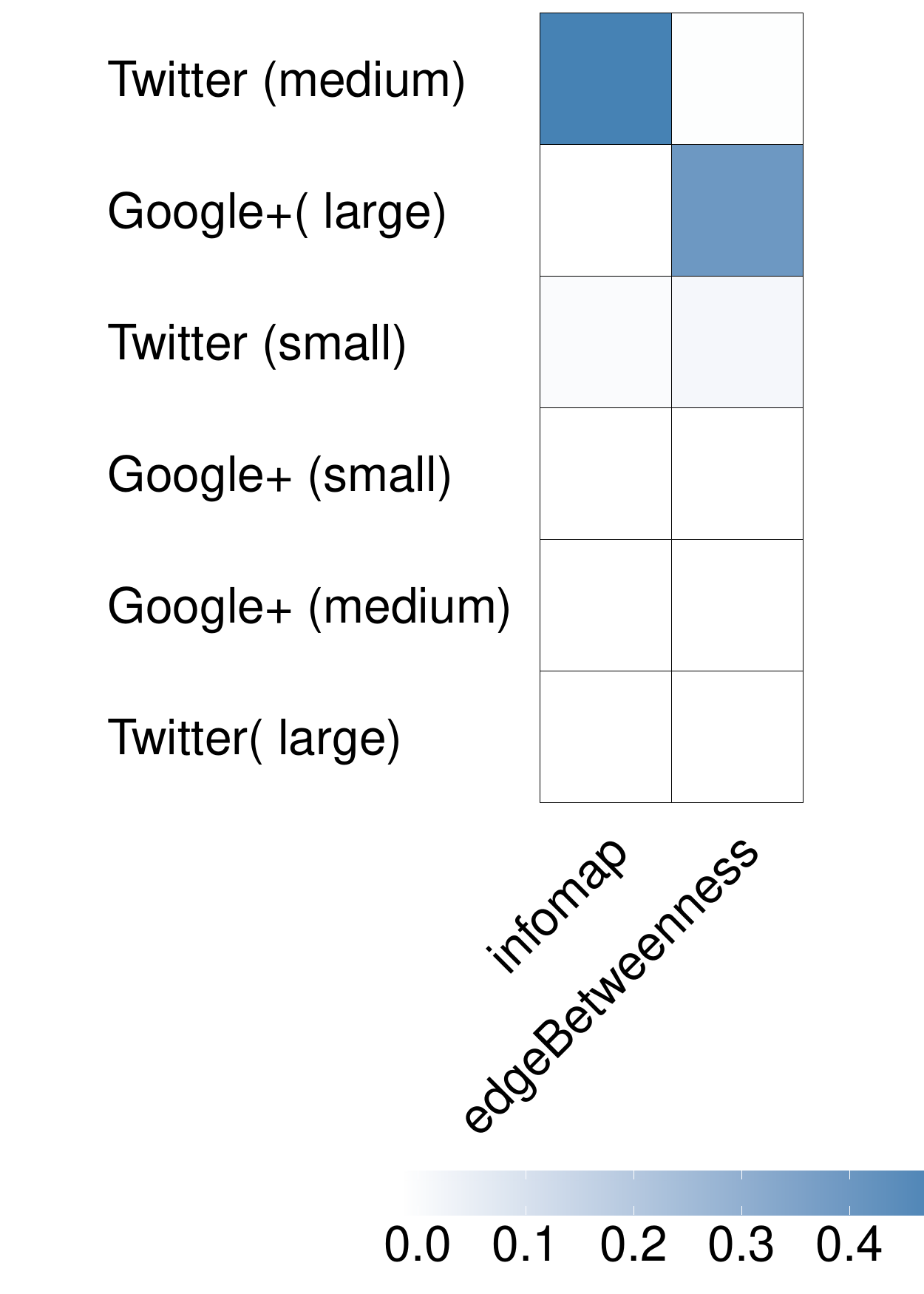} &
	\includegraphics[width=0.95\linewidth]{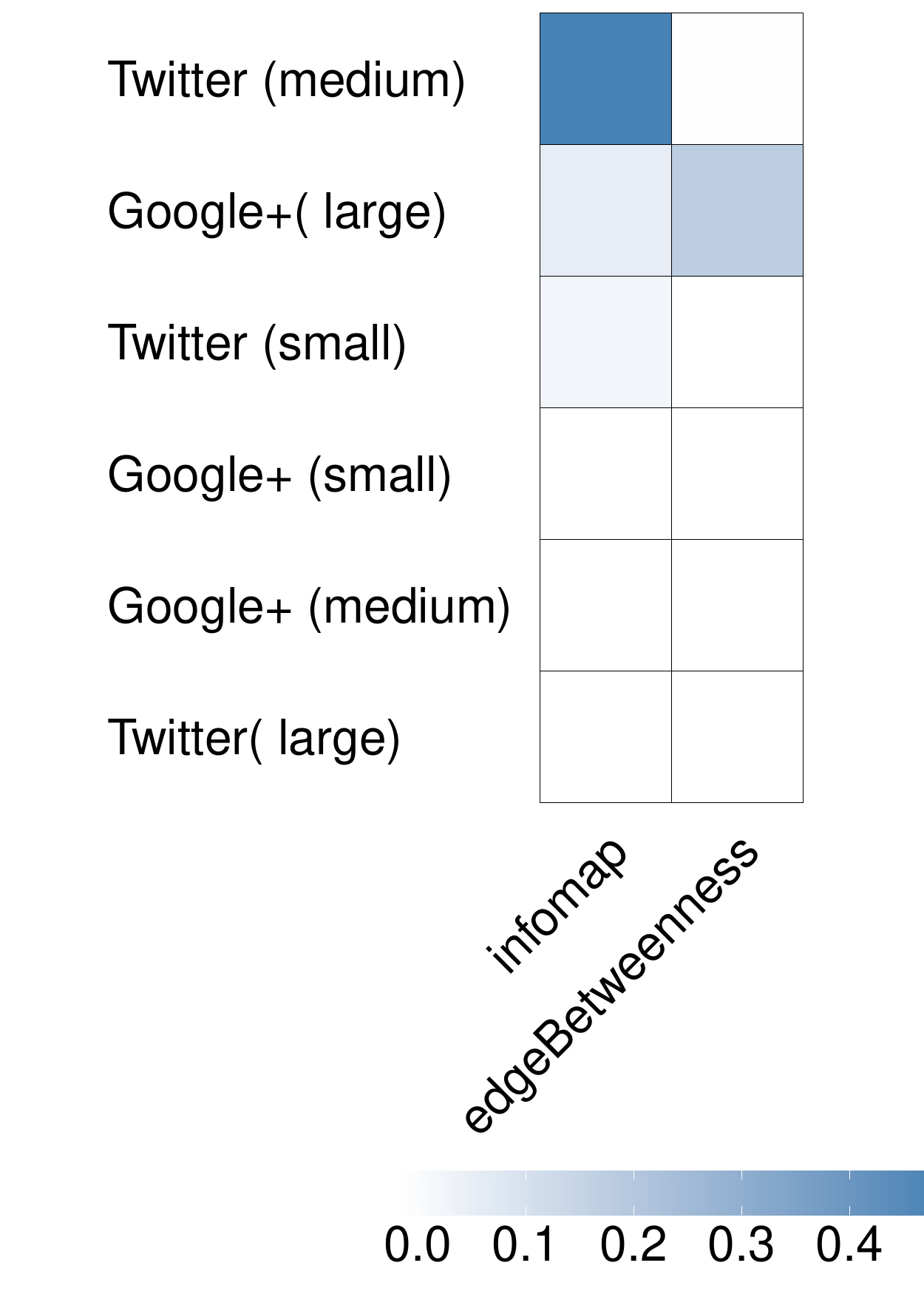} \\
	\end{tabular}
	\caption{\fontsize{10}{10}\selectfont{Average concealment-measure value in each experiment. The results for the directed networks (namely the fragments of Twitter and Google+) are presented separately.}}
	\label{fig:community-heatmaps}
\end{figure}
%%%%%%%%%%%%%%%%%%%%%%%%%%%%%%%%%%%%%%%%%%%%%%%%%%%%%%%%%%%%%%%%%%%%%%%%%%%%%%%

\end{document}